\documentclass[a4paper,10.5pt]{article}
\usepackage{tikz}
\usepackage[compat=1.1.0]{tikz-feynman}
\usepackage{jheppub} 
                     
\usepackage{verbatim}
\usepackage[T1]{fontenc} 
\usepackage{amsmath}
\usepackage{braket}
\usepackage{float}
\usepackage{slashed}
\usepackage{graphicx} 
\usepackage{tabularx}
\usepackage{booktabs} 
\usepackage{braket}
\usepackage{mathtools}
\usepackage{caption,subcaption}
\usepackage{axodraw4j}
\usepackage{pstricks}
\usepackage{color}
\usepackage{braket}
\usepackage{aligned-overset}
\usepackage{multirow}
\usepackage{breqn}
\usepackage{listings}
\usepackage{autobreak}
\usepackage{biblatex}
\allowdisplaybreaks[1]
\DeclareRobustCommand*{\ora}{\overset{\leftrightarrow}}

\title{\boldmath Electroweak Form Factor in Sudakov and Threshold Regimes with Effective Field Theories}

\author[a]{B. Assi,}
\author[a]{B.A. Kniehl}

\affiliation[a]{II. Institut f\"{u}r Theoretische Physik, Universit\"{a}t Hamburg, 22761 Hamburg, Germany}

\abstract{We compute the massive gauge and scalar corrections to form factors in both the Sudakov and threshold regimes up to and including two-loop orders. The corrections are calculated for processes involving two external fermions and scalars in the spontaneously broken SU(N)-Higgs model, examining a range of composite operators. Our results are general, so we discuss how our form factors are mappable from our model to the Standard Model and beyond. The effective theory formalism deployed in our work extends previous studies based on infrared evolution equations, which either neglect scalar contributions or are restricted to the Sudakov regime.}
\begin{document} 
\maketitle
\flushbottom
\newpage
\section{Introduction}
\label{sec:intro}
The often addressed form factor is a crucial building block in the perturbative analysis of scattering processes occurring at the LHC and future colliders \cite{chiu2009factorization,chiesa2013electroweak}. It is also the simplest amplitude which can be used to study the infrared (IR) structure of the Standard Model and beyond. For reference, the QCD form factors of massless quarks have been evaluated through the three-loop approximation \cite{gehrmann2010calculation} and even to four loops in the leading-colour approximation \cite{von2017quark}. On the other hand for massive quarks in QCD three-loop results are available thus far \cite{bernreuther2005two,blumlein2019heavy,ablinger2018heavy}. In our study, we consider massive gauge and Higgs corrections to the form factor for scalar, fermion and mixed external particles on a range of operators, checking and extending the results of \cite{chiu2008electroweak} to two-loop orders. Furthermore, unlike previous work which focus solely on the Sudakov regime relevant for Hadron collider (LHC) studies, we also consider the threshold regime appropriate for future high precision colliders \cite{brambilla2005effective,ablinger2018heavy,manohar1997heavy}. 

When considering the Sudakov regime, one tends to refer to partonic LHC processes since their centre of mass-energy is $\sqrt{s}\sim 14\text{ TeV}$. This energy scale is more than an order of magnitude larger than the masses of the massive electroweak (EW) bosons. Radiative corrections to scattering processes depend on the ratio of mass scales, and such corrections at high energy depend on factors of the form $\log{(s/M_{W/Z}^2)}$. Such radiative corrections are further enhanced by two logarithmic powers in exclusive processes perturbatively at each order, and they are often referred to as Sudakov (double) logarithms. Electro-weak Sudakov corrections are significant at LHC energies, as each double logarithmic contribution is of the $\mathcal{O}(10^{-1})$. Thus, fixed-order perturbation theory breaks down, and one needs to employ resummation at all orders. Double logarithms even appear for inclusive processes, e.g. the total $e^+e^-$ cross-section at large angles, since the colliding particles are not EW gauge singlets \cite{ciafaloni2000bloch}. The literature on EW Sudakov effects in most cases focuses on employing infrared (IR) evolution equations to deal with computations \cite{ciafaloni2000electroweak,fadin2000resummation,kuhn2000summing,feucht2004two,jantzen2005two,denner2001one,hori2000electroweak,jantzen2006two}. The Sudakov logarithm, $\log{(s/M_{W,Z}^2)}$, can be seen as an IR logarithm in EW theory, as it diverges as $M_{W,Z}\rightarrow 0$. In an effective field theory (EFT) formalism, IR logarithms in the original theory are convertible to ultraviolet (UV) logarithms in the effective theory, and then summable using standard renormalisation group (RG) techniques. The effective theories needed are soft-collinear effective theory (SCET) and heavy particle effective theory (HPET) for both fermions and scalars \cite{bauer2001effective,bauer2001invariant}, which have been previously used to study high energy EW Sudakov corrections \cite{chiu2008electroweak,chiu2008electroweak0}, and to perform resummation. This paper studies high energy EW Sudakov corrections using SCET and HPET, expanding on previous work employing the EFT formalism by other authors \cite{chiu2008electroweak,chiu2008electroweak0,chiu2008electroweak2}. 

 With regards to studies in the high energy (Sudakov) regime, an impressive level of accuracy has been achieved with many observables. For instance, the uncertainties on the predictions of the inclusive production cross-section of a top quark pair, now are at around $3-5\%$ for a fixed top quark mass of $m_t = 172.5 \text{ GeV}$ \cite{czakon2013total}. On the other hand, while these precise measurements provide a firm ground for testing the predictions within the SM, beyond the Standard Model (BSM) physics scenarios can just as well hide between said small uncertainties. To find a hint of BSM physics or to rule some hypotheses out, we need more precision than what can be provided by the LHC, and indeed, a future high precision collider which operates at threshold energies along with theoretical studies can achieve that \cite{martinez2003multi,simon2016impact}. In the threshold regime, the processes we consider have a centre of mass energy, $\sqrt{s}$, near equal to the sum of the on-shell masses of the particles produced.  Radiative corrections to scattering processes at threshold depend on the large on-shell external particle masses, as well as EW masses which are significant and, again, must be taken into account. We note further that in the threshold case, we take the gauge and Higgs masses to be IR as in the Sudakov case. Although there is extensive literature on QCD corrections at threshold, there is much more that needs to be achieved when considering EW and even BSM physics. The effective theory we employ at threshold is HPET along with standard RG techniques to perform logarithmic resummation. 

In essence, we generalise previous results in a gauge-invariant fashion to massive scalars and fermions, including radiative corrections due to the Higgs sector. Moreover, we study the threshold regime, as in previous works only the Sudakov regime was considered. Our results are computed without assuming that the Higgs and EW gauge bosons are degenerate in mass, as in previous calculations \cite{jantzen2005two,jantzen2006two}, and we take the EFT analysis to two-loop orders to match the highest precision IR evolution results.  We discuss the form factor computation in detail, checking and expounding on previous results. Although the form factor itself is not of direct relevance to collider processes, it still allows us to illustrate the EFT method for operators involving two external particles. More crucially, the form factor is known to be a building block for a vast array of processes. For example, our findings here can be used to compute corrections to processes relevant for the LHC and beyond, such as di-jet production, $\bar{t}t$ production, squark pair production, or DM production in various models, which all involve operators with four external particles \cite{chiu2008electroweak2,beenakker2010supersymmetric,ovanesyan2015heavy,ciafaloni2011weak}. Previous and future results on processes can be obtained from the computations given in this paper by summing over all pairs of external particles with the appropriate replacements of group-theoretic factors. The reason being is that the model we study, SU(N)-Higgs theory with spontaneous symmetry breaking (SSB), is selected for its generality. Moreover, the various set of composite operators, we look into allows future studies to be derived from our results. To illustrate such derivations, we apply our formalism to EW corrections in the SM for the case of light quarks, leptons and the top quark as external particles.
\section{Full and Effective Theory Formalism}
\label{sec:FTEFTForm}
We begin by outlining the full theory employed in detail as well as including a primer for the effective theories which includes SCET, and HPET for both scalars and fermions.
\subsection{SU(N)-Higgs Theory and the Standard Model}
\label{ssec:SUNHiggs}
Our calculation is set in a spontaneously broken SU(2) gauge model, however we keep our results quite general, i.e. not substituting numerical colour factors and sticking with composite operators so that our results are more conveniently mapped to more specific models for pheonmenological studies. In particular, with regards to the SM, the mapping of our model to the SM has been studied in detail previously \cite{jantzen2006two,chiu2008electroweak,chiu2008electroweak2}, the sole difference being that in the SM, the isospin SU(2) group for left-handed fermions is mixed with the hypercharge U(1) group through the mass eigen-states of the $Z$-boson and the photon. In our model the electroweak fields, $W^{\pm}$ and $Z$, are replaced with neutral SU(2) gauge bosons, $W^a:a=\lbrace 1,2,3\rbrace$, with identical mass, $M = M_W$. The generators of SU(N) in the fundamental representation are labelled by $T^a:a=\lbrace 1,\ldots,N^2-1 \rbrace$. The Lie algebra results in structure constants $f^{abc}$ with Casimir operators for the fundamental and adjoint representations given by, $C_F=(N^2-1)/2N$ and $C_A=N$, respectively. Moreover, we take the convention, $\text{tr}(T^aT^b)=T_F\delta^{ab}$, and even in the specific case of $N=2$ for SU(2), we remain with the general symbols rather than the specific values, which makes our results easily convertible, in particular for the case of the hypercharge U(1) gauge group. In the specific case of SU(2), the group generators are $T^a=\sigma^a/2:a=\lbrace 1,2,3 \rbrace$ where $\sigma^a$ are the Pauli matrices, and $f^{abc} = \epsilon^{abc}$. With the above specifications we may now state the SU(2)-Higgs Lagrangian in the t'Hooft-Feynman gauge,
\begin{equation}
    \mathcal{L}= \mathcal{L}_{\psi}+ \mathcal{L}_{\chi}+ \mathcal{L}_{YM}+ \mathcal{L}_{GF}+ \mathcal{L}_{gh}+ \mathcal{L}_{Higgs}+ \mathcal{L}_{Yuk}.
\end{equation}
The Lagrangian is split into a few parts; $\mathcal{L}_{\psi}$ and $\mathcal{L}_{\chi}$ which describe the fermions and scalars (external particles), respectively; $\mathcal{L}_{YM}$ and $\mathcal{L}_{GF}$ corresponds to the massive Yang Mills (YM) and gauge-fixing (GF) terms, respectively; $\mathcal{L}_{gh}$ describes the Faddeev-Poppov (FP) ghost fields; $\mathcal{L}_{Higgs}$ corresponds to the free Higgs Lagrangian which induces SSB and lastly, $\mathcal{L}_{Yuk}$ entails the Yukawa interaction terms which provide mass to the external fermions and scalars.
\paragraph{Fermions/Scalars}
Let $\psi_i(x)$ and $\chi_i(x)$ correspond to Fermions and scalar fields with subscripts labelling fields as we consider different incoming outgoing external states for generality. The Dirac and scalar Lagrangians then have the following form,
\begin{align} 
    & \mathcal{L}_{\psi}=\bar{\psi_i}i\slashed{D}\psi_i, \quad \mathcal{L}_{\chi}=D_{\mu}\chi_i^{\dagger}D^{\mu}\chi_i,
\end{align}
where $D_{\mu}=\partial_\mu-igW^a_{\mu}T^a$, $W^a(x)$ is the gauge field as previously defined and $g$ corresponds to the $SU(N)_W$ gauge coupling.
\paragraph{YM and Gauge-Fixing}
 The Yang-Mills and gauge-fixing Lagrangians have the usual form,
\begin{align}
    & \mathcal{L}_{YM}=-\frac{1}{4}F_{\mu\nu}^aF^{\mu\nu,a}, \quad \mathcal{L}_{GF}=-\frac{1}{2\xi_W}F_W^2,
\end{align}
such that $F_{\mu\nu}^a=\partial_{\mu}W^a_{\nu}-\partial_{\nu}W^a_{\mu}+gf^{abc}W^b_{\mu}W^c_{\nu}$ and $F_W=(\partial^{\mu}W^{a}_{\mu}-\xi_W M_W\phi^a)T^a$ where $\phi^a$ is the Goldstone boson field and $\xi_W$ the linear t'Hooft gauge fixing parameter.
\paragraph{FP-ghosts}
The gauge-fixing Lagrangian, $\mathcal{L}_{GF}$, involves the unphysical components of gauge fields. In order to compensate for their effects, one introduces the Lagrangian,
\begin{align}
    & \mathcal{L}_{gh}=-i(\partial^{\mu}\Bar{c}^a)D_{\mu}^{ab}c^b-\xi_Wm_W^2\Bar{c}^ac^a,
\end{align}
with FP-ghosts, $c^a(x), \bar{c}^a(x)$, and $D^{ab}_{\mu}=\partial_{\mu}\delta^{ab}+gf^{abc}W_{\mu}^c$.
\paragraph{Higgs and Yukawa}
The minimal Higgs sector consists of a single complex scalar field, $\Phi(x)$, which is coupled to the gauge fields with a covariant derivative and has a self-coupling, resulting in the Lagrangian,
\begin{align}
    & \mathcal{L}_{H}=(D_{\mu}\Phi)^{\dagger}D^{\mu}\Phi-V(|\Phi|^2),
\end{align}
with the Higgs potential, $V(|\Phi|^2)=\frac{\lambda}{2}(|\Phi|^2-v^2/2)^2$. The potential is constructed in such a way that it gives rise to spontaneous symmetry breaking. Meaning the parameters, $\lambda$ and $v$, are chosen in such a way that the potential minimum occurs for a non-vanishing Higgs field. 
More specifically, the theory is constructed such that the classical ground state of the scalar field satisfies, 
\begin{equation}
    |\langle \Phi\rangle|^2=\frac{v^2}{2}\neq 0.
\end{equation}
In perturbation theory one has to expand around the ground state and the Higgs field is
written as
\begin{equation}
    \Phi=\frac{1}{\sqrt{2}}\left((H+v)+i\phi^aT^a\right),
    \label{eqn:phiexp}
\end{equation}  
where $H$ and $\phi^a$ have zero vacuum expectation value and are real. The field, $H$, is the physical Higgs field and $\phi^a$ are the Goldstone bosons which illustrate the unphysical degrees of freedom. Inserting \eqref{eqn:phiexp} back into the full Lagrangian, $\mathcal{L}$, provides mass to the Higgs field and W-boson,
\begin{equation}
    M_H=\sqrt{\frac{\lambda}{2}}v \quad\text{and}\quad M_W=\frac{gv}{2},
\end{equation}
respectively. As for fermion and scalar masses, these arise from the Yukawa-like interactions in the Lagrangian,
\begin{align}
    \mathcal{L}_{Yuk}=-y_{f,i}\Bar{\psi_i}\Phi\psi_i-y_{s,i}\chi_i^{\dagger}\Phi\chi_i+h.c.,
\label{eqn:yuklag}
\end{align}
where $y_{f,i}$ and $y_{s,i}$ are the Yukawa couplings for the fermions and scalars, respectively. After spontaneous symmetry breaking, i.e. inserting \eqref{eqn:phiexp} back into \eqref{eqn:yuklag}, results in mass terms for said fermions and scalars,
\begin{align}
    \mathcal{L}_{Yuk}=-\sqrt{2}(y_{f,i}\Bar{\psi_i}\psi_i+y_{s,i}\chi_i^{\dagger}\chi_i)(H+v),
\end{align}
therefore we can re-write,
\begin{subequations}
\label{eqn:yukdefs}
\begin{align}
   & m_{\psi}=\sqrt{2}vy_{f}\Rightarrow y_{f}=\frac{g}{2\sqrt{2}}\frac{m_{\psi}}{M_W}\equiv \frac{g}{2\sqrt{2}}Y_{f}, \\&
       m_{\chi}^2=\sqrt{2}vy_{s}\Rightarrow y_{s}=\frac{g}{2\sqrt{2}}\frac{m_{\chi}^2}{M_W}\equiv \frac{g}{2\sqrt{2}}Y_{s},
\end{align}
\end{subequations}
and in this notation the Lagrangian becomes,
\begin{equation}
    \mathcal{L}_{Yuk}=-m_{\psi_i}\Bar{\psi_i}\psi_i-m_{\chi_i}^2\chi_i^{\dagger}\chi_i-\frac{g}{2}Y_{f,i}H\Bar{\psi_i}\psi_i-\frac{g}{2}Y_{s,i}H\chi_i^{\dagger}\chi_i,
\end{equation}
where $h_{\psi,\chi}$ is conventionally used in Feynman rules, as given in Appendix \ref{sec:frules}, which we attain by expanding each term in the full Lagrangian.
\subsection{Heavy Particle Effective Theory}
\label{ssec:HQET}
In the case of fermions we deploy heavy quark effective theory (HQET), which we describe briefly in this section but refer to other works for more detail \cite{georgi1990effective,georgi1991heavy,manohar2007heavy}. HQET is used in calculations involving a bound state of a heavy quark with mass $m\gg\Lambda_{QCD}$, and light quarks with mass smaller than the colour confinement scale, $\Lambda_{QCD}$. The energy scale of the interactions between the light and heavy quark is of order $\Lambda_{QCD}$, which is small compared to the mass of the heavy quark. The momentum, $p$, of the system can therefore be decomposed in the following way,
\begin{equation}
    p^{\mu}=mv^{\mu}+k^{\mu},
    \label{eqn:hqetmom}
\end{equation}
 such that $v$ is the velocity of the heavy quark, which is usually normalised such that $v^2=1$, and $k$ is the small residual momentum corresponding to light quark interactions. More precisely, the first part of \eqref{eqn:hqetmom} represents the energy of the heavy quark and is approximately conserved in interactions. The second part corresponds to a parameterisation of the remaining momentum, which is due to the motion of the light quarks and interactions between the light and heavy quarks, such that,
 \begin{equation}
     |k|\sim\mathcal{O}(\Lambda_{QCD})\quad\text{and}\quad m\gg\Lambda_{QCD}
 \end{equation}
A hierarchy of scales is thus present, whence one can organize an effective theory founded upon hierarchy. An interesting feature of HQET, as will be seeing below, is that its propagating degrees
of freedom are massless and the propagating degrees of freedom carry the residual momentum, $k$. We now outline the derivation of the HQET Lagrangian for a quark coupled to our SU(N) gauge and Higgs fields. Our starting point is the Lagrangian,
\begin{equation}
    \mathcal{L}=\bar{\psi}(i\slashed{D}-m)\psi-\frac{g}{2}Y_{f}H\Bar{\psi}\psi,
\end{equation}
such that $D^{\mu}=\partial^{\mu}-igW^{\mu}$ and $Y_{f}$ is the Yukawa coupling as previously defined. Next, following the pedagogical derivation in \cite{schwartz2014quantum}, by introducing the projection operators,
\begin{equation}
    P_{\pm}=\frac{1\pm \slashed{v}}{2},
    \label{eqn:hqet1}
\end{equation}
and two eigen-functions of these operators,
\begin{subequations}
\begin{align}
    & h_{f}=e^{imv\cdot x}P_+\psi,\\
    & H_{f}=e^{imv\cdot x}P_-\psi.
\end{align}
\end{subequations}
This allows us to decompose the spinor field as follows,
\begin{equation}
\psi=\frac{1+\slashed{v}}{2}\psi+\frac{1-\slashed{v}}{2}\psi=e^{-imv\cdot x}(h_{f}+H_{f}),
\label{eqn:hqet2}
\end{equation}
 where the field and anti-field are given by $H_{f}$ and $h_{f}$, respectively and they satisfy the relations, $\slashed{v}h_f=h_f$ and $\slashed{v}H_f=-H_f$. The details of the external states of the heavy fields are explained in previous work \cite{manohar2007heavy}. Now, substituting  into \eqref{eqn:hqet2} into \eqref{eqn:hqet1}, using some simple gamma matrix and projection operator identities, and integrating out the anti-field, $H_{f}$, using its equation of motion, we arrive at our HQET Lagrangian,
\begin{equation}
\mathcal{L}_{HQET}=\Bar{h}_{f}iv\cdot D h_{f} -\frac{g}{2}Y_{f}H\Bar{h}_{f}h_{f}+\mathcal{O}(1/m),
\end{equation}
where we neglect terms of $\mathcal{O}(1/m)$ in our derivation as they are heavily suppressed. The heavy quark propagator is thus,
\begin{equation}
    S(k)=\frac{1}{k\cdot v+i\delta}\frac{1+\slashed{v}}{2},
\end{equation}
where $k$ is the residual momentum defined earlier and the vertex coupling is given in Appendix \ref{sec:frules}.

As for the derivation of the heavy-field limit of a real scalar, or spin-0, field, it is very similar to the fermionic, or spin-1/2, derivation. One starts by considering the Lagrangian of a complex scalar field, $\chi$, with mass, $m$, coupled again to to our SU(N) gauge and Higgs fields,
\begin{equation}
    \mathcal{L}=D_{\mu}\chi^{\dagger}D^{\mu}\chi-m^2\chi^{\dagger}\chi-\frac{g}{2}Y_{s}H\chi^{\dagger}\chi
    \label{eqn:hsetL}
\end{equation}
Motivated by earlier studies \cite{braaten2018classical,namjoo2018relativistic,guth2015dark}, we then decompose the scalar field in the following way,
\begin{equation}
    \chi=\frac{e^{-imv\cdot x}}{\sqrt{2}}\left(h_s+H_s\right),
    \label{eqn:hset1}
\end{equation}
where again, $H_s$ is the anti-field containing the heavy modes, which needs to be integrated out.
More specifically,
\begin{subequations}
\label{eqn:hset2}
\begin{align}
    &h_s=\frac{e^{imv\cdot x}}{\sqrt{2m}}\left(iv\cdot\partial+m\right)\chi\\&
    H_s=\frac{e^{imv\cdot x}}{\sqrt{2m}}\left(-iv\cdot\partial+m\right)\chi,
\end{align}
\end{subequations}
and plugging \eqref{eqn:hset2} into \eqref{eqn:hset1} gives,
\begin{align}
    iv\cdot\partial H_s=(2m+iv\cdot\partial)h_s.
    \label{eqn:hset3}
\end{align}
Hence, substituting \eqref{eqn:hset1} and \eqref{eqn:hset3} into the Lagrangian, \eqref{eqn:hsetL}, and using the equation of motion, one obtains the heavy scalar effective theory (HSET) Lagrangian in our model,
\begin{equation}
    \mathcal{L}_{HSET}=h_s^{\dagger}iv\cdot D h_s-\frac{g}{2}Y_{s}H{h}_{s}^{\dagger}h_{s}+\mathcal{O}(1/m).
\end{equation}
We again neglect terms of $\mathcal{O}(1/m)$ in our derivation as they are heavily suppressed. The heavy scalar propagator is thus,
\begin{equation}
    S(k)=\frac{1}{k\cdot v+i\delta},
\end{equation}
where $k$ is again the residual momentum and the vertex coupling is also given in Appendix \ref{sec:frules}. Comparing the HSET Feynman rules with the the full theory ones, we see that they are related by simply decomposing the momenta as in \eqref{eqn:hqetmom} and dividing by $2m$.
\subsection{Soft-collinear effective theory}
\label{ssec:SCET}
SCET is an effective theory for high-energy particles, with some energy of $\mathcal{O}(Q)$, where $Q$ is a large scale that characterises the scattering process under consideration. SCET preserves the modes of the full theory which have an invariant mass much smaller than $Q^2$. The SCET fields and Lagrangian depend on the null vectors, $n$ and $\bar{n}$, where $n=(1,\boldsymbol{n})$ and $\bar{n}=(1,-\boldsymbol{n})$. The three-vector, $\boldsymbol{n}$, is chosen to be a unit vector, thus, $\bar{n}\cdot n=2$. 

When calculating the Sudakov form factor, we work in the so-called Breit frame, with $n$ chosen to be along the $p_2$ direction and $\bar{n}$ is then along the $p_1$ direction. The momentum transfer, $q=p_2-p_1$, then has time component, $q^0=0$. We work with light-cone components, which for a four-vector, $p$, are defined by $p^+\equiv n\cdot p$ and $p^-\equiv\bar{n}\cdot p$. In our problem, $p_1^-=p_{1{\perp}}=p_2^+=p_{2{\perp}}=0$, and $Q^2=p_1^+p_2^-$, which is reflected in our Feynman rules, see Appendix \ref{sec:frules}. When a fermion moves in a direction close to $n$, it is describable by an $n$-collinear SCET field, $\xi_{n,p}(x)$, where $p$ is a label momentum, and has components $\bar{n}\cdot p$ and $p_{\perp}$ \cite{bauer2001effective,bauer2001invariant}. Kinematically, the field, $\xi_{n,p}(x)$, describes a particle (both on or off-shell) with $2E=\bar{n}\cdot p$ and $p^2\ll Q^2$. The SCET power counting is then as follows,
\begin{equation}
    p^-\sim Q,\quad p^+\sim Q^2 \lambda,\quad p_{\perp}\sim Q \lambda,
\end{equation}
where $\lambda$ is the parameter used for power counting in the EFT expansion. The total momentum of the SCET field, $\xi_{n,p}(x)$, is $p+k$, where as in HPET, $k$ is the residual momentum, except in this effective theory, $k$ is of order $Q\lambda^2$, and is obtained from a Fourier transform of the position vector, $x$. Note that the label momentum, $p$, only contributes to the minus and perpendicular components of the total momentum.

On the other hand, the gauge field in the effective theory is represented in many ways: Labelled $n$-collinear fields, $W_{n,p}(x)$, and $\bar{n}$-collinear fields, $W_{\bar{n},p}(x)$, and unlabelled ultrasoft (US) fields, $W(x)$, which are analogous to the soft and US fields introduced in NRQCD \cite{pineda1998effective,luke2000renormalization}. The $n$-collinear field contains gauge fields with momentum near the $n$-direction, and momentum scaling given by,
\begin{equation}
     \bar{n}\cdot p\sim Q,\quad n\cdot p\sim Q^2 \lambda,\quad p_{\perp}\sim Q \lambda,
\end{equation}
and the $\bar{n}$-collinear fields contain gluons moving near the
$\bar{n}$-direction, with momentum scaling,
\begin{equation}
     n\cdot p\sim Q,\quad \bar{n}\cdot p\sim Q^2 \lambda,\quad p_{\perp}\sim Q \lambda.
\end{equation}
Lastly, the ultrasoft field represents gauge bosons with all momentum components scaling as $Q \lambda^2$. There are is a particularly useful identity we quote that holds for the SCET fermion field,
\begin{equation}
    \frac{n\bar{n}}{4}\xi_{n,p}=\xi_{n,p},
    \label{eqn:0maker}
\end{equation}
where
\begin{equation}
    P_n=\frac{n\bar{n}}{4}, \quad P_{\bar{n}}=\frac{\bar{n}n}{4}, \quad P_{\bar{n}}+ P_{n}=1
\end{equation}
are projection operators. The leading order fermion Lagrangian is \cite{bauer2001effective},
\begin{equation}
    \bar{\xi}_{n,p}\frac{\bar{n}}{4}\left(in\cdot D+\frac{p_{\perp}^2}{2\bar{n}\cdot p}\right){\xi}_{n,p}\frac{\bar{n}}{4}
    \label{eqn:scetf}
\end{equation}
where $iD^{\mu}=i\partial^{\mu}+gW^{\mu}$ is the ultra-soft covariant derivative, and we neglect terms involving the collinear gauge field. The fermionic SCET propagator is then given by,
\begin{equation}
    S(p)=\frac{\slashed{n}}{2}\frac{\bar{n}\cdot p}{p^2}
\end{equation}
SCET knows about the large momentum scale, $Q$, through labels, $\bar{n}\cdot p_2$ and ${n}\cdot p_1$, attached to the fields, $\xi_{n,p_2}$ and $\xi_{\bar{n},p_1}$, for the outgoing and incoming particles, respectively. As a result, SCET anomalous dimensions can depend on $Q$. However, there are no modes in SCET which couple $\bar{n}\cdot p_2$ to ${n}\cdot p_1$, so that SCET does not contain modes with off-shellness of $\mathcal{O}(Q^2)$, which are of course present in the full theory. We also require SCET fields for scalar particles, such as Higgs-like fields for instance. Let $\Phi_{n,p}$ be the scalar analogue of $\xi_{n,p}$ for fermions, which describes the $n$-collinear field for a scalar particle moving in a direction near $n$. One normalises the SCET field, $\Phi_{n,p}$, in the same way as the full theory field, $\phi$, producing scalar particles with unit amplitude. The scalar field kinetic energy term in the Lagrangian then becomes,
\begin{equation}
    D_{\mu}\phi^{\dagger}D^{\mu}\phi\rightarrow \Phi_{n,p}^{\dagger}\left((\bar{n}\cdot p)(in\cdot D)+p_{\perp}^2\right)\Phi_{n,p}
\end{equation}
in SCET. It is also convenient to re-define the scalar field as follows,
\begin{equation}
   \phi_{n,p}=\sqrt{\bar{n}\cdot p}\Phi_{n,p}
\end{equation}
in terms of which the kinetic term becomes,
\begin{equation}
\mathcal{L}=\phi_{n,p}^{\dagger}\left(in\cdot D+\frac{p_{\perp}^2}{(\bar{n}\cdot p)}\right)\phi_{n,p}
\end{equation}
and then has the same normalisation as the fermion Lagrangian in \eqref{eqn:scetf}. The re-scaled scalar propagator is given by,
\begin{equation}
    \frac{1}{p^2}\rightarrow \frac{\bar{n}\cdot p}{p^2 }.
\end{equation}
Hence, $\phi_{n,p}$ as defined, produces scalar particles moving in the $n$-direction with amplitude, $\sqrt{\bar{n}\cdot p}$.

\section{The Form Factor}
\label{sec:outline}

The physical quantity we consider in this work is the form factor in the Euclidean region, defined as the amplitude, $F_E(Q^2)=\bra{p_2}\mathcal{O}\ket{p_1}$ for the scattering of on-shell particles $p_i^2 = m_i^2$ by an operator $\mathcal{O}$, with $Q^2=-(p_2-p_1)^2>0$. The time-like form factor is given by an analytic continuation, $F(s)=F_E(-s-i0^+)$, implying $\log{Q^2/\mu^2}\rightarrow\log{s/\mu^2}-i\pi$. We will compute $F_E(Q^2)$ for fermion scattering with, $\mathcal{O}=\bar{\psi}\gamma^{\mu}\psi,\bar{\psi}\psi,\bar{\psi}\sigma^{\mu\nu}\psi$, scalar scattering with, $\mathcal{O}=\chi^{\dagger}\chi,i(D^{\mu}\chi^{\dagger}\chi-\chi^{\dagger}D^{\mu}\chi)$, and mixed scattering with, $\mathcal{O}=\bar{\psi}\chi,\chi^{\dagger}\psi$. All operators are taken to be gauge singlets and thus the external particles have the same gauge quantum numbers, but differing mass. We then compute the form factor, $F_E(Q^2)$, by employing a sequence of effective theories inspired by previous studies \cite{chiu2008electroweak,chiu2008electroweak0}. In both the threshold and Sudakov regimes we consider, there are various widely separated scales and we switch to the relevant theory as we shift between scales. 

To illustrate the matching, let us consider the Sudakov regime. At scales higher than $Q^2$, the model is the original Higgs-gauge theory or the so-called full theory in EFT terminology. As one shifts to scales below $\mathcal{O}(Q^2)$, we transition to an effective field theory (SCET) where degrees of freedom with off-shellness of $\mathcal{O}(Q^2)$ are integrated out. The full and effective theory share identical infrared (IR) physics but differ in their ultraviolet (UV) behaviour. To ensure that operators in the full and effective theories have the same on-shell matrix elements, one must introduce a so-called multiplicative matching coefficient. If the full theory is matched onto SCET at $\mu \sim Q$ then the matching coefficient selected,
\begin{equation}
    \bra{p_2}\mathcal{O}(\mu)\ket{p_1}=\exp{[C(\mu)]} \bra{p_2}\mathcal{\tilde{O}}(\mu)\ket{p_1},
\end{equation}
where $\exp{[C(\mu)]}$ is the matching coefficient at $\mu \sim Q$ which is in exponential form for convenience and $\mathcal{\tilde{O}}(\mu)$ is the effective theory version of the full theory operator, $\mathcal{O}(\mu)$. The matching coefficient is independent of IR physics and is computable if perturbation theory is valid at $\mu\sim Q$. In general, a single operator, $\mathcal{O}$, can match onto a set of operators, $\tilde{\mathcal{O}}_i$ in the EFT with identical quantum numbers \cite{georgi1992d,hiller2004more}. 
The matching coefficient $C(\mu)$ contains logarithms, $\log{\mu^2/Q^2}$, and logarithms are not large if $\mu \sim Q$. Although we choose $\mu=Q$, any value of $\mathcal{O}(Q)$ may be chosen as well, and all physical observables do not depend on the renormalisation scale, $\mu$. The convention we follow is to pick the coefficient, $c(\mu)$, of $\mathcal{O}$ in the full theory, to be unity at $\mu=Q$. Our choice then provides the normalisation for $F_E(Q^2)$, and $c(Q)=\exp{[C(Q)]}$ is the coefficient of $\mathcal{O}$ in SCET at $\mu=Q$. Moreover, to do RGE for $c(\mu)$ between scales we use the usual equation,
\begin{equation}
    \mu\frac{dc(\mu)}{d\mu}=\gamma(\mu)c(\mu),
\end{equation}
such that $\gamma(\mu)$ is the anomalous dimension of $\tilde{\mathcal{O}}$ in the effective theory. We then repeat these steps of matching and RGE as we shift between well-separated energy scales, integrating out the appropriate degrees of freedom along the way. The EFT approach is superior to IR evolution as it divides a multi-scale calculation into multiple single-scale pieces which are simpler to work with. One can then trivially identify so-called universal quantities which are independent of scale. Lastly, we re-state that in an EFT calculation, the IR divergences in the theory above a matching scale match the UV divergences in the theory below the matching scale. Thus, with regards to most of our results presented here, having checked the above and below UV-IR matching, we need only provide the physically relevant finite parts.

For reference, our notation is as follows, we use $a(\mu) \equiv \alpha(\mu)/(4\pi)$, and for applications to the SM, $a_i(\mu) \equiv \alpha_i(\mu)/(4\pi)$ where $i =\lbrace s, 2, 1 \rbrace$ for the QCD, SU(2) and U(1) couplings. Hypercharge is taken to be normalised such that $Q = T_3 + Y$. Our various Logarithms are denoted by $\mathcal{L}_A \equiv \log{A^2/\mu^2}$, for $A = Q,M_{W,H},m_{1,2}$. $C_A$, $C_F$ and $T_f$ are the SU(N) Casimir operators and index for external particles. 
\section{Renormalisation}
\subsection{Field Renormalisation}
The on-shell renormalization of the external fermion/scalar fields in our form factor expansions require the multiplication of the vertex corrections by a factor of $Z$, where $Z^{1/2}$ is the fermion/scalar wave function renormalization (WFR) constant. The factor, $Z$, is determined by the fermion/scalar self-energy corrections $\Sigma$ at on-shell momentum $p^2=m^2$, in specific ways we will describe below. In a perturbative expansion with the external fields we study, $\lbrace \psi,\chi,h_f,h_s,\xi_{n,p},\phi_{n,p}\rbrace$, letting $\lbrace I,J\rbrace$ denote these fields such that $V_{IJ}$ and $Z_{IJ}=\sqrt{Z_IZ_J}$ correspond to the vertex and wavefunction contributions, we have,
\begin{align}
    &V_{IJ}=1+aV^{(1)}_{IJ}+a^2V^{(2)}_{IJ}+\mathcal{O}(a^3), \\
    &Z_I=1+a\delta Z_I^{(1)}+a^2\delta Z_I^{(2)}+\mathcal{O}(a^3).
\end{align}
Therefore, the WFR is given by,
\begin{equation*}
    Z_{IJ}=1+\frac{a}{2}\left(\delta Z_I^{(1)}+\delta Z_J^{(1)}\right)+\frac{a^2}{2}\left(\delta Z_I^{(2)}+\delta Z_J^{(2)}+\frac{1}{2}\delta Z_I^{(1)}\delta Z_J^{(1)}-\frac{1}{4}(\delta Z{_I^{(1)}})^2-\frac{1}{4}(\delta Z_J^{(1)})^2\right).
\end{equation*}
Whence, the total form factor, $F_{IJ}=V_{IJ}Z_{IJ}$, up to order $\alpha^2$, can be written as follows,
\begin{align}
    F_{IJ}=&1+a\left\lbrace V_{IJ}^{(1)}+\frac{1}{2}\left( \delta Z_I^{(1)}+\delta Z_J^{(1)} \right) \right\rbrace + a^2\left\lbrace V_{IJ}^{(2)}+\frac{1}{2}\left( \delta Z_I^{(2)}+\delta Z_J^{(2)} \right) \right. \nonumber \\ & \left. +\frac{1}{2}\left( \delta Z_I^{(1)}+\delta Z_J^{(1)} \right)V_{IJ}^{(1)}+\frac{1}{4}\delta Z_I^{(1)}\delta Z_J^{(1)}-\frac{1}{8}(\delta Z_I^{(1)})^2-\frac{1}{8}(\delta Z_J^{(1)})^2 \right\rbrace.
    \label{eqn:ffac12}
\end{align}
With the above notation we may now discuss how to obtain the WFR constant, $Z_I$, for the spin-$\lbrace 0,1/2 \rbrace$ fields we study. In all cases, the wavefunction corrections are garnered from self-energy amplitudes, denoted by $\tilde{\Sigma}_I$, which are quadratic matrices both in the spinor and in the isospin space \cite{denner1993techniques,schwartz2014quantum}. Moreover, we note here that the collinear correction to the particle propagator is the same as in the full theory \cite{bauer2001effective,chiu2008electroweak}. Therefore the wavefunction corrections are the same as in the full theory/non-collinear case. Whence, we only need to outline how to obtain the wavefunction contributions to the form factors for the full theory and HPET fields.

\begin{figure}[tb]
\begin{center}
\scalebox{.9}{
\fcolorbox{white}{white}{
  \begin{picture}(451,107) (17,0)
    \SetWidth{0.5}
    \SetColor{Black}
    \Text(56.117,60.419)[lb]{{\Black{$(a)$}}}
    \Text(149.256,60.249)[lb]{{\Black{$(b)$}}}
    \Text(243.174,60.639)[lb]{{\Black{$(c)$}}}
    \Text(150.035,9.978)[lb]{{\Black{$(g)$}}}
    \Text(337.872,60.639)[lb]{{\Black{$(d)$}}}
    \Text(431.011,60.249)[lb]{{\Black{$(e)$}}}
    \Text(54.948,9.419)[lb]{{\Black{$(f)$}}}
    \Text(242.785,-7.338)[lb]{{\Black{$(h)$}}}
    \SetWidth{2.5}
    \Line[arrow,arrowpos=0.5,arrowlength=7.5,arrowwidth=2,arrowinset=0.2](18.706,75.992)(44.036,75.992)
    \Line[arrow,arrowpos=0.5,arrowlength=7.5,arrowwidth=2,arrowinset=0.2](44.036,75.992)(68.977,75.992)
    \Line[arrow,arrowpos=0.5,arrowlength=7.5,arrowwidth=2,arrowinset=0.2](68.977,75.992)(93.139,75.992)
    \Line[arrow,arrowpos=0.5,arrowlength=7.5,arrowwidth=2,arrowinset=0.2](393.21,26.11)(418.541,26.11)
    \Line[arrow,arrowpos=0.5,arrowlength=7.5,arrowwidth=2,arrowinset=0.2](418.541,26.11)(443.482,26.11)
    \Line[arrow,arrowpos=0.5,arrowlength=7.5,arrowwidth=2,arrowinset=0.2](443.482,26.11)(467.643,26.11)
    \SetWidth{2.0}
    \Arc[dash,dashsize=3.897,clock](430.697,25.923)(23.075,178.568,113.06)
    \Arc[dash,dashsize=3.897,clock](431.869,26.524)(22.135,71.738,-0.063)
    \SetWidth{1.0}
    \Arc[arrow,arrowpos=0.5,arrowlength=5,arrowwidth=2,arrowinset=0.2](430.232,47.933)(7.891,147,507)
    \SetWidth{2.5}
    \Line[arrow,arrowpos=0.5,arrowlength=7.5,arrowwidth=2,arrowinset=0.2](206.153,76.382)(231.483,76.382)
    \Line[arrow,arrowpos=0.5,arrowlength=7.5,arrowwidth=2,arrowinset=0.2](231.483,76.382)(256.424,76.382)
    \Line[arrow,arrowpos=0.5,arrowlength=7.5,arrowwidth=2,arrowinset=0.2](256.424,76.382)(280.586,76.382)
    \SetWidth{2.0}
    \Arc[dash,dashsize=3.897,clock](243.639,75.805)(23.075,178.568,113.06)
    \Arc[dash,dashsize=1.169,arrow,arrowpos=0.5,arrowlength=6.667,arrowwidth=2,arrowinset=0.2](243.954,97.426)(8.034,141,501)
    \Arc[dash,dashsize=3.897,clock](247.851,79.629)(18.218,77.648,-10.268)
    \SetWidth{2.5}
    \Line[arrow,arrowpos=0.5,arrowlength=7.5,arrowwidth=2,arrowinset=0.2](18.706,25.331)(44.036,25.331)
    \Line[arrow,arrowpos=0.5,arrowlength=7.5,arrowwidth=2,arrowinset=0.2](44.036,25.331)(68.977,25.331)
    \Line[arrow,arrowpos=0.5,arrowlength=7.5,arrowwidth=2,arrowinset=0.2](68.977,25.331)(93.139,25.331)
    \SetWidth{2.0}
    \Arc[dash,dashsize=3.897,clock](56.117,18.771)(19.455,160.295,19.705)
    \Arc[dash,dashsize=3.897,clock](56.423,40.538)(9.345,-155.293,-382.103)
    \SetWidth{2.5}
    \Line[arrow,arrowpos=0.5,arrowlength=7.5,arrowwidth=2,arrowinset=0.2](300.461,76.382)(319.556,76.382)
    \Line[arrow,arrowpos=0.5,arrowlength=7.5,arrowwidth=2,arrowinset=0.2](319.556,76.382)(338.262,76.382)
    \Line[arrow,arrowpos=0.5,arrowlength=7.5,arrowwidth=2,arrowinset=0.2](338.262,76.382)(356.968,76.382)
    \Line[arrow,arrowpos=0.5,arrowlength=7.5,arrowwidth=2,arrowinset=0.2](356.968,76.382)(375.284,76.382)
    \SetWidth{2.0}
    \Arc[dash,dashsize=3.897,clock](336.672,79.629)(21.709,-170.356,-368.602)
    \Line[dash,dashsize=3.897](337.482,101.323)(337.482,76.382)
    \SetWidth{2.5}
    \Line[arrow,arrowpos=0.5,arrowlength=7.5,arrowwidth=2,arrowinset=0.2](393.6,75.992)(412.695,75.992)
    \Line[arrow,arrowpos=0.5,arrowlength=7.5,arrowwidth=2,arrowinset=0.2](412.695,75.992)(431.401,75.992)
    \Line[arrow,arrowpos=0.5,arrowlength=7.5,arrowwidth=2,arrowinset=0.2](431.401,75.992)(450.106,75.992)
    \Line[arrow,arrowpos=0.5,arrowlength=7.5,arrowwidth=2,arrowinset=0.2](450.106,75.992)(468.422,75.992)
    \SetWidth{2.0}
    \Arc[dash,dashsize=3.897,clock](431.476,75.805)(23.075,178.568,113.06)
    \Arc[dash,dashsize=3.897,clock](431.869,76.406)(22.135,71.738,-0.063)
    \Arc[dash,dashsize=3.897](431.011,98.595)(8.034,141,501)
    \SetWidth{2.5}
    \Line[arrow,arrowpos=0.5,arrowlength=7.5,arrowwidth=2,arrowinset=0.2](113.403,26.11)(132.499,26.11)
    \Line[arrow,arrowpos=0.5,arrowlength=7.5,arrowwidth=2,arrowinset=0.2](132.499,26.11)(151.205,26.11)
    \Line[arrow,arrowpos=0.5,arrowlength=7.5,arrowwidth=2,arrowinset=0.2](151.205,26.11)(169.91,26.11)
    \Line[arrow,arrowpos=0.5,arrowlength=7.5,arrowwidth=2,arrowinset=0.2](169.91,26.11)(188.226,26.11)
    \SetWidth{2.0}
    \Arc[dash,dashsize=3.897,clock](150.394,29.747)(21.709,-170.356,-368.602)
    \Arc[dash,dashsize=3.897,clock](150.425,28.984)(15.851,-169.553,-370.447)
    \SetWidth{2.5}
    \Line[arrow,arrowpos=0.5,arrowlength=7.5,arrowwidth=2,arrowinset=0.2](206.153,26.5)(225.248,26.5)
    \Line[arrow,arrowpos=0.5,arrowlength=7.5,arrowwidth=2,arrowinset=0.2](225.248,26.5)(243.954,26.5)
    \Line[arrow,arrowpos=0.5,arrowlength=7.5,arrowwidth=2,arrowinset=0.2](243.954,26.5)(262.66,26.5)
    \Line[arrow,arrowpos=0.5,arrowlength=7.5,arrowwidth=2,arrowinset=0.2](262.66,26.5)(280.976,26.5)
    \SetWidth{2.0}
    \Arc[dash,dashsize=3.897,clock](252.917,29.374)(15.851,-169.553,-370.447)
    \Arc[dash,dashsize=3.897](233.393,23.272)(15.499,167.98,370.551)
    \Text(336.703,10.978)[lb]{{\Black{$(i)$}}}
    \SetWidth{2.5}
    \Line[arrow,arrowpos=0.5,arrowlength=7.5,arrowwidth=2,arrowinset=0.2](300.461,26.5)(325.791,26.5)
    \Line[arrow,arrowpos=0.5,arrowlength=7.5,arrowwidth=2,arrowinset=0.2](325.791,26.5)(350.732,26.5)
    \Line[arrow,arrowpos=0.5,arrowlength=7.5,arrowwidth=2,arrowinset=0.2](350.732,26.5)(374.894,26.5)
    \SetWidth{2.0}
    \Arc[dash,dashsize=3.897,clock](319.166,27.975)(12.171,-173.038,-366.962)
    \Arc[dash,dashsize=3.897,clock](356.18,27.265)(12.874,-176.592,-361.672)
    \SetWidth{2.5}
    \Line[arrow,arrowpos=0.5,arrowlength=7.5,arrowwidth=2,arrowinset=0.2](112.624,76.382)(137.955,76.382)
    \Line(137.955,76.382)(163.675,76.382)
    \Line[arrow,arrowpos=0.5,arrowlength=7.5,arrowwidth=2,arrowinset=0.2](163.675,76.382)(188.616,76.382)
    \Text(429.452,11.757)[lb]{{\Black{$(j)$}}}
    \SetWidth{1.0}
    \GluonArc[clock](56.271,80.467)(20.903,-167.639,-371.27){2.923}{15}
    \GluonArc[clock](55.922,63.575)(25.278,150.579,29.421){2.923}{11}
    \GluonArc[clock](135.772,79.431)(16.496,-167.968,-370.651){2.923}{12}
    \GluonArc[clock](168.546,79.102)(16.016,-170.223,-369.777){2.923}{11}
  \end{picture}
}}
\end{center}
    \caption{Two-loop wave-function correction graphs, arrowed lines represent all incoming-outgoing particles, dashed lines correspond to bosonic propagators. $(a),(b)$ are seagull terms and only occur with scalar propagators, $(h)$-$(j)$ and $(c)$-$(g)$ represent Abelian and non-Abelian corrections, respectively.}
    \label{fig:wfr}
\end{figure}
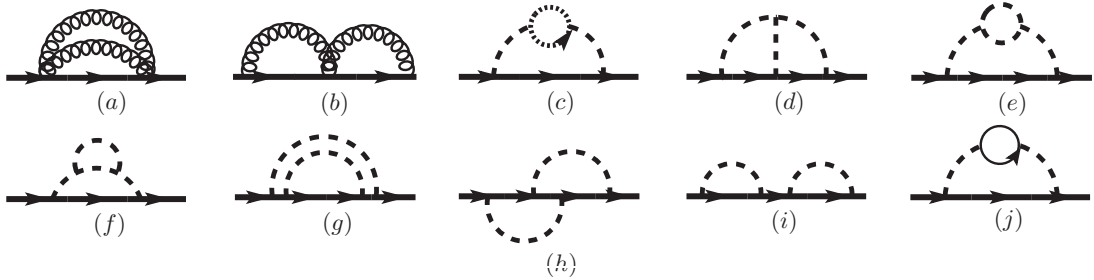

\paragraph{Scalar field:}
For massive scalars of momentum, $p$, and mass, $m$, the self-energy amplitudes, as shown in figure \ref{fig:wfr} are of the form,
\begin{equation}
    \tilde\Sigma_{\chi}=-i\Sigma_{\chi}(p^2)\boldsymbol{1}.
\end{equation}
From this we may extract the WFR contributions in the following way, 
\begin{equation}
    \delta Z_{\chi}=\frac{i}{4}\text{tr}(\partial_{p^2}\tilde\Sigma_{\chi}|_{p^2=m^2}).
\end{equation}
The massless case is identical except one takes $p^2=0$ instead.
\paragraph{Fermion field:}
In the case of fermions of momentum, $p$, and mass, $m$, the self-energy amplitudes are of the form,
\begin{equation}
    \tilde\Sigma_{\psi}=-i\left(\Sigma_{\psi}^V(p^2)\slashed{p}+\Sigma_{\psi}^S(p^2)m\right)\boldsymbol{1},
\end{equation}
where the super-scripts, $V$ and $S$, denote vector and scalar contributions, respectively. From this we may extract the WFR contributions,
\begin{equation}
    \delta Z_{\psi}=\left\lbrace\Sigma_{\psi}^V(m^2)+2m^2\partial_{p^2}\left(\Sigma_{\psi}^V(p^2)+\Sigma_{\psi}^S(p^2)\right)|_{p^2=m^2}\right\rbrace.
\end{equation}
The massless case simplifies as $p^2=0$ instead and the terms proportional to $m^2$ vanish.
\paragraph{Heavy fields:}
Lastly, for heavy scalars and fermions, $h$, of momentum, $p$, and velocity, $v$, the self-energy amplitudes are of the form,
\begin{align}
    &\tilde\Sigma_{h}=-i\left\lbrace\Sigma_{h}^F(v\cdot p)+\Sigma_{h}^R(M_B)\right\rbrace\boldsymbol{1},
\end{align}
for bosons of mass, $M$, coupling to the heavy fields, as in our case. We thus have an additional contribution to the heavy field
residual mass term, $\delta m$, along with the usual wave function contribution,
\begin{align}
    &\delta Z_{h}=i\partial_{v\cdot p}\tilde\Sigma_{h}|_{v\cdot p=0} \\&
    \delta m_I=-i\tilde\Sigma_{h}(v\cdot p=0).
\end{align}
The residual shift in the heavy particle mass, $\delta m$, is non-analytic in the boson mass squared, these arise from loop integrals which diverge as an odd power of loop momenta, $l$. Such non-analytic contributions are known to occur in mass corrections to particles with $l\cdot v$ propagators \cite{jenkins1991baryon,hoang2000charm}. Such integrals are finite, but non-analytic, in dimensional regularization. There are two mass parameters for the heavy particle Lagrangian, the expansion parameter, $m_0$, and the residual mass term, $\delta m$. The two parameters are not independent; one can make the redefinition $m_0 \rightarrow m_0 + \Delta m$, $\delta m\rightarrow \delta m-\Delta m$. A particularly convenient choice is to adjust $m_0$ so that the residual mass term $\delta m$ vanishes, when picking $m_0$ such that $\delta m = 0$ this choice is known as the pole mass \cite{bigi1994pole}, and we will follow this practice here. 

\begin{table}
\small
\centering
\begin{tabular}{|c|c|c|c|c|} 
 \hline
 Field & $m$ & $M$ & $\delta Z^{(1)}$ & $\delta Z^{(2)}$ \\ [0.75ex] 
 \hline\hline
 $\psi$ & $0$ & $0$ & 
\resizebox{.23\hsize}{!}{ $\begin{aligned}[t] \frac{C_A}{2}\left(C_F+\frac{Y_f^2}{8}\right)\left\lbrace\frac{2}{\varepsilon _{\text{IR}}}-\frac{2}{\varepsilon _{\text{UV}}}\right\rbrace \end{aligned}$}
 & 
{ $\begin{aligned}[t] F^{(0,0)}_{\psi}\end{aligned}$}
 \\[0.75ex] \hline
 $\psi$ & $0$ & $M$ & 
\resizebox{.42\hsize}{!}{$\begin{aligned}[t] &\frac{C_A}{2}\left\lbrace C_F\left(\frac{2}{-\varepsilon _{\text{UV}}}+1+2\mathcal{L}_{M_W}\right)+\frac{Y_f^2}{8}\left(-\frac{2}{\varepsilon _{\text{UV}}}-1+2\mathcal{L}_{M_H}\right)\right\rbrace\end{aligned}$} & 
{ $\begin{aligned}[t] F^{(0,M)}_{\psi} \end{aligned}$}
  \\[0.75ex]  \hline
 $\psi$ &  $m$ & $0$ & 
\resizebox{.47\hsize}{!}{$\begin{aligned}[t]&\frac{C_A}{2}\left\lbrace C_F\left(-\frac{2}{\varepsilon _{\text{UV}}}-\frac{4}{\varepsilon _{\text{IR}}}-4+3\mathcal{L}_{m}\right)+\frac{Y_f^2}{8}\left(-\frac{2}{\varepsilon _{\text{UV}}}+\frac{8}{\varepsilon _{\text{IR}}}+14-6\mathcal{L}_{m}\right)\right\rbrace\end{aligned}$}
 & 
{ $\begin{aligned}[t] F^{(m,0)}_{\psi}\end{aligned}$}
 \\[0.75ex] \hline
 $\psi$ & $m$ & $M$ & \resizebox{.59\hsize}{!}{$\begin{aligned}[t]&\frac{C_A}{2}\left\lbrace C_F\left(-\frac{2}{\varepsilon _{\text{UV}}}-8+2\mathcal{L}_{M_W}-P(m/M_W)\right)+  \frac{Y_f^2}{8}\left(-\frac{2}{\varepsilon _{\text{UV}}}+14+2\mathcal{L}_{M_H}-P^'(m/M_H)\right)\right\rbrace\end{aligned}$}
 & 
{ $\begin{aligned}[t] F^{(m,M)}_{\psi}\end{aligned}$}
 \\[0.75ex] \hline
 $\chi$ & $0$ & $0$ &  
\resizebox{.15\hsize}{!}{$\begin{aligned}[t]&\frac{C_AC_F}{2}\left\lbrace\frac{4}{\varepsilon _{\text{UV}}}-\frac{4}{\varepsilon _{\text{IR}}}\right\rbrace\end{aligned}$}
 & 
{$\begin{aligned}[t] F^{(0,0)}_{\chi} \end{aligned}$}
 \\[0.75ex] \hline
 $\chi$ & $0$ & $M$ & 
\resizebox{.28\hsize}{!}{$\begin{aligned}[t]&\frac{C_A}{2}\left\lbrace-\frac{Y_s^2}{4M_H^2}+C_F\left(\frac{4}{\varepsilon _{\text{UV}}}+3-4\mathcal{L}_{M_W}\right)\right\rbrace\end{aligned}$}
 & 
{$\begin{aligned}[t] F^{(m,0)}_{\chi} \end{aligned}$}
 \\[0.75ex] \hline
 $\chi$ & $m$ & $0$ & 
\resizebox{.35\hsize}{!}{$\begin{aligned}[t]&\frac{C_A}{2}\left\lbrace C_F\left(\frac{4}{\varepsilon _{\text{UV}}}-\frac{4}{\varepsilon _{\text{IR}}}\right)+\frac{Y_s^2}{2m^2}\left(\frac{1}{2\varepsilon _{\text{IR}}}+1-\mathcal{L}_m\right)\right\rbrace\end{aligned}$}
 & 
{$\begin{aligned}[t] F^{(0,M)}_{\chi} \end{aligned}$}
 \\[0.75ex] \hline
 $\chi$ & $m$ & $M$ & 
\resizebox{.4\hsize}{!}{$\begin{aligned}[t]&\frac{C_A}{2}\left\lbrace-\frac{Y_s^2}{2M_H^2}S^'(m/M_H)+ C_F\left(\frac{4}{\varepsilon _{\text{UV}}}-4\mathcal{L}_{M_W}+S(m/M_W)\right)\right\rbrace\end{aligned}$}
 & 
{$\begin{aligned}[t]F^{(m,M)}_{\chi} \end{aligned}$}
 \\[0.75ex] \hline
 $h_{f,s}$ &  & $0$ & 
\resizebox{.33\hsize}{!}{$\begin{aligned}[t]&\frac{C_A}{2}\left\lbrace C_F\left(\frac{4}{\varepsilon _{\text{UV}}}-\frac{4}{\varepsilon _{\text{IR}}}\right)+\frac{Y_s^2}{2}\left(-\frac{2}{\varepsilon _{\text{UV}}}+\frac{2}{\varepsilon _{\text{IR}}}\right)\right\rbrace\end{aligned}$}
 & 
{$\begin{aligned}[t]F^{(0)}_{h} \end{aligned}$}
 \\[0.75ex] \hline 
 $h_{f,s}$ &  & $M$ & 
\resizebox{.35\hsize}{!}{$\begin{aligned}[t]&\frac{C_A}{2}\left\lbrace C_F\left(\frac{4}{\varepsilon _{\text{UV}}}-4\mathcal{L}_{M_W}\right)+\frac{Y_s^2}{2}\left(-\frac{2}{\varepsilon _{\text{UV}}}+2\mathcal{L}_{M_H}\right)\right\rbrace\end{aligned}$}
 & 
{$\begin{aligned}[t] F^{(M)}_{h} \end{aligned}$}
 \\ [1ex] 
 \hline
\end{tabular}
\caption{Contributions to on-shell wavefunction renormalization. The exchanged boson masses are $M=M_{W,H}$ where $\mathcal{L}_M=\log{M^2/\mu^2}$, and the external particle (fermion or scalar) mass is $m$. The two-loop wave-function corrections, $F_I^{(i,j)}$, and the parametric integral functions, $P, P^'$ and $S,S'$, are given in Appendix \ref{sec:wfr2}.}
\label{table:wfr}
\end{table}
\subsection{Mass and Coupling Renormalisation}
Our loop calculations up to two loop order have been performed using the unrenormalized Feynman rules. Introducing now the renormalized coupling constant and mass instead of the bare couplings does not change the two loop results at $\mathcal{O}(\alpha^2)$. However, the coupling and mass in the one-loop result have to be regarded as the bare parameters and must be replaced by the renormalised ones. In our work we employ the $\overline{\text{MS}}$ scheme for the coupling renormalisation and the on-shell scheme for mass renormalisation. In the on-shell scheme, the square of the physical, renormalized mass is defined to be the real part of the pole of the propagator. In the case of coupling renormalisation the replacement can be applied naively as shown below. However, in the case of mass renormalisation, say given a mass $M$, with replacement (we denote the bare quantities with index, $0$),
\begin{equation}
    M_0^2=M^2+\delta M^2+\mathcal{O}(\alpha^2),
\end{equation}
in which $\delta M^2$ corresponds to the mass contribution, the masses to be renormalisaed often appear in terms of the form $(\mu^2/M^2)^{\epsilon}$ or in powers of logarithms. Thus the substitutions at one-loop are,
\begin{align}
    &\left(\frac{M^2}{\mu^2}\right)^{\epsilon}\rightarrow \left(\frac{M^2}{\mu^2}\right)^{\epsilon}\left(1+\epsilon \frac{\delta M^2}{M^2}\right)+\mathcal{O}(\alpha^2),\\&
    \mathcal{L}^n_M=\mathcal{L}^n_M+n\mathcal{L}^{n-1}_M\frac{\delta M^2}{M^2}+\mathcal{O}(\alpha^2),
\end{align}
 which, when applied provides corrections of $\mathcal{O}(\alpha^2)$. For the particles we are considering below, the renormalized quantities and renormalization constants are defined as follows,
\begin{subequations}
\label{eqn:rensubs}
\begin{align}
\label{eqn:rensubs1}
    &\alpha_0=(1+\delta Z_{\alpha})\alpha \\
    &M_{W,0}^2=M_W^2+\delta M_W^2\\
    &M_{H,0}^2=M_H^2+\delta M_H^2\\
    &m_{\chi,0}^2=m_{\chi}^2+\delta m_{\chi}^2\\
    &m_{\psi,0}=m_{\psi}+\delta m_{\psi},
\end{align}
\end{subequations}
  where the subscripts $\psi$ and $\chi$ indicate that the masses belong to fermion and scalar fields, respectively, that appear externally in the form factor.

\subsubsection{Coupling Renormalisation}
According to the prescription of the $\overline{\text{MS}}$ scheme, the unrenormalized coupling constant $\alpha_0$ is replaced by the renormalized coupling $\alpha$ via,
\begin{equation}
\alpha_0=(1+\delta Z_{\alpha})\alpha=\alpha\left(1-\frac{\alpha}{4\pi}\frac{\beta_0}{\varepsilon _{\text{UV}}}\right)+\mathcal{O}(\alpha^3),
\label{eqn:aren}
\end{equation}
such that $\beta_0$ is the leading (one-loop) coefficient of the renormalisation group $\beta$-function. We note that $\beta_0$ has the following form,
\begin{equation}
    \beta_0=\frac{11}{3}C_A-\frac{4}{3}T_fn_f-\frac{1}{6},
\end{equation}
where the terms proportional to $C_A$ and $n_f$ correspond to the non-Abelian and fermionic contributions, respectively, while the last term corresponds to a Higgs contribution. Thus by applying the substitution \eqref{eqn:aren} to our  one-loop form factors, we get additional contributions of order $\alpha^2$.
\subsubsection{Gauge Mass Renormalisation}
As this is the first case of mass renormalisation we consider we will discuss this in detail, at the amplitude level. The relation between the bare gauge boson mass, $M_{W,0}$, and the renormalized mass, $M_W$, is determined by the gauge boson self-energy corrections, which have the form,
\begin{equation}
    \tilde\Pi^{\mu\nu,ab}(p)=i\delta^{ab}g^{\mu\nu}p^2\Pi(p^2)\boldsymbol{1}+\text{terms}\propto p^{\mu}p^{\nu}.
\end{equation}
After extracting $\Pi(p^2)$ from the amplitudes with the help of the projection operator, $P_{\mu\nu}=g_{\mu\nu}-\frac{p_{\mu}p_{\nu}}{p^2}$, the renormalised mass is given by setting $\delta M_W^2=-M_W^2\Pi(M_W^2)$, and we may check various contributions at one-loop, up to $\mathcal{O}(\varepsilon)$, where $\epsilon$ are UV divergences. 
The results up to $\mathcal{O}(\varepsilon)^2$, needed for mass renormalisation contributing at two-loop orders is provided in Appendix \ref{sec:massren}. 
We begin with the self-energy contributions from the fermion loop,
\begin{equation}
    \Pi(M_W^2)_{n_f}=-\frac{4a}{9}T_fn_f\left\lbrace5+3i\pi+\frac{3}{\varepsilon }-3\mathcal{L}_{M_W}\right\rbrace+\mathcal{O}(\varepsilon),
\end{equation}
where as we stated before, $a(\mu)=\alpha(\mu)/4\pi$. Next, we have contributions from the non-Abelian gauge boson and ghost field loops,
\begin{equation}
    \Pi(M_W^2)_{WW,cc}=\frac{a}{9}C_A\left\lbrace82-12\sqrt{2}\pi+\frac{51}{\varepsilon }-51\mathcal{L}_{M_W}\right\rbrace+\mathcal{O}(\varepsilon),
\end{equation}
from the loop with gauge and Higgs boson,
\begin{equation}
    \Pi(M_W^2)_{WH}=a\left\lbrace-2-\frac{1}{\varepsilon }+\mathcal{L}_{M_W}+\frac{sr}{M_W}\log{w}+r^2\log{r}\right\rbrace+\mathcal{O}(\varepsilon),
\end{equation}
where we define $r=M_H/M_W$, $s=\sqrt{M_H^2-4M_W^2}$ and $w=\frac{2M_W}{M_H+s}$, and finally a contribution from the loops with Higgs and Goldstone bosons,
\begin{align}
    &\Pi(M_W^2)_{\phi\phi}=\frac{a}{72}C_A\left\lbrace34-3\sqrt{3}+\frac{15}{\varepsilon }-15\mathcal{L}_{M_W}\right\rbrace+\mathcal{O}(\varepsilon), \\&
    \Pi(M_W^2)_{H\phi}=a\left\lbrace\frac{1}{18}\left(5+\frac{3}{\varepsilon }-3\mathcal{L}_{M_W}\right)+\frac{r^2}{2}\left(\log{r}+\frac{3}{2}+\frac{1}{2\varepsilon }-\frac{1}{3}\mathcal{L}_{M_H}\right)\right.\nonumber\\&\left.\quad\quad\quad\quad\quad-\frac{r^4}{12}-\frac{r^4}{2}\log{r}+\frac{r^5s}{12M_W}\log{w}-\frac{r^3s}{3M_W}\log{w}+\frac{r^6}{12}\log{r}\right\rbrace+\mathcal{O}(\varepsilon).
\end{align}
Thus combining all terms provides one with the gauge boson mass correction in the replacement rules. Note that in the above contributions there are no terms from massive fermion and scalar loops, this makes sense as in the EFT formalism the scale where the bosons are no longer IR, i.e. where there masses are no longer zero, is the same scale where the fermions and scalars are taken to be UV or static. Moreover, we note that the self-energy diagrams with tadpoles have been omitted. They do not depend on the momentum of the gauge boson, so their contribution to the mass renormalization cancels exactly the corresponding vertex correction and field renormalization diagrams which are also dropped out.
\subsubsection{Higgs Mass Renormalisation}
As we were explicit in the previous section and broke down each contribution we will be brief now as the above still applies and we simply state the correction. The relation between the bare Higgs mass, $M_{H_0}$, and the renormalized mass, $M_H$, is determined by the Higgs self-energy corrections, $ \tilde\Sigma(p^2)=i\Sigma(p^2)\boldsymbol{1}$. Extracting $\Sigma(p^2)$ gives the renormalized mass by setting $\delta M_H^2=\Sigma(M_H^2)$, which has the following form after combining all contributions,
\begin{align}
    \delta M_H^2=&aC_AC_F\frac{M_W}{64r}\left\lbrace-2M_Wr(r^4-16r^2+36)+M_Wr(r^4-16r^2+48)\left(\mathcal{L}_{M_W}  -\frac{1}{\varepsilon }\right)  \nonumber \right.\\& \left. -s(r^4-16r^2+56)\log{\left(\frac{r(s-M_H)}{2M_W}+1\right)}\right\rbrace-a r^4\frac{9 M_W^2}{32}\left\lbrace2-\frac{\pi}{\sqrt{3}} \nonumber \right.\\& \left. +\frac{1}{\varepsilon }-\mathcal{L}_{M_W}-\log{r}\right\rbrace+\mathcal{O}(\varepsilon),
\end{align}
 up to $\mathcal{O}(\epsilon)$, where $\epsilon$ are UV divergences. The results up to $\mathcal{O}(\epsilon^2)$, needed for mass renormalisation contributing at two-loop orders is provided in Appendix \ref{sec:massren}. 
 Whence the above provide us with the Higgs mass correction at two loop order. Note again the lack of contributions from massive fermion and scalar loops due to the corrections being applied at a scale where fermions and scalars are integrated out. Moreover, we note that the self-energy diagrams with tadpoles have been omitted for the same reason previously described.
\subsubsection{Fermion and Scalar Mass Renormalisation}
Lastly we discuss the mass renormalisation of the massive external fermion and scalar fields we consider. These masses appear and the corrections contribute at two loop order in the threshold regime at the scale where the Higgs and gauge masses are taken to be IR and vanish. We begin with the scalar contributions; the relation between the bare scalar mass, $m_{\chi_0}$, and the renormalised mass, $m_{\chi}$, is determined by the scalar self-energy corrections, $ \tilde\Sigma(p^2)=i\Sigma(p^2)\boldsymbol{1}$. Extracting $\Sigma(p^2)$ gives the renormalised mass by setting $\delta m_{\chi}^2=\Sigma(m_{\chi}^2)$, which has the following form after combining all contributions,
\begin{equation}
    \delta m_{\chi}^2=a e^{\gamma_E  \varepsilon }\left(\frac{\mu ^2}{m_{\chi}^2}\right)^{\varepsilon } C_A\left\lbrace C_F m_{\chi}^2\frac{(2 \varepsilon -3)\Gamma(\varepsilon-1 )}{2 \varepsilon -1}+Y_s^2\frac{ \Gamma (\varepsilon )}{4-8 \varepsilon }\right\rbrace,
\end{equation}
where $\epsilon$ are UV divergences. Note that the dimensions of the second term match the first by definition of $Y_s$ in \eqref{eqn:yukdefs}. On the other hand, the relation between the bare fermion mass, $m_{\psi_0}$, and the renormalised mass, $m_{\psi}$, is determined by the fermion self-energy corrections, 
\begin{equation}
\tilde\Sigma(p^2)=i\left(\Sigma^V(p^2)\slashed{p}+\Sigma^S(p^2)m_{\psi}\right)\boldsymbol{1},
\end{equation}
where the superscripts, $S$ and $V$, label the scalar and vector contributions. Extracting $\Sigma^{S,V}(p^2)$ gives the renormalized mass by setting $\delta m_{\psi}=m_{\psi}\left(\Sigma^V(m_{\chi}^2)+\Sigma^S(m_{\chi}^2)\right)$, which has the following form after combining all contributions,
\begin{equation}
    \delta m_{\psi}=a e^{\gamma_E  \varepsilon }\left(\frac{\mu ^2}{m_{\psi}^2}\right)^{\varepsilon }  C_A\left\lbrace C_F\frac{(2 \varepsilon -3)\Gamma(\varepsilon )}{2 \varepsilon -1} -\frac{Y_f^2}{8} \left(\Gamma (\varepsilon
   -1)+\frac{4 \Gamma (\varepsilon )}{1-2 \varepsilon }\right) \right\rbrace,
\end{equation}
where again $\epsilon$ are UV divergences. In this case dimensions hold since $Y_f$ is dimensionless as shown in \eqref{eqn:yukdefs}. Note that the expansion up to $\mathcal{O}(\varepsilon)^3$ are needed for mass renormalisation. Now we have all the one-loop terms that arise in our problem which, when replacement rules are applied, contribute at the two-loop level. 
\subsection{Operator Renormalisation}
Composite operators like ours require subsequent subtractions beyond wave-function renormalisation \cite{manohar2007heavy}. This holds for both full and effective theory operators, to illustrate, let us take, for instance, the bare heavy-light fermion operator from HPET,
\begin{align}
    \mathcal{O}^{(0)}=\Bar{\psi}^{(0)}\Gamma h_f^{(0)}=\sqrt{Z_fZ_h}\Bar{\psi}\Gamma h_f,
\end{align}
where $\Gamma$ is an arbitrary Dirac matrix. The renormalised composite operator is then,
\begin{align}
    \mathcal{O}=Z_{\mathcal{O}}^{-1}\mathcal{O}^{(0)}=&\frac{\sqrt{Z_fZ_h}}{Z_{\mathcal{O}}}\Bar{\psi}\Gamma h_f \nonumber \\& =\Bar{\psi}\Gamma h_f+\text{counter term},
    \label{eqn:oprenorm1}
\end{align}
such that the additional operator, $Z_{\mathcal{O}}$, is determinable by computing a Green’s function with an insertion of $\mathcal{O}$. Therefore, $Z_{\mathcal{O}}$ can be found by taking the one particle irreducible Green's function of $\bar{\psi}$, $h_f$ and $\mathcal{O}$, where the counter term in \eqref{eqn:oprenorm1} contributes,
\begin{equation}
    \left(\frac{\sqrt{Z_fZ_h}}{Z_{\mathcal{O}}}-1\right)\Gamma,
        \label{eqn:oprenorm2}
\end{equation}
to this time-ordered product. The vertex contribution also provides a UV divergent contribution to the time-ordered product. Consequently, the counter term, \eqref{eqn:oprenorm1}, must eliminate the divergences present in the vertex contribution and thus, \eqref{eqn:oprenorm1} must be finite as $\varepsilon _{\text{UV}}\rightarrow 0$. Plugging in the wave function contributions, $\sqrt{Z_fZ_h}$ then gives $Z_{\mathcal{O}}$ by the finiteness requirement. We then may obtain the anomalous dimension of the composite operator, 
\begin{equation}
    \gamma_{\mathcal{O}}=\frac{\mu}{Z_{\mathcal{O}}}\left(\frac{dZ_{\mathcal{O}}}{d\mu}\right),
\end{equation}
from the renormalisation constant,
\begin{align}
    &Z_{\mathcal{O}}=1+\delta Z_{\mathcal{O}}= 1-\frac{1}{\varepsilon _{\text{UV}}}\gamma_{\mathcal{O}} 
    \label{eqn:oprenorm}
\end{align}
Note in this case that the renormalisation of $\mathcal{O}$ is independent of the gamma matrix of choice, $\Gamma$, in the composite operator. This is a consequence of heavy fermion spin symmetry and light fermion
chiral symmetry. In fact, this independence holds for all our effective operators in the threshold regime as they include operators with heavy/light fermions/scalars \cite{manohar2007heavy}.  On the other hand, in the full theory as well as SCET the gamma matrix plays a role and $Z_{\mathcal{O}}$ varies for different operators. In particular, in the full theory for both scalars and fermions, the scalar and tensor currents require renormalisation while the vector currents, at all orders, do not, meaning $\delta Z_{\mathcal{O}}$ is null \cite{schwartz2014quantum}.
\section{Radiative Corrections in Sudakov Limit}
\label{sec:ffsud}
In this section, we calculate the form factor, $\log{F_E(Q^2)}$, in the large $Q^2$, or Sudakov, limit. We perform calculations up to two-loop order, extending previous studies and refraining from including computational details which have been presented in other works \cite{chiu2008electroweak,jantzen2006two}. 


\subsection{Massless External Particles}
\label{ssec:esud}
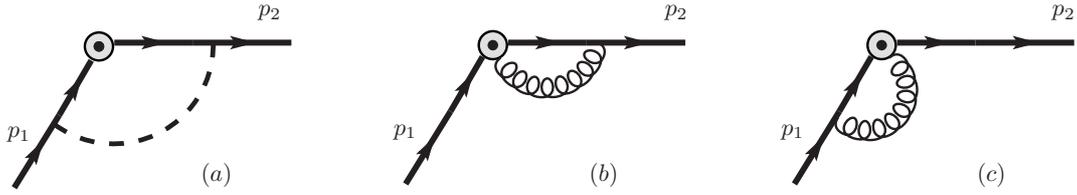
\begin{figure}[tb]
\begin{center}
\scalebox{.9}{
\fcolorbox{white}{white}{
  \begin{picture}(451,82) (19,-16)
    \SetWidth{1.0}
    \SetColor{Black}
    \GOval(55.811,43.586)(5.847,5.847)(0){0.882}
    \SetWidth{2.5}
    \Line[arrow,arrowpos=0.5,arrowlength=7.5,arrowwidth=2,arrowinset=0.2](20.73,-15.415)(39.334,14.883)
    \Line[arrow,arrowpos=0.5,arrowlength=7.5,arrowwidth=2,arrowinset=0.2](39.334,14.883)(52.622,37.739)
    \Line[arrow,arrowpos=0.5,arrowlength=7.5,arrowwidth=2,arrowinset=0.2](62.19,45.181)(94.614,45.181)
    \Line[arrow,arrowpos=0.5,arrowlength=7.5,arrowwidth=2,arrowinset=0.2](94.614,45.181)(135.542,45.181)
    \SetWidth{2.0}
    \Arc[dash,dashsize=6.378](61.924,44.134)(41.207,-126.856,1.455)
    \SetWidth{1.0}
    \Vertex(55.811,43.586){2.192}
    \Text(18.072,3.721)[lb]{{\Black{$p_1$}}}
    \Text(122.254,54.748)[lb]{{\Black{$p_2$}}}
    \Text(98.334,-14.883)[lb]{{\Black{$(a)$}}}
    \Vertex(55.811,43.586){2.192}
    \GOval(218.994,44.118)(5.847,5.847)(0){0.882}
    \SetWidth{2.5}
    \Line[arrow,arrowpos=0.5,arrowlength=7.5,arrowwidth=2,arrowinset=0.2](225.372,45.181)(257.796,45.181)
    \Line[arrow,arrowpos=0.5,arrowlength=7.5,arrowwidth=2,arrowinset=0.2](257.796,45.181)(298.724,45.181)
    \Line[arrow,arrowpos=0.5,arrowlength=7.5,arrowwidth=2,arrowinset=0.2](201.453,16.478)(214.741,39.334)
    \Line[arrow,arrowpos=0.5,arrowlength=7.5,arrowwidth=2,arrowinset=0.2](182.849,-13.82)(201.453,16.478)
    \SetWidth{1.0}
    \Vertex(218.994,44.118){2.192}
    \Text(291.283,54.748)[lb]{{\Black{$p_2$}}}
    \Text(178.597,4.784)[lb]{{\Black{$p_1$}}}
    \Text(259.922,-14.883)[lb]{{\Black{$(b)$}}}
    \GOval(380.049,44.118)(5.847,5.847)(0){0.882}
    \SetWidth{2.5}
    \Line[arrow,arrowpos=0.5,arrowlength=7.5,arrowwidth=2,arrowinset=0.2](386.428,45.181)(418.852,45.181)
    \Line[arrow,arrowpos=0.5,arrowlength=7.5,arrowwidth=2,arrowinset=0.2](418.852,45.181)(459.78,45.181)
    \Line[arrow,arrowpos=0.5,arrowlength=7.5,arrowwidth=2,arrowinset=0.2](362.509,16.478)(375.797,39.334)
    \Line[arrow,arrowpos=0.5,arrowlength=7.5,arrowwidth=2,arrowinset=0.2](343.905,-13.82)(362.509,16.478)
    \SetWidth{1.0}
    \Vertex(380.049,44.118){2.192}
    \Text(452.339,54.748)[lb]{{\Black{$p_2$}}}
    \Text(339.653,4.784)[lb]{{\Black{$p_1$}}}
    \Text(420.978,-14.883)[lb]{{\Black{$(c)$}}}

    \GluonArc(241.303,47.07)(20.831,-156.615,-5.204){3.987}{8}
    \GluonArc[clock](374.201,24.955)(16.535,56.87,-137.669){3.987}{8}
  \end{picture}
}}
\end{center}
    \caption{One-loop vertex corrections, bulls-eye represents composite operator, arrowed lines represent all incoming-outgoing particles we consider, dashed lines correspond to bosonic propagators. $(b),(c)$ only exists with the operator, $\mathcal{O}=i\phi^{\dagger}\ora{D}_{\mu}\phi$, and EFT equivalents, $\tilde{\mathcal{O}}$. }
    \label{fig:verts1}
\end{figure}

Let us begin by considering the case of massless external particles in a fair amount of detail. The limit we consider is thus, $Q^2\gg M^2 \gg m^2$, where $M$ and $m$ denote the bosonic and external masses, respectively. Schematically, in this case, the matching and running steps can be illustrated as follows,
\begin{equation*}
    \mathcal{O}\xleftrightarrow[m,M=0]{\mu\sim Q} e^C \tilde{\mathcal{O}}_1  \xrightarrow[]{\phantom{11}\gamma_1\phantom{11}}      e^C\tilde{\mathcal{O}}_1\xleftrightarrow[m=0]{\mu\sim M}e^{C+D} \tilde{\mathcal{O}}_2,
\end{equation*}
where $C$ and $D$ are multiplicative matching coefficients, $\gamma_1$ the effective theory anomalous dimension and $\tilde{\mathcal{O}}_{1,2}$ the effective theory operators at each scale. At scale, $\mu>Q$, we use the full theory, and at scale, $\mu<Q$, we match down to SCET with Wilson coefficient, $c(\mu)$. The RGE of $c(\mu)$ is given by,
\begin{equation}
    \mu\frac{dc(\mu)}{d\mu}=\gamma_F(a(\mu))c(\mu),
\end{equation}
where $\gamma_F(\mu)$ is the full theory anomalous dimension  for a composite operator, $\mathcal{O}$. The full theory is matched onto SCET at a scale $\mu\sim Q$. The effective theory has modes with off-shellness of $\mathcal{O}(Q)$ integrated out, so the matching coefficient depends on $\mathcal{L}_Q$, and these logarithms are not large if $\mu\sim Q$. The operator, $\mathcal{O}$ in the full theory matches to the operator, $\tilde{\mathcal{O}}$, in SCET. More specifically,
\begin{subequations}
\begin{align}
    &\bar{\psi}\Gamma\psi\rightarrow e^C(\bar{\xi}_{n,p_2}W_n)\Gamma (W^{\dagger}_{\bar{n}}\xi_{\bar{n},p_1}), \\ &
    \chi^{\dagger}\chi\rightarrow e^C({\Phi}^{\dagger}_{n,p_2}W_n) (W^{\dagger}_{\bar{n}}\Phi_{\bar{n},p_1}),\\ &
    i\chi^{\dagger}\overset{\leftrightarrow}{D}_{\mu}\chi\rightarrow e^C({\Phi}^{\dagger}_{n,p_2}W_n)[i\mathcal{D}_1+i\mathcal{D}_2]_{\mu}(W^{\dagger}_{\bar{n}}\Phi_{\bar{n},p_1}),\\ &
    \bar{\psi}\chi\rightarrow e^C(\bar{\xi}_{n,p_2}W_n) (W^{\dagger}_{\bar{n}}\Phi_{\bar{n},p_1}),
\end{align}
\label{eqn:matchoc}
\end{subequations}
where $i\mathcal{D}_1=\mathcal{P}+g\left(n\cdot A_{\bar{n},q}\right)\frac{\bar{n}}{2}$,  $i\mathcal{D}_2=\mathcal{P}^{\dagger}+g\left(\bar{n}\cdot A_{n,-q}\right)\frac{n}{2}$,  $\mathcal{P}$ are label operators in SCET and $W_n$ is a Wilson line containing $n$-collinear gauge fields obtained by integrating over a path in the $\bar{n}$-direction \cite{bauer2001effective}. Of course $C(\mu)$ differs for each operator and we have written the multiplicative matching coefficient as $\exp{[C(\mu)]}$ rather than $C(\mu)$ for convenience. As is well-known, the matching coefficient can be computed as the finite part of the full theory matrix element, evaluated on-shell, with all IR scales, which in our case are the gauge and Higgs boson masses, are set to zero \cite{manohar1997heavy,manohar1977effective,manohar2003deep}. To illustrate the computation let us consider the one-loop result. The full and effective theory graphs to be evaluated are those in figure \ref{fig:verts1}, except in SCET the external lines are both taken to be collinear and graphs $(b)$ and $(c)$ are no longer identical. 
After combining the vertex graphs with the wavefunction and tree-level graphs, one obtains the value of the full and effective theory matrix elements, $\bra{p_2}\mathcal{O}\ket{p_1}$ and $\bra{p_2}\tilde{\mathcal{O}}\ket{p_1}$, respectively. The gauge and external particle masses are IR scales and can be set to zero in the matching, thus leading to scaleless integrals, for the EFT and wavefunction contributions. One then combines the vertex and wave-function contributions as prescribed in \eqref{eqn:ffac12} to obtain the one and two-loop order results. Moreover, as the masses are zero there are no two-loop contributions from mass renormalisation, only coupling renormalisation contributes. Scaleless integrals are set to zero in dimensional regularization, so the EFT matrix element is equal to its tree-level value. The full theory and EFT operators, $\mathcal{O}$ and $\tilde{\mathcal{O}}$, are normalised to have the same tree-level value \cite{manohar1997heavy}, thus,
\begin{equation}
\exp{[C(\mu)]}=\frac{\bra{p_2}\mathcal{O}\ket{p_1}}{\bra{p_2}\mathcal{O}\ket{p_1}}_{tree}.
\end{equation}
\begin{table}
\footnotesize
\centering
\begin{tabular}{|c|c|c|c|c|} 
 \hline
 $\mathcal{O}$ & $\gamma_F^{(1)}$ & $C^{(1)}(\mu)$ & $C^{(2)}(\mu)$ & $\gamma_1^{(1)}$ \\ [0.75ex] 
 \hline\hline
 $\bar{\psi}\psi$ & 
\resizebox{.12\hsize}{!}{$-3C_AC_F+\frac{C_A}{8}Y_f^2$}
 & 
\resizebox{.3\hsize}{!}{ $\begin{aligned}[t]\frac{C_AC_F}{6}\left\lbrace-6\mathcal{L}_Q^2+\pi^2-12\right\rbrace+\frac{C_AY_f^2}{4}\left\lbrace\mathcal{L}_Q-2\right\rbrace\end{aligned}$}
 & $V^{(Q)}_1$& 
\resizebox{.12\hsize}{!}{$C_AC_F(4\mathcal{L}_Q-6)$} \\[0.75ex] \hline
 $\bar{\psi}\gamma^{\mu}\psi$& 
\resizebox{.05\hsize}{!}{$-\frac{C_A}{4}Y_f^2$}
 &  
\resizebox{.4\hsize}{!}{ $\frac{C_AC_F}{6}\left\lbrace-6\mathcal{L}_Q\left(\mathcal{L}_Q-3\right)\mathcal{L}_Q+\pi^2-48\right\rbrace-\frac{C_AY_f^2}{8}\left\lbrace\mathcal{L}_Q-1\right\rbrace$}
 & $V^{(Q)}_2$& 
\resizebox{.12\hsize}{!}{$C_AC_F(4\mathcal{L}_Q-6)$} \\[0.75ex]  \hline
 $\bar{\psi}\sigma^{\mu\nu}\psi$& 
\resizebox{.1\hsize}{!}{$C_AC_F-\frac{C_A}{8}Y_f^2$}
 & 
\resizebox{.35\hsize}{!}{ $\frac{C_AC_F}{6}\left\lbrace-6\mathcal{L}_Q\left(\mathcal{L}_Q-4\right)\mathcal{L}_Q+\pi^2-48\right\rbrace+\frac{C_AY_f^2}{4}$}
 & $V^{(Q)}_3$ & 
\resizebox{.12\hsize}{!}{$C_AC_F(4\mathcal{L}_Q-6)$}\\[0.75ex] \hline
 $\chi^{\dagger}\chi$ &  
\resizebox{.07\hsize}{!}{$-3C_AC_F$}
  &   
\resizebox{.28\hsize}{!}{ $\frac{C_AC_F}{6}\left\lbrace-6\mathcal{L}_Q\left(\mathcal{L}_Q-1\right)\mathcal{L}_Q+\pi^2-12\right\rbrace$}
 & $V^{(Q)}_4$ & 
\resizebox{.12\hsize}{!}{$C_AC_F(4\mathcal{L}_Q-8)$}\\ [0.75ex] \hline
 $i\chi^{\dagger}\overset{\leftrightarrow}{D}_{\mu}\chi$&  
\resizebox{.01\hsize}{!}{$0$}  &  
\resizebox{.28\hsize}{!}{ $\frac{C_AC_F}{6}\left\lbrace-6\mathcal{L}_Q\left(\mathcal{L}_Q-4\right)\mathcal{L}_Q+\pi^2-48\right\rbrace$}
 & $V^{(Q)}_5$& 
\resizebox{.12\hsize}{!}{$C_AC_F(4\mathcal{L}_Q-8)$} \\ [0.75ex] \hline
 $\bar{\psi}\chi$, $\chi^{\dagger}\psi$&  
\resizebox{.13\hsize}{!}{$-\frac{3}{2}C_AC_F-\frac{C_A}{16}Y_f^2$}
  & 
\resizebox{.28\hsize}{!}{ $\frac{C_AC_F}{6}\left\lbrace-6\mathcal{L}_Q\left(\mathcal{L}_Q-3\right)\mathcal{L}_Q+\pi^2-36\right\rbrace$}
 & $V^{(Q)}_6$ & 
\resizebox{.12\hsize}{!}{$C_AC_F(4\mathcal{L}_Q-7)$}\\ [1ex] 
 \hline
\end{tabular}
\caption{Matching corrections, $C(\mu)$, to the Sudakov form-factor at $\mu\sim Q$. $V^{(Q)}_i$ are two loop vertex corrections, given in Appendix \ref{sec:vertcorr}. $\gamma_F$ and $\gamma_1$ are the full theory and SCET anomalous dimension, $a\equiv \alpha/(4\pi)$, and $\mathcal{L}_Q\equiv \log{Q^2/\mu^2}$. }
\label{tab:expc}
\end{table}
When computing the one loop graphs for $\mathcal{O}$, $\exp{[C(\mu)]}$ is given by the on-shell full theory matrix element, normalised by its tree-level value. The particle masses are all much smaller than $Q^2$, and only contribute $M^2/Q^2$ (where $M$ correspeonds to the gauge and Higgs masses) power corrections at the large scale, $Q$, which are being neglected. The one-loop values of $C(\mu)$ for the other cases are computed similarly, and are given in table \ref{tab:expc}, where in the loop expansion, $C(\mu)=aC^{(1)}(\mu)+a^2C^{(1)}(\mu)+\mathcal{O}(a^3)$. Large logarithms do not appear if the matching scale $\mu\sim Q$, in this work we choose $\mu=Q$ and the RGE of $c(\mu)$ in the EFT is given by the anomalous dimension, $\gamma_{1}$, of $\tilde{O}$ in SCET. The full thoery anomalous dimension, $\gamma_F$, of $\tilde{O}$ is also given in \ref{tab:expc}, we avoid presenting the two loop result as this has been previously found for a number of operators \cite{machacek1985two}. On the other hand, the SCET anomalous dimension, $\gamma_{1}$, is used to evolve $c(\mu)$ from $\mu=Q\rightarrow M$. As previously defined, the anomalous dimension is given by the UV counter terms for the SCET graphs, and can depend on $Q$, the largest scale. UV divergences are independent of IR properties and $\gamma_1$ is linear in $\log{\mu^2/Q^2}$ to all order \cite{manohar2003deep,bauer2004shape}, so one can always write,
\begin{equation}
    \gamma_1(\mu)=A(\alpha(\mu))\log{\frac{\mu^2}{Q^2}}+B(\alpha(\mu)).
\end{equation}
The loop expansion of the anomalous dimension, $\gamma_1= a\gamma_1^{(1)}+a^2\gamma_2^{(2)}+\mathcal{O}(a^3)$, is given for each operator in table \ref{tab:expc}. By inspection, the SCET anomalous dimension, $\gamma_1$, depends solely on the external fields for the operators, as in it is equal for the three fermion and two scalar operators, respectively, as well as being the average of the two field's result for the mixed operator. The reason being that the effective theory anomalous dimension depends on the IR divergence of the full theory graph, and the IR divergence is independent of the vertex factors.

\begin{figure}[b]
\begin{center}
\scalebox{0.9}{
\fcolorbox{white}{white}{
  \begin{picture}(451,201) (-3,-14)
    \SetWidth{2.0}
    \SetColor{Black}
    \Arc[dash,dashsize=4.734](28.994,177.113)(21.789,-115.179,-9.365)
    \Arc[dash,dashsize=4.734](29.585,176.78)(41.544,-118.964,-4.434)
    \SetWidth{2.5}
    \Line[arrow,arrowpos=0.5,arrowlength=7.5,arrowwidth=2,arrowinset=0.2](47.731,173.568)(66.271,173.568)
    \SetWidth{1.0}
    \GOval(28.796,173.173)(3.55,3.55)(0){0.882}
    \Vertex(28.796,173.173){1.674}
    \SetWidth{2.5}
    \Line[arrow,arrowpos=0.5,arrowlength=7.5,arrowwidth=2,arrowinset=0.2](19.329,157)(26.824,170.017)
    \Line[arrow,arrowpos=0.5,arrowlength=7.5,arrowwidth=2,arrowinset=0.2](9.862,141.221)(18.935,156.605)
    \Line[arrow,arrowpos=0.5,arrowlength=7.5,arrowwidth=2,arrowinset=0.2](0,124.259)(9.467,140.826)
    \Line[arrow,arrowpos=0.5,arrowlength=7.5,arrowwidth=2,arrowinset=0.2](33.136,173.568)(47.731,173.568)
    \Line[arrow,arrowpos=0.5,arrowlength=7.5,arrowwidth=2,arrowinset=0.2](66.271,173.568)(91.517,173.568)
    \SetWidth{2.0}
    \Arc[dash,dashsize=4.734](108.066,169.286)(30.361,-108.131,8.86)
    \SetWidth{2.5}
    \Line[arrow,arrowpos=0.5,arrowlength=7.5,arrowwidth=2,arrowinset=0.2](137.671,173.962)(156.211,173.962)
    \SetWidth{1.0}
    \GOval(118.736,173.568)(3.55,3.55)(0){0.882}
    \Vertex(118.736,173.568){1.674}
    \SetWidth{2.5}
    \Line[arrow,arrowpos=0.5,arrowlength=7.5,arrowwidth=2,arrowinset=0.2](109.269,157.394)(116.764,170.412)
    \Line[arrow,arrowpos=0.5,arrowlength=7.5,arrowwidth=2,arrowinset=0.2](99.801,141.615)(108.874,157)
    \Line[arrow,arrowpos=0.5,arrowlength=7.5,arrowwidth=2,arrowinset=0.2](89.94,124.653)(99.407,141.221)
    \Line[arrow,arrowpos=0.5,arrowlength=7.5,arrowwidth=2,arrowinset=0.2](123.075,173.962)(137.671,173.962)
    \Line[arrow,arrowpos=0.5,arrowlength=7.5,arrowwidth=2,arrowinset=0.2](156.211,173.962)(181.457,173.962)
    \SetWidth{2.0}
    \Arc[dash,dashsize=4.734](125.475,188.555)(36.538,-117.72,-23.541)
    \SetWidth{1.0}
    \Vertex(206.703,173.568){1.674}
    \SetWidth{2.0}
    \Arc[dash,dashsize=4.734](208.281,173.173)(39.061,-123.048,1.157)
    \Arc[dash,dashsize=4.734,clock](232.344,174.378)(10.659,-177.765,-362.235)
    \SetWidth{2.5}
    \Line[arrow,arrowpos=0.5,arrowlength=7.5,arrowwidth=2,arrowinset=0.2](312.422,173.962)(330.962,173.962)
    \SetWidth{1.0}
    \GOval(293.487,173.568)(3.55,3.55)(0){0.882}
    \Vertex(293.487,173.568){1.674}
    \SetWidth{2.5}
    \Line[arrow,arrowpos=0.5,arrowlength=7.5,arrowwidth=2,arrowinset=0.2](284.02,157.394)(291.515,170.412)
    \Line[arrow,arrowpos=0.5,arrowlength=7.5,arrowwidth=2,arrowinset=0.2](274.552,141.615)(283.625,157)
    \Line[arrow,arrowpos=0.5,arrowlength=7.5,arrowwidth=2,arrowinset=0.2](264.69,124.653)(274.158,141.221)
    \Line[arrow,arrowpos=0.5,arrowlength=7.5,arrowwidth=2,arrowinset=0.2](297.826,173.962)(312.422,173.962)
    \Line[arrow,arrowpos=0.5,arrowlength=7.5,arrowwidth=2,arrowinset=0.2](330.962,173.962)(356.208,173.962)
    \SetWidth{2.0}
    \Arc[dash,dashsize=4.734](292.737,174.746)(39.021,-118.433,-1.152)
    \Arc[dash,dashsize=4.734,clock](332.203,168.931)(17.26,164.417,16.947)
    \Text(65.482,124.259)[lb]{{\Black{$(a)$}}}
    \Text(154.238,123.864)[lb]{{\Black{$(b)$}}}
    \Text(246.15,123.075)[lb]{{\Black{$(c)$}}}
    \Text(336.879,123.47)[lb]{{\Black{$(d)$}}}
    \SetWidth{2.5}
    \Line[arrow,arrowpos=0.5,arrowlength=7.5,arrowwidth=2,arrowinset=0.2](402.756,173.568)(421.296,173.568)
    \SetWidth{1.0}
    \GOval(383.821,173.173)(3.55,3.55)(0){0.882}
    \Vertex(383.821,173.173){1.674}
    \SetWidth{2.5}
    \Line[arrow,arrowpos=0.5,arrowlength=7.5,arrowwidth=2,arrowinset=0.2](374.354,157)(381.849,170.017)
    \Line[arrow,arrowpos=0.5,arrowlength=7.5,arrowwidth=2,arrowinset=0.2](364.886,141.221)(373.959,156.605)
    \Line[arrow,arrowpos=0.5,arrowlength=7.5,arrowwidth=2,arrowinset=0.2](355.025,124.259)(364.492,140.826)
    \Line[arrow,arrowpos=0.5,arrowlength=7.5,arrowwidth=2,arrowinset=0.2](388.16,173.568)(402.756,173.568)
    \Line[arrow,arrowpos=0.5,arrowlength=7.5,arrowwidth=2,arrowinset=0.2](421.296,173.568)(446.542,173.568)
    \SetWidth{2.0}
    \Arc[dash,dashsize=4.734](378.626,170.973)(34.535,-114.879,-68.442)
    \Arc[dash,dashsize=4.734](393.754,177.554)(34.088,-48.887,-6.716)
    \SetWidth{1.0}
    \Arc[arrow,arrowpos=0.5,arrowlength=5,arrowwidth=2,arrowinset=0.2](402.756,143.982)(13.778,156,516)
    \Text(431.158,126.231)[lb]{{\Black{$(e)$}}}
    \SetWidth{2.5}
    \Line[arrow,arrowpos=0.5,arrowlength=7.5,arrowwidth=2,arrowinset=0.2](44.575,102.168)(63.115,102.168)
    \SetWidth{1.0}
    \GOval(25.641,101.774)(3.55,3.55)(0){0.882}
    \Vertex(25.641,101.774){1.674}
    \SetWidth{2.5}
    \Line[arrow,arrowpos=0.5,arrowlength=7.5,arrowwidth=2,arrowinset=0.2](16.173,85.6)(23.668,98.618)
    \Line[arrow,arrowpos=0.5,arrowlength=7.5,arrowwidth=2,arrowinset=0.2](6.706,69.821)(15.779,85.206)
    \Line[arrow,arrowpos=0.5,arrowlength=7.5,arrowwidth=2,arrowinset=0.2](-3.156,52.859)(6.312,69.427)
    \Line[arrow,arrowpos=0.5,arrowlength=7.5,arrowwidth=2,arrowinset=0.2](29.98,102.168)(44.575,102.168)
    \Line[arrow,arrowpos=0.5,arrowlength=7.5,arrowwidth=2,arrowinset=0.2](63.115,102.168)(88.362,102.168)
    \SetWidth{2.0}
    \Arc[dash,dashsize=4.734](24.063,104.635)(40.312,-116.752,-3.508)
    \Arc[dash,dashsize=4.734](56.737,93.802)(14.761,145.477,259.482)
    \Text(59.565,52.859)[lb]{{\Black{$(f)$}}}
    \SetWidth{2.5}
    \Line[arrow,arrowpos=0.5,arrowlength=7.5,arrowwidth=2,arrowinset=0.2](133.726,101.774)(152.266,101.774)
    \SetWidth{1.0}
    \GOval(114.791,101.379)(3.55,3.55)(0){0.882}
    \Vertex(114.791,101.379){1.674}
    \SetWidth{2.5}
    \Line[arrow,arrowpos=0.5,arrowlength=7.5,arrowwidth=2,arrowinset=0.2](105.324,85.206)(112.819,98.223)
    \Line[arrow,arrowpos=0.5,arrowlength=7.5,arrowwidth=2,arrowinset=0.2](95.857,69.427)(104.929,84.811)
    \Line[arrow,arrowpos=0.5,arrowlength=7.5,arrowwidth=2,arrowinset=0.2](85.995,52.465)(95.462,69.033)
    \Line[arrow,arrowpos=0.5,arrowlength=7.5,arrowwidth=2,arrowinset=0.2](119.13,101.774)(133.726,101.774)
    \Line[arrow,arrowpos=0.5,arrowlength=7.5,arrowwidth=2,arrowinset=0.2](152.266,101.774)(177.512,101.774)
    \SetWidth{2.0}
    \Arc[dash,dashsize=4.734](114.988,103.887)(40.489,-119.473,-2.992)
    \Arc[dash,dashsize=4.734](134.026,69.415)(11.205,-155.076,27.304)
    \Text(156.211,53.254)[lb]{{\Black{$(g)$}}}
    \SetWidth{2.5}
    \Line[arrow,arrowpos=0.5,arrowlength=7.5,arrowwidth=2,arrowinset=0.2](222.482,101.774)(241.022,101.774)
    \SetWidth{1.0}
    \GOval(203.547,101.379)(3.55,3.55)(0){0.882}
    \Vertex(203.547,101.379){1.674}
    \SetWidth{2.5}
    \Line[arrow,arrowpos=0.5,arrowlength=7.5,arrowwidth=2,arrowinset=0.2](194.08,85.206)(201.575,98.223)
    \Line[arrow,arrowpos=0.5,arrowlength=7.5,arrowwidth=2,arrowinset=0.2](184.613,69.427)(193.686,84.811)
    \Line[arrow,arrowpos=0.5,arrowlength=7.5,arrowwidth=2,arrowinset=0.2](174.751,52.465)(184.218,69.033)
    \Line[arrow,arrowpos=0.5,arrowlength=7.5,arrowwidth=2,arrowinset=0.2](207.887,101.774)(222.482,101.774)
    \Line[arrow,arrowpos=0.5,arrowlength=7.5,arrowwidth=2,arrowinset=0.2](241.022,101.774)(266.268,101.774)
    \SetWidth{2.0}
    \Arc[dash,dashsize=4.734](196.762,84.338)(20.959,-128.121,-48.51)
    \Arc[dash,dashsize=4.734](213.48,105.76)(34.088,-48.887,-6.716)
    \Arc[dash,dashsize=1.105,arrow,arrowpos=0.5,arrowlength=6.667,arrowwidth=2,arrowinset=0.2](222.876,72.977)(12.605,160,520)
    \Text(240.233,53.254)[lb]{{\Black{$(h)$}}}
    \SetWidth{2.5}
    \Line[arrow,arrowpos=0.5,arrowlength=7.5,arrowwidth=2,arrowinset=0.2](312.027,102.168)(330.567,102.168)
    \SetWidth{1.0}
    \GOval(293.092,101.774)(3.55,3.55)(0){0.882}
    \Vertex(293.092,101.774){1.674}
    \SetWidth{2.5}
    \Line[arrow,arrowpos=0.5,arrowlength=7.5,arrowwidth=2,arrowinset=0.2](283.625,85.6)(291.12,98.618)
    \Line[arrow,arrowpos=0.5,arrowlength=7.5,arrowwidth=2,arrowinset=0.2](274.158,69.821)(283.231,85.206)
    \Line[arrow,arrowpos=0.5,arrowlength=7.5,arrowwidth=2,arrowinset=0.2](264.296,52.859)(273.763,69.427)
    \Line[arrow,arrowpos=0.5,arrowlength=7.5,arrowwidth=2,arrowinset=0.2](297.432,102.168)(312.027,102.168)
    \Line[arrow,arrowpos=0.5,arrowlength=7.5,arrowwidth=2,arrowinset=0.2](330.567,102.168)(355.813,102.168)
    \SetWidth{2.0}
    \Arc[dash,dashsize=4.734](303.814,105.76)(34.088,-48.887,-6.716)
    \Arc[dash,dashsize=4.734](288.243,102.798)(37.62,-113.29,-68.28)
    \Arc[dash,dashsize=4.734](313.999,73.766)(12.586,148,508)
    \Text(331.356,52.465)[lb]{{\Black{$(i)$}}}
    \SetWidth{1.0}
    \Vertex(381.454,101.774){1.674}
    \Text(426.029,53.648)[lb]{{\Black{$(j)$}}}
    \SetWidth{2.5}
    \Line[arrow,arrowpos=0.5,arrowlength=7.5,arrowwidth=2,arrowinset=0.2](400.389,102.168)(418.929,102.168)
    \SetWidth{1.0}
    \GOval(381.454,101.774)(3.55,3.55)(0){0.882}
    \SetWidth{2.5}
    \Line[arrow,arrowpos=0.5,arrowlength=7.5,arrowwidth=2,arrowinset=0.2](371.987,85.6)(379.482,98.618)
    \Line[arrow,arrowpos=0.5,arrowlength=7.5,arrowwidth=2,arrowinset=0.2](362.519,69.821)(371.592,85.206)
    \Line[arrow,arrowpos=0.5,arrowlength=7.5,arrowwidth=2,arrowinset=0.2](352.658,52.859)(362.125,69.427)
    \Line[arrow,arrowpos=0.5,arrowlength=7.5,arrowwidth=2,arrowinset=0.2](385.793,102.168)(400.389,102.168)
    \Line[arrow,arrowpos=0.5,arrowlength=7.5,arrowwidth=2,arrowinset=0.2](418.929,102.168)(444.175,102.168)
    \SetWidth{2.0}
    \Arc[dash,dashsize=4.734,clock](375.537,92.87)(27.889,18.618,-119.673)
    \SetWidth{1.0}
    \GluonArc[clock](400.623,97.095)(20.138,155.881,14.592){2.959}{10}
    \GOval(26.035,35.108)(3.55,3.55)(0){0.882}
    \SetWidth{2.5}
    \Line[arrow,arrowpos=0.5,arrowlength=7.5,arrowwidth=2,arrowinset=0.2](44.97,35.502)(63.51,35.502)
    \SetWidth{1.0}
    \Vertex(26.035,35.108){1.674}
    \SetWidth{2.5}
    \Line[arrow,arrowpos=0.5,arrowlength=7.5,arrowwidth=2,arrowinset=0.2](16.568,18.935)(24.063,31.952)
    \Line[arrow,arrowpos=0.5,arrowlength=7.5,arrowwidth=2,arrowinset=0.2](7.1,3.156)(16.173,18.54)
    \Line[arrow,arrowpos=0.5,arrowlength=7.5,arrowwidth=2,arrowinset=0.2](-2.761,-13.807)(6.706,2.761)
    \Line[arrow,arrowpos=0.5,arrowlength=7.5,arrowwidth=2,arrowinset=0.2](30.374,35.502)(44.97,35.502)
    \Line[arrow,arrowpos=0.5,arrowlength=7.5,arrowwidth=2,arrowinset=0.2](63.51,35.502)(88.756,35.502)
    \SetWidth{2.0}
    \Arc[dash,dashsize=4.734,clock](19.425,26.131)(25.539,-10.001,-117.848)
    \Text(62.327,-14.595)[lb]{{\Black{$(k)$}}}
    \SetWidth{1.0}
    \GluonArc(45.333,45.499)(21.439,-142.16,-27.792){2.959}{9}
    \SetWidth{2.5}
    \Line[arrow,arrowpos=0.5,arrowlength=7.5,arrowwidth=2,arrowinset=0.2](133.331,35.897)(151.872,35.897)
    \SetWidth{1.0}
    \GOval(114.397,35.502)(3.55,3.55)(0){0.882}
    \Vertex(114.397,35.502){1.674}
    \SetWidth{2.5}
    \Line[arrow,arrowpos=0.5,arrowlength=7.5,arrowwidth=2,arrowinset=0.2](104.929,19.329)(112.424,32.347)
    \Line[arrow,arrowpos=0.5,arrowlength=7.5,arrowwidth=2,arrowinset=0.2](95.462,3.55)(104.535,18.935)
    \Line[arrow,arrowpos=0.5,arrowlength=7.5,arrowwidth=2,arrowinset=0.2](85.6,-13.412)(95.068,3.156)
    \Line[arrow,arrowpos=0.5,arrowlength=7.5,arrowwidth=2,arrowinset=0.2](118.736,35.897)(133.331,35.897)
    \Line[arrow,arrowpos=0.5,arrowlength=7.5,arrowwidth=2,arrowinset=0.2](151.872,35.897)(177.118,35.897)
    \SetWidth{2.0}
    \Arc[dash,dashsize=4.734,clock](115.975,33.133)(37.97,4.174,-124.848)
    \Text(154.238,-14.595)[lb]{{\Black{$(l)$}}}
    \SetWidth{1.0}
    \GluonArc[clock](125.99,39.626)(10.032,-176.724,-381.822){2.959}{7}
    \Vertex(381.454,101.774){1.674}
    \SetWidth{2.5}
    \Line[arrow,arrowpos=0.5,arrowlength=7.5,arrowwidth=2,arrowinset=0.2](225.638,173.962)(244.178,173.962)
    \SetWidth{1.0}
    \GOval(206.703,173.568)(3.55,3.55)(0){0.882}
    \SetWidth{2.5}
    \Line[arrow,arrowpos=0.5,arrowlength=7.5,arrowwidth=2,arrowinset=0.2](197.236,157.394)(204.731,170.412)
    \Line[arrow,arrowpos=0.5,arrowlength=7.5,arrowwidth=2,arrowinset=0.2](187.769,141.615)(196.841,157)
    \Line[arrow,arrowpos=0.5,arrowlength=7.5,arrowwidth=2,arrowinset=0.2](177.907,124.653)(187.374,141.221)
    \Line[arrow,arrowpos=0.5,arrowlength=7.5,arrowwidth=2,arrowinset=0.2](211.042,173.962)(225.638,173.962)
    \Line[arrow,arrowpos=0.5,arrowlength=7.5,arrowwidth=2,arrowinset=0.2](244.178,173.962)(269.424,173.962)
    \Line[arrow,arrowpos=0.5,arrowlength=7.5,arrowwidth=2,arrowinset=0.2](313.999,35.897)(332.54,35.897)
    \SetWidth{1.0}
    \GOval(295.065,35.502)(3.55,3.55)(0){0.882}
    \SetWidth{2.5}
    \Line[arrow,arrowpos=0.5,arrowlength=7.5,arrowwidth=2,arrowinset=0.2](285.597,19.329)(293.092,32.347)
    \Line[arrow,arrowpos=0.5,arrowlength=7.5,arrowwidth=2,arrowinset=0.2](276.13,3.55)(285.203,18.935)
    \Line[arrow,arrowpos=0.5,arrowlength=7.5,arrowwidth=2,arrowinset=0.2](266.268,-13.412)(275.736,3.156)
    \Line[arrow,arrowpos=0.5,arrowlength=7.5,arrowwidth=2,arrowinset=0.2](299.404,35.897)(313.999,35.897)
    \Line[arrow,arrowpos=0.5,arrowlength=7.5,arrowwidth=2,arrowinset=0.2](332.54,35.897)(357.786,35.897)
    \SetWidth{1.0}
    \Vertex(295.065,35.502){1.674}
    \GluonArc(297.355,36.077)(42.68,-122.295,0.287){2.959}{19}
    \GluonArc[clock](323.861,38.941)(15.296,-11.479,-168.521){2.959}{9}
    \SetWidth{2.5}
    \Line[arrow,arrowpos=0.5,arrowlength=7.5,arrowwidth=2,arrowinset=0.2](402.756,37.08)(421.296,37.08)
    \SetWidth{1.0}
    \GOval(383.821,36.686)(3.55,3.55)(0){0.882}
    \SetWidth{2.5}
    \Line[arrow,arrowpos=0.5,arrowlength=7.5,arrowwidth=2,arrowinset=0.2](374.354,20.513)(381.849,33.53)
    \Line[arrow,arrowpos=0.5,arrowlength=7.5,arrowwidth=2,arrowinset=0.2](364.886,4.734)(373.959,20.118)
    \Line[arrow,arrowpos=0.5,arrowlength=7.5,arrowwidth=2,arrowinset=0.2](355.025,-12.229)(364.492,4.339)
    \Line[arrow,arrowpos=0.5,arrowlength=7.5,arrowwidth=2,arrowinset=0.2](388.16,37.08)(402.756,37.08)
    \Line[arrow,arrowpos=0.5,arrowlength=7.5,arrowwidth=2,arrowinset=0.2](421.296,37.08)(446.542,37.08)
    \SetWidth{1.0}
    \Vertex(383.821,36.686){1.674}
    \GluonArc(386.111,36.866)(42.68,-122.295,0.287){2.959}{19}
    \GluonArc[clock](395.704,43.775)(33.538,-12.894,-132.212){2.959}{14}
    \Text(241.811,-13.412)[lb]{{\Black{$(m)$}}}
    \Text(335.695,-14.595)[lb]{{\Black{$(n)$}}}
    \Vertex(206.703,173.568){1.674}
    \SetWidth{2.5}
    \Line[arrow,arrowpos=0.5,arrowlength=7.5,arrowwidth=2,arrowinset=0.2](222.088,35.897)(240.628,35.897)
    \SetWidth{1.0}
    \GOval(203.153,35.502)(3.55,3.55)(0){0.882}
    \SetWidth{2.5}
    \Line[arrow,arrowpos=0.5,arrowlength=7.5,arrowwidth=2,arrowinset=0.2](193.686,19.329)(201.181,32.347)
    \Line[arrow,arrowpos=0.5,arrowlength=7.5,arrowwidth=2,arrowinset=0.2](184.218,3.55)(193.291,18.935)
    \Line[arrow,arrowpos=0.5,arrowlength=7.5,arrowwidth=2,arrowinset=0.2](174.356,-13.412)(183.824,3.156)
    \Line[arrow,arrowpos=0.5,arrowlength=7.5,arrowwidth=2,arrowinset=0.2](207.492,35.897)(222.088,35.897)
    \Line[arrow,arrowpos=0.5,arrowlength=7.5,arrowwidth=2,arrowinset=0.2](240.628,35.897)(265.874,35.897)
    \SetWidth{1.0}
    \Vertex(203.153,35.502){1.674}
    \GluonArc(205.838,36.077)(42.68,-122.295,0.287){2.959}{19}
    \GluonArc(227.073,38.384)(22.763,-164.62,-6.272){2.959}{13}
    \Text(429.974,-14.595)[lb]{{\Black{$(o)$}}}
  \end{picture}
}}
\end{center}
    \caption{Two-loop vertex correction graphs, $(a)$-$(d)$ are Abelian corrections; $(f)$-$(i)$ are non-Abelian,  $(j)$-$(m)$ only exists with the operator, $\mathcal{O}=i\phi^{\dagger}\ora{D}_{\mu}\phi$ and EFT equivalents, $(m)$-$(o)$ are seagull terms and occur only for scalar fields. }
    \label{fig:verts2}
\end{figure}
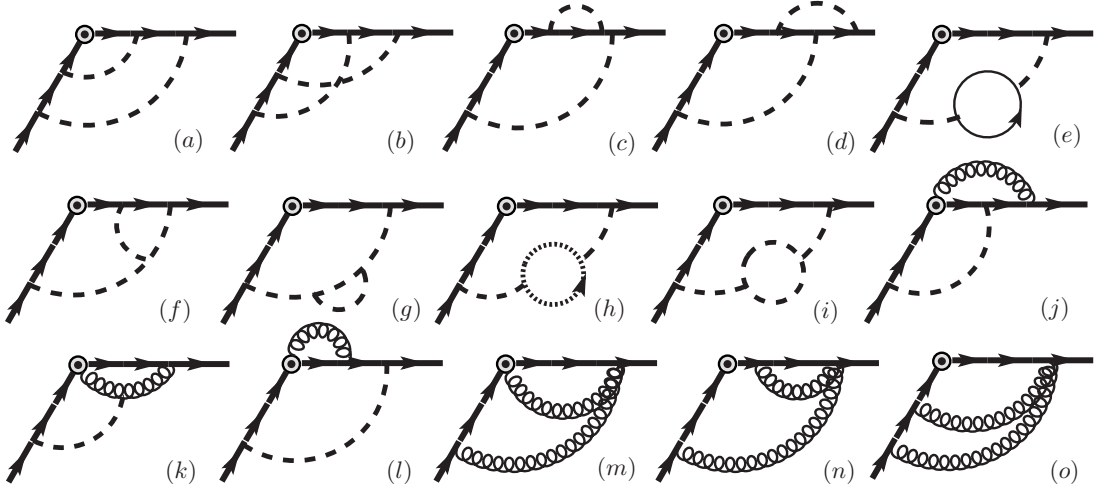

The next step matching step occurs at the lower scale, $\mu\sim M$, where the massive bosons are integrated out. The matching is done from SCET with massive bosons ($\mu> M$), to SCET without massive bosons ($\mu< M$). In our model, this is a free theory, so there is no need for propagating bosonic modes below $M$. The matching coefficient at $\mu\sim M$ is given by $d(\mu)=\exp{[D(\mu)]}$ in table \ref{tab:expd} and is found from the SCET vertex and wave-function corrections. More specifically, one matches in the following way,
\begin{subequations}
\begin{align}
    &e^{C}(\bar{\xi}_{n,p_2}W_n)\Gamma (W^{\dagger}_{\bar{n}}\xi_{\bar{n},p_1})\rightarrow e^{C+D}\bar{\xi}_{n,p_2}\Gamma \xi_{\bar{n},p_1} , \\ &
    e^{C}({\Phi}^{\dagger}_{n,p_2}W_n) (W^{\dagger}_{\bar{n}}\Phi_{\bar{n},p_1})\rightarrow    e^{C+D}{\Phi}^{\dagger}_{n,p_2} \Phi_{\bar{n},p_1},\\ &
    e^{C}({\Phi}^{\dagger}_{n,p_2}W_n)[i\mathcal{D}_1+i\mathcal{D}_2]_{\mu}(W^{\dagger}_{\bar{n}}\Phi_{\bar{n},p_1})\rightarrow e^{C+D}{\Phi}^{\dagger}_{n,p_2}i(\mathcal{P}^{\dagger}+\mathcal{P})_{\mu}\Phi_{\bar{n},p_1},\\ &
   e^{C}(\bar{\xi}_{n,p_2}W_n) (W^{\dagger}_{\bar{n}}\Phi_{\bar{n},p_1})\rightarrow e^{C+D}\bar{\xi}_{n,p_2} \Phi_{\bar{n},p_1}.
\end{align}
\label{eqn:matchod}
\end{subequations}
As for the results, although we calculate up to two loops fully for $c(\mu)$, we do not calculate the two-loop vertex contribute to $d(\mu)$ due to the complexity of massive SCET integrals. \begin{table}[tb]
\centering
\footnotesize
\begin{tabular}{|c|c|} 
 \hline
 $\mathcal{O}$ & $\Delta C^{(2)}(\mu)$  \\
 \hline\hline
 $\bar{\psi}\psi$ & 
\resizebox{.82\hsize}{!}{ $\begin{aligned}[t] &\frac{C_A C_F}{36}\left\lbrace-2 \mathcal{L}_Q^3+\left(\pi ^2-12\right)\mathcal{L}_Q-28 \zeta_3+24\right\rbrace\left(C_A \left(3
   C_A+22\right)-8 n_f T_f-4\right)+\frac{C_A Y_f^2}{288}\left\lbrace6\mathcal{L}_Q \left(\mathcal{L}_Q \left(-2 C_A C_F \log \mathcal{L}_Q+C_A\left(6 C_F+22\right) -8 n_f T_f-1\right)   \right.\right. \\ &\left.\left. +C_A \left(\left(\pi ^2-36\right)C_F-88\right)+32 n_f T_f+4\right)-168 \zeta_3 C_A C_F+48 C_A \left(9C_F+22\right)-2 \pi ^2 C_A \left(3 C_F+11\right)+\left(\pi ^2-48\right) \left(8 n_fT_f+1\right)\right\rbrace\end{aligned}$}
  \\ \hline
 $\bar{\psi}\gamma^{\mu}\psi$ &  \resizebox{.82\hsize}{!}{ $\begin{aligned}[t] &\frac{C_A C_F}{72}\left\lbrace2 \mathcal{L}_Q \left(\left(9-2 \mathcal{L}_Q \right) \mathcal{L}_Q +\pi
   ^2-48\right)-56 \zeta_3-3 \pi ^2+192\right\rbrace \left(C_A
   \left(3 C_A+22\right)-8 n_f T_f-4\right) + \frac{C_A Y_f^2}{576}\left\lbrace6\mathcal{L}_Q^2 \left(2 C_A \left(6 C_F-11\right)+8 n_f
   T_f+1\right)  \right. \\ &\left.  +12 \mathcal{L}_Q
   \left(C_A \left(\left(\pi ^2-42\right) C_F+22\right)-8
   n_f T_f-1\right)-24 C_A C_F \mathcal{L}_Q^3-12 \left(28 \zeta_3-90+\pi ^2\right) C_A C_F+22 \left(\pi ^2-12\right)
   C_A-\left(\pi ^2-12\right) \left(8 n_f
   T_f+1\right)\right\rbrace \end{aligned}$}
  \\\hline
 $\bar{\psi}\sigma^{\mu\nu}\psi$ &  \resizebox{.72\hsize}{!}{ $\begin{aligned}[t] &\frac{C_A C_F}{36}\left\lbrace\mathcal{L}_Q
   \left(-2 \left(\mathcal{L}_Q-6\right)
  \mathcal{L}_Q+\pi ^2-48\right)-28
   \zeta_3-2 \pi ^2+84\right\rbrace \left(C_A \left(3
   C_A+22\right)-8 n_f T_f-4\right)   +\frac{C_AY_f^2}{48}\left\lbrace\mathcal{L}_Q
   \left(-2 C_A C_F \left(\mathcal{L}_Q \right.\right.\right.\\ &\left.\left.\left.- 6\right) \mathcal{L}_Q+C_A
   \left(\left(\pi ^2-36\right) C_F+44\right)-2 \left(8
   n_f T_f+1\right)\right)-28 \zeta_3 C_A C_F-2 C_A
   \left(\left(\pi ^2-24\right) C_F+66\right)+48 n_f
   T_f+6\right\rbrace \end{aligned}$}
  \\\hline
 $\chi^{\dagger}\chi$  &  
\resizebox{.5\hsize}{!}{$\begin{aligned}[t] &\frac{C_A C_F }{72}\left\lbrace2 \mathcal{L}_Q \left(\left(3-2 \mathcal{L}_Q\right) \mathcal{L}_Q+\pi^2-12\right)-56 \zeta_3-\pi ^2+48\right\rbrace \left(-6 C_A^2+22
   C_A-8 n_f T_f+5\right)\end{aligned}$}
  \\\hline
 $i\chi^{\dagger}\overset{\leftrightarrow}{D}_{\mu}\chi$  &  
\resizebox{.5\hsize}{!}{$\begin{aligned}[t] &\frac{C_A C_F}{36}\left\lbrace\mathcal{L}_Q
   \left(-2 \left(\mathcal{L}_Q-6\right)
   \mathcal{L}_Q+\pi ^2-48\right)-28
   \zeta_3-2 \pi ^2+96\right\rbrace \left(-6 C_A^2+22 C_A-8
   n_f T_f+5\right)\end{aligned}$}
 \\\hline
 $\bar{\psi}\chi$, $\chi^{\dagger}\psi$  &  
\resizebox{.6\hsize}{!}{$\begin{aligned}[t] &\frac{C_A C_F }{576}\left\lbrace-2 \mathcal{L}_Q \left(\left(9-2 \mathcal{L}_Q\right) \mathcal{L}_Q+\pi
   ^2-36\right)+56 \zeta_3+3 \pi ^2-144\right\rbrace \left(C_A
   \left(-12 C_A+3 Yf^2-176\right)+64
   n_f T_f+20\right)\end{aligned}$}
 \\[1ex]
 \hline
\end{tabular}
\caption{Mass and coupling corrections to matching which contribute at two-loop order, $\Delta C^{(2)}(\mu)$ is the correction at $\mu\sim Q$.  }
\label{tab:corrcd}
\end{table}
Hence, mass and coupling renormalisation is not necessary as they only affect the next order, nonetheless, we present these sub-divergent $\mathcal{O}(a^2)$ contributions for both $c(\mu)$ and $d(\mu)$ in tables \ref{tab:corrcd} and \ref{tab:expd}. Moreover, by inspection of table \ref{table:wfr}, we do not include the collinear correction to the particle propagator for each case as it is the same as in the full theory \cite{bauer2001effective}. The ultrasoft correction vanishes, so the wavefunction corrections are the same as in the full theory and we have these up to two loops. For a more detailed description on the specific one-loop SCET integrals, we point to previous work \cite{chiu2008electroweak,becher2015introduction}, and it would be interesting to calculate the SCET vertex contributions at two-loop to have a complete account at this order. The above matching steps are identical at each order, and the two-loop vertex and wavefunction graphs we calculated are shown in figures \ref{fig:verts2} and \ref{fig:wfr}. 
\begin{table}
\footnotesize
\centering
\begin{tabular}{|c|c|c|} 
 \hline
 $\mathcal{O}$ & $D^{(1)}(\mu)$ & $\Delta D^{(2)}(\mu)$  \\ [0.75ex]
 \hline\hline
 $\bar{\psi}\Gamma\psi$ & 
\resizebox{.33\hsize}{!}{ $\frac{C_AC_F}{6}\left\lbrace-6\mathcal{L}_{M_W}^2+12\mathcal{L}_{M_W}\mathcal{L}_Q-18\mathcal{L}_{M_W}+27-5\pi^2\right\rbrace$}
 &  
\resizebox{.42\hsize}{!}{ $\begin{aligned}[t] -\frac{C_A C_F}{72 M_W^6}&\left\lbrace M_W^6
   \left(\left(99 \sqrt{3} \pi -690\right) C_A+32 (5+3 i \pi )
   n_f T_f+124\right) \right. \\ & \left.+3 M_W^4\left(\mathcal{L}_{M_W}\left(M_W^2 \left(141 C_A-32n_f T_f-20\right)+10 M_H^2\right)-8 M_H^2\mathcal{L}_{M_H}\right) \right. \\ & \left. -54 M_W^4M_H^2+6 M_W^2 M_H^4 -6 M_H s \left(12 M_W^4-4M_W^2 M_H^2+M_H^4\right) \log (w) \right. \\ & \left. +6 \left(8 M_W^4M_H^2-6 M_W^2 M_H^4+M_H^6\right) \log{(M_W/M_H)}\right\rbrace\left(2 \mathcal{L}_{M_W}-2\mathcal{L}_{Q}+3\right) \end{aligned}$}
  \\[0.75ex] \hline
 {$\chi^{\dagger}\chi$, $i\chi^{\dagger}\overset{\leftrightarrow}{D}_{\mu}\chi$}
  &  
\resizebox{.33\hsize}{!}{ $\begin{aligned}\frac{C_AC_F}{6}\left\lbrace-6\mathcal{L}_{M_W}^2+12\mathcal{L}_{M_W}\mathcal{L}_Q-24\mathcal{L}_{M_W}+21-5\pi^2\right\rbrace\end{aligned}$}
 & 
\resizebox{.42\hsize}{!}{ $\begin{aligned}[t] -\frac{C_A C_F}{36
   M_W^6}&\left\lbrace3 M_W^4 \left(\mathcal{L}_{M_W} \left(M_W^2 \left(141 C_A-32 n_f
   T_f-20\right)+10 M_H^2\right) -8 M_H^2 \mathcal{L}_{M_H}\right) \right. \\ & \left. +M_W^6
   \left(\left(99 \sqrt{3} \pi -690\right) C_A  +32 (5+3 i \pi )
   n_f T_f+124\right) +6 M_H^4 M_W^2 \right. \\ & \left.  -54
   M_H^2 M_W^4 -  6 M_H s \left(M_H^4-4
   M_H^2 M_W^2+12 M_W^4\right) \log (w) \right. \\ & \left. +6
   \left(M_H^6-6 M_H^4 M_W^2+8 M_H^2
   M_W^4\right) \log{(M_W/M_H)}\right\rbrace\left(\mathcal{L}_{M_W}-\mathcal{L}_{Q}+2\right) \end{aligned}$}
\\ [0.75ex] \hline
 $\bar{\psi}\chi$, $\chi^{\dagger}\psi$  &  
\resizebox{.33\hsize}{!}{ $\frac{C_AC_F}{6}\left\lbrace-6\mathcal{L}_{M_W}^2+12\mathcal{L}_{M_W}\mathcal{L}_Q-21\mathcal{L}_{M_W}+24-5\pi^2\right\rbrace$}
 & 
\resizebox{.42\hsize}{!}{ $\begin{aligned}[t] -\frac{C_A C_F}{144
   M_W^6}&\left\lbrace3 M_W^4 \left(\mathcal{L}_{M_W} \left(M_W^2 \left(141 C_A-32 n_f
   T_f-20\right)+10 M_H^2\right)-8 M_H^2\mathcal{L}_{M_H}\right) \right. \\ & \left. +M_W^6
   \left(\left(99 \sqrt{3} \pi -690\right) C_A+32 (5+3 i \pi )
   n_f T_f+124\right)+6 M_H^4 M_W^2 \right. \\ & \left. -54
   M_H^2 M_W^4-6 M_H s \left(M_H^4-4
   M_H^2 M_W^2+12 M_W^4\right) \log (w) \right. \\ & \left. +6
   \left(M_H^6-6 M_H^4 M_W^2+8 M_H^2
   M_W^4\right) \log{(M_W/M_H)}\right\rbrace\left(4 \mathcal{L}_{M_W}-4 \mathcal{L}_{Q}+7\right) \end{aligned}$}
 \\ [1ex] 
 \hline
\end{tabular}
\caption{SCET contribution to the Sudakov form-factor at $\mu\sim M$, one-loop matching coefficient, $D^{(1)}(\mu)$, two-loop mass and coupling renormalisation correction, $\Delta D^{(2)}(\mu)$.  }
\label{tab:expd}
\end{table}
Furthermore, we note that both in the massive and massless external particle cases of SCET, their is no Higgs contributions in the vertex corrections. This is because the fermion Yukawa vertex vanishes, as by construction, \eqref{eqn:0maker} implies that,
\begin{equation}
    \bar{\xi}_{n,p}{\xi}_{n,p}=\bar{\xi}_{n,p}\frac{\slashed{\bar{n}}\slashed{n}}{4}\frac{\slashed{n}\slashed{\bar{n}}}{4}\xi_{n,p}=0,
\end{equation}
using the identity, $\slashed{n}\slashed{n}=n^2=0$. Moreover the tri-scalar couplings have dimension of mass and Higgs exchange corrections to the scalar operators are suppressed by powers of $Y_s/Q$, which is sub-leading in SCET power counting and we drop such terms. This is easily seen when using the re-scaled fields, $\phi_{n,p}$, which have a propagator of identical form to those of fermions. Then the Yukawa coupling becomes,
\begin{align}
    Y_sH\chi^{\dagger}\chi\rightarrow Y_sH\Phi_{n,p}^{\dagger}\Phi_{n,p}=\frac{Y_s}{\bar{n}\cdot p}H\phi_{n,p}^{\dagger}\phi_{n,p},
\end{align}
which is $\mathcal{O}(1/Q)$ as $\bar{n}\cdot p$ is of order $\mathcal{O}(Q)$ which suppresses any graph at each tri-scalar coupling. Thus, the only scalar graphs which appear are the matching at $Q$, which are full theory graphs as well as scalar contributions to the wave-function renormalisation in the effective theories. 
\subsection{Massive External Particles}
\label{ssec:mesud}
Thus far our results have been computed for external particles with masses, $m_{1,2}$, much smaller than the bosonic masses. In this section, we consider the Sudakov regime for massive external particles, extending previous results. Thus, we are primarily interested in the limits, $Q\gg m_{1,2} \gg M$, although we will discuss other cases that can be studied as well, in particular one that can be applied for LHC studies of the top quark.

There are two cases to consider, $Q\gg m_2 \gg  m_1 \gg M$ and $Q\gg m_2 \sim  m_1 \gg M$, we begin with the former. Again, the Sudakov form-factor can be computed using a sequence of effective field theories \cite{leibovich2003comment}. One begins as in the massless external particle case by matching the full theory onto SCET with a single massive particle at the scale, $\mu\sim Q$. The same operators are matched to as in \eqref{eqn:matchoc}, except now the $n$-collinear SCET field, $\xi_{n,p_2}$, is taken to have mass, $m_2$. Again, this matching is independent of IR scales much smaller than $Q$, and thus is given by $\exp{[C(\mu)]}$, as presented in table \ref{tab:expc}. The next step is to run the operator from the scale $Q$ to $m_2$, which can be done with the anomalous dimension, $\gamma_1$, given in \ref{tab:expc}, as the anomalous dimension is also independent of the lower mass scales. The matching steps that follow lie at scales $\mu=m_2$, $\mu=m_1$ and $\mu=M$. Schematically, the matching and running steps can be illustrated as follows,
\small{
\begin{equation*}
        e^C\tilde{\mathcal{O}}_1 \xleftrightarrow[m_1,M=0]{\mu\sim m_2} e^{C+R}\tilde{\mathcal{O}}_2
        \xrightarrow[]{\phantom{1}\gamma_2\phantom{1}} e^{C+R} \tilde{\mathcal{O}}_2\xleftrightarrow[M=0]{\mu\sim m_1} e^{C+R+T} \tilde{\mathcal{O}}_3 \xrightarrow[]{\phantom{1}\gamma_3\phantom{1}}   e^{C+R+T} \tilde{\mathcal{O}}_3\xleftrightarrow[M\neq 0]{\mu\sim M}e^{C+R+T+U} \tilde{\mathcal{O}}_4,
\end{equation*}  }
where the exponents are multiplicative matching coefficients, $\gamma_{i}$ the effective theory anomalous dimensions and $\tilde{\mathcal{O}}_{i}$ the effective theory operators at each scale. 

Firstly, for the matching step at $\mu=m_2$, one switches from SCET to a new EFT with the massive particle described by a heavy field \cite{manohar2007heavy}, $h_{f,s}$, with a velocity, $v_2$, such that $v_2^2=1$. Whereas, the other particle remaining massless continues to be described by the $\bar{n}$-collinear SCET field, $\xi_{\bar{n},p_1}$. The fermionic operators, for instance, are then given by $\bar{h}_{f,1}\Gamma W^{\dagger}_{\bar{n}}\xi_{\bar{n},p_1}$, and similarly for other operators \cite{fleming2008jets}. The matching correction at $\mu=m_2$ is then given by the difference between the vertex graphs in figures \ref{fig:verts1} and \ref{fig:verts2}, for the corresponding external particles in the effective theories above and below $m_2$. More specifically, in the fermion example, the difference between graphs where $\xi_{n,p_2}$ and $h_{f,2}$, for the particle with mass, $m_2$. Note that in the theory below $m_2$ there are no graphs with collinear Wilson lines associated with $h_{f,s}$ and thus such corrections do not appear. The graphs in the theory above $m_2$ are evaluated with bosonic masses set to zero, as $m_2\gg M$, and on-shell at $p_2^2=m_2^2$. Below $m_2$ the graphs in the effective theory are evaluated at $M=0$ as well, at the on-shell point, $k_2\cdot v_2=0$ where $k_2$ is the residual momentum of the heavy particle. 

As for the wave-function graphs, the $\xi_{\bar{n},p1}$ and HQET graphs both vanish on-shell. Hence, the matching is given by the vertex correction and the on-shell wavefunction graph for $\xi_{n,p_2}$, the results of which are discussed in detail in previous work \cite{chiu2008electroweak}, and show in tables \ref{tab:expr} and \ref{table:wfr}, respectively.
\begin{table}[tb]
\footnotesize
\centering
\scalebox{.77}{
\begin{tabular}{|c|c|c|c|c|} 
 \hline
 $\mathcal{O}$ & $R^{(1)}(\mu)$ & $\gamma_2^{(1)}(\mu)$ & $T^{(1)}(\mu)$ & $\gamma_3^{(1)}(\mu)$\\ [0.75ex]
 \hline\hline
 $\bar{\psi}_2\Gamma\psi_1$ & 
\resizebox{.29\hsize}{!}{ $\begin{aligned}[t]\frac{C_AC_F}{2}\left\lbrace\mathcal{L}_{m_2}^2-\mathcal{L}_{m_2}+\frac{\pi^2}{6}+4\right\rbrace\end{aligned}$}
 & 
\resizebox{.23\hsize}{!}{ $\begin{aligned}[t] C_AC_F\left\lbrace4\mathcal{L}_{Q}-2\mathcal{L}_{m_2}-5\right\rbrace\end{aligned}$}
 & 
\resizebox{.29\hsize}{!}{ $\begin{aligned}[t]\frac{C_AC_F}{2}\left\lbrace\mathcal{L}_{m_1}^2-\mathcal{L}_{m_1}+\frac{\pi^2}{6}+4\right\rbrace\end{aligned}$}
 & 
\resizebox{.19\hsize}{!}{ $\begin{aligned}[t]4C_AC_F\left\lbrace w h(w)-1\right\rbrace\end{aligned}$}
  \\[0.75ex]  
 $\chi^{\dagger}_2\chi_1$, $i\chi^{\dagger}_2\overset{\leftrightarrow}{D}_{\mu}\chi_1$  &   
\resizebox{.29\hsize}{!}{ $\begin{aligned}[t]\frac{C_AC_F}{2}\left\lbrace\mathcal{L}_{m_2}^2-2\mathcal{L}_{m_2}+\frac{\pi^2}{6}+4\right\rbrace\end{aligned}$}
 & 
\resizebox{.23\hsize}{!}{ $\begin{aligned}[t] C_AC_F\left\lbrace4\mathcal{L}_{Q}-2\mathcal{L}_{m_2}-6\right\rbrace\end{aligned}$}
 & 
\resizebox{.29\hsize}{!}{ $\begin{aligned}[t]\frac{C_AC_F}{2}\left\lbrace\mathcal{L}_{m_1}^2-2\mathcal{L}_{m_1}+\frac{\pi^2}{6}+4\right\rbrace\end{aligned}$}
 & 
\resizebox{.19\hsize}{!}{ $\begin{aligned}[t]4C_AC_F\left\lbrace w h(w)-1\right\rbrace\end{aligned}$}
  \\[0.75ex] 
 $\bar{\psi}_2\chi_1$  & 
\resizebox{.29\hsize}{!}{ $\begin{aligned}[t]\frac{C_AC_F}{2}\left\lbrace\mathcal{L}_{m_2}^2-\mathcal{L}_{m_2}+\frac{\pi^2}{6}+4\right\rbrace\end{aligned}$}
 & 
\resizebox{.23\hsize}{!}{ $\begin{aligned}[t] C_AC_F\left\lbrace4\mathcal{L}_{Q}-2\mathcal{L}_{m_2}-6\right\rbrace\end{aligned}$}
 & 
\resizebox{.29\hsize}{!}{ $\begin{aligned}[t]\frac{C_AC_F}{2}\left\lbrace\mathcal{L}_{m_1}^2-2\mathcal{L}_{m_1}+\frac{\pi^2}{6}+4\right\rbrace\end{aligned}$}
 & 
\resizebox{.19\hsize}{!}{ $\begin{aligned}[t]4C_AC_F\left\lbrace w h(w)-1\right\rbrace\end{aligned}$}
   \\[0.75ex] 
 $\chi^{\dagger}_2\psi_1$  &  
\resizebox{.29\hsize}{!}{ $\begin{aligned}[t]\frac{C_AC_F}{2}\left\lbrace\mathcal{L}_{m_2}^2-2\mathcal{L}_{m_2}+\frac{\pi^2}{6}+4\right\rbrace\end{aligned}$}
 & 
\resizebox{.23\hsize}{!}{ $\begin{aligned}[t] C_AC_F\left\lbrace4\mathcal{L}_{Q}-2\mathcal{L}_{m_2}-5\right\rbrace\end{aligned}$}
 & 
\resizebox{.29\hsize}{!}{ $\begin{aligned}[t]\frac{C_AC_F}{2}\left\lbrace\mathcal{L}_{m_1}^2-\mathcal{L}_{m_1}+\frac{\pi^2}{6}+4\right\rbrace\end{aligned}$}
 & 
\resizebox{.19\hsize}{!}{ $\begin{aligned}[t]4C_AC_F\left\lbrace w h(w)-1\right\rbrace\end{aligned}$}
    \\[1ex] 
 \hline
\end{tabular}}
\caption{Matching and running results for $Q\gg m_2 \gg m_1 \gg M$. $R$ and $T$ are the matching coefficients at $\mu\sim m_2$ and $\mu\sim m_1$, $\gamma_2$ is the anomalous dimension between $m_2$ and $m_1$, $\gamma_3$ is the anomalous dimension between $m_1$ and $M$. $R$ and $T$ only depend on whether the light particle is a fermion or scalar.  }
\label{tab:expr}
\end{table}
We proceed then with next matching step with coefficient, $\exp{[T(\mu)]}$, at the scale of the lower particle mass, $\mu\sim m_1$. At this scale, the theory above $m_1$ is SCET with heavy field for particle with mass, $m_2$, and the theory below $m_1$, the $\bar{n}$-collinear SCET field, $\xi_{\bar{n},p_1}$ is replaced by the heavy field, $h_{f,s}$, with velocity, $v_1$, such that $v_1^2=1$ and $v_1\cdot v_2=w$. The fermionic operators, for example, are then given by $\bar{h}_{f,2}\Gamma{h}_{f,1}$ instead of $\bar{h}_{f,2}\Gamma W^{\dagger}_{\bar{n}}\xi_{\bar{n},p_1}$. 

In the theory below $m_1$ there are no vertex corrections due to collinear Wilson lines, as there are no collinear wilson lines, $W$, associated with heavy fields. The matching contribution is again given by the difference of the vertex and wavefunction graphs in the theories above and below $m_1$, setting all scales less than $m_1$ to zero. Note that in the theory below $m_1$ we have scaleless integrals which vanish trivially and thus, the sole non-zero contributions come from vertex contributions above $m_1$ and the $\bar{n}$-collinear wavefunction graph. Conveniently, these are the same graphs that contribute to the matching condition at $m_2$, so $T$ is given by $R$ with $m_2 \rightarrow m_1$, and is presented in table \ref{tab:expr}.

\begin{table}[tb]
\footnotesize
\centering
\scalebox{0.695}{
\begin{tabular}{|c|c|c|} 
 \hline
 $\mathcal{O}$ & $U^{(1)}(\mu)$ & $U^{(2)}(\mu)$ \\ [0.75ex] 
 \hline\hline
 $\bar{\psi}\Gamma\psi$ &  
\resizebox{.23\hsize}{!}{ $\begin{aligned}[t]&\frac{C_A}{4}\left\lbrace8C_F(w h(w)-1)\mathcal{L}_{M_W} \right. \\& \left. -Y_f^2(h(w)-1)\mathcal{L}_{M_H}\right\rbrace\end{aligned}$}
& \resizebox{.85\hsize}{!}{$\begin{aligned}[t]&(V^{(M)}_1+F^{(M)}_{h}) +\frac{2}{3} w C_A^2 C_F^2 \left(12 (\mathcal{L}_{M_W}-1) \mathcal{L}_{M_W}+\pi ^2\right) h(w)+\frac{1}{24} C_A^2Y_f^4 \left(12 (\mathcal{L}_{M_H}-1) \mathcal{L}_{M_H}+\pi ^2\right) h(w) \\& - \frac{1}{6} C_A^2 C_FY_f^2 h(w) \left(6 \mathcal{L}_{M_W} ((w+1) \mathcal{L}_{M_H}-2)+3 (w+1) \mathcal{L}_{M_W}^2+3 \mathcal{L}_{M_H} ((w+1)\mathcal{L}_{M_H}-4 w)+\pi ^2 (w+1)\right) \end{aligned}$}
 \\[0.75ex]\hline 
 $\chi^{\dagger}\chi$, $i\chi^{\dagger}\overset{\leftrightarrow}{D}_{\mu}\chi$
 &  
\resizebox{.23\hsize}{!}{ $\begin{aligned}[t]&\frac{C_A}{4}\left\lbrace8C_F(w h(w)-1)\mathcal{L}_{M_W}  \right. \\& \left. -Y_s^2(h(w)-1)\mathcal{L}_{M_H}\right\rbrace \end{aligned}$} 
 & \resizebox{.85\hsize}{!}{$\begin{aligned}[t]& (V^{(M)}_2+F^{(M)}_{h})+\frac{2}{3} w C_A^2 C_F^2 \left(12 (\mathcal{L}_{M_W}-1) \mathcal{L}_{M_W}+\pi ^2\right) h(w)+\frac{1}{24} C_A^2Y_s^4 \left(12 (\mathcal{L}_{M_H}-1) \mathcal{L}_{M_H}+\pi ^2\right) h(w) \\&  -\frac{1}{6} C_A^2 C_FY_s^2 h(w) \left(6 \mathcal{L}_{M_W} ((w+1) \mathcal{L}_{M_H}-2)+3 (w+1) \mathcal{L}_{M_W}^2+3 \mathcal{L}_{M_H} ((w+1)\mathcal{L}_{M_H}-4 w)+\pi ^2 (w+1)\right)\end{aligned}$}\\[0.75ex]\hline
 $\chi^{\dagger}\psi$, $\bar{\psi}\chi$ &  
\resizebox{.31\hsize}{!}{ $\begin{aligned}[t]&\frac{C_A}{4}\left\lbrace 8C_F(w h(w)-1)\mathcal{L}_{M_W} \right. \\& \left. -\left(Y_fY_sh(w)-\frac{1}{2}(Y_f^2+Y_s^2)\right)\mathcal{L}_{M_H}\right\rbrace\end{aligned}$}
 &\resizebox{.92\hsize}{!}{$\begin{aligned}[t]& (V^{(M)}_3+F^{(M)}_{h})+\frac{2}{3} w C_A^2 C_F^2 \left(12 (\mathcal{L}_{M_W}-1) \mathcal{L}_{M_W}+\pi ^2\right) h(w)+\frac{1}{48} C_A^2 (Y_f^3Y_s+Y_s^3Y_f)\left(12 (\mathcal{L}_{M_H}-1) \mathcal{L}_{M_H}+\pi ^2\right) \\&  -\frac{1}{6} C_A^2 C_FY_fY_s h(w) \left(3 \left(2 \mathcal{L}_{M_W} (\mathcal{L}_{M_H}-2)+\mathcal{L}_{M_W}^2+\mathcal{L}_{M_H}^2\right)+\pi
   ^2\right) -\frac{1}{12} w C_A^2 C_F(Y_f^2+Y_s^2) h(w) \left(3 (\mathcal{L}_{M_W}+\mathcal{L}_{M_H})^2 \right.\\&\left. -12 \mathcal{L}_{M_H}+\pi ^2\right)  h(w)\end{aligned}$}\\ [1ex] 
 \hline
\end{tabular}}
\caption{One and two-loop matching contribution, $U^{(1,2)}$, at $\mu \sim M$. $V^{(M)}_i$ and $F_I$ are two loop vertex and wave-function corrections, given in Appendices \ref{sec:vertcorr} and \ref{sec:wfr2}. }
\label{tab:exp2u}
\end{table}
The final contributions needed are the anomalous dimension, $\gamma_3$, for the running between $m_1$ and $M$, and the matching condition, $\exp{[U(\mu)]}$, at $M$. These can be computed from the HPET graphs, evaluated on-shell, but now with bosonic masses, $M$, included as they are no longer IR. The one-loop contributions are presented in the last column of table \ref{tab:expr}, which by inspection are independent of the composite operator. Whence, they are identical and depend solely on whether the external particles are fermions or scalars, and their dependence lies in the appropriate Yukawa factors. The function, 
\begin{equation}
    h(w)=\frac{\log{(w+\sqrt{w^2-1})}}{\sqrt{w^2-1}},
    \label{eqn:romega}
\end{equation}
is the well-known factor which occurs in the velocity-dependent anomalous dimension in HQET \cite{manohar2007heavy}. Note further that in the Sudakov regime, the Higgs contribution in $\exp{U}$ is sub-leading as, $Q^2\sim m_1m_2 w$, and in this limit,
\begin{equation}
    h(w)\sim \frac{\log{w}}{w},
\end{equation}
thus the gauge contribution dominates in the Sudakov regime. We will see later on that in the threshold regime, the Higgs and gauge contributions turn out to be on equal footing. We also present the two-loop contribution to the matching contribution, $\exp{[U(\mu)]}$, in table \ref{tab:exp2u} which combines the vertex and wave-function contribution listed in table \ref{table:wfr}. 

As for remaining two-loop contributions, we present the mass and coupling renormalisation, which contribute at two-loop order for each matching coefficient in table \ref{tab:corrrtu}.
\begin{table}[b]
\footnotesize
\centering
\scalebox{0.79}{
\begin{tabular}{|c|c|c|c|} 
 \hline
 $\mathcal{O}$ & $\Delta R^{(2)}$ & $\Delta T^{(2)}$ & $\Delta U^{(2)}$ \\ [0.75ex]
 \hline\hline
 $\bar{\psi}_2\Gamma\psi_1$ & 
\resizebox{.47\hsize}{!}{ $\begin{aligned}[t]\frac{C_A^2C_F}{2}\left\lbrace C_F\left(1-2 \mathcal{L}_{m_2}\right)\left(4-3\mathcal{L}_{m_2}\right)-\frac{Y_f^2}{8}\left(1-2 \mathcal{L}_{m_2}\right)\left(7-3\mathcal{L}_{m_2}\right)\right\rbrace\end{aligned}$}
 & 
\resizebox{.47\hsize}{!}{ $\begin{aligned}[t]\frac{C_A^2C_F}{2}\left\lbrace C_F\left(1-2 \mathcal{L}_{m_1}\right)\left(4-3\mathcal{L}_{m_1}\right)-\frac{Y_f^2}{8}\left(1-2 \mathcal{L}_{m_1}\right)\left(7-3\mathcal{L}_{m_1}\right)\right\rbrace\end{aligned}$}
 &  
\resizebox{.065\hsize}{!}{ $\begin{aligned}[t] \Delta U^{(2)}_1 \end{aligned}$}
  \\[0.75ex]  \hline
 $\chi^{\dagger}_2\chi_1$, $i\chi^{\dagger}_2\overset{\leftrightarrow}{D}_{\mu}\chi_1$  &   
\resizebox{.47\hsize}{!}{ $\begin{aligned}[t]C_A^2C_F\left\lbrace C_F\left(1- \mathcal{L}_{m_2}\right)\left(7-3\mathcal{L}_{m_2}\right)-\frac{Y_s^2}{4m_2^2}\left(1-2 \mathcal{L}_{m_2}\right)\left(2-\mathcal{L}_{m_2}\right)\right\rbrace\end{aligned}$}
 & 
\resizebox{.47\hsize}{!}{ $\begin{aligned}[t]C_A^2C_F\left\lbrace C_F\left(1- \mathcal{L}_{m_1}\right)\left(7-3\mathcal{L}_{m_1}\right)-\frac{Y_s^2}{4m_1^2}\left(1-2 \mathcal{L}_{m_1}\right)\left(2-\mathcal{L}_{m_1}\right)\right\rbrace\end{aligned}$}
 &  
\resizebox{.065\hsize}{!}{ $\begin{aligned}[t] \Delta U^{(2)}_2 \end{aligned}$}
 \\[0.75ex] \hline
 $\bar{\psi}_2\chi_1$  & 
\resizebox{.47\hsize}{!}{ $\begin{aligned}[t]\frac{C_A^2C_F}{2}\left\lbrace C_F\left(1-2 \mathcal{L}_{m_2}\right)\left(4-3\mathcal{L}_{m_2}\right)-\frac{Y_f^2}{8}\left(1-2 \mathcal{L}_{m_2}\right)\left(7-3\mathcal{L}_{m_2}\right)\right\rbrace\end{aligned}$}
& 
\resizebox{.47\hsize}{!}{ $\begin{aligned}[t]C_A^2C_F\left\lbrace C_F\left(1- \mathcal{L}_{m_1}\right)\left(7-3\mathcal{L}_{m_1}\right)-\frac{Y_s^2}{4m_1^2}\left(1-2 \mathcal{L}_{m_1}\right)\left(2-\mathcal{L}_{m_1}\right)\right\rbrace\end{aligned}$}
 &  
\resizebox{.065\hsize}{!}{ $\begin{aligned}[t] \Delta U^{(2)}_3 \end{aligned}$}
 \\[0.75ex] \hline
 $\chi^{\dagger}_2\psi_1$  & 
\resizebox{.47\hsize}{!}{ $\begin{aligned}[t]C_A^2C_F\left\lbrace C_F\left(1- \mathcal{L}_{m_2}\right)\left(7-3\mathcal{L}_{m_2}\right)-\frac{Y_s^2}{4m_2^2}\left(1-2 \mathcal{L}_{m_2}\right)\left(2-\mathcal{L}_{m_2}\right)\right\rbrace\end{aligned}$}
 & 
\resizebox{.47\hsize}{!}{ $\begin{aligned}[t]\frac{C_A^2C_F}{2}\left\lbrace C_F\left(1-2 \mathcal{L}_{m_1}\right)\left(4-3\mathcal{L}_{m_1}\right)-\frac{Y_f^2}{8}\left(1-2 \mathcal{L}_{m_1}\right)\left(7-3\mathcal{L}_{m_1}\right)\right\rbrace\end{aligned}$}
 & 
\resizebox{.065\hsize}{!}{ $\begin{aligned}[t] \Delta U^{(2)}_3 \end{aligned}$}
   \\[1ex] 
 \hline
\end{tabular}}
\caption{Matching and contribution due to mass and coupling renormalisation for $Q\gg m_2 \gg m_1 \gg M$. $\Delta R^{(2)}$, $\Delta T^{(2)}$ and $\Delta U^{(2)}$ are the two-loop order matching coefficients at $\mu\sim m_2$, $\mu\sim m_1$ and $\mu\sim M$. The contributions, $\Delta U^{(2)}_i$, are presented in Appendix \ref{sec:massren}. }
\label{tab:corrrtu}
\end{table}
The situation is similar in the case $Q\gg m_2\sim m_1 \gg M$, which is why we left this for last. The evolution down to the scale $m_1\sim m_2$ is the same as for the case where $m_i=0$. The $n$ and $\bar{n}$ collinear graphs at the scale $m_1\sim m_2$ are independent of each other, so the matching is imply given by the sum of $R$ and $T$ at $m_2$ and $m_1$ respectively. Below $m_1\sim m_2$ the matching and running is identical to the previous case with anomalous dimension, $\gamma_3$, and matching coefficient, $\exp{[U(\mu)]}$. Lastly, if $m_2=m_1$, then the case is identical to $m_2\sim m_1$, except one sets $m_2=m_1$ in all matching and running contributions.

\paragraph{Further Cases:} We note, as considered in previous work \cite{chiu2008electroweak}, that there are other cases one can compare for complete generality, in particular, one case resonates with regard to top quark physics in the high energy regime. The Sudakov limit being, $Q\gg m_1\sim m_2 \sim M$, which involves one running step with $\gamma_1$, as the running from $Q$ is independent of the IR scales, and two matching steps. The matching at $\mu\sim Q$ is represented by the usual, $\exp{[C(\mu)]}$. On the other hand, the matching at $\mu\sim m_{1,2}\sim M$ is the same as for the massless case, except the matching condition, $\exp{[D(\mu)]}$, now involves massive collinear propagators, which modifies the matching in the following way,
\begin{equation}
    D(m_1,m_2)=D(m_{1,2}=0)+(f_2(z_2)-\tilde{f}_2(z_2)/2)+(f_1(z_1)-\tilde{f}_1(z_1)/2),
\end{equation}
where $z_i=m_i/M_W$, $f_{1,2}$ corresponds to the massive collinear contributions,
\begin{equation}
f_i(z_i)\equiv I_{n}(m_i)-I_{n}(0),
\end{equation}
where $I_{n}$ is the collinear vertex contribution and,
\begin{equation}
\tilde{f}_i(z_i)\equiv\delta Z_i(m_i,M)-\delta Z_i(0,M),
\end{equation}
is the difference between the wave-function contribution with all mass scales non-zero and the external mass scales set to zero from table \ref{table:wfr}. Both the vertex and wave-function contributions depend solely on whether the corresponding particle is a fermion or a scalar. More specifically, 
${f}_i(z_i)$ maps to $f_F(z_i)$ and $f_S(z_i)$ in the case of fermions and scalars, respectively, and are given by,
\begin{subequations}
\begin{align}
    &f_F(z)=2+\left(\frac{1}{z^2}-2\right)\log{z^2}+\frac{2\sqrt{1-4z^2}}{z^2}\tanh^{-1}{\sqrt{1-4z^2}}+\frac{1}{2}\log^2{(z^2)}-2(\tanh^{-1}{\sqrt{1-4z^2}})^2
\\&
    f_S(z)=1-\left(1-\frac{1}{2z^2}\right)\log{z^2}+\frac{\sqrt{1-4z^2}}{z^2}\tanh^{-1}{\sqrt{1-4z^2}}+\frac{1}{2}\log^2{(z^2)}-2(\tanh^{-1}{\sqrt{1-4z^2}})^2
\end{align}
\end{subequations}
as was also found in \cite{chiu2008electroweak}. Thus, now that we have considered cases of interest in the Sudakov limit, we can shift to studying counterparts in the threshold limit. 

\section{Radiative Corrections in Threshold Limit}
\label{sec:methre}
In this section, we calculate the form factor, $\log{F_E(Q^2)}$, in the opposite limit, i.e. small $Q^2$ and large $m^2$, or threshold regime. Evidently, at threshold, the masses of the external particles are then taken to be the largest scale, resulting in two cases to consider, $m_1\sim m_2 \gg M \gg Q$ and $m_2 \gg m_1 \gg M \gg Q$. These cases have not been studied previously, and we provide the form factor up to and including two-loop order, which is computed using a sequence of effective field theories.

We begin by noting that at scales higher than $m^2$, the theory is the original Higgs-gauge theory, or so-called full theory. Moving to scales below $m^2$, we transition to HPET where degrees of freedom of off-shellness on the order $m^2$ are integrated out. More specifically, let us commence with the simpler case, $m_1\sim m_2\gg M^2 \gg Q^2$, where $m_{1,2}$ and $M$ denote the external particle and bosonic masses, respectively. Schematically, we then have the following matching and running steps, illustrated as follows,
\begin{equation*}
    \mathcal{O}\xleftrightarrow[Q,M=0]{\mu\sim m_{1,2}} e^B \tilde{\mathcal{O}}_1\xrightarrow[]{\phantom{11}\gamma_3\phantom{11}}   e^B\tilde{\mathcal{O}}_1 \xleftrightarrow[Q=0]{\mu\sim M}e^{B+U} \tilde{\mathcal{O}}_2,
\end{equation*}
where $B$ and $U$ are multiplicative matching coefficients, $\gamma_3$, is the effective theory anomalous dimension, and $\tilde{\mathcal{O}}_{1,2}$ the effective theory operators at each scale. At the scale $\mu>m_{1,2}$, we employ the full theory graphs and below, at $\mu<m_{1,2}$, we match down to HPET with matching coefficient, $b(\mu)$, and RGE given by,
\begin{equation}
    \mu\frac{d b(\mu)}{d \mu}=\gamma_F(a(\mu))b(\mu),
\end{equation}
where $\gamma_F$ is the full theory anomalous dimension for operator, $\mathcal{O}$, and is independent of energetic regime as given in table \ref{tab:expc}. The full theory is then matched onto HPET at $\mu\sim m_{1,2}$. The matching coefficient then depends on logarithms, $\mathcal{L}_{m_{1,2}}$, which are not divergent if $\mu\sim m_{1,2}$. The matching is done between full and effective theory operators as follows,
\begin{subequations}
\begin{align}
    &\bar{\psi}_2\Gamma\psi_1\rightarrow e^B \bar{h}_{f,2}\Gamma h_{f,1}, \\ &
    \chi^{\dagger}_2\chi_1\rightarrow e^B {h}^{\dagger}_{s,2} h_{s,1},\\ &
    i\chi^{\dagger}_2\overset{\leftrightarrow}{D}_{\mu}\chi_1\rightarrow e^B {h}^{\dagger}_{s,2}[v_1+v_2]_{\mu} h_{s,1},\\ &
    \bar{\psi}_2\chi_1\rightarrow e^B \bar{h}_{f,2} h_{s,1},\quad{\chi}^{\dagger}_2\psi_1\rightarrow e^B{h}^{\dagger}_{s,2} h_{s,1}.
\end{align}
\label{eqn:matchob}
\end{subequations}
We can then calculate the matching coefficient, $\exp{[B(\mu)]}$, as the full theory vertex and wave-function corrections with IR scales, $M$ and $Q$, set to zero. The results of which at one and two loop order are given in table \ref{tab:expb}. Note that for the two-loop results, since $m_1\sim m_2$ and we want to evaluate the master integrals analytically, this can only be achieved with master integrals at a single scale, whence, we expand the two-loop contributions about the difference of $m_1$ and $m_2$ to first order. This is an accurate representation as the scale we are considering is where $m_1\sim m_2$ and although we chose to expand to first order as is conventionally done one can expand to any order and perform the single-scale two-loop master integrals as they are independent of expansion order. As for the remaining two-loop contributions, we present the mass and coupling renormalisation contributions at two-loop order in table \ref{tab:expb}.

\begin{table}[tb]
\footnotesize
\centering
\scalebox{0.61}{
\begin{tabular}{|c|c|c|c|} 
 \hline
 $\mathcal{O}$ & $B^{(1)}(\mu)$ & $B^{(2)}(\mu)$ & $\Delta B^{(2)}(\mu)$  \\ [0.75ex] 
 \hline\hline
 $\bar{\psi}_2\psi_1$ & 
\resizebox{.52\hsize}{!}{ $\begin{aligned}[t]&\frac{C_A C_F \mathcal{L}_{m_2}^2 \left(m_1^2+m_2^2\right)}{2 m_- m_+}  -\frac{C_A C_F\mathcal{L}_{m_2} \left(m_1^2-4 m_1 m_2-m_2^2\right)}{2 m_-m_+} \\& -\frac{C_A C_F \mathcal{L}_{m_1} \left(m_1^2 \mathcal{L}_{m_1}+m_2^2\mathcal{L}_{m_1}+m_1^2+4 m_1 m_2-m_2^2\right)}{2 m_- m_+} \\&    -\frac{Y_f^2 C_A \left(5 m_1^2 \mathcal{L}_{m_1}-9 m_2^2 \mathcal{L}_{m_1}-4 m_1 m_2\mathcal{L}_{m_1}^2+8 m_1 m_2 \mathcal{L}_{m_1}-6 m_1^2+6 m_2^2\right)}{16 m_-m_+} \\& -\frac{Y_f^2 C_A \mathcal{L}_{m_2} \left(9 m_1^2-8 m_1 m_2-5m_2^2\right)}{16 m_- m_+}  -\frac{Y_f^2 m_1 m_2 C_A \mathcal{L}_{m_2}^2}{4m_- m_+}\end{aligned}$}
 & 
\resizebox{.7\hsize}{!}{ $\begin{aligned}[t]&(V^{(m_{1,2})}_1+\frac{1}{2}F^{(m_2)}_{\psi}+\frac{1}{2}F^{(m_1)}_{\psi})+\\&\frac{C_A^2 C_F^2 [60 m_1^2 \mathcal{L}_{m_1}^2-80 m_1^2 \mathcal{L}_{m_1}-12 \mathcal{L}_{m_1}^2+28\mathcal{L}_{m_1}+5 \pi ^2 m_1^2+104 m_1^2-\pi ^2-36]}{2 m_1^2}+\\& \frac{C_A^2 C_F^2[12 m_1^2 \mathcal{L}_{m_1}^2+176 m_1^2 \mathcal{L}_{m_1}+60 \mathcal{L}_{m_1}^2-188 \mathcal{L}_{m_1}+\pi^2 m_1^2-64 m_1^2+5 \pi ^2+236] (m_2-m_1)}{4 m_1^3}-\\& 
\frac{C_A^2 C_FY_f^2 [228 m_1^2 \mathcal{L}_{m_1}^2-292 m_1^2 \mathcal{L}_{m_1}-60\mathcal{L}_{m_1}^2+128\mathcal{L}_{m_1}+19 \pi ^2 m_1^2+386 m_1^2-5 \pi^2-160]}{16 m_1^2}- \\& \frac{C_A^2 C_FY_f^2[12 m_1^2 \mathcal{L}_{m_1}^2+764 m_1^2 \mathcal{L}_{m_1}+252 \mathcal{L}_{m_1}^2-768\mathcal{L}_{m_1}+ \pi ^2 m_1^2-366 m_1^2+21 \pi^2+912] (m_2-m_1)}{32m_1^3} + \\&
\frac{C_A^2 Y_f^4 [72 m_1^2 \mathcal{L}_{m_1}^2-144 m_1^2 \mathcal{L}_{m_1}-36 \mathcal{L}_{m_1}^2+72\mathcal{L}_{m_1}+6 \pi ^2 m_1^2+176 m_1^2-3 \pi ^2-88]}{64 m_1^2}- \\& \frac{C_A^2 Y_f^4[36 m_1^2 \mathcal{L}_{m_1}^2-324 m_1^2 \mathcal{L}_{m_1}-108 \mathcal{L}_{m_1}^2+360 \mathcal{L}_{m_1}+3\pi ^2 m_1^2+322 m_1^2-9 \pi ^2-408] (m_2-m_1)}{128 m_1^3}  \end{aligned}$}
 & $\Delta B^{(2)}_1$ \\[0.75ex]\hline
 $\bar{\psi}_2\gamma^{\mu}\psi_1$ &  
\resizebox{.52\hsize}{!}{$\begin{aligned}[t]& -\frac{C_A C_F \mathcal{L}_{m_2} \left(m_1^2-4 m_1 m_2+5 m_2^2\right)}{2 m_-m_+} \\& -\frac{C_A C_F}{2 m_- m_+} \left\lbrace M_1^2 \mathcal{L}_{m_1}^2-5 m_1^2 \mathcal{L}_{m_1}+m_2^2\mathcal{L}_{m_1}^2-m_2^2 \mathcal{L}_{m_1}-2 m_1 m_2 \mathcal{L}_{m_1}^2 \right.\\&\left.+4 m_1 m_2\mathcal{L}_{m_1}+6 m_1^2-6 m_2^2\right\rbrace \\& +\frac{m_- C_A C_F\mathcal{L}_{m_2}^2}{2 m_+}  -\frac{Y_f^2C_A \mathcal{L}_{m_2} \left(9 m_1^3+9 m_1^2 m_2-3 m_1 m_2^2-11m_2^3\right)}{16 m_- m_+^2} \\& -\frac{Y_f^2 C_A}{16 m_- m_+^2} \left\lbrace11 m_1^3 \mathcal{L}_{m_1}+11 m_1^2 m_2 \mathcal{L}_{m_1}-9m_2^3 \mathcal{L}_{m_1}-17 m_1 m_2^2 \mathcal{L}_{m_1}  -18 m_1^3 \right.\\&\left. -26 m_1^2m_2 +18 m_1 m_2^2+26 m_2^3\right)] \end{aligned}$}
 & 
\resizebox{.76\hsize}{!}{  $\begin{aligned}[t]&(V^{(m_{1,2})}_2+\frac{1}{2}F^{(m_2)}_{\psi}+\frac{1}{2}F^{(m_1)}_{\psi}) -\frac{C_A^2 C_F^2 \left(12 m_1^2 \mathcal{L}_{m_1}^2-28 m_1^2 \mathcal{L}_{m_1}+12 \mathcal{L}_{m_1}^2-16\mathcal{L}_{m_1}+\pi ^2 m_1^2+30 m_1^2+\pi ^2+16\right)}{4 m_1(m_2-m_1)} \\&- \frac{C_A^2 C_F^2 \left(36 m_1^2 \mathcal{L}_{m_1}^2+36 m_1^2\mathcal{L}_{m_1}+36 \mathcal{L}_{m_1}^2+3 \pi ^2 m_1^2-74 m_1^2+3 \pi ^2+40\right)}{8m_1^2}+ \\& \frac{C_A^2 C_FY_f^2 \left(12 m_1^2 \mathcal{L}_{m_1}^2-16 m_1^2 \mathcal{L}_{m_1}+12 \mathcal{L}_{m_1}^2-4\mathcal{L}_{m_1}+\pi ^2 m_1^2+10 m_1^2+\pi^2+2\right)}{16 m_1(m_2-m_1)}+ \\& \frac{C_A^2 C_FY_f^2 \left(36 m_1^2 \mathcal{L}_{m_1}^2+72 m_1^2\mathcal{L}_{m_1}+36 \mathcal{L}_{m_1}^2+36 \mathcal{L}_{m_1}+3 \pi ^2 m_1^2-74 m_1^2+3 \pi^2+22\right)}{32 m_1^2} -\\& \frac{C_A^2 C_FY_f^2 \left(324 m_1^2 \mathcal{L}_{m_1}^2-1512 m_1^2\mathcal{L}_{m_1}+900 \mathcal{L}_{m_1}^2-1284 \mathcal{L}_{m_1}+27 \pi ^2m_1^2+1278 m_1^2+75 \pi^2+1202\right) (m_2-m_1)}{192 m_1^3} + \\& \frac{C_A^2 Y_f^4 \left(36 m_1^2 \mathcal{L}_{m_1}^2-108 m_1^2 \mathcal{L}_{m_1}+36\mathcal{L}_{m_1}^2-72\mathcal{L}_{m_1}+3 \pi ^2 m_1^2+124 m_1^2+3 \pi ^2+70\right)}{256 m_1(m_2-m_1)}+ \\& \frac{3 C_A^2 \left(36 m_1^2 \mathcal{L}_{m_1}^2+12 m_1^2 \mathcal{L}_{m_1}+36\mathcal{L}_{m_1}^2-24 \mathcal{L}_{m_1}+3 \pi ^2 m_1^2-80m_1^2+3 \pi ^2+46\right)}{512m_1^2} + \\& \frac{C_A^2 \left(324 m_1^2 \mathcal{L}_{m_1}^2+324 m_1^2 \mathcal{L}_{m_1}-252\mathcal{L}_{m_1}^2+840 \mathcal{L}_{m_1}+27 \pi ^2m_1^2-36m_1^2-21 \pi ^2-778\right)(m_2-m_1)}{1024 m_1^3}
     \end{aligned}$}
 & $\Delta B^{(2)}_2$ \\[0.75ex]  \hline
 $\bar{\psi}_2\sigma^{\mu\nu}\psi_1$ & 
\resizebox{.55\hsize}{!}{ $\begin{aligned}[t]&\frac{C_A C_F \mathcal{L}_{m_2}^2 \left(m_1^2+m_2^2\right)}{2 m_- m_+}  -\frac{C_A C_F\mathcal{L}_{m_2} \left(m_1^2-4 m_1 m_2+7 m_2^2\right)}{2 m_-m_+} \\& -\frac{C_A C_F \left(m_1^2 \mathcal{L}_{m_1}^2-7 m_1^2 \mathcal{L}_{m_1}+m_2^2\mathcal{L}_{m_1}^2-m_2^2 \mathcal{L}_{m_1}+4 m_1 m_2 \mathcal{L}_{m_1}+4 m_1^2-4m_2^2\right)}{2 m_- m_+} \\& -\frac{Y_f^2 C_A \left(9 m_1^2 \mathcal{L}_{m_1}-9 m_2^2 \mathcal{L}_{m_1}-4 m_1 m_2\mathcal{L}_{m_1}^2+8 m_1 m_2 \mathcal{L}_{m_1}-14 m_1^2+14 m_2^2\right)}{16 m_-m_+} \\& -\frac{Y_f^2 C_A \mathcal{L}_{m_2} \left(9 m_1^2-8 m_1 m_2-9m_2^2\right)}{16 m_- m_+} -\frac{Y_f^2 m_1 m_2 C_A \mathcal{L}_{m_2}^2}{4m_- m_+}
   \end{aligned}$}
 &
\resizebox{.77\hsize}{!}{ $\begin{aligned}[t]&(V^{(m_{1,2})}_3+\frac{1}{2}F^{(m_2)}_{\psi}+\frac{1}{2}F^{(m_1)}_{\psi})+\frac{C_A^2 C_F^2 \left(12 m_1^2\mathcal{L}_{m_1}^2-58 m_1^2\mathcal{L}_{m_1}-12 \mathcal{L}_{m_1}^2+34 \mathcal{L}_{m_1}+\pi ^2 m_1^2+80 m_1^2-\pi ^2-55\right)}{2m_1^2} \\& -\frac{C_A^2 C_F^2 \left(6 m_1^2 \mathcal{L}_{m_1}+6 \mathcal{L}_{m_1}-10 m_1^2-7\right)}{m_1(m_2-m_1)}+ \frac{C_A^2 C_FY_f^2 \left(12 m_1^2 \mathcal{L}_{m_1}^2-4 m_1^2 \mathcal{L}_{m_1}+12 \mathcal{L}_{m_1}^2+8\mathcal{L}_{m_1}+\pi ^2m_1^2-4 m_1^2+\pi ^2-6\right)}{16 m_1(m_2-m_1)} \\& + \frac{C_A^2 C_F^2 \left(12 m_1^2 \mathcal{L}_{m_1}^2-22 m_1^2 \mathcal{L}_{m_1}+60\mathcal{L}_{m_1}^2-194 \mathcal{L}_{m_1}+\pi ^2 m_1^2-10 m_1^2+5 \pi ^2+263\right)(m_2-m_1)}{4 m_1^3} -\\& \frac{C_A^2 C_FY_f^2\left(84 m_1^2 \mathcal{L}_{m_1}^2-388 m_1^2\mathcal{L}_{m_1}-108 \mathcal{L}_{m_1}^2+216 \mathcal{L}_{m_1}+7 \pi ^2 m_1^2+548 m_1^2-9 \pi^2-342\right)}{32 m_1^2} - \\& \frac{C_A^2 C_FY_f^2 \left(36 m_1^2\mathcal{L}_{m_1}^2+132m_1^2\mathcal{L}_{m_1}+1476\mathcal{L}_{m_1}^2-4392 \mathcal{L}_{m_1}+3 \pi ^2 m_1^2-936 m_1^2+123 \pi^2+5378\right) (m_2-m_1)}{192 m_1^3} -\\& -\frac{C_A^2Y_f^4 \left(36 m_1^2 \mathcal{L}_{m_1}^2-108 m_1^2 \mathcal{L}_{m_1}+36 \mathcal{L}_{m_1}^2-72\mathcal{L}_{m_1}+3 \pi ^2 m_1^2+124 m_1^2+3 \pi ^2+70\right)}{128 m_1(m_2-m_1)}+ \\& \frac{C_A^2Y_f^4 \left(36 m_1^2 \mathcal{L}_{m_1}^2-180 m_1^2 \mathcal{L}_{m_1}-108\mathcal{L}_{m_1}^2+72 \mathcal{L}_{m_1}+3 \pi ^2 m_1^2+376 m_1^2-9 \pi ^2-138\right)}{256m_1^2} + \\& \frac{C_A^2 \left(180 m_1^2 \mathcal{L}_{m_1}^2-396 m_1^2 \mathcal{L}_{m_1}-396\mathcal{L}_{m_1}^2+1128 \mathcal{L}_{m_1}+15 \pi^2 m_1^2+548 m_1^2-33 \pi ^2-1130\right)(m_2-m_1)}{512 m_1^3}
     \end{aligned}$}
 & $\Delta B^{(2)}_3$ \\[0.75ex] \hline
 $\chi^{\dagger}_2\chi_1$  &   
\resizebox{.48\hsize}{!}{ $\begin{aligned}[t]& \frac{C_A C_F \mathcal{L}_{m_2}^2 \left(m_1^2+m_2^2\right)}{2 m_- m_+}-\frac{m_2^2 C_A C_F \mathcal{L}_{m_2}}{m_- m_+}-\frac{Y_s^2 C_A \mathcal{L}_{m_2}}{8 m_2^2}  -\frac{Y_s^2 C_A\mathcal{L}_{m_2}^2}{16 m_- m_+} \\& -\frac{C_A C_F\left(m_1^2 \mathcal{L}_{m_1}^2-2 m_1^2 \mathcal{L}_{m_1}+m_2^2 \mathcal{L}_{m_1}^2-6 m_1^2+6m_2^2\right)}{2 m_- m_+} \\& + \frac{Y_s^2 C_A \left(m_1^2 m_2^2 \mathcal{L}_{m_1}^2-2 m_1^2 m_2^2\mathcal{L}_{m_1}+2 m_2^4 \mathcal{L}_{m_1}+2 m_1^4-2 m_2^4\right)}{16 m_1^2 m_2^2m_- m_+} \end{aligned}$}
 & 
\resizebox{.73\hsize}{!}{ $\begin{aligned}[t]&(V^{(m_{1,2})}_4+\frac{1}{2}F^{(m_2)}_{\chi}+\frac{1}{2}F^{(m_1)}_{\chi})+\frac{C_A^2 C_F^2 \left(24 m_1^2 \mathcal{L}_{m_1}^2-24 m_1^2 \mathcal{L}_{m_1}-12 \mathcal{L}_{m_1}^2+12\mathcal{L}_{m_1}+2 \pi ^2 m_1^2+24 m_1^2-\pi ^2-12\right)}{3 m_1^2}+\\& \frac{2 C_A^2 C_F^2\left(18 m_1^2 \mathcal{L}_{m_1}+12 \mathcal{L}_{m_1}^2-24 \mathcal{L}_{m_1}-6 m_1^2+\pi ^2+18\right)(m_2-m_1)}{3 m_1^3}+ \\& \frac{C_A^2 C_FY_s^2 \left(12 m_1^2 \mathcal{L}_{m_1}^2-36 m_1^2 \mathcal{L}_{m_1}-12 \mathcal{L}_{m_1}^2+32\mathcal{L}_{m_1}+\pi ^2 m_1^2+30 m_1^2-\pi ^2-38\right) (m_2-m_1)}{8m_1^5}- \\& \frac{C_A^2 C_FY_s^2 \left(36 m_1^2 \mathcal{L}_{m_1}^2-48 m_1^2 \mathcal{L}_{m_1}-12\mathcal{L}_{m_1}^2+24 \mathcal{L}_{m_1}+3 \pi ^2 m_1^2+60 m_1^2-\pi ^2-30\right)}{24 m_1^4} +\\& \frac{C_A^2Y_s^4 \left(12 \mathcal{L}_{m_1}^2-12 \mathcal{L}_{m_1}+\pi ^2+12\right)}{192 m_1^4}- \frac{C_A^2Y_s^4\left(12 \mathcal{L}_{m_1}^2-24 \mathcal{L}_{m_1}+\pi ^2+18\right) (m_2-m_1)}{96 m_1^5}
     \end{aligned}$}
 & $\Delta B^{(2)}_4$ \\ [0.75ex] \hline
 $i\chi^{\dagger}_2\overset{\leftrightarrow}{D}_{\mu}\chi_1$  &  
\resizebox{.55\hsize}{!}{ $\begin{aligned}[t] & -\frac{2 m_2^2 C_A C_F \mathcal{L}_{m_2} \left(m_1^2+8 m_2^2\right)}{m_- m_+\left(m_1^2+m_2^2\right)} +\frac{C_A C_F \mathcal{L}_{m_2}^2\left(m_1^2+m_2^2\right)}{2 m_- m_+} \\& -\frac{C_A C_F \left(m_1^4\mathcal{L}_{m_1}^2-32 m_1^4 \mathcal{L}_{m_1}+2 m_1^2 m_2^2 \mathcal{L}_{m_1}^2-4 m_1^2m_2^2 \mathcal{L}_{m_1}+m_2^4 \mathcal{L}_{m_1}^2+18 m_1^4-18 m_2^4\right)}{2 m_-m_+ \left(m_1^2+m_2^2\right)} \\& -\frac{Y_s^2 C_A \mathcal{L}_{m_2} \left(m_1^2-2 m_2^2\right)}{2 m_2^2 m_-m_+}-\frac{Y_s^2 C_A \mathcal{L}_{m_2}^2}{16 m_- m_+} \\& +\frac{Y_s^2 C_A}{16 m_1^2 m_2^2 m_- m_+\left(m_1^2+m_2^2\right)} \left\lbrace M_1^4 m_2^2 \mathcal{L}_{m_1}^2-16 m_1^4m_2^2 \mathcal{L}_{m_1} +m_1^2 m_2^4 \mathcal{L}_{m_1}^2-8 m_1^2 m_2^4\mathcal{L}_{m_1}\right. \\& \left. +8 m_2^6 \mathcal{L}_{m_1}+8 m_1^6-8 m_1^4 m_2^2+8 m_1^2m_2^4-8 m_2^6\right\rbrace \end{aligned}$}
 & 
\resizebox{.66\hsize}{!}{ $\begin{aligned}[t]&(V^{(m_{1,2})}_5+\frac{1}{2}F^{(m_2)}_{\chi}+\frac{1}{2}F^{(m_1)}_{\chi})+\\& \frac{C_A^2 C_F^2 \left(36 m_1^2 \mathcal{L}_{m_1}^2-120 m_1^2 \mathcal{L}_{m_1}+60 \mathcal{L}_{m_1}^2-72\mathcal{L}_{m_1}+3 \pi ^2 m_1^2+108 m_1^2+5 \pi ^2+48\right) (m_2-m_1)}{12m_1^3}\\& -    \frac{C_A^2 C_F Y_s^2 \left(12 m_1^2 \mathcal{L}_{m_1}^2-54 m_1^2 \mathcal{L}_{m_1}+72 \mathcal{L}_{m_1}^2-132\mathcal{L}_{m_1}+\pi ^2 m_1^2+78 m_1^2+6 \pi ^2+126\right) (m_2-m_1)}{48m_1^5} \\& - \frac{C_A^2 C_FY_s^2 \left(12 m_1^2 \mathcal{L}_{m_1}^2-24 m_1^2 \mathcal{L}_{m_1}-36\mathcal{L}_{m_1}^2+36 \mathcal{L}_{m_1}+\pi ^2 m_1^2+24 m_1^2-3 \pi ^2-36\right)}{96 m_1^4}   \\& +  \frac{C_A^2 \left(12 m_1^2 \mathcal{L}_{m_1}^2-36 m_1^2 \mathcal{L}_{m_1}-12 \mathcal{L}_{m_1}^2+24\mathcal{L}_{m_1}+\pi ^2 m_1^2+48 m_1^2-\pi ^2-30\right)}{384 m_1^6}  \\& +  \frac{C_A^2 \left(12m_1^2 \mathcal{L}_{m_1}^2-42 m_1^2 \mathcal{L}_{m_1}-24 \mathcal{L}_{m_1}^2+60 \mathcal{L}_{m_1}+\pi ^2m_1^2+57 m_1^2-2 \pi ^2-72\right) (m_2-m_1)}{192 m_1^7} \\& -\frac{C_A^2 C_F^2 \left(12 m_1^2 \mathcal{L}_{m_1}^2-12 m_1^2 \mathcal{L}_{m_1}+12\mathcal{L}_{m_1}^2+\pi ^2 m_1^2+6 m_1^2+\pi ^2\right)}{12 m_1^2}
     \end{aligned}$}
 & $\Delta B^{(2)}_5$\\ [0.75ex] \hline
$\begin{aligned}[t]& \chi^{\dagger}_2\psi_1, \\& \bar{\psi}_2\chi_1  (1\leftrightarrow2 ) \end{aligned}$
   & 
\resizebox{.55\hsize}{!}{ $\begin{aligned}[t]& \frac{C_A C_F \mathcal{L}_{m_2}^2 \left(m_1^2+m_2^2\right)}{2 m_- m_+} -\frac{2 m_2^3 C_A C_F \mathcal{L}_{m_2}}{m_-
   m_+^2}-\frac{Y_f Y_s m_1 C_A \mathcal{L}_{m_2}^2}{8 m_- m_+}  \\& -\frac{C_A C_F}{2 m_- m_+^2} \left\lbrace M_1^3 \mathcal{L}_{m_1}^2-3 m_1^3 \mathcal{L}_{m_1}+m_1^2 m_2 \mathcal{L}_{m_1}^2-3m_1^2 m_2 \mathcal{L}_{m_1}\right. \\ & \left.+m_2^3 \mathcal{L}_{m_1}^2-m_2^3 \mathcal{L}_{m_1}+m_1
   m_2^2 \mathcal{L}_{m_1}^2+3 m_1 m_2^2 \mathcal{L}_{m_1}-4 m_1^3+4 m_1
   m_2^2\right\rbrace\\& +\frac{C_A}{16 m_2^2 m_- m_+^2} \left\lbrace-9 Y_f^2 m_1^3 m_2^2 \mathcal{L}_{m_1}-9 Y_f^2 m_1^2 m_2^3\mathcal{L}_{m_1}+9 Y_f^2 m_2^5 \mathcal{L}_{m_1}+9 Y_f^2 m_1 m_2^4 \mathcal{L}_{m_1}\right. \\ & \left.+9Y_f^2 m_1^3 m_2^2+9 Y_f^2 m_1^2 m_2^3  -9 Y_f^2 m_1m_2^4-9 Y_f^2 m_2^5  +2 Y_f Y_s m_1^2 m_2^2 \mathcal{L}_{m_1}^2+2Y_f Y_s m_1 m_2^3 \mathcal{L}_{m_1}^2\right. \\ & \left.-4 Y_f Y_s m_1 m_2^3\mathcal{L}_{m_1}-4 Y_f Y_s m_1^2 m_2^2+4 Y_f Y_s m_2^4+2
   Y_s^2 m_1^3+2 Y_s^2 m_1^2 m_2-2 Y_s^2 m_1 m_2^2-2
   Y_s^2 m_2^3\right\rbrace \\& +\frac{Y_s C_A \mathcal{L}_{m_2}
   \left(2 Y_f m_1 m_2^3-Y_s m_1^3-Y_s m_1^2 m_2+Y_s
   m_1 m_2^2+Y_s m_2^3\right)}{8 m_2^2 m_-
   m_+^2} \\& \end{aligned}$}
 & 
\resizebox{.77\hsize}{!}{ $\begin{aligned}[t]&(V^{(m_{1,2})}_6+\frac{1}{2}F^{(m_2)}_{\chi}+\frac{1}{2}F^{(m_1)}_{\psi})+\frac{C_A^2 C_F^2 \left(300 m_1^2 \mathcal{L}_{m_1}^2-600 m_1^2 \mathcal{L}_{m_1}-60 \mathcal{L}_{m_1}^2+108\mathcal{L}_{m_1}+25 \pi ^2 m_1^2+744 m_1^2-5 \pi ^2-132\right)}{24 m_1^2}+\\& \frac{C_A^2C_F^2 \left(60 m_1^2 \mathcal{L}_{m_1}^2+312 m_1^2 \mathcal{L}_{m_1}+180 \mathcal{L}_{m_1}^2-432\mathcal{L}_{m_1}+5 \pi ^2 m_1^2-144 m_1^2+15 \pi ^2+474\right) (m_2-m_1)}{24m_1^3}+ \\& \frac{C_A^2 C_F Y_fY_s\left(120 m_1^2 \mathcal{L}_{m_1}^2-288 m_1^2 \mathcal{L}_{m_1}-60 \mathcal{L}_{m_1}^2+162\mathcal{L}_{m_1}+10 \pi ^2 m_1^2+228 m_1^2-5 \pi ^2-171\right) (m_2-m_1)}{96m_1^4}- \\& \frac{C_A^2 C_F Y_fY_s \left(180 m_1^2 \mathcal{L}_{m_1}^2-264 m_1^2 \mathcal{L}_{m_1}-60\mathcal{L}_{m_1}^2+108 \mathcal{L}_{m_1}+15 \pi ^2 m_1^2+312 m_1^2-5 \pi ^2-132\right)}{192m_1^3} +\\& \frac{C_A^2 C_F Y_s^2 \left(108 m_1^2 \mathcal{L}_{m_1}^2-396 m_1^2 \mathcal{L}_{m_1}-60 \mathcal{L}_{m_1}^2+156\mathcal{L}_{m_1}+9 \pi ^2 m_1^2+468 m_1^2-5 \pi ^2-186\right) (m_2-m_1)}{96m_1^5} -\\& \frac{C_A^2 C_F Y_s^2 \left(60 m_1^2 \mathcal{L}_{m_1}^2-132 m_1^2 \mathcal{L}_{m_1}-12\mathcal{L}_{m_1}^2+24 \mathcal{L}_{m_1}+5 \pi ^2 m_1^2+168 m_1^2-\pi ^2-30\right)}{96 m_1^4}- \\& \frac{C_A^2 C_FY_f^2 \left(180 m_1^2 \mathcal{L}_{m_1}^2-396 m_1^2 \mathcal{L}_{m_1}-36 \mathcal{L}_{m_1}^2+72\mathcal{L}_{m_1}+15 \pi ^2 m_1^2+494 m_1^2-3 \pi ^2-88\right)}{64 m_1^2}- \\& \frac{C_A^2 C_F Y_f^2\left(36 m_1^2 \mathcal{L}_{m_1}^2+36 m_1^2 \mathcal{L}_{m_1}+108 \mathcal{L}_{m_1}^2-252 \mathcal{L}_{m_1}+3\pi ^2 m_1^2-2 m_1^2+9 \pi ^2+300\right) (m_2-m_1)}{64 m_1^3}+\\&\frac{C_A^2Y_f^3Y_s \left(108 m_1^2 \mathcal{L}_{m_1}^2-180 m_1^2 \mathcal{L}_{m_1}-36 \mathcal{L}_{m_1}^2+72\mathcal{L}_{m_1}+9 \pi ^2 m_1^2+210 m_1^2-3 \pi ^2-88\right)}{512 m_1^3} - \\ & \frac{C_A^2 Y_f^3Y_s\left(72 m_1^2 \mathcal{L}_{m_1}^2-144 m_1^2 \mathcal{L}_{m_1}-36 \mathcal{L}_{m_1}^2+90 \mathcal{L}_{m_1}+6\pi ^2 m_1^2+140 m_1^2-3 \pi ^2-106\right) (m_2-m_1)}{256 m_1^4}+ \\& \frac{C_A^2 Y_s^3Y_f \left(36 m_1^2 \mathcal{L}_{m_1}^2-60 m_1^2 \mathcal{L}_{m_1}-12 \mathcal{L}_{m_1}^2+24 \mathcal{L}_{m_1}+3 \pi ^2 m_1^2+72 m_1^2-\pi ^2-30\right)}{768 m_1^5} - \\& \frac{C_A^2 Y_s^3Y_f\left(60 m_1^2 \mathcal{L}_{m_1}^2-144 m_1^2 \mathcal{L}_{m_1}-24 \mathcal{L}_{m_1}^2+66 \mathcal{L}_{m_1}+5\pi ^2 m_1^2+150 m_1^2-2 \pi ^2-78\right) (m_2-m_1)}{384 m_1^6}
     \end{aligned}$}
 & $\Delta B^{(2)}_6$\\ [1ex] \hline
\end{tabular}}
\caption{Matching corrections, $B(\mu)$, to the threshold form-factor at $\mu\sim m_{1,2}$, $a\equiv \alpha/(4\pi)$, $\mathcal{L}_{m_{1,2}}\equiv \log{m_{1,2}^2/\mu^2}$, and $m_{\pm}=m_1\pm m_2$. $V^{(m_{1,2})}_i$ and $F_I$ are two loop vertex and wave-function corrections given in Appendices \ref{sec:vertcorr} and \ref{sec:wfr2}. $\Delta B^{(2)}$ is the two-loop order contribution from mass and coupling renormalisation given in Appendix \ref{sec:massren}. }
\label{tab:expb}
\end{table}
What remains is the anomalous dimension, $\gamma_3$, between $m_{1,2}$ and $M$, and the matching coefficient,  $\exp{[U(\mu)]}$, at $\mu\sim M$. These have been computed in the previously in tables \ref{tab:corrrtu} and \ref{tab:exp2u}. Again, these contributions are found by computing graphs in figures \ref{fig:verts2} and \ref{fig:wfr}, evaluated on-shell, with bosonic masses, $M$, included and external lines taken to be heavy with incoming and outgoing velocities, $v_1$ and $v_2$, reespectively. The difference here being that in the threshold limit, 
\begin{equation}
    w\sim \frac{m_1^2+m_2^2}{2m_1m_2}\sim \mathcal{O}(1),
\end{equation}
since we take $m_1\sim m_2$, and thus, $h(w)\sim \mathcal{O}(1)$, by inspection of \eqref{eqn:romega}. Whence, the sub-leading Higgs contribution which was sub-leading in the Sudakov regime becomes of the same order as the gauge contribution in the threshold regime. The remaining contributions at two-loop order from mass and coupling renormalisation were presented previously in table \ref{tab:corrrtu}.
\begin{table}[tb]
\footnotesize
\centering
\scalebox{0.67}{
\begin{tabular}{|c|c|c|c|c|} 
 \hline
 $\mathcal{O}$ & $\tilde{B}^{(1)}(\mu)$ & $\tilde{B}^{(2)}(\mu)$ & $\tilde{\gamma}^{(1)}_3(\mu)$ & $\Delta \tilde{B}^{(2)}(\mu)$ \\ [0.75ex] 
 \hline\hline
 $\bar{\psi}_2\psi_1$ & 
\resizebox{.25\hsize}{!}{ $\begin{aligned}[t]& -\frac{1}{48} C_A \left(3 \mathcal{L}_{m_2} \left(8 C_F \mathcal{L}_{m_2}+8 C_F+5 Y_f^2\right) \right. \\& \left. +4 \left(24+\pi^2\right) C_F+9Y_f^2\right)\end{aligned}$}
 & 
\resizebox{.43\hsize}{!}{ $\begin{aligned}[t]& (V^{(m_2)}_1+\frac{1}{2}F^{(m_2,0)}_{\psi}+\frac{1}{2}F^{(0,0)}_{\psi}) \\& + \frac{1}{6} C_A^2 C_F^2 \left(3 \mathcal{L}_{m_2} \left(4 \left(\mathcal{L}_{m_2}-2\right) \mathcal{L}_{m_2}+\pi^2+16\right)+6 \zeta_3-2 \pi ^2-48\right) \\& -\frac{1}{32} C_A^2 C_FY_f^2 \left(2 \mathcal{L}_{m_2} \left(6 \mathcal{L}_{m_2} \left(2 \mathcal{L}_{m_2}+1\right)+3\pi ^2-46\right)+12 \zeta_3+\pi ^2+158\right) \\& + \frac{1}{64} C_A^2Y_f^4 \left(36 \left(\mathcal{L}_{m_2}-3\right) \mathcal{L}_{m_2}+3 \pi ^2+142\right)
   \end{aligned}$}
 & 
\resizebox{.14\hsize}{!}{ $\begin{aligned}[t]& -\frac{C_A}{8} \left(40C_F-Y_f^2\right)\end{aligned}$}
 & 
\resizebox{.4\hsize}{!}{ $\begin{aligned}[t]& 
\frac{1}{24} C_A C_F \left\lbrace-2 \mathcal{L}_{m_2} \left(3 \mathcal{L}_{m_2} \left(\beta_0-48 C_A
   C_F\right)+120 C_A C_F+\left(24+\pi ^2\right) \beta_0 \right.\right. \\& \left.\left. +2 \beta_0
   \mathcal{L}_{m_2}^2\right)+24 \left(24+\pi ^2\right) C_A C_F-\left(\pi ^2-96\right) \beta_0-8
   \beta_0 \zeta_3\right\rbrace \\& -\frac{1}{192} Y_f^2 C_A \left\lbrace6 \mathcal{L}_{m_2} \left(\mathcal{L}_{m_2} \left(48 C_A C_F+5 \beta_0\right)-208 C_A C_F+6 \beta_0\right) \right.\\&\left. +24 \left(46+\pi ^2\right) C_A C_F+\left(5 \pi
   ^2-132\right) \beta_0\right\rbrace  +\frac{1}{32} Y_f^4 (13-15 \mathcal{L}_{m_2}) C_A^2  \end{aligned}$}
  \\[0.75ex]\hline
 $\bar{\psi}_2\gamma^{\mu}\psi_1$ &  
\resizebox{.25\hsize}{!}{$\begin{aligned}[t]& \frac{1}{48} C_A \left(-3 \mathcal{L}_{m_2} \left(8 C_F \mathcal{L}_{m_2}-40 C_F+11 Y_f^2\right)\right. \\& \left. -4\left(60+\pi ^2\right) C_F+27 Y_f^2\right) \end{aligned}$}
 & 
\resizebox{.43\hsize}{!}{ $\begin{aligned}[t]& (V^{(m_2)}_2+\frac{1}{2}F^{(m_2,0)}_{\psi}+\frac{1}{2}F^{(0,0)}_{\psi}) \\& + \frac{1}{12} C_A^2 C_F^2 \left(6 \mathcal{L}_{m_2} \left(4 \mathcal{L}_{m_2}^2-2 \mathcal{L}_{m_2}+\pi^2+26\right)+12 \zeta_3-\pi ^2-246\right)\\& + \frac{1}{32} C_A^2 C_FY_f^2 \left(-6 \mathcal{L}_{m_2} \left(4 \mathcal{L}_{m_2}^2-2 \mathcal{L}_{m_2}+\pi^2+30\right)+\pi ^2+262-12\zeta_3\right) \\& - \frac{1}{256} C_A^2Y_f^4 \left(36 \left(\mathcal{L}_{m_2}-3\right) \mathcal{L}_{m_2}+3 \pi ^2+124\right) \end{aligned}$}
 & 
\resizebox{.13\hsize}{!}{ $\begin{aligned}[t]& -\frac{C_A}{4}  \left(8 C_F+Y_f^2\right)\end{aligned}$}
 &  
\resizebox{.4\hsize}{!}{ $\begin{aligned}[t]& 
\frac{1}{24} C_A C_F \left\lbrace-2 \mathcal{L}_{m_2} \left(-3 \mathcal{L}_{m_2} \left(48 C_A C_F+5 \beta_0\right)+552 C_A C_F+\left(60+\pi ^2\right) \beta_0 \right.\right.\\&\left.\left.  +2 \beta_0\mathcal{L}_{m_2}^2\right) +24 \left(66+\pi ^2\right) C_A C_F+\left(264+5 \pi ^2\right) \beta_0-8\beta_0 \zeta_3\right\rbrace \\& -\frac{1}{192} Y_f^2 C_A \left\lbrace6 \mathcal{L}_{m_2} \left(48 C_A C_F \mathcal{L}_{m_2}-496 C_A C_F-18\beta_0+11 \beta_0 \mathcal{L}_{m_2}\right) \right.\\&\left. +24 \left(166+\pi ^2\right) C_AC_F+\left(180+11 \pi ^2\right) \beta_0\right\rbrace + \frac{1}{32} Y_f^4 (52-33 \mathcal{L}_{m_2}) C_A^2 \end{aligned}$}
  \\[0.75ex]  \hline
 $\bar{\psi}_2\sigma^{\mu\nu}\psi_1$ & 
\resizebox{.25\hsize}{!}{ $\begin{aligned}[t]& \frac{1}{48} C_A \left(-3 \mathcal{L}_{m_2} \left(8 C_F \mathcal{L}_{m_2}-56 C_F+9 Y_f^2\right) \right. \\& \left. -4 \left(48+\pi^2\right) C_F+15 Y_f^2\right)
   \end{aligned}$}
 & 
\resizebox{.4\hsize}{!}{ $\begin{aligned}[t]& (V^{(m_2)}_3+\frac{1}{2}F^{(m_2,0)}_{\psi}+\frac{1}{2}F^{(0,0)}_{\psi}) \\& +\frac{1}{6} C_A^2 C_F^2 \left(3 \mathcal{L}_{m_2} \left(4 \left(\mathcal{L}_{m_2}-2\right) \mathcal{L}_{m_2}+\pi^2+28\right)+6 \zeta_3-2 \pi ^2-108\right)  \\&  -\frac{1}{32} C_A^2 C_FY_f^2 \left(2 \mathcal{L}_{m_2} \left(6 \mathcal{L}_{m_2} \left(2 \mathcal{L}_{m_2}+3\right)+3\pi ^2-50\right) \right. \\& \left. +3 \left(4 \zeta_3+50+\pi ^2\right)\right) +  \frac{3}{128} C_A^2Y_f^4 \left(36 \left(\mathcal{L}_{m_2}-3\right) \mathcal{L}_{m_2}+3 \pi ^2+136\right) \end{aligned}$}
 & 
\resizebox{.13\hsize}{!}{ $\begin{aligned}[t]& -\frac{C_A}{8} \left(8 C_F+Y_f^2\right)  \end{aligned}$}
 &  
\resizebox{.42\hsize}{!}{ $\begin{aligned}[t]& 
\frac{1}{24} C_A C_F \left\lbrace-2 \mathcal{L}_{m_2} \left(-3 \mathcal{L}_{m_2} \left(48 C_A C_F+7 \beta_0\right)+696 C_A C_F+\left(48+\pi ^2\right) \beta_0 \right.\right.\\&\left.\left. +2 \beta_0
   \mathcal{L}_{m_2}^2\right) +24 \left(68+\pi ^2\right) C_A C_F+\left(96+7 \pi ^2\right) \beta_0-8\beta_0 \zeta_3\right\rbrace \\& -\frac{1}{64} Y_f^2 C_A \left\lbrace2 \mathcal{L}_{m_2} \left(48 C_A C_F \mathcal{L}_{m_2}-496 C_A C_F-10\beta_0+9 \beta_0 \mathcal{L}_{m_2}\right) \right.\\&\left.+8 \left(154+\pi ^2\right) C_A C_F+\left(4+3 \pi^2\right) \beta_0\right\rbrace +\frac{3}{32} Y_f^4 (13-9 \mathcal{L}_{m_2}) C_A^2  \end{aligned}$}
  \\[0.75ex] \hline
 $\chi^{\dagger}_2\chi_1$  &   
\resizebox{.25\hsize}{!}{ $\begin{aligned}[t]& -\frac{1}{12} C_A C_F \left(6 (\mathcal{L}_{m_2}-2) \mathcal{L}_{m_2}+\pi ^2-12\right)\\ & +\frac{Y_s^2 C_A \left(-12 \mathcal{L}_{m_2}+6 \log ^2\left(m_2^2\right)+\pi ^2+12\right)}{96m_2^2} \end{aligned}$}
 & 
\resizebox{.4\hsize}{!}{ $\begin{aligned}[t]& (V^{(m_2)}_4+\frac{1}{2}F^{(m_2,0)}_{\chi}+\frac{1}{2}F^{(0,0)}_{\chi}) \\& +\frac{1}{6} C_A^2 C_F^2 \left(2 \mathcal{L}_{m_2} \left(4 \mathcal{L}_{m_2}^2+6 \mathcal{L}_{m_2}+\pi^2-12\right)+4 \zeta_3+\pi ^2+24\right)+\\& \frac{C_A^2 C_FY_s^2 \left(-\mathcal{L}_{m_2} \left(4 \mathcal{L}_{m_2}^2+\pi ^2-3\right)-6+\psi ^{(2)}(1)\right)}{12 m_2^2} \\& +  \frac{C_A^2Y_s^4 \left(2 \mathcal{L}_{m_2} \left(4 \mathcal{L}_{m_2}^2-6 \mathcal{L}_{m_2}+\pi^2+12\right)+4 \zeta_3-\pi ^2-24\right)}{384 m_2^4} \end{aligned}$}
 & 
\resizebox{.06\hsize}{!}{ $\begin{aligned}[t]& C_AC_F \end{aligned}$}
 &  
\resizebox{.4\hsize}{!}{ $\begin{aligned}[t]& 
\frac{1}{12} C_A C_F \left\lbrace6 \mathcal{L}_{m_2}^2 \left(12 C_A C_F+\beta_0\right)-\mathcal{L}_{m_2}
   \left(240 C_A C_F+\left(\pi ^2-12\right) \beta_0\right) \right.\\&\left. +6 \left(38+\pi ^2\right) C_A C_F-2\beta_0 \mathcal{L}_{m_2}^3+\beta_0 \left(\pi ^2-4 (\zeta_3+9)\right)\right\rbrace \\& +\frac{Y_s^2 C_A}{96 m_2^2}\left\lbrace\mathcal{L}_{m_2} \left(-6 \mathcal{L}_{m_2} \left(20 C_A C_F+\beta_0\right)+384 C_A C_F+\left(12+\pi ^2\right) \beta_0 \right.\right. \\& \left.\left. +2 \beta_0\mathcal{L}_{m_2}^2\right)  -10 \left(42+\pi ^2\right) C_A C_F-\left(24+\pi ^2\right) \beta_0+4\beta_0 \zeta_3\right\rbrace \\&  +\frac{Y_s^4 \left(12 (\mathcal{L}_{m_2}-3) \mathcal{L}_{m_2}+\pi ^2+42\right) C_A^2}{192m_2^4} +\frac{1}{32} Y_f^4 (13-15 \mathcal{L}_{m_2}) C_A^2  \end{aligned}$}
   \\ [0.75ex] \hline
 $i\chi^{\dagger}_2\overset{\leftrightarrow}{D}_{\mu}\chi_1$  &  
\resizebox{.23\hsize}{!}{ $\begin{aligned}[t] & -\frac{1}{12} C_A C_F \left(6 (\mathcal{L}_{m_2}-8) \mathcal{L}_{m_2}+\pi ^2+51\right) \\& +\frac{Y_s^2 C_A \left(6 (\mathcal{L}_{m_2}-4) \mathcal{L}_{m_2}+\pi ^2+12\right)}{96 m_2^2}\end{aligned}$}
 & 
\resizebox{.44\hsize}{!}{ $\begin{aligned}[t]& (V^{(m_2)}_5+\frac{1}{2}F^{(m_2,0)}_{\chi}+\frac{1}{2}F^{(0,0)}_{\chi}) \\& +\frac{1}{12} C_A^2 C_F^2 \left(4 \mathcal{L}_{m_2} \left(\mathcal{L}_{m_2} \left(4 \mathcal{L}_{m_2}-9\right)+\pi ^2+15\right)+8 \zeta_3-3 \pi ^2-54\right)+\\& \frac{C_A^2 C_FY_s^2 \left(-4 \mathcal{L}_{m_2} \left(8 \mathcal{L}_{m_2}^2-9 \mathcal{L}_{m_2}+2 \pi^2+24\right)+3 \pi ^2+8 (15+\psi ^{(2)}(1))\right)}{96 m_2^2} \\& + \frac{C_A^2Y_s^4 \left(4 \mathcal{L}_{m_2}^3+\left(\pi ^2-6\right) \mathcal{L}_{m_2}+2 (\zeta_3+6)\right)}{192 m_2^4} \end{aligned}$}
 & 
\resizebox{.065\hsize}{!}{ $\begin{aligned}[t]& 4 C_A C_F\end{aligned}$}
 &  
\resizebox{.42\hsize}{!}{ $\begin{aligned}[t]& 
\frac{1}{24} C_A C_F \left\lbrace-2 \mathcal{L}_{m_2} \left(-24 \mathcal{L}_{m_2} \left(3 C_A C_F+\beta_0\right)+456 C_A C_F+\left(51+\pi ^2\right) \beta_0 \right.\right. \\& \left.\left. +2 \beta_0\mathcal{L}_{m_2}^2\right)  +6 \left(223+2 \pi ^2\right) C_A C_F+\left(153+8 \pi ^2\right) \beta_0-8\beta_0 \zeta_3\right\rbrace \\&  +\frac{Y_s^2 C_A}{96 m_2^2} \left\lbrace-2 \left(\left(387+5 \pi ^2\right) C_A C_F+\pi ^2 \beta_0\right) +\mathcal{L}_{m_2} \left(2 \mathcal{L}_{m_2} \left(\beta_0 \mathcal{L}_{m_2}\right.\right.\right. \\& \left.\left.\left.-6 \left(10 C_AC_F+\beta_0\right)\right)  +600 C_A C_F+\left(12+\pi ^2\right) \beta_0\right)  +4\beta_0 \zeta_3\right\rbrace \\& +\frac{Y_s^4 \left(12 (\mathcal{L}_{m_2}-4) \mathcal{L}_{m_2}+\pi ^2+54\right) C_A^2}{192m_2^4}  \end{aligned}$}
   \\ [0.75ex] \hline
 $\chi^{\dagger}_2\psi_1$  & 
\resizebox{.25\hsize}{!}{ $\begin{aligned}[t]&-\frac{1}{12} C_A C_F \left(6 (\mathcal{L}_{m_2}-2) \mathcal{L}_{m_2}+\pi ^2-24\right) \\& -\frac{Y_s^2 C_A (\mathcal{L}_{m_2}-1)}{8 m_2^2} \\& +\frac{Y_f Y_s C_A \left(6 \mathcal{L}_{m_2}^2+\pi ^2-12\right)}{48 m_2}
\end{aligned}$}
 & 
\resizebox{.4\hsize}{!}{ $\begin{aligned}[t]& (V^{(m_2)}_6+\frac{1}{2}F^{(m_2,0)}_{\chi}+\frac{1}{2}F^{(0,0)}_{\psi}) \\& +\frac{1}{3} C_A^2 C_F^2 \left(\mathcal{L}_{m_2} \left(4 \mathcal{L}_{m_2} \left(\mathcal{L}_{m_2}+3\right)+\pi^2-24\right)+2 \zeta_3+\pi ^2+24\right)- \\& \frac{C_A^2 C_F Y_fY_s\left(2 \mathcal{L}_{m_2} \left(4 \mathcal{L}_{m_2}^2+6 \mathcal{L}_{m_2}+\pi^2-12\right)+4 \zeta_3+\pi ^2+24\right)}{24 m_2} \\& - \frac{C_A^2 C_FY_s^2 \left(2 \mathcal{L}_{m_2} \left(4 \mathcal{L}_{m_2}^2+6 \mathcal{L}_{m_2}+\pi^2-24\right)+4 \zeta_3+\pi ^2+72\right)}{48 m_2^2}+ \\& \frac{C_A^2Y_s^3Y_f \left(4 \mathcal{L}_{m_2}^3+\left(\pi ^2-6\right) \mathcal{L}_{m_2}+2 (\zeta_3+6)\right)}{96 m_2^3} \end{aligned}$}
 & 
\resizebox{.06\hsize}{!}{ $\begin{aligned}[t]& C_AC_F\end{aligned}$}
 &  
\resizebox{.4\hsize}{!}{ $\begin{aligned}[t]& 
\frac{1}{12} C_A C_F \left\lbrace6 \mathcal{L}_{m_2}^2 \left(12 C_A C_F+\beta_0\right)-\mathcal{L}_{m_2}
   \left(240 C_A C_F+\left(\pi ^2-24\right) \beta_0\right) \right.\\&\left. +6 \left(32+\pi ^2\right) C_A C_F-2\beta_0 \mathcal{L}_{m_2}^3+\beta_0 \left(-4 \zeta_3-72+\pi ^2\right)\right\rbrace \\&
    -\frac{Y_s^2 C_A}{96 m_2^2}  \left\lbrace6 \mathcal{L}_{m_2} \left(\mathcal{L}_{m_2} \left(8 C_A C_F+\beta_0\right) -2\left(18 C_A C_F+\beta_0\right)\right)+4 \left(54 \right.\right.\\&\left.\left. +\pi ^2\right) C_A C_F  +\left(24+\pi ^2\right)\beta_0\right\rbrace +\frac{Y_f Y_s C_A}{48 m_2} \left\lbrace\mathcal{L}_{m_2} \left(168 C_A C_F+\left(\pi ^2-12\right) \beta_0 \right)\right.\\&\left.-72 C_A C_F \mathcal{L}_{m_2}^2-6 \left(24+\pi ^2\right) C_A C_F+2 \beta_0
   \mathcal{L}_{m_2}^3+4 \beta_0 (\zeta_3+9)\right\rbrace \\& \frac{Y_f Y_s^3 \left(12 (\mathcal{L}_{m_2}-2) \mathcal{L}_{m_2}+\pi ^2+18\right) \left(2 C_A
   C_F+1\right)}{96 m_2^3}  -\frac{Y_s^4 (2 \mathcal{L}_{m_2}-3) C_A^2}{32 m_2^4} \end{aligned}$}
  \\ [0.75ex] \hline
 $\bar{\psi}_2\chi_1$  & 
\resizebox{.25\hsize}{!}{ $\begin{aligned}[t]&  -\frac{1}{12} C_A C_F \left(6 (\mathcal{L}_{m_2}-5) \mathcal{L}_{m_2}+\pi ^2+48\right) \\& -\frac{Y_f Y_s C_A}{4 m_2}-\frac{9}{16} Y_f^2 C_A (\mathcal{L}_{m_2}-1)
\end{aligned}$}
 & 
\resizebox{.43\hsize}{!}{ $\begin{aligned}[t]& (V^{(m_2)}_7+\frac{1}{2}F^{(m_2,0)}_{\psi}+\frac{1}{2}F^{(0,0)}_{\chi}) \\& + \frac{1}{12} C_A^2 C_F^2 \left(6 \mathcal{L}_{m_2} \left(2 \mathcal{L}_{m_2} \left(2 \mathcal{L}_{m_2}-7\right)+\pi ^2+36\right)+12 \zeta_3-7 \pi ^2-264\right) \\& -\frac{C_A^2 C_F Y_fY_s \left(4 \mathcal{L}_{m_2} \left(3 \mathcal{L}_{m_2}-10\right)+\pi ^2+56\right)}{16 m_2} \\& +  \frac{1}{16} C_A^2 C_FY_f^2 \left(-\mathcal{L}_{m_2} \left(12 \left(\mathcal{L}_{m_2}-3\right) \mathcal{L}_{m_2}  +3 \pi^2+88\right)-6 \zeta_3+3 \pi ^2+104\right) \\&+ \frac{C_A^2Y_f^3Y_s \left(36 \left(\mathcal{L}_{m_2}-3\right) \mathcal{L}_{m_2}+3 \pi ^2+142\right)}{128 m_2} \end{aligned}$}
 & 
\resizebox{.06\hsize}{!}{ $\begin{aligned}[t]& C_A C_F \end{aligned}$}
 &  
\resizebox{.4\hsize}{!}{ $\begin{aligned}[t]& 
\frac{1}{24} C_A C_F \left\lbrace-2 \mathcal{L}_{m_2} \left(-3 \mathcal{L}_{m_2} \left(48 C_A C_F+5 \beta_0\right)+552 C_A C_F+\left(48+\pi ^2\right) \beta_0 \right.\right.\\&\left.\left. +2 \beta_0\mathcal{L}_{m_2}^2\right) +24 \left(60+\pi ^2\right) C_A C_F+\left(192+5 \pi ^2\right) \beta_0-8\beta_0 \zeta_3\right\rbrace \\& -\frac{1}{64} Y_f^2 C_A \left\lbrace2 \mathcal{L}_{m_2} \left(48 C_A C_F \mathcal{L}_{m_2}  -448 C_A C_F-18\beta_0+9 \beta_0 \mathcal{L}_{m_2}\right) \right.\\&\left.+8 \left(152+\pi ^2\right) C_A C_F+\left(68+3\pi ^2\right) \beta_0\right\rbrace+ \frac{9}{32} Y_f^4 (5-3 \mathcal{L}_{m_2}) C_A^2 \\& + \frac{Y_f Y_s C_A \left(-\mathcal{L}_{m_2} \left(6 C_AC_F+\beta_0\right)+19 C_A C_F+3\beta_0\right)}{4 m_2} \\& +\frac{Y_f^3 Y_s (3 \mathcal{L}_{m_2}-11) C_A^2}{16 m_2}   \end{aligned}$}
  \\ [1ex] 
 \hline
\end{tabular}}
\caption{Matching and running, $\tilde{B}(\mu)$ and $\tilde{\gamma}_3(\mu)$, to the threshold form-factor for $m_2\gg m_1\gg M\gg Q$ at $\mu\sim m_2$, $a\equiv \alpha/(4\pi)$, and $\mathcal{L}_{m_{2}}\equiv \log{m_{2}^2/\mu^2}$. $\Delta \tilde{B}^{(2)}$ are the mass and coupling renormalisation contributions contributing at two-loop order. $V^{(m_2)}_i$ and $F_I$ are two loop vertex and wave-function corrections, given in Appendices \ref{sec:vertcorr} and \ref{sec:wfr2}. }
\label{tab:expbt}
\end{table}
Finally, we consider the slightly more involved, $m_2\gg m_1\gg M^2 \gg Q^2$ case, where $m_{1,2}$ and $M$ denote the external particle and bosonic masses, respectively. Schematically, we then have following matching and running steps, illustrated as follows,
\begin{equation*}
    \mathcal{O} \xleftrightarrow[Q,M,m_1=0]{\mu\sim m_{2}} e^{\tilde{B}} \tilde{\mathcal{O}}_1 \xrightarrow[]{\phantom{11}\tilde{\gamma}_3\phantom{11}}  e^{\tilde{B}} \xleftrightarrow[Q,M=0]{\mu\sim m_{1}} e^{\tilde{B}+G} \tilde{\mathcal{O}}_2 \xrightarrow[]{\phantom{11}\gamma_3\phantom{11}}    e^{\tilde{B}+G}\tilde{\mathcal{O}}_2 \xleftrightarrow[Q=0]{\mu\sim M}e^{\tilde{B}+G+U} \tilde{\mathcal{O}}_3,
\end{equation*}
where $\tilde{B}$, $G$ and $U$ are multiplicative matching coefficients, $\gamma_3$ and $\tilde{\gamma}_3$, are the effective theory anomalous dimensions, and $\tilde{\mathcal{O}}_{1,2,3}$ the effective theory operators at each scale. At the scale $\mu>m_2$, we employ the full theory graphs and below, at $\mu<m_2$, we match down to an effective theory with a single heavy field of mass, $m_2$. Thus, the effective theory operator is given by the full theory operators with particle $2$ represented by a heavy field, $h_{f,s}$, for instance in the fermionic case we have, $\bar{h}_{f,2}\Gamma \psi_1$, and similarly for the other operators. 
\begin{table}[tb]
\footnotesize
\centering
\scalebox{0.69}{
\begin{tabular}{|c|c|c|c|} 
 \hline
 $\mathcal{O}$ & $G^{(1)}(\mu)$ & $G^{(2)}(\mu)$ & $\Delta G^{(2)}(\mu)$  \\ [0.75ex] 
 \hline\hline
 $\bar{\psi}_2\Gamma\psi_1$ & 
\resizebox{.32\hsize}{!}{ $\begin{aligned}[t]& \frac{1}{16} C_A \left(\mathcal{L}_{m_1} \left(16 C_F-7 Y_f^2\right)-16 C_F+5 Y_f^2\right)\end{aligned}$}
 & 
\resizebox{.42\hsize}{!}{ $\begin{aligned}[t]& (V^{(m_1)}_1+\frac{1}{2}F^{(m_1,0)}_{\psi}+\frac{1}{2}F^{(0)}_{h}) \\&-\frac{1}{8} C_A^2 C_F^2 \left(4 \mathcal{L}_{m_1} (3 \mathcal{L}_{m_1}-10)+\pi ^2+56\right)+ \\& \frac{C_A^2 C_F Y_f^2}{96 m_1^2} \left\lbrace-2 \mathcal{L}_{m_1} \left(2 \mathcal{L}_{m_1} (2 \mathcal{L}_{m_1}-9)+\pi ^2+48\right)\right.\\&\left. -4 \zeta_3+3 \pi ^2+120\right\rbrace -\frac{1}{256} C_A^2Y_f^4 \left(36 (\mathcal{L}_{m_1}-3) \mathcal{L}_{m_1}+3 \pi ^2+142\right)
\end{aligned}$}
 & 
\resizebox{.4\hsize}{!}{ $\begin{aligned}[t]& \frac{1}{12} C_A C_F \left\lbrace6 \mathcal{L}_{m_1} \left(\beta_0 \mathcal{L}_{m_1}-2 \left(12 C_AC_F+\beta_0\right)\right) \right.\\&\left.+168 C_A C_F+\left(24+\pi ^2\right) \beta_0\right\rbrace +\frac{1}{192} Y_f^2 C_A \left\lbrace6 \mathcal{L}_{m_1} \left(216 C_A C_F\right.\right.\\&\left.\left. +10 \beta_0-7 \text{$\beta$0} \mathcal{L}_{m_1}\right) -1512 C_A C_F-\left(108+7 \pi ^2\right) \beta_0\right\rbrace \\&+\frac{1}{32} Y_f^4 (32-21 \mathcal{L}_{m_1}) C_A^2\end{aligned}$}
 \\[0.75ex]\hline
 $\chi^{\dagger}_2\chi_1$  &   
\resizebox{.35\hsize}{!}{ $\begin{aligned}[t]& \frac{1}{24} C_A C_F \left(6 \mathcal{L}_{m_1}^2+\pi ^2+48\right)  -\frac{Y_s^2 C_A (\mathcal{L}_{m_1}-1)}{8 m_1^2} \end{aligned}$}
 & 
\resizebox{.4\hsize}{!}{ $\begin{aligned}[t]& (V^{(m_1)}_2+\frac{1}{2}F^{(m_1,0)}_{\chi}+\frac{1}{2}F^{(0)}_{h}) \\&-\frac{1}{6} C_A^2 C_F^2 \left\lbrace\mathcal{L}_{m_1} \left(4 (\mathcal{L}_{m_1}-3) \mathcal{L}_{m_1}+\pi ^2+24\right) \right.\\&\left. +2 \zeta_3-\pi ^2-24\right\rbrace  +  \frac{C_A^2 C_FY_s^2}{96 m_1^2} \left\lbrace-2 \mathcal{L}_{m_1} \left(2 \mathcal{L}_{m_1} (2 \mathcal{L}_{m_1}-9) +\pi ^2 \right.\right.\\&\left.\left.+48\right)-4 \zeta_3+3 \pi ^2+120\right\rbrace
\end{aligned}$}
  & 
\resizebox{.4\hsize}{!}{ $\begin{aligned}[t]& \frac{1}{24} C_A C_F \left\lbrace-144 C_A C_F \mathcal{L}_{m_1}+2 \beta_0\mathcal{L}_{m_1}^3+\left(48+\pi ^2\right) \beta_0 \mathcal{L}_{m_1} \right.\\&\left. +168 C_A C_F+4\beta_0 (\zeta_3-24)\right\rbrace +\frac{Y_s^2 C_A}{96 m_1^2} \left\lbrace6 \mathcal{L}_{m_1} \left(20 C_A C_F+2 \beta_0  \right.\right.\\&\left.\left.-\beta_0\mathcal{L}_{m_1}\right)  -168 C_A C_F-\left(24+\pi ^2\right) \beta_0\right\rbrace -\frac{Y_s^4 (2 \mathcal{L}_{m_1}-3) C_A^2}{32 m_1^4} \end{aligned}$}
 \\ [0.75ex] \hline
 $i\chi^{\dagger}_2\overset{\leftrightarrow}{D}_{\mu}\chi_1$  &  
\resizebox{.26\hsize}{!}{ $\begin{aligned}[t]& \frac{1}{12} C_A C_F \left\lbrace6 (\mathcal{L}_{m_1}-2) \mathcal{L}_{m_1}+\pi ^2+48\right\rbrace\\&-\frac{Y_s^2 C_A (\mathcal{L}_{m_1}-1)}{8 m_1^2}\end{aligned}$}
 & 
\resizebox{.4\hsize}{!}{ $\begin{aligned}[t]& (V^{(m_1)}_3+\frac{1}{2}F^{(m_1,0)}_{\chi}+\frac{1}{2}F^{(0)}_{h}) \\&+\frac{1}{3} C_A^2 C_F^2 \left\lbrace\mathcal{L}_{m_1} \left(4 (\mathcal{L}_{m_1}-3) \mathcal{L}_{m_1}+\pi ^2+24\right)\right.\\&\left.+2 \zeta_3-\pi ^2-24\right\rbrace+ \frac{C_A^2 C_F}{48 m_1^2}\left\lbrace-2 \mathcal{L}_{m_1} \left(2 \mathcal{L}_{m_1} (2 \mathcal{L}_{m_1}-9)+\pi ^2\right.\right.\\&\left.\left.+48\right)-4 \zeta_3+3 \pi ^2+120\right\rbrace
\end{aligned}$}
  & 
\resizebox{.4\hsize}{!}{ $\begin{aligned}[t]& \frac{1}{12} C_A C_F \left\lbrace\mathcal{L}_{m_1} \left(\left(48+\pi ^2\right) \beta_0-72 C_A C_F\right)+84C_A C_F+\right.\\&\left. 2 \beta_0 \mathcal{L}_{m_1}^3  -6 \beta_0 \mathcal{L}_{m_1}^2-\beta_0 \left(-4\zeta_3+96+\pi ^2\right)\right\rbrace \\& +\frac{Y_s^2 C_A}{96 m_1^2} \left\lbrace6 \mathcal{L}_{m_1} \left(20 C_A C_F+2 \beta_0-\beta_0\mathcal{L}_{m_1}\right)  -168 C_A C_F-\left(24 \right.\right.\\&\left.\left.+\pi ^2\right) \beta_0\right\rbrace  -\frac{Y_s^4 (2 \mathcal{L}_{m_1}-3) C_A^2}{32 m_1^4}\end{aligned}$}
 \\ [0.75ex] \hline
 $\chi^{\dagger}_2\psi_1$  & 
\resizebox{.32\hsize}{!}{ $\begin{aligned}[t]& \frac{1}{16} C_A \left\lbrace\mathcal{L}_{m_1} \left(16 C_F+Y_f (2 Y_s-9 Y_f)\right)-16C_F\right.\\&\left.+Y_f (9 Y_f-4 Y_s)\right\rbrace \end{aligned}$}
 & 
\resizebox{.4\hsize}{!}{ $\begin{aligned}[t]& (V^{(m_1)}_4+\frac{1}{2}F^{(m_1,0)}_{\psi}+\frac{1}{2}F^{(0)}_{h}) \\&-\frac{1}{12} C_A^2 C_F^2 \left(12 (\mathcal{L}_{m_1}-2) \mathcal{L}_{m_1}  +\pi ^2+24\right)+ \frac{1}{48} C_A^2 C_F Y_f Y_s\left(\pi ^2 \right.\\&\left. + 24 + 12 (\mathcal{L}_{m_1}-2) \mathcal{L}_{m_1}\right)+ \frac{C_A^2 C_FY_s^2}{96 m_1^2} \left\lbrace12 (\mathcal{L}_{m_1}-3) \mathcal{L}_{m_1}+\pi ^2\right. \\&\left. +48\right\rbrace- \frac{C_A^2Y_s^3Y_f \left(12 (\mathcal{L}_{m_1}-3) \mathcal{L}_{m_1}+\pi ^2+48\right)}{384 m_1^2}
\end{aligned}$}
 & 
\resizebox{.47\hsize}{!}{ $\begin{aligned}[t]& \frac{1}{12} C_A C_F \left\lbrace6 \mathcal{L}_{m_1} \left(\beta_0 \mathcal{L}_{m_1}-2 \left(12 C_A C_F+\beta_0\right)\right)+\left(24+\pi ^2\right) \beta_0 \right.\\&\left.+168 C_A C_F\right\rbrace  +\frac{1}{64} Y_f^2 C_A \left\lbrace6 \mathcal{L}_{m_1} \left(88 C_A C_F+6 \beta_0-3 \beta_0\mathcal{L}_{m_1}\right)-\left(68+3 \pi ^2\right) \beta_0 \right.\\&\left. -664 C_A C_F\right\rbrace  + \frac{1}{96} Y_f Y_s C_A \left\lbrace6 \mathcal{L}_{m_1} \left(\beta_0 \mathcal{L}_{m_1}-4 \left(6C_A C_F+\beta_0\right)\right) +240 C_A C_F \right.\\&\left. +\left(48+\pi ^2\right) \beta_0\right\rbrace  +\frac{1}{32} Y_f^3 Y_s (6 \mathcal{L}_{m_1}-13) C_A^2+\frac{9}{32} Y_f^4 (5-3 \mathcal{L}_{m_1}) C_A^2 \end{aligned}$}
 \\ [0.75ex] \hline
 $\bar{\psi}_2\chi_1$  & 
\resizebox{.27\hsize}{!}{ $\begin{aligned}[t]& \frac{C_A }{24 m_1^2}\left\lbrace6 m_1^2 C_F \mathcal{L}_{m_1}^2+\left(48+\pi ^2\right) m_1^2 C_F \right.\\&\left. -3 Y_s^2(\mathcal{L}_{m_1}-1)\right\rbrace \end{aligned}$}
 & 
\resizebox{.4\hsize}{!}{ $\begin{aligned}[t]& (V^{(m_1)}_5+\frac{1}{2}F^{(m_1,0)}_{\chi}+\frac{1}{2}F^{(0)}_{h}) \\&+\frac{1}{12} C_A^2 C_F^2 \left(3 \mathcal{L}_{m_1} \left(4 (\mathcal{L}_{m_1}-5) \mathcal{L}_{m_1}+\pi ^2+56\right)\right.\\&\left.+6\zeta_3-5 \pi ^2-216\right) + \frac{1}{64} C_A^2 C_FY_f^2 \left\lbrace-2 \mathcal{L}_{m_1} \left(6 \mathcal{L}_{m_1} (2 \mathcal{L}_{m_1}-9)\right.\right.\\&\left.\left.+3 \pi ^2+142\right) +9\pi ^2+350+6 \psi ^{(2)}(1)\right\rbrace
\end{aligned}$}
 & 
\resizebox{.42\hsize}{!}{ $\begin{aligned}[t]& \frac{1}{24} C_A C_F \left\lbrace-144 C_A C_F \mathcal{L}_{m_1}+168 C_A C_F+2 \beta_0\mathcal{L}_{m_1}^3+\left(48+\pi ^2\right) \beta_0 \mathcal{L}_{m_1}\right.\\&\left.+4\beta_0 (\zeta_3-24)\right\rbrace  +\frac{Y_s^2 C_A}{96 m_1^2}  \left\lbrace6 \mathcal{L}_{m_1} \left(20 C_A C_F+2 \beta_0-\beta_0\mathcal{L}_{m_1}\right)\right.\\&\left.-168 C_A C_F-\left(24+\pi ^2\right) \beta_0\right\rbrace -\frac{Y_s^4 (2 \mathcal{L}_{m_1}-3) C_A^2}{32 m_1^4}\end{aligned}$}
 \\ [1ex] 
 \hline
\end{tabular}}
\caption{Matching corrections, $G(\mu)$, to the threshold form-factor for $m_1\gg m_2\gg M\gg Q$ at $\mu\sim m_1$, $a\equiv \alpha/(4\pi)$, and $\mathcal{L}_{m_{1}}\equiv \log{m_{1}^2/\mu^2}$. $\Delta \tilde{G}^{(2)}$ are the mass and coupling renormalisation contributions contributing at two-loop order. $V^{(m_1)}_i$ and $F_I$ are two loop vertex and wave-function corrections, given in Appendices \ref{sec:vertcorr} and \ref{sec:wfr2}. }
\label{tab:expg}
\end{table}
We can then calculate the matching coefficient, $\exp{[\tilde{B}(\mu)]}$, as the full theory vertex and wave-function corrections with IR scales, $m_1$, $M$ and $Q$, set to zero. The results of the vertex and wave-function contributions, $\exp{\tilde{B}(\mu)}$, as well as the anomalous dimension, $\tilde{\gamma}_3$, between $m_2$ and $m_1$ are given in table \ref{tab:expbt}. Moreover, the coupling and mass renormalisation corrections that contribute at two-loop order are also given, in table \ref{tab:expbt}. What remains then is to evaluate the matching at $\mu\sim m_1$ as the final matching and running, $\exp{[U(\mu)]}$ and $\gamma_3$, at $M$ is identical to the previous case. The theory above, $\mu>m_1$, is the effective theory with particle $2$ taken to be a heavy field and the theory below, $\mu<m_1$, is heavy particle effective theory where both particles $1$ and $2$ are taken to be heavy and the IR scale being the bosonic masses are set to zero. The theory below $m_1$ is scaleless and thus does not contribute to the matching but the theory above $m_1$ is one of two scales, $m_1$ and $w'=p_1 \cdot v_2$. However, $w'$ is integrated out at leading order in the threshold limit as,
\begin{equation}
    w'=p_1\cdot v_2= \frac{m_1^2+m_2^2-Q^2}{2m_2}\sim \frac{m_2}{2},
\end{equation}
and thus, we obtained the matching and wave-function contributions, $\exp{[G(\mu)]}$, with logarithms of a single scale, $m_1$, and these are presented  up to two-loops in table \ref{tab:expg}. As for the coupling and mass renormalisation corrections that also contribute at two-loop order, we present these results in table \ref{tab:expg}. With the above results, due to their generality one can map our results to operators in models that are similar to the SU(N)-Higgs model we discuss here, including those with spontaneous symmetry breaking at a certain scale.
\section{Application to the Standard Model}
The results we have obtained for our SU(N)-Higgs model can now be used to compute results for the SM. We dedicate this treatment to illustrating how the mapping from our to other models of a similar type, which may exhibit SSB, can be achieved. When considering the SM, one must select the correct coupling constants with care, since, it is a chiral gauge theory, and our model is vector-like.
One can then obtain results for more than two external particles by combining the two-particle results computed in this paper with the appropriate gauge theory factors included. We focus here on
how our results can be used to calculate the radiative corrections to quark and charged lepton production by gauge-invariant currents, $\bar{Q}_i\gamma_{\mu}P_LQ_i$ and $\bar{L}\gamma_{\mu}P_LL$, respectively, where $Q_i$ is the quark doublet for generation, $i=u,c,t$, with only the top quark mass, $m_t$, taken to be a non-zero fermion mass.
\subsection{Light Quarks}
Let us begin by considering the representation of light quarks in the SM \cite{schwartz2014quantum}. The first generation of the quark doublet in the mass eigenstate basis is given by,
\begin{equation}
    Q_u=\left[ \begin{array}{c} u \\ d' \end{array} \right] = \begin{bmatrix} t \\ V_{ud}d+V_{us}s+V_{ub}b \end{bmatrix}.
    \label{eqn:udoub}
\end{equation}
At the scale, $Q\gg m_q$, in the full electro-weak theory, the operator coefficient is assumed to be unity. For the first generation, all quark masses and Yukawa couplings can be neglected, and so the matching is given by combining the gauge boson contributions computed earlier.
The operator in SCET at the scale $Q$ is,
\begin{equation}
    \bar{Q}_u\gamma_{\mu}P_LQ_u\rightarrow c(Q)\left[\bar{\xi}^{(Q_u)}_{n,p_2}W_n\right]\gamma_{\mu}P_L\left[ W_{\bar{n}}^{\dagger}\xi^{(Q_u)}_{\bar{n},p_1}\right],
    \label{eqn:udoub2}
\end{equation}
where $\xi^{(Q_u)}$ represents the left-handed EW quark doublet of \eqref{eqn:udoub} in SCET. Thus, the matching condition, $c(\mu)$ at the scale $\mu=Q$ with $\mathcal{L}_Q=0$ is,
\begin{equation}
    \log{c(Q)}=a_{EW}(Q)\log{c^{(1)}(Q)}+a_{EW}(Q)^2\log{c^{(2)}(Q)}+\mathcal{O}(a_{EW}^3),
\end{equation}
where,
\begin{equation}
    a_{EW}(\mu)=\left(\frac{\alpha_s(\mu)}{4\pi}\frac{4}{3}+\frac{\alpha_2(\mu)}{4\pi}\frac{3}{4}+\frac{\alpha_1(\mu)}{4\pi}\frac{1}{36}\right).
\end{equation}
The gauge couplings have been multiplied by the corresponding gauge factors, $C_F$, which are $4/3$ for an SU(3) triplet, $3/4$ for an SU(2) doublet, and $1/36$ for $Y = 1/6$. Moreover, the electroweak couplings, renormalized at $\mu = M_Z$, are given by \cite{denner1993techniques},
\begin{align}
    &\alpha_1(M_Z)=\frac{\alpha_{em}(M_Z)}{\cos^2{\theta_W}}, \nonumber \\ & \alpha_2(M_Z)=\frac{\alpha_{em}(M_Z)}{\sin^2{\theta_W}},
\end{align}
and their values at $\mu\sim Q$ are obtained by the usual $\beta$-functions of the SM. The theory below $Q$ is SCET with an $\text{SU(3)}\times\text{SU(2)}\times\text{U(1)}$ gauge symmetry. In this regime, the SCET current in \eqref{eqn:udoub2} is multiplicatively renormalised with the anomalous dimension $\gamma(\mu)$ given by,
\begin{equation}
    \log{\gamma(\mu)}=a_{EW}(\mu){\gamma^{(1)}_1(\mu)}+a_{EW}(\mu)^2{\gamma^{(2)}_1(\mu)}+\mathcal{O}(a_{EW}^3).
\end{equation}
The anomalous dimension, $\gamma(\mu)$, is used to run $c(\mu)$ down to a scale of order the gauge boson mass. One can integrate out the weak gauge bosons sequentially, by first integrating out the $Z$-boson at $\mu=M_Z$ , followed by the $W$-boson at $\mu=M_W$. This is not a good choice to use for the SM, as $M_W/M_Z$ is not negligible, and summing powers of $M_W/M_Z$ is more important than summing $\alpha\log^2{M_W/M_Z}$ terms. Instead, we integrate out the $W$ and $Z$ at a common scale, chosen to be $\mu=M_Z$. In this way, we match directly from an $\text{SU(3)}\times \text{SU(2)}\times \text{U(1)}$ gauge theory onto a $\text{SU(3)}\times  \text{U(1)}_{em}$ gauge theory of gluons and photons, which lacks the complications of an intermediate stage of broken EW symmetry where $Z$ is integrated out, but not the $W$. Moreover, the Higgs corrections for light particles are sub-leading as the Yukawa coupling is proportional to the light mass and thus, are suppressed. At the scale $\mu=M_Z$, integrating out the $W$ and $Z$ bosons give a matching correction to the SCET operator,
\begin{align}
    \left[\bar{\xi}^{(Q_u)}_{n,p_2}W_n\right]\gamma_{\mu}P_L\left[ W_{\bar{n}}^{\dagger}\xi^{(Q_u)}_{\bar{n},p_1}\right] \rightarrow & d^{(u)}\left[\bar{\xi}^{(u)}_{n,p_2}W_n\right]\gamma_{\mu}P_L\left[ W_{\bar{n}}^{\dagger}\xi^{(u)}_{\bar{n},p_1}\right] + \nonumber \\& d^{(d')}\left[\bar{\xi}^{(d')}_{n,p_2}W_n\right]\gamma_{\mu}P_L\left[ W_{\bar{n}}^{\dagger}\xi^{(d')}_{\bar{n},p_1}\right].
    \label{eqn:uddoub1}
\end{align}
Since the EW symmetry is broken, the $u$ and$\phantom{x}d'$ parts of the operator get different matching corrections. The matching corrections are as follows,
\begin{align}
    \log{d^{(u)}}(M_Z)=& a_1\log{d^{(1)}(M_W)}+a_1^2\log{d^{(2)}(M_W)}+\mathcal{O}(a_{1}^3)\nonumber \\& + a_2\log{d^{(1)}(M_Z)}+a_2^2\log{d^{(2)}(M_Z)}+\mathcal{O}(a_{2}^3)
\end{align}
where the terms proportional to $a_1$ and $a_2$ correspond the $Z$ and $W$ contributions, respectively, and,
\begin{align}
    & a_1=\frac{\alpha_{em}}{4\pi\sin^2{\theta_W}\cos^2{\theta_W}}\left(\frac{1}{2}-\frac{2}{3}\sin^2{\theta_W}\right)^2 \\ &
    a_2=\frac{\alpha_{em}}{4\pi\sin^2{\theta_W}\cos^2{\theta_W}}\left(\frac{1}{2}\right)^2.
\end{align}
Below $M_Z$, the operators in \eqref{eqn:udoub} are multiplicatively renormalised, with anomalous dimensions,
\begin{align}
   & \gamma^{(u)}(\mu)=\tilde{a}_1(\mu){\gamma^{(1)}_1(\mu)}+\tilde{a}_1(\mu)^2{\gamma^{(2)}_1(\mu)}+\mathcal{O}(\tilde{a}_1^3), \\ &
   \gamma^{(d')}(\mu)=\tilde{a}_2(\mu){\gamma^{(1)}_1(\mu)}+\tilde{a}_2(\mu)^2{\gamma^{(2)}_1(\mu)}+\mathcal{O}(\tilde{a}_2^3)
\end{align}
such that
\begin{equation}
    \tilde{a}_1(\mu)=\left\lbrace \frac{\alpha_s(\mu)}{4\pi}\frac{4}{3}+\frac{\alpha_{em}(\mu)}{4\pi}\frac{4}{9} \right\rbrace \quad\text{and}\quad \tilde{a}_2(\mu)=\left\lbrace \frac{\alpha_s(\mu)}{4\pi}\frac{4}{3}+\frac{\alpha_{em}(\mu)}{4\pi}\frac{1}{9} \right\rbrace,
\end{equation}
for the $u$ and $d'$ terms. The final result for the operator at a low scale is thus,
\begin{align}
        \left[\bar{\xi}^{(Q_u)}_{n,p_2}W_n\right]\gamma_{\mu}P_L\left[ W_{\bar{n}}^{\dagger}\xi^{(Q_u)}_{\bar{n},p_1}\right] \rightarrow & c^{(u)}\left[\bar{\xi}^{(u)}_{n,p_2}W_n\right]\gamma_{\mu}P_L\left[ W_{\bar{n}}^{\dagger}\xi^{(u)}_{\bar{n},p_1}\right] + \nonumber \\& c^{(d')}\left[\bar{\xi}^{(d')}_{n,p_2}W_n\right]\gamma_{\mu}P_L\left[ W_{\bar{n}}^{\dagger}\xi^{(d')}_{\bar{n},p_1}\right],
        \label{eqn:uddoub2}
\end{align}
with
\begin{align}
    &\log{c^{u}}(\mu)=\log{c(Q)}+\int_Q^{M_Z}\frac{d\mu}{\mu}\gamma(\mu)+\log{{d^{(u)}}(M_Z)}+\int_{M_Z}^{\mu}\frac{d\mu}{\mu}\gamma^{(u)}(\mu) \\&
    \log{c^{d}}(\mu)=\log{c(Q)}+\int_Q^{M_Z}\frac{d\mu}{\mu}\gamma(\mu)+\log{{d^{(d')}}(M_Z)}+\int_{M_Z}^{\mu}\frac{d\mu}{\mu}\gamma^{(d')}(\mu).
    \label{eqn:lqmatch}
\end{align}
The EFT operator in \eqref{eqn:uddoub2} can then be used to compute processes such as dijet production using SCET \cite{bauer2003enhanced}. For jet production, the renormalisation scale, $\mu$, would be chosen to be of order the jet invariant mass, or $30 \text{ GeV}$ for jets at the LHC.
\subsection{Leptons}
The computation for the radiative corrections to the lepton current, $\bar{L}\gamma_{\mu}P_L L$, where $L$ is the lepton doublet,
\begin{equation}
        L=\left( \begin{array}{c} \nu \\ l \end{array} \right),
\end{equation}
is identical to that for the quark doublet, aside from a few replacements. The full theory operator at the low scale, $\mu$, is,
\begin{align}
      \bar{L}\gamma_{\mu}P_LL\rightarrow & c^{(\nu)}\left[\bar{\xi}^{(\nu)}_{n,p_2}W_n\right]\gamma_{\mu}P_L\left[ W_{\bar{n}}^{\dagger}\xi^{(\nu)}_{\bar{n},p_1}\right] + \nonumber \\& c^{(l)}\left[\bar{\xi}^{(l)}_{n,p_2}W_n\right]\gamma_{\mu}P_L\left[ W_{\bar{n}}^{\dagger}\xi^{(l)}_{\bar{n},p_1}\right],
\end{align}
with the coefficients given by \eqref{eqn:lqmatch} and replacements $u\rightarrow \nu$, $d'\rightarrow l$, along with different gauge theory factors which implies the following coupling replacements,
\begin{align}
    &a_{EW}(\mu)\rightarrow a_{EW}'(\mu)=\left(\frac{\alpha_2(\mu)}{4\pi}\frac{3}{4}+\frac{\alpha_1(\mu)}{4\pi}\frac{1}{4}\right),\\&
    a_1\rightarrow a_1'=\frac{\alpha_{em}}{4\pi\sin^2{\theta_W}\cos^2{\theta_W}}\left(\frac{1}{2}\right)^2,\\&
    a_2\rightarrow a_2'=\frac{\alpha_{em}}{4\pi\sin^2{\theta_W}\cos^2{\theta_W}}\left(-\frac{1}{2}+\sin^2{\theta_W}\right)^2,\\&
    \tilde{a}_1(\mu)\rightarrow \tilde{a}_1'(\mu)=0, \\&  
    \tilde{a}_2(\mu)\rightarrow \tilde{a}_2'(\mu)=\frac{\alpha_{em}(\mu)}{4\pi},
\end{align}
which provides us with the leptonic equivalent of the previous result.
\subsection{Top Quarks}
In this subsection, we show how our results can be used to compute the radiative corrections to $t\bar{t}$-production by a gauge-invariant vector current $\bar{Q}_t\gamma_\mu P_LQ_t$, where $Q_t$ is the left-handed quark doublet in the SM. We may write the quark doublet in the mass eigenstate basis,
\begin{equation}
        Q_t=\left(\begin{array}{c} t \\ b' \end{array} \right)=\left[ \begin{array}{c} t \\ V_{td}d+V_{ts}s+V_{tb}b \end{array} \right].
        \label{eqn:tdoub}
\end{equation}
We will neglect all quark masses other than $m_t$. This example is useful as it illustrates how to use the fermion mass and Higgs exchange contributions computed in our model. We will examine both the Sudakov and threshold regimes in this case as they are both available to us in this example.
\subsubsection{Sudakov Regime}
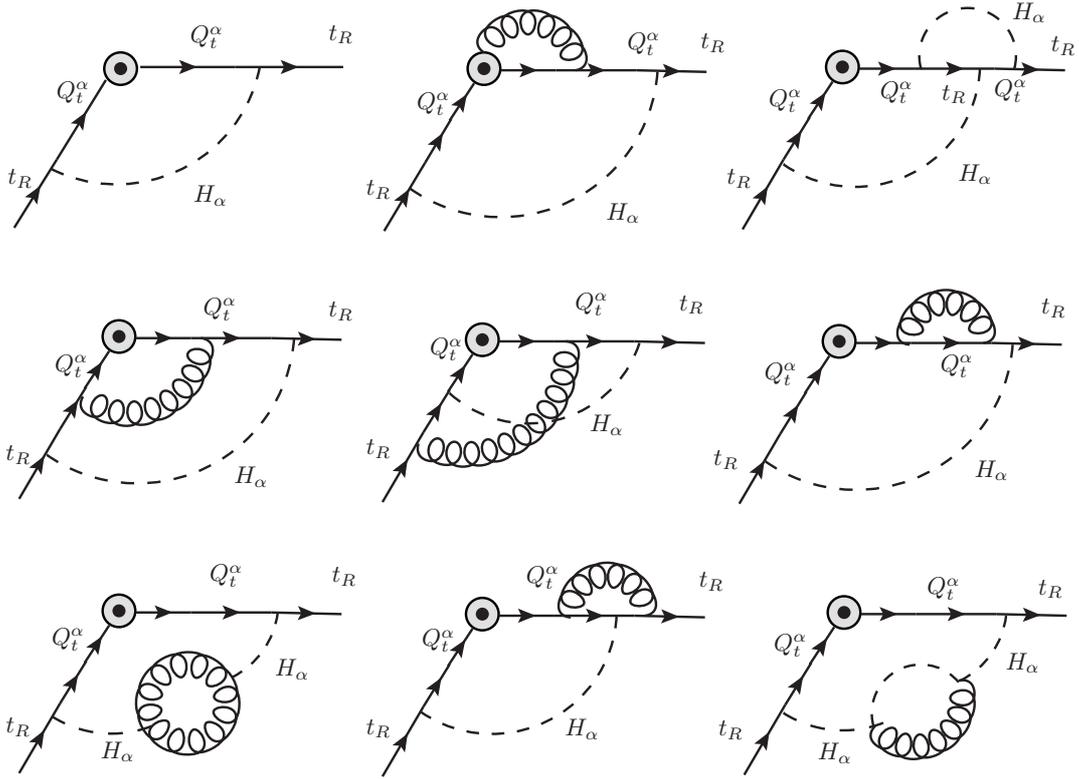
\begin{figure}[tb]
\begin{center}
\scalebox{0.91}{
\fcolorbox{white}{white}{
  \begin{picture}(451,327) (3,-3)
    \SetWidth{1.0}
    \SetColor{Black}
    \Vertex(48.279,289.675){3.353}
    \GOval(48.279,289.675)(6.705,6.705)(0){0.882}
    \Line[arrow,arrowpos=0.5,arrowlength=5,arrowwidth=2,arrowinset=0.2](4.694,223.961)(22.128,252.124)
    \Line[arrow,arrowpos=0.5,arrowlength=5,arrowwidth=2,arrowinset=0.2](22.128,252.124)(42.244,284.981)
    \Line[arrow,arrowpos=0.5,arrowlength=5,arrowwidth=2,arrowinset=0.2](56.326,290.345)(95.217,290.345)
    \Line[arrow,arrowpos=0.5,arrowlength=5,arrowwidth=2,arrowinset=0.2](95.217,290.345)(139.473,290.345)
    \Arc[dash,dashsize=5.364](46.616,301.637)(59.736,-116.335,-10.896)
    \Text(2.012,241.396)[lb]{{\Black{$t_R$}}}
    \Text(22.128,276.264)[lb]{{\Black{$Q_t^{\alpha}$}}}
    \Text(77.112,299.062)[lb]{{\Black{$Q_t^{\alpha}$}}}
    \Text(134.109,301.074)[lb]{{\Black{$t_R$}}}
    \Text(78.454,234.02)[lb]{{\Black{$H_{\alpha}$}}}
    \Vertex(48.279,289.675){2.682}
    \GOval(197.14,289.675)(6.705,6.705)(0){0.882}
    \Vertex(197.14,289.675){2.682}
    \Line[arrow,arrowpos=0.5,arrowlength=5,arrowwidth=2,arrowinset=0.2](183.058,268.217)(193.116,283.64)
    \Line[arrow,arrowpos=0.5,arrowlength=5,arrowwidth=2,arrowinset=0.2](173,250.783)(183.729,268.888)
    \Line[arrow,arrowpos=0.5,arrowlength=5,arrowwidth=2,arrowinset=0.2](156.237,222.62)(173,250.783)
    \Line[arrow,arrowpos=0.5,arrowlength=5,arrowwidth=2,arrowinset=0.2](203.845,289.004)(226.644,289.004)
    \Line[arrow,arrowpos=0.5,arrowlength=5,arrowwidth=2,arrowinset=0.2](226.644,289.004)(259.5,289.004)
    \Line[arrow,arrowpos=0.5,arrowlength=5,arrowwidth=2,arrowinset=0.2](259.5,289.004)(288.334,288.334)
    \Arc[dash,dashsize=5.364](203.996,292.706)(64.37,-125.119,-3.895)
    \GluonArc[clock](215.277,289.446)(19.418,159.079,-1.305){5.029}{6}
    \Text(286.992,297.721)[lb]{{\Black{$t_R$}}}
    \Text(256.818,296.38)[lb]{{\Black{$Q_t^{\alpha}$}}}
    \Text(170.318,271.57)[lb]{{\Black{$Q_t^{\alpha}$}}}
    \Text(149.531,236.702)[lb]{{\Black{$t_R$}}}
    \Text(248.101,226.644)[lb]{{\Black{$H_{\alpha}$}}}
    \GOval(343.989,290.345)(6.705,6.705)(0){0.882}
    \Vertex(343.989,290.345){2.682}
    \Line[arrow,arrowpos=0.5,arrowlength=5,arrowwidth=2,arrowinset=0.2](329.907,268.888)(339.965,284.31)
    \Line[arrow,arrowpos=0.5,arrowlength=5,arrowwidth=2,arrowinset=0.2](319.849,252.124)(330.578,269.558)
    \Line[arrow,arrowpos=0.5,arrowlength=5,arrowwidth=2,arrowinset=0.2](303.086,223.291)(319.849,252.124)
    \Line[arrow,arrowpos=0.5,arrowlength=5,arrowwidth=2,arrowinset=0.2](350.694,289.675)(373.493,289.675)
    \Line[arrow,arrowpos=0.5,arrowlength=5,arrowwidth=2,arrowinset=0.2](373.493,289.675)(406.349,289.675)
    \Line[arrow,arrowpos=0.5,arrowlength=5,arrowwidth=2,arrowinset=0.2](406.349,289.675)(435.183,289.004)
    \Arc[dash,dashsize=5.364](349.222,291.982)(51.145,-125.051,-2.586)
    \Arc[dash,dashsize=5.364,clock](395.285,294.903)(19.813,-164.699,-375.301)
    \GOval(47.609,179.035)(6.705,6.705)(0){0.882}
    \Vertex(47.609,179.035){2.682}
    \Line[arrow,arrowpos=0.5,arrowlength=5,arrowwidth=2,arrowinset=0.2](33.527,157.578)(43.585,173)
    \Line[arrow,arrowpos=0.5,arrowlength=5,arrowwidth=2,arrowinset=0.2](23.469,140.814)(34.198,158.248)
    \Line[arrow,arrowpos=0.5,arrowlength=5,arrowwidth=2,arrowinset=0.2](6.705,111.981)(23.469,140.814)
    \Line[arrow,arrowpos=0.5,arrowlength=5,arrowwidth=2,arrowinset=0.2](54.314,178.364)(77.112,178.364)
    \Line[arrow,arrowpos=0.5,arrowlength=5,arrowwidth=2,arrowinset=0.2](77.112,178.364)(109.969,178.364)
    \Line[arrow,arrowpos=0.5,arrowlength=5,arrowwidth=2,arrowinset=0.2](109.969,178.364)(138.802,177.694)
    \GOval(196.469,177.694)(6.705,6.705)(0){0.882}
    \Vertex(196.469,177.694){2.682}
    \Line[arrow,arrowpos=0.5,arrowlength=5,arrowwidth=2,arrowinset=0.2](182.388,156.237)(192.446,171.659)
    \Line[arrow,arrowpos=0.5,arrowlength=5,arrowwidth=2,arrowinset=0.2](172.33,139.473)(183.058,156.907)
    \Line[arrow,arrowpos=0.5,arrowlength=5,arrowwidth=2,arrowinset=0.2](155.566,110.64)(172.33,139.473)
    \Line[arrow,arrowpos=0.5,arrowlength=5,arrowwidth=2,arrowinset=0.2](203.175,177.023)(225.973,177.023)
    \Line[arrow,arrowpos=0.5,arrowlength=5,arrowwidth=2,arrowinset=0.2](225.973,177.023)(258.83,177.023)
    \Line[arrow,arrowpos=0.5,arrowlength=5,arrowwidth=2,arrowinset=0.2](258.83,177.023)(287.663,176.353)
    \Arc[dash,dashsize=5.364](55.135,182.737)(64.37,-125.119,-3.895)
    \GluonArc(51.967,176.353)(29.238,-130.815,3.945){5.029}{8}
    \GluonArc(189.152,172.102)(41.805,-116.777,6.761){5.029}{11}
    \Arc[dash,dashsize=5.364](215.269,190.959)(47.856,-132.305,-17.771)
    \GOval(342.648,177.023)(6.705,6.705)(0){0.882}
    \Vertex(342.648,177.023){2.682}
    \Line[arrow,arrowpos=0.5,arrowlength=5,arrowwidth=2,arrowinset=0.2](328.566,155.566)(338.624,170.989)
    \Line[arrow,arrowpos=0.5,arrowlength=5,arrowwidth=2,arrowinset=0.2](318.508,138.802)(329.237,156.237)
    \Line[arrow,arrowpos=0.5,arrowlength=5,arrowwidth=2,arrowinset=0.2](301.744,109.969)(318.508,138.802)
    \Line[arrow,arrowpos=0.5,arrowlength=5,arrowwidth=2,arrowinset=0.2](349.353,176.353)(372.151,176.353)
    \Line[arrow,arrowpos=0.5,arrowlength=5,arrowwidth=2,arrowinset=0.2](372.151,176.353)(405.008,176.353)
    \Line[arrow,arrowpos=0.5,arrowlength=5,arrowwidth=2,arrowinset=0.2](405.008,176.353)(433.841,175.682)
    \Arc[dash,dashsize=5.364](349.504,180.055)(64.37,-125.119,-3.895)
    \GluonArc[clock](386.524,177.994)(15.22,-171.263,-366.19){5.029}{6}
    \Text(385.562,162.942)[lb]{{\Black{$Q_t^{\alpha}$}}}
    \Text(427.136,187.752)[lb]{{\Black{$t_R$}}}
    \Text(292.357,124.721)[lb]{{\Black{$t_R$}}}
    \Text(313.144,159.589)[lb]{{\Black{$Q_t^{\alpha}$}}}
    \GOval(47.609,65.713)(6.705,6.705)(0){0.882}
    \Vertex(47.609,65.713){2.682}
    \Line[arrow,arrowpos=0.5,arrowlength=5,arrowwidth=2,arrowinset=0.2](33.527,44.256)(43.585,59.678)
    \Line[arrow,arrowpos=0.5,arrowlength=5,arrowwidth=2,arrowinset=0.2](23.469,27.492)(34.198,44.926)
    \Line[arrow,arrowpos=0.5,arrowlength=5,arrowwidth=2,arrowinset=0.2](6.705,-1.341)(23.469,27.492)
    \Line[arrow,arrowpos=0.5,arrowlength=5,arrowwidth=2,arrowinset=0.2](54.314,65.043)(77.112,65.043)
    \Line[arrow,arrowpos=0.5,arrowlength=5,arrowwidth=2,arrowinset=0.2](77.112,65.043)(109.969,65.043)
    \Line[arrow,arrowpos=0.5,arrowlength=5,arrowwidth=2,arrowinset=0.2](109.969,65.043)(138.802,64.372)
    \GOval(196.469,64.372)(6.705,6.705)(0){0.882}
    \Vertex(196.469,64.372){2.682}
    \Line[arrow,arrowpos=0.5,arrowlength=5,arrowwidth=2,arrowinset=0.2](182.388,42.915)(192.446,58.337)
    \Line[arrow,arrowpos=0.5,arrowlength=5,arrowwidth=2,arrowinset=0.2](172.33,26.151)(183.058,43.585)
    \Line[arrow,arrowpos=0.5,arrowlength=5,arrowwidth=2,arrowinset=0.2](155.566,-2.682)(172.33,26.151)
    \Line[arrow,arrowpos=0.5,arrowlength=5,arrowwidth=2,arrowinset=0.2](203.175,63.702)(225.973,63.702)
    \Line[arrow,arrowpos=0.5,arrowlength=5,arrowwidth=2,arrowinset=0.2](225.973,63.702)(258.83,63.702)
    \Line[arrow,arrowpos=0.5,arrowlength=5,arrowwidth=2,arrowinset=0.2](258.83,63.702)(287.663,63.031)
    \Arc[dash,dashsize=5.364](200.361,66.009)(51.145,-125.051,-2.586)
    \GluonArc[clock](247.722,65.343)(15.22,-171.263,-366.19){5.029}{6}
    \GOval(344.659,65.043)(6.705,6.705)(0){0.882}
    \Vertex(344.659,65.043){2.682}
    \Line[arrow,arrowpos=0.5,arrowlength=5,arrowwidth=2,arrowinset=0.2](330.578,43.585)(340.636,59.008)
    \Line[arrow,arrowpos=0.5,arrowlength=5,arrowwidth=2,arrowinset=0.2](320.52,26.822)(331.248,44.256)
    \Line[arrow,arrowpos=0.5,arrowlength=5,arrowwidth=2,arrowinset=0.2](303.756,-2.012)(320.52,26.822)
    \Line[arrow,arrowpos=0.5,arrowlength=5,arrowwidth=2,arrowinset=0.2](351.365,64.372)(374.163,64.372)
    \Line[arrow,arrowpos=0.5,arrowlength=5,arrowwidth=2,arrowinset=0.2](374.163,64.372)(407.02,64.372)
    \Line[arrow,arrowpos=0.5,arrowlength=5,arrowwidth=2,arrowinset=0.2](407.02,64.372)(435.853,63.702)
    \Arc[dash,dashsize=5.364](344.243,61.539)(45.215,-122.651,-68.585)
    \Arc[dash,dashsize=5.364](376.465,67.924)(34.835,-63.022,-6.962)
    \Text(392.938,242.737)[lb]{{\Black{$H_{\alpha}$}}}
    \Text(415.066,309.791)[lb]{{\Black{$H_{\alpha}$}}}
    \Text(431.159,296.38)[lb]{{\Black{$t_R$}}}
    \Text(407.69,276.934)[lb]{{\Black{$Q_t^{\alpha}$}}}
    \Text(360.752,276.264)[lb]{{\Black{$Q_t^{\alpha}$}}}
    \Text(386.233,276.264)[lb]{{\Black{$t_R$}}}
    \Text(297.721,241.396)[lb]{{\Black{$t_R$}}}
    \Text(315.155,272.911)[lb]{{\Black{$Q_t^{\alpha}$}}}
    \Text(235.361,188.423)[lb]{{\Black{$Q_t^{\alpha}$}}}
    \Text(175.682,169.647)[lb]{{\Black{$Q_t^{\alpha}$}}}
    \Text(149.531,129.415)[lb]{{\Black{$t_R$}}}
    \Text(278.275,189.093)[lb]{{\Black{$t_R$}}}
    \Text(134.109,186.411)[lb]{{\Black{$t_R$}}}
    \Text(1.341,126.733)[lb]{{\Black{$t_R$}}}
    \Text(21.457,162.942)[lb]{{\Black{$Q_t^{\alpha}$}}}
    \Text(82.477,186.411)[lb]{{\Black{$Q_t^{\alpha}$}}}
    \Arc[dash,dashsize=5.364](44.51,60.198)(45.215,-122.651,-68.585)
    \Arc[dash,dashsize=5.364](78.073,69.265)(34.835,-63.022,-6.962)
    \GluonArc(75.771,27.492)(16.121,-17,343){5.029}{12}
    \Arc[dash,dashsize=5.364,clock](376.09,23.446)(20.084,-164.578,-320.543)
    \GluonArc(373.628,29.078)(19.59,-145.935,23.471){5.029}{7}
    \Text(399.644,120.027)[lb]{{\Black{$H_{\alpha}$}}}
    \Text(241.396,138.802)[lb]{{\Black{$H_{\alpha}$}}}
    \Text(95.217,117.345)[lb]{{\Black{$H_{\alpha}$}}}
    \Text(40.233,3.353)[lb]{{\Black{$H_{\alpha}$}}}
    \Text(111.981,37.55)[lb]{{\Black{$H_{\alpha}$}}}
    \Text(231.337,12.74)[lb]{{\Black{$H_{\alpha}$}}}
    \Text(335.272,2.682)[lb]{{\Black{$H_{\alpha}$}}}
    \Text(412.384,40.233)[lb]{{\Black{$H_{\alpha}$}}}
    \Text(380.198,70.407)[lb]{{\Black{$Q_t^{\alpha}$}}}
    \Text(317.167,46.938)[lb]{{\Black{$Q_t^{\alpha}$}}}
    \Text(294.368,12.74)[lb]{{\Black{$t_R$}}}
    \Text(425.795,71.078)[lb]{{\Black{$t_R$}}}
    \Text(286.322,74.43)[lb]{{\Black{$t_R$}}}
    \Text(149.531,13.411)[lb]{{\Black{$t_R$}}}
    \Text(215.244,75.101)[lb]{{\Black{$Q_t^{\alpha}$}}}
    \Text(172.33,48.95)[lb]{{\Black{$Q_t^{\alpha}$}}}
    \Text(85.159,75.771)[lb]{{\Black{$Q_t^{\alpha}$}}}
    \Text(20.116,49.62)[lb]{{\Black{$Q_t^{\alpha}$}}}
    \Text(135.45,76.442)[lb]{{\Black{$t_R$}}}
    \Text(1.341,14.081)[lb]{{\Black{$t_R$}}}
  \end{picture}}
}
\end{center}
    \caption{Vertex contributions to matching, $c_R(\mu)$, at one and two loop order. Higgs exchanges cause $\bar{Q}_t\gamma_{\mu}P_LQ_t$ to mix with $\bar{t}_R\gamma_{\mu}P_R t$ and the index, $\alpha$, is the fundamental SU(2) and index and is summed over.}
    \label{fig:trgraph}
\end{figure}

In the Sudakov regime, the operator in SCET at the scale, $\mu\sim Q$, is as follows,
\begin{align}
     \bar{Q}_t\gamma_{\mu}P_LQ_t\rightarrow & c_{L}(Q)\left[\bar{\xi}^{(Q_t)}_{n,p_2}W_n\right]\gamma_{\mu}P_L\left[ w_{\bar{n}}^{\dagger}\xi^{(Q_t)}_{\bar{n},p_1}\right] + \nonumber \\& c_R(Q)\left[\bar{\xi}^{(t)}_{n,p_2}W_n\right]\gamma_{\mu}P_R\left[ w_{\bar{n}}^{\dagger}\xi^{(t)}_{\bar{n},p_1}\right],
     \label{eqn:tdoub1}
\end{align}
where $\xi^{(Q_t)}$ and $\xi^{(t)}$ represent the left-handed and right-handed $t$-quark doublet, \eqref{eqn:tdoub}, and singlet, $t_R$, respectively, in SCET with gauge indices suppressed. The reason $t_R$ appears in this case is due to Higgs exchange graphs which are chiral in the SM, and have been computed in our model value which is a vector-like theory, thus when mapping to the SM we must plaster on the fact that the Yukawa vertex flips the fermion chirality. Practically, the Higgs exchange vertex correction mixes the $Q_L$ operator with the $t_R$ operator. The matching condition at the scale $\mu=Q$ is then given by $c_{L/R}(Q)$, where one splits the left and right handed contributions of $c(Q)$ which now has non-zero Yukawa coupings.  Hence, $c_R(Q)$ includes all terms which arise from Higgs exchange graphs of type illustrated in figure \ref{fig:trgraph}; and the remaining graphs contribute to $c_L(Q)$.
Note further that one must include appropriate factors of two for terms in $c_{L/R}$ arising from summing over each closed SU(2) index loop, i.e. because both the Higgs and $Q_t$ are SU(2) doublets. As for the wavefunction correction, the $t_L$ and $t_R$ field renormalisation contributions which include Higgs exchange must also include appropriate factors of two from loops with SU(2) index summation. 

The theory below $Q$ is SCET with an $\text{SU(3)}\times\text{SU(2)}\times\text{U(1)}$ gauge symmetry. In this regime the two operators in \eqref{eqn:tdoub1} are multiplicatively renormalised with anomalous dimensions (again splitting chiral contributions in the same way as for matching),
\begin{align}
     \frac{dc_L(\mu)}{d\mu}=\gamma_L(\mu)c_L \quad \text{and}\quad \frac{dc_R(\mu)}{d\mu}=\gamma_R(\mu)c_R .
\end{align}
At this scale, the Higgs vertex graph, which causes $c_L/c_R$ mixing, is $1/Q^2$ suppressed. The anomalous dimension $\gamma$ is as usual used to run $c_L$  and $c_R$ down to a scale of order $m_t$. At $\mu\sim m_t$ there are several different methods one can use. Since $\lbrace m_t,M_{W,Z},M_H\rbrace$ are not widely separated, one can integrate them all out together. In this way, one goes directly from an $\text{SU(3)}\times\text{SU(2)}\times\text{U(1)}$ invariant theory to a $\text{SU(2)}\times\text{U(1)}_{em}$ gauge theory, with broken $SU(2)\times U(1)$ symmetry and no EW gauge bosons. This procedure keeps the entire mass dependence on the four mass scales. At the scale $\mu=m_t$ the $t$-quark SCET field is replaced by the heavy quark field $t_v$, whereas the $b'$ quark in the SCET field $\xi^{(Q_t)}$ remains a SCET field, $\xi^{(b')}$. The operator matching is,
\begin{align}
    & \left[\bar{\xi}^{(Q_t)}_{n,p_2}W_n\right]\gamma_{\mu}P_L\left[ W_{\bar{n}}^{\dagger}\xi^{(Q_t)}_{\bar{n},p_1}\right]\rightarrow a_1\left[\bar{\xi}^{(b')}_{n,p_2}W_n\right]\gamma_{\mu}P_L\left[ W_{\bar{n}}^{\dagger}\xi^{(b')}_{\bar{n},p_1}\right]+a_2\bar{t}_{v_2}t_{v_1}  \\&
    \left[\bar{\xi}^{(t)}_{n,p_2}W_n\right]\gamma_{\mu}P_R\left[ W_{\bar{n}}^{\dagger}\xi^{(t)}_{\bar{n},p_1}\right]\rightarrow  a_3\bar{t}_{v_2}t_{v_1},
\end{align}
where the matching coefficients, $a_i$, and are obtained using the results of the results from section \ref{ssec:mesud}. All running couplings are renormalized at $\mu=m_t$. The expressions are given by adding the contributions due to the $W/Z$,  $g,\gamma$ and $H,h^{0,+}$, where $h^{0,+}$ are the unphysical Higgs scalars present in $R_{\xi}=1$ gauge.  Lastly, below $\mu=m_t$, the $\bar{t}_{v_2}t_{v_1}$ operator has the anomalous dimension, 
\begin{equation}
    \gamma_3=a\gamma_3^{(1)}+a^2\gamma_3^{(2)}+\mathcal{O}(a^3),\quad  a=\left(\frac{4}{3}\frac{\alpha_s}{4\pi}+\frac{4}{9}\frac{\alpha_{em}}{4\pi}\right)
\end{equation}
from the fourth column of table \ref{tab:expr}, with the given group theory factor replacements. The radiative corrections to the $\bar{t}t$ operator can then be combined with known methods to obtain $t$-quark decay distributions. The QCD corrections (the $\alpha_s$ terms) have already been included in the analysis of previous work \cite{fleming2008jets}. The new results in this paper are the additional two-loop EW radiative corrections, including Higgs effects.
\subsubsection{Threshold Regime}
In the Threshold regime at $\mu\sim m_t$, and $m_t$ is the largest scale in the problem. Although the scales, $\lbrace m_t,M_{W,Z},M_H\rbrace$, are not widely separated, one can no longer integrate them all out together as in the Sudakov regime. One can integrate out the scales $m_t$, $M_{W,Z}$ and $M_H$ in various ways, e.g. one can integrate out each particle at a scale $\mu$ equal to its mass, or integrate out one or more particles simultaneously at some common value of $\mu$, as was done in Sudakov regime. The most experimentally relevant way to integrate out the relevant scales is to first integrate out the top quark as $m_t\sim \text{ 172 GeV}$ and $m_t>m_H>M_{W,Z}\gg m_b$, which leads to an effective theory that breaks $SU(2)\times U(1)$ invariance as the $b'$ quark remains along with dynamical $W/Z$ bosons. From which one integrates out the Higgs first \cite{dittmaier1996integrating}, as $M_H\sim \text{ 125 GeV}$, and then the $W/Z$ bosons at a common scale, $M_Z\sim \text{ 81 GeV}$. Otherwise if one wants to avoid breaking $SU(2)\times U(1)$ invariance, one is free to integrate out both $t$ and $b'$ at a common scale, $\mu \sim m_t$. Then one can either integrate all massive bosons at a common scale, $M_Z$ or separate into integrating out the Higgs first then the $W/Z$ at common scale. We consider the former here to illustrate and leave further analyses and numerical comparisons to upcoming work in the SM and beyond, as there is the further case of heavy-light currents to consider as well.

Integrating out both $t$ and $b'$ at the common scale $\mu=m_t$, below $m_t$ the fields are replaced by their heavy quark counterparts $t_v$ and $b_v$, respecitively. The operator matching is,
\begin{align}
    &\bar{Q}_t\gamma_{\mu}P_LQ_t\rightarrow b_1\bar{t}_{v_2}t_{v_1}+b_2\bar{b}_{v_2}b_{v_1}  \quad \text{and} \quad
    \bar{t}_R\gamma_{\mu}P_Rt_R\rightarrow  b_3\bar{t}_{v_2}t_{v_1},
\end{align}
where the matching coefficients, $b_i$, are obtained using the matching coefficient, $b(m_t)$, from section \ref{sec:methre} with the appropriate graphs and group theory factors for each part, adding the contributions due to both $b'$ and $t$ individually. All running couplings are then to be renormalized at $\mu=m_t$.  As in the Sudakov case, below $\mu=m_t$, the $\bar{t}_{v_2}t_{v_1}$ operator has the anomalous dimension for running from $m_t\rightarrow M_Z$ and is given by, 
\begin{equation}
    \gamma_3=a\gamma_3^{(1)}+a^2\gamma_3^{(2)}+\mathcal{O}(a^3),\quad  a=\left(\frac{4}{3}\frac{\alpha_s}{4\pi}+\frac{4}{9}\frac{\alpha_{em}}{4\pi}\right)
\end{equation}
from the fourth column of table \ref{tab:expr}, with all running coupling renormalised at $\mu=m_t$. The remaining quantities needed are the matching contributions at $\mu=M_Z$, where the operators $\bar{t}_{v_2}{t}_{v_1}$ and $\bar{b}_{v_2}{b}_{v_1}$ above $M_Z$ are matched to their counterparts below $M_Z$ with massive bosons integrated out,
\begin{equation}
    \bar{t}_{v_2}{t}_{v_1}\rightarrow u_1 \bar{t}_{v_2}{t}_{v_1}  \quad \text{and} \quad \bar{b}_{v_2}{b}_{v_1}\rightarrow u_2 \bar{b}_{v_2}{b}_{v_1}
\end{equation}
where the matching coefficients, $u_i$, are obtained using the matching coefficient, $u(M_Z)$, from combining contributions from tables in sections \ref{ssec:mesud} and \ref{sec:methre} with all running coupling renormalised at $\mu=M_Z$. The expressions are given by adding individual contributions due to each massive boson in the SM.
\section{Technical Calculation}
The Feynman diagrams we needed were generated using \texttt{QGRAF} \cite{nogueira1993automatic}, the output of which was then processed using FORM \cite{vermaseren2000new} to express the diagrams in terms of a linear combination of a set of scalar integrals. We then reduced these integrals to a much smaller set of so-called master integrals (MIs) using integration-by-parts identities (IBPs) \cite{chetyrkin1980new}, with the help of \texttt{LiteRed} \cite{lee2012presenting} and home-grown tools. Our master integrals in some cases are dependent on two mass scales taken to be not widely separated, either the external particle masses or the bosonic masses, respectively. One can perform these integrals numerically but to obtain analytic results we expand such amplitudes in the mass difference to NLO in said difference, leading to single scale integrals. Once the integrals are maximally reduced, all that remains is to evaluate the master integrals. As these procedures are well-known, we refrain from delving into too much detail.

\begin{figure}[tb!]
\centering
\begin{center}
\scalebox{0.58}{
\fcolorbox{white}{white}{
  \begin{picture}(725,632) (26,-14)
    \SetWidth{1.0}
    \SetColor{Black}
    \Vertex(51,535){6.403}
    \Line[dash,dashsize=2,arrow,arrowpos=0.5,arrowlength=5,arrowwidth=2,arrowinset=0.2](53,536)(177,615)
    \Line[dash,dashsize=2,arrow,arrowpos=0.5,arrowlength=5,arrowwidth=2,arrowinset=0.2,flip](52,532)(177,454)
    \Vertex(229,535){6.403}
    \Line[dash,dashsize=2](231,536)(355,615)
    \Line[dash,dashsize=2](230,532)(355,454)
    \Line[dash,dashsize=2,arrow,arrowpos=0.5,arrowlength=5,arrowwidth=2,arrowinset=0.2,flip](101,566)(101,501)
    \Line[dash,dashsize=2,arrow,arrowpos=0.5,arrowlength=5,arrowwidth=2,arrowinset=0.2](141,591)(142,476)
    \Line[dash,dashsize=2](270,560)(320,475)
    \Line[dash,dashsize=2](271,505)(318,592)
    \Text(104,431)[lb]{{\Black{$(a_1)$}}}
    \Text(280,431)[lb]{{\Black{$(a_2)$}}}
    \Vertex(407,537){6.403}
    \Line(408,537)(530,616)
    \Line(409,534)(529,456)
    \Line[dash,dashsize=2](458,568)(458,502)
    \Line[dash,dashsize=2](499,596)(499,475)
    \Text(462,432)[lb]{{\Black{$(b_1)$}}}
    \Vertex(584,536){6.403}
    \Line(585,536)(707,615)
    \Line(586,533)(706,455)
    \Line[dash,dashsize=2](626,562)(677,473)
    \Line[dash,dashsize=2](628,505)(679,596)
    \Text(640,432)[lb]{{\Black{$(b_2)$}}}
    \Line(143,368)(176,390)
    \Line(142,251)(177,229)
    \Line(114,348)(114,270)
    \Text(114,207)[lb]{{\Black{$(b_3)$}}}
    \Vertex(230,312){6.403}
    \Line(231,312)(353,391)
    \Line[dash,dashsize=2](281,343)(281,277)
    \Line[dash,dashsize=2](322,371)(322,250)
    \Text(281,206)[lb]{{\Black{$(c_1)$}}}
    \Vertex(407,312){6.403}
    \Line(408,312)(530,391)
    \Line[dash,dashsize=2](449,338)(500,249)
    \Line[dash,dashsize=2](451,281)(502,372)
    \Text(461,206)[lb]{{\Black{$(c_2)$}}}
    \Line(141,251)(142,366)
    \Line[dash,dashsize=2](114,348)(142,366)
    \Line[dash,dashsize=2](114,269)(142,251)
    \Line[dash,dashsize=2](229,310)(352,230)
    \Line[dash,dashsize=2](406,308)(528,231)
    \Vertex(582,310){6.403}
    \Line[double,sep=2](583,310)(705,389)
    \Line[dash,dashsize=2](638,345)(638,272)
    \Line[dash,dashsize=2](678,370)(678,246)
    \Text(633,205)[lb]{{\Black{$(d_1)$}}}
    \Line(582,309)(705,230)
    \Line[dash,dashsize=2,arrow,arrowpos=0.5,arrowlength=5,arrowwidth=2,arrowinset=0.2](27,534)(44,534)
    \Line[dash,dashsize=10,arrow,arrowpos=0.5,arrowlength=5,arrowwidth=2,arrowinset=0.2](74,517)(80,514)
    \Line[dash,dashsize=10,arrow,arrowpos=0.5,arrowlength=5,arrowwidth=2,arrowinset=0.2](81,553)(74,549)
    \Line[dash,dashsize=10,arrow,arrowpos=0.5,arrowlength=5,arrowwidth=2,arrowinset=0.2](156,601)(160,604)
    \Line[dash,dashsize=10,arrow,arrowpos=0.5,arrowlength=5,arrowwidth=2,arrowinset=0.2](165,461)(161,464)
    \Line[dash,dashsize=2,arrow,arrowpos=0.5,arrowlength=5,arrowwidth=2,arrowinset=0.2](206,536)(222,536)
    \Line[dash,dashsize=2,arrow,arrowpos=0.5,arrowlength=5,arrowwidth=2,arrowinset=0.2](253,549)(248,545)
    \Line[dash,dashsize=2,arrow,arrowpos=0.5,arrowlength=5,arrowwidth=2,arrowinset=0.2](253,517)(259,514)
    \Line[dash,dashsize=2,arrow,arrowpos=0.5,arrowlength=5,arrowwidth=2,arrowinset=0.2](292,492)(288,495)
    \Line[dash,dashsize=2,arrow,arrowpos=0.5,arrowlength=5,arrowwidth=2,arrowinset=0.2](336,467)(331,469)
    \Line[dash,dashsize=2,arrow,arrowpos=0.5,arrowlength=5,arrowwidth=2,arrowinset=0.2](293,522)(297,515)
    \Line[dash,dashsize=2,arrow,arrowpos=0.5,arrowlength=5,arrowwidth=2,arrowinset=0.2](293,544)(298,552)
    \Line[dash,dashsize=2,arrow,arrowpos=0.5,arrowlength=5,arrowwidth=2,arrowinset=0.2](289,572)(296,577)
    \Line[dash,dashsize=2,arrow,arrowpos=0.5,arrowlength=5,arrowwidth=2,arrowinset=0.2](331,599)(337,603)
    \Line[dash,dashsize=2,arrow,arrowpos=0.5,arrowlength=5,arrowwidth=2,arrowinset=0.2](436,555)(432,552)
    \Line[dash,dashsize=2,arrow,arrowpos=0.5,arrowlength=5,arrowwidth=2,arrowinset=0.2](431,520)(436,515)
    \Line[dash,dashsize=2,arrow,arrowpos=0.5,arrowlength=5,arrowwidth=2,arrowinset=0.2](481,487)(475,492)
    \Line[dash,dashsize=2,arrow,arrowpos=0.5,arrowlength=5,arrowwidth=2,arrowinset=0.2](473,579)(479,583)
    \Line[dash,dashsize=2,arrow,arrowpos=0.5,arrowlength=5,arrowwidth=2,arrowinset=0.2](511,603)(518,608)
    \Line[dash,dashsize=2,arrow,arrowpos=0.5,arrowlength=5,arrowwidth=2,arrowinset=0.2](520,463)(512,468)
    \Line[dash,dashsize=2,arrow,arrowpos=0.5,arrowlength=5,arrowwidth=2,arrowinset=0.2](499,538)(499,531)
    \Line[dash,dashsize=2,arrow,arrowpos=0.5,arrowlength=5,arrowwidth=2,arrowinset=0.2](459,532)(459,538)
    \Line[dash,dashsize=2,arrow,arrowpos=0.5,arrowlength=5,arrowwidth=2,arrowinset=0.2](649,541)(652,545)
    \Line[dash,dashsize=2,arrow,arrowpos=0.5,arrowlength=5,arrowwidth=2,arrowinset=0.2](651,517)(655,510)
    \Line[dash,dashsize=2,arrow,arrowpos=0.5,arrowlength=5,arrowwidth=2,arrowinset=0.2](611,552)(606,549)
    \Line[dash,dashsize=2,arrow,arrowpos=0.5,arrowlength=5,arrowwidth=2,arrowinset=0.2](646,576)(655,581)
    \Line[dash,dashsize=2,arrow,arrowpos=0.5,arrowlength=5,arrowwidth=2,arrowinset=0.2](690,604)(697,608)
    \Line[dash,dashsize=2,arrow,arrowpos=0.5,arrowlength=5,arrowwidth=2,arrowinset=0.2](691,464)(685,469)
    \Line[dash,dashsize=2,arrow,arrowpos=0.5,arrowlength=5,arrowwidth=2,arrowinset=0.2](652,489)(645,496)
    \Line[dash,dashsize=2,arrow,arrowpos=0.5,arrowlength=5,arrowwidth=2,arrowinset=0.2](606,520)(613,518)
    \Line[dash,dashsize=2,arrow,arrowpos=0.5,arrowlength=5,arrowwidth=2,arrowinset=0.2](386,536)(403,536)
    \Line[dash,dashsize=2,arrow,arrowpos=0.5,arrowlength=5,arrowwidth=2,arrowinset=0.2](561,535)(581,535)
    \Line[dash,dashsize=2,arrow,arrowpos=0.5,arrowlength=5,arrowwidth=2,arrowinset=0.2](207,311)(228,311)
    \Line[dash,dashsize=2,arrow,arrowpos=0.5,arrowlength=5,arrowwidth=2,arrowinset=0.2](386,312)(405,312)
    \Line[dash,dashsize=2,arrow,arrowpos=0.5,arrowlength=5,arrowwidth=2,arrowinset=0.2](561,310)(577,310)
    \Line[dash,dashsize=2,arrow,arrowpos=0.5,arrowlength=5,arrowwidth=2,arrowinset=0.2](114,303)(115,314)
    \Line[dash,dashsize=2,arrow,arrowpos=0.5,arrowlength=5,arrowwidth=2,arrowinset=0.2](142,314)(142,300)
    \Line[dash,dashsize=2,arrow,arrowpos=0.5,arrowlength=5,arrowwidth=2,arrowinset=0.2](164,237)(156,242)
    \Line[dash,dashsize=2,arrow,arrowpos=0.5,arrowlength=5,arrowwidth=2,arrowinset=0.2](156,376)(165,383)
    \Line[dash,dashsize=2,arrow,arrowpos=0.5,arrowlength=5,arrowwidth=2,arrowinset=0.2](126,355)(130,358)
    \Line[dash,dashsize=2,arrow,arrowpos=0.5,arrowlength=5,arrowwidth=2,arrowinset=0.2](128,259)(124,262)
    \Line[dash,dashsize=2,arrow,arrowpos=0.5,arrowlength=5,arrowwidth=2,arrowinset=0.2](282,306)(282,317)
    \Line[dash,dashsize=2,arrow,arrowpos=0.5,arrowlength=5,arrowwidth=2,arrowinset=0.2](305,260)(298,265)
    \Line[dash,dashsize=2,arrow,arrowpos=0.5,arrowlength=5,arrowwidth=2,arrowinset=0.2](341,237)(337,240)
    \Line[dash,dashsize=2,arrow,arrowpos=0.5,arrowlength=5,arrowwidth=2,arrowinset=0.2](322,316)(322,306)
    \Line[dash,dashsize=2,arrow,arrowpos=0.5,arrowlength=5,arrowwidth=2,arrowinset=0.2](298,355)(306,360)
    \Line[dash,dashsize=2,arrow,arrowpos=0.5,arrowlength=5,arrowwidth=2,arrowinset=0.2](333,378)(341,382)
    \Line[dash,dashsize=2,arrow,arrowpos=0.5,arrowlength=5,arrowwidth=2,arrowinset=0.2](260,330)(252,325)
    \Line[dash,dashsize=2,arrow,arrowpos=0.5,arrowlength=5,arrowwidth=2,arrowinset=0.2](255,294)(261,289)
    \Line[dash,dashsize=2,arrow,arrowpos=0.5,arrowlength=5,arrowwidth=2,arrowinset=0.2](433,328)(427,324)
    \Line[dash,dashsize=2,arrow,arrowpos=0.5,arrowlength=5,arrowwidth=2,arrowinset=0.2](467,351)(478,358)
    \Line[dash,dashsize=2,arrow,arrowpos=0.5,arrowlength=5,arrowwidth=2,arrowinset=0.2](514,381)(523,386)
    \Line[dash,dashsize=2,arrow,arrowpos=0.5,arrowlength=5,arrowwidth=2,arrowinset=0.2](517,237)(513,240)
    \Line[dash,dashsize=2,arrow,arrowpos=0.5,arrowlength=5,arrowwidth=2,arrowinset=0.2](478,263)(474,266)
    \Line[dash,dashsize=2,arrow,arrowpos=0.5,arrowlength=5,arrowwidth=2,arrowinset=0.2](427,296)(432,292)
    \Line[dash,dashsize=2,arrow,arrowpos=0.5,arrowlength=5,arrowwidth=2,arrowinset=0.2,flip](476,293)(473,298)
    \Line[dash,dashsize=2,arrow,arrowpos=0.5,arrowlength=5,arrowwidth=2,arrowinset=0.2,flip](475,320)(472,315)
    \Line[dash,dashsize=2,arrow,arrowpos=0.5,arrowlength=5,arrowwidth=2,arrowinset=0.2,flip](639,307)(639,316)
    \Line[dash,dashsize=2,arrow,arrowpos=0.5,arrowlength=5,arrowwidth=2,arrowinset=0.2](679,314)(679,304)
    \Line[dash,dashsize=2,arrow,arrowpos=0.5,arrowlength=5,arrowwidth=2,arrowinset=0.2,flip](605,294)(612,290)
    \Line[dash,dashsize=2,arrow,arrowpos=0.5,arrowlength=5,arrowwidth=2,arrowinset=0.2](659,260)(654,264)
    \Line[dash,dashsize=2,arrow,arrowpos=0.5,arrowlength=5,arrowwidth=2,arrowinset=0.2](698,234)(692,238)
    \Line[dash,dashsize=2,arrow,arrowpos=0.5,arrowlength=5,arrowwidth=2,arrowinset=0.2](688,378)(694,382)
    \Line[dash,dashsize=2,arrow,arrowpos=0.5,arrowlength=5,arrowwidth=2,arrowinset=0.2](650,353)(656,358)
    \Line[dash,dashsize=2,arrow,arrowpos=0.5,arrowlength=5,arrowwidth=2,arrowinset=0.2,flip](615,331)(610,328)
    \Text(117,477)[lb]{{\Black{$1$}}}
    \Text(71,504)[lb]{{\Black{$2$}}}
    \Text(112,532)[lb]{{\Black{$3$}}}
    \Text(154,536)[lb]{{\Black{$4$}}}
    \Text(72,562)[lb]{{\Black{$5$}}}
    \Text(116,592)[lb]{{\Black{$6$}}}
    \Text(288,586)[lb]{{\Black{$7$}}}
    \Text(253,563)[lb]{{\Black{$5$}}}
    \Text(305,546)[lb]{{\Black{$3$}}}
    \Text(312,515)[lb]{{\Black{$4$}}}
    \Text(286,484)[lb]{{\Black{$1$}}}
    \Text(250,507)[lb]{{\Black{$2$}}}
    \Text(474,480)[lb]{{\Black{$1$}}}
    \Text(431,504)[lb]{{\Black{$2$}}}
    \Text(430,569)[lb]{{\Black{$5$}}}
    \Text(478,597)[lb]{{\Black{$6$}}}
    \Text(468,537)[lb]{{\Black{$3$}}}
    \Text(514,537)[lb]{{\Black{$4$}}}
    \Text(666,550)[lb]{{\Black{$3$}}}
    \Text(669,510)[lb]{{\Black{$4$}}}
    \Text(642,480)[lb]{{\Black{$1$}}}
    \Text(600,507)[lb]{{\Black{$2$}}}
    \Text(605,563)[lb]{{\Black{$5$}}}
    \Text(652,590)[lb]{{\Black{$7$}}}
    \Text(427,340)[lb]{{\Black{$5$}}}
    \Text(478,372)[lb]{{\Black{$7$}}}
    \Text(296,371)[lb]{{\Black{$6$}}}
    \Text(251,342)[lb]{{\Black{$5$}}}
    \Text(300,250)[lb]{{\Black{$1$}}}
    \Text(254,283)[lb]{{\Black{$2$}}}
    \Text(295,309)[lb]{{\Black{$3$}}}
    \Text(342,310)[lb]{{\Black{$4$}}}
    \Text(122,308)[lb]{{\Black{$3$}}}
    \Text(159,304)[lb]{{\Black{$4$}}}
    \Text(125,248)[lb]{{\Black{$1$}}}
    \Text(78,276)[lb]{{\Black{$2$}}}
    \Text(126,370)[lb]{{\Black{$6$}}}
    \Text(491,327)[lb]{{\Black{$3$}}}
    \Text(492,290)[lb]{{\Black{$4$}}}
    \Text(470,252)[lb]{{\Black{$1$}}}
    \Text(426,282)[lb]{{\Black{$2$}}}
    \Text(611,277)[lb]{{\Black{$1$}}}
    \Text(660,247)[lb]{{\Black{$2$}}}
    \Text(646,311)[lb]{{\Black{$3$}}}
    \Text(693,312)[lb]{{\Black{$4$}}}
    \Text(604,341)[lb]{{\Black{$5$}}}
    \Text(652,370)[lb]{{\Black{$6$}}}
    \Line[dash,dashsize=2,arrow,arrowpos=0.5,arrowlength=5,arrowwidth=2,arrowinset=0.2](28,312)(49,312)
    \Vertex(56,312){6.403}
    \Line[arrow,arrowpos=0.5,arrowlength=5,arrowwidth=2,arrowinset=0.2](115,348)(59,313)
    \Line[arrow,arrowpos=0.5,arrowlength=5,arrowwidth=2,arrowinset=0.2](56,311)(115,267)

    \Vertex(586,87){6.403}
    \Line[double,sep=2](587,87)(709,166)
    \Line[double,sep=2](588,84)(708,6)
    \Text(652,-18)[lb]{{\Black{$(e_3)$}}}
    \Line[dash,dashsize=2,arrow,arrowpos=0.5,arrowlength=5,arrowwidth=2,arrowinset=0.2](562,88)(581,87)
    \Line[dash,dashsize=2,arrow,arrowpos=0.5,arrowlength=5,arrowwidth=2,arrowinset=0.2,flip](640,122)(634,117)
    \Line[dash,dashsize=2,arrow,arrowpos=0.5,arrowlength=5,arrowwidth=2,arrowinset=0.2](697,158)(702,161)
    \Line[dash,dashsize=2,arrow,arrowpos=0.5,arrowlength=5,arrowwidth=2,arrowinset=0.2](638,51)(632,54)
    \Line[dash,dashsize=2,arrow,arrowpos=0.5,arrowlength=5,arrowwidth=2,arrowinset=0.2](704,9)(700,12)
    \Text(629,135)[lb]{{\Black{$6$}}}
    \Text(695,47)[lb]{{\Black{$3$}}}
    \Text(626,40)[lb]{{\Black{$1$}}}
    \Text(691,129)[lb]{{\Black{$3$}}}
    \Vertex(231,87){6.403}
    \Line[double,sep=2](232,87)(354,166)
    \Line[double,sep=2](233,84)(353,6)
    \Text(287,-19)[lb]{{\Black{$(e_1)$}}}
    \Line(273,113)(273,58)
    \Line(321,144)(320,28)
    \Line[dash,dashsize=2,arrow,arrowpos=0.5,arrowlength=5,arrowwidth=2,arrowinset=0.2](210,87)(227,87)
    \Line[dash,dashsize=2,arrow,arrowpos=0.5,arrowlength=5,arrowwidth=2,arrowinset=0.2](342,14)(337,18)
    \Line[dash,dashsize=2,arrow,arrowpos=0.5,arrowlength=5,arrowwidth=2,arrowinset=0.2](302,39)(299,42)
    \Line[dash,dashsize=2,arrow,arrowpos=0.5,arrowlength=5,arrowwidth=2,arrowinset=0.2,flip](250,73)(256,70)
    \Line[dash,dashsize=2,arrow,arrowpos=0.5,arrowlength=5,arrowwidth=2,arrowinset=0.2,flip](274,81)(274,86)
    \Line[dash,dashsize=2,arrow,arrowpos=0.5,arrowlength=5,arrowwidth=2,arrowinset=0.2](321,90)(321,82)
    \Line[dash,dashsize=2,arrow,arrowpos=0.5,arrowlength=5,arrowwidth=2,arrowinset=0.2](292,126)(296,129)
    \Line[dash,dashsize=2,arrow,arrowpos=0.5,arrowlength=5,arrowwidth=2,arrowinset=0.2](330,151)(335,154)
    \Line[dash,dashsize=2,arrow,arrowpos=0.5,arrowlength=5,arrowwidth=2,arrowinset=0.2,flip](258,105)(253,100)
    \Text(249,113)[lb]{{\Black{$5$}}}
    \Text(298,141)[lb]{{\Black{$6$}}}
    \Text(281,83)[lb]{{\Black{$3$}}}
    \Text(334,84)[lb]{{\Black{$4$}}}
    \Text(292,30)[lb]{{\Black{$1$}}}
    \Text(247,62)[lb]{{\Black{$2$}}}
    \Vertex(412,87){6.403}
    \Line[double,sep=2](413,87)(535,166)
    \Line[double,sep=2](414,84)(534,6)
    \Text(478,-18)[lb]{{\Black{$(e_2)$}}}
    \Line[dash,dashsize=2,arrow,arrowpos=0.5,arrowlength=5,arrowwidth=2,arrowinset=0.2](388,88)(407,87)
    \Line[dash,dashsize=2,arrow,arrowpos=0.5,arrowlength=5,arrowwidth=2,arrowinset=0.2,flip](442,106)(436,101)
    \Line[dash,dashsize=2,arrow,arrowpos=0.5,arrowlength=5,arrowwidth=2,arrowinset=0.2](474,126)(480,130)
    \Line[dash,dashsize=2,arrow,arrowpos=0.5,arrowlength=5,arrowwidth=2,arrowinset=0.2](523,158)(528,161)
    \Line[dash,dashsize=2,arrow,arrowpos=0.5,arrowlength=5,arrowwidth=2,arrowinset=0.2](464,51)(458,54)
    \Line[dash,dashsize=2,arrow,arrowpos=0.5,arrowlength=5,arrowwidth=2,arrowinset=0.2](530,9)(526,12)
    \Text(427,111)[lb]{{\Black{$6$}}}
    \Text(476,141)[lb]{{\Black{$5$}}}
    \Line[arrow,arrowpos=0.5,arrowlength=5,arrowwidth=2,arrowinset=0.2](504,143)(504,93)
    \Line[arrow,arrowpos=0.5,arrowlength=5,arrowwidth=2,arrowinset=0.2](504,93)(504,25)
    \Line[arrow,arrowpos=0.5,arrowlength=5,arrowwidth=2,arrowinset=0.2](504,95)(460,117)
    \Text(475,94)[lb]{{\Black{$3$}}}
    \Text(516,66)[lb]{{\Black{$4$}}}
    \Text(452,40)[lb]{{\Black{$1$}}}
    \Text(517,120)[lb]{{\Black{$7$}}}
    \Vertex(54,86){6.403}
    \Line[double,sep=2](55,86)(177,165)
    \Line[dash,dashsize=2](95,109)(150,21)
    \Line[dash,dashsize=2](96,57)(154,150)
    \Text(120,-19)[lb]{{\Black{$(d_2)$}}}
    \Line(56,86)(175,3)
    \Line[dash,dashsize=2,arrow,arrowpos=0.5,arrowlength=5,arrowwidth=2,arrowinset=0.2](32,87)(50,87)
    \Line[dash,dashsize=2,arrow,arrowpos=0.5,arrowlength=5,arrowwidth=2,arrowinset=0.2,flip](77,100)(71,96)
    \Line[dash,dashsize=2,arrow,arrowpos=0.5,arrowlength=5,arrowwidth=2,arrowinset=0.2](115,125)(123,129)
    \Line[dash,dashsize=2,arrow,arrowpos=0.5,arrowlength=5,arrowwidth=2,arrowinset=0.2](162,155)(169,158)
    \Line[dash,dashsize=2,arrow,arrowpos=0.5,arrowlength=5,arrowwidth=2,arrowinset=0.2](118,72)(123,64)
    \Line[dash,dashsize=2,arrow,arrowpos=0.5,arrowlength=5,arrowwidth=2,arrowinset=0.2,flip](120,94)(124,101)
    \Line[dash,dashsize=2,arrow,arrowpos=0.5,arrowlength=5,arrowwidth=2,arrowinset=0.2](168,8)(162,12)
    \Line[dash,dashsize=2,arrow,arrowpos=0.5,arrowlength=5,arrowwidth=2,arrowinset=0.2](126,37)(118,42)
    \Line[dash,dashsize=2,arrow,arrowpos=0.5,arrowlength=5,arrowwidth=2,arrowinset=0.2,flip](73,73)(81,68)
    \Text(133,99)[lb]{{\Black{$3$}}}
    \Text(135,68)[lb]{{\Black{$4$}}}
    \Text(115,28)[lb]{{\Black{$1$}}}
    \Text(67,60)[lb]{{\Black{$2$}}}
    \Text(71,114)[lb]{{\Black{$5$}}}
    \Text(121,145)[lb]{{\Black{$7$}}}
    \Line[arrow,arrowpos=0.5,arrowlength=5,arrowwidth=2,arrowinset=0.2](681,147)(681,109)
    \Arc[dash,dashsize=2,arrow,arrowpos=0.5,arrowlength=5,arrowwidth=2,arrowinset=0.2](682,88)(21.213,135,495)
    \Line[arrow,arrowpos=0.5,arrowlength=5,arrowwidth=2,arrowinset=0.2](683,66)(683,23)
    \Text(712,89)[lb]{{\Black{$8$}}}
    \Text(652,88)[lb]{{\Black{$9$}}}
    \Text(80,342)[lb]{{\Black{$5$}}}
  \end{picture}}
}
\end{center}
   \caption{Topologies required for the calculation of two-loop form factors. Solid lines represent massive particles, double lines represent heavy particles, dashed lines correspond to massless propagators. Arrows represent direction of momentum. $(a_i)$ represent topologies of MIs at $\mu\sim Q$ (Sudakov), $(b_i)$ represent topologies of MIs at $\mu\sim m_{1,2}$ (threshold), $(c_i)$ represent topologies of MIs at $\mu\sim m_{2}$ (threshold), $(d_i)$ represent topologies of MIs at $\mu\sim m_{1}$ (threshold) and $(e_i)$ represent topologies of MIs at $\mu\sim M$ (Sudakov/threshold). }
    \label{fig:top1}
\end{figure}
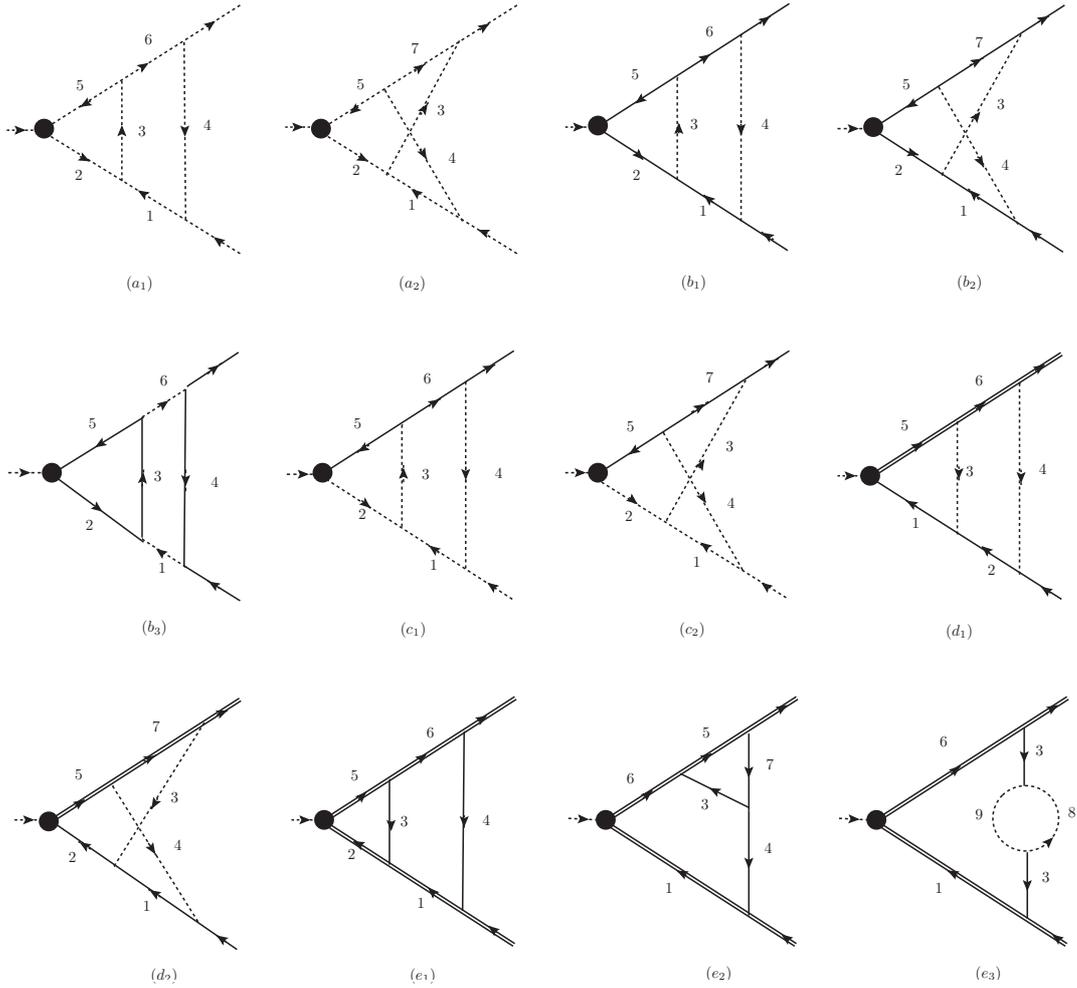

We focus here on the calculation of MIs of the two-loop vertex and wave-function contributions. The full theory integrals have been computed analytically \cite{ablinger2018heavy}. The result was found as a Laurent expansion in the dimensional parameter, $\epsilon$, using the differential equation method \cite{kotikov1991differential,remiddi1997differential}. We evaluate our MIs in the effective theory, HPET, which are not known analytically, in the same fashion. One requires the two-loop master integrals up to a sufficiently high order in $\epsilon$ to obtain $\mathcal{O}(\epsilon^2)$ accuracy in the form factors. As we are still evaluating the heavy-heavy and heavy-light master integrals, we will keep the details of this calculation for upcoming work. Instead, we briefly outline our classification scheme for reference and present the results we do have explicitly. The master integrals are classified according to their underlying topology. 

We begin by distinguishing between the vertex topologies for external full theory fields displayed in figures \ref{fig:top1} (a-c). The master integrals for all topologies can be expressed in terms of a single integral family with seven propagators given by,
\begin{equation}
    J^{(s)}_{\lbrace\nu^{(m)}_1\cdots\nu^{(m)}_7\rbrace}=\left[(4\pi)^{2-\epsilon}e^{\gamma_E\epsilon}\right]^2\int\slashed{d}^Dl_1\slashed{d}^Dl_2\frac{1}{D_1^{\nu_1}(m)\cdots D_7^{\nu_7}(m)},
\end{equation}
where $l_i:i=1,2$ are the loop momenta, $s$ is the scale in the EFT formalism at which the MIs play a role, and,
\begin{align}
    & \quad D_1(m)=l_1^2-m^2, \quad D_2(m)=l_2^2-m^2, \quad D_3(m)=(l_1+l_2)^2-m^2, \nonumber \\& \quad D_4(m)=(l_1-p_1)^2-m^2,  \quad D_5(m)=(l_2-q)^2-m^2, \\& \quad D_6(m)=(l_1+q)^2-m^2, \quad D_7(m)=(l_1+l_2-q-p_1)^2-m^2. 
\end{align}
Here the $p_i:i=1,2$ are the external momenta, which are taken to be on-shell ($p_i^2=m_i^2$) and $q=p_2-p_1$ is the usual transfer momentum. We therefore label the MIs by the exponents, $\nu_1\ldots\nu_7$ of the denominators, $D_1\ldots D_7$. Note the single mass scale in our denominators, this arises from the fact that for integrals involving two mass scales or more, we expand our results in the difference of mass scales up to NLO. For instance, for graphs that include propagators of both $W$ and Higgs bosons, we expand about $\Delta M=M_W-M_H$, assuming them to be not widely separated. This is done purely so we can present our results analytically as any number of scales can be handled numerically, moreover our choice to expand to NLO is for presentability as there is no issue in expanding the amplitudes to higher orders in $\Delta M$ computationally.

In the Sudakov regime the vertex matching contributions at the scale $\mu \sim Q$, $V^{(Q)}_i$, has all mass scales set to zero as they are taken to be IR, and thus $m=0$ in cases below, in which case we have MIs with topology given by figures \ref{fig:top1} (a). Post-reduction one is left with the following master integrals,
\small{
\begin{subequations}
\begin{align}
    & J^{(Q)}_{1010100}=\frac{Q^2 \left(13 \epsilon -4 \epsilon  \log \left(\frac{Q^2}{\mu ^2}\right)+2\right)}{8 \pi ^4 \epsilon }, \\&
   J^{(Q)}_{1100110}=\frac{\epsilon  \left(2 \left(\pi ^2-72\right) \epsilon -3\right)+6 \epsilon  \mathcal{L}_Q \left(\epsilon -4 \epsilon  \mathcal{L}_Q+4\right)-12}{12 \pi ^4 \epsilon ^2}, \\&
    J^{(Q)}_{1100101}=-\frac{\epsilon  \left(\left(114+\pi ^2\right) \epsilon +30\right)+12 \epsilon  \mathcal{L}_Q \left(-5 \epsilon +\epsilon  \mathcal{L}_Q-1\right)+6}{12 \pi ^4 \epsilon ^2}, \\&
    J^{(Q)}_{1111101}=\frac{59 \pi ^4 \epsilon ^4+120 \pi ^2 \epsilon ^2-80 \epsilon  \mathcal{L}_Q \left(3 \pi ^2 \epsilon ^2+\epsilon  \mathcal{L}_Q \left(-3 \pi^2 \epsilon ^2+\epsilon  \mathcal{L}_Q \left(\epsilon  \mathcal{L}_Q-2\right)+3\right)-3\right)}{120 \pi ^4 \epsilon ^4 Q^2} \nonumber \\& \quad\quad\quad\quad\quad + \frac{83 \epsilon ^3 \zeta_3 \left(1-2\epsilon  \mathcal{L}_Q\right)-3}{3 \pi ^4 \epsilon ^4 Q^2},
\end{align}
\end{subequations}}
where $\mathcal{L}_Q\equiv \log{Q^2/\mu^2}$ and $\zeta_n$ denotes the Riemann $\zeta$-function,
\begin{equation}
    \zeta_m=\sum_{k=1}^{\infty}\frac{1}{k^n},\quad n\geq 2:n\in\mathbb{N},
\end{equation}
and these integrals have been verified from previous work \cite{van1986dimensional}. 
\begin{figure}[htp]
\centering
\begin{center}
\scalebox{0.4}{
\fcolorbox{white}{white}{
  \begin{picture}(307,224) (112,-10)
    \SetWidth{2}
    \SetColor{Black}
    \Line[arrow,arrowpos=0.5,arrowlength=10,arrowwidth=5,arrowinset=0.2](113,95)(160,95)
    \Arc(265,95)(102.786,162,522)
    \Line[arrow,arrowpos=0.5,arrowlength=10,arrowwidth=5,arrowinset=0.2](369,95)(418,95)
    \Line[arrow,arrowpos=0.5,arrowlength=10,arrowwidth=5,arrowinset=0.2](265,198)(266,-9)
    \Line[arrow,arrowpos=0.5,arrowlength=10,arrowwidth=5,arrowinset=0.2](185,160)(192,167)
    \Line[arrow,arrowpos=0.5,arrowlength=10,arrowwidth=5,arrowinset=0.2](341,164)(347,158)
    \Line[arrow,arrowpos=0.5,arrowlength=10,arrowwidth=5,arrowinset=0.2](347,33)(342,26)
    \Line[arrow,arrowpos=0.5,arrowlength=10,arrowwidth=5,arrowinset=0.2](192,22)(187,28)
    \Text(160,190)[lb]{\huge{\Black{$1$}}}
    \Text(368,193)[lb]{\huge{\Black{$2$}}}
    \Text(160,-1)[lb]{\huge{\Black{$3$}}}
    \Text(368,0)[lb]{\huge{\Black{$4$}}}
    \Text(231,96)[lb]{\huge{\Black{$5$}}}
  \end{picture}}
}
\end{center}
   \caption{Full theory self-energy topology. Arrows represent momentum direction. The MIs associated to other topologies are subsets of the MIs required for this topology.}
    \label{fig:top2}
\end{figure}
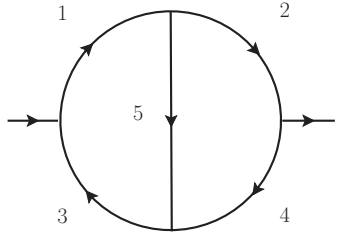
On the other hand, in the threshold regime, the full theory MIs have topologies represented by figures \ref{fig:top1} (b,c). Due to the threshold limit, $Q\rightarrow 0$, the MIs are further reduced down to two-loop self-energy topologies, as shown in figure \ref{fig:top2}, and analytic expressions can be obtained from the standard \texttt{ON-SHELL2} package \cite{fleischer2000shell2}. The master integrals can therefore be expressed in terms of a single integral family with five propagators given by,
\begin{equation}
    J^{(s)}_{\lbrace\nu_1^{(m)}\cdots\nu_5^{(m)}\rbrace}=\left[(4\pi)^{2-\epsilon}e^{\gamma_E\epsilon}\right]^2\int\slashed{d}^Dl_1\slashed{d}^Dl_2\frac{1}{D_1^{\nu_1}(m)\cdots D_5^{\nu_5}(m)},
\end{equation}
where $l_i:i=1,2$ are the loop momenta, $s$ is the scale in the EFT formalism at which the MIs play a role, and,
\begin{align}
    & \quad D_1(m)=l_1^2-m^2, \quad D_2(m)=l_2^2-m^2, \quad D_3(m)=(l_1-p)^2-m^2, \nonumber \\& \quad D_4(m)=(l_2-p)^2-m^2,  \quad D_5(m)=(l_1-l_2)^2-m^2.
\end{align}
To begin with, the vertex matching contributions at the scale $\mu \sim m_2\gg m_1,M$, for matching from the full theory to heavy-light operators is represented by $V^{(m_2)}_i$. Post-reduction one is left with the following master integrals,
\small{
\begin{subequations}
\begin{align}
    & J^{(m_2)}_{1^{(m)}1^{(m)}000}=-\frac{m_2^4 (2 \epsilon \mathcal{L}_{m_2} (\epsilon \mathcal{L}_{m_2}-2 \epsilon-1)+\epsilon (\epsilon(\zeta_2+3)+2)+1)}{ \epsilon^2},
    \\&
   J^{(m_2)}_{11001^{(m)}}=-\frac{m_2^2 (2 \epsilon \mathcal{L}_{m_2} (\epsilon \mathcal{L}_{m_2}-3 \epsilon-1)+\epsilon (\epsilon (3\zeta_2+7)+3)+1)}{2  \epsilon^2}, 
   \\&
    J^{(m_2)}_{11^{(m)}100}=-\frac{m_2^2 (2 \epsilon \mathcal{L}_{m_2} (\epsilon \mathcal{L}_{m_2}-3 \epsilon-1)+\epsilon (7\epsilon+3)+1)}{ \epsilon^2}, 
    \\&
    J^{(m_2)}_{01^{(m)}101}=-\frac{m_2^2 (4 \epsilon \mathcal{L}_{m_2} (2 \epsilon \mathcal{L}_{m_2}-5 \epsilon-2)+\epsilon (\epsilon (20\zeta_2+11)+10)+4)}{8  \epsilon^2},
    \\&
    J^{(m_2)}_{011^{(m)}02^{(m)}}=-\frac{-6 \epsilon \mathcal{L}_{m_2} (\epsilon (-\mathcal{L}_{m_2})+\epsilon+1)+\epsilon (\epsilon (6S_1+\zeta_2-3)+3)+3}{6  \epsilon^2},
    \\&
    J^{(m_2)}_{01^{(m)}1^{(m)}01^{(m)}}=-\frac{m_2^2 (4 \epsilon \mathcal{L}_{m_2} (6 \epsilon \mathcal{L}_{m_2}-17 \epsilon-6)+\epsilon (\epsilon (12\zeta_2+59)+34)+12)}{8  \epsilon^2},
    \\&
    J^{(m_2)}_{01101}=\frac{1}{8 } m_2^2 \left(\frac{2}{\epsilon}-4 \mathcal{L}_{m_2}+13\right),
    \\&
    J^{(m_2)}_{1^{(m)}001^{(m)}1}=-\frac{m_2^2 (4 \epsilon \mathcal{L}_{m_2} (4 \epsilon \mathcal{L}_{m_2}-11 \epsilon-4)+\epsilon (\epsilon (12S_1+8 \zeta_2+35)+22)+8)}{8  \epsilon^2},
    \\&
    J^{(m_2)}_{11110}=\frac{2 \epsilon \mathcal{L}_{m_2} (\epsilon (-\mathcal{L}_{m_2})+4 \epsilon+1)+\epsilon (\epsilon(\zeta_2-12)-4)-1}{ \epsilon^2},
    \\&
    J^{(m_2)}_{11^{(m)}111^{(m)}}=\frac{\frac{27}{2}S_1 S_2+i \pi  \zeta_2-3 \zeta_3}{ m_2^2},
\end{align}
\end{subequations}}
where $S_1=\frac{\pi}{\sqrt{3}}$ and $S_2=\frac{4}{9\sqrt{3}}Cl_2(\frac{\pi}{3})$ such that $Cl_2$ denotes the second order Clusen function,
\begin{equation}
   Cl_2(\theta)=-\int_0^{\theta}\log{|2\sin{(\theta/2)}|}d\theta:\quad 0<\theta<2\pi.
\end{equation}
Next, we consider the vertex contributions, $V^{(m)}_i$, at the scale $\mu \sim m_{1,2}\gg M$, for matching from the full theory to heavy-heavy operators. After reduction we have the following master integrals,
\small{
\begin{subequations}
\begin{align}
    & J^{(m)}_{001^{(m)}1^{(m)}0}=-\frac{m^4 (2 \epsilon \mathcal{L}_{m} (\epsilon \mathcal{L}_{m}-2 \epsilon-1)+\epsilon (\epsilon(\zeta_2+3)+2)+1)}{ \epsilon^2},
    \\&
   J^{(m)}_{1001^{(m)}1}=-\frac{m^2 (4 \epsilon \mathcal{L}_{m} (2 \epsilon \mathcal{L}_{m}-5 \epsilon-2)+\epsilon (\epsilon (20\zeta_2+11)+10)+4)}{8 \epsilon^2}, 
   \\&
    J^{(m)}_{01^{(m)}1^{(m)}01^{(m)}}=\frac{m^2 (-2 \epsilon \mathcal{L}_{m} (\epsilon \mathcal{L}_{m}+\epsilon (S_1-3)-1)}{\epsilon^2} \\& \quad\quad\quad\quad\quad\quad\quad\quad +\frac{\epsilon(-\epsilon (S_1 (\log{(3)}-3)-9 S_2+\zeta_2+7)+S_1-3)-1)}{\epsilon^2}.
\end{align}
\end{subequations}}
As for the full theory two-loop wave-function contributions, present in Appendix \ref{sec:wfr2}, it is well-known that they map to MIs illustrated by figure \ref{fig:top2}, and thus we refrain from going into detail. 

\begin{figure}[htp]
\centering
\begin{center}
\scalebox{0.48}{
\fcolorbox{white}{white}{
  \begin{picture}(868,189) (79,-20)
    \SetWidth{1.0}
    \SetColor{Black}
    \Line[arrow,arrowpos=0.5,arrowlength=5,arrowwidth=2,arrowinset=0.2,double,sep=2](80,45)(145,45)
    \Line[arrow,arrowpos=0.5,arrowlength=5,arrowwidth=2,arrowinset=0.2,double,sep=2](145,45)(210,45)
    \Line[arrow,arrowpos=0.5,arrowlength=5,arrowwidth=2,arrowinset=0.2,double,sep=2](210,45)(273,45)
    \Line[arrow,arrowpos=0.5,arrowlength=5,arrowwidth=2,arrowinset=0.2,double,sep=2](273,45)(337,45)
    \Line[arrow,arrowpos=0.5,arrowlength=5,arrowwidth=2,arrowinset=0.2](129,44)(208,125)
    \Line[arrow,arrowpos=0.5,arrowlength=5,arrowwidth=2,arrowinset=0.2,flip](208,124)(288,44)
    \Line[arrow,arrowpos=0.5,arrowlength=5,arrowwidth=2,arrowinset=0.2](208,125)(209,44)
    \Text(167,21)[lb]{\Large{\Black{$1$}}}
    \Text(246,20)[lb]{\Large{\Black{$2$}}}
    \Text(255,109)[lb]{\Large{\Black{$3$}}}
    \Text(224,73)[lb]{\Large{\Black{$4$}}}
    \Text(160,108)[lb]{\Large{\Black{$5$}}}
    \Line[arrow,arrowpos=0.5,arrowlength=5,arrowwidth=2,arrowinset=0.2,double,sep=2](385,45)(450,45)
    \Line[arrow,arrowpos=0.5,arrowlength=5,arrowwidth=2,arrowinset=0.2,double,sep=2](450,45)(513,45)
    \Line[arrow,arrowpos=0.5,arrowlength=5,arrowwidth=2,arrowinset=0.2,double,sep=2](513,45)(578,44)
    \Line[arrow,arrowpos=0.5,arrowlength=5,arrowwidth=2,arrowinset=0.2,double,sep=2](578,44)(641,44)
    \Arc[arrow,arrowpos=0.5,arrowlength=5,arrowwidth=2,arrowinset=0.2,clock](494.5,28.173)(68.357,166.613,13.387)
    \Arc[arrow,arrowpos=0.5,arrowlength=5,arrowwidth=2,arrowinset=0.2,flip](555.779,61.73)(65.937,-165.302,-16.503)
    \Text(455,23)[lb]{\Large{\Black{$1$}}}
    \Text(581,23)[lb]{\Large{\Black{$2$}}}
    \Text(493,116)[lb]{\Large{\Black{$5$}}}
    \Text(559,-25)[lb]{\Large{\Black{$3$}}}
    \Text(527,23)[lb]{\Large{\Black{$6$}}}
    \Arc[arrow,arrowpos=0.5,arrowlength=5,arrowwidth=2,arrowinset=0.2,clock](813.362,38.666)(73.489,176.619,109.36)
    \Arc[dash,dashsize=2,arrow,arrowpos=0.785,arrowlength=5,arrowwidth=2,arrowinset=0.2,flip](817,108)(26.926,165,525)
    \Arc[arrow,arrowpos=0.5,arrowlength=5,arrowwidth=2,arrowinset=0.2,clock](825.189,38.811)(72.932,74.238,3.293)
    \Text(817,23)[lb]{\Large{\Black{$1$}}}
    \Text(736,94)[lb]{\Large{\Black{$5$}}}
    \Text(897,93)[lb]{\Large{\Black{$5$}}}
    \Text(816,148)[lb]{\Large{\Black{$7$}}}
    \Text(817,69)[lb]{\Large{\Black{$8$}}}
    \Line[arrow,arrowpos=0.5,arrowlength=5,arrowwidth=2,arrowinset=0.2,double,sep=2](690,44)(771,44)
    \Line[arrow,arrowpos=0.5,arrowlength=5,arrowwidth=2,arrowinset=0.2,double,sep=2](771,44)(865,44)
    \Line[arrow,arrowpos=0.5,arrowlength=5,arrowwidth=2,arrowinset=0.2,double,sep=2](865,44)(946,44)
  \end{picture}}
}
\end{center}
   \caption{Heavy field self-energy topologies. The MIs associated to other topologies are subsets of the MIs required for topologies illustrated.}
    \label{fig:top3}
\end{figure}
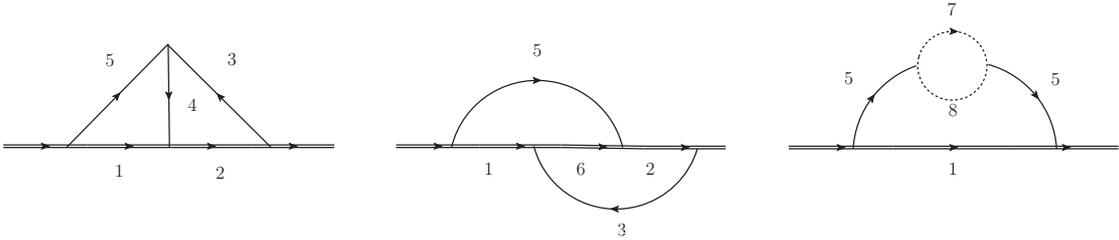

With regards to the effective theory MIs, the HPET vertex and wave-function contributions have MIs with topologies represented by figures \ref{fig:top1} (d,e) and \ref{fig:top3}, respectively, the analytic expressions of which we will present in upcoming work \cite{assi}. Instead we present our results in an attached ancillary file, in general dimension, $d$, with unevaluated master integrals, defined by the notation below. We begin with considering the heavy-light currents in figures \ref{fig:top1} (d), the master integrals of which can be expressed in terms of a single integral family with seven propagators given by,
\begin{equation}
    R^{(s)}_{\nu_1\ldots\nu_7}=\left[(4\pi)^{2-\epsilon}e^{\gamma_E\epsilon}\right]^2\int\slashed{d}^Dl_1\slashed{d}^Dl_2\frac{1}{D_1^{\nu_1}\ldots D_7^{\nu_7}},
\end{equation}
where $v_2$ is the heavy particle velocity, $p_1$ and $m_1$ are the full theory field momentum and mass, $p_1\cdot v_2\equiv w'$, and thus,
\begin{align}
    & \quad D_1(m)=(l_1+p_1)^2-m^2, \quad D_2(m)=(l_2+p_1)^2-m^2, \quad D_3=(l_1-l_2)^2, \quad D_4=l_2^2, \nonumber \\&  \quad D_5=l_1\cdot v_2,  \quad D_6=l_2\cdot v_2,  \quad D_7=(l_1-l_2)\cdot v_2+w'.
\end{align}
Similarly, for the heavy-heavy vertex contributions, all sub-topologies can be mapped to the largest unique two that are shown in figure \ref{fig:top1} (e). We can again express all master integrals in terms of nine propagators given by,
\begin{equation}
    K^{(s)}_{\nu_1\ldots\nu_9}=\left[(4\pi)^{2-\epsilon}e^{\gamma_E\epsilon}\right]^2\int\slashed{d}^Dl_1\slashed{d}^Dl_2\frac{1}{D_1^{\nu_1}\ldots D_9^{\nu_9}},
\end{equation}
where $v_{1,2}$ are the heavy particle velocities, $v_1\cdot v_2\equiv w$, $M$ is the mass of exchanged bosons, and
\begin{align}
    & \quad D_1=l_2\cdot v_1, \quad D_2=l_1\cdot v_1, \quad D_3=(l_1-l_2)^2-M^2,\quad  D_4=l_1^2-M^2, \nonumber \\& \quad D_5(M)=l_1\cdot v_2,\quad D_6(M)=l_2\cdot v_2, \quad D_7(M)=l_2^2-M^2 , \quad D_8=l_2^2, \quad D_9=l_1^2 
\end{align}
Finally, we examine the wave-function contributions of which all topologies are mapped to those shown in figure \ref{fig:top3}. In this case we can express all MIs in terms of six propagators given by,
\begin{equation}
    L^{(s)}_{\nu_1\ldots\nu_8}=\left[(4\pi)^{2-\epsilon}e^{\gamma_E\epsilon}\right]^2\int\slashed{d}^Dl_1\slashed{d}^Dl_2\frac{1}{D_1^{\nu_1}\ldots D_8^{\nu_8}},
\end{equation}
where $p$ and $v$ are the heavy particle residual momentum and velocity, $M$ is the mass of exchanged bosons, and
\begin{align}
    & \quad D_1=(p-l_1)\cdot v, \quad D_2=(p+l_2)\cdot v, \quad D_3(M)=l_2^2-M^2, \quad D_4(M)=(l_1+l_2)^2-m^2, \nonumber \\& \quad D_5(M)=l_1^2-M^2,  \quad D_6=(p+l_2-l_1)\cdot v,  \quad D_7=l_2^2,  \quad D_8=(l_1-l_2)^2.
\end{align}
\newpage
\section{Discussion}
Both the massive and massless form factors are indispensable building blocks to a broad set of observables in both high and low energy regimes. Precisely studying these factors are crucial for shedding light on mysteries that remain in the standard model and beyond, such as the physical structure of the top quark, aspects of mass generation and the nature of dark matter. Our broad set of composite operators, choice of model for applicability, as well as consideration of two critical energetic regimes is emblematic of the breadth of the problem at hand. Our two-loop results are not complete, as we have not considered all possible regimes, for instance, the low-energy regime, nor have we calculated the vertex corrections at two-loop orders for SCET graphs. Continuing to map this space at two-loops and beyond is essential for our predicting power to be able to match the high precision potential of a future electron-positron collider and the LHC in its upcoming high luminosity operating phase. 

At this point, the effective theory formalism is a central front of attack when it comes to tackling such complex problems by breaking them downscale by scale. By application to the SM, we have begun extending the work on EW corrections to high energy processes beyond NLO, as stated in the latest review \cite{denner2020electroweak}. Moreover, we are mapping other parts of the energetic landscape, aside from just the Sudakov regime, which itself opens the door for further investigation. Beyond the SM, the generality of the model and operators studied means that our results can be applied to BSM models by replacement of the proper coupling and group theory factors, which would be very interesting to examine further. For instance, one can apply our results to various models of dark matter \cite{ciafaloni2011weak,ovanesyan2015heavy}, where weak corrections are significant for indirect detection.
\appendix
\section{Feynman Rules}
\label{sec:frules}
This appendix lists the Feynman rules of the vertices which are needed for the calculation of the form factor, as they follow from the Lagrangian of the spontaneously broken SU(2) gauge model described in section \ref{ssec:SUNHiggs}. The gauge boson fields of mass $M = M_W$ are $W^a_{\mu}: a = 1, 2, 3$ (with Lorentz vector index $\mu$). To each $W^a$ corresponds a ghost field $c$ (and antighost $\bar{c}$ ) and a Goldstone boson $\phi^a$, one of the unphysical components of the Higgs field. In the Feynman-t'Hooft gauge used by us, one sets $M_{\phi} = M_W$ . The fields, $\psi$ and $\chi$ denote fermions and complex scalars, respectively.  Lastly, $g$ is the weak SU(2) coupling, and the labelling, $\lbrace 1,2 \rbrace$ differentiates between the particles on the grounds of mass, if two of the same kind exist in a vertex. Vertices which apply beyond two-loops are omitted here but should be included if one wants to venture beyond.
\subsection{Gauge boson self and Ghost couplings}
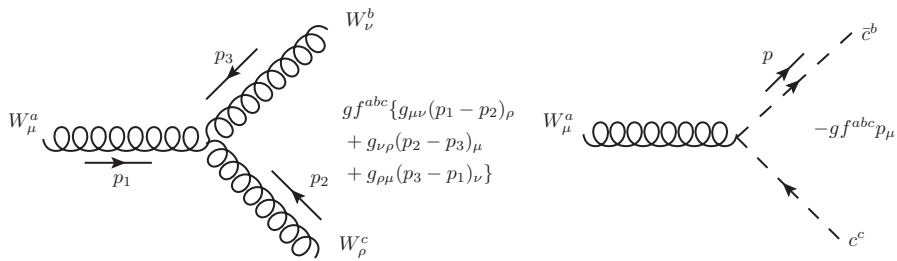
\begin{figure}[H]
\centering
\begin{center}
\scalebox{0.75}{
\fcolorbox{white}{white}{
  \begin{picture}(451,131) (21,7)
    \SetWidth{0.5}
    \SetColor{Black}
    \Text(19.068,69.916)[lb]{{\Black{$W^a_{\mu}$}}}
    \Text(187.502,122.353)[lb]{{\Black{$W^b_{\nu}$}}}
    \Text(184.324,7.945)[lb]{{\Black{$W^c_{\rho}$}}}
    \SetWidth{1.0}
    \Line[arrow,arrowpos=0.5,arrowlength=5,arrowwidth=2,arrowinset=0.2](57.204,55.615)(90.573,55.615)
    \Line[arrow,arrowpos=0.5,arrowlength=5,arrowwidth=2,arrowinset=0.2](174.79,27.013)(150.16,51.642)
    \Line[arrow,arrowpos=0.5,arrowlength=5,arrowwidth=2,arrowinset=0.2](143.01,112.819)(117.586,87.395)
    \Text(122.353,105.668)[lb]{{\Black{$p_3$}}}
    \Text(71.505,43.697)[lb]{{\Black{$p_1$}}}
    \Text(170.023,43.697)[lb]{{\Black{$p_2$}}}
    \Text(185.913,43.697)[lb]{{\Black{$\begin{aligned}[t] &gf^{abc}\lbrace g_{\mu\nu}(p_1-p_2)_{\rho}\\&+g_{\nu\rho}(p_2-p_3)_{\mu}\\&+ g_{\rho\mu}(p_3-p_1)_{\nu} \rbrace \end{aligned}$}}}
    \Line[dash,dashsize=7.945,arrow,arrowpos=0.5,arrowlength=5,arrowwidth=2,arrowinset=0.2](381.36,68.327)(438.564,120.764)
    \Text(286.02,69.916)[lb]{{\Black{$W^a_{\mu}$}}}
    \Text(444.92,117.586)[lb]{{\Black{$\Bar{c}^b$}}}
    \Text(438.564,14.246)[lb]{{\Black{$c^{c}$}}}
    \Text(421.085,67.532)[lb]{{\Black{$-gf^{abc}p_{\mu}$}}}
    \Line[dash,dashsize=7.945,arrow,arrowpos=0.5,arrowlength=5,arrowwidth=2,arrowinset=0.2](432.208,17.479)(381.36,68.327)
    \Line[arrow,arrowpos=0.5,arrowlength=5,arrowwidth=2,arrowinset=0.2](395.661,91.367)(414.729,110.435)
    \Text(395.661,105.668)[lb]{{\Black{$p$}}}
    \Gluon(119.175,67.532)(36.547,67.532){5.959}{8}
    \Gluon(117.586,65.149)(171.612,7.945){5.959}{8}
    \Gluon(119.175,67.532)(177.968,122.353){5.959}{8}
    \Gluon(381.36,69.916)(305.088,69.916){5.959}{8}
  \end{picture}}
}
\end{center}
    \label{fig:my_label}
\end{figure}
\subsection{Bosonic couplings to fermions/scalars}
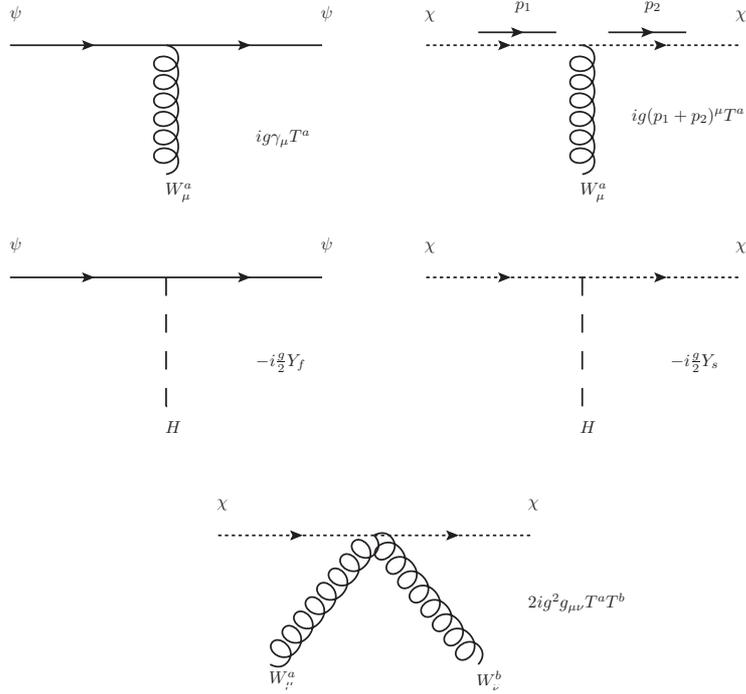
\begin{figure}[H]
\centering
\begin{center}
\scalebox{0.65}{
\fcolorbox{white}{white}{
  \begin{picture}(451,408) (13,-19)
    \SetWidth{1.0}
    \SetColor{Black}
    \Line[arrow,arrowpos=0.5,arrowlength=5,arrowwidth=2,arrowinset=0.2](14.918,348.713)(104.427,348.713)
    \Line[arrow,arrowpos=0.5,arrowlength=5,arrowwidth=2,arrowinset=0.2](104.427,348.713)(193.937,348.713)
    \Text(104.427,259.204)[lb]{\normalsize{\Black{$W^a_{\mu}$}}}
    \Text(14.918,363.631)[lb]{{\Black{$\psi$}}}
    \Text(193.937,363.631)[lb]{{\Black{$\psi$}}}
    \Text(156.641,290.905)[lb]{{\Black{$i g \gamma_{\mu}T^a$}}}
    \Line[arrow,arrowpos=0.5,arrowlength=5,arrowwidth=2,arrowinset=0.2](104.427,214.449)(193.937,214.449)
    \Line[arrow,arrowpos=0.5,arrowlength=5,arrowwidth=2,arrowinset=0.2](14.918,214.449)(104.427,214.449)
    \Text(193.937,229.367)[lb]{{\Black{$\psi$}}}
    \Text(14.918,229.367)[lb]{{\Black{$\psi$}}}
    \Line[dash,dashsize=9.324](104.427,139.858)(104.427,214.449)
    \Text(104.427,124.94)[lb]{{\Black{$H$}}}
    \Text(156.641,160.371)[lb]{{\Black{$-i\frac{g}{2}Y_f$}}}
    \Line[dash,dashsize=9.324](343.119,214.449)(343.119,139.858)
    \Text(372.955,303.958)[lb]{{\Black{$i g (p_1+p_2)^{\mu}T^a$}}}
    \Text(395.332,160.371)[lb]{{\Black{$-i\frac{g}{2}Y_s$}}}
    \Text(253.609,363.631)[lb]{{\Black{$\chi$}}}
    \Text(432.628,363.631)[lb]{{\Black{$\chi$}}}
    \Text(253.609,229.367)[lb]{{\Black{$\chi$}}}
    \Text(432.628,229.367)[lb]{{\Black{$\chi$}}}
    \Text(343.119,259.204)[lb]{{\Black{$W^a_{\mu}$}}}
    \Text(343.119,124.94)[lb]{{\Black{$H$}}}
    \Line[dash,dashsize=1.865,arrow,arrowpos=0.5,arrowlength=5,arrowwidth=2,arrowinset=0.2](253.609,348.713)(343.119,348.713)
    \Line[dash,dashsize=1.865,arrow,arrowpos=0.5,arrowlength=5,arrowwidth=2,arrowinset=0.2](343.119,348.713)(432.628,348.713)
    \Line[dash,dashsize=1.865,arrow,arrowpos=0.5,arrowlength=5,arrowwidth=2,arrowinset=0.2](343.119,214.449)(432.628,214.449)
    \Line[dash,dashsize=1.865,arrow,arrowpos=0.5,arrowlength=5,arrowwidth=2,arrowinset=0.2](253.609,214.449)(343.119,214.449)
    \Text(380.414,369.225)[lb]{{\Black{$p_2$}}}
    \Text(305.823,369.225)[lb]{{\Black{$p_1$}}}
    \Line[arrow,arrowpos=0.5,arrowlength=5,arrowwidth=2,arrowinset=0.2](283.446,356.172)(328.2,356.172)
    \Line[arrow,arrowpos=0.5,arrowlength=5,arrowwidth=2,arrowinset=0.2](358.037,356.172)(402.791,356.172)
    \Line[dash,dashsize=1.865,arrow,arrowpos=0.5,arrowlength=5,arrowwidth=2,arrowinset=0.2](134.264,65.267)(223.773,65.267)
    \Line[dash,dashsize=1.865,arrow,arrowpos=0.5,arrowlength=5,arrowwidth=2,arrowinset=0.2](223.773,65.267)(313.282,65.267)
    \Text(134.264,80.185)[lb]{{\Black{$\chi$}}}
    \Text(313.282,80.185)[lb]{{\Black{$\chi$}}}
    \Gluon(223.773,65.267)(164.1,-9.324){6.993}{8}
    \Gluon(223.773,65.267)(283.446,-9.324){6.993}{8}
    \Text(313.282,20.513)[lb]{{\Black{$2ig^2g_{\mu\nu}T^aT^b$}}}
    \Text(164.1,-24.242)[lb]{\normalsize{\Black{$W^a_{\mu}$}}}
    \Text(283.446,-24.242)[lb]{\normalsize{\Black{$W^b_{\nu}$}}}
    \Gluon(104.427,348.713)(104.427,274.122){6.993}{6}
    \Gluon(343.119,348.713)(343.119,274.122){6.993}{6}
  \end{picture}}
}
\end{center}
    \label{fig:my_label}
\end{figure}
\subsection{Gauge boson coupling to Higgs and Goldstone bosons}
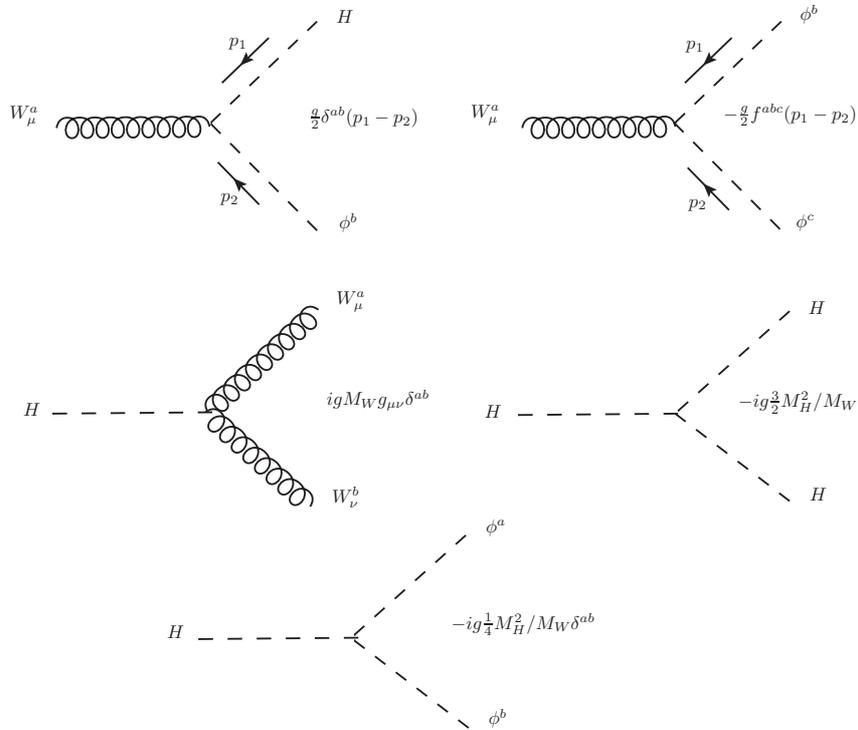
\begin{figure}[H]
\centering
\begin{center}
\scalebox{0.7}{
\fcolorbox{white}{white}{
  \begin{picture}(451,396) (20,-12)
    \SetWidth{1.0}
    \SetColor{Black}
    \Line[dash,dashsize=7.338](125.477,313.325)(181.978,367.624)
    \Line[dash,dashsize=7.338](125.477,312.591)(181.978,255.356)
    \Text(18.345,312.591)[lb]{{\Black{$W^a_{\mu}$}}}
    \Text(192.985,368.358)[lb]{{\Black{$H$}}}
    \Text(194.452,255.356)[lb]{{\Black{$\phi^b$}}}
    \Line[arrow,arrowpos=0.5,arrowlength=5,arrowwidth=2,arrowinset=0.2](156.295,358.819)(131.347,334.604)
    \Line[arrow,arrowpos=0.5,arrowlength=5,arrowwidth=2,arrowinset=0.2](151.159,269.298)(129.146,292.045)
    \Text(135.75,353.683)[lb]{{\Black{$p_1$}}}
    \Text(131.347,270.032)[lb]{{\Black{$p_2$}}}
    \Text(178.309,311.123)[lb]{{\Black{$\frac{g}{2}\delta^{ab}(p_1-p_2)$}}}
    \Line[dash,dashsize=7.338](372.761,313.325)(429.262,367.624)
    \Line[dash,dashsize=7.338](372.027,313.325)(428.528,256.09)
    \Text(441.003,367.624)[lb]{{\Black{$\phi^b$}}}
    \Text(438.801,256.09)[lb]{{\Black{$\phi^c$}}}
    \Text(400.047,312.591)[lb]{{\Black{$-\frac{g}{2}f^{abc}(p_1-p_2)$}}}
    \Text(264.895,313.325)[lb]{{\Black{$W^a_{\mu}$}}}
    \Line[dash,dashsize=7.338](41.092,157.029)(126.21,157.763)
    \Text(25.682,156.295)[lb]{{\Black{$H$}}}
    \Text(192.985,212.797)[lb]{{\Black{$W^a_{\mu}$}}}
    \Text(190.783,107.132)[lb]{{\Black{$W^b_{\nu}$}}}
    \Text(187.848,159.231)[lb]{{\Black{$igM_Wg_{\mu\nu}\delta^{ab}$}}}
    \Line[dash,dashsize=7.338](289.11,156.295)(374.229,157.029)
    \Text(272.233,155.562)[lb]{{\Black{$H$}}}
    \Line[dash,dashsize=7.338](374.229,158.497)(433.665,212.063)
    \Line[dash,dashsize=7.338](372.761,157.029)(433.665,110.067)
    \Text(445.405,211.329)[lb]{{\Black{$H$}}}
    \Text(446.139,110.801)[lb]{{\Black{$H$}}}
    \Line[dash,dashsize=7.338](118.873,35.955)(203.991,36.689)
    \Line[dash,dashsize=7.338](201.79,38.157)(261.226,91.723)
    \Line[dash,dashsize=7.338](201.79,35.222)(262.694,-11.741)
    \Text(272.967,91.723)[lb]{{\Black{$\phi^a$}}}
    \Text(274.434,-11.741)[lb]{{\Black{$\phi^b$}}}
    \Text(102.729,36.689)[lb]{{\Black{$H$}}}
    \Text(407.982,157.763)[lb]{{\Black{$-ig\frac{3}{2}M_H^2/M_W$}}}
    \Text(254.622,38.157)[lb]{{\Black{$-ig\frac{1}{4}M_H^2/M_W\delta^{ab}$}}}
    \Line[arrow,arrowpos=0.5,arrowlength=5,arrowwidth=2,arrowinset=0.2](402.846,358.819)(377.897,334.604)
    \Line[arrow,arrowpos=0.5,arrowlength=5,arrowwidth=2,arrowinset=0.2](401.378,266.363)(379.365,289.11)
    \Text(381.566,267.096)[lb]{{\Black{$p_2$}}}
    \Text(380.099,352.215)[lb]{{\Black{$p_1$}}}
    \Gluon(43.293,310.39)(124.743,311.123){5.503}{9}
    \Gluon(291.311,310.39)(372.761,311.123){5.503}{9}
    \Gluon(124.009,157.763)(182.712,212.797){5.503}{9}
    \Gluon(122.541,157.763)(178.309,107.132){5.503}{8}
  \end{picture}}
}
\end{center}
    \label{fig:my_label}
\end{figure}
\subsection{Effective theory couplings}
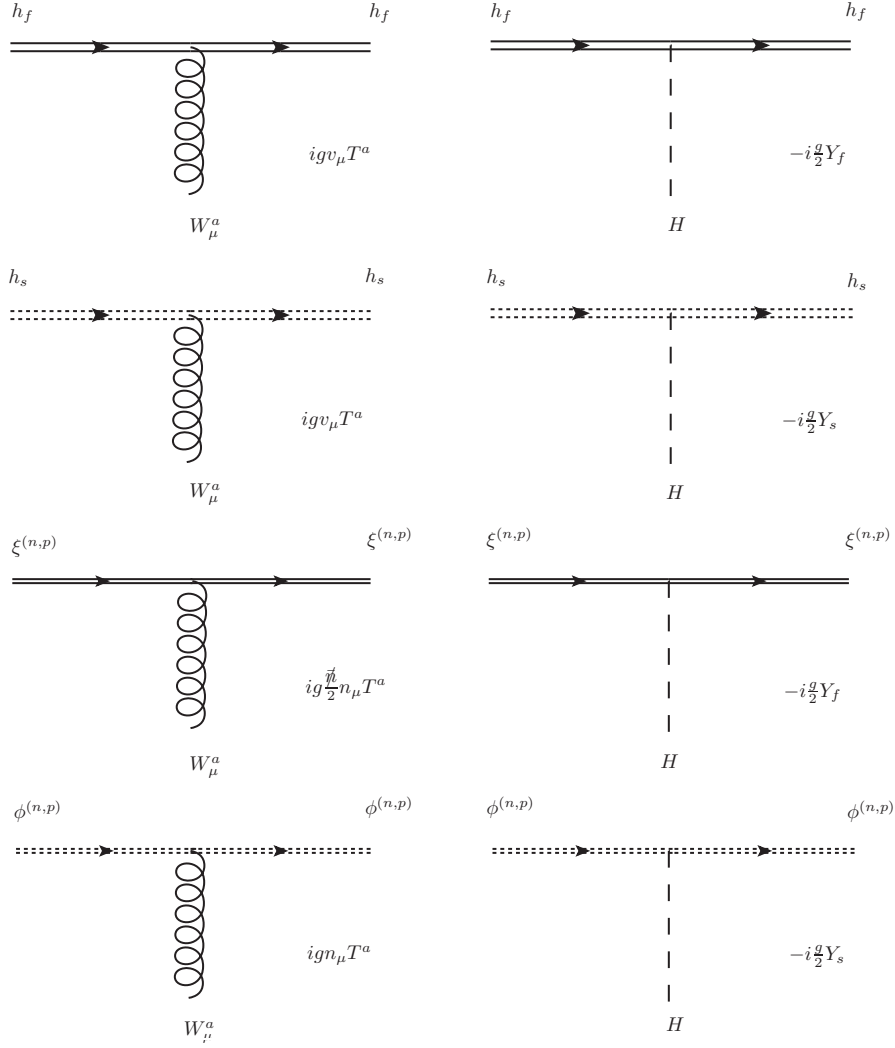
\begin{figure}[H]
\centering
\begin{center}
\scalebox{0.75}{
\fcolorbox{white}{white}{
  \begin{picture}(451,528) (44,-11)
    \SetWidth{1.0}
    \SetColor{Black}
    \Line[arrow,arrowpos=0.5,arrowlength=6.591,arrowwidth=2.636,arrowinset=0.2,double,sep=4](45.687,482.977)(135.196,482.977)
    \Line[arrow,arrowpos=0.5,arrowlength=6.591,arrowwidth=2.636,arrowinset=0.2,double,sep=4](135.196,482.977)(224.705,482.977)
    \Line[arrow,arrowpos=0.5,arrowlength=6.591,arrowwidth=2.636,arrowinset=0.2,double,sep=4](284.378,483.909)(373.887,483.909)
    \Line[arrow,arrowpos=0.5,arrowlength=6.591,arrowwidth=2.636,arrowinset=0.2,double,sep=4](373.887,483.909)(463.397,483.909)
    \Line[dash,dashsize=9.324](373.887,483.909)(373.887,408.386)
    \Line[dash,dashsize=1.865,arrow,arrowpos=0.5,arrowlength=6.591,arrowwidth=2.636,arrowinset=0.2,double,sep=4](45.687,348.713)(135.196,348.713)
    \Line[dash,dashsize=1.865,arrow,arrowpos=0.5,arrowlength=6.591,arrowwidth=2.636,arrowinset=0.2,double,sep=4](135.196,348.713)(224.705,348.713)
    \Line[dash,dashsize=1.865,arrow,arrowpos=0.5,arrowlength=6.591,arrowwidth=2.636,arrowinset=0.2,double,sep=4](284.378,349.645)(373.887,349.645)
    \Line[dash,dashsize=1.865,arrow,arrowpos=0.5,arrowlength=6.591,arrowwidth=2.636,arrowinset=0.2,double,sep=4](374.82,349.645)(464.329,349.645)
    \Line[dash,dashsize=9.324](373.887,349.645)(373.887,274.122)
    \Line[arrow,arrowpos=0.5,arrowlength=5,arrowwidth=2,arrowinset=0.2,double,sep=2](46.619,215.382)(136.129,215.382)
    \Line[arrow,arrowpos=0.5,arrowlength=5,arrowwidth=2,arrowinset=0.2,double,sep=2](135.196,215.382)(224.705,215.382)
    \Line[arrow,arrowpos=0.5,arrowlength=5,arrowwidth=2,arrowinset=0.2,double,sep=2](283.446,215.382)(372.955,215.382)
    \Line[arrow,arrowpos=0.5,arrowlength=5,arrowwidth=2,arrowinset=0.2,double,sep=2](372.955,215.382)(462.464,215.382)
    \Line[dash,dashsize=9.324](372.955,215.382)(372.955,139.858)
    \Line[dash,dashsize=1.865,arrow,arrowpos=0.5,arrowlength=5,arrowwidth=2,arrowinset=0.2,double,sep=2](48.484,80.185)(137.993,80.185)
    \Line[dash,dashsize=1.865,arrow,arrowpos=0.5,arrowlength=5,arrowwidth=2,arrowinset=0.2,double,sep=2](135.196,80.185)(224.705,80.185)
    \Line[dash,dashsize=1.865,arrow,arrowpos=0.5,arrowlength=5,arrowwidth=2,arrowinset=0.2,double,sep=2](285.311,80.185)(374.82,80.185)
    \Line[dash,dashsize=1.865,arrow,arrowpos=0.5,arrowlength=5,arrowwidth=2,arrowinset=0.2,double,sep=2](375.752,80.185)(465.261,80.185)
    \Line[dash,dashsize=9.324](372.955,80.185)(372.955,4.662)
    \Text(135.196,387.873)[lb]{{\Black{$W^a_{\mu}$}}}
    \Text(135.196,254.542)[lb]{{\Black{$W^a_{\mu}$}}}
    \Text(135.196,117.481)[lb]{{\Black{$W^a_{\mu}$}}}
    \Text(132.399,-15.851)[lb]{{\Black{$W^a_{\mu}$}}}
    \Text(373.887,392.535)[lb]{{\Black{$H$}}}
    \Text(372.955,257.339)[lb]{{\Black{$H$}}}
    \Text(370.158,122.143)[lb]{{\Black{$H$}}}
    \Text(372.955,-10.256)[lb]{{\Black{$H$}}}
    \Text(46.619,496.963)[lb]{{\Black{$h_f$}}}
    \Text(224.705,496.963)[lb]{{\Black{$h_f$}}}
    \Text(285.311,496.963)[lb]{{\Black{$h_f$}}}
    \Text(462.464,497.895)[lb]{{\Black{$h_f$}}}
    \Text(44.755,364.564)[lb]{{\Black{$h_s$}}}
    \Text(222.841,365.496)[lb]{{\Black{$h_s$}}}
    \Text(283.446,364.564)[lb]{{\Black{$h_s$}}}
    \Text(463.397,362.699)[lb]{{\Black{$h_s$}}}
    \Text(46.619,228.435)[lb]{{\Black{$\xi^{(n,p)}$}}}
    \Text(223.773,232.165)[lb]{{\Black{$\xi^{(n,p)}$}}}
    \Text(283.446,232.165)[lb]{{\Black{$\xi^{(n,p)}$}}}
    \Text(462.464,232.165)[lb]{{\Black{$\xi^{(n,p)}$}}}
    \Text(47.552,95.104)[lb]{{\Black{$\phi^{(n,p)}$}}}
    \Text(222.841,96.968)[lb]{{\Black{$\phi^{(n,p)}$}}}
    \Text(283.446,96.968)[lb]{{\Black{$\phi^{(n,p)}$}}}
    \Text(463.397,95.104)[lb]{{\Black{$\phi^{(n,p)}$}}}
    \Text(194.869,425.169)[lb]{{\Black{$igv_{\mu}T^a$}}}
    \Text(191.139,291.837)[lb]{{\Black{$igv_{\mu}T^a$}}}
    \Text(434.493,425.169)[lb]{{\Black{$-i\frac{g}{2}Y_f$}}}
    \Text(430.763,290.905)[lb]{{\Black{$-i\frac{g}{2}Y_s$}}}
    \Text(193.004,156.641)[lb]{{\Black{$ig\frac{\slashed{\Bar{n}}}{2}n_{\mu}T^a$}}}
    \Text(193.937,23.31)[lb]{{\Black{$ign_{\mu}T^a$}}}
    \Text(431.695,154.776)[lb]{{\Black{$-i\frac{g}{2}Y_f$}}}
    \Text(434.493,22.377)[lb]{{\Black{$-i\frac{g}{2}Y_s$}}}
    \Gluon(135.196,482.977)(134.264,409.318){6.993}{6}
    \Gluon(134.264,348.713)(133.331,275.054){6.993}{6}
    \Gluon(136.129,215.382)(135.196,141.723){6.993}{6}
    \Gluon(135.196,80.185)(134.264,6.527){6.993}{6}
  \end{picture}}
}
\end{center}
    \label{fig:my_label}
\end{figure}
In the effective theory vertices, solid (dashed) lines correspond to fermions (scalars) and widely (thinly) spaced lines correspond to heavy (co-linear) particles. As for the co-linear vertices we do not distinguish between soft/Wilson line couplings as they are identical up to the order we are considering.
\section{Mass Renormalisation}
\label{sec:massren}
\subsection{Matching at $\mu\sim m_{1,2}$:}
\begin{tiny}
\begin{align}
\begin{autobreak}
\Delta B^{(2)}_1=
-\frac{3 C_A m_+^2 \mathcal{L}_{m_1}^3 C_F^3}{2 m_-^2}+3 C_A \mathcal{L}_{m_1}^3 C_F^3-
\frac{3 C_A m_-^2 \mathcal{L}_{m_1}^3 C_F^3}{2 m_+^2}-
\frac{3 C_A m_+^2 \mathcal{L}_{m_2}^3 C_F^3}{2 m_-^2}+3 C_A \mathcal{L}_{m_2}^3 C_F^3-
\frac{3 C_A m_-^2 \mathcal{L}_{m_2}^3 C_F^3}{2 m_+^2}+
\frac{C_A m_+^2 \mathcal{L}_{m_1}^2 C_F^3}{2 m_-^2}+
\frac{3 C_A m_+^2 \mathcal{L}_{m_1}^2 C_F^3}{2 \varepsilon m_-^2}-4 C_A \mathcal{L}_{m_1}^2 C_F^3+
\frac{12 C_A m_+ \mathcal{L}_{m_1}^2 C_F^3}{m_-}-
\frac{3 C_A \mathcal{L}_{m_1}^2 C_F^3}{\varepsilon }+
\frac{12 C_A m_- \mathcal{L}_{m_1}^2 C_F^3}{m_+}+
\frac{7 C_A m_-^2 \mathcal{L}_{m_1}^2 C_F^3}{2 m_+^2}+
\frac{3 C_A m_-^2 \mathcal{L}_{m_1}^2 C_F^3}{2 \varepsilon m_+^2}+
\frac{C_A m_+^2 \mathcal{L}_{m_2}^2 C_F^3}{2 m_-^2}+
\frac{3 C_A m_+^2 \mathcal{L}_{m_2}^2 C_F^3}{2 \varepsilon m_-^2}-4 C_A \mathcal{L}_{m_2}^2 C_F^3-
\frac{12 C_A m_+ \mathcal{L}_{m_2}^2 C_F^3}{m_-}+
\frac{3 C_A m_+^2 \mathcal{L}_{m_1} \mathcal{L}_{m_2}^2 C_F^3}{2 m_-^2}-3 C_A \mathcal{L}_{m_1} \mathcal{L}_{m_2}^2 C_F^3+
\frac{3 C_A m_-^2 \mathcal{L}_{m_1} \mathcal{L}_{m_2}^2 C_F^3}{2 m_+^2}-
\frac{3 C_A \mathcal{L}_{m_2}^2 C_F^3}{\varepsilon }-
\frac{12 C_A m_- \mathcal{L}_{m_2}^2 C_F^3}{m_+}+
\frac{7 C_A m_-^2 \mathcal{L}_{m_2}^2 C_F^3}{2 m_+^2}+
\frac{3 C_A m_-^2 \mathcal{L}_{m_2}^2 C_F^3}{2 \varepsilon m_+^2}-16 C_A C_F^3+12 C_A \mathcal{L}_{m_1} C_F^3-
\frac{4 C_A m_+ \mathcal{L}_{m_1} C_F^3}{m_-}-
\frac{12 C_A m_+ \mathcal{L}_{m_1} C_F^3}{\varepsilon m_-}-
\frac{28 C_A m_- \mathcal{L}_{m_1} C_F^3}{m_+}-
\frac{12 C_A m_- \mathcal{L}_{m_1} C_F^3}{\varepsilon m_+}+
\frac{3 C_A m_+^2 \mathcal{L}_{m_1}^2 \mathcal{L}_{m_2} C_F^3}{2 m_-^2}-3 C_A \mathcal{L}_{m_1}^2 \mathcal{L}_{m_2} C_F^3+
\frac{3 C_A m_-^2 \mathcal{L}_{m_1}^2 \mathcal{L}_{m_2} C_F^3}{2 m_+^2}+12 C_A \mathcal{L}_{m_2} C_F^3+
\frac{4 C_A m_+ \mathcal{L}_{m_2} C_F^3}{m_-}+
\frac{12 C_A m_+ \mathcal{L}_{m_2} C_F^3}{\varepsilon m_-}-
\frac{C_A m_+^2 \mathcal{L}_{m_1} \mathcal{L}_{m_2} C_F^3}{m_-^2}-
\frac{3 C_A m_+^2 \mathcal{L}_{m_1} \mathcal{L}_{m_2} C_F^3}{\varepsilon m_-^2}+8 C_A \mathcal{L}_{m_1} \mathcal{L}_{m_2} C_F^3+
\frac{6 C_A \mathcal{L}_{m_1} \mathcal{L}_{m_2} C_F^3}{\varepsilon }-
\frac{7 C_A m_-^2 \mathcal{L}_{m_1} \mathcal{L}_{m_2} C_F^3}{m_+^2}-
\frac{3 C_A m_-^2 \mathcal{L}_{m_1} \mathcal{L}_{m_2} C_F^3}{\varepsilon m_+^2}+
\frac{28 C_A m_- \mathcal{L}_{m_2} C_F^3}{m_+}+
\frac{12 C_A m_- \mathcal{L}_{m_2} C_F^3}{\varepsilon m_+}-
\frac{12 C_A C_F^3}{\varepsilon }-
\frac{3 m_+^2 \mathcal{L}_{m_1}^3 C_F^2}{4 m_-^2}-
\frac{3 m_-^2 \mathcal{L}_{m_1}^3 C_F^2}{4 m_+^2}+
\frac{3}{2} \mathcal{L}_{m_1}^3 C_F^2-
\frac{3 m_+^2 \mathcal{L}_{m_2}^3 C_F^2}{4 m_-^2}-
\frac{3 m_-^2 \mathcal{L}_{m_2}^3 C_F^2}{4 m_+^2}+
\frac{3}{2} \mathcal{L}_{m_2}^3 C_F^2+
\frac{3 m_+^2 \mathcal{L}_{m_1}^2 C_F^2}{4 \varepsilon m_-^2}+
\frac{m_+^2 \mathcal{L}_{m_1}^2 C_F^2}{4 m_-^2}+
\frac{6 m_+ \mathcal{L}_{m_1}^2 C_F^2}{m_-}-
\frac{3 \mathcal{L}_{m_1}^2 C_F^2}{2 \varepsilon }+
\frac{6 m_- \mathcal{L}_{m_1}^2 C_F^2}{m_+}+
\frac{3 m_-^2 \mathcal{L}_{m_1}^2 C_F^2}{4 \varepsilon m_+^2}+
\frac{7 m_-^2 \mathcal{L}_{m_1}^2 C_F^2}{4 m_+^2}-2 \mathcal{L}_{m_1}^2 C_F^2+
\frac{3 m_+^2 \mathcal{L}_{m_2}^2 C_F^2}{4 \varepsilon m_-^2}+
\frac{m_+^2 \mathcal{L}_{m_2}^2 C_F^2}{4 m_-^2}-
\frac{6 m_+ \mathcal{L}_{m_2}^2 C_F^2}{m_-}+
\frac{3 m_+^2 \mathcal{L}_{m_1} \mathcal{L}_{m_2}^2 C_F^2}{4 m_-^2}+
\frac{3 m_-^2 \mathcal{L}_{m_1} \mathcal{L}_{m_2}^2 C_F^2}{4 m_+^2}-
\frac{3}{2} \mathcal{L}_{m_1} \mathcal{L}_{m_2}^2 C_F^2-
\frac{3 \mathcal{L}_{m_2}^2 C_F^2}{2 \varepsilon }-
\frac{6 m_- \mathcal{L}_{m_2}^2 C_F^2}{m_+}+
\frac{3 m_-^2 \mathcal{L}_{m_2}^2 C_F^2}{4 \varepsilon m_+^2}+
\frac{7 m_-^2 \mathcal{L}_{m_2}^2 C_F^2}{4 m_+^2}-2 \mathcal{L}_{m_2}^2 C_F^2-
\frac{6 m_+ \mathcal{L}_{m_1} C_F^2}{\varepsilon m_-}-
\frac{2 m_+ \mathcal{L}_{m_1} C_F^2}{m_-}-
\frac{6 m_- \mathcal{L}_{m_1} C_F^2}{\varepsilon m_+}-
\frac{14 m_- \mathcal{L}_{m_1} C_F^2}{m_+}+6 \mathcal{L}_{m_1} C_F^2+
\frac{3 m_+^2 \mathcal{L}_{m_1}^2 \mathcal{L}_{m_2} C_F^2}{4 m_-^2}+
\frac{3 m_-^2 \mathcal{L}_{m_1}^2 \mathcal{L}_{m_2} C_F^2}{4 m_+^2}-
\frac{3}{2} \mathcal{L}_{m_1}^2 \mathcal{L}_{m_2} C_F^2+
\frac{6 m_+ \mathcal{L}_{m_2} C_F^2}{\varepsilon m_-}+
\frac{2 m_+ \mathcal{L}_{m_2} C_F^2}{m_-}-
\frac{3 m_+^2 \mathcal{L}_{m_1} \mathcal{L}_{m_2} C_F^2}{2 \varepsilon m_-^2}-
\frac{m_+^2 \mathcal{L}_{m_1} \mathcal{L}_{m_2} C_F^2}{2 m_-^2}+
\frac{3 \mathcal{L}_{m_1} \mathcal{L}_{m_2} C_F^2}{\varepsilon }-
\frac{3 m_-^2 \mathcal{L}_{m_1} \mathcal{L}_{m_2} C_F^2}{2 \varepsilon m_+^2}-
\frac{7 m_-^2 \mathcal{L}_{m_1} \mathcal{L}_{m_2} C_F^2}{2 m_+^2}+4 \mathcal{L}_{m_1} \mathcal{L}_{m_2} C_F^2+
\frac{6 m_- \mathcal{L}_{m_2} C_F^2}{\varepsilon m_+}+
\frac{14 m_- \mathcal{L}_{m_2} C_F^2}{m_+}+6 \mathcal{L}_{m_2} C_F^2-
\frac{6 C_F^2}{\varepsilon }-8 C_F^2-
\frac{C_A m_+ \beta_0 \mathcal{L}_{m_1}^3 C_F}{12 m_-}-
\frac{C_A m_- \beta_0 \mathcal{L}_{m_1}^3 C_F}{12 m_+}+
\frac{C_A m_+ \beta_0 \mathcal{L}_{m_2}^3 C_F}{12 m_-}+
\frac{C_A m_- \beta_0 \mathcal{L}_{m_2}^3 C_F}{12 m_+}-
\frac{1}{4} C_A \beta_0 \mathcal{L}_{m_1}^2 C_F-
\frac{C_A m_+ \beta_0 \mathcal{L}_{m_1}^2 C_F}{4 m_-}+
\frac{C_A m_+ \beta_0 \mathcal{L}_{m_1}^2 C_F}{4 \varepsilon m_-}+
\frac{C_A m_- \beta_0 \mathcal{L}_{m_1}^2 C_F}{4 m_+}+
\frac{C_A m_- \beta_0 \mathcal{L}_{m_1}^2 C_F}{4 \varepsilon m_+}-
\frac{1}{4} C_A \beta_0 \mathcal{L}_{m_2}^2 C_F+
\frac{C_A m_+ \beta_0 \mathcal{L}_{m_2}^2 C_F}{4 m_-}-
\frac{C_A m_+ \beta_0 \mathcal{L}_{m_2}^2 C_F}{4 \varepsilon m_-}-
\frac{C_A m_- \beta_0 \mathcal{L}_{m_2}^2 C_F}{4 m_+}-
\frac{C_A m_- \beta_0 \mathcal{L}_{m_2}^2 C_F}{4 \varepsilon m_+}-
\frac{C_A \beta_0 C_F}{\varepsilon ^2}-
\frac{1}{12} C_A \pi ^2 \beta_0 C_F+
\frac{C_A m_+ \beta_0 \mathcal{L}_{m_1} C_F}{2 \varepsilon m_-}-
\frac{C_A m_+ \beta_0 \mathcal{L}_{m_1} C_F}{2 \varepsilon ^2 m_-}+
\frac{C_A \beta_0 \mathcal{L}_{m_1} C_F}{2 \varepsilon }-
\frac{2 C_A m_- \beta_0 \mathcal{L}_{m_1} C_F}{m_+}-
\frac{C_A m_- \beta_0 \mathcal{L}_{m_1} C_F}{2 \varepsilon m_+}-
\frac{C_A m_- \beta_0 \mathcal{L}_{m_1} C_F}{2 \varepsilon ^2 m_+}-
\frac{C_A m_+ \pi ^2 \beta_0 \mathcal{L}_{m_1} C_F}{24 m_-}-
\frac{C_A m_- \pi ^2 \beta_0 \mathcal{L}_{m_1} C_F}{24 m_+}-
\frac{C_A m_+ \beta_0 \mathcal{L}_{m_2} C_F}{2 \varepsilon m_-}+
\frac{C_A m_+ \beta_0 \mathcal{L}_{m_2} C_F}{2 \varepsilon ^2 m_-}+
\frac{C_A \beta_0 \mathcal{L}_{m_2} C_F}{2 \varepsilon }+
\frac{2 C_A m_- \beta_0 \mathcal{L}_{m_2} C_F}{m_+}+
\frac{C_A m_- \beta_0 \mathcal{L}_{m_2} C_F}{2 \varepsilon m_+}+
\frac{C_A m_- \beta_0 \mathcal{L}_{m_2} C_F}{2 \varepsilon ^2 m_+}+
\frac{C_A m_+ \pi ^2 \beta_0 \mathcal{L}_{m_2} C_F}{24 m_-}+
\frac{C_A m_- \pi ^2 \beta_0 \mathcal{L}_{m_2} C_F}{24 m_+}-
\frac{3 C_A C_F^2 Y_f^2 \mathcal{L}_{m_1}^2 m_1^4}{m_-^2 m_+^2}-
\frac{3 C_F Y_f^2 \mathcal{L}_{m_1}^2 m_1^4}{2 m_-^2 m_+^2}+
\frac{3 C_A C_F^2 Y_f^2 \mathcal{L}_{m_2}^2 m_1^4}{m_-^2 m_+^2}+
\frac{3 C_F Y_f^2 \mathcal{L}_{m_2}^2 m_1^4}{2 m_-^2 m_+^2}-
\frac{9 C_A Y_f^2 \beta_0 \mathcal{L}_{m_2}^2 m_1^4}{32 m_-^2 m_+^2}+
\frac{3 C_A C_F^2 m_2 Y_f^2 \mathcal{L}_{m_1}^3 m_1^3}{m_-^2 m_+^2}+
\frac{3 C_F m_2 Y_f^2 \mathcal{L}_{m_1}^3 m_1^3}{2 m_-^2 m_+^2}+
\frac{C_A m_2 Y_f^2 \beta_0 \mathcal{L}_{m_1}^3 m_1^3}{12 m_-^2 m_+^2}+
\frac{3 C_A C_F^2 m_2 Y_f^2 \mathcal{L}_{m_2}^3 m_1^3}{m_-^2 m_+^2}+
\frac{3 C_F m_2 Y_f^2 \mathcal{L}_{m_2}^3 m_1^3}{2 m_-^2 m_+^2}-
\frac{C_A m_2 Y_f^2 \beta_0 \mathcal{L}_{m_2}^3 m_1^3}{12 m_-^2 m_+^2}-
\frac{35 C_A C_F^2 m_2 Y_f^2 \mathcal{L}_{m_1}^2 m_1^3}{2 m_-^2 m_+^2}-
\frac{35 C_F m_2 Y_f^2 \mathcal{L}_{m_1}^2 m_1^3}{4 m_-^2 m_+^2}-
\frac{3 C_A C_F^2 m_2 Y_f^2 \mathcal{L}_{m_1}^2 m_1^3}{\varepsilon m_-^2 m_+^2}-
\frac{3 C_F m_2 Y_f^2 \mathcal{L}_{m_1}^2 m_1^3}{2 \varepsilon m_-^2 m_+^2}+
\frac{13 C_A C_F^2 m_2 Y_f^2 \mathcal{L}_{m_2}^2 m_1^3}{2 m_-^2 m_+^2}+
\frac{13 C_F m_2 Y_f^2 \mathcal{L}_{m_2}^2 m_1^3}{4 m_-^2 m_+^2}-
\frac{3 C_A C_F^2 m_2 Y_f^2 \mathcal{L}_{m_2}^2 m_1^3}{\varepsilon m_-^2 m_+^2}-
\frac{3 C_F m_2 Y_f^2 \mathcal{L}_{m_2}^2 m_1^3}{2 \varepsilon m_-^2 m_+^2}+
\frac{C_A m_2 Y_f^2 \beta_0 \mathcal{L}_{m_2}^2 m_1^3}{4 m_-^2 m_+^2}+
\frac{C_A m_2 Y_f^2 \beta_0 \mathcal{L}_{m_2}^2 m_1^3}{4 \varepsilon m_-^2 m_+^2}-
\frac{3 C_A^2 C_F m_2 Y_f^2 \mathcal{L}_{m_1} \mathcal{L}_{m_2}^2 m_1^3}{2 m_-^2 m_+^2}-
\frac{3 C_A^2 C_F m_2 Y_f^2 \mathcal{L}_{m_1}^2 \mathcal{L}_{m_2} m_1^3}{2 m_-^2 m_+^2}+
\frac{11 C_A^2 C_F m_2 Y_f^2 \mathcal{L}_{m_1} \mathcal{L}_{m_2} m_1^3}{2 m_-^2 m_+^2}+
\frac{3 C_A^2 C_F m_2 Y_f^2 \mathcal{L}_{m_1} \mathcal{L}_{m_2} m_1^3}{\varepsilon m_-^2 m_+^2}+
\frac{3 C_A C_F^2 m_2^2 Y_f^2 \mathcal{L}_{m_1}^3 m_1^2}{m_-^2 m_+^2}+
\frac{3 C_F m_2^2 Y_f^2 \mathcal{L}_{m_1}^3 m_1^2}{2 m_-^2 m_+^2}+
\frac{3 C_A C_F^2 m_2^2 Y_f^2 \mathcal{L}_{m_2}^3 m_1^2}{m_-^2 m_+^2}+
\frac{3 C_F m_2^2 Y_f^2 \mathcal{L}_{m_2}^3 m_1^2}{2 m_-^2 m_+^2}-
\frac{13 C_A C_F^2 m_2^2 Y_f^2 \mathcal{L}_{m_1}^2 m_1^2}{m_-^2 m_+^2}-
\frac{13 C_F m_2^2 Y_f^2 \mathcal{L}_{m_1}^2 m_1^2}{2 m_-^2 m_+^2}-
\frac{3 C_A C_F^2 m_2^2 Y_f^2 \mathcal{L}_{m_1}^2 m_1^2}{\varepsilon m_-^2 m_+^2}-
\frac{3 C_F m_2^2 Y_f^2 \mathcal{L}_{m_1}^2 m_1^2}{2 \varepsilon m_-^2 m_+^2}-
\frac{9 C_A Y_f^2 \beta_0 \mathcal{L}_{m_1}^2 m_1^2}{32 m_- m_+}-
\frac{13 C_A C_F^2 m_2^2 Y_f^2 \mathcal{L}_{m_2}^2 m_1^2}{m_-^2 m_+^2}-
\frac{13 C_F m_2^2 Y_f^2 \mathcal{L}_{m_2}^2 m_1^2}{2 m_-^2 m_+^2}-
\frac{3 C_A C_F^2 m_2^2 Y_f^2 \mathcal{L}_{m_2}^2 m_1^2}{\varepsilon m_-^2 m_+^2}-
\frac{3 C_F m_2^2 Y_f^2 \mathcal{L}_{m_2}^2 m_1^2}{2 \varepsilon m_-^2 m_+^2}+
\frac{9 C_A m_2^2 Y_f^2 \beta_0 \mathcal{L}_{m_2}^2 m_1^2}{16 m_-^2 m_+^2}-
\frac{3 C_A^2 C_F m_2^2 Y_f^2 \mathcal{L}_{m_1} \mathcal{L}_{m_2}^2 m_1^2}{2 m_-^2 m_+^2}+
\frac{31 C_A C_F^2 Y_f^2 \mathcal{L}_{m_1} m_1^2}{m_- m_+}+
\frac{31 C_F Y_f^2 \mathcal{L}_{m_1} m_1^2}{2 m_- m_+}+
\frac{3 C_A C_F^2 Y_f^2 \mathcal{L}_{m_1} m_1^2}{\varepsilon m_- m_+}+
\frac{3 C_F Y_f^2 \mathcal{L}_{m_1} m_1^2}{2 \varepsilon m_- m_+}+
\frac{5 C_A Y_f^2 \beta_0 \mathcal{L}_{m_1} m_1^2}{16 m_- m_+}+
\frac{9 C_A Y_f^2 \beta_0 \mathcal{L}_{m_1} m_1^2}{16 \varepsilon m_- m_+}+
\frac{5 C_A C_F^2 Y_f^2 \mathcal{L}_{m_2} m_1^2}{m_- m_+}+
\frac{5 C_F Y_f^2 \mathcal{L}_{m_2} m_1^2}{2 m_- m_+}-
\frac{3 C_A C_F^2 Y_f^2 \mathcal{L}_{m_2} m_1^2}{\varepsilon m_- m_+}-
\frac{3 C_F Y_f^2 \mathcal{L}_{m_2} m_1^2}{2 \varepsilon m_- m_+}-
\frac{3 C_A^2 C_F m_2^2 Y_f^2 \mathcal{L}_{m_1}^2 \mathcal{L}_{m_2} m_1^2}{2 m_-^2 m_+^2}+
\frac{9 C_A Y_f^2 \beta_0 \mathcal{L}_{m_2} m_1^2}{16 m_- m_+}+
\frac{9 C_A Y_f^2 \beta_0 \mathcal{L}_{m_2} m_1^2}{16 \varepsilon m_- m_+}+
\frac{13 C_A^2 C_F m_2^2 Y_f^2 \mathcal{L}_{m_1} \mathcal{L}_{m_2} m_1^2}{m_-^2 m_+^2}+
\frac{3 C_A^2 C_F m_2^2 Y_f^2 \mathcal{L}_{m_1} \mathcal{L}_{m_2} m_1^2}{\varepsilon m_-^2 m_+^2}+
\frac{3 C_A C_F^2 m_2^3 Y_f^2 \mathcal{L}_{m_1}^3 m_1}{m_-^2 m_+^2}+
\frac{3 C_F m_2^3 Y_f^2 \mathcal{L}_{m_1}^3 m_1}{2 m_-^2 m_+^2}-
\frac{C_A m_2^3 Y_f^2 \beta_0 \mathcal{L}_{m_1}^3 m_1}{12 m_-^2 m_+^2}+
\frac{3 C_A C_F^2 m_2^3 Y_f^2 \mathcal{L}_{m_2}^3 m_1}{m_-^2 m_+^2}+
\frac{3 C_F m_2^3 Y_f^2 \mathcal{L}_{m_2}^3 m_1}{2 m_-^2 m_+^2}+
\frac{C_A m_2^3 Y_f^2 \beta_0 \mathcal{L}_{m_2}^3 m_1}{12 m_-^2 m_+^2}+
\frac{13 C_A C_F^2 m_2^3 Y_f^2 \mathcal{L}_{m_1}^2 m_1}{2 m_-^2 m_+^2}+
\frac{13 C_F m_2^3 Y_f^2 \mathcal{L}_{m_1}^2 m_1}{4 m_-^2 m_+^2}-
\frac{3 C_A C_F^2 m_2^3 Y_f^2 \mathcal{L}_{m_1}^2 m_1}{\varepsilon m_-^2 m_+^2}-
\frac{3 C_F m_2^3 Y_f^2 \mathcal{L}_{m_1}^2 m_1}{2 \varepsilon m_-^2 m_+^2}-
\frac{C_A m_2 Y_f^2 \beta_0 \mathcal{L}_{m_1}^2 m_1}{4 m_- m_+}-
\frac{C_A m_2 Y_f^2 \beta_0 \mathcal{L}_{m_1}^2 m_1}{4 \varepsilon m_- m_+}-
\frac{35 C_A C_F^2 m_2^3 Y_f^2 \mathcal{L}_{m_2}^2 m_1}{2 m_-^2 m_+^2}-
\frac{35 C_F m_2^3 Y_f^2 \mathcal{L}_{m_2}^2 m_1}{4 m_-^2 m_+^2}-
\frac{3 C_A C_F^2 m_2^3 Y_f^2 \mathcal{L}_{m_2}^2 m_1}{\varepsilon m_-^2 m_+^2}-
\frac{3 C_F m_2^3 Y_f^2 \mathcal{L}_{m_2}^2 m_1}{2 \varepsilon m_-^2 m_+^2}-
\frac{C_A m_2^3 Y_f^2 \beta_0 \mathcal{L}_{m_2}^2 m_1}{4 m_-^2 m_+^2}-
\frac{C_A m_2^3 Y_f^2 \beta_0 \mathcal{L}_{m_2}^2 m_1}{4 \varepsilon m_-^2 m_+^2}-
\frac{3 C_A^2 C_F m_2^3 Y_f^2 \mathcal{L}_{m_1} \mathcal{L}_{m_2}^2 m_1}{2 m_-^2 m_+^2}+
\frac{22 C_A C_F^2 m_2 Y_f^2 \mathcal{L}_{m_1} m_1}{m_- m_+}+
\frac{11 C_F m_2 Y_f^2 \mathcal{L}_{m_1} m_1}{m_- m_+}+
\frac{12 C_A C_F^2 m_2 Y_f^2 \mathcal{L}_{m_1} m_1}{\varepsilon m_- m_+}+
\frac{6 C_F m_2 Y_f^2 \mathcal{L}_{m_1} m_1}{\varepsilon m_- m_+}+
\frac{C_A m_2 Y_f^2 \beta_0 \mathcal{L}_{m_1} m_1}{m_- m_+}+
\frac{C_A m_2 Y_f^2 \beta_0 \mathcal{L}_{m_1} m_1}{2 \varepsilon m_- m_+}+
\frac{C_A m_2 Y_f^2 \beta_0 \mathcal{L}_{m_1} m_1}{2 \varepsilon ^2 m_- m_+}+
\frac{C_A m_2 \pi ^2 Y_f^2 \beta_0 \mathcal{L}_{m_1} m_1}{24 m_- m_+}-
\frac{22 C_A C_F^2 m_2 Y_f^2 \mathcal{L}_{m_2} m_1}{m_- m_+}-
\frac{11 C_F m_2 Y_f^2 \mathcal{L}_{m_2} m_1}{m_- m_+}-
\frac{12 C_A C_F^2 m_2 Y_f^2 \mathcal{L}_{m_2} m_1}{\varepsilon m_- m_+}-
\frac{6 C_F m_2 Y_f^2 \mathcal{L}_{m_2} m_1}{\varepsilon m_- m_+}-
\frac{3 C_A^2 C_F m_2^3 Y_f^2 \mathcal{L}_{m_1}^2 \mathcal{L}_{m_2} m_1}{2 m_-^2 m_+^2}-
\frac{C_A m_2 Y_f^2 \beta_0 \mathcal{L}_{m_2} m_1}{m_- m_+}-
\frac{C_A m_2 Y_f^2 \beta_0 \mathcal{L}_{m_2} m_1}{2 \varepsilon m_- m_+}-
\frac{C_A m_2 Y_f^2 \beta_0 \mathcal{L}_{m_2} m_1}{2 \varepsilon ^2 m_- m_+}-
\frac{C_A m_2 \pi ^2 Y_f^2 \beta_0 \mathcal{L}_{m_2} m_1}{24 m_- m_+}+
\frac{11 C_A^2 C_F m_2^3 Y_f^2 \mathcal{L}_{m_1} \mathcal{L}_{m_2} m_1}{2 m_-^2 m_+^2}+
\frac{3 C_A^2 C_F m_2^3 Y_f^2 \mathcal{L}_{m_1} \mathcal{L}_{m_2} m_1}{\varepsilon m_-^2 m_+^2}-42 C_A C_F^2 Y_f^2-21 C_F Y_f^2-
\frac{18 C_A C_F^2 Y_f^2}{\varepsilon }-
\frac{9 C_F Y_f^2}{\varepsilon }+
\frac{3 C_A C_F^2 m_2^4 Y_f^2 \mathcal{L}_{m_1}^2}{m_-^2 m_+^2}+
\frac{3 C_F m_2^4 Y_f^2 \mathcal{L}_{m_1}^2}{2 m_-^2 m_+^2}+
\frac{9 C_A m_2^2 Y_f^2 \beta_0 \mathcal{L}_{m_1}^2}{32 m_- m_+}-
\frac{3 C_A C_F^2 m_2^4 Y_f^2 \mathcal{L}_{m_2}^2}{m_-^2 m_+^2}-
\frac{3 C_F m_2^4 Y_f^2 \mathcal{L}_{m_2}^2}{2 m_-^2 m_+^2}-
\frac{9 C_A m_2^4 Y_f^2 \beta_0 \mathcal{L}_{m_2}^2}{32 m_-^2 m_+^2}-
\frac{9}{8} C_A Y_f^2 \beta_0-
\frac{7 C_A Y_f^2 \beta_0}{8 \varepsilon }-
\frac{9 C_A Y_f^2 \beta_0}{8 \varepsilon ^2}-
\frac{3}{32} C_A \pi ^2 Y_f^2 \beta_0-
\frac{5 C_A C_F^2 m_2^2 Y_f^2 \mathcal{L}_{m_1}}{m_- m_+}-
\frac{5 C_F m_2^2 Y_f^2 \mathcal{L}_{m_1}}{2 m_- m_+}+
\frac{3 C_A C_F^2 m_2^2 Y_f^2 \mathcal{L}_{m_1}}{\varepsilon m_- m_+}+
\frac{3 C_F m_2^2 Y_f^2 \mathcal{L}_{m_1}}{2 \varepsilon m_- m_+}-
\frac{9 C_A m_2^2 Y_f^2 \beta_0 \mathcal{L}_{m_1}}{16 m_- m_+}-
\frac{9 C_A m_2^2 Y_f^2 \beta_0 \mathcal{L}_{m_1}}{16 \varepsilon m_- m_+}-
\frac{31 C_A C_F^2 m_2^2 Y_f^2 \mathcal{L}_{m_2}}{m_- m_+}-
\frac{31 C_F m_2^2 Y_f^2 \mathcal{L}_{m_2}}{2 m_- m_+}-
\frac{3 C_A C_F^2 m_2^2 Y_f^2 \mathcal{L}_{m_2}}{\varepsilon m_- m_+}-
\frac{3 C_F m_2^2 Y_f^2 \mathcal{L}_{m_2}}{2 \varepsilon m_- m_+}-
\frac{5 C_A m_2^2 Y_f^2 \beta_0 \mathcal{L}_{m_2}}{16 m_- m_+}-
\frac{9 C_A m_2^2 Y_f^2 \beta_0 \mathcal{L}_{m_2}}{16 \varepsilon m_- m_+}+
\frac{3 m_1^3 m_2 Y_f^4 C_A^2 \mathcal{L}_{m_1}^2}{16 \varepsilon m_-^2 m_+^2}+
\frac{3 m_1^3 m_2 Y_f^4 C_A^2 \mathcal{L}_{m_2}^2}{16 \varepsilon m_-^2 m_+^2}-
\frac{3 m_1^3 m_2 Y_f^4 C_A^2 \mathcal{L}_{m_1} \mathcal{L}_{m_2}}{8 \varepsilon m_-^2 m_+^2}-
\frac{3 m_1^3 m_2 Y_f^4 C_A^2 \mathcal{L}_{m_1}^3}{16 m_-^2 m_+^2}-
\frac{3 m_1^3 m_2 Y_f^4 C_A^2 \mathcal{L}_{m_2}^3}{16 m_-^2 m_+^2}+
\frac{11 m_1^3 m_2 Y_f^4 C_A^2 \mathcal{L}_{m_1}^2}{8 m_-^2 m_+^2}+
\frac{3 m_1^3 m_2 Y_f^4 C_A^2 \mathcal{L}_{m_1} \mathcal{L}_{m_2}^2}{16 m_-^2 m_+^2}-
\frac{m_1^3 m_2 Y_f^4 C_A^2 \mathcal{L}_{m_2}^2}{8 m_-^2 m_+^2}+
\frac{3 m_1^3 m_2 Y_f^4 C_A^2 \mathcal{L}_{m_1}^2 \mathcal{L}_{m_2}}{16 m_-^2 m_+^2}-
\frac{5 m_1^3 m_2 Y_f^4 C_A^2 \mathcal{L}_{m_1} \mathcal{L}_{m_2}}{4 m_-^2 m_+^2}-
\frac{3 m_1^2 m_2^2 Y_f^4 C_A^2 \mathcal{L}_{m_1}^2}{16 m_-^2 m_+^2}-
\frac{3 m_1^2 m_2^2 Y_f^4 C_A^2 \mathcal{L}_{m_2}^2}{16 m_-^2 m_+^2}+
\frac{3 m_1^2 m_2^2 Y_f^4 C_A^2 \mathcal{L}_{m_1} \mathcal{L}_{m_2}}{8 m_-^2 m_+^2}-
\frac{27 m_1^2 Y_f^4 C_A^2 \mathcal{L}_{m_2}}{32 m_- m_+}-
\frac{15 m_1^2 Y_f^4 C_A^2 \mathcal{L}_{m_1}}{32 m_- m_+}+
\frac{3 m_1 m_2^3 Y_f^4 C_A^2 \mathcal{L}_{m_1}^2}{16 \varepsilon m_-^2 m_+^2}+
\frac{3 m_1 m_2^3 Y_f^4 C_A^2 \mathcal{L}_{m_2}^2}{16 \varepsilon m_-^2 m_+^2}-
\frac{3 m_1 m_2^3 Y_f^4 C_A^2 \mathcal{L}_{m_1} \mathcal{L}_{m_2}}{8 \varepsilon m_-^2 m_+^2}-
\frac{3 m_1 m_2^3 Y_f^4 C_A^2 \mathcal{L}_{m_1}^3}{16 m_-^2 m_+^2}-
\frac{3 m_1 m_2^3 Y_f^4 C_A^2 \mathcal{L}_{m_2}^3}{16 m_-^2 m_+^2}-
\frac{m_1 m_2^3 Y_f^4 C_A^2 \mathcal{L}_{m_1}^2}{8 m_-^2 m_+^2}+
\frac{3 m_1 m_2^3 Y_f^4 C_A^2 \mathcal{L}_{m_1} \mathcal{L}_{m_2}^2}{16 m_-^2 m_+^2}+
\frac{11 m_1 m_2^3 Y_f^4 C_A^2 \mathcal{L}_{m_2}^2}{8 m_-^2 m_+^2}+
\frac{3 m_1 m_2^3 Y_f^4 C_A^2 \mathcal{L}_{m_1}^2 \mathcal{L}_{m_2}}{16 m_-^2 m_+^2}-
\frac{5 m_1 m_2^3 Y_f^4 C_A^2 \mathcal{L}_{m_1} \mathcal{L}_{m_2}}{4 m_-^2 m_+^2}+
\frac{27 m_2^2 Y_f^4 C_A^2 \mathcal{L}_{m_1}}{32 m_- m_+}-
\frac{3 m_1 m_2 Y_f^4 C_A^2 \mathcal{L}_{m_1}}{4 \varepsilon m_- m_+}+
\frac{3 m_1 m_2 Y_f^4 C_A^2 \mathcal{L}_{m_2}}{4 \varepsilon m_- m_+}-
\frac{5 m_1 m_2 Y_f^4 C_A^2 \mathcal{L}_{m_1}}{2 m_- m_+}+
\frac{5 m_1 m_2 Y_f^4 C_A^2 \mathcal{L}_{m_2}}{2 m_- m_+}+
\frac{15 m_2^2 Y_f^4 C_A^2 \mathcal{L}_{m_2}}{32 m_- m_+}+
\frac{21 Y_f^4 C_A^2}{32 \varepsilon }+
\frac{29}{16} Y_f^4 C_A^2
\end{autobreak}
\end{align}
\end{tiny}

\begin{tiny}
\begin{align}
\begin{autobreak}
\Delta B^{(2)}_2=
3 C_A \mathcal{L}_{m_1}^3 C_F^3-
\frac{3 C_A m_-^2 \mathcal{L}_{m_1}^3 C_F^3}{m_+^2}+3 C_A \mathcal{L}_{m_2}^3 C_F^3-
\frac{3 C_A m_-^2 \mathcal{L}_{m_2}^3 C_F^3}{m_+^2}+
\frac{3 C_A m_+^2 \mathcal{L}_{m_1}^2 C_F^3}{4 m_-^2}-
\frac{17}{2} C_A \mathcal{L}_{m_1}^2 C_F^3-
\frac{3 C_A \mathcal{L}_{m_1}^2 C_F^3}{\varepsilon }+
\frac{24 C_A m_- \mathcal{L}_{m_1}^2 C_F^3}{m_+}+
\frac{31 C_A m_-^2 \mathcal{L}_{m_1}^2 C_F^3}{4 m_+^2}+
\frac{3 C_A m_-^2 \mathcal{L}_{m_1}^2 C_F^3}{\varepsilon m_+^2}+
\frac{3 C_A m_+^2 \mathcal{L}_{m_2}^2 C_F^3}{4 m_-^2}-
\frac{17}{2} C_A \mathcal{L}_{m_2}^2 C_F^3-3 C_A \mathcal{L}_{m_1} \mathcal{L}_{m_2}^2 C_F^3+
\frac{3 C_A m_-^2 \mathcal{L}_{m_1} \mathcal{L}_{m_2}^2 C_F^3}{m_+^2}-
\frac{3 C_A \mathcal{L}_{m_2}^2 C_F^3}{\varepsilon }-
\frac{24 C_A m_- \mathcal{L}_{m_2}^2 C_F^3}{m_+}+
\frac{31 C_A m_-^2 \mathcal{L}_{m_2}^2 C_F^3}{4 m_+^2}+
\frac{3 C_A m_-^2 \mathcal{L}_{m_2}^2 C_F^3}{\varepsilon m_+^2}+68 C_A C_F^3-24 C_A \mathcal{L}_{m_1} C_F^3-
\frac{6 C_A m_+ \mathcal{L}_{m_1} C_F^3}{m_-}-
\frac{62 C_A m_- \mathcal{L}_{m_1} C_F^3}{m_+}-
\frac{24 C_A m_- \mathcal{L}_{m_1} C_F^3}{\varepsilon m_+}-3 C_A \mathcal{L}_{m_1}^2 \mathcal{L}_{m_2} C_F^3+
\frac{3 C_A m_-^2 \mathcal{L}_{m_1}^2 \mathcal{L}_{m_2} C_F^3}{m_+^2}-24 C_A \mathcal{L}_{m_2} C_F^3+
\frac{6 C_A m_+ \mathcal{L}_{m_2} C_F^3}{m_-}-
\frac{3 C_A m_+^2 \mathcal{L}_{m_1} \mathcal{L}_{m_2} C_F^3}{2 m_-^2}+17 C_A \mathcal{L}_{m_1} \mathcal{L}_{m_2} C_F^3+
\frac{6 C_A \mathcal{L}_{m_1} \mathcal{L}_{m_2} C_F^3}{\varepsilon }-
\frac{31 C_A m_-^2 \mathcal{L}_{m_1} \mathcal{L}_{m_2} C_F^3}{2 m_+^2}-
\frac{6 C_A m_-^2 \mathcal{L}_{m_1} \mathcal{L}_{m_2} C_F^3}{\varepsilon m_+^2}+
\frac{62 C_A m_- \mathcal{L}_{m_2} C_F^3}{m_+}+
\frac{24 C_A m_- \mathcal{L}_{m_2} C_F^3}{\varepsilon m_+}+
\frac{24 C_A C_F^3}{\varepsilon }-
\frac{3 m_-^2 \mathcal{L}_{m_1}^3 C_F^2}{2 m_+^2}+
\frac{3}{2} \mathcal{L}_{m_1}^3 C_F^2-
\frac{3 m_-^2 \mathcal{L}_{m_2}^3 C_F^2}{2 m_+^2}+
\frac{3}{2} \mathcal{L}_{m_2}^3 C_F^2+
\frac{3 m_+^2 \mathcal{L}_{m_1}^2 C_F^2}{8 m_-^2}-
\frac{3 \mathcal{L}_{m_1}^2 C_F^2}{2 \varepsilon }+
\frac{12 m_- \mathcal{L}_{m_1}^2 C_F^2}{m_+}+
\frac{3 m_-^2 \mathcal{L}_{m_1}^2 C_F^2}{2 \varepsilon m_+^2}+
\frac{31 m_-^2 \mathcal{L}_{m_1}^2 C_F^2}{8 m_+^2}-
\frac{17}{4} \mathcal{L}_{m_1}^2 C_F^2+
\frac{3 m_+^2 \mathcal{L}_{m_2}^2 C_F^2}{8 m_-^2}+
\frac{3 m_-^2 \mathcal{L}_{m_1} \mathcal{L}_{m_2}^2 C_F^2}{2 m_+^2}-
\frac{3}{2} \mathcal{L}_{m_1} \mathcal{L}_{m_2}^2 C_F^2-
\frac{3 \mathcal{L}_{m_2}^2 C_F^2}{2 \varepsilon }-
\frac{12 m_- \mathcal{L}_{m_2}^2 C_F^2}{m_+}+
\frac{3 m_-^2 \mathcal{L}_{m_2}^2 C_F^2}{2 \varepsilon m_+^2}+
\frac{31 m_-^2 \mathcal{L}_{m_2}^2 C_F^2}{8 m_+^2}-
\frac{17}{4} \mathcal{L}_{m_2}^2 C_F^2-
\frac{3 m_+ \mathcal{L}_{m_1} C_F^2}{m_-}-
\frac{12 m_- \mathcal{L}_{m_1} C_F^2}{\varepsilon m_+}-
\frac{31 m_- \mathcal{L}_{m_1} C_F^2}{m_+}-12 \mathcal{L}_{m_1} C_F^2+
\frac{3 m_-^2 \mathcal{L}_{m_1}^2 \mathcal{L}_{m_2} C_F^2}{2 m_+^2}-
\frac{3}{2} \mathcal{L}_{m_1}^2 \mathcal{L}_{m_2} C_F^2+
\frac{3 m_+ \mathcal{L}_{m_2} C_F^2}{m_-}-
\frac{3 m_+^2 \mathcal{L}_{m_1} \mathcal{L}_{m_2} C_F^2}{4 m_-^2}+
\frac{3 \mathcal{L}_{m_1} \mathcal{L}_{m_2} C_F^2}{\varepsilon }-
\frac{3 m_-^2 \mathcal{L}_{m_1} \mathcal{L}_{m_2} C_F^2}{\varepsilon m_+^2}-
\frac{31 m_-^2 \mathcal{L}_{m_1} \mathcal{L}_{m_2} C_F^2}{4 m_+^2}+
\frac{17}{2} \mathcal{L}_{m_1} \mathcal{L}_{m_2} C_F^2+
\frac{12 m_- \mathcal{L}_{m_2} C_F^2}{\varepsilon m_+}+
\frac{31 m_- \mathcal{L}_{m_2} C_F^2}{m_+}-12 \mathcal{L}_{m_2} C_F^2+
\frac{12 C_F^2}{\varepsilon }+34 C_F^2-
\frac{C_A m_- \beta_0 \mathcal{L}_{m_1}^3 C_F}{6 m_+}+
\frac{C_A m_- \beta_0 \mathcal{L}_{m_2}^3 C_F}{6 m_+}+
\frac{1}{2} C_A \beta_0 \mathcal{L}_{m_1}^2 C_F+
\frac{C_A m_+ \beta_0 \mathcal{L}_{m_1}^2 C_F}{8 m_-}+
\frac{5 C_A m_- \beta_0 \mathcal{L}_{m_1}^2 C_F}{8 m_+}+
\frac{C_A m_- \beta_0 \mathcal{L}_{m_1}^2 C_F}{2 \varepsilon m_+}+
\frac{1}{2} C_A \beta_0 \mathcal{L}_{m_2}^2 C_F-
\frac{C_A m_+ \beta_0 \mathcal{L}_{m_2}^2 C_F}{8 m_-}-
\frac{5 C_A m_- \beta_0 \mathcal{L}_{m_2}^2 C_F}{8 m_+}-
\frac{C_A m_- \beta_0 \mathcal{L}_{m_2}^2 C_F}{2 \varepsilon m_+}+5 C_A \beta_0 C_F+
\frac{3 C_A \beta_0 C_F}{\varepsilon }+
\frac{2 C_A m_- \beta_0 C_F}{m_+}+
\frac{2 C_A \beta_0 C_F}{\varepsilon ^2}+
\frac{1}{6} C_A \pi ^2 \beta_0 C_F-C_A \beta_0 \mathcal{L}_{m_1} C_F-
\frac{C_A m_+ \beta_0 \mathcal{L}_{m_1} C_F}{4 m_-}-
\frac{C_A m_+ \beta_0 \mathcal{L}_{m_1} C_F}{4 \varepsilon m_-}-
\frac{C_A \beta_0 \mathcal{L}_{m_1} C_F}{\varepsilon }-
\frac{13 C_A m_- \beta_0 \mathcal{L}_{m_1} C_F}{4 m_+}-
\frac{5 C_A m_- \beta_0 \mathcal{L}_{m_1} C_F}{4 \varepsilon m_+}-
\frac{C_A m_- \beta_0 \mathcal{L}_{m_1} C_F}{\varepsilon ^2 m_+}-
\frac{C_A m_-^2 \beta_0 \mathcal{L}_{m_1} C_F}{2 m_+^2}-
\frac{C_A m_- \pi ^2 \beta_0 \mathcal{L}_{m_1} C_F}{12 m_+}-2 C_A \beta_0 \mathcal{L}_{m_2} C_F+
\frac{C_A m_+ \beta_0 \mathcal{L}_{m_2} C_F}{4 m_-}+
\frac{C_A m_+ \beta_0 \mathcal{L}_{m_2} C_F}{4 \varepsilon m_-}-
\frac{C_A \beta_0 \mathcal{L}_{m_2} C_F}{\varepsilon }+
\frac{13 C_A m_- \beta_0 \mathcal{L}_{m_2} C_F}{4 m_+}+
\frac{5 C_A m_- \beta_0 \mathcal{L}_{m_2} C_F}{4 \varepsilon m_+}+
\frac{C_A m_- \beta_0 \mathcal{L}_{m_2} C_F}{\varepsilon ^2 m_+}+
\frac{C_A m_-^2 \beta_0 \mathcal{L}_{m_2} C_F}{2 m_+^2}+
\frac{C_A m_- \pi ^2 \beta_0 \mathcal{L}_{m_2} C_F}{12 m_+}+
\frac{3 C_A C_F^2 Y_f^2 \mathcal{L}_{m_2}^2 m_1^4}{m_- m_+^3}+
\frac{3 C_F Y_f^2 \mathcal{L}_{m_2}^2 m_1^4}{2 m_- m_+^3}-
\frac{9 C_A Y_f^2 \beta_0 \mathcal{L}_{m_2}^2 m_1^4}{32 m_- m_+^3}-
\frac{3 C_A C_F^2 Y_f^2 \mathcal{L}_{m_1}^2 m_1^3}{m_+^3}-
\frac{3 C_F Y_f^2 \mathcal{L}_{m_1}^2 m_1^3}{2 m_+^3}-
\frac{11 C_A Y_f^2 \beta_0 \mathcal{L}_{m_1}^2 m_1^3}{32 m_- m_+^2}+
\frac{5 C_A C_F^2 m_2 Y_f^2 \mathcal{L}_{m_2}^2 m_1^3}{m_- m_+^3}+
\frac{5 C_F m_2 Y_f^2 \mathcal{L}_{m_2}^2 m_1^3}{2 m_- m_+^3}+
\frac{3 C_A C_F^2 m_2 Y_f^2 \mathcal{L}_{m_2}^2 m_1^3}{2 \varepsilon m_- m_+^3}+
\frac{3 C_F m_2 Y_f^2 \mathcal{L}_{m_2}^2 m_1^3}{4 \varepsilon m_- m_+^3}-
\frac{9 C_A m_2 Y_f^2 \beta_0 \mathcal{L}_{m_2}^2 m_1^3}{16 m_- m_+^3}+
\frac{31 C_A C_F^2 Y_f^2 \mathcal{L}_{m_1} m_1^3}{m_- m_+^2}+
\frac{31 C_F Y_f^2 \mathcal{L}_{m_1} m_1^3}{2 m_- m_+^2}+
\frac{3 C_A C_F^2 Y_f^2 \mathcal{L}_{m_1} m_1^3}{\varepsilon m_- m_+^2}+
\frac{3 C_F Y_f^2 \mathcal{L}_{m_1} m_1^3}{2 \varepsilon m_- m_+^2}+
\frac{9 C_A Y_f^2 \beta_0 \mathcal{L}_{m_1} m_1^3}{16 m_- m_+^2}+
\frac{11 C_A Y_f^2 \beta_0 \mathcal{L}_{m_1} m_1^3}{16 \varepsilon m_- m_+^2}+
\frac{5 C_A C_F^2 Y_f^2 \mathcal{L}_{m_2} m_1^3}{m_- m_+^2}+
\frac{5 C_F Y_f^2 \mathcal{L}_{m_2} m_1^3}{2 m_- m_+^2}-
\frac{3 C_A C_F^2 Y_f^2 \mathcal{L}_{m_2} m_1^3}{\varepsilon m_- m_+^2}-
\frac{3 C_F Y_f^2 \mathcal{L}_{m_2} m_1^3}{2 \varepsilon m_- m_+^2}+
\frac{9 C_A Y_f^2 \beta_0 \mathcal{L}_{m_2} m_1^3}{16 m_- m_+^2}+
\frac{9 C_A Y_f^2 \beta_0 \mathcal{L}_{m_2} m_1^3}{16 \varepsilon m_- m_+^2}+
\frac{2 C_A C_F^2 m_2 Y_f^2 \mathcal{L}_{m_1}^2 m_1^2}{m_+^3}+
\frac{C_F m_2 Y_f^2 \mathcal{L}_{m_1}^2 m_1^2}{m_+^3}+
\frac{3 C_A C_F^2 m_2 Y_f^2 \mathcal{L}_{m_1}^2 m_1^2}{2 \varepsilon m_+^3}+
\frac{3 C_F m_2 Y_f^2 \mathcal{L}_{m_1}^2 m_1^2}{4 \varepsilon m_+^3}-
\frac{11 C_A m_2 Y_f^2 \beta_0 \mathcal{L}_{m_1}^2 m_1^2}{32 m_- m_+^2}-
\frac{3 C_A C_F^2 m_2^2 Y_f^2 \mathcal{L}_{m_2}^2 m_1^2}{m_- m_+^3}-
\frac{3 C_F m_2^2 Y_f^2 \mathcal{L}_{m_2}^2 m_1^2}{2 m_- m_+^3}-
\frac{3 C_A m_2^2 Y_f^2 \beta_0 \mathcal{L}_{m_2}^2 m_1^2}{16 m_- m_+^3}+
\frac{14 C_A C_F^2 m_2 Y_f^2 \mathcal{L}_{m_1} m_1^2}{m_- m_+^2}+
\frac{7 C_F m_2 Y_f^2 \mathcal{L}_{m_1} m_1^2}{m_- m_+^2}-
\frac{3 C_A C_F^2 m_2 Y_f^2 \mathcal{L}_{m_1} m_1^2}{\varepsilon m_- m_+^2}-
\frac{3 C_F m_2 Y_f^2 \mathcal{L}_{m_1} m_1^2}{2 \varepsilon m_- m_+^2}+
\frac{17 C_A m_2 Y_f^2 \beta_0 \mathcal{L}_{m_1} m_1^2}{16 m_- m_+^2}+
\frac{11 C_A m_2 Y_f^2 \beta_0 \mathcal{L}_{m_1} m_1^2}{16 \varepsilon m_- m_+^2}+
\frac{22 C_A C_F^2 m_2 Y_f^2 \mathcal{L}_{m_2} m_1^2}{m_- m_+^2}+
\frac{11 C_F m_2 Y_f^2 \mathcal{L}_{m_2} m_1^2}{m_- m_+^2}+
\frac{3 C_A C_F^2 m_2 Y_f^2 \mathcal{L}_{m_2} m_1^2}{\varepsilon m_- m_+^2}+
\frac{3 C_F m_2 Y_f^2 \mathcal{L}_{m_2} m_1^2}{2 \varepsilon m_- m_+^2}+
\frac{9 C_A m_2 Y_f^2 \beta_0 \mathcal{L}_{m_2} m_1^2}{16 m_- m_+^2}+
\frac{9 C_A m_2 Y_f^2 \beta_0 \mathcal{L}_{m_2} m_1^2}{16 \varepsilon m_- m_+^2}-
\frac{5 C_A^2 C_F m_2 Y_f^2 \mathcal{L}_{m_1} \mathcal{L}_{m_2} m_1^2}{m_+^3}-
\frac{3 C_A^2 C_F m_2 Y_f^2 \mathcal{L}_{m_1} \mathcal{L}_{m_2} m_1^2}{2 \varepsilon m_+^3}-
\frac{3 C_A^2 C_F m_2 Y_f^2 \mathcal{L}_{m_1}^3 m_1}{4 m_+^2}-
\frac{3 C_A^2 C_F m_2 Y_f^2 \mathcal{L}_{m_2}^3 m_1}{4 m_+^2}-
\frac{45 C_A C_F^2 Y_f^2 m_1}{m_+}-
\frac{45 C_F Y_f^2 m_1}{2 m_+}-
\frac{18 C_A C_F^2 Y_f^2 m_1}{\varepsilon m_+}-
\frac{9 C_F Y_f^2 m_1}{\varepsilon m_+}+
\frac{11 C_A C_F^2 m_2^2 Y_f^2 \mathcal{L}_{m_1}^2 m_1}{m_+^3}+
\frac{11 C_F m_2^2 Y_f^2 \mathcal{L}_{m_1}^2 m_1}{2 m_+^3}+
\frac{3 C_A C_F^2 m_2^2 Y_f^2 \mathcal{L}_{m_1}^2 m_1}{2 \varepsilon m_+^3}+
\frac{3 C_F m_2^2 Y_f^2 \mathcal{L}_{m_1}^2 m_1}{4 \varepsilon m_+^3}+
\frac{17 C_A m_2^2 Y_f^2 \beta_0 \mathcal{L}_{m_1}^2 m_1}{32 m_- m_+^2}-
\frac{8 C_A C_F^2 m_2^3 Y_f^2 \mathcal{L}_{m_2}^2 m_1}{m_- m_+^3}-
\frac{4 C_F m_2^3 Y_f^2 \mathcal{L}_{m_2}^2 m_1}{m_- m_+^3}-
\frac{3 C_A C_F^2 m_2^3 Y_f^2 \mathcal{L}_{m_2}^2 m_1}{2 \varepsilon m_- m_+^3}-
\frac{3 C_F m_2^3 Y_f^2 \mathcal{L}_{m_2}^2 m_1}{4 \varepsilon m_- m_+^3}+
\frac{7 C_A m_2^3 Y_f^2 \beta_0 \mathcal{L}_{m_2}^2 m_1}{16 m_- m_+^3}+
\frac{3 C_A^2 C_F m_2 Y_f^2 \mathcal{L}_{m_1} \mathcal{L}_{m_2}^2 m_1}{4 m_+^2}-
\frac{2 C_A Y_f^2 \beta_0 m_1}{m_+}-
\frac{9 C_A Y_f^2 \beta_0 m_1}{8 \varepsilon m_+}-
\frac{5 C_A Y_f^2 \beta_0 m_1}{4 \varepsilon ^2 m_+}-
\frac{5 C_A \pi ^2 Y_f^2 \beta_0 m_1}{48 m_+}-
\frac{40 C_A C_F^2 m_2^2 Y_f^2 \mathcal{L}_{m_1} m_1}{m_- m_+^2}-
\frac{20 C_F m_2^2 Y_f^2 \mathcal{L}_{m_1} m_1}{m_- m_+^2}-
\frac{3 C_A C_F^2 m_2^2 Y_f^2 \mathcal{L}_{m_1} m_1}{\varepsilon m_- m_+^2}-
\frac{3 C_F m_2^2 Y_f^2 \mathcal{L}_{m_1} m_1}{2 \varepsilon m_- m_+^2}-
\frac{25 C_A m_2^2 Y_f^2 \beta_0 \mathcal{L}_{m_1} m_1}{16 m_- m_+^2}-
\frac{17 C_A m_2^2 Y_f^2 \beta_0 \mathcal{L}_{m_1} m_1}{16 \varepsilon m_- m_+^2}+
\frac{4 C_A C_F^2 m_2^2 Y_f^2 \mathcal{L}_{m_2} m_1}{m_- m_+^2}+
\frac{2 C_F m_2^2 Y_f^2 \mathcal{L}_{m_2} m_1}{m_- m_+^2}+
\frac{3 C_A C_F^2 m_2^2 Y_f^2 \mathcal{L}_{m_2} m_1}{\varepsilon m_- m_+^2}+
\frac{3 C_F m_2^2 Y_f^2 \mathcal{L}_{m_2} m_1}{2 \varepsilon m_- m_+^2}+
\frac{3 C_A^2 C_F m_2 Y_f^2 \mathcal{L}_{m_1}^2 \mathcal{L}_{m_2} m_1}{4 m_+^2}+
\frac{7 C_A m_2^2 Y_f^2 \beta_0 \mathcal{L}_{m_2} m_1}{16 m_- m_+^2}-
\frac{3 C_A m_2^2 Y_f^2 \beta_0 \mathcal{L}_{m_2} m_1}{16 \varepsilon m_- m_+^2}-
\frac{8 C_A^2 C_F m_2^2 Y_f^2 \mathcal{L}_{m_1} \mathcal{L}_{m_2} m_1}{m_+^3}-
\frac{3 C_A^2 C_F m_2^2 Y_f^2 \mathcal{L}_{m_1} \mathcal{L}_{m_2} m_1}{2 \varepsilon m_+^3}-
\frac{51 C_A C_F^2 m_2 Y_f^2}{m_+}-
\frac{51 C_F m_2 Y_f^2}{2 m_+}-
\frac{18 C_A C_F^2 m_2 Y_f^2}{\varepsilon m_+}-
\frac{9 C_F m_2 Y_f^2}{\varepsilon m_+}+
\frac{3 C_A C_F^2 m_2^3 Y_f^2 \mathcal{L}_{m_1}^2}{m_+^3}+
\frac{3 C_F m_2^3 Y_f^2 \mathcal{L}_{m_1}^2}{2 m_+^3}+
\frac{9 C_A m_2^3 Y_f^2 \beta_0 \mathcal{L}_{m_1}^2}{32 m_- m_+^2}+
\frac{3 C_A C_F^2 m_2^4 Y_f^2 \mathcal{L}_{m_2}^2}{m_- m_+^3}+
\frac{3 C_F m_2^4 Y_f^2 \mathcal{L}_{m_2}^2}{2 m_- m_+^3}+
\frac{11 C_A m_2^4 Y_f^2 \beta_0 \mathcal{L}_{m_2}^2}{32 m_- m_+^3}-
\frac{7 C_A m_2 Y_f^2 \beta_0}{2 m_+}-
\frac{13 C_A m_2 Y_f^2 \beta_0}{8 \varepsilon m_+}-
\frac{5 C_A m_2 Y_f^2 \beta_0}{4 \varepsilon ^2 m_+}-
\frac{5 C_A m_2 \pi ^2 Y_f^2 \beta_0}{48 m_+}-
\frac{5 C_A C_F^2 m_2^3 Y_f^2 \mathcal{L}_{m_1}}{m_- m_+^2}-
\frac{5 C_F m_2^3 Y_f^2 \mathcal{L}_{m_1}}{2 m_- m_+^2}+
\frac{3 C_A C_F^2 m_2^3 Y_f^2 \mathcal{L}_{m_1}}{\varepsilon m_- m_+^2}+
\frac{3 C_F m_2^3 Y_f^2 \mathcal{L}_{m_1}}{2 \varepsilon m_- m_+^2}-
\frac{9 C_A m_2^3 Y_f^2 \beta_0 \mathcal{L}_{m_1}}{16 m_- m_+^2}-
\frac{9 C_A m_2^3 Y_f^2 \beta_0 \mathcal{L}_{m_1}}{16 \varepsilon m_- m_+^2}-
\frac{31 C_A C_F^2 m_2^3 Y_f^2 \mathcal{L}_{m_2}}{m_- m_+^2}-
\frac{31 C_F m_2^3 Y_f^2 \mathcal{L}_{m_2}}{2 m_- m_+^2}-
\frac{3 C_A C_F^2 m_2^3 Y_f^2 \mathcal{L}_{m_2}}{\varepsilon m_- m_+^2}-
\frac{3 C_F m_2^3 Y_f^2 \mathcal{L}_{m_2}}{2 \varepsilon m_- m_+^2}-
\frac{17 C_A m_2^3 Y_f^2 \beta_0 \mathcal{L}_{m_2}}{16 m_- m_+^2}-
\frac{11 C_A m_2^3 Y_f^2 \beta_0 \mathcal{L}_{m_2}}{16 \varepsilon m_- m_+^2}-
\frac{27 m_1^5 Y_f^4 C_A^2 \mathcal{L}_{m_2}}{32 m_-^2 m_+^3}-
\frac{33 m_1^5 Y_f^4 C_A^2 \mathcal{L}_{m_1}}{32 m_-^2 m_+^3}-
\frac{39 m_1^4 m_2 Y_f^4 C_A^2 \mathcal{L}_{m_1}}{32 m_-^2 m_+^3}-
\frac{21 m_1^4 m_2 Y_f^4 C_A^2 \mathcal{L}_{m_2}}{32 m_-^2 m_+^3}-
\frac{9 m_1^3 m_2^2 Y_f^4 C_A^2 \mathcal{L}_{m_1}^2}{32 m_-^2 m_+^3}-
\frac{9 m_1^3 m_2^2 Y_f^4 C_A^2 \mathcal{L}_{m_2}^2}{32 m_-^2 m_+^3}+
\frac{45 m_1^3 m_2^2 Y_f^4 C_A^2 \mathcal{L}_{m_1}}{16 m_-^2 m_+^3}+
\frac{15 m_1^3 m_2^2 Y_f^4 C_A^2 \mathcal{L}_{m_2}}{16 m_-^2 m_+^3}+
\frac{9 m_1^3 m_2^2 Y_f^4 C_A^2 \mathcal{L}_{m_1} \mathcal{L}_{m_2}}{16 m_-^2 m_+^3}+
\frac{9 m_1^2 m_2^3 Y_f^4 C_A^2 \mathcal{L}_{m_1}^2}{32 m_-^2 m_+^3}+
\frac{9 m_1^2 m_2^3 Y_f^4 C_A^2 \mathcal{L}_{m_2}^2}{32 m_-^2 m_+^3}+
\frac{33 m_1^2 m_2^3 Y_f^4 C_A^2 \mathcal{L}_{m_1}}{16 m_-^2 m_+^3}+
\frac{27 m_1^2 m_2^3 Y_f^4 C_A^2 \mathcal{L}_{m_2}}{16 m_-^2 m_+^3}-
\frac{9 m_1^2 m_2^3 Y_f^4 C_A^2 \mathcal{L}_{m_1} \mathcal{L}_{m_2}}{16 m_-^2 m_+^3}-
\frac{27 m_2^5 Y_f^4 C_A^2 \mathcal{L}_{m_1}}{32 m_-^2 m_+^3}-
\frac{3 m_1 m_2^4 Y_f^4 C_A^2 \mathcal{L}_{m_1}^2}{16 m_-^2 m_+^3}-
\frac{3 m_1 m_2^4 Y_f^4 C_A^2 \mathcal{L}_{m_2}^2}{16 m_-^2 m_+^3}-
\frac{57 m_1 m_2^4 Y_f^4 C_A^2 \mathcal{L}_{m_1}}{32 m_-^2 m_+^3}-
\frac{3 m_1 m_2^4 Y_f^4 C_A^2 \mathcal{L}_{m_2}}{32 m_-^2 m_+^3}+
\frac{3 m_1 m_2^4 Y_f^4 C_A^2 \mathcal{L}_{m_1} \mathcal{L}_{m_2}}{8 m_-^2 m_+^3}-
\frac{33 m_2^5 Y_f^4 C_A^2 \mathcal{L}_{m_2}}{32 m_-^2 m_+^3}+
\frac{15 m_1 Y_f^4 C_A^2}{16 \varepsilon m_+}+
\frac{97 m_1 Y_f^4 C_A^2}{32 m_+}+
\frac{15 m_2 Y_f^4 C_A^2}{16 \varepsilon m_+}+
\frac{109 m_2 Y_f^4 C_A^2}{32 m_+}
\end{autobreak}
\end{align}
\end{tiny}

\begin{tiny}
\begin{align}
\begin{autobreak}
\Delta B^{(2)}_3=
-\frac{3 C_A m_+^2 \mathcal{L}_{m_1}^3 C_F^3}{2 m_-^2}+3 C_A \mathcal{L}_{m_1}^3 C_F^3-
\frac{3 C_A m_-^2 \mathcal{L}_{m_1}^3 C_F^3}{2 m_+^2}-
\frac{3 C_A m_+^2 \mathcal{L}_{m_2}^3 C_F^3}{2 m_-^2}+3 C_A \mathcal{L}_{m_2}^3 C_F^3-
\frac{3 C_A m_-^2 \mathcal{L}_{m_2}^3 C_F^3}{2 m_+^2}+
\frac{7 C_A m_+^2 \mathcal{L}_{m_1}^2 C_F^3}{2 m_-^2}+
\frac{3 C_A m_+^2 \mathcal{L}_{m_1}^2 C_F^3}{2 \varepsilon m_-^2}-10 C_A \mathcal{L}_{m_1}^2 C_F^3+
\frac{12 C_A m_+ \mathcal{L}_{m_1}^2 C_F^3}{m_-}-
\frac{3 C_A \mathcal{L}_{m_1}^2 C_F^3}{\varepsilon }+
\frac{12 C_A m_- \mathcal{L}_{m_1}^2 C_F^3}{m_+}+
\frac{13 C_A m_-^2 \mathcal{L}_{m_1}^2 C_F^3}{2 m_+^2}+
\frac{3 C_A m_-^2 \mathcal{L}_{m_1}^2 C_F^3}{2 \varepsilon m_+^2}+
\frac{7 C_A m_+^2 \mathcal{L}_{m_2}^2 C_F^3}{2 m_-^2}+
\frac{3 C_A m_+^2 \mathcal{L}_{m_2}^2 C_F^3}{2 \varepsilon m_-^2}-10 C_A \mathcal{L}_{m_2}^2 C_F^3-
\frac{12 C_A m_+ \mathcal{L}_{m_2}^2 C_F^3}{m_-}+
\frac{3 C_A m_+^2 \mathcal{L}_{m_1} \mathcal{L}_{m_2}^2 C_F^3}{2 m_-^2}-3 C_A \mathcal{L}_{m_1} \mathcal{L}_{m_2}^2 C_F^3+
\frac{3 C_A m_-^2 \mathcal{L}_{m_1} \mathcal{L}_{m_2}^2 C_F^3}{2 m_+^2}-
\frac{3 C_A \mathcal{L}_{m_2}^2 C_F^3}{\varepsilon }-
\frac{12 C_A m_- \mathcal{L}_{m_2}^2 C_F^3}{m_+}+
\frac{13 C_A m_-^2 \mathcal{L}_{m_2}^2 C_F^3}{2 m_+^2}+
\frac{3 C_A m_-^2 \mathcal{L}_{m_2}^2 C_F^3}{2 \varepsilon m_+^2}+72 C_A C_F^3-36 C_A \mathcal{L}_{m_1} C_F^3-
\frac{28 C_A m_+ \mathcal{L}_{m_1} C_F^3}{m_-}-
\frac{12 C_A m_+ \mathcal{L}_{m_1} C_F^3}{\varepsilon m_-}-
\frac{52 C_A m_- \mathcal{L}_{m_1} C_F^3}{m_+}-
\frac{12 C_A m_- \mathcal{L}_{m_1} C_F^3}{\varepsilon m_+}+
\frac{3 C_A m_+^2 \mathcal{L}_{m_1}^2 \mathcal{L}_{m_2} C_F^3}{2 m_-^2}-3 C_A \mathcal{L}_{m_1}^2 \mathcal{L}_{m_2} C_F^3+
\frac{3 C_A m_-^2 \mathcal{L}_{m_1}^2 \mathcal{L}_{m_2} C_F^3}{2 m_+^2}-36 C_A \mathcal{L}_{m_2} C_F^3+
\frac{28 C_A m_+ \mathcal{L}_{m_2} C_F^3}{m_-}+
\frac{12 C_A m_+ \mathcal{L}_{m_2} C_F^3}{\varepsilon m_-}-
\frac{7 C_A m_+^2 \mathcal{L}_{m_1} \mathcal{L}_{m_2} C_F^3}{m_-^2}-
\frac{3 C_A m_+^2 \mathcal{L}_{m_1} \mathcal{L}_{m_2} C_F^3}{\varepsilon m_-^2}+20 C_A \mathcal{L}_{m_1} \mathcal{L}_{m_2} C_F^3+
\frac{6 C_A \mathcal{L}_{m_1} \mathcal{L}_{m_2} C_F^3}{\varepsilon }-
\frac{13 C_A m_-^2 \mathcal{L}_{m_1} \mathcal{L}_{m_2} C_F^3}{m_+^2}-
\frac{3 C_A m_-^2 \mathcal{L}_{m_1} \mathcal{L}_{m_2} C_F^3}{\varepsilon m_+^2}+
\frac{52 C_A m_- \mathcal{L}_{m_2} C_F^3}{m_+}+
\frac{12 C_A m_- \mathcal{L}_{m_2} C_F^3}{\varepsilon m_+}+
\frac{36 C_A C_F^3}{\varepsilon }-
\frac{3 m_+^2 \mathcal{L}_{m_1}^3 C_F^2}{4 m_-^2}-
\frac{3 m_-^2 \mathcal{L}_{m_1}^3 C_F^2}{4 m_+^2}+
\frac{3}{2} \mathcal{L}_{m_1}^3 C_F^2-
\frac{3 m_+^2 \mathcal{L}_{m_2}^3 C_F^2}{4 m_-^2}-
\frac{3 m_-^2 \mathcal{L}_{m_2}^3 C_F^2}{4 m_+^2}+
\frac{3}{2} \mathcal{L}_{m_2}^3 C_F^2+
\frac{3 m_+^2 \mathcal{L}_{m_1}^2 C_F^2}{4 \varepsilon m_-^2}+
\frac{7 m_+^2 \mathcal{L}_{m_1}^2 C_F^2}{4 m_-^2}+
\frac{6 m_+ \mathcal{L}_{m_1}^2 C_F^2}{m_-}-
\frac{3 \mathcal{L}_{m_1}^2 C_F^2}{2 \varepsilon }+
\frac{6 m_- \mathcal{L}_{m_1}^2 C_F^2}{m_+}+
\frac{3 m_-^2 \mathcal{L}_{m_1}^2 C_F^2}{4 \varepsilon m_+^2}+
\frac{13 m_-^2 \mathcal{L}_{m_1}^2 C_F^2}{4 m_+^2}-5 \mathcal{L}_{m_1}^2 C_F^2+
\frac{3 m_+^2 \mathcal{L}_{m_2}^2 C_F^2}{4 \varepsilon m_-^2}+
\frac{7 m_+^2 \mathcal{L}_{m_2}^2 C_F^2}{4 m_-^2}-
\frac{6 m_+ \mathcal{L}_{m_2}^2 C_F^2}{m_-}+
\frac{3 m_+^2 \mathcal{L}_{m_1} \mathcal{L}_{m_2}^2 C_F^2}{4 m_-^2}+
\frac{3 m_-^2 \mathcal{L}_{m_1} \mathcal{L}_{m_2}^2 C_F^2}{4 m_+^2}-
\frac{3}{2} \mathcal{L}_{m_1} \mathcal{L}_{m_2}^2 C_F^2-
\frac{3 \mathcal{L}_{m_2}^2 C_F^2}{2 \varepsilon }-
\frac{6 m_- \mathcal{L}_{m_2}^2 C_F^2}{m_+}+
\frac{3 m_-^2 \mathcal{L}_{m_2}^2 C_F^2}{4 \varepsilon m_+^2}+
\frac{13 m_-^2 \mathcal{L}_{m_2}^2 C_F^2}{4 m_+^2}-5 \mathcal{L}_{m_2}^2 C_F^2-
\frac{6 m_+ \mathcal{L}_{m_1} C_F^2}{\varepsilon m_-}-
\frac{14 m_+ \mathcal{L}_{m_1} C_F^2}{m_-}-
\frac{6 m_- \mathcal{L}_{m_1} C_F^2}{\varepsilon m_+}-
\frac{26 m_- \mathcal{L}_{m_1} C_F^2}{m_+}-18 \mathcal{L}_{m_1} C_F^2+
\frac{3 m_+^2 \mathcal{L}_{m_1}^2 \mathcal{L}_{m_2} C_F^2}{4 m_-^2}+
\frac{3 m_-^2 \mathcal{L}_{m_1}^2 \mathcal{L}_{m_2} C_F^2}{4 m_+^2}-
\frac{3}{2} \mathcal{L}_{m_1}^2 \mathcal{L}_{m_2} C_F^2+
\frac{6 m_+ \mathcal{L}_{m_2} C_F^2}{\varepsilon m_-}+
\frac{14 m_+ \mathcal{L}_{m_2} C_F^2}{m_-}-
\frac{3 m_+^2 \mathcal{L}_{m_1} \mathcal{L}_{m_2} C_F^2}{2 \varepsilon m_-^2}-
\frac{7 m_+^2 \mathcal{L}_{m_1} \mathcal{L}_{m_2} C_F^2}{2 m_-^2}+
\frac{3 \mathcal{L}_{m_1} \mathcal{L}_{m_2} C_F^2}{\varepsilon }-
\frac{3 m_-^2 \mathcal{L}_{m_1} \mathcal{L}_{m_2} C_F^2}{2 \varepsilon m_+^2}-
\frac{13 m_-^2 \mathcal{L}_{m_1} \mathcal{L}_{m_2} C_F^2}{2 m_+^2}+10 \mathcal{L}_{m_1} \mathcal{L}_{m_2} C_F^2+
\frac{6 m_- \mathcal{L}_{m_2} C_F^2}{\varepsilon m_+}+
\frac{26 m_- \mathcal{L}_{m_2} C_F^2}{m_+}-18 \mathcal{L}_{m_2} C_F^2+
\frac{18 C_F^2}{\varepsilon }+36 C_F^2-
\frac{C_A m_+ \beta_0 \mathcal{L}_{m_1}^3 C_F}{12 m_-}-
\frac{C_A m_- \beta_0 \mathcal{L}_{m_1}^3 C_F}{12 m_+}+
\frac{C_A m_+ \beta_0 \mathcal{L}_{m_2}^3 C_F}{12 m_-}+
\frac{C_A m_- \beta_0 \mathcal{L}_{m_2}^3 C_F}{12 m_+}+
\frac{3}{4} C_A \beta_0 \mathcal{L}_{m_1}^2 C_F+
\frac{C_A m_+ \beta_0 \mathcal{L}_{m_1}^2 C_F}{4 m_-}+
\frac{C_A m_+ \beta_0 \mathcal{L}_{m_1}^2 C_F}{4 \varepsilon m_-}+
\frac{3 C_A m_- \beta_0 \mathcal{L}_{m_1}^2 C_F}{4 m_+}+
\frac{C_A m_- \beta_0 \mathcal{L}_{m_1}^2 C_F}{4 \varepsilon m_+}+
\frac{3}{4} C_A \beta_0 \mathcal{L}_{m_2}^2 C_F-
\frac{C_A m_+ \beta_0 \mathcal{L}_{m_2}^2 C_F}{4 m_-}-
\frac{C_A m_+ \beta_0 \mathcal{L}_{m_2}^2 C_F}{4 \varepsilon m_-}-
\frac{3 C_A m_- \beta_0 \mathcal{L}_{m_2}^2 C_F}{4 m_+}-
\frac{C_A m_- \beta_0 \mathcal{L}_{m_2}^2 C_F}{4 \varepsilon m_+}+
\frac{2 C_A \beta_0 C_F}{\varepsilon }+
\frac{3 C_A \beta_0 C_F}{\varepsilon ^2}+
\frac{1}{4} C_A \pi ^2 \beta_0 C_F-C_A \beta_0 \mathcal{L}_{m_1} C_F-
\frac{C_A m_+ \beta_0 \mathcal{L}_{m_1} C_F}{2 m_-}-
\frac{C_A m_+ \beta_0 \mathcal{L}_{m_1} C_F}{2 \varepsilon m_-}-
\frac{C_A m_+ \beta_0 \mathcal{L}_{m_1} C_F}{2 \varepsilon ^2 m_-}-
\frac{3 C_A \beta_0 \mathcal{L}_{m_1} C_F}{2 \varepsilon }-
\frac{5 C_A m_- \beta_0 \mathcal{L}_{m_1} C_F}{2 m_+}-
\frac{3 C_A m_- \beta_0 \mathcal{L}_{m_1} C_F}{2 \varepsilon m_+}-
\frac{C_A m_- \beta_0 \mathcal{L}_{m_1} C_F}{2 \varepsilon ^2 m_+}-
\frac{C_A m_+ \pi ^2 \beta_0 \mathcal{L}_{m_1} C_F}{24 m_-}-
\frac{C_A m_- \pi ^2 \beta_0 \mathcal{L}_{m_1} C_F}{24 m_+}-C_A \beta_0 \mathcal{L}_{m_2} C_F+
\frac{C_A m_+ \beta_0 \mathcal{L}_{m_2} C_F}{2 m_-}+
\frac{C_A m_+ \beta_0 \mathcal{L}_{m_2} C_F}{2 \varepsilon m_-}+
\frac{C_A m_+ \beta_0 \mathcal{L}_{m_2} C_F}{2 \varepsilon ^2 m_-}-
\frac{3 C_A \beta_0 \mathcal{L}_{m_2} C_F}{2 \varepsilon }+
\frac{5 C_A m_- \beta_0 \mathcal{L}_{m_2} C_F}{2 m_+}+
\frac{3 C_A m_- \beta_0 \mathcal{L}_{m_2} C_F}{2 \varepsilon m_+}+
\frac{C_A m_- \beta_0 \mathcal{L}_{m_2} C_F}{2 \varepsilon ^2 m_+}+
\frac{C_A m_+ \pi ^2 \beta_0 \mathcal{L}_{m_2} C_F}{24 m_-}+
\frac{C_A m_- \pi ^2 \beta_0 \mathcal{L}_{m_2} C_F}{24 m_+}-
\frac{3 C_A C_F^2 Y_f^2 \mathcal{L}_{m_1}^2 m_1^4}{m_-^2 m_+^2}-
\frac{3 C_F Y_f^2 \mathcal{L}_{m_1}^2 m_1^4}{2 m_-^2 m_+^2}+
\frac{3 C_A C_F^2 Y_f^2 \mathcal{L}_{m_2}^2 m_1^4}{m_-^2 m_+^2}+
\frac{3 C_F Y_f^2 \mathcal{L}_{m_2}^2 m_1^4}{2 m_-^2 m_+^2}-
\frac{9 C_A Y_f^2 \beta_0 \mathcal{L}_{m_2}^2 m_1^4}{32 m_-^2 m_+^2}+
\frac{3 C_A C_F^2 m_2 Y_f^2 \mathcal{L}_{m_1}^3 m_1^3}{m_-^2 m_+^2}+
\frac{3 C_F m_2 Y_f^2 \mathcal{L}_{m_1}^3 m_1^3}{2 m_-^2 m_+^2}+
\frac{C_A m_2 Y_f^2 \beta_0 \mathcal{L}_{m_1}^3 m_1^3}{12 m_-^2 m_+^2}+
\frac{3 C_A C_F^2 m_2 Y_f^2 \mathcal{L}_{m_2}^3 m_1^3}{m_-^2 m_+^2}+
\frac{3 C_F m_2 Y_f^2 \mathcal{L}_{m_2}^3 m_1^3}{2 m_-^2 m_+^2}-
\frac{C_A m_2 Y_f^2 \beta_0 \mathcal{L}_{m_2}^3 m_1^3}{12 m_-^2 m_+^2}-
\frac{35 C_A C_F^2 m_2 Y_f^2 \mathcal{L}_{m_1}^2 m_1^3}{2 m_-^2 m_+^2}-
\frac{35 C_F m_2 Y_f^2 \mathcal{L}_{m_1}^2 m_1^3}{4 m_-^2 m_+^2}-
\frac{3 C_A C_F^2 m_2 Y_f^2 \mathcal{L}_{m_1}^2 m_1^3}{\varepsilon m_-^2 m_+^2}-
\frac{3 C_F m_2 Y_f^2 \mathcal{L}_{m_1}^2 m_1^3}{2 \varepsilon m_-^2 m_+^2}+
\frac{13 C_A C_F^2 m_2 Y_f^2 \mathcal{L}_{m_2}^2 m_1^3}{2 m_-^2 m_+^2}+
\frac{13 C_F m_2 Y_f^2 \mathcal{L}_{m_2}^2 m_1^3}{4 m_-^2 m_+^2}-
\frac{3 C_A C_F^2 m_2 Y_f^2 \mathcal{L}_{m_2}^2 m_1^3}{\varepsilon m_-^2 m_+^2}-
\frac{3 C_F m_2 Y_f^2 \mathcal{L}_{m_2}^2 m_1^3}{2 \varepsilon m_-^2 m_+^2}+
\frac{C_A m_2 Y_f^2 \beta_0 \mathcal{L}_{m_2}^2 m_1^3}{4 m_-^2 m_+^2}+
\frac{C_A m_2 Y_f^2 \beta_0 \mathcal{L}_{m_2}^2 m_1^3}{4 \varepsilon m_-^2 m_+^2}-
\frac{3 C_A^2 C_F m_2 Y_f^2 \mathcal{L}_{m_1} \mathcal{L}_{m_2}^2 m_1^3}{2 m_-^2 m_+^2}-
\frac{3 C_A^2 C_F m_2 Y_f^2 \mathcal{L}_{m_1}^2 \mathcal{L}_{m_2} m_1^3}{2 m_-^2 m_+^2}+
\frac{11 C_A^2 C_F m_2 Y_f^2 \mathcal{L}_{m_1} \mathcal{L}_{m_2} m_1^3}{2 m_-^2 m_+^2}+
\frac{3 C_A^2 C_F m_2 Y_f^2 \mathcal{L}_{m_1} \mathcal{L}_{m_2} m_1^3}{\varepsilon m_-^2 m_+^2}+
\frac{3 C_A C_F^2 m_2^2 Y_f^2 \mathcal{L}_{m_1}^3 m_1^2}{m_-^2 m_+^2}+
\frac{3 C_F m_2^2 Y_f^2 \mathcal{L}_{m_1}^3 m_1^2}{2 m_-^2 m_+^2}+
\frac{3 C_A C_F^2 m_2^2 Y_f^2 \mathcal{L}_{m_2}^3 m_1^2}{m_-^2 m_+^2}+
\frac{3 C_F m_2^2 Y_f^2 \mathcal{L}_{m_2}^3 m_1^2}{2 m_-^2 m_+^2}-
\frac{13 C_A C_F^2 m_2^2 Y_f^2 \mathcal{L}_{m_1}^2 m_1^2}{m_-^2 m_+^2}-
\frac{13 C_F m_2^2 Y_f^2 \mathcal{L}_{m_1}^2 m_1^2}{2 m_-^2 m_+^2}-
\frac{3 C_A C_F^2 m_2^2 Y_f^2 \mathcal{L}_{m_1}^2 m_1^2}{\varepsilon m_-^2 m_+^2}-
\frac{3 C_F m_2^2 Y_f^2 \mathcal{L}_{m_1}^2 m_1^2}{2 \varepsilon m_-^2 m_+^2}-
\frac{9 C_A Y_f^2 \beta_0 \mathcal{L}_{m_1}^2 m_1^2}{32 m_- m_+}-
\frac{13 C_A C_F^2 m_2^2 Y_f^2 \mathcal{L}_{m_2}^2 m_1^2}{m_-^2 m_+^2}-
\frac{13 C_F m_2^2 Y_f^2 \mathcal{L}_{m_2}^2 m_1^2}{2 m_-^2 m_+^2}-
\frac{3 C_A C_F^2 m_2^2 Y_f^2 \mathcal{L}_{m_2}^2 m_1^2}{\varepsilon m_-^2 m_+^2}-
\frac{3 C_F m_2^2 Y_f^2 \mathcal{L}_{m_2}^2 m_1^2}{2 \varepsilon m_-^2 m_+^2}+
\frac{9 C_A m_2^2 Y_f^2 \beta_0 \mathcal{L}_{m_2}^2 m_1^2}{16 m_-^2 m_+^2}-
\frac{3 C_A^2 C_F m_2^2 Y_f^2 \mathcal{L}_{m_1} \mathcal{L}_{m_2}^2 m_1^2}{2 m_-^2 m_+^2}+
\frac{31 C_A C_F^2 Y_f^2 \mathcal{L}_{m_1} m_1^2}{m_- m_+}+
\frac{31 C_F Y_f^2 \mathcal{L}_{m_1} m_1^2}{2 m_- m_+}+
\frac{3 C_A C_F^2 Y_f^2 \mathcal{L}_{m_1} m_1^2}{\varepsilon m_- m_+}+
\frac{3 C_F Y_f^2 \mathcal{L}_{m_1} m_1^2}{2 \varepsilon m_- m_+}+
\frac{5 C_A Y_f^2 \beta_0 \mathcal{L}_{m_1} m_1^2}{16 m_- m_+}+
\frac{9 C_A Y_f^2 \beta_0 \mathcal{L}_{m_1} m_1^2}{16 \varepsilon m_- m_+}+
\frac{5 C_A C_F^2 Y_f^2 \mathcal{L}_{m_2} m_1^2}{m_- m_+}+
\frac{5 C_F Y_f^2 \mathcal{L}_{m_2} m_1^2}{2 m_- m_+}-
\frac{3 C_A C_F^2 Y_f^2 \mathcal{L}_{m_2} m_1^2}{\varepsilon m_- m_+}-
\frac{3 C_F Y_f^2 \mathcal{L}_{m_2} m_1^2}{2 \varepsilon m_- m_+}-
\frac{3 C_A^2 C_F m_2^2 Y_f^2 \mathcal{L}_{m_1}^2 \mathcal{L}_{m_2} m_1^2}{2 m_-^2 m_+^2}+
\frac{9 C_A Y_f^2 \beta_0 \mathcal{L}_{m_2} m_1^2}{16 m_- m_+}+
\frac{9 C_A Y_f^2 \beta_0 \mathcal{L}_{m_2} m_1^2}{16 \varepsilon m_- m_+}+
\frac{13 C_A^2 C_F m_2^2 Y_f^2 \mathcal{L}_{m_1} \mathcal{L}_{m_2} m_1^2}{m_-^2 m_+^2}+
\frac{3 C_A^2 C_F m_2^2 Y_f^2 \mathcal{L}_{m_1} \mathcal{L}_{m_2} m_1^2}{\varepsilon m_-^2 m_+^2}+
\frac{3 C_A C_F^2 m_2^3 Y_f^2 \mathcal{L}_{m_1}^3 m_1}{m_-^2 m_+^2}+
\frac{3 C_F m_2^3 Y_f^2 \mathcal{L}_{m_1}^3 m_1}{2 m_-^2 m_+^2}-
\frac{C_A m_2^3 Y_f^2 \beta_0 \mathcal{L}_{m_1}^3 m_1}{12 m_-^2 m_+^2}+
\frac{3 C_A C_F^2 m_2^3 Y_f^2 \mathcal{L}_{m_2}^3 m_1}{m_-^2 m_+^2}+
\frac{3 C_F m_2^3 Y_f^2 \mathcal{L}_{m_2}^3 m_1}{2 m_-^2 m_+^2}+
\frac{C_A m_2^3 Y_f^2 \beta_0 \mathcal{L}_{m_2}^3 m_1}{12 m_-^2 m_+^2}+
\frac{13 C_A C_F^2 m_2^3 Y_f^2 \mathcal{L}_{m_1}^2 m_1}{2 m_-^2 m_+^2}+
\frac{13 C_F m_2^3 Y_f^2 \mathcal{L}_{m_1}^2 m_1}{4 m_-^2 m_+^2}-
\frac{3 C_A C_F^2 m_2^3 Y_f^2 \mathcal{L}_{m_1}^2 m_1}{\varepsilon m_-^2 m_+^2}-
\frac{3 C_F m_2^3 Y_f^2 \mathcal{L}_{m_1}^2 m_1}{2 \varepsilon m_-^2 m_+^2}-
\frac{C_A m_2 Y_f^2 \beta_0 \mathcal{L}_{m_1}^2 m_1}{4 m_- m_+}-
\frac{C_A m_2 Y_f^2 \beta_0 \mathcal{L}_{m_1}^2 m_1}{4 \varepsilon m_- m_+}-
\frac{35 C_A C_F^2 m_2^3 Y_f^2 \mathcal{L}_{m_2}^2 m_1}{2 m_-^2 m_+^2}-
\frac{35 C_F m_2^3 Y_f^2 \mathcal{L}_{m_2}^2 m_1}{4 m_-^2 m_+^2}-
\frac{3 C_A C_F^2 m_2^3 Y_f^2 \mathcal{L}_{m_2}^2 m_1}{\varepsilon m_-^2 m_+^2}-
\frac{3 C_F m_2^3 Y_f^2 \mathcal{L}_{m_2}^2 m_1}{2 \varepsilon m_-^2 m_+^2}-
\frac{C_A m_2^3 Y_f^2 \beta_0 \mathcal{L}_{m_2}^2 m_1}{4 m_-^2 m_+^2}-
\frac{C_A m_2^3 Y_f^2 \beta_0 \mathcal{L}_{m_2}^2 m_1}{4 \varepsilon m_-^2 m_+^2}-
\frac{3 C_A^2 C_F m_2^3 Y_f^2 \mathcal{L}_{m_1} \mathcal{L}_{m_2}^2 m_1}{2 m_-^2 m_+^2}+
\frac{22 C_A C_F^2 m_2 Y_f^2 \mathcal{L}_{m_1} m_1}{m_- m_+}+
\frac{11 C_F m_2 Y_f^2 \mathcal{L}_{m_1} m_1}{m_- m_+}+
\frac{12 C_A C_F^2 m_2 Y_f^2 \mathcal{L}_{m_1} m_1}{\varepsilon m_- m_+}+
\frac{6 C_F m_2 Y_f^2 \mathcal{L}_{m_1} m_1}{\varepsilon m_- m_+}+
\frac{C_A m_2 Y_f^2 \beta_0 \mathcal{L}_{m_1} m_1}{m_- m_+}+
\frac{C_A m_2 Y_f^2 \beta_0 \mathcal{L}_{m_1} m_1}{2 \varepsilon m_- m_+}+
\frac{C_A m_2 Y_f^2 \beta_0 \mathcal{L}_{m_1} m_1}{2 \varepsilon ^2 m_- m_+}+
\frac{C_A m_2 \pi ^2 Y_f^2 \beta_0 \mathcal{L}_{m_1} m_1}{24 m_- m_+}-
\frac{22 C_A C_F^2 m_2 Y_f^2 \mathcal{L}_{m_2} m_1}{m_- m_+}-
\frac{11 C_F m_2 Y_f^2 \mathcal{L}_{m_2} m_1}{m_- m_+}-
\frac{12 C_A C_F^2 m_2 Y_f^2 \mathcal{L}_{m_2} m_1}{\varepsilon m_- m_+}-
\frac{6 C_F m_2 Y_f^2 \mathcal{L}_{m_2} m_1}{\varepsilon m_- m_+}-
\frac{3 C_A^2 C_F m_2^3 Y_f^2 \mathcal{L}_{m_1}^2 \mathcal{L}_{m_2} m_1}{2 m_-^2 m_+^2}-
\frac{C_A m_2 Y_f^2 \beta_0 \mathcal{L}_{m_2} m_1}{m_- m_+}-
\frac{C_A m_2 Y_f^2 \beta_0 \mathcal{L}_{m_2} m_1}{2 \varepsilon m_- m_+}-
\frac{C_A m_2 Y_f^2 \beta_0 \mathcal{L}_{m_2} m_1}{2 \varepsilon ^2 m_- m_+}-
\frac{C_A m_2 \pi ^2 Y_f^2 \beta_0 \mathcal{L}_{m_2} m_1}{24 m_- m_+}+
\frac{11 C_A^2 C_F m_2^3 Y_f^2 \mathcal{L}_{m_1} \mathcal{L}_{m_2} m_1}{2 m_-^2 m_+^2}+
\frac{3 C_A^2 C_F m_2^3 Y_f^2 \mathcal{L}_{m_1} \mathcal{L}_{m_2} m_1}{\varepsilon m_-^2 m_+^2}-42 C_A C_F^2 Y_f^2-21 C_F Y_f^2-
\frac{18 C_A C_F^2 Y_f^2}{\varepsilon }-
\frac{9 C_F Y_f^2}{\varepsilon }+
\frac{3 C_A C_F^2 m_2^4 Y_f^2 \mathcal{L}_{m_1}^2}{m_-^2 m_+^2}+
\frac{3 C_F m_2^4 Y_f^2 \mathcal{L}_{m_1}^2}{2 m_-^2 m_+^2}+
\frac{9 C_A m_2^2 Y_f^2 \beta_0 \mathcal{L}_{m_1}^2}{32 m_- m_+}-
\frac{3 C_A C_F^2 m_2^4 Y_f^2 \mathcal{L}_{m_2}^2}{m_-^2 m_+^2}-
\frac{3 C_F m_2^4 Y_f^2 \mathcal{L}_{m_2}^2}{2 m_-^2 m_+^2}-
\frac{9 C_A m_2^4 Y_f^2 \beta_0 \mathcal{L}_{m_2}^2}{32 m_-^2 m_+^2}-
\frac{9}{8} C_A Y_f^2 \beta_0-
\frac{7 C_A Y_f^2 \beta_0}{8 \varepsilon }-
\frac{9 C_A Y_f^2 \beta_0}{8 \varepsilon ^2}-
\frac{3}{32} C_A \pi ^2 Y_f^2 \beta_0-
\frac{5 C_A C_F^2 m_2^2 Y_f^2 \mathcal{L}_{m_1}}{m_- m_+}-
\frac{5 C_F m_2^2 Y_f^2 \mathcal{L}_{m_1}}{2 m_- m_+}+
\frac{3 C_A C_F^2 m_2^2 Y_f^2 \mathcal{L}_{m_1}}{\varepsilon m_- m_+}+
\frac{3 C_F m_2^2 Y_f^2 \mathcal{L}_{m_1}}{2 \varepsilon m_- m_+}-
\frac{9 C_A m_2^2 Y_f^2 \beta_0 \mathcal{L}_{m_1}}{16 m_- m_+}-
\frac{9 C_A m_2^2 Y_f^2 \beta_0 \mathcal{L}_{m_1}}{16 \varepsilon m_- m_+}-
\frac{31 C_A C_F^2 m_2^2 Y_f^2 \mathcal{L}_{m_2}}{m_- m_+}-
\frac{31 C_F m_2^2 Y_f^2 \mathcal{L}_{m_2}}{2 m_- m_+}-
\frac{3 C_A C_F^2 m_2^2 Y_f^2 \mathcal{L}_{m_2}}{\varepsilon m_- m_+}-
\frac{3 C_F m_2^2 Y_f^2 \mathcal{L}_{m_2}}{2 \varepsilon m_- m_+}-
\frac{5 C_A m_2^2 Y_f^2 \beta_0 \mathcal{L}_{m_2}}{16 m_- m_+}-
\frac{9 C_A m_2^2 Y_f^2 \beta_0 \mathcal{L}_{m_2}}{16 \varepsilon m_- m_+}+
\frac{3 m_1^3 m_2 Y_f^4 C_A^2 \mathcal{L}_{m_1}^2}{16 \varepsilon m_-^2 m_+^2}+
\frac{3 m_1^3 m_2 Y_f^4 C_A^2 \mathcal{L}_{m_2}^2}{16 \varepsilon m_-^2 m_+^2}-
\frac{3 m_1^3 m_2 Y_f^4 C_A^2 \mathcal{L}_{m_1} \mathcal{L}_{m_2}}{8 \varepsilon m_-^2 m_+^2}-
\frac{3 m_1^3 m_2 Y_f^4 C_A^2 \mathcal{L}_{m_1}^3}{16 m_-^2 m_+^2}-
\frac{3 m_1^3 m_2 Y_f^4 C_A^2 \mathcal{L}_{m_2}^3}{16 m_-^2 m_+^2}+
\frac{11 m_1^3 m_2 Y_f^4 C_A^2 \mathcal{L}_{m_1}^2}{8 m_-^2 m_+^2}+
\frac{3 m_1^3 m_2 Y_f^4 C_A^2 \mathcal{L}_{m_1} \mathcal{L}_{m_2}^2}{16 m_-^2 m_+^2}-
\frac{m_1^3 m_2 Y_f^4 C_A^2 \mathcal{L}_{m_2}^2}{8 m_-^2 m_+^2}+
\frac{3 m_1^3 m_2 Y_f^4 C_A^2 \mathcal{L}_{m_1}^2 \mathcal{L}_{m_2}}{16 m_-^2 m_+^2}-
\frac{5 m_1^3 m_2 Y_f^4 C_A^2 \mathcal{L}_{m_1} \mathcal{L}_{m_2}}{4 m_-^2 m_+^2}-
\frac{27 m_1^2 Y_f^4 C_A^2 \mathcal{L}_{m_2}}{32 m_- m_+}-
\frac{27 m_1^2 Y_f^4 C_A^2 \mathcal{L}_{m_1}}{32 m_- m_+}+
\frac{3 m_1 m_2^3 Y_f^4 C_A^2 \mathcal{L}_{m_1}^2}{16 \varepsilon m_-^2 m_+^2}+
\frac{3 m_1 m_2^3 Y_f^4 C_A^2 \mathcal{L}_{m_2}^2}{16 \varepsilon m_-^2 m_+^2}-
\frac{3 m_1 m_2^3 Y_f^4 C_A^2 \mathcal{L}_{m_1} \mathcal{L}_{m_2}}{8 \varepsilon m_-^2 m_+^2}-
\frac{3 m_1 m_2^3 Y_f^4 C_A^2 \mathcal{L}_{m_1}^3}{16 m_-^2 m_+^2}-
\frac{3 m_1 m_2^3 Y_f^4 C_A^2 \mathcal{L}_{m_2}^3}{16 m_-^2 m_+^2}-
\frac{m_1 m_2^3 Y_f^4 C_A^2 \mathcal{L}_{m_1}^2}{8 m_-^2 m_+^2}+
\frac{3 m_1 m_2^3 Y_f^4 C_A^2 \mathcal{L}_{m_1} \mathcal{L}_{m_2}^2}{16 m_-^2 m_+^2}+
\frac{11 m_1 m_2^3 Y_f^4 C_A^2 \mathcal{L}_{m_2}^2}{8 m_-^2 m_+^2}+
\frac{3 m_1 m_2^3 Y_f^4 C_A^2 \mathcal{L}_{m_1}^2 \mathcal{L}_{m_2}}{16 m_-^2 m_+^2}-
\frac{5 m_1 m_2^3 Y_f^4 C_A^2 \mathcal{L}_{m_1} \mathcal{L}_{m_2}}{4 m_-^2 m_+^2}+
\frac{27 m_2^2 Y_f^4 C_A^2 \mathcal{L}_{m_1}}{32 m_- m_+}-
\frac{3 m_1 m_2 Y_f^4 C_A^2 \mathcal{L}_{m_1}}{4 \varepsilon m_- m_+}+
\frac{3 m_1 m_2 Y_f^4 C_A^2 \mathcal{L}_{m_2}}{4 \varepsilon m_- m_+}-
\frac{5 m_1 m_2 Y_f^4 C_A^2 \mathcal{L}_{m_1}}{2 m_- m_+}+
\frac{5 m_1 m_2 Y_f^4 C_A^2 \mathcal{L}_{m_2}}{2 m_- m_+}+
\frac{27 m_2^2 Y_f^4 C_A^2 \mathcal{L}_{m_2}}{32 m_- m_+}+
\frac{27 Y_f^4 C_A^2}{32 \varepsilon }+
\frac{21}{8} Y_f^4 C_A^2
\end{autobreak}
\end{align}
\end{tiny}

\begin{tiny}
\begin{align}
\begin{autobreak}
\Delta B^{(2)}_4=
-\frac{3 C_A m_+^2 \mathcal{L}_{m_1}^3 C_F^3}{4 m_-^2}+
\frac{3}{2} C_A \mathcal{L}_{m_1}^3 C_F^3-
\frac{3 C_A m_-^2 \mathcal{L}_{m_1}^3 C_F^3}{4 m_+^2}-
\frac{3 C_A m_+^2 \mathcal{L}_{m_2}^3 C_F^3}{4 m_-^2}+
\frac{3}{2} C_A \mathcal{L}_{m_2}^3 C_F^3-
\frac{3 C_A m_-^2 \mathcal{L}_{m_2}^3 C_F^3}{4 m_+^2}+
\frac{17 C_A m_+^2 \mathcal{L}_{m_1}^2 C_F^3}{8 m_-^2}+
\frac{3 C_A m_+^2 \mathcal{L}_{m_1}^2 C_F^3}{4 \varepsilon m_-^2}-
\frac{17}{4} C_A \mathcal{L}_{m_1}^2 C_F^3+
\frac{6 C_A m_+ \mathcal{L}_{m_1}^2 C_F^3}{m_-}-
\frac{3 C_A \mathcal{L}_{m_1}^2 C_F^3}{2 \varepsilon }+
\frac{6 C_A m_- \mathcal{L}_{m_1}^2 C_F^3}{m_+}+
\frac{17 C_A m_-^2 \mathcal{L}_{m_1}^2 C_F^3}{8 m_+^2}+
\frac{3 C_A m_-^2 \mathcal{L}_{m_1}^2 C_F^3}{4 \varepsilon m_+^2}+
\frac{17 C_A m_+^2 \mathcal{L}_{m_2}^2 C_F^3}{8 m_-^2}+
\frac{3 C_A m_+^2 \mathcal{L}_{m_2}^2 C_F^3}{4 \varepsilon m_-^2}-
\frac{17}{4} C_A \mathcal{L}_{m_2}^2 C_F^3-
\frac{6 C_A m_+ \mathcal{L}_{m_2}^2 C_F^3}{m_-}+
\frac{3 C_A m_+^2 \mathcal{L}_{m_1} \mathcal{L}_{m_2}^2 C_F^3}{4 m_-^2}-
\frac{3}{2} C_A \mathcal{L}_{m_1} \mathcal{L}_{m_2}^2 C_F^3+
\frac{3 C_A m_-^2 \mathcal{L}_{m_1} \mathcal{L}_{m_2}^2 C_F^3}{4 m_+^2}-
\frac{3 C_A \mathcal{L}_{m_2}^2 C_F^3}{2 \varepsilon }-
\frac{6 C_A m_- \mathcal{L}_{m_2}^2 C_F^3}{m_+}+
\frac{17 C_A m_-^2 \mathcal{L}_{m_2}^2 C_F^3}{8 m_+^2}+
\frac{3 C_A m_-^2 \mathcal{L}_{m_2}^2 C_F^3}{4 \varepsilon m_+^2}-4 C_A C_F^3-6 C_A \mathcal{L}_{m_1} C_F^3-
\frac{17 C_A m_+ \mathcal{L}_{m_1} C_F^3}{m_-}-
\frac{6 C_A m_+ \mathcal{L}_{m_1} C_F^3}{\varepsilon m_-}-
\frac{17 C_A m_- \mathcal{L}_{m_1} C_F^3}{m_+}-
\frac{6 C_A m_- \mathcal{L}_{m_1} C_F^3}{\varepsilon m_+}+
\frac{3 C_A m_+^2 \mathcal{L}_{m_1}^2 \mathcal{L}_{m_2} C_F^3}{4 m_-^2}-
\frac{3}{2} C_A \mathcal{L}_{m_1}^2 \mathcal{L}_{m_2} C_F^3+
\frac{3 C_A m_-^2 \mathcal{L}_{m_1}^2 \mathcal{L}_{m_2} C_F^3}{4 m_+^2}-6 C_A \mathcal{L}_{m_2} C_F^3+
\frac{17 C_A m_+ \mathcal{L}_{m_2} C_F^3}{m_-}+
\frac{6 C_A m_+ \mathcal{L}_{m_2} C_F^3}{\varepsilon m_-}-
\frac{17 C_A m_+^2 \mathcal{L}_{m_1} \mathcal{L}_{m_2} C_F^3}{4 m_-^2}-
\frac{3 C_A m_+^2 \mathcal{L}_{m_1} \mathcal{L}_{m_2} C_F^3}{2 \varepsilon m_-^2}+
\frac{17}{2} C_A \mathcal{L}_{m_1} \mathcal{L}_{m_2} C_F^3+
\frac{3 C_A \mathcal{L}_{m_1} \mathcal{L}_{m_2} C_F^3}{\varepsilon }-
\frac{17 C_A m_-^2 \mathcal{L}_{m_1} \mathcal{L}_{m_2} C_F^3}{4 m_+^2}-
\frac{3 C_A m_-^2 \mathcal{L}_{m_1} \mathcal{L}_{m_2} C_F^3}{2 \varepsilon m_+^2}+
\frac{17 C_A m_- \mathcal{L}_{m_2} C_F^3}{m_+}+
\frac{6 C_A m_- \mathcal{L}_{m_2} C_F^3}{\varepsilon m_+}+
\frac{6 C_A C_F^3}{\varepsilon }-
\frac{3 m_+^2 \mathcal{L}_{m_1}^3 C_F^2}{8 m_-^2}-
\frac{3 m_-^2 \mathcal{L}_{m_1}^3 C_F^2}{8 m_+^2}+
\frac{3}{4} \mathcal{L}_{m_1}^3 C_F^2-
\frac{3 m_+^2 \mathcal{L}_{m_2}^3 C_F^2}{8 m_-^2}-
\frac{3 m_-^2 \mathcal{L}_{m_2}^3 C_F^2}{8 m_+^2}+
\frac{3}{4} \mathcal{L}_{m_2}^3 C_F^2+
\frac{3 m_+^2 \mathcal{L}_{m_1}^2 C_F^2}{8 \varepsilon m_-^2}+
\frac{17 m_+^2 \mathcal{L}_{m_1}^2 C_F^2}{16 m_-^2}+
\frac{3 m_+ \mathcal{L}_{m_1}^2 C_F^2}{m_-}-
\frac{3 \mathcal{L}_{m_1}^2 C_F^2}{4 \varepsilon }+
\frac{3 m_- \mathcal{L}_{m_1}^2 C_F^2}{m_+}+
\frac{3 m_-^2 \mathcal{L}_{m_1}^2 C_F^2}{8 \varepsilon m_+^2}+
\frac{17 m_-^2 \mathcal{L}_{m_1}^2 C_F^2}{16 m_+^2}-
\frac{17}{8} \mathcal{L}_{m_1}^2 C_F^2+
\frac{3 m_+^2 \mathcal{L}_{m_2}^2 C_F^2}{8 \varepsilon m_-^2}+
\frac{17 m_+^2 \mathcal{L}_{m_2}^2 C_F^2}{16 m_-^2}-
\frac{3 m_+ \mathcal{L}_{m_2}^2 C_F^2}{m_-}+
\frac{3 m_+^2 \mathcal{L}_{m_1} \mathcal{L}_{m_2}^2 C_F^2}{8 m_-^2}+
\frac{3 m_-^2 \mathcal{L}_{m_1} \mathcal{L}_{m_2}^2 C_F^2}{8 m_+^2}-
\frac{3}{4} \mathcal{L}_{m_1} \mathcal{L}_{m_2}^2 C_F^2-
\frac{3 \mathcal{L}_{m_2}^2 C_F^2}{4 \varepsilon }-
\frac{3 m_- \mathcal{L}_{m_2}^2 C_F^2}{m_+}+
\frac{3 m_-^2 \mathcal{L}_{m_2}^2 C_F^2}{8 \varepsilon m_+^2}+
\frac{17 m_-^2 \mathcal{L}_{m_2}^2 C_F^2}{16 m_+^2}-
\frac{17}{8} \mathcal{L}_{m_2}^2 C_F^2-
\frac{3 m_+ \mathcal{L}_{m_1} C_F^2}{\varepsilon m_-}-
\frac{17 m_+ \mathcal{L}_{m_1} C_F^2}{2 m_-}-
\frac{3 m_- \mathcal{L}_{m_1} C_F^2}{\varepsilon m_+}-
\frac{17 m_- \mathcal{L}_{m_1} C_F^2}{2 m_+}-3 \mathcal{L}_{m_1} C_F^2+
\frac{3 m_+^2 \mathcal{L}_{m_1}^2 \mathcal{L}_{m_2} C_F^2}{8 m_-^2}+
\frac{3 m_-^2 \mathcal{L}_{m_1}^2 \mathcal{L}_{m_2} C_F^2}{8 m_+^2}-
\frac{3}{4} \mathcal{L}_{m_1}^2 \mathcal{L}_{m_2} C_F^2+
\frac{3 m_+ \mathcal{L}_{m_2} C_F^2}{\varepsilon m_-}+
\frac{17 m_+ \mathcal{L}_{m_2} C_F^2}{2 m_-}-
\frac{3 m_+^2 \mathcal{L}_{m_1} \mathcal{L}_{m_2} C_F^2}{4 \varepsilon m_-^2}-
\frac{17 m_+^2 \mathcal{L}_{m_1} \mathcal{L}_{m_2} C_F^2}{8 m_-^2}+
\frac{3 \mathcal{L}_{m_1} \mathcal{L}_{m_2} C_F^2}{2 \varepsilon }-
\frac{3 m_-^2 \mathcal{L}_{m_1} \mathcal{L}_{m_2} C_F^2}{4 \varepsilon m_+^2}-
\frac{17 m_-^2 \mathcal{L}_{m_1} \mathcal{L}_{m_2} C_F^2}{8 m_+^2}+
\frac{17}{4} \mathcal{L}_{m_1} \mathcal{L}_{m_2} C_F^2+
\frac{3 m_- \mathcal{L}_{m_2} C_F^2}{\varepsilon m_+}+
\frac{17 m_- \mathcal{L}_{m_2} C_F^2}{2 m_+}-3 \mathcal{L}_{m_2} C_F^2+
\frac{3 C_F^2}{\varepsilon }-2 C_F^2-
\frac{C_A m_+ \beta_0 \mathcal{L}_{m_1}^3 C_F}{12 m_-}-
\frac{C_A m_- \beta_0 \mathcal{L}_{m_1}^3 C_F}{12 m_+}+
\frac{C_A m_+ \beta_0 \mathcal{L}_{m_2}^3 C_F}{12 m_-}+
\frac{C_A m_- \beta_0 \mathcal{L}_{m_2}^3 C_F}{12 m_+}+
\frac{1}{4} C_A \beta_0 \mathcal{L}_{m_1}^2 C_F+
\frac{C_A m_+ \beta_0 \mathcal{L}_{m_1}^2 C_F}{8 m_-}+
\frac{C_A m_+ \beta_0 \mathcal{L}_{m_1}^2 C_F}{4 \varepsilon m_-}+
\frac{C_A m_- \beta_0 \mathcal{L}_{m_1}^2 C_F}{8 m_+}+
\frac{C_A m_- \beta_0 \mathcal{L}_{m_1}^2 C_F}{4 \varepsilon m_+}+
\frac{1}{4} C_A \beta_0 \mathcal{L}_{m_2}^2 C_F-
\frac{C_A m_+ \beta_0 \mathcal{L}_{m_2}^2 C_F}{8 m_-}-
\frac{C_A m_+ \beta_0 \mathcal{L}_{m_2}^2 C_F}{4 \varepsilon m_-}-
\frac{C_A m_- \beta_0 \mathcal{L}_{m_2}^2 C_F}{8 m_+}-
\frac{C_A m_- \beta_0 \mathcal{L}_{m_2}^2 C_F}{4 \varepsilon m_+}-7 C_A \beta_0 C_F-
\frac{3 C_A \beta_0 C_F}{\varepsilon }+
\frac{C_A \beta_0 C_F}{\varepsilon ^2}+
\frac{1}{12} C_A \pi ^2 \beta_0 C_F+
\frac{3}{2} C_A \beta_0 \mathcal{L}_{m_1} C_F-
\frac{C_A m_+ \beta_0 \mathcal{L}_{m_1} C_F}{4 m_-}-
\frac{C_A m_+ \beta_0 \mathcal{L}_{m_1} C_F}{4 \varepsilon m_-}-
\frac{C_A m_+ \beta_0 \mathcal{L}_{m_1} C_F}{2 \varepsilon ^2 m_-}-
\frac{C_A \beta_0 \mathcal{L}_{m_1} C_F}{2 \varepsilon }-
\frac{C_A m_- \beta_0 \mathcal{L}_{m_1} C_F}{4 m_+}-
\frac{C_A m_- \beta_0 \mathcal{L}_{m_1} C_F}{4 \varepsilon m_+}-
\frac{C_A m_- \beta_0 \mathcal{L}_{m_1} C_F}{2 \varepsilon ^2 m_+}-
\frac{C_A m_+ \pi ^2 \beta_0 \mathcal{L}_{m_1} C_F}{24 m_-}-
\frac{C_A m_- \pi ^2 \beta_0 \mathcal{L}_{m_1} C_F}{24 m_+}+
\frac{3}{2} C_A \beta_0 \mathcal{L}_{m_2} C_F+
\frac{C_A m_+ \beta_0 \mathcal{L}_{m_2} C_F}{4 m_-}+
\frac{C_A m_+ \beta_0 \mathcal{L}_{m_2} C_F}{4 \varepsilon m_-}+
\frac{C_A m_+ \beta_0 \mathcal{L}_{m_2} C_F}{2 \varepsilon ^2 m_-}-
\frac{C_A \beta_0 \mathcal{L}_{m_2} C_F}{2 \varepsilon }+
\frac{C_A m_- \beta_0 \mathcal{L}_{m_2} C_F}{4 m_+}+
\frac{C_A m_- \beta_0 \mathcal{L}_{m_2} C_F}{4 \varepsilon m_+}+
\frac{C_A m_- \beta_0 \mathcal{L}_{m_2} C_F}{2 \varepsilon ^2 m_+}+
\frac{C_A m_+ \pi ^2 \beta_0 \mathcal{L}_{m_2} C_F}{24 m_-}+
\frac{C_A m_- \pi ^2 \beta_0 \mathcal{L}_{m_2} C_F}{24 m_+}+
\frac{C_A C_F^2 Y_s^2 \mathcal{L}_{m_2}^2 m_1^4}{m_2^2 m_-^2 m_+^2}+
\frac{C_F Y_s^2 \mathcal{L}_{m_2}^2 m_1^4}{2 m_2^2 m_-^2 m_+^2}-
\frac{C_A Y_s^2 \beta_0 \mathcal{L}_{m_2}^2 m_1^4}{16 m_2^2 m_-^2 m_+^2}+
\frac{17 C_A C_F^2 Y_s^2 \mathcal{L}_{m_1}^3 m_1^2}{24 m_-^2 m_+^2}+
\frac{17 C_F Y_s^2 \mathcal{L}_{m_1}^3 m_1^2}{48 m_-^2 m_+^2}+
\frac{C_A Y_s^2 \beta_0 \mathcal{L}_{m_1}^3 m_1^2}{48 m_-^2 m_+^2}+
\frac{25 C_A C_F^2 Y_s^2 \mathcal{L}_{m_2}^3 m_1^2}{24 m_-^2 m_+^2}+
\frac{25 C_F Y_s^2 \mathcal{L}_{m_2}^3 m_1^2}{48 m_-^2 m_+^2}-
\frac{C_A Y_s^2 \beta_0 \mathcal{L}_{m_2}^3 m_1^2}{48 m_-^2 m_+^2}-
\frac{31 C_A C_F^2 Y_s^2 \mathcal{L}_{m_1}^2 m_1^2}{8 m_-^2 m_+^2}-
\frac{31 C_F Y_s^2 \mathcal{L}_{m_1}^2 m_1^2}{16 m_-^2 m_+^2}-
\frac{5 C_A C_F^2 Y_s^2 \mathcal{L}_{m_1}^2 m_1^2}{8 \varepsilon m_-^2 m_+^2}-
\frac{5 C_F Y_s^2 \mathcal{L}_{m_1}^2 m_1^2}{16 \varepsilon m_-^2 m_+^2}-
\frac{11 C_A C_F^2 Y_s^2 \mathcal{L}_{m_2}^2 m_1^2}{8 m_-^2 m_+^2}-
\frac{11 C_F Y_s^2 \mathcal{L}_{m_2}^2 m_1^2}{16 m_-^2 m_+^2}-
\frac{9 C_A C_F^2 Y_s^2 \mathcal{L}_{m_2}^2 m_1^2}{8 \varepsilon m_-^2 m_+^2}-
\frac{9 C_F Y_s^2 \mathcal{L}_{m_2}^2 m_1^2}{16 \varepsilon m_-^2 m_+^2}+
\frac{C_A Y_s^2 \beta_0 \mathcal{L}_{m_2}^2 m_1^2}{8 m_-^2 m_+^2}+
\frac{C_A Y_s^2 \beta_0 \mathcal{L}_{m_2}^2 m_1^2}{16 \varepsilon m_-^2 m_+^2}-
\frac{7 C_A^2 C_F Y_s^2 \mathcal{L}_{m_1} \mathcal{L}_{m_2}^2 m_1^2}{16 m_-^2 m_+^2}-
\frac{C_A C_F^2 Y_s^2 \mathcal{L}_{m_2} m_1^2}{2 m_2^2 m_- m_+}-
\frac{C_F Y_s^2 \mathcal{L}_{m_2} m_1^2}{4 m_2^2 m_- m_+}-
\frac{C_A C_F^2 Y_s^2 \mathcal{L}_{m_2} m_1^2}{\varepsilon m_2^2 m_- m_+}-
\frac{C_F Y_s^2 \mathcal{L}_{m_2} m_1^2}{2 \varepsilon m_2^2 m_- m_+}-
\frac{7 C_A^2 C_F Y_s^2 \mathcal{L}_{m_1}^2 \mathcal{L}_{m_2} m_1^2}{16 m_-^2 m_+^2}+
\frac{C_A Y_s^2 \beta_0 \mathcal{L}_{m_2} m_1^2}{8 m_2^2 m_- m_+}+
\frac{C_A Y_s^2 \beta_0 \mathcal{L}_{m_2} m_1^2}{8 \varepsilon m_2^2 m_- m_+}+
\frac{17 C_A^2 C_F Y_s^2 \mathcal{L}_{m_1} \mathcal{L}_{m_2} m_1^2}{8 m_-^2 m_+^2}+
\frac{7 C_A^2 C_F Y_s^2 \mathcal{L}_{m_1} \mathcal{L}_{m_2} m_1^2}{8 \varepsilon m_-^2 m_+^2}+
\frac{25 C_A C_F^2 m_2^2 Y_s^2 \mathcal{L}_{m_1}^3}{24 m_-^2 m_+^2}+
\frac{25 C_F m_2^2 Y_s^2 \mathcal{L}_{m_1}^3}{48 m_-^2 m_+^2}-
\frac{C_A m_2^2 Y_s^2 \beta_0 \mathcal{L}_{m_1}^3}{48 m_-^2 m_+^2}+
\frac{17 C_A C_F^2 m_2^2 Y_s^2 \mathcal{L}_{m_2}^3}{24 m_-^2 m_+^2}+
\frac{17 C_F m_2^2 Y_s^2 \mathcal{L}_{m_2}^3}{48 m_-^2 m_+^2}+
\frac{C_A m_2^2 Y_s^2 \beta_0 \mathcal{L}_{m_2}^3}{48 m_-^2 m_+^2}+
\frac{C_A C_F^2 Y_s^2}{2 m_2^2}+
\frac{C_F Y_s^2}{4 m_2^2}+
\frac{C_A C_F^2 Y_s^2}{4 \varepsilon m_2^2}+
\frac{C_F Y_s^2}{8 \varepsilon m_2^2}+
\frac{C_A C_F^2 Y_s^2}{2 \varepsilon ^2 m_2^2}+
\frac{C_F Y_s^2}{4 \varepsilon ^2 m_2^2}+
\frac{C_A C_F^2 \pi ^2 Y_s^2}{12 m_2^2}+
\frac{C_F \pi ^2 Y_s^2}{24 m_2^2}-
\frac{11 C_A C_F^2 m_2^2 Y_s^2 \mathcal{L}_{m_1}^2}{8 m_-^2 m_+^2}-
\frac{11 C_F m_2^2 Y_s^2 \mathcal{L}_{m_1}^2}{16 m_-^2 m_+^2}-
\frac{9 C_A C_F^2 m_2^2 Y_s^2 \mathcal{L}_{m_1}^2}{8 \varepsilon m_-^2 m_+^2}-
\frac{9 C_F m_2^2 Y_s^2 \mathcal{L}_{m_1}^2}{16 \varepsilon m_-^2 m_+^2}-
\frac{C_A Y_s^2 \beta_0 \mathcal{L}_{m_1}^2}{16 m_- m_+}-
\frac{C_A Y_s^2 \beta_0 \mathcal{L}_{m_1}^2}{16 \varepsilon m_- m_+}-
\frac{31 C_A C_F^2 m_2^2 Y_s^2 \mathcal{L}_{m_2}^2}{8 m_-^2 m_+^2}-
\frac{31 C_F m_2^2 Y_s^2 \mathcal{L}_{m_2}^2}{16 m_-^2 m_+^2}-
\frac{5 C_A C_F^2 m_2^2 Y_s^2 \mathcal{L}_{m_2}^2}{8 \varepsilon m_-^2 m_+^2}-
\frac{5 C_F m_2^2 Y_s^2 \mathcal{L}_{m_2}^2}{16 \varepsilon m_-^2 m_+^2}-
\frac{C_A m_2^2 Y_s^2 \beta_0 \mathcal{L}_{m_2}^2}{16 m_-^2 m_+^2}-
\frac{C_A m_2^2 Y_s^2 \beta_0 \mathcal{L}_{m_2}^2}{16 \varepsilon m_-^2 m_+^2}-
\frac{7 C_A^2 C_F m_2^2 Y_s^2 \mathcal{L}_{m_1} \mathcal{L}_{m_2}^2}{16 m_-^2 m_+^2}-
\frac{C_A Y_s^2 \beta_0}{4 m_2^2}-
\frac{C_A Y_s^2 \beta_0}{8 \varepsilon m_2^2}-
\frac{C_A Y_s^2 \beta_0}{8 \varepsilon ^2 m_2^2}-
\frac{C_A \pi ^2 Y_s^2 \beta_0}{96 m_2^2}+
\frac{25 C_A C_F^2 Y_s^2 \mathcal{L}_{m_1}}{4 m_- m_+}+
\frac{25 C_F Y_s^2 \mathcal{L}_{m_1}}{8 m_- m_+}+
\frac{7 C_A C_F^2 Y_s^2 \mathcal{L}_{m_1}}{4 \varepsilon m_- m_+}+
\frac{7 C_F Y_s^2 \mathcal{L}_{m_1}}{8 \varepsilon m_- m_+}-
\frac{C_A C_F^2 Y_s^2 \mathcal{L}_{m_1}}{4 \varepsilon ^2 m_- m_+}-
\frac{C_F Y_s^2 \mathcal{L}_{m_1}}{8 \varepsilon ^2 m_- m_+}-
\frac{C_A C_F^2 \pi ^2 Y_s^2 \mathcal{L}_{m_1}}{24 m_- m_+}-
\frac{C_F \pi ^2 Y_s^2 \mathcal{L}_{m_1}}{48 m_- m_+}+
\frac{C_A Y_s^2 \beta_0 \mathcal{L}_{m_1}}{8 m_- m_+}+
\frac{C_A Y_s^2 \beta_0 \mathcal{L}_{m_1}}{8 \varepsilon m_- m_+}+
\frac{C_A Y_s^2 \beta_0 \mathcal{L}_{m_1}}{8 \varepsilon ^2 m_- m_+}+
\frac{C_A \pi ^2 Y_s^2 \beta_0 \mathcal{L}_{m_1}}{96 m_- m_+}-
\frac{25 C_A C_F^2 Y_s^2 \mathcal{L}_{m_2}}{4 m_- m_+}-
\frac{25 C_F Y_s^2 \mathcal{L}_{m_2}}{8 m_- m_+}-
\frac{7 C_A C_F^2 Y_s^2 \mathcal{L}_{m_2}}{4 \varepsilon m_- m_+}-
\frac{7 C_F Y_s^2 \mathcal{L}_{m_2}}{8 \varepsilon m_- m_+}+
\frac{C_A C_F^2 Y_s^2 \mathcal{L}_{m_2}}{4 \varepsilon ^2 m_- m_+}+
\frac{C_F Y_s^2 \mathcal{L}_{m_2}}{8 \varepsilon ^2 m_- m_+}+
\frac{C_A C_F^2 \pi ^2 Y_s^2 \mathcal{L}_{m_2}}{24 m_- m_+}+
\frac{C_F \pi ^2 Y_s^2 \mathcal{L}_{m_2}}{48 m_- m_+}-
\frac{7 C_A^2 C_F m_2^2 Y_s^2 \mathcal{L}_{m_1}^2 \mathcal{L}_{m_2}}{16 m_-^2 m_+^2}-
\frac{C_A Y_s^2 \beta_0 \mathcal{L}_{m_2}}{8 m_- m_+}-
\frac{C_A Y_s^2 \beta_0 \mathcal{L}_{m_2}}{8 \varepsilon m_- m_+}-
\frac{C_A Y_s^2 \beta_0 \mathcal{L}_{m_2}}{8 \varepsilon ^2 m_- m_+}-
\frac{C_A \pi ^2 Y_s^2 \beta_0 \mathcal{L}_{m_2}}{96 m_- m_+}+
\frac{17 C_A^2 C_F m_2^2 Y_s^2 \mathcal{L}_{m_1} \mathcal{L}_{m_2}}{8 m_-^2 m_+^2}+
\frac{7 C_A^2 C_F m_2^2 Y_s^2 \mathcal{L}_{m_1} \mathcal{L}_{m_2}}{8 \varepsilon m_-^2 m_+^2}+
\frac{C_A C_F^2 Y_s^2}{2 m_1^2}+
\frac{C_F Y_s^2}{4 m_1^2}+
\frac{C_A C_F^2 Y_s^2}{4 \varepsilon m_1^2}+
\frac{C_F Y_s^2}{8 \varepsilon m_1^2}+
\frac{C_A C_F^2 Y_s^2}{2 \varepsilon ^2 m_1^2}+
\frac{C_F Y_s^2}{4 \varepsilon ^2 m_1^2}+
\frac{C_A C_F^2 \pi ^2 Y_s^2}{12 m_1^2}+
\frac{C_F \pi ^2 Y_s^2}{24 m_1^2}+
\frac{C_A C_F^2 m_2^4 Y_s^2 \mathcal{L}_{m_1}^2}{m_-^2 m_+^2 m_1^2}+
\frac{C_F m_2^4 Y_s^2 \mathcal{L}_{m_1}^2}{2 m_-^2 m_+^2 m_1^2}+
\frac{C_A m_2^2 Y_s^2 \beta_0 \mathcal{L}_{m_1}^2}{16 m_- m_+ m_1^2}-
\frac{C_A Y_s^2 \beta_0}{4 m_1^2}-
\frac{C_A Y_s^2 \beta_0}{8 \varepsilon m_1^2}-
\frac{C_A Y_s^2 \beta_0}{8 \varepsilon ^2 m_1^2}-
\frac{C_A \pi ^2 Y_s^2 \beta_0}{96 m_1^2}+
\frac{C_A C_F^2 m_2^2 Y_s^2 \mathcal{L}_{m_1}}{2 m_- m_+ m_1^2}+
\frac{C_F m_2^2 Y_s^2 \mathcal{L}_{m_1}}{4 m_- m_+ m_1^2}+
\frac{C_A C_F^2 m_2^2 Y_s^2 \mathcal{L}_{m_1}}{\varepsilon m_- m_+ m_1^2}+
\frac{C_F m_2^2 Y_s^2 \mathcal{L}_{m_1}}{2 \varepsilon m_- m_+ m_1^2}-
\frac{C_A m_2^2 Y_s^2 \beta_0 \mathcal{L}_{m_1}}{8 m_- m_+ m_1^2}-
\frac{C_A m_2^2 Y_s^2 \beta_0 \mathcal{L}_{m_1}}{8 \varepsilon m_- m_+ m_1^2}+
\frac{m_2^2 Y_s^4 C_A^2 \mathcal{L}_{m_1}}{16 m_1^4 m_- m_+}-
\frac{m_1^2 Y_s^4 C_A^2 \mathcal{L}_{m_2}}{16 m_2^4 m_- m_+}-
\frac{m_2^2 Y_s^4 C_A^2 \mathcal{L}_{m_1}^2}{16 m_1^2 m_-^2 m_+^2}-
\frac{m_1^2 Y_s^4 C_A^2 \mathcal{L}_{m_2}^2}{16 m_2^2 m_-^2 m_+^2}-
\frac{Y_s^4 C_A^2 \mathcal{L}_{m_1}}{16 \varepsilon m_1^2 m_- m_+}-
\frac{3 Y_s^4 C_A^2 \mathcal{L}_{m_1}}{16 m_1^2 m_- m_+}-
\frac{Y_s^4 C_A^2 \mathcal{L}_{m_1} \mathcal{L}_{m_2}}{16 \varepsilon m_-^2 m_+^2}+
\frac{Y_s^4 C_A^2 \mathcal{L}_{m_1} \mathcal{L}_{m_2}^2}{32 m_-^2 m_+^2}+
\frac{Y_s^4 C_A^2 \mathcal{L}_{m_1}^2 \mathcal{L}_{m_2}}{32 m_-^2 m_+^2}-
\frac{Y_s^4 C_A^2 \mathcal{L}_{m_1} \mathcal{L}_{m_2}}{8 m_-^2 m_+^2}+
\frac{Y_s^4 C_A^2 \mathcal{L}_{m_1}^2}{32 \varepsilon m_-^2 m_+^2}-
\frac{Y_s^4 C_A^2 \mathcal{L}_{m_1}^3}{32 m_-^2 m_+^2}+
\frac{Y_s^4 C_A^2 \mathcal{L}_{m_1}^2}{8 m_-^2 m_+^2}+
\frac{Y_s^4 C_A^2 \mathcal{L}_{m_2}}{16 \varepsilon m_2^2 m_- m_+}+
\frac{3 Y_s^4 C_A^2 \mathcal{L}_{m_2}}{16 m_2^2 m_- m_+}+
\frac{Y_s^4 C_A^2 \mathcal{L}_{m_2}^2}{32 \varepsilon m_-^2 m_+^2}-
\frac{Y_s^4 C_A^2 \mathcal{L}_{m_2}^3}{32 m_-^2 m_+^2}+
\frac{Y_s^4 C_A^2 \mathcal{L}_{m_2}^2}{8 m_-^2 m_+^2}+
\frac{Y_s^4 C_A^2}{32 \varepsilon m_1^4}+
\frac{3 Y_s^4 C_A^2}{32 m_1^4}-
\frac{Y_s^4 C_A^2}{16 \varepsilon m_1^2 m_2^2}-
\frac{Y_s^4 C_A^2}{32 \varepsilon ^2 m_1^2 m_2^2}-
\frac{Y_s^4 C_A^2}{8 m_1^2 m_2^2}-
\frac{\pi ^2 Y_s^4 C_A^2}{192 m_1^2 m_2^2}+
\frac{Y_s^4 C_A^2}{32 \varepsilon m_2^4}+
\frac{3 Y_s^4 C_A^2}{32 m_2^4}
\end{autobreak}
\end{align}
\end{tiny}

\begin{tiny}

\end{tiny}
\subsection{Matching at $\mu\sim M$:}
The one-loop bosonic mass renormalisation contributions are non-trivial as they involve two mass scales, $M_H$ and $M_W$, and are required to $\mathcal{O}(\varepsilon^2)$ to provide the correct contributions at two-loop order. 
\begin{tiny}
\begin{align}
\begin{autobreak}
\frac{\delta M_W^2}{aM_W^2}=
-\frac{1}{48} \varepsilon \log ^2(r (r-s)) r^6-
\frac{1}{48} \varepsilon \log ^2(r (r+s))r^6+
\frac{1}{48} \varepsilon \log ^2(u) r^6+
\frac{1}{48} \varepsilon \log ^2(v)r^6-
\frac{1}{9} \varepsilon \log \left(r^2\right) r^6+
\frac{1}{24} \varepsilon \mathcal{L}_{M_W} \log\left(r^2\right) r^6-
\frac{1}{24} \log \left(r^2\right) r^6+
\frac{1}{24} \varepsilon \log \left(r^2\right) \log (r (r-s)) r^6+
\frac{1}{24} \varepsilon \log \left(r^2\right) \log (r (r+s)) r^6-
\frac{1}{24} \varepsilon \log \left(\frac{s-r}{2 s}\right) \log (r (r+s)) r^6-
\frac{1}{24} \varepsilon \log (r (r-s)) \log \left(\frac{r+s}{2 s}\right) r^6+
\frac{1}{24} \varepsilon \log \left(-\frac{u}{2 r s}\right) \log (v) r^6+
\frac{1}{24} \varepsilon \log (u) \log \left(\frac{v}{2 r s}\right) r^6-
\frac{1}{24} \varepsilon \text{Li}_2\left(\frac{r+s}{2 s}\right) r^6+
\frac{1}{24} \varepsilon \text{Li}_2\left(-\frac{u}{2 r s}\right) r^6+
\frac{1}{24} \varepsilon \text{Li}_2\left(\frac{v}{2 r s}\right) r^6+
\frac{1}{48} \varepsilon s \log ^2(r (r-s)) r^5-
\frac{1}{48} \varepsilon s \log ^2(r (r+s)) r^5-
\frac{1}{48} \varepsilon s \log ^2(u) r^5+
\frac{1}{48} \varepsilon s \log ^2(v) r^5-
\frac{1}{24} \varepsilon s \log \left(r^2\right) \log (r (r-s)) r^5-
\frac{1}{9} \varepsilon s \log (s-r) r^5-
\frac{1}{24} s \log (s-r) r^5+
\frac{1}{24} \varepsilon s \mathcal{L}_{M_W} \log (s-r) r^5+
\frac{1}{9} \varepsilon s \log (r+s) r^5+
\frac{1}{24} s \log (r+s) r^5-
\frac{1}{24} \varepsilon s \mathcal{L}_{M_W} \log (r+s) r^5+
\frac{1}{24} \varepsilon s \log \left(r^2\right) \log (r (r+s)) r^5-
\frac{1}{24} \varepsilon s \log \left(\frac{s-r}{2 s}\right) \log (r (r+s)) r^5+
\frac{1}{24} \varepsilon s \log (r (r-s)) \log \left(\frac{r+s}{2 s}\right) r^5+
\frac{1}{144} \varepsilon s \log (64) \log (u) r^5+
\frac{1}{9} \varepsilon s \log (-v) r^5+
\frac{1}{24} s \log (-v) r^5-
\frac{1}{24} \varepsilon s \mathcal{L}_{M_W} \log (-v) r^5+
\frac{1}{24} \varepsilon s \log \left(-\frac{u}{2 r s}\right) \log (v) r^5-
\frac{1}{9} \varepsilon s \log \left(\frac{v}{s}\right) r^5-
\frac{1}{24} s \log \left(\frac{v}{s}\right) r^5+
\frac{1}{24} \varepsilon s \mathcal{L}_{M_W} \log \left(\frac{v}{s}\right) r^5-
\frac{1}{24} \varepsilon s \log (u) \log \left(\frac{v}{r s}\right) r^5-
\frac{1}{24} \varepsilon s \text{Li}_2\left(\frac{r+s}{2 s}\right) r^5-
\frac{1}{24} \varepsilon s \text{Li}_2\left(-\frac{u}{2 r s}\right) r^5+
\frac{1}{24} \varepsilon s \text{Li}_2\left(\frac{v}{2 r s}\right) r^5-
\frac{1}{24} \varepsilon \mathcal{L}_{M_H}^2 r^4+
\frac{1}{24} \varepsilon \mathcal{L}_{M_W}^2 r^4+
\frac{1}{24} \varepsilon \log ^2\left(r^2\right) r^4+
\frac{1}{8} \varepsilon \log ^2(r (r-s)) r^4+
\frac{1}{8} \varepsilon \log ^2(r (r+s)) r^4-
\frac{1}{8} \varepsilon \log ^2(u) r^4-
\frac{1}{8} \varepsilon \log ^2(v) r^4+
\frac{11 \varepsilon r^4}{36}+
\frac{5}{36} \varepsilon \mathcal{L}_{M_H} r^4+
\frac{1}{12} \mathcal{L}_{M_H} r^4-
\frac{2}{9} \varepsilon \mathcal{L}_{M_W} r^4-
\frac{1}{12} \mathcal{L}_{M_W} r^4+
\frac{4}{9} \varepsilon \log \left(r^2\right) r^4-
\frac{1}{6} \varepsilon \mathcal{L}_{M_W} \log \left(r^2\right) r^4+
\frac{1}{6} \log \left(r^2\right) r^4-
\frac{1}{4} \varepsilon \log \left(r^2\right) \log (r (r-s)) r^4-
\frac{1}{4} \varepsilon \log \left(r^2\right) \log (r (r+s)) r^4+
\frac{1}{4} \varepsilon \log \left(\frac{s-r}{2 s}\right) \log (r (r+s)) r^4+
\frac{1}{4} \varepsilon \log (r (r-s)) \log \left(\frac{r+s}{2 s}\right) r^4+
\frac{1}{12} \varepsilon \log (2) \log (u) r^4-
\frac{1}{4} \varepsilon \log \left(-\frac{u}{2 r s}\right) \log (v) r^4-
\frac{1}{6} \varepsilon \log (u) \log \left(\frac{v}{2 r s}\right) r^4-
\frac{1}{12} \varepsilon \log (u) \log \left(\frac{v}{r s}\right) r^4-
\frac{1}{24} \varepsilon u \text{Li}_2\left(\frac{s-r}{2 s}\right) r^4+
\frac{1}{4} \varepsilon \text{Li}_2\left(\frac{r+s}{2 s}\right) r^4-
\frac{1}{4} \varepsilon \text{Li}_2\left(-\frac{u}{2 r s}\right) r^4-
\frac{1}{4} \varepsilon \text{Li}_2\left(\frac{v}{2 r s}\right) r^4+
\frac{1}{144} \varepsilon \pi ^2 r^4+\frac{r^4}{12}-
\frac{1}{12} \varepsilon s \log ^2(r (r-s)) r^3+
\frac{1}{12} \varepsilon s \log ^2(r (r+s)) r^3+
\frac{1}{12} \varepsilon s \log ^2(u) r^3-
\frac{1}{12} \varepsilon s \log ^2(v) r^3+
\frac{1}{6} \varepsilon s \log \left(r^2\right) \log (r (r-s)) r^3+
\frac{4}{9} \varepsilon s \log (s-r) r^3+
\frac{1}{6} s \log (s-r) r^3-
\frac{1}{6} \varepsilon s \mathcal{L}_{M_W} \log (s-r) r^3-
\frac{4}{9} \varepsilon s \log (r+s) r^3-
\frac{1}{6} s \log (r+s) r^3+
\frac{1}{6} \varepsilon s \mathcal{L}_{M_W} \log (r+s) r^3-
\frac{1}{6} \varepsilon s \log \left(r^2\right) \log (r (r+s)) r^3+
\frac{1}{6} \varepsilon s \log \left(\frac{s-r}{2 s}\right) \log (r (r+s)) r^3-
\frac{1}{6} \varepsilon s \log (r (r-s)) \log \left(\frac{r+s}{2 s}\right) r^3-
\frac{1}{36} \varepsilon s \log (64) \log (u) r^3-
\frac{4}{9} \varepsilon s \log (-v) r^3-
\frac{1}{6} s \log (-v) r^3+
\frac{1}{6} \varepsilon s \mathcal{L}_{M_W} \log (-v) r^3-
\frac{1}{6} \varepsilon s \log \left(-\frac{u}{2 r s}\right) \log (v) r^3+
\frac{4}{9} \varepsilon s \log \left(\frac{v}{s}\right) r^3+
\frac{1}{6} s \log \left(\frac{v}{s}\right) r^3-
\frac{1}{6} \varepsilon s \mathcal{L}_{M_W} \log \left(\frac{v}{s}\right) r^3+
\frac{1}{6} \varepsilon s \log (u) \log \left(\frac{v}{r s}\right) r^3+
\frac{1}{6} \varepsilon s \text{Li}_2\left(\frac{r+s}{2 s}\right) r^3+
\frac{1}{6} \varepsilon s \text{Li}_2\left(-\frac{u}{2 r s}\right) r^3-
\frac{1}{6} \varepsilon s \text{Li}_2\left(\frac{v}{2 r s}\right) r^3-
\frac{1}{8} \varepsilon \mathcal{L}_{M_W}^2 r^2-
\frac{1}{6} \varepsilon \log ^2\left(r^2\right) r^2-
\frac{5}{12} \varepsilon \log ^2(r (r-s)) r^2-
\frac{5}{12} \varepsilon \log ^2(r (r+s)) r^2+
\frac{5}{12} \varepsilon \log ^2(u) r^2+
\frac{5}{12} \varepsilon \log ^2(v) r^2-
\frac{7 \varepsilon r^2}{4}+
\frac{3}{4} \varepsilon \mathcal{L}_{M_W} r^2+
\frac{1}{4} \mathcal{L}_{M_W} r^2
-\varepsilon \log \left(r^2\right) r^2+
\frac{1}{2} \varepsilon \mathcal{L}_{M_W} \log \left(r^2\right) r^2-
\frac{1}{2} \log \left(r^2\right) r^2+
\frac{5}{6} \varepsilon \log \left(r^2\right) \log (r (r-s)) r^2+
\frac{5}{6} \varepsilon \log \left(r^2\right) \log (r (r+s)) r^2-
\frac{5}{6} \varepsilon \log \left(\frac{s-r}{2 s}\right) \log (r (r+s)) r^2-
\frac{5}{6} \varepsilon \log (r (r-s)) \log \left(\frac{r+s}{2 s}\right) r^2-
\frac{1}{3} \varepsilon \log (2) \log (u) r^2+
\frac{5}{6} \varepsilon \log \left(-\frac{u}{2 r s}\right) \log (v) r^2+
\frac{1}{2} \varepsilon \log (u) \log \left(\frac{v}{2 r s}\right) r^2+
\frac{1}{3} \varepsilon \log (u) \log \left(\frac{v}{r s}\right) r^2+
\frac{1}{6} \varepsilon u \text{Li}_2\left(\frac{s-r}{2 s}\right) r^2-
\frac{5}{6} \varepsilon \text{Li}_2\left(\frac{r+s}{2 s}\right) r^2+
\frac{5}{6} \varepsilon \text{Li}_2\left(-\frac{u}{2 r s}\right) r^2+
\frac{5}{6} \varepsilon \text{Li}_2\left(\frac{v}{2 r s}\right) r^2-\frac{r^2}{4 \varepsilon }-
\frac{7}{144} \varepsilon \pi ^2 r^2
-\frac{3 r^2}{4}+
\frac{1}{4} \varepsilon s \log ^2(r (r-s)) r-
\frac{1}{4} \varepsilon s \log ^2(r (r+s)) r-
\frac{1}{4} \varepsilon s \log ^2(u) r+
\frac{1}{4} \varepsilon s \log ^2(v) r-
\frac{1}{2} \varepsilon s \log \left(r^2\right) \log (r (r-s)) r-\varepsilon s \log (s-r) r-
\frac{1}{2} s \log (s-r) r+
\frac{1}{2} \varepsilon s \mathcal{L}_{M_W} \log (s-r) r+\varepsilon s \log (r+s) r+
\frac{1}{2} s \log (r+s) r-
\frac{1}{2} \varepsilon s \mathcal{L}_{M_W} \log (r+s) r+
\frac{1}{2} \varepsilon s \log \left(r^2\right) \log (r (r+s)) r-
\frac{1}{2} \varepsilon s \log \left(\frac{s-r}{2 s}\right) \log (r (r+s)) r+
\frac{1}{2} \varepsilon s \log (r (r-s)) \log \left(\frac{r+s}{2 s}\right) r+
\frac{1}{12} \varepsilon s \log (64) \log (u) r+\varepsilon s \log (-v) r+
\frac{1}{2} s \log (-v) r-
\frac{1}{2} \varepsilon s \mathcal{L}_{M_W} \log (-v) r+
\frac{1}{2} \varepsilon s \log \left(-\frac{u}{2 r s}\right) \log (v) r-\varepsilon s \log \left(\frac{v}{s}\right) r-
\frac{1}{2} s \log \left(\frac{v}{s}\right) r+
\frac{1}{2} \varepsilon s \mathcal{L}_{M_W} \log \left(\frac{v}{s}\right) r-
\frac{1}{2} \varepsilon s \log (u) \log \left(\frac{v}{r s}\right) r-
\frac{1}{2} \varepsilon s \text{Li}_2\left(\frac{r+s}{2 s}\right) r-
\frac{1}{2} \varepsilon s \text{Li}_2\left(-\frac{u}{2 r s}\right) r+
\frac{1}{2} \varepsilon s \text{Li}_2\left(\frac{v}{2 r s}\right) r-
\frac{47}{16} C_A \varepsilon \mathcal{L}_{M_W}^2+
\frac{5}{12} \varepsilon \mathcal{L}_{M_W}^2+
\frac{8}{3} \varepsilon n_f T_f \mathcal{L}_{M_W}^2+
\frac{1}{2} \varepsilon \log ^2\left(r^2\right)+
\frac{1}{2} \varepsilon \log ^2(r (r-s))+
\frac{1}{2} \varepsilon \log ^2(r (r+s))-
\frac{1}{2} \varepsilon \log ^2(u)-
\frac{1}{2} \varepsilon \log ^2(v)-
\frac{115 C_A}{12}-
\frac{623 C_A \varepsilon }{36}+
\frac{197 \varepsilon }{54}+
\frac{112}{27} \varepsilon n_f T_f+
\frac{4 n_f T_f}{3 \varepsilon }+
\frac{20 n_f T_f}{9}-
\frac{7}{9} \varepsilon n_f \pi ^2 T_f-
\frac{20}{9} i \varepsilon n_f \pi T_f-
\frac{4}{3} i n_f \pi T_f+
\frac{47}{8} C_A \mathcal{L}_{M_W}+
\frac{115}{12} C_A \varepsilon \mathcal{L}_{M_W}-
\frac{31}{18} \varepsilon \mathcal{L}_{M_W}-
\frac{11}{8} \sqrt{3} C_A \varepsilon \pi \mathcal{L}_{M_W}-
\frac{5 \mathcal{L}_{M_W}}{6}-
\frac{40}{9} \varepsilon n_f T_f \mathcal{L}_{M_W}-
\frac{8}{3} n_f T_f\mathcal{L}_{M_W}+
\frac{8}{3} i \varepsilon n_f \pi T_f \mathcal{L}_{M_W}-
\varepsilon \log \left(r^2\right) \log (r (r-s))
-\varepsilon \log \left(r^2\right) \log (r (r+s))+
\varepsilon \log \left(\frac{s-r}{2 s}\right)\log (r (r+s))+
\varepsilon \log (r (r-s)) \log \left(\frac{r+s}{2 s}\right)+
\frac{20}{9} \varepsilon n_f T_f \log \left(\mu ^2\right) -
\frac{8}{3} \varepsilon n_f T_f \mathcal{L}_{M_W} \log \left(\mu ^2\right)
+\varepsilon \log (2) \log (u)
-\varepsilon \log \left(-\frac{u}{2 r s}\right) \log (v)-
\varepsilon \log (u) \log \left(\frac{v}{r s}\right)-
\frac{1}{2} \varepsilon u \text{Li}_2\left(\frac{s-r}{2 s}\right)+
\varepsilon \text{Li}_2\left(\frac{r+s}{2 s}\right)-
\varepsilon \text{Li}_2\left(-\frac{u}{2 r s}\right)-
\varepsilon \text{Li}_2\left(\frac{v}{2 r s}\right)-
\frac{47 C_A}{8 \varepsilon }+
\frac{5}{6 \varepsilon }-
\frac{33}{8} i \sqrt{3} C_A \varepsilon \text{Li}_2\left(\frac{1}{6} \left(3+i \sqrt{3}\right)\right)+
\frac{33}{8} i \sqrt{3} C_A \varepsilon \text{Li}_2\left(\frac{1}{6} \left(3-i \sqrt{3}\right)\right)-
\frac{11}{16} \sqrt{3} C_A \varepsilon \pi \log (3)-
\frac{5}{6} C_A \varepsilon \pi ^2+
\frac{11 \varepsilon \pi ^2}{72}+
\frac{11}{8} \sqrt{3} C_A \pi +
\frac{8 C_A \varepsilon \pi }{\sqrt{3}}+
\frac{31}{18} 
\end{autobreak}
\end{align}
\end{tiny}

\begin{tiny}
\begin{align}
\begin{autobreak}
\frac{\delta M_H^2}{aM_H^2}=
\frac{9}{64} \varepsilon \mathcal{L}_{M_H}^2 r^2
+\frac{1}{128} C_A C_F \varepsilon \mathcal{L}_{M_W}^2 r^2
+\frac{1}{256} C_A C_F \varepsilon \log ^2(r (r-s)) r^2
-\frac{1}{256} C_A C_F \varepsilon \log ^2(r (s-r)) r^2
-\frac{1}{256} C_A C_F \varepsilon \log ^2(-r (r+s)) r^2
+\frac{1}{256} C_A C_F \varepsilon \log ^2(r (r+s)) r^2
+\frac{1}{32} C_A C_F r^2
+\frac{1}{16} C_A C_F \varepsilon r^2
+\frac{9 \varepsilon r^2}{8}
-\frac{9}{16} \varepsilon \mathcal{L}_{M_H} r^2
+\frac{3}{32} \sqrt{3} \varepsilon \pi \mathcal{L}_{M_H} r^2
-\frac{9}{32} \mathcal{L}_{M_H} r^2
-\frac{1}{64} C_A C_F \mathcal{L}_{M_W} r^2
-\frac{1}{32} C_A C_F \varepsilon \mathcal{L}_{M_W} r^2
+\frac{1}{128} C_A C_F \varepsilon \log (2) \log (r (s-r)) r^2
-\frac{1}{128} C_A C_F \varepsilon \log \left(\frac{s-r}{2 s}\right) \log (-r (r+s)) r^2
+\frac{1}{128} C_A C_F \varepsilon \log \left(\frac{s-r}{2 s}\right) \log (r (r+s)) r^2
+\frac{1}{128} C_A C_F \varepsilon \log (r (r-s)) \log \left(\frac{r+s}{2 s}\right) r^2
-\frac{1}{128} C_A C_F \varepsilon \log (r (s-r)) \log \left(\frac{r+s}{s}\right) r^2
+\frac{C_A C_F r^2}{64 \varepsilon }
+\frac{9 r^2}{32 \varepsilon }
+\frac{9}{32} i \sqrt{3} \varepsilon \text{Li}_2\left(\frac{1}{6} \left(3+i \sqrt{3}\right)\right) r^2
-\frac{9}{32} i \sqrt{3} \varepsilon \text{Li}_2\left(\frac{1}{6} \left(3-i \sqrt{3}\right)\right) r^2
+\frac{3}{64} \sqrt{3} \varepsilon \pi \log (3) r^2
+\frac{1}{384} C_A C_F \varepsilon \pi ^2 r^2
+\frac{3}{64} \varepsilon \pi ^2 r^2
-\frac{3}{16} \sqrt{3} \varepsilon \pi r^2
-\frac{3}{32} \sqrt{3} \pi r^2
+\frac{9 r^2}{16}
+\frac{1}{256} C_A C_F \varepsilon s \log ^2(r (r-s)) r
+\frac{1}{256} C_A C_F \varepsilon s \log ^2(r (s-r)) r
-\frac{1}{256} C_A C_F \varepsilon s \log ^2(-r (r+s)) r
-\frac{1}{256} C_A C_F \varepsilon s \log ^2(r (r+s)) r
-\frac{1}{64} C_A C_F s \log (s-r) r
-\frac{1}{32} C_A C_F \varepsilon s \log (s-r) r
+\frac{1}{64} C_A C_F \varepsilon s \mathcal{L}_{M_W} \log (s-r) r
+\frac{1}{64} C_A C_F s \log (r+s) r
+\frac{1}{32} C_A C_F \varepsilon s \log (r+s) r
-\frac{1}{64} C_A C_F \varepsilon s \mathcal{L}_{M_W} \log (r+s) r
-\frac{1}{128} C_A C_F \varepsilon s \log \left(\frac{s-r}{2 s}\right) \log (-r (r+s)) r
-\frac{1}{128} C_A C_F \varepsilon s \log \left(\frac{s-r}{2 s}\right) \log (r (r+s)) r
+\frac{1}{128} C_A C_F \varepsilon s \log (r (r-s)) \log \left(\frac{r+s}{2 s}\right) r
+\frac{1}{128} C_A C_F \varepsilon s \log (r (s-r)) \log \left(\frac{r+s}{2 s}\right) r
+\frac{1}{64} C_A C_F \varepsilon s \text{Li}_2\left(\frac{s-r}{2 s}\right) r
-\frac{1}{64} C_A C_F \varepsilon s \text{Li}_2\left(\frac{r+s}{2 s}\right) r
-\frac{1}{8} C_A C_F \varepsilon \mathcal{L}_{M_W}^2
-\frac{1}{16} C_A C_F \varepsilon \log ^2(r (r-s))
+\frac{1}{16} C_A C_F \varepsilon \log ^2(r (s-r))
+\frac{1}{16} C_A C_F \varepsilon \log ^2(-r (r+s))
-\frac{1}{16} C_A C_F \varepsilon \log ^2(r (r+s))
-\frac{C_A C_F}{2}-C_A C_F \varepsilon 
+\frac{1}{4} C_A C_F \mathcal{L}_{M_W}
+\frac{1}{2} C_A C_F \varepsilon \mathcal{L}_{M_W}
-\frac{1}{8} C_A C_F \varepsilon \log (2) \log (r (s-r))
+\frac{1}{8} C_A C_F \varepsilon \log \left(\frac{s-r}{2 s}\right) \log (-r (r+s))
-\frac{1}{8} C_A C_F \varepsilon \log \left(\frac{s-r}{2 s}\right) \log (r (r+s))
-\frac{1}{8} C_A C_F \varepsilon \log (r (r-s)) \log \left(\frac{r+s}{2 s}\right)
+\frac{1}{8} C_A C_F \varepsilon \log (r (s-r)) \log \left(\frac{r+s}{s}\right)
-\frac{C_A C_F}{4 \varepsilon }
-\frac{1}{24} C_A C_F \varepsilon \pi ^2
-\frac{C_A C_F \varepsilon s \log ^2(r (r-s))}{16 r}
-\frac{C_A C_F \varepsilon s \log ^2(r (s-r))}{16 r}
+\frac{C_A C_F \varepsilon s \log ^2(-r (r+s))}{16 r}
+\frac{C_A C_F \varepsilon s \log ^2(r (r+s))}{16 r}
+\frac{C_A C_F s \log (s-r)}{4 r}
+\frac{C_A C_F \varepsilon s \log (s-r)}{2 r}
-\frac{C_A C_F \varepsilon s \mathcal{L}_{M_W} \log (s-r)}{4 r}
-\frac{C_A C_F s \log (r+s)}{4 r}
-\frac{C_A C_F \varepsilon s \log (r+s)}{2 r}
+\frac{C_A C_F \varepsilon s \mathcal{L}_{M_W} \log (r+s)}{4 r}
+\frac{C_A C_F \varepsilon s \log \left(\frac{s-r}{2 s}\right) \log (-r (r+s))}{8 r}
+\frac{C_A C_F \varepsilon s \log \left(\frac{s-r}{2 s}\right) \log (r (r+s))}{8 r}
-\frac{C_A C_F \varepsilon s \log (r (r-s)) \log \left(\frac{r+s}{2 s}\right)}{8 r}
-\frac{C_A C_F \varepsilon s \log (r (s-r)) \log \left(\frac{r+s}{2 s}\right)}{8 r}
-\frac{C_A C_F \varepsilon s \text{Li}_2\left(\frac{s-r}{2 s}\right)}{4 r}
+\frac{C_A C_F \varepsilon s \text{Li}_2\left(\frac{r+s}{2 s}\right)}{4 r}
+\frac{3 C_A C_F \varepsilon \mathcal{L}_{M_W}^2}{8 r^2}
+\frac{7 C_A C_F \varepsilon \log ^2(r (r-s))}{32 r^2}
-\frac{7 C_A C_F \varepsilon \log ^2(r (s-r))}{32 r^2}
-\frac{7 C_A C_F \varepsilon \log ^2(-r (r+s))}{32 r^2}
+\frac{7 C_A C_F \varepsilon \log ^2(r (r+s))}{32 r^2}
+\frac{9 C_A C_F}{8 r^2}
+\frac{19 C_A C_F \varepsilon }{8 r^2}
-\frac{3 C_A C_F \mathcal{L}_{M_W}}{4 r^2}
-\frac{9 C_A C_F \varepsilon \mathcal{L}_{M_W}}{8 r^2}
+\frac{7 C_A C_F \varepsilon \log (2) \log (r (s-r))}{16 r^2}
-\frac{7 C_A C_F \varepsilon \log \left(\frac{s-r}{2 s}\right) \log (-r (r+s))}{16 r^2}
+\frac{7 C_A C_F \varepsilon \log \left(\frac{s-r}{2 s}\right) \log (r (r+s))}{16 r^2}
+\frac{7 C_A C_F \varepsilon \log (r (r-s)) \log \left(\frac{r+s}{2 s}\right)}{16 r^2}
-\frac{7 C_A C_F \varepsilon \log (r (s-r)) \log \left(\frac{r+s}{s}\right)}{16 r^2}
+\frac{3 C_A C_F}{4 \varepsilon r^2}
+\frac{13 C_A C_F \varepsilon \pi ^2}{96 r^2}
+\frac{7 C_A C_F \varepsilon s \log ^2(r (r-s))}{32 r^3}
+\frac{7 C_A C_F \varepsilon s \log ^2(r (s-r))}{32 r^3}
-\frac{7 C_A C_F \varepsilon s \log ^2(-r (r+s))}{32 r^3}
-\frac{7 C_A C_F \varepsilon s \log ^2(r (r+s))}{32 r^3}
-\frac{7 C_A C_F s \log (s-r)}{8 r^3}
-\frac{5 C_A C_F \varepsilon s \log (s-r)}{4 r^3}
+\frac{7 C_A C_F \varepsilon s \mathcal{L}_{M_W} \log (s-r)}{8 r^3}
+\frac{7 C_A C_F s \log (r+s)}{8 r^3}
+\frac{5 C_A C_F \varepsilon s \log (r+s)}{4 r^3}
-\frac{7 C_A C_F \varepsilon s \mathcal{L}_{M_W} \log (r+s)}{8 r^3}
-\frac{7 C_A C_F \varepsilon s \log \left(\frac{s-r}{2 s}\right) \log (-r (r+s))}{16 r^3}
-\frac{7 C_A C_F \varepsilon s \log \left(\frac{s-r}{2 s}\right) \log (r (r+s))}{16 r^3}
+\frac{7 C_A C_F \varepsilon s \log (r (r-s)) \log \left(\frac{r+s}{2 s}\right)}{16 r^3}
+\frac{7 C_A C_F \varepsilon s \log (r (s-r)) \log \left(\frac{r+s}{2 s}\right)}{16 r^3}
+\frac{7 C_A C_F \varepsilon s \text{Li}_2\left(\frac{s-r}{2 s}\right)}{8 r^3}
-\frac{7 C_A C_F \varepsilon s \text{Li}_2\left(\frac{r+s}{2 s}\right)}{8 r^3}
\end{autobreak}
\end{align}
\end{tiny}
where $r=M_H/M_W$, $s=\sqrt{r^2-4}$, $u=r^2-rs-2$ and $v=r^2+rs-2$ and $\text{Li}_2(z)$ is the dilogarithm function,
\begin{equation}
    \text{Li}_2(z)=\sum_{k=1}^{\infty}\frac{z^k}{k^2}:\quad z\in\mathbb{C}.
\end{equation}
These corrections are required for two-loop order matching contributions for the heavy-heavy graphs at the scale $\mu\sim M_{W,H}$, we provide these contributions here.
\begin{tiny}
\begin{align}
\begin{autobreak}
\Delta U^{(2)}_1=
-\frac{1}{12} C_A C_F \log \left(r^2\right) r^6
+\frac{1}{12} C_A C_F w \log \left(r^2\right) r(w) r^6
-\frac{1}{12} C_A C_F s \log (s-r) r^5
+\frac{1}{12} C_A C_F s \log (r+s) r^5
+\frac{1}{12} C_A C_F s \log (-v) r^5
-\frac{1}{12} C_A C_F s \log \left(\frac{v}{s}\right) r^5
+\frac{1}{12} C_A C_F s w \log (s-r) r(w) r^5
-\frac{1}{12} C_A C_F s w \log (r+s) r(w) r^5
-\frac{1}{12} C_A C_F s w \log (-v) r(w) r^5
+\frac{1}{12} C_A C_F s w \log \left(\frac{v}{s}\right) r(w) r^5
+\frac{1}{6} C_A C_F r^4
+\frac{1}{6} C_A C_F \mathcal{L}_{M_H} r^4
-\frac{1}{6} C_A C_F \mathcal{L}_{M_W} r^4
+\frac{1}{3} C_A C_F \log \left(r^2\right) r^4
-\frac{1}{6} C_A C_F w r(w) r^4
-\frac{1}{6} C_A C_F w \mathcal{L}_{M_H} r(w) r^4
+\frac{1}{6} C_A C_F w \mathcal{L}_{M_W} r(w) r^4
-\frac{1}{3} C_A C_F w \log \left(r^2\right) r(w) r^4
+\frac{1}{3} C_A C_F s \log (s-r) r^3
-\frac{1}{3} C_A C_F s \log (r+s) r^3
-\frac{1}{3} C_A C_F s \log (-v) r^3
+\frac{1}{3} C_A C_F s \log \left(\frac{v}{s}\right) r^3
-\frac{1}{3} C_A C_F s w \log (s-r) r(w) r^3
+\frac{1}{3} C_A C_F s w \log (r+s) r(w) r^3
+\frac{1}{3} C_A C_F s w \log (-v) r(w) r^3
-\frac{1}{3} C_A C_F s w \log \left(\frac{v}{s}\right) r(w) r^3
-\frac{1}{16} C_A C_F^2 Y_f^2 r^2
-\frac{9}{16} C_A Y_f^2 r^2
-\frac{1}{32} C_F Y_f^2 r^2
-\frac{C_A C_F^2 Y_f^2 r^2}{64 \varepsilon }
-\frac{9 C_A Y_f^2 r^2}{64 \varepsilon }
-\frac{C_F Y_f^2 r^2}{128 \varepsilon }
-\frac{5}{2} C_A C_F r^2
+\frac{1}{64} C_A C_F^2 Y_f^2 \mathcal{L}_{M_H} r^2
+\frac{9}{32} C_A Y_f^2 \mathcal{L}_{M_H} r^2
+\frac{1}{128} C_F Y_f^2 \mathcal{L}_{M_H} r^2
+\frac{1}{64} C_A C_F^2 Y_f^2 \mathcal{L}_{M_W} r^2
+\frac{1}{128} C_F Y_f^2 \mathcal{L}_{M_W} r^2
+C_A C_F \mathcal{L}_{M_W} r^2
-C_A C_F \log \left(r^2\right) r^2
+\frac{1}{32} C_A C_F^2 Y_f^2 r(w) r^2
+\frac{9}{32} C_A Y_f^2 r(w) r^2
+\frac{1}{64} C_F Y_f^2 r(w) r^2
+\frac{C_A C_F^2 Y_f^2 r(w) r^2}{64 \varepsilon }
+\frac{9 C_A Y_f^2 r(w) r^2}{64 \varepsilon }
+\frac{C_F Y_f^2 r(w) r^2}{128 \varepsilon }
+\frac{3}{2} C_A C_F w r(w) r^2
+\frac{C_A C_F w r(w) r^2}{2 \varepsilon }
-\frac{1}{64} C_A C_F^2 Y_f^2 \mathcal{L}_{M_H} r(w) r^2
-\frac{9}{32} C_A Y_f^2 \mathcal{L}_{M_H} r(w) r^2
-\frac{1}{128} C_F Y_f^2 \mathcal{L}_{M_H} r(w) r^2
-\frac{1}{64} C_A C_F^2 Y_f^2 \mathcal{L}_{M_W} r(w) r^2
-\frac{1}{128} C_F Y_f^2 \mathcal{L}_{M_W} r(w) r^2
-C_A C_F w \mathcal{L}_{M_W} r(w) r^2
+C_A C_F w \log \left(r^2\right) r(w) r^2
-\frac{3}{64} \sqrt{3} C_A Y_f^2 \pi r(w) r^2
-\frac{C_A C_F r^2}{2 \varepsilon }
+\frac{3}{64} \sqrt{3} C_A Y_f^2 \pi r^2
+\frac{1}{64} C_A C_F^2 Y_f^2 s \log (s-r) r
+\frac{1}{128} C_F Y_f^2 s \log (s-r) r-C_A C_F s \log (s-r) r
-\frac{1}{64} C_A C_F^2 Y_f^2 s \log (r+s) r
-\frac{1}{128} C_F Y_f^2 s \log (r+s) r
+C_A C_F s \log (r+s) r+C_A C_F s \log (-v) r-C_A C_F s \log \left(\frac{v}{s}\right) r
-\frac{1}{64} C_A C_F^2 Y_f^2 s \log (s-r) r(w) r
-\frac{1}{128} C_F Y_f^2 s \log (s-r) r(w) r
+C_A C_F s w \log (s-r) r(w) r
+\frac{1}{64} C_A C_F^2 Y_f^2 s \log (r+s) r(w) r
+\frac{1}{128} C_F Y_f^2 s \log (r+s) r(w) r
-C_A C_F s w \log (r+s) r(w) r-C_A C_F s w \log (-v) r(w) r+C_A C_F s w \log \left(\frac{v}{s}\right) r(w) r
-\frac{256}{3} C_A C_F^2+C_A C_F^2 Y_f^2
+\frac{C_F Y_f^2}{2}
+\frac{C_A C_F^2 Y_f^2}{4 \varepsilon }
+\frac{C_F Y_f^2}{8 \varepsilon }
+\frac{1}{4} C_A Y_f^2 \beta_0 \mathcal{L}_{M_H}^2-C_A C_F \beta_0 \mathcal{L}_{M_W}^2
+\frac{61 C_A C_F}{9}
-\frac{128 C_F}{3}
+\frac{88}{9} C_A C_F n_f T_f
+\frac{8 C_A C_F n_f T_f}{3 \varepsilon }
-\frac{8}{3} i C_A C_F n_f \pi T_f
+\frac{C_A Y_f^2 \beta_0}{2 \varepsilon ^2}
-\frac{2 C_A C_F \beta_0}{\varepsilon ^2}
+\frac{1}{24} C_A Y_f^2 \pi ^2 \beta_0
-\frac{1}{6} C_A C_F \pi ^2 \beta_0
-\frac{1}{4} C_A C_F^2 Y_f^2 \mathcal{L}_{M_H}
-\frac{1}{8} C_F Y_f^2 \mathcal{L}_{M_H}
-C_A Y_f^2 \beta_0 \mathcal{L}_{M_H}
-\frac{C_A Y_f^2 \beta_0 \mathcal{L}_{M_H}}{2 \varepsilon }
+47 C_A C_F^2 \mathcal{L}_{M_W}
-\frac{1}{4} C_A C_F^2 Y_f^2 \mathcal{L}_{M_W}
-\frac{1}{8} C_F Y_f^2 \mathcal{L}_{M_W}
-\frac{10}{3} C_A C_F \mathcal{L}_{M_W}
+\frac{47}{2} C_F \mathcal{L}_{M_W}
-\frac{16}{3} C_A C_F n_f T_f \mathcal{L}_{M_W}
+4 C_A C_F \beta_0 \mathcal{L}_{M_W}
+\frac{2 C_A C_F \beta_0 \mathcal{L}_{M_W}}{\varepsilon }
-\frac{1}{2} C_A C_F^2 Y_f^2 r(w)
-\frac{1}{4} C_F Y_f^2 r(w)
-\frac{C_A C_F^2 Y_f^2 r(w)}{4 \varepsilon }
-\frac{C_F Y_f^2 r(w)}{8 \varepsilon }
-\frac{1}{4} C_A Y_f^2 \beta_0 \mathcal{L}_{M_H}^2 r(w)
+C_A C_F w \beta_0 \mathcal{L}_{M_W}^2 r(w)
+\frac{115}{3} C_A C_F^2 w r(w)
-\frac{31}{9} C_A C_F w r(w)
+\frac{115}{6} C_F w r(w)
-\frac{40}{9} C_A C_F n_f T_f w r(w)
-\frac{8 C_A C_F n_f T_f w r(w)}{3 \varepsilon }
+\frac{8}{3} i C_A C_F n_f \pi T_f w r(w)
+\frac{47 C_A C_F^2 w r(w)}{2 \varepsilon }
-\frac{5 C_A C_F w r(w)}{3 \varepsilon }
+\frac{47 C_F w r(w)}{4 \varepsilon }
-\frac{11}{2} \sqrt{3} C_A C_F^2 \pi w r(w)
-\frac{11}{4} \sqrt{3} C_F \pi w r(w)
-\frac{C_A Y_f^2 \beta_0 r(w)}{2 \varepsilon ^2}
+\frac{2 C_A C_F w \beta_0 r(w)}{\varepsilon ^2}
+\frac{1}{6} C_A C_F \pi ^2 w \beta_0 r(w)
-\frac{1}{24} C_A Y_f^2 \pi ^2 \beta_0 r(w)
+\frac{1}{4} C_A C_F^2 Y_f^2 \mathcal{L}_{M_H} r(w)
+\frac{1}{8} C_F Y_f^2 \mathcal{L}_{M_H} r(w)
+\frac{C_A Y_f^2 \beta_0 \mathcal{L}_{M_H} r(w)}{2 \varepsilon }
+\frac{1}{4} C_A C_F^2 Y_f^2 \mathcal{L}_{M_W} r(w)
+\frac{1}{8} C_F Y_f^2 \mathcal{L}_{M_W} r(w)
-47 C_A C_F^2 w \mathcal{L}_{M_W} r(w)
+\frac{10}{3} C_A C_F w \mathcal{L}_{M_W} r(w)
-\frac{47}{2} C_F w \mathcal{L}_{M_W} r(w)
+\frac{16}{3} C_A C_F n_f T_f w \mathcal{L}_{M_W} r(w)
-\frac{2 C_A C_F w \beta_0 \mathcal{L}_{M_W} r(w)}{\varepsilon }
-\frac{47 C_A C_F^2}{2 \varepsilon }
+\frac{5 C_A C_F}{3 \varepsilon }
-\frac{47 C_F}{4 \varepsilon }
+\frac{11}{2} \sqrt{3} C_A C_F^2 \pi 
+\frac{11}{4} \sqrt{3} C_F \pi 
-\frac{C_A C_F^2 Y_f^2 s \log (s-r)}{4 r}
-\frac{C_F Y_f^2 s \log (s-r)}{8 r}
+\frac{C_A C_F^2 Y_f^2 s \log (r+s)}{4 r}
+\frac{C_F Y_f^2 s \log (r+s)}{8 r}
+\frac{C_A C_F^2 Y_f^2 s \log (s-r) r(w)}{4 r}
+\frac{C_F Y_f^2 s \log (s-r) r(w)}{8 r}
-\frac{C_A C_F^2 Y_f^2 s \log (r+s) r(w)}{4 r}
-\frac{C_F Y_f^2 s \log (r+s) r(w)}{8 r}
-\frac{21 C_A C_F^2 Y_f^2}{8 r^2}
-\frac{21 C_F Y_f^2}{16 r^2}
-\frac{3 C_A C_F^2 Y_f^2}{4 \varepsilon r^2}
-\frac{3 C_F Y_f^2}{8 \varepsilon r^2}
+\frac{3 C_A C_F^2 Y_f^2 \mathcal{L}_{M_H}}{4 r^2}
+\frac{3 C_F Y_f^2 \mathcal{L}_{M_H}}{8 r^2}
+\frac{3 C_A C_F^2 Y_f^2 \mathcal{L}_{M_W}}{4 r^2}
+\frac{3 C_F Y_f^2 \mathcal{L}_{M_W}}{8 r^2}
+\frac{9 C_A C_F^2 Y_f^2 r(w)}{8 r^2}
+\frac{9 C_F Y_f^2 r(w)}{16 r^2}
+\frac{3 C_A C_F^2 Y_f^2 r(w)}{4 \varepsilon r^2}
+\frac{3 C_F Y_f^2 r(w)}{8 \varepsilon r^2}
-\frac{3 C_A C_F^2 Y_f^2 \mathcal{L}_{M_H} r(w)}{4 r^2}
-\frac{3 C_F Y_f^2 \mathcal{L}_{M_H} r(w)}{8 r^2}
-\frac{3 C_A C_F^2 Y_f^2 \mathcal{L}_{M_W} r(w)}{4 r^2}
-\frac{3 C_F Y_f^2 \mathcal{L}_{M_W} r(w)}{8 r^2}
+\frac{7 C_A C_F^2 Y_f^2 s \log (s-r)}{8 r^3}
+\frac{7 C_F Y_f^2 s \log (s-r)}{16 r^3}
-\frac{7 C_A C_F^2 Y_f^2 s \log (r+s)}{8 r^3}
-\frac{7 C_F Y_f^2 s \log (r+s)}{16 r^3}
-\frac{7 C_A C_F^2 Y_f^2 s \log (s-r) r(w)}{8 r^3}
-\frac{7 C_F Y_f^2 s \log (s-r) r(w)}{16 r^3}
+\frac{7 C_A C_F^2 Y_f^2 s \log (r+s) r(w)}{8 r^3}
+\frac{7 C_F Y_f^2 s \log (r+s) r(w)}{16 r^3}
\end{autobreak}
\end{align}
\end{tiny}

\begin{tiny}
\begin{align}
\begin{autobreak}
\Delta U^{(2)}_2=
-\frac{1}{12} C_A C_F \log \left(r^2\right) r^6
+\frac{1}{12} C_A C_F w \log \left(r^2\right) r(w) r^6
-\frac{1}{12} C_A C_F s \log (s-r) r^5
+\frac{1}{12} C_A C_F s \log (r+s) r^5
+\frac{1}{12} C_A C_F s \log (-v) r^5
-\frac{1}{12} C_A C_F s \log \left(\frac{v}{s}\right) r^5
+\frac{1}{12} C_A C_F s w \log (s-r) r(w) r^5
-\frac{1}{12} C_A C_F s w \log (r+s) r(w) r^5
-\frac{1}{12} C_A C_F s w \log (-v) r(w) r^5
+\frac{1}{12} C_A C_F s w \log \left(\frac{v}{s}\right) r(w) r^5
+\frac{1}{6} C_A C_F r^4
+\frac{1}{6} C_A C_F \mathcal{L}_{M_H} r^4
-\frac{1}{6} C_A C_F \mathcal{L}_{M_W} r^4
+\frac{1}{3} C_A C_F \log \left(r^2\right) r^4
-\frac{1}{6} C_A C_F w r(w) r^4
-\frac{1}{6} C_A C_F w \mathcal{L}_{M_H} r(w) r^4
+\frac{1}{6} C_A C_F w \mathcal{L}_{M_W} r(w) r^4
-\frac{1}{3} C_A C_F w \log \left(r^2\right) r(w) r^4
+\frac{1}{3} C_A C_F s \log (s-r) r^3
-\frac{1}{3} C_A C_F s \log (r+s) r^3
-\frac{1}{3} C_A C_F s \log (-v) r^3
+\frac{1}{3} C_A C_F s \log \left(\frac{v}{s}\right) r^3
-\frac{1}{3} C_A C_F s w \log (s-r) r(w) r^3
+\frac{1}{3} C_A C_F s w \log (r+s) r(w) r^3
+\frac{1}{3} C_A C_F s w \log (-v) r(w) r^3
-\frac{1}{3} C_A C_F s w \log \left(\frac{v}{s}\right) r(w) r^3
-\frac{1}{16} C_A C_F^2 Y_s^2 r^2
-\frac{9}{16} C_A Y_s^2 r^2
-\frac{1}{32} C_F Y_s^2 r^2
-\frac{C_A C_F^2 Y_s^2 r^2}{64 \varepsilon }
-\frac{9 C_A Y_s^2 r^2}{64 \varepsilon }
-\frac{C_F Y_s^2 r^2}{128 \varepsilon }
-\frac{5}{2} C_A C_F r^2
+\frac{1}{64} C_A C_F^2 Y_s^2 \mathcal{L}_{M_H} r^2
+\frac{9}{32} C_A Y_s^2 \mathcal{L}_{M_H} r^2
+\frac{1}{128} C_F Y_s^2 \mathcal{L}_{M_H} r^2
+\frac{1}{64} C_A C_F^2 Y_s^2 \mathcal{L}_{M_W} r^2
+\frac{1}{128} C_F Y_s^2 \mathcal{L}_{M_W} r^2
+C_A C_F \mathcal{L}_{M_W} r^2-C_A C_F \log \left(r^2\right) r^2
+\frac{1}{32} C_A C_F^2 Y_s^2 r(w) r^2
+\frac{9}{32} C_A Y_s^2 r(w) r^2
+\frac{1}{64} C_F Y_s^2 r(w) r^2
+\frac{C_A C_F^2 Y_s^2 r(w) r^2}{64 \varepsilon }
+\frac{9 C_A Y_s^2 r(w) r^2}{64 \varepsilon }
+\frac{C_F Y_s^2 r(w) r^2}{128 \varepsilon }
+\frac{3}{2} C_A C_F w r(w) r^2
+\frac{C_A C_F w r(w) r^2}{2 \varepsilon }
-\frac{1}{64} C_A C_F^2 Y_s^2 \mathcal{L}_{M_H} r(w) r^2
-\frac{9}{32} C_A Y_s^2 \mathcal{L}_{M_H} r(w) r^2
-\frac{1}{128} C_F Y_s^2 \mathcal{L}_{M_H} r(w) r^2
-\frac{1}{64} C_A C_F^2 Y_s^2 \mathcal{L}_{M_W} r(w) r^2
-\frac{1}{128} C_F Y_s^2 \mathcal{L}_{M_W} r(w) r^2
-C_A C_F w \mathcal{L}_{M_W} r(w) r^2
+C_A C_F w \log \left(r^2\right) r(w) r^2
-\frac{3}{64} \sqrt{3} C_A Y_s^2 \pi r(w) r^2
-\frac{C_A C_F r^2}{2 \varepsilon }
+\frac{3}{64} \sqrt{3} C_A Y_s^2 \pi r^2
+\frac{1}{64} C_A C_F^2 Y_s^2 s \log (s-r) r
+\frac{1}{128} C_F Y_s^2 s \log (s-r) r-C_A C_F s \log (s-r) r
-\frac{1}{64} C_A C_F^2 Y_s^2 s \log (r+s) r
-\frac{1}{128} C_F Y_s^2 s \log (r+s) r+C_A C_F s \log (r+s) r
+C_A C_F s \log (-v) r-C_A C_F s \log \left(\frac{v}{s}\right) r
-\frac{1}{64} C_A C_F^2 Y_s^2 s \log (s-r) r(w) r
-\frac{1}{128} C_F Y_s^2 s \log (s-r) r(w) r+C_A C_F s w \log (s-r) r(w) r
+\frac{1}{64} C_A C_F^2 Y_s^2 s \log (r+s) r(w) r
+\frac{1}{128} C_F Y_s^2 s \log (r+s) r(w) r-C_A C_F s w \log (r+s) r(w) r
-C_A C_F s w \log (-v) r(w) r+C_A C_F s w \log \left(\frac{v}{s}\right) r(w) r
-\frac{256}{3} C_A C_F^2+C_A C_F^2 Y_s^2
+\frac{C_F Y_s^2}{2}
+\frac{C_A C_F^2 Y_s^2}{4 \varepsilon }
+\frac{C_F Y_s^2}{8 \varepsilon }
+\frac{1}{4} C_A Y_s^2 \beta_0 \mathcal{L}_{M_H}^2
-C_A C_F \beta_0 \mathcal{L}_{M_W}^2
+\frac{61 C_A C_F}{9}
-\frac{128 C_F}{3}
+\frac{88}{9} C_A C_F n_f T_f
+\frac{8 C_A C_F n_f T_f}{3 \varepsilon }
-\frac{8}{3} i C_A C_F n_f \pi T_f
+\frac{C_A Y_s^2 \beta_0}{2 \varepsilon ^2}
-\frac{2 C_A C_F \beta_0}{\varepsilon ^2}
+\frac{1}{24} C_A Y_s^2 \pi ^2 \beta_0
-\frac{1}{6} C_A C_F \pi ^2 \beta_0
-\frac{1}{4} C_A C_F^2 Y_s^2 \mathcal{L}_{M_H}
-\frac{1}{8} C_F Y_s^2 \mathcal{L}_{M_H}-C_A Y_s^2 \beta_0 \mathcal{L}_{M_H}
-\frac{C_A Y_s^2 \beta_0 \mathcal{L}_{M_H}}{2 \varepsilon }
+47 C_A C_F^2 \mathcal{L}_{M_W}
-\frac{1}{4} C_A C_F^2 Y_s^2 \mathcal{L}_{M_W}
-\frac{1}{8} C_F Y_s^2 \mathcal{L}_{M_W}
-\frac{10}{3} C_A C_F \mathcal{L}_{M_W}
+\frac{47}{2} C_F \mathcal{L}_{M_W}
-\frac{16}{3} C_A C_F n_f T_f \mathcal{L}_{M_W}+4 C_A C_F \beta_0 \mathcal{L}_{M_W}
+\frac{2 C_A C_F \beta_0 \mathcal{L}_{M_W}}{\varepsilon }
-\frac{1}{2} C_A C_F^2 Y_s^2 r(w)
-\frac{1}{4} C_F Y_s^2 r(w)
-\frac{C_A C_F^2 Y_s^2 r(w)}{4 \varepsilon }
-\frac{C_F Y_s^2 r(w)}{8 \varepsilon }
-\frac{1}{4} C_A Y_s^2 \beta_0 \mathcal{L}_{M_H}^2 r(w)
+C_A C_F w \beta_0 \mathcal{L}_{M_W}^2 r(w)
+\frac{115}{3} C_A C_F^2 w r(w)
-\frac{31}{9} C_A C_F w r(w)
+\frac{115}{6} C_F w r(w)
-\frac{40}{9} C_A C_F n_f T_f w r(w)
-\frac{8 C_A C_F n_f T_f w r(w)}{3 \varepsilon }
+\frac{8}{3} i C_A C_F n_f \pi T_f w r(w)
+\frac{47 C_A C_F^2 w r(w)}{2 \varepsilon }
-\frac{5 C_A C_F w r(w)}{3 \varepsilon }
+\frac{47 C_F w r(w)}{4 \varepsilon }
-\frac{11}{2} \sqrt{3} C_A C_F^2 \pi w r(w)
-\frac{11}{4} \sqrt{3} C_F \pi w r(w)
-\frac{C_A Y_s^2 \beta_0 r(w)}{2 \varepsilon ^2}
+\frac{2 C_A C_F w \beta_0 r(w)}{\varepsilon ^2}
+\frac{1}{6} C_A C_F \pi ^2 w \beta_0 r(w)
-\frac{1}{24} C_A Y_s^2 \pi ^2 \beta_0 r(w)
+\frac{1}{4} C_A C_F^2 Y_s^2 \mathcal{L}_{M_H} r(w)
+\frac{1}{8} C_F Y_s^2 \mathcal{L}_{M_H} r(w)
+\frac{C_A Y_s^2 \beta_0 \mathcal{L}_{M_H} r(w)}{2 \varepsilon }
+\frac{1}{4} C_A C_F^2 Y_s^2 \mathcal{L}_{M_W} r(w)
+\frac{1}{8} C_F Y_s^2 \mathcal{L}_{M_W} r(w)
-47 C_A C_F^2 w \mathcal{L}_{M_W} r(w)
+\frac{10}{3} C_A C_F w \mathcal{L}_{M_W} r(w)
-\frac{47}{2} C_F w \mathcal{L}_{M_W} r(w)
+\frac{16}{3} C_A C_F n_f T_f w \mathcal{L}_{M_W} r(w)
-\frac{2 C_A C_F w \beta_0 \mathcal{L}_{M_W} r(w)}{\varepsilon }
-\frac{47 C_A C_F^2}{2 \varepsilon }
+\frac{5 C_A C_F}{3 \varepsilon }
-\frac{47 C_F}{4 \varepsilon }
+\frac{11}{2} \sqrt{3} C_A C_F^2 \pi 
+\frac{11}{4} \sqrt{3} C_F \pi 
-\frac{C_A C_F^2 Y_s^2 s \log (s-r)}{4 r}
-\frac{C_F Y_s^2 s \log (s-r)}{8 r}
+\frac{C_A C_F^2 Y_s^2 s \log (r+s)}{4 r}
+\frac{C_F Y_s^2 s \log (r+s)}{8 r}
+\frac{C_A C_F^2 Y_s^2 s \log (s-r) r(w)}{4 r}
+\frac{C_F Y_s^2 s \log (s-r) r(w)}{8 r}
-\frac{C_A C_F^2 Y_s^2 s \log (r+s) r(w)}{4 r}
-\frac{C_F Y_s^2 s \log (r+s) r(w)}{8 r}
-\frac{21 C_A C_F^2 Y_s^2}{8 r^2}
-\frac{21 C_F Y_s^2}{16 r^2}
-\frac{3 C_A C_F^2 Y_s^2}{4 \varepsilon r^2}
-\frac{3 C_F Y_s^2}{8 \varepsilon r^2}
+\frac{3 C_A C_F^2 Y_s^2 \mathcal{L}_{M_H}}{4 r^2}
+\frac{3 C_F Y_s^2 \mathcal{L}_{M_H}}{8 r^2}
+\frac{3 C_A C_F^2 Y_s^2 \mathcal{L}_{M_W}}{4 r^2}
+\frac{3 C_F Y_s^2 \mathcal{L}_{M_W}}{8 r^2}
+\frac{9 C_A C_F^2 Y_s^2 r(w)}{8 r^2}
+\frac{9 C_F Y_s^2 r(w)}{16 r^2}
+\frac{3 C_A C_F^2 Y_s^2 r(w)}{4 \varepsilon r^2}
+\frac{3 C_F Y_s^2 r(w)}{8 \varepsilon r^2}
-\frac{3 C_A C_F^2 Y_s^2 \mathcal{L}_{M_H} r(w)}{4 r^2}
-\frac{3 C_F Y_s^2 \mathcal{L}_{M_H} r(w)}{8 r^2}
-\frac{3 C_A C_F^2 Y_s^2 \mathcal{L}_{M_W} r(w)}{4 r^2}
-\frac{3 C_F Y_s^2 \mathcal{L}_{M_W} r(w)}{8 r^2}
+\frac{7 C_A C_F^2 Y_s^2 s \log (s-r)}{8 r^3}
+\frac{7 C_F Y_s^2 s \log (s-r)}{16 r^3}
-\frac{7 C_A C_F^2 Y_s^2 s \log (r+s)}{8 r^3}
-\frac{7 C_F Y_s^2 s \log (r+s)}{16 r^3}
-\frac{7 C_A C_F^2 Y_s^2 s \log (s-r) r(w)}{8 r^3}
-\frac{7 C_F Y_s^2 s \log (s-r) r(w)}{16 r^3}
+\frac{7 C_A C_F^2 Y_s^2 s \log (r+s) r(w)}{8 r^3}
+\frac{7 C_F Y_s^2 s \log (r+s) r(w)}{16 r^3}
\end{autobreak}
\end{align}
\end{tiny}

\begin{tiny}
\begin{align}
\begin{autobreak}
\Delta U^{(2)}_3=
-\frac{1}{12} C_A C_F \log \left(r^2\right) r^6
+\frac{1}{12} C_A C_F w \log \left(r^2\right) r(w) r^6
-\frac{1}{12} C_A C_F s \log (s-r) r^5
+\frac{1}{12} C_A C_F s \log (r+s) r^5
+\frac{1}{12} C_A C_F s \log (-v) r^5
-\frac{1}{12} C_A C_F s \log \left(\frac{v}{s}\right) r^5
+\frac{1}{12} C_A C_F s w \log (s-r) r(w) r^5
-\frac{1}{12} C_A C_F s w \log (r+s) r(w) r^5
-\frac{1}{12} C_A C_F s w \log (-v) r(w) r^5
+\frac{1}{12} C_A C_F s w \log \left(\frac{v}{s}\right) r(w) r^5
+\frac{1}{6} C_A C_F r^4
+\frac{1}{6} C_A C_F \mathcal{L}_{M_H} r^4
-\frac{1}{6} C_A C_F \mathcal{L}_{M_W} r^4
+\frac{1}{3} C_A C_F \log \left(r^2\right) r^4
-\frac{1}{6} C_A C_F w r(w) r^4
-\frac{1}{6} C_A C_F w \mathcal{L}_{M_H} r(w) r^4
+\frac{1}{6} C_A C_F w \mathcal{L}_{M_W} r(w) r^4
-\frac{1}{3} C_A C_F w \log \left(r^2\right) r(w) r^4
+\frac{1}{3} C_A C_F s \log (s-r) r^3
-\frac{1}{3} C_A C_F s \log (r+s) r^3
-\frac{1}{3} C_A C_F s \log (-v) r^3
+\frac{1}{3} C_A C_F s \log \left(\frac{v}{s}\right) r^3
-\frac{1}{3} C_A C_F s w \log (s-r) r(w) r^3
+\frac{1}{3} C_A C_F s w \log (r+s) r(w) r^3
+\frac{1}{3} C_A C_F s w \log (-v) r(w) r^3
-\frac{1}{3} C_A C_F s w \log \left(\frac{v}{s}\right) r(w) r^3
-\frac{1}{32} C_A C_F^2 Y_f^2 r^2
-\frac{9}{32} C_A Y_f^2 r^2
-\frac{1}{64} C_F Y_f^2 r^2
-\frac{C_A C_F^2 Y_f^2 r^2}{128 \varepsilon }
-\frac{9 C_A Y_f^2 r^2}{128 \varepsilon }
-\frac{C_F Y_f^2 r^2}{256 \varepsilon }
-\frac{1}{32} C_A C_F^2 Y_s^2 r^2
-\frac{9}{32} C_A Y_s^2 r^2
-\frac{1}{64} C_F Y_s^2 r^2
-\frac{C_A C_F^2 Y_s^2 r^2}{128 \varepsilon }
-\frac{9 C_A Y_s^2 r^2}{128 \varepsilon }
-\frac{C_F Y_s^2 r^2}{256 \varepsilon }
-\frac{5}{2} C_A C_F r^2
+\frac{1}{128} C_A C_F^2 Y_f^2 \mathcal{L}_{M_H} r^2
+\frac{9}{64} C_A Y_f^2 \mathcal{L}_{M_H} r^2
+\frac{1}{256} C_F Y_f^2 \mathcal{L}_{M_H} r^2
+\frac{1}{128} C_A C_F^2 Y_s^2 \mathcal{L}_{M_H} r^2
+\frac{9}{64} C_A Y_s^2 \mathcal{L}_{M_H} r^2
+\frac{1}{256} C_F Y_s^2 \mathcal{L}_{M_H} r^2
+\frac{1}{128} C_A C_F^2 Y_f^2 \mathcal{L}_{M_W} r^2
+\frac{1}{256} C_F Y_f^2 \mathcal{L}_{M_W} r^2
+\frac{1}{128} C_A C_F^2 Y_s^2 \mathcal{L}_{M_W} r^2
+\frac{1}{256} C_F Y_s^2 \mathcal{L}_{M_W} r^2
+C_A C_F \mathcal{L}_{M_W} r^2-C_A C_F \log \left(r^2\right) r^2
+\frac{1}{32} C_A C_F^2 Y_f Y_s r(w) r^2
+\frac{9}{32} C_A Y_f Y_s r(w) r^2
+\frac{1}{64} C_F Y_f Y_s r(w) r^2
+\frac{C_A C_F^2 Y_f Y_s r(w) r^2}{64 \varepsilon }
+\frac{9 C_A Y_f Y_s r(w) r^2}{64 \varepsilon }
+\frac{C_F Y_f Y_s r(w) r^2}{128 \varepsilon }
+\frac{3}{2} C_A C_F w r(w) r^2
+\frac{C_A C_F w r(w) r^2}{2 \varepsilon }
-\frac{1}{64} C_A C_F^2 Y_f Y_s \mathcal{L}_{M_H} r(w) r^2
-\frac{9}{32} C_A Y_f Y_s \mathcal{L}_{M_H} r(w) r^2
-\frac{1}{128} C_F Y_f Y_s \mathcal{L}_{M_H} r(w) r^2
-\frac{1}{64} C_A C_F^2 Y_f Y_s \mathcal{L}_{M_W} r(w) r^2
-\frac{1}{128} C_F Y_f Y_s \mathcal{L}_{M_W} r(w) r^2
-C_A C_F w \mathcal{L}_{M_W} r(w) r^2
+C_A C_F w \log \left(r^2\right) r(w) r^2
-\frac{3}{64} \sqrt{3} C_A Y_f Y_s \pi r(w) r^2
-\frac{C_A C_F r^2}{2 \varepsilon }
+\frac{3}{128} \sqrt{3} C_A Y_f^2 \pi r^2
+\frac{3}{128} \sqrt{3} C_A Y_s^2 \pi r^2
+\frac{1}{128} C_A C_F^2 Y_f^2 s \log (s-r) r
+\frac{1}{256} C_F Y_f^2 s \log (s-r) r
+\frac{1}{128} C_A C_F^2 Y_s^2 s \log (s-r) r
+\frac{1}{256} C_F Y_s^2 s \log (s-r) r-C_A C_F s \log (s-r) r
-\frac{1}{128} C_A C_F^2 Y_f^2 s \log (r+s) r
-\frac{1}{256} C_F Y_f^2 s \log (r+s) r
-\frac{1}{128} C_A C_F^2 Y_s^2 s \log (r+s) r
-\frac{1}{256} C_F Y_s^2 s \log (r+s) r
+C_A C_F s \log (r+s) r+C_A C_F s \log (-v) r
-C_A C_F s \log \left(\frac{v}{s}\right) r
-\frac{1}{64} C_A C_F^2 Y_f Y_s s \log (s-r) r(w) r
-\frac{1}{128} C_F Y_f Y_s s \log (s-r) r(w) r+C_A C_F s w \log (s-r) r(w) r
+\frac{1}{64} C_A C_F^2 Y_f Y_s s \log (r+s) r(w) r
+\frac{1}{128} C_F Y_f Y_s s \log (r+s) r(w) r-C_A C_F s w \log (r+s) r(w) r
-C_A C_F s w \log (-v) r(w) r
+C_A C_F s w \log \left(\frac{v}{s}\right) r(w) r
-\frac{256}{3} C_A C_F^2
+\frac{1}{2} C_A C_F^2 Y_f^2
+\frac{C_F Y_f^2}{4}
+\frac{C_A C_F^2 Y_f^2}{8 \varepsilon }
+\frac{C_F Y_f^2}{16 \varepsilon }
+\frac{1}{2} C_A C_F^2 Y_s^2
+\frac{C_F Y_s^2}{4}
+\frac{C_A C_F^2 Y_s^2}{8 \varepsilon }
+\frac{C_F Y_s^2}{16 \varepsilon }
+\frac{1}{8} C_A Y_f^2 \beta_0 \mathcal{L}_{M_H}^2
+\frac{1}{8} C_A Y_s^2 \beta_0 \mathcal{L}_{M_H}^2
-C_A C_F \beta_0 \mathcal{L}_{M_W}^2
+\frac{61 C_A C_F}{9}
-\frac{128 C_F}{3}
+\frac{88}{9} C_A C_F n_f T_f
+\frac{8 C_A C_F n_f T_f}{3 \varepsilon }
-\frac{8}{3} i C_A C_F n_f \pi T_f
+\frac{C_A Y_f^2 \beta_0}{4 \varepsilon ^2}
+\frac{C_A Y_s^2 \beta_0}{4 \varepsilon ^2}
-\frac{2 C_A C_F \beta_0}{\varepsilon ^2}
+\frac{1}{48} C_A Y_f^2 \pi ^2 \beta_0
+\frac{1}{48} C_A Y_s^2 \pi ^2 \beta_0
-\frac{1}{6} C_A C_F \pi ^2 \beta_0
-\frac{1}{8} C_A C_F^2 Y_f^2 \mathcal{L}_{M_H}
-\frac{1}{16} C_F Y_f^2 \mathcal{L}_{M_H}
-\frac{1}{8} C_A C_F^2 Y_s^2 \mathcal{L}_{M_H}
-\frac{1}{16} C_F Y_s^2 \mathcal{L}_{M_H}
-\frac{1}{2} C_A Y_f^2 \beta_0 \mathcal{L}_{M_H}
-\frac{C_A Y_f^2 \beta_0 \mathcal{L}_{M_H}}{4 \varepsilon }
-\frac{1}{2} C_A Y_s^2 \beta_0 \mathcal{L}_{M_H}
-\frac{C_A Y_s^2 \beta_0 \mathcal{L}_{M_H}}{4 \varepsilon }
+47 C_A C_F^2 \mathcal{L}_{M_W}
-\frac{1}{8} C_A C_F^2 Y_f^2 \mathcal{L}_{M_W}
-\frac{1}{16} C_F Y_f^2 \mathcal{L}_{M_W}
-\frac{1}{8} C_A C_F^2 Y_s^2 \mathcal{L}_{M_W}
-\frac{1}{16} C_F Y_s^2 \mathcal{L}_{M_W}
-\frac{10}{3} C_A C_F \mathcal{L}_{M_W}
+\frac{47}{2} C_F \mathcal{L}_{M_W}
-\frac{16}{3} C_A C_F n_f T_f \mathcal{L}_{M_W}
+4 C_A C_F \beta_0 \mathcal{L}_{M_W}
+\frac{2 C_A C_F \beta_0 \mathcal{L}_{M_W}}{\varepsilon }
-\frac{1}{4} C_A Y_f Y_s \beta_0 \mathcal{L}_{M_H}^2 r(w)
+C_A C_F w \beta_0 \mathcal{L}_{M_W}^2 r(w)
-\frac{1}{2} C_A C_F^2 Y_f Y_s r(w)
-\frac{1}{4} C_F Y_f Y_s r(w)
-\frac{C_A C_F^2 Y_f Y_s r(w)}{4 \varepsilon }
-\frac{C_F Y_f Y_s r(w)}{8 \varepsilon }
+\frac{115}{3} C_A C_F^2 w r(w)
-\frac{31}{9} C_A C_F w r(w)
+\frac{115}{6} C_F w r(w)
-\frac{40}{9} C_A C_F n_f T_f w r(w)
-\frac{8 C_A C_F n_f T_f w r(w)}{3 \varepsilon }
+\frac{8}{3} i C_A C_F n_f \pi T_f w r(w)
+\frac{47 C_A C_F^2 w r(w)}{2 \varepsilon }
-\frac{5 C_A C_F w r(w)}{3 \varepsilon }
+\frac{47 C_F w r(w)}{4 \varepsilon }
-\frac{11}{2} \sqrt{3} C_A C_F^2 \pi w r(w)
-\frac{11}{4} \sqrt{3} C_F \pi w r(w)
-\frac{C_A Y_f Y_s \beta_0 r(w)}{2 \varepsilon ^2}
+\frac{2 C_A C_F w \beta_0 r(w)}{\varepsilon ^2}
+\frac{1}{6} C_A C_F \pi ^2 w \beta_0 r(w)
-\frac{1}{24} C_A Y_f Y_s \pi ^2 \beta_0 r(w)
+\frac{1}{4} C_A C_F^2 Y_f Y_s \mathcal{L}_{M_H} r(w)
+\frac{1}{8} C_F Y_f Y_s \mathcal{L}_{M_H} r(w)
+\frac{C_A Y_f Y_s \beta_0 \mathcal{L}_{M_H} r(w)}{2 \varepsilon }
+\frac{1}{4} C_A C_F^2 Y_f Y_s \mathcal{L}_{M_W} r(w)
+\frac{1}{8} C_F Y_f Y_s \mathcal{L}_{M_W} r(w)
-47 C_A C_F^2 w \mathcal{L}_{M_W} r(w)
+\frac{10}{3} C_A C_F w \mathcal{L}_{M_W} r(w)
-\frac{47}{2} C_F w \mathcal{L}_{M_W} r(w)
+\frac{16}{3} C_A C_F n_f T_f w \mathcal{L}_{M_W} r(w)
-\frac{2 C_A C_F w \beta_0 \mathcal{L}_{M_W} r(w)}{\varepsilon }
-\frac{47 C_A C_F^2}{2 \varepsilon }
+\frac{5 C_A C_F}{3 \varepsilon }
-\frac{47 C_F}{4 \varepsilon }
+\frac{11}{2} \sqrt{3} C_A C_F^2 \pi 
+\frac{11}{4} \sqrt{3} C_F \pi 
-\frac{C_A C_F^2 Y_f^2 s \log (s-r)}{8 r}
-\frac{C_F Y_f^2 s \log (s-r)}{16 r}
-\frac{C_A C_F^2 Y_s^2 s \log (s-r)}{8 r}
-\frac{C_F Y_s^2 s \log (s-r)}{16 r}
+\frac{C_A C_F^2 Y_f^2 s \log (r+s)}{8 r}
+\frac{C_F Y_f^2 s \log (r+s)}{16 r}
+\frac{C_A C_F^2 Y_s^2 s \log (r+s)}{8 r}
+\frac{C_F Y_s^2 s \log (r+s)}{16 r}
+\frac{C_A C_F^2 Y_f Y_s s \log (s-r) r(w)}{4 r}
+\frac{C_F Y_f Y_s s \log (s-r) r(w)}{8 r}
-\frac{C_A C_F^2 Y_f Y_s s \log (r+s) r(w)}{4 r}
-\frac{C_F Y_f Y_s s \log (r+s) r(w)}{8 r}
-\frac{21 C_A C_F^2 Y_f^2}{16 r^2}
-\frac{21 C_F Y_f^2}{32 r^2}
-\frac{3 C_A C_F^2 Y_f^2}{8 \varepsilon r^2}
-\frac{3 C_F Y_f^2}{16 \varepsilon r^2}
-\frac{21 C_A C_F^2 Y_s^2}{16 r^2}
-\frac{21 C_F Y_s^2}{32 r^2}
-\frac{3 C_A C_F^2 Y_s^2}{8 \varepsilon r^2}
-\frac{3 C_F Y_s^2}{16 \varepsilon r^2}
+\frac{3 C_A C_F^2 Y_f^2 \mathcal{L}_{M_H}}{8 r^2}
+\frac{3 C_F Y_f^2 \mathcal{L}_{M_H}}{16 r^2}
+\frac{3 C_A C_F^2 Y_s^2 \mathcal{L}_{M_H}}{8 r^2}
+\frac{3 C_F Y_s^2 \mathcal{L}_{M_H}}{16 r^2}
+\frac{3 C_A C_F^2 Y_f^2 \mathcal{L}_{M_W}}{8 r^2}
+\frac{3 C_F Y_f^2 \mathcal{L}_{M_W}}{16 r^2}
+\frac{3 C_A C_F^2 Y_s^2 \mathcal{L}_{M_W}}{8 r^2}
+\frac{3 C_F Y_s^2 \mathcal{L}_{M_W}}{16 r^2}
+\frac{9 C_A C_F^2 Y_f Y_s r(w)}{8 r^2}
+\frac{9 C_F Y_f Y_s r(w)}{16 r^2}
+\frac{3 C_A C_F^2 Y_f Y_s r(w)}{4 \varepsilon r^2}
+\frac{3 C_F Y_f Y_s r(w)}{8 \varepsilon r^2}
-\frac{3 C_A C_F^2 Y_f Y_s \mathcal{L}_{M_H} r(w)}{4 r^2}
-\frac{3 C_F Y_f Y_s \mathcal{L}_{M_H} r(w)}{8 r^2}
-\frac{3 C_A C_F^2 Y_f Y_s \mathcal{L}_{M_W} r(w)}{4 r^2}
-\frac{3 C_F Y_f Y_s \mathcal{L}_{M_W} r(w)}{8 r^2}
+\frac{7 C_A C_F^2 Y_f^2 s \log (s-r)}{16 r^3}
+\frac{7 C_F Y_f^2 s \log (s-r)}{32 r^3}
+\frac{7 C_A C_F^2 Y_s^2 s \log (s-r)}{16 r^3}
+\frac{7 C_F Y_s^2 s \log (s-r)}{32 r^3}
-\frac{7 C_A C_F^2 Y_f^2 s \log (r+s)}{16 r^3}
-\frac{7 C_F Y_f^2 s \log (r+s)}{32 r^3}
-\frac{7 C_A C_F^2 Y_s^2 s \log (r+s)}{16 r^3}
-\frac{7 C_F Y_s^2 s \log (r+s)}{32 r^3}
-\frac{7 C_A C_F^2 Y_f Y_s s \log (s-r) r(w)}{8 r^3}
-\frac{7 C_F Y_f Y_s s \log (s-r) r(w)}{16 r^3}
+\frac{7 C_A C_F^2 Y_f Y_s s \log (r+s) r(w)}{8 r^3}
+\frac{7 C_F Y_f Y_s s \log (r+s) r(w)}{16 r^3}
\end{autobreak}
\end{align}
\end{tiny}
\section{Vertex Corrections}
\label{sec:vertcorr}
\subsection{Matching at $\mu\sim Q$:}
\begin{tiny}
\begin{align}
\begin{autobreak}
V^{(Q)}_1=
\frac{1}{96} C_A \mathcal{L}_Q^4 Y_f^4
-\frac{5}{12} C_A \mathcal{L}_Q^3 Y_f^4
-\frac{C_A \mathcal{L}_Q^3 Y_f^4}{48 \epsilon }
+\frac{105}{128} C_A \mathcal{L}_Q^2 Y_f^4
+\frac{5 C_A \mathcal{L}_Q^2 Y_f^4}{8 \epsilon }
+\frac{C_A \mathcal{L}_Q^2 Y_f^4}{32 \epsilon ^2}
-\frac{2463 C_A Y_f^4}{512}
-\frac{329}{256} C_A \mathcal{L}_Q Y_f^4
-\frac{105 C_A \mathcal{L}_Q Y_f^4}{128 \epsilon }
-\frac{5 C_A \mathcal{L}_Q Y_f^4}{8 \epsilon ^2}
-\frac{C_A \mathcal{L}_Q Y_f^4}{32 \epsilon ^3}
+\frac{3}{128} C_A \pi ^2 \mathcal{L}_Q Y_f^4
+\frac{329 C_A Y_f^4}{512 \epsilon }
+\frac{105 C_A Y_f^4}{256 \epsilon ^2}
+\frac{5 C_A Y_f^4}{16 \epsilon ^3}
+\frac{C_A Y_f^4}{64 \epsilon ^4}
-\frac{83}{192} C_A \zeta_3 Y_f^4
-\frac{5}{256} C_A \pi ^2 Y_f^4
-\frac{3 C_A \pi ^2 Y_f^4}{256 \epsilon }
+\frac{1}{4} C_A C_F \mathcal{L}_Q^4 Y_f^2
-\frac{17}{2} C_A C_F \mathcal{L}_Q^3 Y_f^2
-\frac{C_A C_F \mathcal{L}_Q^3 Y_f^2}{2 \epsilon }
+\frac{19}{4} C_A C_F^2 Y_f^2
+\frac{1}{2} C_A C_F^2 \mathcal{L}_Q^2 Y_f^2
+\frac{333}{16} C_A C_F \mathcal{L}_Q^2 Y_f^2
+\frac{1}{4} C_F \mathcal{L}_Q^2 Y_f^2
+\frac{51 C_A C_F \mathcal{L}_Q^2 Y_f^2}{4 \epsilon }
+\frac{3 C_A C_F \mathcal{L}_Q^2 Y_f^2}{4 \epsilon ^2}
+\frac{1}{12} C_A C_F \pi ^2 \mathcal{L}_Q^2 Y_f^2
-\frac{167}{2} C_A C_F Y_f^2
+\frac{19 C_F Y_f^2}{8}
-\frac{5}{2} C_A C_F^2 \mathcal{L}_Q Y_f^2
+\frac{2501}{16} C_A C_F \mathcal{L}_Q Y_f^2
-\frac{5}{4} C_F \mathcal{L}_Q Y_f^2
-\frac{C_A C_F^2 \mathcal{L}_Q Y_f^2}{2 \epsilon }
-\frac{333 C_A C_F \mathcal{L}_Q Y_f^2}{16 \epsilon }
-\frac{C_F \mathcal{L}_Q Y_f^2}{4 \epsilon }
-\frac{51 C_A C_F \mathcal{L}_Q Y_f^2}{4 \epsilon ^2}
-\frac{3 C_A C_F \mathcal{L}_Q Y_f^2}{4 \epsilon ^3}
-\frac{9}{8} C_A C_F \pi ^2 \mathcal{L}_Q Y_f^2
-\frac{C_A C_F \pi ^2 \mathcal{L}_Q Y_f^2}{12 \epsilon }
+\frac{5 C_A C_F^2 Y_f^2}{4 \epsilon }
-\frac{2501 C_A C_F Y_f^2}{32 \epsilon }
+\frac{5 C_F Y_f^2}{8 \epsilon }
+\frac{C_A C_F^2 Y_f^2}{4 \epsilon ^2}
+\frac{333 C_A C_F Y_f^2}{32 \epsilon ^2}
+\frac{C_F Y_f^2}{8 \epsilon ^2}
+\frac{51 C_A C_F Y_f^2}{8 \epsilon ^3}
+\frac{3 C_A C_F Y_f^2}{8 \epsilon ^4}
+\frac{83}{12} C_A C_F \zeta_3 Y_f^2
+\frac{1}{24} C_A C_F^2 \pi ^2 Y_f^2
+\frac{47}{48} C_A C_F \pi ^2 Y_f^2
+\frac{1}{48} C_F \pi ^2 Y_f^2
+\frac{9 C_A C_F \pi ^2 Y_f^2}{16 \epsilon }
+\frac{C_A C_F \pi ^2 Y_f^2}{24 \epsilon ^2}-6 C_A C_F^2 \mathcal{L}_Q^4+2 C_F \mathcal{L}_Q^4
-\frac{157}{6} C_A C_F^2 \mathcal{L}_Q^3
+\frac{1}{18} C_A C_F \mathcal{L}_Q^3
-\frac{133}{12} C_F \mathcal{L}_Q^3
-\frac{2}{9} C_A C_F n_f T_f \mathcal{L}_Q^3
+\frac{12 C_A C_F^2 \mathcal{L}_Q^3}{\epsilon }
-\frac{4 C_F \mathcal{L}_Q^3}{\epsilon }
-\frac{78349}{216} C_A C_F^2
+\frac{650}{3} C_A C_F^2 \mathcal{L}_Q^2
-\frac{2}{9} C_A C_F \mathcal{L}_Q^2
+\frac{263}{6} C_F \mathcal{L}_Q^2
-\frac{1}{9} C_A C_F n_f T_f \mathcal{L}_Q^2
+\frac{C_A C_F n_f T_f \mathcal{L}_Q^2}{3 \epsilon }
+\frac{157 C_A C_F^2 \mathcal{L}_Q^2}{4 \epsilon }
-\frac{C_A C_F \mathcal{L}_Q^2}{12 \epsilon }
+\frac{133 C_F \mathcal{L}_Q^2}{8 \epsilon }
-\frac{18 C_A C_F^2 \mathcal{L}_Q^2}{\epsilon ^2}
+\frac{6 C_F \mathcal{L}_Q^2}{\epsilon ^2}+C_A C_F^2 \pi ^2 \mathcal{L}_Q^2
-\frac{5}{3} C_F \pi ^2 \mathcal{L}_Q^2
+\frac{1129 C_A C_F}{648}
+\frac{362507 C_F}{432}
-\frac{1615}{162} C_A C_F n_f T_f
+\frac{8 C_A C_F n_f T_f}{27 \epsilon }
-\frac{C_A C_F n_f T_f}{18 \epsilon ^2}
+\frac{C_A C_F n_f T_f}{6 \epsilon ^3}
-\frac{4}{27} C_A C_F n_f \pi ^2 T_f
+\frac{C_A C_F n_f \pi ^2 T_f}{36 \epsilon }
-\frac{5759}{18} C_A C_F^2 \mathcal{L}_Q
+\frac{35}{54} C_A C_F \mathcal{L}_Q
-\frac{31147}{72} C_F \mathcal{L}_Q
-\frac{16}{27} C_A C_F n_f T_f \mathcal{L}_Q
+\frac{C_A C_F n_f T_f \mathcal{L}_Q}{9 \epsilon }
-\frac{C_A C_F n_f T_f \mathcal{L}_Q}{3 \epsilon ^2}
-\frac{1}{18} C_A C_F n_f \pi ^2 T_f \mathcal{L}_Q
-\frac{650 C_A C_F^2 \mathcal{L}_Q}{3 \epsilon }
+\frac{2 C_A C_F \mathcal{L}_Q}{9 \epsilon }
-\frac{263 C_F \mathcal{L}_Q}{6 \epsilon }
-\frac{157 C_A C_F^2 \mathcal{L}_Q}{4 \epsilon ^2}
+\frac{C_A C_F \mathcal{L}_Q}{12 \epsilon ^2}
-\frac{133 C_F \mathcal{L}_Q}{8 \epsilon ^2}
+\frac{18 C_A C_F^2 \mathcal{L}_Q}{\epsilon ^3}
-\frac{6 C_F \mathcal{L}_Q}{\epsilon ^3}
+\frac{17}{8} C_A C_F^2 \pi ^2 \mathcal{L}_Q
+\frac{1}{72} C_A C_F \pi ^2 \mathcal{L}_Q
+\frac{93}{16} C_F \pi ^2 \mathcal{L}_Q
-\frac{C_A C_F^2 \pi ^2 \mathcal{L}_Q}{\epsilon }
+\frac{5 C_F \pi ^2 \mathcal{L}_Q}{3 \epsilon }
+\frac{5759 C_A C_F^2}{36 \epsilon }
-\frac{35 C_A C_F}{108 \epsilon }
+\frac{31147 C_F}{144 \epsilon }
+\frac{325 C_A C_F^2}{3 \epsilon ^2}
-\frac{C_A C_F}{9 \epsilon ^2}
+\frac{263 C_F}{12 \epsilon ^2}
+\frac{157 C_A C_F^2}{8 \epsilon ^3}
-\frac{C_A C_F}{24 \epsilon ^3}
+\frac{133 C_F}{16 \epsilon ^3}
-\frac{9 C_A C_F^2}{\epsilon ^4}
+\frac{3 C_F}{\epsilon ^4}
-\frac{415 C_F \zeta_3}{6}
+\frac{166}{3} C_F \mathcal{L}_Q \zeta_3
-\frac{83 C_F \zeta_3}{3 \epsilon }
-\frac{59 C_F \pi ^4}{120}
+\frac{389}{144} C_A C_F^2 \pi ^2
+\frac{7}{432} C_A C_F \pi ^2
-\frac{3835 C_F \pi ^2}{288}
-\frac{17 C_A C_F^2 \pi ^2}{16 \epsilon }
-\frac{C_A C_F \pi ^2}{144 \epsilon }
-\frac{93 C_F \pi ^2}{32 \epsilon }
+\frac{C_A C_F^2 \pi ^2}{2 \epsilon ^2}
-\frac{5 C_F \pi ^2}{6 \epsilon ^2}
\end{autobreak}
\end{align}
\end{tiny}

\begin{tiny}
\begin{align}
\begin{autobreak}
V^{(Q)}_2=
\frac{1}{96} C_A \mathcal{L}_Q^4 Y_f^4
-\frac{31}{48} C_A \mathcal{L}_Q^3 Y_f^4
-\frac{C_A \mathcal{L}_Q^3 Y_f^4}{48 \epsilon }
+\frac{67}{64} C_A \mathcal{L}_Q^2 Y_f^4
+\frac{31 C_A \mathcal{L}_Q^2 Y_f^4}{32 \epsilon }
+\frac{C_A \mathcal{L}_Q^2 Y_f^4}{32 \epsilon ^2}
+\frac{1}{192} C_A \pi ^2 \mathcal{L}_Q^2 Y_f^4
+\frac{823 C_A Y_f^4}{512}
+\frac{2595}{256} C_A \mathcal{L}_Q Y_f^4
-\frac{67 C_A \mathcal{L}_Q Y_f^4}{64 \epsilon }
-\frac{31 C_A \mathcal{L}_Q Y_f^4}{32 \epsilon ^2}
-\frac{C_A \mathcal{L}_Q Y_f^4}{32 \epsilon ^3}
-\frac{17}{384} C_A \pi ^2 \mathcal{L}_Q Y_f^4
-\frac{C_A \pi ^2 \mathcal{L}_Q Y_f^4}{192 \epsilon }
-\frac{2595 C_A Y_f^4}{512 \epsilon }
+\frac{67 C_A Y_f^4}{128 \epsilon ^2}
+\frac{31 C_A Y_f^4}{64 \epsilon ^3}
+\frac{C_A Y_f^4}{64 \epsilon ^4}
+\frac{83}{192} C_A \zeta_3 Y_f^4
+\frac{55}{768} C_A \pi ^2 Y_f^4
+\frac{17 C_A \pi ^2 Y_f^4}{768 \epsilon }
+\frac{C_A \pi ^2 Y_f^4}{384 \epsilon ^2}
-\frac{2}{3} C_A C_F \mathcal{L}_Q^4 Y_f^2
+\frac{91}{9} C_A C_F \mathcal{L}_Q^3 Y_f^2
+\frac{4 C_A C_F \mathcal{L}_Q^3 Y_f^2}{3 \epsilon }
-\frac{37}{32} C_A C_F^2 Y_f^2
-\frac{1}{4} C_A C_F^2 \mathcal{L}_Q^2 Y_f^2
+\frac{1493}{144} C_A C_F \mathcal{L}_Q^2 Y_f^2
-\frac{1}{8} C_F \mathcal{L}_Q^2 Y_f^2
-\frac{91 C_A C_F \mathcal{L}_Q^2 Y_f^2}{6 \epsilon }
-\frac{2 C_A C_F \mathcal{L}_Q^2 Y_f^2}{\epsilon ^2}
-\frac{1}{8} C_A C_F \pi ^2 \mathcal{L}_Q^2 Y_f^2
+\frac{1244593 C_A C_F Y_f^2}{5184}
-\frac{37 C_F Y_f^2}{64}
+\frac{7}{8} C_A C_F^2 \mathcal{L}_Q Y_f^2
-\frac{91409}{864} C_A C_F \mathcal{L}_Q Y_f^2
+\frac{7}{16} C_F \mathcal{L}_Q Y_f^2
+\frac{C_A C_F^2 \mathcal{L}_Q Y_f^2}{4 \epsilon }
-\frac{1493 C_A C_F \mathcal{L}_Q Y_f^2}{144 \epsilon }
+\frac{C_F \mathcal{L}_Q Y_f^2}{8 \epsilon }
+\frac{91 C_A C_F \mathcal{L}_Q Y_f^2}{6 \epsilon ^2}
+\frac{2 C_A C_F \mathcal{L}_Q Y_f^2}{\epsilon ^3}
+\frac{163}{144} C_A C_F \pi ^2 \mathcal{L}_Q Y_f^2
+\frac{C_A C_F \pi ^2 \mathcal{L}_Q Y_f^2}{8 \epsilon }
-\frac{7 C_A C_F^2 Y_f^2}{16 \epsilon }
+\frac{91409 C_A C_F Y_f^2}{1728 \epsilon }
-\frac{7 C_F Y_f^2}{32 \epsilon }
-\frac{C_A C_F^2 Y_f^2}{8 \epsilon ^2}
+\frac{1493 C_A C_F Y_f^2}{288 \epsilon ^2}
-\frac{C_F Y_f^2}{16 \epsilon ^2}
-\frac{91 C_A C_F Y_f^2}{12 \epsilon ^3}
-\frac{C_A C_F Y_f^2}{\epsilon ^4}
-\frac{415}{24} C_A C_F \zeta_3 Y_f^2
-\frac{1}{48} C_A C_F^2 \pi ^2 Y_f^2
-\frac{3041}{864} C_A C_F \pi ^2 Y_f^2
-\frac{1}{96} C_F \pi ^2 Y_f^2
-\frac{163 C_A C_F \pi ^2 Y_f^2}{288 \epsilon }
-\frac{C_A C_F \pi ^2 Y_f^2}{16 \epsilon ^2}
-\frac{20}{3} C_A C_F^2 \mathcal{L}_Q^4-2 C_F \mathcal{L}_Q^4
-\frac{1783}{18} C_A C_F^2 \mathcal{L}_Q^3
+\frac{1}{18} C_A C_F \mathcal{L}_Q^3
-\frac{269}{12} C_F \mathcal{L}_Q^3
-\frac{2}{9} C_A C_F n_f T_f \mathcal{L}_Q^3
+\frac{40 C_A C_F^2 \mathcal{L}_Q^3}{3 \epsilon }
+\frac{4 C_F \mathcal{L}_Q^3}{\epsilon }
-\frac{239309 C_A C_F^2}{5184}
+\frac{43897}{72} C_A C_F^2 \mathcal{L}_Q^2
-\frac{25}{72} C_A C_F \mathcal{L}_Q^2
+\frac{20131}{48} C_F \mathcal{L}_Q^2
+\frac{17}{9} C_A C_F n_f T_f \mathcal{L}_Q^2
+\frac{C_A C_F n_f T_f \mathcal{L}_Q^2}{3 \epsilon }
+\frac{1783 C_A C_F^2 \mathcal{L}_Q^2}{12 \epsilon }
-\frac{C_A C_F \mathcal{L}_Q^2}{12 \epsilon }
+\frac{269 C_F \mathcal{L}_Q^2}{8 \epsilon }
-\frac{20 C_A C_F^2 \mathcal{L}_Q^2}{\epsilon ^2}
-\frac{6 C_F \mathcal{L}_Q^2}{\epsilon ^2}+C_A C_F^2 \pi ^2 \mathcal{L}_Q^2
-\frac{5}{2} C_F \pi ^2 \mathcal{L}_Q^2
-\frac{283 C_A C_F}{5184}
+\frac{8829121 C_F}{3456}
+\frac{2683}{324} C_A C_F n_f T_f
+\frac{259 C_A C_F n_f T_f}{54 \epsilon }
+\frac{17 C_A C_F n_f T_f}{18 \epsilon ^2}
+\frac{C_A C_F n_f T_f}{6 \epsilon ^3}
+\frac{1}{54} C_A C_F n_f \pi ^2 T_f
+\frac{C_A C_F n_f \pi ^2 T_f}{36 \epsilon }
+\frac{201881}{432} C_A C_F^2 \mathcal{L}_Q
+\frac{631}{432} C_A C_F \mathcal{L}_Q
-\frac{283861}{288} C_F \mathcal{L}_Q
-\frac{259}{27} C_A C_F n_f T_f \mathcal{L}_Q
-\frac{17 C_A C_F n_f T_f \mathcal{L}_Q}{9 \epsilon }
-\frac{C_A C_F n_f T_f \mathcal{L}_Q}{3 \epsilon ^2}
-\frac{1}{18} C_A C_F n_f \pi ^2 T_f \mathcal{L}_Q
-\frac{43897 C_A C_F^2 \mathcal{L}_Q}{72 \epsilon }
+\frac{25 C_A C_F \mathcal{L}_Q}{72 \epsilon }
-\frac{20131 C_F \mathcal{L}_Q}{48 \epsilon }
-\frac{1783 C_A C_F^2 \mathcal{L}_Q}{12 \epsilon ^2}
+\frac{C_A C_F \mathcal{L}_Q}{12 \epsilon ^2}
-\frac{269 C_F \mathcal{L}_Q}{8 \epsilon ^2}
+\frac{20 C_A C_F^2 \mathcal{L}_Q}{\epsilon ^3}
+\frac{6 C_F \mathcal{L}_Q}{\epsilon ^3}
-\frac{427}{72} C_A C_F^2 \pi ^2 \mathcal{L}_Q
+\frac{1}{72} C_A C_F \pi ^2 \mathcal{L}_Q
+\frac{535}{48} C_F \pi ^2 \mathcal{L}_Q
-\frac{C_A C_F^2 \pi ^2 \mathcal{L}_Q}{\epsilon }
+\frac{5 C_F \pi ^2 \mathcal{L}_Q}{2 \epsilon }
-\frac{201881 C_A C_F^2}{864 \epsilon }
-\frac{631 C_A C_F}{864 \epsilon }
+\frac{283861 C_F}{576 \epsilon }
+\frac{43897 C_A C_F^2}{144 \epsilon ^2}
-\frac{25 C_A C_F}{144 \epsilon ^2}
+\frac{20131 C_F}{96 \epsilon ^2}
+\frac{1783 C_A C_F^2}{24 \epsilon ^3}
-\frac{C_A C_F}{24 \epsilon ^3}
+\frac{269 C_F}{16 \epsilon ^3}
-\frac{10 C_A C_F^2}{\epsilon ^4}
-\frac{3 C_F}{\epsilon ^4}
-\frac{581 C_F \zeta_3}{3}
+\frac{166}{3} C_F \mathcal{L}_Q \zeta_3
-\frac{83 C_F \zeta_3}{3 \epsilon }
-\frac{59 C_F \pi ^4}{120}
+\frac{1999}{864} C_A C_F^2 \pi ^2
+\frac{5}{864} C_A C_F \pi ^2
-\frac{20507 C_F \pi ^2}{576}
+\frac{427 C_A C_F^2 \pi ^2}{144 \epsilon }
-\frac{C_A C_F \pi ^2}{144 \epsilon }
-\frac{535 C_F \pi ^2}{96 \epsilon }
+\frac{C_A C_F^2 \pi ^2}{2 \epsilon ^2}
-\frac{5 C_F \pi ^2}{4 \epsilon ^2}
\end{autobreak}  
\end{align}
\end{tiny}

\begin{tiny}
\begin{align}
\begin{autobreak}
V^{(Q)}_3=
\frac{1}{192} C_A \mathcal{L}_Q^4 Y_f^4
-\frac{9}{32} C_A \mathcal{L}_Q^3 Y_f^4
-\frac{C_A \mathcal{L}_Q^3 Y_f^4}{96 \epsilon }
+\frac{201}{256} C_A \mathcal{L}_Q^2 Y_f^4
+\frac{27 C_A \mathcal{L}_Q^2 Y_f^4}{64 \epsilon }
+\frac{C_A \mathcal{L}_Q^2 Y_f^4}{64 \epsilon ^2}
-\frac{1075 C_A Y_f^4}{512}
-\frac{99}{128} C_A \mathcal{L}_Q Y_f^4
-\frac{201 C_A \mathcal{L}_Q Y_f^4}{256 \epsilon }
-\frac{27 C_A \mathcal{L}_Q Y_f^4}{64 \epsilon ^2}
-\frac{C_A \mathcal{L}_Q Y_f^4}{64 \epsilon ^3}
+\frac{1}{48} C_A \pi ^2 \mathcal{L}_Q Y_f^4
+\frac{99 C_A Y_f^4}{256 \epsilon }
+\frac{201 C_A Y_f^4}{512 \epsilon ^2}
+\frac{27 C_A Y_f^4}{128 \epsilon ^3}
+\frac{C_A Y_f^4}{128 \epsilon ^4}
-\frac{83}{192} C_A \zeta_3 Y_f^4
-\frac{1}{24} C_A \pi ^2 Y_f^4
-\frac{C_A \pi ^2 Y_f^4}{96 \epsilon }
+\frac{3}{8} C_A C_F \mathcal{L}_Q^4 Y_f^2
-\frac{33}{4} C_A C_F \mathcal{L}_Q^3 Y_f^2
-\frac{3 C_A C_F \mathcal{L}_Q^3 Y_f^2}{4 \epsilon }
+\frac{19}{4} C_A C_F^2 Y_f^2
+\frac{1}{2} C_A C_F^2 \mathcal{L}_Q^2 Y_f^2
-\frac{209}{32} C_A C_F \mathcal{L}_Q^2 Y_f^2
+\frac{1}{4} C_F \mathcal{L}_Q^2 Y_f^2
+\frac{99 C_A C_F \mathcal{L}_Q^2 Y_f^2}{8 \epsilon }
+\frac{9 C_A C_F \mathcal{L}_Q^2 Y_f^2}{8 \epsilon ^2}
+\frac{1}{8} C_A C_F \pi ^2 \mathcal{L}_Q^2 Y_f^2
-\frac{8079}{64} C_A C_F Y_f^2
+\frac{19 C_F Y_f^2}{8}
-\frac{5}{2} C_A C_F^2 \mathcal{L}_Q Y_f^2
+\frac{809}{4} C_A C_F \mathcal{L}_Q Y_f^2
-\frac{5}{4} C_F \mathcal{L}_Q Y_f^2
-\frac{C_A C_F^2 \mathcal{L}_Q Y_f^2}{2 \epsilon }
+\frac{209 C_A C_F \mathcal{L}_Q Y_f^2}{32 \epsilon }
-\frac{C_F \mathcal{L}_Q Y_f^2}{4 \epsilon }
-\frac{99 C_A C_F \mathcal{L}_Q Y_f^2}{8 \epsilon ^2}
-\frac{9 C_A C_F \mathcal{L}_Q Y_f^2}{8 \epsilon ^3}-2 C_A C_F \pi ^2 \mathcal{L}_Q Y_f^2
-\frac{C_A C_F \pi ^2 \mathcal{L}_Q Y_f^2}{8 \epsilon }
+\frac{5 C_A C_F^2 Y_f^2}{4 \epsilon }
-\frac{809 C_A C_F Y_f^2}{8 \epsilon }
+\frac{5 C_F Y_f^2}{8 \epsilon }
+\frac{C_A C_F^2 Y_f^2}{4 \epsilon ^2}
-\frac{209 C_A C_F Y_f^2}{64 \epsilon ^2}
+\frac{C_F Y_f^2}{8 \epsilon ^2}
+\frac{99 C_A C_F Y_f^2}{16 \epsilon ^3}
+\frac{9 C_A C_F Y_f^2}{16 \epsilon ^4}
+\frac{581}{24} C_A C_F \zeta_3 Y_f^2
+\frac{1}{24} C_A C_F^2 \pi ^2 Y_f^2
+\frac{101}{24} C_A C_F \pi ^2 Y_f^2
+\frac{1}{48} C_F \pi ^2 Y_f^2
+\frac{C_A C_F \pi ^2 Y_f^2}{\epsilon }
+\frac{C_A C_F \pi ^2 Y_f^2}{16 \epsilon ^2}
-\frac{19}{3} C_A C_F^2 \mathcal{L}_Q^4-3 C_F \mathcal{L}_Q^4
-\frac{665}{6} C_A C_F^2 \mathcal{L}_Q^3
+\frac{1}{18} C_A C_F \mathcal{L}_Q^3
-\frac{149}{12} C_F \mathcal{L}_Q^3
-\frac{2}{9} C_A C_F n_f T_f \mathcal{L}_Q^3
+\frac{38 C_A C_F^2 \mathcal{L}_Q^3}{3 \epsilon }
+\frac{6 C_F \mathcal{L}_Q^3}{\epsilon }
-\frac{5689}{54} C_A C_F^2
+\frac{6377}{12} C_A C_F^2 \mathcal{L}_Q^2
-\frac{2}{9} C_A C_F \mathcal{L}_Q^2
+\frac{1144}{3} C_F \mathcal{L}_Q^2
-\frac{1}{9} C_A C_F n_f T_f \mathcal{L}_Q^2
+\frac{C_A C_F n_f T_f \mathcal{L}_Q^2}{3 \epsilon }
+\frac{665 C_A C_F^2 \mathcal{L}_Q^2}{4 \epsilon }
-\frac{C_A C_F \mathcal{L}_Q^2}{12 \epsilon }
+\frac{149 C_F \mathcal{L}_Q^2}{8 \epsilon }
-\frac{19 C_A C_F^2 \mathcal{L}_Q^2}{\epsilon ^2}
-\frac{9 C_F \mathcal{L}_Q^2}{\epsilon ^2}+C_A C_F^2 \pi ^2 \mathcal{L}_Q^2
-\frac{17}{6} C_F \pi ^2 \mathcal{L}_Q^2
+\frac{1129 C_A C_F}{648}
+\frac{267173 C_F}{108}
-\frac{3235}{162} C_A C_F n_f T_f
-\frac{46 C_A C_F n_f T_f}{27 \epsilon }
-\frac{C_A C_F n_f T_f}{18 \epsilon ^2}
+\frac{C_A C_F n_f T_f}{6 \epsilon ^3}
-\frac{4}{27} C_A C_F n_f \pi ^2 T_f
+\frac{C_A C_F n_f \pi ^2 T_f}{36 \epsilon }
+\frac{7088}{9} C_A C_F^2 \mathcal{L}_Q
+\frac{35}{54} C_A C_F \mathcal{L}_Q
-\frac{71539}{72} C_F \mathcal{L}_Q
+\frac{92}{27} C_A C_F n_f T_f \mathcal{L}_Q
+\frac{C_A C_F n_f T_f \mathcal{L}_Q}{9 \epsilon }
-\frac{C_A C_F n_f T_f \mathcal{L}_Q}{3 \epsilon ^2}
-\frac{1}{18} C_A C_F n_f \pi ^2 T_f \mathcal{L}_Q
-\frac{6377 C_A C_F^2 \mathcal{L}_Q}{12 \epsilon }
+\frac{2 C_A C_F \mathcal{L}_Q}{9 \epsilon }
-\frac{1144 C_F \mathcal{L}_Q}{3 \epsilon }
-\frac{665 C_A C_F^2 \mathcal{L}_Q}{4 \epsilon ^2}
+\frac{C_A C_F \mathcal{L}_Q}{12 \epsilon ^2}
-\frac{149 C_F \mathcal{L}_Q}{8 \epsilon ^2}
+\frac{19 C_A C_F^2 \mathcal{L}_Q}{\epsilon ^3}
+\frac{9 C_F \mathcal{L}_Q}{\epsilon ^3}
-\frac{55}{8} C_A C_F^2 \pi ^2 \mathcal{L}_Q
+\frac{1}{72} C_A C_F \pi ^2 \mathcal{L}_Q
+\frac{205}{16} C_F \pi ^2 \mathcal{L}_Q
-\frac{C_A C_F^2 \pi ^2 \mathcal{L}_Q}{\epsilon }
+\frac{17 C_F \pi ^2 \mathcal{L}_Q}{6 \epsilon }
-\frac{3544 C_A C_F^2}{9 \epsilon }
-\frac{35 C_A C_F}{108 \epsilon }
+\frac{71539 C_F}{144 \epsilon }
+\frac{6377 C_A C_F^2}{24 \epsilon ^2}
-\frac{C_A C_F}{9 \epsilon ^2}
+\frac{572 C_F}{3 \epsilon ^2}
+\frac{665 C_A C_F^2}{8 \epsilon ^3}
-\frac{C_A C_F}{24 \epsilon ^3}
+\frac{149 C_F}{16 \epsilon ^3}
-\frac{19 C_A C_F^2}{2 \epsilon ^4}
-\frac{9 C_F}{2 \epsilon ^4}
-\frac{664 C_F \zeta_3}{3}
+\frac{166}{3} C_F \mathcal{L}_Q \zeta_3
-\frac{83 C_F \zeta_3}{3 \epsilon }
-\frac{59 C_F \pi ^4}{120}
+\frac{413}{144} C_A C_F^2 \pi ^2
+\frac{7}{432} C_A C_F \pi ^2
-\frac{11347 C_F \pi ^2}{288}
+\frac{55 C_A C_F^2 \pi ^2}{16 \epsilon }
-\frac{C_A C_F \pi ^2}{144 \epsilon }
-\frac{205 C_F \pi ^2}{32 \epsilon }
+\frac{C_A C_F^2 \pi ^2}{2 \epsilon ^2}
-\frac{17 C_F \pi ^2}{12 \epsilon ^2}
\end{autobreak}
\end{align}
\end{tiny}

\begin{tiny}
\begin{align}
\begin{autobreak}
V^{(Q)}_4=
-\frac{4 n_f C_A C_F T_f \mathcal{L}_Q^2}{3 \epsilon }
+\frac{32 n_f C_A C_F T_f \mathcal{L}_Q}{9 \epsilon }
+\frac{4 n_f C_A C_F T_f \mathcal{L}_Q}{3 \epsilon ^2}
+\frac{8}{9} n_f C_A C_F T_f \mathcal{L}_Q^3
-\frac{32}{9} n_f C_A C_F T_f \mathcal{L}_Q^2
+\frac{280}{27} n_f C_A C_F T_f \mathcal{L}_Q
+\frac{2}{9} \pi ^2 n_f C_A C_F T_f \mathcal{L}_Q
-\frac{140 n_f C_A C_F T_f}{27 \epsilon }
-\frac{16 n_f C_A C_F T_f}{9 \epsilon ^2}
-\frac{2 n_f C_A C_F T_f}{3 \epsilon ^3}
-\frac{\pi ^2 n_f C_A C_F T_f}{9 \epsilon }
+\frac{2258}{81} n_f C_A C_F T_f
+\frac{7}{27} \pi ^2 n_f C_A C_F T_f
+\frac{1579 C_A C_F^2}{72 \epsilon }
-\frac{2047 C_A C_F}{216 \epsilon }
+\frac{677 C_A C_F^2}{24 \epsilon ^2}
-\frac{155 C_A C_F}{72 \epsilon ^2}
+\frac{69 C_A C_F^2}{8 \epsilon ^3}
-\frac{C_A C_F}{24 \epsilon ^3}
-\frac{8 C_A C_F^2}{\epsilon ^4}
-\frac{59 \pi ^2 C_A C_F^2}{48 \epsilon }
-\frac{\pi ^2 C_A C_F}{144 \epsilon }
+\frac{\pi ^2 C_A C_F^2}{2 \epsilon ^2}
+\frac{32 C_A C_F^2 \mathcal{L}_Q^3}{3 \epsilon }
+\frac{69 C_A C_F^2 \mathcal{L}_Q^2}{4 \epsilon }
-\frac{C_A C_F \mathcal{L}_Q^2}{12 \epsilon }
-\frac{16 C_A C_F^2 \mathcal{L}_Q^2}{\epsilon ^2}
-\frac{677 C_A C_F^2 \mathcal{L}_Q}{12 \epsilon }
+\frac{155 C_A C_F \mathcal{L}_Q}{36 \epsilon }
-\frac{69 C_A C_F^2 \mathcal{L}_Q}{4 \epsilon ^2}
+\frac{C_A C_F \mathcal{L}_Q}{12 \epsilon ^2}
+\frac{16 C_A C_F^2 \mathcal{L}_Q}{\epsilon ^3}
-\frac{\pi ^2 C_A C_F^2 \mathcal{L}_Q}{\epsilon }
-\frac{16}{3} C_A C_F^2 \mathcal{L}_Q^4
-\frac{23}{2} C_A C_F^2 \mathcal{L}_Q^3
+\frac{1}{18} C_A C_F \mathcal{L}_Q^3
+\frac{677}{12} C_A C_F^2 \mathcal{L}_Q^2
-\frac{155}{36} C_A C_F \mathcal{L}_Q^2+\pi ^2 C_A C_F^2 \mathcal{L}_Q^2
-\frac{1579}{36} C_A C_F^2 \mathcal{L}_Q
+\frac{2047}{108} C_A C_F \mathcal{L}_Q
+\frac{59}{24} \pi ^2 C_A C_F^2 \mathcal{L}_Q
+\frac{1}{72} \pi ^2 C_A C_F \mathcal{L}_Q
-\frac{3823}{54} C_A C_F^2
-\frac{5141 C_A C_F}{162}
+\frac{29}{36} \pi ^2 C_A C_F^2
-\frac{35}{108} \pi ^2 C_A C_F
+\frac{20767 C_F}{144 \epsilon }
-\frac{1201 C_F}{48 \epsilon ^2}
+\frac{253 C_F}{16 \epsilon ^3}
+\frac{6 C_F}{\epsilon ^4}
-\frac{81 \pi ^2 C_F}{32 \epsilon }
-\frac{2 \pi ^2 C_F}{3 \epsilon ^2}
-\frac{8 C_F \mathcal{L}_Q^3}{\epsilon }
+\frac{253 C_F \mathcal{L}_Q^2}{8 \epsilon }
+\frac{12 C_F \mathcal{L}_Q^2}{\epsilon ^2}
+\frac{1201 C_F \mathcal{L}_Q}{24 \epsilon }
-\frac{253 C_F \mathcal{L}_Q}{8 \epsilon ^2}
-\frac{12 C_F \mathcal{L}_Q}{\epsilon ^3}
+\frac{4 \pi ^2 C_F \mathcal{L}_Q}{3 \epsilon }
+\frac{166}{3} \zeta_3 C_F \mathcal{L}_Q+4 C_F \mathcal{L}_Q^4
-\frac{253}{12} C_F \mathcal{L}_Q^3
-\frac{1201}{24} C_F \mathcal{L}_Q^2
-\frac{4}{3} \pi ^2 C_F \mathcal{L}_Q^2
-\frac{20767}{72} C_F \mathcal{L}_Q
+\frac{81}{16} \pi ^2 C_F \mathcal{L}_Q
-\frac{83 \zeta_3 C_F}{3 \epsilon }
-\frac{166 \zeta_3 C_F}{3}
+\frac{32033 C_F}{108}
-\frac{59 \pi ^4 C_F}{120}
-\frac{269 \pi ^2 C_F}{36}
\end{autobreak}
\end{align}
\end{tiny}

\begin{tiny}
\begin{align}
\begin{autobreak}
V^{(Q)}_5=
-\frac{4 n_f C_A C_F T_f \mathcal{L}_Q^2}{3 \epsilon }
+\frac{68 n_f C_A C_F T_f \mathcal{L}_Q}{9 \epsilon }
+\frac{4 n_f C_A C_F T_f \mathcal{L}_Q}{3 \epsilon ^2}
+\frac{8}{9} n_f C_A C_F T_f \mathcal{L}_Q^3
-\frac{68}{9} n_f C_A C_F T_f \mathcal{L}_Q^2
+\frac{802}{27} n_f C_A C_F T_f \mathcal{L}_Q
+\frac{2}{9} \pi ^2 n_f C_A C_F T_f \mathcal{L}_Q
-\frac{401 n_f C_A C_F T_f}{27 \epsilon }
-\frac{34 n_f C_A C_F T_f}{9 \epsilon ^2}
-\frac{2 n_f C_A C_F T_f}{3 \epsilon ^3}
-\frac{\pi ^2 n_f C_A C_F T_f}{9 \epsilon }
-\frac{1361}{162} n_f C_A C_F T_f
-\frac{2}{27} \pi ^2 n_f C_A C_F T_f
+\frac{7649 C_A C_F^2}{144 \epsilon }
-\frac{953 C_A C_F}{432 \epsilon }
+\frac{301 C_A C_F^2}{12 \epsilon ^2}
-\frac{5 C_A C_F}{18 \epsilon ^2}
-\frac{69 C_A C_F^2}{\epsilon ^3}
+\frac{477 C_A^2 C_F}{16 \epsilon ^3}
-\frac{C_A C_F}{24 \epsilon ^3}
-\frac{20 C_A C_F^2}{\epsilon ^4}
+\frac{6 C_A^2 C_F}{\epsilon ^4}
+\frac{37 \pi ^2 C_A C_F^2}{48 \epsilon }
-\frac{\pi ^2 C_A C_F}{144 \epsilon }
+\frac{\pi ^2 C_A C_F^2}{2 \epsilon ^2}
+\frac{32 C_A C_F^2 \mathcal{L}_Q^3}{3 \epsilon }
-\frac{75 C_A C_F^2 \mathcal{L}_Q^2}{4 \epsilon }
-\frac{C_A C_F \mathcal{L}_Q^2}{12 \epsilon }
-\frac{16 C_A C_F^2 \mathcal{L}_Q^2}{\epsilon ^2}
-\frac{301 C_A C_F^2 \mathcal{L}_Q}{6 \epsilon }
+\frac{5 C_A C_F \mathcal{L}_Q}{9 \epsilon }
+\frac{75 C_A C_F^2 \mathcal{L}_Q}{4 \epsilon ^2}
+\frac{C_A C_F \mathcal{L}_Q}{12 \epsilon ^2}
+\frac{40 C_A C_F^2 \mathcal{L}_Q}{\epsilon ^3}
-\frac{12 C_A^2 C_F \mathcal{L}_Q}{\epsilon ^3}
-\frac{\pi ^2 C_A C_F^2 \mathcal{L}_Q}{\epsilon }
-\frac{16}{3} C_A C_F^2 \mathcal{L}_Q^4
+\frac{25}{2} C_A C_F^2 \mathcal{L}_Q^3
+\frac{1}{18} C_A C_F \mathcal{L}_Q^3
+\frac{301}{6} C_A C_F^2 \mathcal{L}_Q^2
-\frac{5}{9} C_A C_F \mathcal{L}_Q^2+\pi ^2 C_A C_F^2 \mathcal{L}_Q^2
-\frac{7649}{72} C_A C_F^2 \mathcal{L}_Q
+\frac{953}{216} C_A C_F \mathcal{L}_Q
-\frac{37}{24} \pi ^2 C_A C_F^2 \mathcal{L}_Q
+\frac{1}{72} \pi ^2 C_A C_F \mathcal{L}_Q
+\frac{62051}{864} C_A C_F^2
-\frac{20039 C_A C_F}{2592}
+\frac{41}{144} \pi ^2 C_A C_F^2
-\frac{5}{432} \pi ^2 C_A C_F
+\frac{59201 C_F}{288 \epsilon }
+\frac{116 C_F}{3 \epsilon ^2}
-\frac{355 \pi ^2 C_F}{96 \epsilon }
-\frac{2 \pi ^2 C_F}{3 \epsilon ^2}
-\frac{8 C_F \mathcal{L}_Q^3}{\epsilon }
+\frac{477 C_F \mathcal{L}_Q^2}{8 \epsilon }
+\frac{12 C_F \mathcal{L}_Q^2}{\epsilon ^2}
-\frac{232 C_F \mathcal{L}_Q}{3 \epsilon }
-\frac{477 C_F \mathcal{L}_Q}{8 \epsilon ^2}
+\frac{4 \pi ^2 C_F \mathcal{L}_Q}{3 \epsilon }
+\frac{166}{3} \zeta_3 C_F \mathcal{L}_Q+4 C_F \mathcal{L}_Q^4
-\frac{159}{4} C_F \mathcal{L}_Q^3
+\frac{232}{3} C_F \mathcal{L}_Q^2
-\frac{4}{3} \pi ^2 C_F \mathcal{L}_Q^2
-\frac{59201}{144} C_F \mathcal{L}_Q
+\frac{355}{48} \pi ^2 C_F \mathcal{L}_Q
-\frac{83 \zeta_3 C_F}{3 \epsilon }
-\frac{332 \zeta_3 C_F}{3}
+\frac{1265063 C_F}{1728}
-\frac{59 \pi ^4 C_F}{120}
-\frac{4375 \pi ^2 C_F}{288}
\end{autobreak}
\end{align}
\end{tiny}

\begin{tiny}
\begin{align}
\begin{autobreak}
V^{(Q)}_6=
\frac{11}{8} C_A^3 \mathcal{L}_Q^4
-\frac{1}{48} C_A^2 Y_f^2 \mathcal{L}_Q^4
+\frac{1}{48} Y_f^2 \mathcal{L}_Q^4
-\frac{11}{8} C_A \mathcal{L}_Q^4
-\frac{35}{6} C_A^2 C_F \mathcal{L}_Q^4
+\frac{35}{6} C_F \mathcal{L}_Q^4
-\frac{427}{24} C_A^3 \mathcal{L}_Q^3
+\frac{1}{36} C_A^2 \mathcal{L}_Q^3
-\frac{7}{24} C_A^2 Y_f^2 \mathcal{L}_Q^3
+\frac{C_A^2 Y_f^2 \mathcal{L}_Q^3}{24 \epsilon }
-\frac{Y_f^2 \mathcal{L}_Q^3}{24 \epsilon }
+\frac{7}{24} Y_f^2 \mathcal{L}_Q^3
+\frac{427}{24} C_A \mathcal{L}_Q^3
+\frac{58}{3} C_A^2 C_F \mathcal{L}_Q^3
-\frac{58}{3} C_F \mathcal{L}_Q^3
-\frac{11 C_A^3 \mathcal{L}_Q^3}{4 \epsilon }
+\frac{11 C_A \mathcal{L}_Q^3}{4 \epsilon }
+\frac{35 C_A^2 C_F \mathcal{L}_Q^3}{3 \epsilon }
-\frac{35 C_F \mathcal{L}_Q^3}{3 \epsilon }
-\frac{\mathcal{L}_Q^3}{36}
+\frac{4543}{96} C_A^3 \mathcal{L}_Q^2
-\frac{7}{36} C_A^2 \mathcal{L}_Q^2
+\frac{107}{64} C_A^2 Y_f^2 \mathcal{L}_Q^2
+\frac{7 C_A^2 Y_f^2 \mathcal{L}_Q^2}{16 \epsilon }
-\frac{7 Y_f^2 \mathcal{L}_Q^2}{16 \epsilon }
-\frac{C_A^2 Y_f^2 \mathcal{L}_Q^2}{16 \epsilon ^2}
+\frac{Y_f^2 \mathcal{L}_Q^2}{16 \epsilon ^2}
-\frac{107}{64} Y_f^2 \mathcal{L}_Q^2
-\frac{4543}{96} C_A \mathcal{L}_Q^2
+\frac{103}{8} C_A^2 C_F \mathcal{L}_Q^2
-\frac{103}{8} C_F \mathcal{L}_Q^2
-\frac{1}{3} C_A^2 n_f T_f \mathcal{L}_Q^2
+\frac{1}{3} n_f T_f \mathcal{L}_Q^2
+\frac{427 C_A^3 \mathcal{L}_Q^2}{16 \epsilon }
-\frac{C_A^2 \mathcal{L}_Q^2}{24 \epsilon }
-\frac{427 C_A \mathcal{L}_Q^2}{16 \epsilon }
-\frac{29 C_A^2 C_F \mathcal{L}_Q^2}{\epsilon }
+\frac{29 C_F \mathcal{L}_Q^2}{\epsilon }
+\frac{\mathcal{L}_Q^2}{24 \epsilon }
+\frac{33 C_A^3 \mathcal{L}_Q^2}{8 \epsilon ^2}
-\frac{33 C_A \mathcal{L}_Q^2}{8 \epsilon ^2}
-\frac{35 C_A^2 C_F \mathcal{L}_Q^2}{2 \epsilon ^2}
+\frac{35 C_F \mathcal{L}_Q^2}{2 \epsilon ^2}
-\frac{3}{4} C_A^3 \pi ^2 \mathcal{L}_Q^2
+\frac{3}{4} C_A \pi ^2 \mathcal{L}_Q^2+2 C_A^2 C_F \pi ^2 \mathcal{L}_Q^2-2 C_F \pi ^2 \mathcal{L}_Q^2
+\frac{7 \mathcal{L}_Q^2}{36}
-\frac{9865}{144} C_A^3 \mathcal{L}_Q
+\frac{77}{108} C_A^2 \mathcal{L}_Q
+\frac{623}{64} C_A^2 Y_f^2 \mathcal{L}_Q
-\frac{107 C_A^2 Y_f^2 \mathcal{L}_Q}{64 \epsilon }
+\frac{107 Y_f^2 \mathcal{L}_Q}{64 \epsilon }
-\frac{7 C_A^2 Y_f^2 \mathcal{L}_Q}{16 \epsilon ^2}
+\frac{7 Y_f^2 \mathcal{L}_Q}{16 \epsilon ^2}
+\frac{C_A^2 Y_f^2 \mathcal{L}_Q}{16 \epsilon ^3}
-\frac{Y_f^2 \mathcal{L}_Q}{16 \epsilon ^3}
-\frac{1}{32} C_A^2 \pi ^2 Y_f^2 \mathcal{L}_Q
+\frac{1}{32} \pi ^2 Y_f^2 \mathcal{L}_Q
-\frac{623}{64} Y_f^2 \mathcal{L}_Q
+\frac{9865}{144} C_A \mathcal{L}_Q
+\frac{1153}{4} C_A^2 C_F \mathcal{L}_Q
-\frac{1153}{4} C_F \mathcal{L}_Q
+\frac{11}{9} C_A^2 n_f T_f \mathcal{L}_Q
+\frac{C_A^2 n_f T_f \mathcal{L}_Q}{3 \epsilon }
-\frac{n_f T_f \mathcal{L}_Q}{3 \epsilon }
-\frac{11}{9} n_f T_f \mathcal{L}_Q
-\frac{4543 C_A^3 \mathcal{L}_Q}{96 \epsilon }
+\frac{7 C_A^2 \mathcal{L}_Q}{36 \epsilon }
+\frac{4543 C_A \mathcal{L}_Q}{96 \epsilon }
-\frac{103 C_A^2 C_F \mathcal{L}_Q}{8 \epsilon }
+\frac{103 C_F \mathcal{L}_Q}{8 \epsilon }
-\frac{7 \mathcal{L}_Q}{36 \epsilon }
-\frac{427 C_A^3 \mathcal{L}_Q}{16 \epsilon ^2}
+\frac{C_A^2 \mathcal{L}_Q}{24 \epsilon ^2}
+\frac{427 C_A \mathcal{L}_Q}{16 \epsilon ^2}
+\frac{29 C_A^2 C_F \mathcal{L}_Q}{\epsilon ^2}
-\frac{29 C_F \mathcal{L}_Q}{\epsilon ^2}
-\frac{\mathcal{L}_Q}{24 \epsilon ^2}
-\frac{33 C_A^3 \mathcal{L}_Q}{8 \epsilon ^3}
+\frac{33 C_A \mathcal{L}_Q}{8 \epsilon ^3}
+\frac{35 C_A^2 C_F \mathcal{L}_Q}{2 \epsilon ^3}
-\frac{35 C_F \mathcal{L}_Q}{2 \epsilon ^3}
+\frac{83}{3} C_A^3 \zeta_3 \mathcal{L}_Q
-\frac{83}{3} C_A \zeta_3 \mathcal{L}_Q
-\frac{166}{3} C_A^2 C_F \zeta_3 \mathcal{L}_Q
+\frac{166}{3} C_F \zeta_3 \mathcal{L}_Q
+\frac{81}{32} C_A^3 \pi ^2 \mathcal{L}_Q
+\frac{1}{144} C_A^2 \pi ^2 \mathcal{L}_Q
-\frac{81}{32} C_A \pi ^2 \mathcal{L}_Q
-\frac{71}{12} C_A^2 C_F \pi ^2 \mathcal{L}_Q
+\frac{71}{12} C_F \pi ^2 \mathcal{L}_Q
+\frac{3 C_A^3 \pi ^2 \mathcal{L}_Q}{4 \epsilon }
-\frac{3 C_A \pi ^2 \mathcal{L}_Q}{4 \epsilon }
-\frac{2 C_A^2 C_F \pi ^2 \mathcal{L}_Q}{\epsilon }
+\frac{2 C_F \pi ^2 \mathcal{L}_Q}{\epsilon }
-\frac{1}{144} \pi ^2 \mathcal{L}_Q
-\frac{77 \mathcal{L}_Q}{108}
+\frac{684445 C_A^3}{1728}
+\frac{205 C_A^2}{1296}
+\frac{1147}{128} C_A^2 Y_f^2
-\frac{623 C_A^2 Y_f^2}{128 \epsilon }
+\frac{623 Y_f^2}{128 \epsilon }
+\frac{107 C_A^2 Y_f^2}{128 \epsilon ^2}
-\frac{107 Y_f^2}{128 \epsilon ^2}
+\frac{7 C_A^2 Y_f^2}{32 \epsilon ^3}
-\frac{7 Y_f^2}{32 \epsilon ^3}
-\frac{C_A^2 Y_f^2}{32 \epsilon ^4}
+\frac{Y_f^2}{32 \epsilon ^4}
-\frac{7}{96} C_A^2 \pi ^2 Y_f^2
+\frac{C_A^2 \pi ^2 Y_f^2}{64 \epsilon }
-\frac{\pi ^2 Y_f^2}{64 \epsilon }
+\frac{7 \pi ^2 Y_f^2}{96}
-\frac{1147 Y_f^2}{128}
-\frac{684445 C_A}{1728}
-\frac{5179}{8} C_A^2 C_F
+\frac{5179 C_F}{8}
-\frac{56}{27} C_A^2 n_f T_f
-\frac{11 C_A^2 n_f T_f}{18 \epsilon }
+\frac{11 n_f T_f}{18 \epsilon }
-\frac{C_A^2 n_f T_f}{6 \epsilon ^2}
+\frac{n_f T_f}{6 \epsilon ^2}
+\frac{56 n_f T_f}{27}
-\frac{1}{36} C_A^2 n_f \pi ^2 T_f
+\frac{1}{36} n_f \pi ^2 T_f
+\frac{9865 C_A^3}{288 \epsilon }
-\frac{77 C_A^2}{216 \epsilon }
-\frac{9865 C_A}{288 \epsilon }
-\frac{1153 C_A^2 C_F}{8 \epsilon }
+\frac{1153 C_F}{8 \epsilon }
+\frac{77}{216 \epsilon }
+\frac{4543 C_A^3}{192 \epsilon ^2}
-\frac{7 C_A^2}{72 \epsilon ^2}
-\frac{4543 C_A}{192 \epsilon ^2}
+\frac{103 C_A^2 C_F}{16 \epsilon ^2}
-\frac{103 C_F}{16 \epsilon ^2}
+\frac{7}{72 \epsilon ^2}
+\frac{427 C_A^3}{32 \epsilon ^3}
-\frac{C_A^2}{48 \epsilon ^3}
-\frac{427 C_A}{32 \epsilon ^3}
-\frac{29 C_A^2 C_F}{2 \epsilon ^3}
+\frac{29 C_F}{2 \epsilon ^3}
+\frac{1}{48 \epsilon ^3}
+\frac{33 C_A^3}{16 \epsilon ^4}
-\frac{33 C_A}{16 \epsilon ^4}
-\frac{35 C_A^2 C_F}{4 \epsilon ^4}
+\frac{35 C_F}{4 \epsilon ^4}
-\frac{83 C_A^3 \zeta_3}{2}
+\frac{83 C_A \zeta_3}{2}+83 C_A^2 C_F \zeta_3-83 C_F \zeta_3
-\frac{83 C_A^3 \zeta_3}{6 \epsilon }
+\frac{83 C_A \zeta_3}{6 \epsilon }
+\frac{83 C_A^2 C_F \zeta_3}{3 \epsilon }
-\frac{83 C_F \zeta_3}{3 \epsilon }
-\frac{59}{240} C_A^3 \pi ^4
+\frac{59 C_A \pi ^4}{240}
+\frac{59}{120} C_A^2 C_F \pi ^4
-\frac{59 C_F \pi ^4}{120}
-\frac{3829}{576} C_A^3 \pi ^2
+\frac{1}{864} C_A^2 \pi ^2
+\frac{3829 C_A \pi ^2}{576}+12 C_A^2 C_F \pi ^2-12 C_F \pi ^2
-\frac{81 C_A^3 \pi ^2}{64 \epsilon }
-\frac{C_A^2 \pi ^2}{288 \epsilon }
+\frac{81 C_A \pi ^2}{64 \epsilon }
+\frac{71 C_A^2 C_F \pi ^2}{24 \epsilon }
-\frac{71 C_F \pi ^2}{24 \epsilon }
+\frac{\pi ^2}{288 \epsilon }
-\frac{3 C_A^3 \pi ^2}{8 \epsilon ^2}
+\frac{3 C_A \pi ^2}{8 \epsilon ^2}
+\frac{C_A^2 C_F \pi ^2}{\epsilon ^2}
-\frac{C_F \pi ^2}{\epsilon ^2}
-\frac{\pi ^2}{864}
-\frac{205}{1296}
\end{autobreak}
\end{align}
\end{tiny}

\subsection{Matching at $\mu\sim m_{1,2}$:}
\begin{tiny}
\begin{align}
\begin{autobreak}
V^{(m_{1,2})}_1=
\frac{n_f C_A C_F T_f \mathcal{L}_{m_2}}{2 \varepsilon }-
\frac{1}{2} n_f C_A C_F T_f \mathcal{L}_{m_2}^2-
\frac{n_f C_A C_FT_f}{4 \varepsilon ^2}-
\frac{5}{4} n_f \zeta_2 C_A C_F T_f-
\frac{1}{2} n_f C_A C_F T_f-
\frac{45 C_A C_F^2}{8\varepsilon }+
\frac{3 C_A C_F}{4 \varepsilon }+
\frac{23 C_A C_F^2}{8 \varepsilon ^2}+
\frac{C_A C_F}{4 \varepsilon ^2}-
\frac{7 C_AC_F^2}{8 \varepsilon ^3}-
\frac{7 C_A C_F^2 \mathcal{L}_{m_2}^2}{4 \varepsilon }-
\frac{23 C_A C_F^2 \mathcal{L}_{m_2}}{4 \varepsilon}-
\frac{C_A C_F \mathcal{L}_{m_2}}{2 \varepsilon }+
\frac{7 C_A C_F^2 \mathcal{L}_{m_2}}{4 \varepsilon ^2}-
\frac{6 Y_f^2 C_A C_F \mathcal{L}_{m_2}^2}{\varepsilon }-
\frac{3 Y_f^2 C_A C_F^2 \mathcal{L}_{m_2}}{2 \varepsilon }-
\frac{Y_f^2 C_A C_F \mathcal{L}_{m_2}}{\varepsilon }+
\frac{6 Y_f^2 C_A C_F \mathcal{L}_{m_2}}{\varepsilon ^2}+
\frac{3}{2} Y_f^2 C_A C_F^2 \mathcal{L}_{m_2}^2+16 Y_f^2 C_A C_F \mathcal{L}_{m_2}^2-
\frac{13}{2} Y_f^2 C_A C_F^2 \mathcal{L}_{m_2}-
\frac{21}{2} Y_f^2 C_A C_F \mathcal{L}_{m_2}+
\frac{81}{8} C_A C_F^2 \mathcal{L}_{m_2}^2+
\frac{1}{2} C_A C_F \mathcal{L}_{m_2}^2+
\frac{103}{16} C_A C_F^2 \mathcal{L}_{m_2}-
\frac{3}{2} C_A C_F \mathcal{L}_{m_2}+
\frac{13 Y_f^2 C_A C_F^2}{4 \varepsilon }-
\frac{3 Y_f^2 C_A C_F}{\varepsilon }+
\frac{3 Y_f^2 C_A C_F^2}{4 \varepsilon ^2}+
\frac{Y_f^2 C_A C_F}{2 \varepsilon ^2}-
\frac{3 Y_f^2 C_A C_F}{\varepsilon ^3}-
\frac{15 Y_f^2 \zeta_2 C_A C_F}{\varepsilon }+
\frac{15}{4} Y_f^2 \zeta_2 C_A C_F^2+40 Y_f^2 \zeta_2 C_A C_F+
\frac{47}{4} Y_f^2 C_A C_F^2-
\frac{13}{8} Y_f^2 C_A C_F-
\frac{35 \zeta_2 C_A C_F^2}{8 \varepsilon }+
\frac{405}{16} \zeta_2 C_A C_F^2+
\frac{5}{4} \zeta_2 C_A C_F-
\frac{3977}{64} C_A C_F^2+
\frac{9 C_A C_F}{4}+
\frac{3 Y_f^4 C_A \mathcal{L}_{m_2}^2}{4 \varepsilon }-
\frac{3 Y_f^4 C_A \mathcal{L}_{m_2}}{4 \varepsilon ^2}-
\frac{15}{8} Y_f^4 C_A \mathcal{L}_{m_2}^2+
\frac{7}{4} Y_f^4 C_A \mathcal{L}_{m_2}+
\frac{5 Y_f^4 C_A}{32 \varepsilon }+
\frac{3 Y_f^4 C_A}{8 \varepsilon ^3}+
\frac{15 Y_f^4 \zeta_2 C_A}{8 \varepsilon }-
\frac{75}{16} Y_f^4 \zeta_2 C_A-
\frac{59 Y_f^4 C_A}{64}-
\frac{237 C_F}{16 \varepsilon }-
\frac{57 C_F}{16 \varepsilon ^2}-
\frac{55 C_F}{16 \varepsilon ^3}-
\frac{55 C_F \mathcal{L}_{m_2}^2}{8 \varepsilon }+
\frac{57 C_F \mathcal{L}_{m_2}}{8 \varepsilon }+
\frac{55 C_F \mathcal{L}_{m_2}}{8 \varepsilon ^2}-
\frac{3 Y_f^2 C_F \mathcal{L}_{m_2}}{4 \varepsilon }+
\frac{3}{4} Y_f^2 C_F \mathcal{L}_{m_2}^2-
\frac{13}{4} Y_f^2 C_F \mathcal{L}_{m_2}+
\frac{161}{16} C_F \mathcal{L}_{m_2}^2+
\frac{343}{32} C_F \mathcal{L}_{m_2}+
\frac{13 Y_f^2 C_F}{8 \varepsilon }+
\frac{3 Y_f^2 C_F}{8 \varepsilon ^2}+
\frac{15}{8} Y_f^2 \zeta_2 C_F+
\frac{47 Y_f^2 C_F}{8}-
\frac{275 \zeta_2 C_F}{16 \varepsilon }+
\frac{805 \zeta_2 C_F}{32}-
\frac{6489 C_F}{128}
\end{autobreak}
\end{align}
\end{tiny}

\begin{tiny}
\begin{align}
\begin{autobreak}
V^{(m_{1,2})}_2=
-\frac{9}{32}C_A\mathcal{L}_{m_2}^2Y_f^4
-\frac{C_A\mathcal{L}_{m_2}^2Y_f^4}{32\varepsilon}
-\frac{219C_AY_f^4}{512}
-\frac{45}{64}C_A\zeta_2Y_f^4
-\frac{5C_A\zeta_2Y_f^4}{64\varepsilon}
+\frac{41}{128}C_A\mathcal{L}_{m_2}Y_f^4
+\frac{23C_A\mathcal{L}_{m_2}Y_f^4}{64\varepsilon}
+\frac{C_A\mathcal{L}_{m_2}Y_f^4}{32\varepsilon^2}
-\frac{13C_AY_f^4}{64\varepsilon}
-\frac{23C_AY_f^4}{128\varepsilon^2}
-\frac{C_AY_f^4}{64\varepsilon^3}
+\frac{19}{8}C_AC_F^2Y_f^2
+\frac{1}{4}C_AC_F^2\mathcal{L}_{m_2}^2Y_f^2
+\frac{33}{8}C_AC_F\mathcal{L}_{m_2}^2Y_f^2
+\frac{1}{8}C_F\mathcal{L}_{m_2}^2Y_f^2
-\frac{C_AC_F\mathcal{L}_{m_2}^2Y_f^2}{4\varepsilon}
+\frac{1129}{64}C_AC_FY_f^2
+\frac{19C_FY_f^2}{16}
+\frac{5}{8}C_AC_F^2\zeta_2Y_f^2
+\frac{165}{16}C_AC_F\zeta_2Y_f^2
+\frac{5}{16}C_F\zeta_2Y_f^2
-\frac{5C_AC_F\zeta_2Y_f^2}{8\varepsilon}
-\frac{5}{4}C_AC_F^2\mathcal{L}_{m_2}Y_f^2
-\frac{137}{16}C_AC_F\mathcal{L}_{m_2}Y_f^2
-\frac{5}{8}C_F\mathcal{L}_{m_2}Y_f^2
-\frac{C_AC_F^2\mathcal{L}_{m_2}Y_f^2}{4\varepsilon}
-\frac{7C_AC_F\mathcal{L}_{m_2}Y_f^2}{2\varepsilon}
-\frac{C_F\mathcal{L}_{m_2}Y_f^2}{8\varepsilon}
+\frac{C_AC_F\mathcal{L}_{m_2}Y_f^2}{4\varepsilon^2}
+\frac{5C_AC_F^2Y_f^2}{8\varepsilon}
+\frac{63C_AC_FY_f^2}{16\varepsilon}
+\frac{5C_FY_f^2}{16\varepsilon}
+\frac{C_AC_F^2Y_f^2}{8\varepsilon^2}
+\frac{7C_AC_FY_f^2}{4\varepsilon^2}
+\frac{C_FY_f^2}{16\varepsilon^2}
-\frac{C_AC_FY_f^2}{8\varepsilon^3}
+\frac{C_AC_FM_W\mathcal{L}_{m_2}^2Y_f}{4m_-}
+\frac{C_AC_FM_W\mathcal{L}_{m_2}^2Y_f}{2\varepsilon m_-}
+\frac{175C_AC_FM_WY_f}{32m_-}
+\frac{7C_AC_FM_WY_f}{4\varepsilon m_-}
+\frac{3C_AC_FM_WY_f}{4\varepsilon^2m_-}
+\frac{C_AC_FM_WY_f}{4\varepsilon^3m_-}
+\frac{5C_AC_FM_W\zeta_2Y_f}{8m_-}
+\frac{5C_AC_FM_W\zeta_2Y_f}{4\varepsilon m_-}
-\frac{17C_AC_FM_W\mathcal{L}_{m_2}Y_f}{8m_-}
-\frac{3C_AC_FM_W\mathcal{L}_{m_2}Y_f}{2\varepsilon m_-}
-\frac{C_AC_FM_W\mathcal{L}_{m_2}Y_f}{2\varepsilon^2 m_-}
-\frac{495}{4}C_AC_F^2
-\frac{141}{16}C_AC_F^2\mathcal{L}_{m_2}^2
-\frac{7}{16}C_AC_F\mathcal{L}_{m_2}^2
-\frac{53}{32}C_F\mathcal{L}_{m_2}^2
+\frac{C_AC_F^2m_2\mathcal{L}_{m_2}^2}{m_1-m_2}
+\frac{C_Fm_2\mathcal{L}_{m_2}^2}{2m_-}
+\frac{2C_AC_F^2m_2\mathcal{L}_{m_2}^2}{\varepsilon m_-}
+\frac{C_Fm_2\mathcal{L}_{m_2}^2}{\varepsilon m_-}
-\frac{1}{4}C_AC_Fn_fT_f\mathcal{L}_{m_2}^2
-\frac{3C_AC_F^2\mathcal{L}_{m_2}^2}{\varepsilon}
-\frac{4C_F\mathcal{L}_{m_2}^2}{\varepsilon}
-\frac{41C_AC_F}{16}
-\frac{1503C_F}{32}
+\frac{175C_AC_F^2m_2}{8m_-}
+\frac{175C_Fm_2}{16m_-}
+\frac{7C_AC_F^2m_2}{\varepsilon m_-}
+\frac{7C_Fm_2}{2\varepsilon m_-}
+\frac{3C_AC_F^2m_2}{\varepsilon^2 m_-}
+\frac{3C_Fm_2}{2\varepsilon^2 m_-}
+\frac{C_AC_F^2m_2}{\varepsilon^3m_-}
+\frac{C_Fm_2}{2\varepsilon^3m_-}
-\frac{85}{32}C_AC_Fn_fT_f
-\frac{11C_AC_Fn_fT_f}{32\varepsilon}
-\frac{C_AC_Fn_fT_f}{8\varepsilon^2}
-\frac{705}{32}C_AC_F^2\zeta_2
-\frac{35}{32}C_AC_F\zeta_2
-\frac{265C_F\zeta_2}{64}
+\frac{5C_AC_F^2m_2\zeta_2}{2m_-}
+\frac{5C_Fm_2\zeta_2}{4m_-}
+\frac{5C_AC_F^2m_2\zeta_2}{\varepsilon m_-}
+\frac{5C_Fm_2\zeta_2}{2\varepsilon m_-}
-\frac{5}{8}C_AC_Fn_fT_f\zeta_2
-\frac{15C_AC_F^2\zeta_2}{2\varepsilon}
-\frac{10C_F\zeta_2}{\varepsilon}
+\frac{989}{16}C_AC_F^2\mathcal{L}_{m_2}
+\frac{19}{16}C_AC_F\mathcal{L}_{m_2}
+\frac{833}{32}C_F\mathcal{L}_{m_2}
-\frac{17C_AC_F^2m_2\mathcal{L}_{m_2}}{2m_-}
-\frac{17C_Fm_2\mathcal{L}_{m_2}}{4m_-}
-\frac{6C_AC_F^2m_2\mathcal{L}_{m_2}}{\varepsilon m_-}
-\frac{3C_Fm_2\mathcal{L}_{m_2}}{\varepsilon m_-}
-\frac{2C_AC_F^2m_2\mathcal{L}_{m_2}}{\varepsilon^2m_-}
-\frac{C_Fm_2\mathcal{L}_{m_2}}{\varepsilon^2m_-}
+\frac{11}{16}C_AC_Fn_fT_f\mathcal{L}_{m_2}
+\frac{C_AC_Fn_fT_f\mathcal{L}_{m_2}}{4\varepsilon}
+\frac{261C_AC_F^2\mathcal{L}_{m_2}}{16\varepsilon}
+\frac{7C_AC_F\mathcal{L}_{m_2}}{16\varepsilon}
+\frac{373C_F\mathcal{L}_{m_2}}{32\varepsilon}
+\frac{3C_AC_F^2\mathcal{L}_{m_2}}{\varepsilon^2}
+\frac{4C_F\mathcal{L}_{m_2}}{\varepsilon^2}
-\frac{1121C_AC_F^2}{32\varepsilon}
-\frac{19C_AC_F}{32\varepsilon}
-\frac{1185C_F}{64\varepsilon}
-\frac{261C_AC_F^2}{32\varepsilon^2}
-\frac{7C_AC_F}{32\varepsilon^2}
-\frac{373C_F}{64\varepsilon^2}
-\frac{3C_AC_F^2}{2\varepsilon^3}
-\frac{2C_F}{\varepsilon^3}
\end{autobreak}
\end{align}
\end{tiny}

\begin{tiny}
\begin{align}
\begin{autobreak}
V^{(m_{1,2})}_3=
\frac{n_f C_A C_F T_f \mathcal{L}_{m_2}}{2 \varepsilon }
-\frac{1}{2} n_f C_A C_F T_f \mathcal{L}_{m_2}^2+n_f C_A C_F T_f \mathcal{L}_{m_2}
-\frac{n_f C_A C_F T_f}{2 \varepsilon }
-\frac{n_f C_A C_F T_f}{4 \varepsilon ^2}
-\frac{5}{4} n_f \zeta_2 C_A C_F T_f-n_f C_A C_F T_f
-\frac{24 C_A C_F^2}{\varepsilon }
+\frac{3 C_A C_F}{4 \varepsilon }
-\frac{47 C_A C_F^2}{8 \varepsilon ^2}
+\frac{C_A C_F}{4 \varepsilon ^2}
-\frac{3 C_A C_F^2}{4 \varepsilon ^3}
-\frac{3 C_A C_F^2 \mathcal{L}_{m_2}^2}{2 \varepsilon }
+\frac{47 C_A C_F^2 \mathcal{L}_{m_2}}{4 \varepsilon }
-\frac{C_A C_F \mathcal{L}_{m_2}}{2 \varepsilon }
+\frac{3 C_A C_F^2 \mathcal{L}_{m_2}}{2 \varepsilon ^2}
-\frac{6 Y_f^2 C_A C_F \mathcal{L}_{m_2}^2}{\varepsilon }
-\frac{3 Y_f^2 C_A C_F^2 \mathcal{L}_{m_2}}{2 \varepsilon }
+\frac{2 Y_f^2 C_A C_F \mathcal{L}_{m_2}}{\varepsilon }
+\frac{6 Y_f^2 C_A C_F \mathcal{L}_{m_2}}{\varepsilon ^2}
+\frac{3}{2} Y_f^2 C_A C_F^2 \mathcal{L}_{m_2}^2+13 Y_f^2 C_A C_F \mathcal{L}_{m_2}^2
-\frac{13}{2} Y_f^2 C_A C_F^2 \mathcal{L}_{m_2}
-\frac{17}{2} Y_f^2 C_A C_F \mathcal{L}_{m_2}-8 C_A C_F^2 \mathcal{L}_{m_2}^2
+\frac{1}{2} C_A C_F \mathcal{L}_{m_2}^2
+\frac{351}{8} C_A C_F^2 \mathcal{L}_{m_2}
-\frac{3}{2} C_A C_F \mathcal{L}_{m_2}
+\frac{13 Y_f^2 C_A C_F^2}{4 \varepsilon }
-\frac{4 Y_f^2 C_A C_F}{\varepsilon }
+\frac{3 Y_f^2 C_A C_F^2}{4 \varepsilon ^2}
-\frac{Y_f^2 C_A C_F}{\varepsilon ^2}
-\frac{3 Y_f^2 C_A C_F}{\varepsilon ^3}
-\frac{15 Y_f^2 \zeta_2 C_A C_F}{\varepsilon }
+\frac{15}{4} Y_f^2 \zeta_2 C_A C_F^2
+\frac{65}{2} Y_f^2 \zeta_2 C_A C_F
+\frac{47}{4} Y_f^2 C_A C_F^2
+\frac{7}{8} Y_f^2 C_A C_F
-\frac{15 \zeta_2 C_A C_F^2}{4 \varepsilon }-20 \zeta_2 C_A C_F^2
+\frac{5}{4} \zeta_2 C_A C_F
-\frac{3165}{32} C_A C_F^2
+\frac{9 C_A C_F}{4}
+\frac{3 Y_f^4 C_A \mathcal{L}_{m_2}^2}{4 \varepsilon }
-\frac{3 Y_f^4 C_A \mathcal{L}_{m_2}}{4 \varepsilon ^2}
-\frac{15}{8} Y_f^4 C_A \mathcal{L}_{m_2}^2
+\frac{7}{4} Y_f^4 C_A \mathcal{L}_{m_2}
+\frac{5 Y_f^4 C_A}{32 \varepsilon }
+\frac{3 Y_f^4 C_A}{8 \varepsilon ^3}
+\frac{15 Y_f^4 \zeta_2 C_A}{8 \varepsilon }
-\frac{75}{16} Y_f^4 \zeta_2 C_A
-\frac{59 Y_f^4 C_A}{64}
-\frac{26 C_F}{\varepsilon }
-\frac{95 C_F}{16 \varepsilon ^2}
-\frac{27 C_F}{8 \varepsilon ^3}
-\frac{27 C_F \mathcal{L}_{m_2}^2}{4 \varepsilon }
+\frac{95 C_F \mathcal{L}_{m_2}}{8 \varepsilon }
+\frac{27 C_F \mathcal{L}_{m_2}}{4 \varepsilon ^2}
-\frac{3 Y_f^2 C_F \mathcal{L}_{m_2}}{4 \varepsilon }
+\frac{3}{4} Y_f^2 C_F \mathcal{L}_{m_2}^2
-\frac{13}{4} Y_f^2 C_F \mathcal{L}_{m_2}+5 C_F \mathcal{L}_{m_2}^2
+\frac{535}{16} C_F \mathcal{L}_{m_2}
+\frac{13 Y_f^2 C_F}{8 \varepsilon }
+\frac{3 Y_f^2 C_F}{8 \varepsilon ^2}
+\frac{15}{8} Y_f^2 \zeta_2 C_F
+\frac{47 Y_f^2 C_F}{8}
-\frac{135 \zeta_2 C_F}{8 \varepsilon }
+\frac{25 \zeta_2 C_F}{2}
-\frac{4293 C_F}{64}
\end{autobreak}  
\end{align}
\end{tiny}

\begin{tiny}
\begin{align}
\begin{autobreak}
V^{(m_{1,2})}_4=
-\frac{27 C_A C_F^2 Y_s^2 M_H^4}{512 m_- m_2^3 M_W^2}
-\frac{243 C_A Y_s^2 M_H^4}{512 m_- m_2^3 M_W^2}
-\frac{27 C_F Y_s^2 M_H^4}{1024 m_- m_2^3 M_W^2}
-\frac{C_A C_F^2 Y_s^2 M_H^4}{256 \varepsilon m_- m_2^3 M_W^2}
-\frac{9 C_A Y_s^2 M_H^4}{256 \varepsilon m_- m_2^3 M_W^2}
-\frac{C_F Y_s^2 M_H^4}{512 \varepsilon m_- m_2^3 M_W^2}
+\frac{C_A C_F^2 Y_s^2 M_H^4}{512 \varepsilon ^2 m_- m_2^3 M_W^2}
+\frac{9 C_A Y_s^2 M_H^4}{512 \varepsilon ^2 m_- m_2^3 M_W^2}
+\frac{C_F Y_s^2 M_H^4}{1024 \varepsilon ^2 m_- m_2^3 M_W^2}
+\frac{31 C_A C_F^2 Y_s^2 M_H^4}{1024 m_2^4 M_W^2}
+\frac{279 C_A Y_s^2 M_H^4}{1024 m_2^4 M_W^2}
+\frac{31 C_F Y_s^2 M_H^4}{2048 m_2^4 M_W^2}
+\frac{C_A C_F^2 Y_s^2 M_H^4}{128 \varepsilon m_2^4 M_W^2}
+\frac{9 C_A Y_s^2 M_H^4}{128 \varepsilon m_2^4 M_W^2}
+\frac{C_F Y_s^2 M_H^4}{256 \varepsilon m_2^4 M_W^2}
-\frac{5 C_A C_F^2 Y_s^2 M_H^4}{1024 \varepsilon ^2 m_2^4 M_W^2}
-\frac{45 C_A Y_s^2 M_H^4}{1024 \varepsilon ^2 m_2^4 M_W^2}
-\frac{5 C_F Y_s^2 M_H^4}{2048 \varepsilon ^2 m_2^4 M_W^2}
+\frac{C_A C_F^2 Y_s^2 \mathcal{L}_{m_2}^2 M_H^4}{256 m_- m_2^3 M_W^2}
+\frac{9 C_A Y_s^2 \mathcal{L}_{m_2}^2 M_H^4}{256 m_- m_2^3 M_W^2}
+\frac{C_F Y_s^2 \mathcal{L}_{m_2}^2 M_H^4}{512 m_- m_2^3 M_W^2}
-\frac{5 C_A C_F^2 Y_s^2 \mathcal{L}_{m_2}^2 M_H^4}{512 m_2^4 M_W^2}
-\frac{45 C_A Y_s^2 \mathcal{L}_{m_2}^2 M_H^4}{512 m_2^4 M_W^2}
-\frac{5 C_F Y_s^2 \mathcal{L}_{m_2}^2 M_H^4}{1024 m_2^4 M_W^2}
+\frac{5 C_A C_F^2 Y_s^2 \zeta_2 M_H^4}{512 m_- m_2^3 M_W^2}
+\frac{45 C_A Y_s^2 \zeta_2 M_H^4}{512 m_- m_2^3 M_W^2}
+\frac{5 C_F Y_s^2 \zeta_2 M_H^4}{1024 m_- m_2^3 M_W^2}
-\frac{25 C_A C_F^2 Y_s^2 \zeta_2 M_H^4}{1024 m_2^4 M_W^2}
-\frac{225 C_A Y_s^2 \zeta_2 M_H^4}{1024 m_2^4 M_W^2}
-\frac{25 C_F Y_s^2 \zeta_2 M_H^4}{2048 m_2^4 M_W^2}
+\frac{C_A C_F^2 Y_s^2 \mathcal{L}_{m_2} M_H^4}{128 m_- m_2^3 M_W^2}
+\frac{9 C_A Y_s^2 \mathcal{L}_{m_2} M_H^4}{128 m_- m_2^3 M_W^2}
+\frac{C_F Y_s^2 \mathcal{L}_{m_2} M_H^4}{256 m_- m_2^3 M_W^2}
-\frac{C_A C_F^2 Y_s^2 \mathcal{L}_{m_2} M_H^4}{256 \varepsilon m_- m_2^3 M_W^2}
-\frac{9 C_A Y_s^2 \mathcal{L}_{m_2} M_H^4}{256 \varepsilon m_- m_2^3 M_W^2}
-\frac{C_F Y_s^2 \mathcal{L}_{m_2} M_H^4}{512 \varepsilon m_- m_2^3 M_W^2}
-\frac{C_A C_F^2 Y_s^2 \mathcal{L}_{m_2} M_H^4}{64 m_2^4 M_W^2}
-\frac{9 C_A Y_s^2 \mathcal{L}_{m_2} M_H^4}{64 m_2^4 M_W^2}
-\frac{C_F Y_s^2 \mathcal{L}_{m_2} M_H^4}{128 m_2^4 M_W^2}
+\frac{5 C_A C_F^2 Y_s^2 \mathcal{L}_{m_2} M_H^4}{512 \varepsilon m_2^4 M_W^2}
+\frac{45 C_A Y_s^2 \mathcal{L}_{m_2} M_H^4}{512 \varepsilon m_2^4 M_W^2}
+\frac{5 C_F Y_s^2 \mathcal{L}_{m_2} M_H^4}{1024 \varepsilon m_2^4 M_W^2}
+\frac{2193 C_A Y_s^3 M_H^2}{1024 m_2^4 M_W}
+\frac{141 C_A Y_s^3 M_H^2}{128 \varepsilon m_2^4 M_W}
+\frac{123 C_A Y_s^3 M_H^2}{256 \varepsilon ^2 m_2^4 M_W}
-\frac{9 C_A Y_s^3 M_H^2}{128 \varepsilon ^3 m_2^4 M_W}
-\frac{2943 C_A m_1 Y_s^3 M_H^2}{2048 m_2^5 M_W}
-\frac{195 C_A m_1 Y_s^3 M_H^2}{256 \varepsilon m_2^5 M_W}
-\frac{21 C_A m_1 Y_s^3 M_H^2}{64 \varepsilon ^2 m_2^5 M_W}
+\frac{15 C_A m_1 Y_s^3 M_H^2}{256 \varepsilon ^3 m_2^5 M_W}
+\frac{15 C_A Y_s^3 M_H^2}{128 m_- m_2^5 M_W}
+\frac{3 C_A Y_s^3 M_H^2}{16 \varepsilon m_- m_2^5 M_W}
+\frac{9 C_A Y_s^3 M_H^2}{128 \varepsilon ^2 m_- m_2^5 M_W}
+\frac{3 C_A Y_s^3 M_H^2}{128 \varepsilon ^3 m_- m_2^5 M_W}
-\frac{165 C_A Y_s^3 M_H^2}{256 m_2^6 M_W}
-\frac{33 C_A Y_s^3 M_H^2}{32 \varepsilon m_2^6 M_W}
-\frac{99 C_A Y_s^3 M_H^2}{256 \varepsilon ^2 m_2^6 M_W}
-\frac{33 C_A Y_s^3 M_H^2}{256 \varepsilon ^3 m_2^6 M_W}
+\frac{21 C_A Y_s^3 \mathcal{L}_{m_2}^2 M_H^2}{16 m_2^4 M_W}
-\frac{9 C_A Y_s^3 \mathcal{L}_{m_2}^2 M_H^2}{64 \varepsilon m_2^4 M_W}
-\frac{243 C_A m_1 Y_s^3 \mathcal{L}_{m_2}^2 M_H^2}{256 m_2^5 M_W}
+\frac{15 C_A m_1 Y_s^3 \mathcal{L}_{m_2}^2 M_H^2}{128 \varepsilon m_2^5 M_W}
+\frac{3 C_A Y_s^3 \mathcal{L}_{m_2}^2 M_H^2}{64 \varepsilon m_- m_2^5 M_W}
-\frac{33 C_A Y_s^3 \mathcal{L}_{m_2}^2 M_H^2}{128 \varepsilon m_2^6 M_W}
+\frac{105 C_A Y_s^3 \zeta_2 M_H^2}{32 m_2^4 M_W}
-\frac{45 C_A Y_s^3 \zeta_2 M_H^2}{128 \varepsilon m_2^4 M_W}
-\frac{1215 C_A m_1 Y_s^3 \zeta_2 M_H^2}{512 m_2^5 M_W}
+\frac{75 C_A m_1 Y_s^3 \zeta_2 M_H^2}{256 \varepsilon m_2^5 M_W}
+\frac{3 C_A Y_s^3 \zeta_2 M_H^2}{128 \varepsilon m_- m_2^5 M_W}
-\frac{33 C_A Y_s^3 \zeta_2 M_H^2}{256 \varepsilon m_2^6 M_W}
-\frac{663 C_A Y_s^3 \mathcal{L}_{m_2} M_H^2}{256 m_2^4 M_W}
-\frac{123 C_A Y_s^3 \mathcal{L}_{m_2} M_H^2}{128 \varepsilon m_2^4 M_W}
+\frac{9 C_A Y_s^3 \mathcal{L}_{m_2} M_H^2}{64 \varepsilon ^2 m_2^4 M_W}
+\frac{945 C_A m_1 Y_s^3 \mathcal{L}_{m_2} M_H^2}{512 m_2^5 M_W}
+\frac{21 C_A m_1 Y_s^3 \mathcal{L}_{m_2} M_H^2}{32 \varepsilon m_2^5 M_W}
-\frac{15 C_A m_1 Y_s^3 \mathcal{L}_{m_2} M_H^2}{128 \varepsilon ^2 m_2^5 M_W}
-\frac{3 C_A Y_s^3 \mathcal{L}_{m_2} M_H^2}{64 m_- m_2^5 M_W}
-\frac{9 C_A Y_s^3 \mathcal{L}_{m_2} M_H^2}{64 \varepsilon m_- m_2^5 M_W}
-\frac{3 C_A Y_s^3 \mathcal{L}_{m_2} M_H^2}{64 \varepsilon ^2 m_- m_2^5 M_W}
+\frac{33 C_A Y_s^3 \mathcal{L}_{m_2} M_H^2}{128 m_2^6 M_W}
+\frac{99 C_A Y_s^3 \mathcal{L}_{m_2} M_H^2}{128 \varepsilon m_2^6 M_W}
+\frac{33 C_A Y_s^3 \mathcal{L}_{m_2} M_H^2}{128 \varepsilon ^2 m_2^6 M_W}
+\frac{339 C_A Y_s^4}{256 m_2^4}
+\frac{99 C_A Y_s^4}{256 \varepsilon m_2^4}
-\frac{3 C_A Y_s^4}{256 \varepsilon ^2 m_2^4}
-\frac{267 C_A m_1 Y_s^4}{256 m_2^5}
-\frac{75 C_A m_1 Y_s^4}{256 \varepsilon m_2^5}
-\frac{3 C_A m_1 Y_s^4}{256 \varepsilon ^2 m_2^5}
-\frac{1449 C_A Y_s^4}{1024 m_2^6}
-\frac{289 C_A Y_s^4}{1024 \varepsilon m_2^6}
-\frac{921 C_A Y_s^4}{4096 \varepsilon ^2 m_2^6}
-\frac{27 C_A Y_s^4}{1024 \varepsilon ^3 m_2^6}
+\frac{1315 C_A m_1 Y_s^4}{1024 m_2^7}
+\frac{339 C_A m_1 Y_s^4}{1024 \varepsilon m_2^7}
+\frac{915 C_A m_1 Y_s^4}{4096 \varepsilon ^2 m_2^7}
+\frac{9 C_A m_1 Y_s^4}{1024 \varepsilon ^3 m_2^7}
-\frac{119}{2} C_A C_F^2
-\frac{13 C_A C_F^2 M_W^2 Y_s^2}{4 m_- m_2^3}
-\frac{13 C_F M_W^2 Y_s^2}{8 m_- m_2^3}
-\frac{5 C_A C_F^2 M_W^2 Y_s^2}{16 \varepsilon m_- m_2^3}
-\frac{5 C_F M_W^2 Y_s^2}{32 \varepsilon m_- m_2^3}
+\frac{C_A C_F^2 M_W^2 Y_s^2}{8 \varepsilon ^2 m_- m_2^3}
+\frac{C_F M_W^2 Y_s^2}{16 \varepsilon ^2 m_- m_2^3}
+\frac{27 C_A C_F^2 M_W^2 Y_s^2}{16 m_2^4}
+\frac{27 C_F M_W^2 Y_s^2}{32 m_2^4}
+\frac{21 C_A C_F^2 M_W^2 Y_s^2}{32 \varepsilon m_2^4}
+\frac{21 C_F M_W^2 Y_s^2}{64 \varepsilon m_2^4}
-\frac{5 C_A C_F^2 M_W^2 Y_s^2}{16 \varepsilon ^2 m_2^4}
-\frac{5 C_F M_W^2 Y_s^2}{32 \varepsilon ^2 m_2^4}
+\frac{3 C_A C_F^2 Y_s^2}{4 m_2^2}
+\frac{231 C_A C_F Y_s^2}{64 m_2^2}
+\frac{3 C_F Y_s^2}{8 m_2^2}
+\frac{3 C_A C_F^2 Y_s^2}{8 \varepsilon m_2^2}
-\frac{5 C_A C_F Y_s^2}{16 \varepsilon m_2^2}
+\frac{3 C_F Y_s^2}{16 \varepsilon m_2^2}
+\frac{3 C_A C_F^2 Y_s^2}{16 \varepsilon ^2 m_2^2}
-\frac{11 C_A C_F Y_s^2}{16 \varepsilon ^2 m_2^2}
+\frac{3 C_F Y_s^2}{32 \varepsilon ^2 m_2^2}
-\frac{3 C_A C_F Y_s^2}{8 \varepsilon ^3 m_2^2}
-\frac{C_A C_F^2 m_1 Y_s^2}{4 m_2^3}
-\frac{19 C_A C_F m_1 Y_s^2}{16 m_2^3}
-\frac{C_F m_1 Y_s^2}{8 m_2^3}
-\frac{C_A C_F^2 m_1 Y_s^2}{8 \varepsilon m_2^3}
+\frac{13 C_A C_F m_1 Y_s^2}{16 \varepsilon m_2^3}
-\frac{C_F m_1 Y_s^2}{16 \varepsilon m_2^3}
-\frac{C_A C_F^2 m_1 Y_s^2}{16 \varepsilon ^2 m_2^3}
+\frac{23 C_A C_F m_1 Y_s^2}{16 \varepsilon ^2 m_2^3}
-\frac{C_F m_1 Y_s^2}{32 \varepsilon ^2 m_2^3}
-\frac{565 C_A C_F Y_s^2}{64 m_2^4}
-\frac{287 C_A C_F Y_s^2}{64 \varepsilon m_2^4}
-\frac{709 C_A C_F Y_s^2}{128 \varepsilon ^2 m_2^4}
+\frac{41 C_A C_F Y_s^2}{64 \varepsilon ^3 m_2^4}
-\frac{3 C_A C_F Y_s^2}{16 \varepsilon ^4 m_2^4}
-\frac{3 C_A Y_s^4 \mathcal{L}_{m_2}^2}{128 m_2^4}
-\frac{3 C_A m_1 Y_s^4 \mathcal{L}_{m_2}^2}{128 m_2^5}
-\frac{705 C_A Y_s^4 \mathcal{L}_{m_2}^2}{2048 m_2^6}
-\frac{27 C_A Y_s^4 \mathcal{L}_{m_2}^2}{512 \varepsilon m_2^6}
+\frac{843 C_A m_1 Y_s^4 \mathcal{L}_{m_2}^2}{2048 m_2^7}
+\frac{9 C_A m_1 Y_s^4 \mathcal{L}_{m_2}^2}{512 \varepsilon m_2^7}
-\frac{11}{4} C_A C_F^2 \mathcal{L}_{m_2}^2
+\frac{C_A C_F^2 M_W^2 Y_s^2 \mathcal{L}_{m_2}^2}{4 m_- m_2^3}
+\frac{C_F M_W^2 Y_s^2 \mathcal{L}_{m_2}^2}{8 m_- m_2^3}
-\frac{5 C_A C_F^2 M_W^2 Y_s^2 \mathcal{L}_{m_2}^2}{8 m_2^4}
-\frac{5 C_F M_W^2 Y_s^2 \mathcal{L}_{m_2}^2}{16 m_2^4}
+\frac{3 C_A C_F^2 Y_s^2 \mathcal{L}_{m_2}^2}{8 m_2^2}
+\frac{C_A C_F Y_s^2 \mathcal{L}_{m_2}^2}{2 m_2^2}
+\frac{3 C_F Y_s^2 \mathcal{L}_{m_2}^2}{16 m_2^2}
-\frac{3 C_A C_F Y_s^2 \mathcal{L}_{m_2}^2}{4 \varepsilon m_2^2}
-\frac{C_A C_F^2 m_1 Y_s^2 \mathcal{L}_{m_2}^2}{8 m_2^3}
+\frac{23 C_A C_F m_1 Y_s^2 \mathcal{L}_{m_2}^2}{8 m_2^3}
-\frac{C_F m_1 Y_s^2 \mathcal{L}_{m_2}^2}{16 m_2^3}
-\frac{963 C_A C_F Y_s^2 \mathcal{L}_{m_2}^2}{64 m_2^4}
+\frac{65 C_A C_F Y_s^2 \mathcal{L}_{m_2}^2}{32 \varepsilon m_2^4}
-\frac{3 C_A C_F Y_s^2 \mathcal{L}_{m_2}^2}{8 \varepsilon ^2 m_2^4}
-\frac{15}{4} C_A C_F \mathcal{L}_{m_2}^2
+\frac{69}{8} C_F \mathcal{L}_{m_2}^2+C_A C_F n_f T_f \mathcal{L}_{m_2}^2
-\frac{191 C_A C_F M_W Y_s \mathcal{L}_{m_2}^2}{16 m_2^2}
+\frac{11 C_A C_F M_W Y_s \mathcal{L}_{m_2}^2}{8 \varepsilon m_2^2}
-\frac{2 C_A C_F^2 \mathcal{L}_{m_2}^2}{\varepsilon }
-\frac{7 C_F \mathcal{L}_{m_2}^2}{\varepsilon }
+\frac{43 C_A C_F}{8}
-\frac{447 C_F}{8}+7 C_A C_F n_f T_f
+\frac{2 C_A C_F n_f T_f}{\varepsilon }
+\frac{C_A C_F n_f T_f}{2 \varepsilon ^2}
-\frac{2931 C_A C_F M_W Y_s}{128 m_2^2}
-\frac{85 C_A C_F M_W Y_s}{8 \varepsilon m_2^2}
-\frac{17 C_A C_F M_W Y_s}{4 \varepsilon ^2 m_2^2}
+\frac{11 C_A C_F M_W Y_s}{16 \varepsilon ^3 m_2^2}
-\frac{15 C_A Y_s^4 \zeta_2}{256 m_2^4}
-\frac{15 C_A m_1 Y_s^4 \zeta_2}{256 m_2^5}
+\frac{1239 C_A Y_s^4 \zeta_2}{4096 m_2^6}
-\frac{135 C_A Y_s^4 \zeta_2}{1024 \varepsilon m_2^6}
+\frac{195 C_A m_1 Y_s^4 \zeta_2}{4096 m_2^7}
+\frac{45 C_A m_1 Y_s^4 \zeta_2}{1024 \varepsilon m_2^7}
-\frac{55}{8} C_A C_F^2 \zeta_2
+\frac{5 C_A C_F^2 M_W^2 Y_s^2 \zeta_2}{8 m_- m_2^3}
+\frac{5 C_F M_W^2 Y_s^2 \zeta_2}{16 m_- m_2^3}
-\frac{25 C_A C_F^2 M_W^2 Y_s^2 \zeta_2}{16 m_2^4}
-\frac{25 C_F M_W^2 Y_s^2 \zeta_2}{32 m_2^4}
+\frac{15 C_A C_F^2 Y_s^2 \zeta_2}{16 m_2^2}
+\frac{5 C_A C_F Y_s^2 \zeta_2}{4 m_2^2}
+\frac{15 C_F Y_s^2 \zeta_2}{32 m_2^2}
-\frac{15 C_A C_F Y_s^2 \zeta_2}{8 \varepsilon m_2^2}
-\frac{5 C_A C_F^2 m_1 Y_s^2 \zeta_2}{16 m_2^3}
+\frac{115 C_A C_F m_1 Y_s^2 \zeta_2}{16 m_2^3}
-\frac{5 C_F m_1 Y_s^2 \zeta_2}{32 m_2^3}
-\frac{4379 C_A C_F Y_s^2 \zeta_2}{128 m_2^4}
+\frac{289 C_A C_F Y_s^2 \zeta_2}{64 \varepsilon m_2^4}
-\frac{15 C_A C_F Y_s^2 \zeta_2}{16 \varepsilon ^2 m_2^4}
-\frac{75}{8} C_A C_F \zeta_2
+\frac{345 C_F \zeta_2}{16}
+\frac{5}{2} C_A C_F n_f T_f \zeta_2
-\frac{955 C_A C_F M_W Y_s \zeta_2}{32 m_2^2}
+\frac{55 C_A C_F M_W Y_s \zeta_2}{16 \varepsilon m_2^2}
-\frac{5 C_A C_F^2 \zeta_2}{\varepsilon }
-\frac{35 C_F \zeta_2}{2 \varepsilon }
-\frac{99 C_A Y_s^4 \mathcal{L}_{m_2}}{128 m_2^4}
+\frac{3 C_A Y_s^4 \mathcal{L}_{m_2}}{128 \varepsilon m_2^4}
+\frac{75 C_A m_1 Y_s^4 \mathcal{L}_{m_2}}{128 m_2^5}
+\frac{3 C_A m_1 Y_s^4 \mathcal{L}_{m_2}}{128 \varepsilon m_2^5}
+\frac{131 C_A Y_s^4 \mathcal{L}_{m_2}}{256 m_2^6}
+\frac{921 C_A Y_s^4 \mathcal{L}_{m_2}}{2048 \varepsilon m_2^6}
+\frac{27 C_A Y_s^4 \mathcal{L}_{m_2}}{512 \varepsilon ^2 m_2^6}
-\frac{165 C_A m_1 Y_s^4 \mathcal{L}_{m_2}}{256 m_2^7}
-\frac{915 C_A m_1 Y_s^4 \mathcal{L}_{m_2}}{2048 \varepsilon m_2^7}
-\frac{9 C_A m_1 Y_s^4 \mathcal{L}_{m_2}}{512 \varepsilon ^2 m_2^7}
+\frac{93}{4} C_A C_F^2 \mathcal{L}_{m_2}
+\frac{5 C_A C_F^2 M_W^2 Y_s^2 \mathcal{L}_{m_2}}{8 m_- m_2^3}
+\frac{5 C_F M_W^2 Y_s^2 \mathcal{L}_{m_2}}{16 m_- m_2^3}
-\frac{C_A C_F^2 M_W^2 Y_s^2 \mathcal{L}_{m_2}}{4 \varepsilon m_- m_2^3}
-\frac{C_F M_W^2 Y_s^2 \mathcal{L}_{m_2}}{8 \varepsilon m_- m_2^3}
-\frac{21 C_A C_F^2 M_W^2 Y_s^2 \mathcal{L}_{m_2}}{16 m_2^4}
-\frac{21 C_F M_W^2 Y_s^2 \mathcal{L}_{m_2}}{32 m_2^4}
+\frac{5 C_A C_F^2 M_W^2 Y_s^2 \mathcal{L}_{m_2}}{8 \varepsilon m_2^4}
+\frac{5 C_F M_W^2 Y_s^2 \mathcal{L}_{m_2}}{16 \varepsilon m_2^4}
-\frac{3 C_A C_F^2 Y_s^2 \mathcal{L}_{m_2}}{4 m_2^2}
-\frac{23 C_A C_F Y_s^2 \mathcal{L}_{m_2}}{16 m_2^2}
-\frac{3 C_F Y_s^2 \mathcal{L}_{m_2}}{8 m_2^2}
-\frac{3 C_A C_F^2 Y_s^2 \mathcal{L}_{m_2}}{8 \varepsilon m_2^2}
+\frac{11 C_A C_F Y_s^2 \mathcal{L}_{m_2}}{8 \varepsilon m_2^2}
-\frac{3 C_F Y_s^2 \mathcal{L}_{m_2}}{16 \varepsilon m_2^2}
+\frac{3 C_A C_F Y_s^2 \mathcal{L}_{m_2}}{4 \varepsilon ^2 m_2^2}
+\frac{C_A C_F^2 m_1 Y_s^2 \mathcal{L}_{m_2}}{4 m_2^3}
-\frac{13 C_A C_F m_1 Y_s^2 \mathcal{L}_{m_2}}{8 m_2^3}
+\frac{C_F m_1 Y_s^2 \mathcal{L}_{m_2}}{8 m_2^3}
+\frac{C_A C_F^2 m_1 Y_s^2 \mathcal{L}_{m_2}}{8 \varepsilon m_2^3}
-\frac{23 C_A C_F m_1 Y_s^2 \mathcal{L}_{m_2}}{8 \varepsilon m_2^3}
+\frac{C_F m_1 Y_s^2 \mathcal{L}_{m_2}}{16 \varepsilon m_2^3}
+\frac{203 C_A C_F Y_s^2 \mathcal{L}_{m_2}}{16 m_2^4}
+\frac{685 C_A C_F Y_s^2 \mathcal{L}_{m_2}}{64 \varepsilon m_2^4}
-\frac{41 C_A C_F Y_s^2 \mathcal{L}_{m_2}}{32 \varepsilon ^2 m_2^4}
+\frac{3 C_A C_F Y_s^2 \mathcal{L}_{m_2}}{8 \varepsilon ^3 m_2^4}
+\frac{3}{4} C_A C_F \mathcal{L}_{m_2}
+\frac{129}{8} C_F \mathcal{L}_{m_2}-4 C_A C_F n_f T_f \mathcal{L}_{m_2}
-\frac{C_A C_F n_f T_f \mathcal{L}_{m_2}}{\varepsilon }
+\frac{801 C_A C_F M_W Y_s \mathcal{L}_{m_2}}{32 m_2^2}
+\frac{17 C_A C_F M_W Y_s \mathcal{L}_{m_2}}{2 \varepsilon m_2^2}
-\frac{11 C_A C_F M_W Y_s \mathcal{L}_{m_2}}{8 \varepsilon ^2 m_2^2}
+\frac{31 C_A C_F^2 \mathcal{L}_{m_2}}{4 \varepsilon }
+\frac{15 C_A C_F \mathcal{L}_{m_2}}{4 \varepsilon }
+\frac{71 C_F \mathcal{L}_{m_2}}{8 \varepsilon }
+\frac{2 C_A C_F^2 \mathcal{L}_{m_2}}{\varepsilon ^2}
+\frac{7 C_F \mathcal{L}_{m_2}}{\varepsilon ^2}
-\frac{115 C_A C_F^2}{8 \varepsilon }
-\frac{3 C_A C_F}{8 \varepsilon }
-\frac{283 C_F}{16 \varepsilon }
-\frac{31 C_A C_F^2}{8 \varepsilon ^2}
-\frac{15 C_A C_F}{8 \varepsilon ^2}
-\frac{71 C_F}{16 \varepsilon ^2}
-\frac{C_A C_F^2}{\varepsilon ^3}
-\frac{7 C_F}{2 \varepsilon ^3}
\end{autobreak}
\end{align}
\end{tiny}

\begin{tiny}
\begin{align}
\begin{autobreak}
V^{(m_{1,2})}_5=
-\frac{27 C_A C_F^2 Y_s^2 M_H^4}{512 m_- m_2^3 M_W^2}
-\frac{243 C_A Y_s^2 M_H^4}{512 m_- m_2^3 M_W^2}
-\frac{27 C_F Y_s^2 M_H^4}{1024 m_- m_2^3 M_W^2}
-\frac{C_A C_F^2 Y_s^2 M_H^4}{256 \varepsilon m_- m_2^3 M_W^2}
-\frac{9 C_A Y_s^2 M_H^4}{256 \varepsilon m_- m_2^3 M_W^2}
-\frac{C_F Y_s^2 M_H^4}{512 \varepsilon m_- m_2^3 M_W^2}
+\frac{C_A C_F^2 Y_s^2 M_H^4}{512 \varepsilon ^2 m_- m_2^3 M_W^2}
+\frac{9 C_A Y_s^2 M_H^4}{512 \varepsilon ^2 m_- m_2^3 M_W^2}
+\frac{C_F Y_s^2 M_H^4}{1024 \varepsilon ^2 m_- m_2^3 M_W^2}
+\frac{17 C_A C_F^2 Y_s^2 M_H^4}{256 m_2^4 M_W^2}
+\frac{153 C_A Y_s^2 M_H^4}{256 m_2^4 M_W^2}
+\frac{17 C_F Y_s^2 M_H^4}{512 m_2^4 M_W^2}
+\frac{3 C_A C_F^2 Y_s^2 M_H^4}{512 \varepsilon m_2^4 M_W^2}
+\frac{27 C_A Y_s^2 M_H^4}{512 \varepsilon m_2^4 M_W^2}
+\frac{3 C_F Y_s^2 M_H^4}{1024 \varepsilon m_2^4 M_W^2}
+\frac{C_A C_F^2 Y_s^2 M_H^4}{256 \varepsilon ^2 m_2^4 M_W^2}
+\frac{9 C_A Y_s^2 M_H^4}{256 \varepsilon ^2 m_2^4 M_W^2}
+\frac{C_F Y_s^2 M_H^4}{512 \varepsilon ^2 m_2^4 M_W^2}
+\frac{C_A C_F^2 Y_s^2 \mathcal{L}_{m_2}^2 M_H^4}{256 m_- m_2^3 M_W^2}
+\frac{9 C_A Y_s^2 \mathcal{L}_{m_2}^2 M_H^4}{256 m_- m_2^3 M_W^2}
+\frac{C_F Y_s^2 \mathcal{L}_{m_2}^2 M_H^4}{512 m_- m_2^3 M_W^2}
+\frac{C_A C_F^2 Y_s^2 \mathcal{L}_{m_2}^2 M_H^4}{128 m_2^4 M_W^2}
+\frac{9 C_A Y_s^2 \mathcal{L}_{m_2}^2 M_H^4}{128 m_2^4 M_W^2}
+\frac{C_F Y_s^2 \mathcal{L}_{m_2}^2 M_H^4}{256 m_2^4 M_W^2}
+\frac{5 C_A C_F^2 Y_s^2 \zeta_2 M_H^4}{512 m_- m_2^3 M_W^2}
+\frac{45 C_A Y_s^2 \zeta_2 M_H^4}{512 m_- m_2^3 M_W^2}
+\frac{5 C_F Y_s^2 \zeta_2 M_H^4}{1024 m_- m_2^3 M_W^2}
+\frac{5 C_A C_F^2 Y_s^2 \zeta_2 M_H^4}{256 m_2^4 M_W^2}
+\frac{45 C_A Y_s^2 \zeta_2 M_H^4}{256 m_2^4 M_W^2}
+\frac{5 C_F Y_s^2 \zeta_2 M_H^4}{512 m_2^4 M_W^2}
+\frac{C_A C_F^2 Y_s^2 \mathcal{L}_{m_2} M_H^4}{128 m_- m_2^3 M_W^2}
+\frac{9 C_A Y_s^2 \mathcal{L}_{m_2} M_H^4}{128 m_- m_2^3 M_W^2}
+\frac{C_F Y_s^2 \mathcal{L}_{m_2} M_H^4}{256 m_- m_2^3 M_W^2}
-\frac{C_A C_F^2 Y_s^2 \mathcal{L}_{m_2} M_H^4}{256 \varepsilon m_- m_2^3 M_W^2}
-\frac{9 C_A Y_s^2 \mathcal{L}_{m_2} M_H^4}{256 \varepsilon m_- m_2^3 M_W^2}
-\frac{C_F Y_s^2 \mathcal{L}_{m_2} M_H^4}{512 \varepsilon m_- m_2^3 M_W^2}
-\frac{3 C_A C_F^2 Y_s^2 \mathcal{L}_{m_2} M_H^4}{256 m_2^4 M_W^2}
-\frac{27 C_A Y_s^2 \mathcal{L}_{m_2} M_H^4}{256 m_2^4 M_W^2}
-\frac{3 C_F Y_s^2 \mathcal{L}_{m_2} M_H^4}{512 m_2^4 M_W^2}
-\frac{C_A C_F^2 Y_s^2 \mathcal{L}_{m_2} M_H^4}{128 \varepsilon m_2^4 M_W^2}
-\frac{9 C_A Y_s^2 \mathcal{L}_{m_2} M_H^4}{128 \varepsilon m_2^4 M_W^2}
-\frac{C_F Y_s^2 \mathcal{L}_{m_2} M_H^4}{256 \varepsilon m_2^4 M_W^2}
+\frac{369 C_A Y_s^3 M_H^2}{4096 m_2^4 M_W}
+\frac{105 C_A Y_s^3 M_H^2}{512 \varepsilon m_2^4 M_W}
-\frac{21 C_A Y_s^3 M_H^2}{512 \varepsilon ^2 m_2^4 M_W}
-\frac{81 C_A Y_s^3 M_H^2}{512 \varepsilon ^3 m_2^4 M_W}
-\frac{1875 C_A m_1 Y_s^3 M_H^2}{4096 m_2^5 M_W}
-\frac{123 C_A m_1 Y_s^3 M_H^2}{512 \varepsilon m_2^5 M_W}
+\frac{3 C_A m_1 Y_s^3 M_H^2}{512 \varepsilon ^2 m_2^5 M_W}
+\frac{51 C_A m_1 Y_s^3 M_H^2}{512 \varepsilon ^3 m_2^5 M_W}
+\frac{15 C_A Y_s^3 M_H^2}{128 m_- m_2^5 M_W}
+\frac{3 C_A Y_s^3 M_H^2}{16 \varepsilon m_- m_2^5 M_W}
+\frac{9 C_A Y_s^3 M_H^2}{128 \varepsilon ^2 m_- m_2^5 M_W}
+\frac{3 C_A Y_s^3 M_H^2}{128 \varepsilon ^3 m_- m_2^5 M_W}
+\frac{51 C_A Y_s^3 M_H^2}{128 m_2^6 M_W}
-\frac{9 C_A Y_s^3 M_H^2}{32 \varepsilon m_2^6 M_W}
-\frac{15 C_A Y_s^3 M_H^2}{128 \varepsilon ^2 m_2^6 M_W}
-\frac{9 C_A Y_s^3 M_H^2}{128 \varepsilon ^3 m_2^6 M_W}
+\frac{363 C_A Y_s^3 \mathcal{L}_{m_2}^2 M_H^2}{512 m_2^4 M_W}
-\frac{81 C_A Y_s^3 \mathcal{L}_{m_2}^2 M_H^2}{256 \varepsilon m_2^4 M_W}
-\frac{249 C_A m_1 Y_s^3 \mathcal{L}_{m_2}^2 M_H^2}{512 m_2^5 M_W}
+\frac{51 C_A m_1 Y_s^3 \mathcal{L}_{m_2}^2 M_H^2}{256 \varepsilon m_2^5 M_W}
+\frac{3 C_A Y_s^3 \mathcal{L}_{m_2}^2 M_H^2}{64 \varepsilon m_- m_2^5 M_W}
+\frac{3 C_A Y_s^3 \mathcal{L}_{m_2}^2 M_H^2}{16 m_2^6 M_W}
-\frac{9 C_A Y_s^3 \mathcal{L}_{m_2}^2 M_H^2}{64 \varepsilon m_2^6 M_W}
+\frac{1815 C_A Y_s^3 \zeta_2 M_H^2}{1024 m_2^4 M_W}
-\frac{405 C_A Y_s^3 \zeta_2 M_H^2}{512 \varepsilon m_2^4 M_W}
-\frac{1245 C_A m_1 Y_s^3 \zeta_2 M_H^2}{1024 m_2^5 M_W}
+\frac{255 C_A m_1 Y_s^3 \zeta_2 M_H^2}{512 \varepsilon m_2^5 M_W}
+\frac{3 C_A Y_s^3 \zeta_2 M_H^2}{128 \varepsilon m_- m_2^5 M_W}
+\frac{3 C_A Y_s^3 \zeta_2 M_H^2}{32 m_2^6 M_W}
-\frac{9 C_A Y_s^3 \zeta_2 M_H^2}{128 \varepsilon m_2^6 M_W}
-\frac{1311 C_A Y_s^3 \mathcal{L}_{m_2} M_H^2}{1024 m_2^4 M_W}
+\frac{21 C_A Y_s^3 \mathcal{L}_{m_2} M_H^2}{256 \varepsilon m_2^4 M_W}
+\frac{81 C_A Y_s^3 \mathcal{L}_{m_2} M_H^2}{256 \varepsilon ^2 m_2^4 M_W}
+\frac{1053 C_A m_1 Y_s^3 \mathcal{L}_{m_2} M_H^2}{1024 m_2^5 M_W}
-\frac{3 C_A m_1 Y_s^3 \mathcal{L}_{m_2} M_H^2}{256 \varepsilon m_2^5 M_W}
-\frac{51 C_A m_1 Y_s^3 \mathcal{L}_{m_2} M_H^2}{256 \varepsilon ^2 m_2^5 M_W}
-\frac{3 C_A Y_s^3 \mathcal{L}_{m_2} M_H^2}{64 m_- m_2^5 M_W}
-\frac{9 C_A Y_s^3 \mathcal{L}_{m_2} M_H^2}{64 \varepsilon m_- m_2^5 M_W}
-\frac{3 C_A Y_s^3 \mathcal{L}_{m_2} M_H^2}{64 \varepsilon ^2 m_- m_2^5 M_W}
-\frac{27 C_A Y_s^3 \mathcal{L}_{m_2} M_H^2}{64 m_2^6 M_W}
+\frac{15 C_A Y_s^3 \mathcal{L}_{m_2} M_H^2}{64 \varepsilon m_2^6 M_W}
+\frac{9 C_A Y_s^3 \mathcal{L}_{m_2} M_H^2}{64 \varepsilon ^2 m_2^6 M_W}
-\frac{111 C_A Y_s^4}{256 m_2^4}
-\frac{3 C_A Y_s^4}{512 \varepsilon m_2^4}
+\frac{13 C_A Y_s^4}{128 \varepsilon ^2 m_2^4}
+\frac{C_A Y_s^4}{32 \varepsilon ^3 m_2^4}
+\frac{151 C_A m_1 Y_s^4}{512 m_2^5}
+\frac{5 C_A m_1 Y_s^4}{512 \varepsilon m_2^5}
-\frac{11 C_A m_1 Y_s^4}{128 \varepsilon ^2 m_2^5}
-\frac{C_A m_1 Y_s^4}{64 \varepsilon ^3 m_2^5}
-\frac{177 C_A Y_s^4}{128 m_2^6}
-\frac{175 C_A Y_s^4}{512 \varepsilon m_2^6}
+\frac{127 C_A Y_s^4}{1024 \varepsilon ^2 m_2^6}
-\frac{29 C_A Y_s^4}{1024 \varepsilon ^3 m_2^6}
+\frac{5 C_A Y_s^4}{256 \varepsilon ^4 m_2^6}
+\frac{297 C_A m_1 Y_s^4}{256 m_2^7}
+\frac{153 C_A m_1 Y_s^4}{512 \varepsilon m_2^7}
-\frac{171 C_A m_1 Y_s^4}{2048 \varepsilon ^2 m_2^7}
-\frac{11 C_A m_1 Y_s^4}{1024 \varepsilon ^3 m_2^7}
-\frac{3 C_A m_1 Y_s^4}{256 \varepsilon ^4 m_2^7}
+\frac{2361}{32} C_A C_F^2
-\frac{13 C_A C_F^2 M_W^2 Y_s^2}{4 m_- m_2^3}
-\frac{13 C_F M_W^2 Y_s^2}{8 m_- m_2^3}
-\frac{5 C_A C_F^2 M_W^2 Y_s^2}{16 \varepsilon m_- m_2^3}
-\frac{5 C_F M_W^2 Y_s^2}{32 \varepsilon m_- m_2^3}
+\frac{C_A C_F^2 M_W^2 Y_s^2}{8 \varepsilon ^2 m_- m_2^3}
+\frac{C_F M_W^2 Y_s^2}{16 \varepsilon ^2 m_- m_2^3}
+\frac{65 C_A C_F^2 M_W^2 Y_s^2}{16 m_2^4}
+\frac{65 C_F M_W^2 Y_s^2}{32 m_2^4}
+\frac{C_A C_F^2 M_W^2 Y_s^2}{4 \varepsilon m_2^4}
+\frac{C_F M_W^2 Y_s^2}{8 \varepsilon m_2^4}
+\frac{C_A C_F^2 M_W^2 Y_s^2}{4 \varepsilon ^2 m_2^4}
+\frac{C_F M_W^2 Y_s^2}{8 \varepsilon ^2 m_2^4}
+\frac{89 C_A C_F^2 Y_s^2}{16 m_2^2}
+\frac{247 C_A C_F Y_s^2}{32 m_2^2}
+\frac{89 C_F Y_s^2}{32 m_2^2}
+\frac{19 C_A C_F^2 Y_s^2}{16 \varepsilon m_2^2}
-\frac{313 C_A C_F Y_s^2}{128 \varepsilon m_2^2}
+\frac{19 C_F Y_s^2}{32 \varepsilon m_2^2}
+\frac{C_A C_F^2 Y_s^2}{16 \varepsilon ^2 m_2^2}
-\frac{133 C_A C_F Y_s^2}{64 \varepsilon ^2 m_2^2}
+\frac{C_F Y_s^2}{32 \varepsilon ^2 m_2^2}
-\frac{3 C_A C_F Y_s^2}{4 \varepsilon ^3 m_2^2}
-\frac{41 C_A C_F^2 m_1 Y_s^2}{16 m_2^3}
-\frac{77 C_A C_F m_1 Y_s^2}{32 m_2^3}
-\frac{41 C_F m_1 Y_s^2}{32 m_2^3}
-\frac{7 C_A C_F^2 m_1 Y_s^2}{16 \varepsilon m_2^3}
+\frac{267 C_A C_F m_1 Y_s^2}{128 \varepsilon m_2^3}
-\frac{7 C_F m_1 Y_s^2}{32 \varepsilon m_2^3}
+\frac{C_A C_F^2 m_1 Y_s^2}{16 \varepsilon ^2 m_2^3}
+\frac{111 C_A C_F m_1 Y_s^2}{64 \varepsilon ^2 m_2^3}
+\frac{C_F m_1 Y_s^2}{32 \varepsilon ^2 m_2^3}
+\frac{3 C_A C_F m_1 Y_s^2}{8 \varepsilon ^3 m_2^3}
-\frac{65 C_A C_F Y_s^2}{32 m_2^4}
-\frac{135 C_A C_F Y_s^2}{32 \varepsilon m_2^4}
-\frac{261 C_A C_F Y_s^2}{32 \varepsilon ^2 m_2^4}
-\frac{C_A C_F Y_s^2}{64 \varepsilon ^3 m_2^4}
-\frac{3 C_A C_F Y_s^2}{16 \varepsilon ^4 m_2^4}
+\frac{3 C_A Y_s^4 \mathcal{L}_{m_2}^2}{64 m_2^4}
+\frac{C_A Y_s^4 \mathcal{L}_{m_2}^2}{16 \varepsilon m_2^4}
-\frac{3 C_A m_1 Y_s^4 \mathcal{L}_{m_2}^2}{32 m_2^5}
-\frac{C_A m_1 Y_s^4 \mathcal{L}_{m_2}^2}{32 \varepsilon m_2^5}
+\frac{57 C_A Y_s^4 \mathcal{L}_{m_2}^2}{128 m_2^6}
-\frac{69 C_A Y_s^4 \mathcal{L}_{m_2}^2}{512 \varepsilon m_2^6}
+\frac{5 C_A Y_s^4 \mathcal{L}_{m_2}^2}{128 \varepsilon ^2 m_2^6}
-\frac{153 C_A m_1 Y_s^4 \mathcal{L}_{m_2}^2}{1024 m_2^7}
+\frac{13 C_A m_1 Y_s^4 \mathcal{L}_{m_2}^2}{512 \varepsilon m_2^7}
-\frac{3 C_A m_1 Y_s^4 \mathcal{L}_{m_2}^2}{128 \varepsilon ^2 m_2^7}+16 C_A C_F^2 \mathcal{L}_{m_2}^2
+\frac{C_A C_F^2 M_W^2 Y_s^2 \mathcal{L}_{m_2}^2}{4 m_- m_2^3}
+\frac{C_F M_W^2 Y_s^2 \mathcal{L}_{m_2}^2}{8 m_- m_2^3}
+\frac{C_A C_F^2 M_W^2 Y_s^2 \mathcal{L}_{m_2}^2}{2 m_2^4}
+\frac{C_F M_W^2 Y_s^2 \mathcal{L}_{m_2}^2}{4 m_2^4}
+\frac{C_A C_F^2 Y_s^2 \mathcal{L}_{m_2}^2}{8 m_2^2}
-\frac{13 C_A C_F Y_s^2 \mathcal{L}_{m_2}^2}{32 m_2^2}
+\frac{C_F Y_s^2 \mathcal{L}_{m_2}^2}{16 m_2^2}
-\frac{3 C_A C_F Y_s^2 \mathcal{L}_{m_2}^2}{2 \varepsilon m_2^2}
+\frac{C_A C_F^2 m_1 Y_s^2 \mathcal{L}_{m_2}^2}{8 m_2^3}
+\frac{51 C_A C_F m_1 Y_s^2 \mathcal{L}_{m_2}^2}{32 m_2^3}
+\frac{C_F m_1 Y_s^2 \mathcal{L}_{m_2}^2}{16 m_2^3}
+\frac{3 C_A C_F m_1 Y_s^2 \mathcal{L}_{m_2}^2}{4 \varepsilon m_2^3}
-\frac{565 C_A C_F Y_s^2 \mathcal{L}_{m_2}^2}{32 m_2^4}
+\frac{23 C_A C_F Y_s^2 \mathcal{L}_{m_2}^2}{32 \varepsilon m_2^4}
-\frac{3 C_A C_F Y_s^2 \mathcal{L}_{m_2}^2}{8 \varepsilon ^2 m_2^4}
+\frac{13}{2} C_F \mathcal{L}_{m_2}^2
-\frac{87 C_A C_F M_W Y_s \mathcal{L}_{m_2}^2}{16 m_2^2}
+\frac{3 C_A C_F M_W Y_s \mathcal{L}_{m_2}^2}{8 \varepsilon m_2^2}
-\frac{2 C_A C_F^2 \mathcal{L}_{m_2}^2}{\varepsilon }
-\frac{3 C_F \mathcal{L}_{m_2}^2}{\varepsilon }
-\frac{51 C_A C_F}{32}
-\frac{887 C_F}{64}
+\frac{11}{8} C_A C_F n_f T_f
+\frac{C_A C_F n_f T_f}{4 \varepsilon }
-\frac{1515 C_A C_F M_W Y_s}{128 m_2^2}
-\frac{41 C_A C_F M_W Y_s}{8 \varepsilon m_2^2}
-\frac{9 C_A C_F M_W Y_s}{4 \varepsilon ^2 m_2^2}
+\frac{3 C_A C_F M_W Y_s}{16 \varepsilon ^3 m_2^2}
+\frac{15 C_A Y_s^4 \zeta_2}{128 m_2^4}
+\frac{5 C_A Y_s^4 \zeta_2}{32 \varepsilon m_2^4}
-\frac{15 C_A m_1 Y_s^4 \zeta_2}{64 m_2^5}
-\frac{5 C_A m_1 Y_s^4 \zeta_2}{64 \varepsilon m_2^5}
+\frac{409 C_A Y_s^4 \zeta_2}{256 m_2^6}
-\frac{413 C_A Y_s^4 \zeta_2}{1024 \varepsilon m_2^6}
+\frac{25 C_A Y_s^4 \zeta_2}{256 \varepsilon ^2 m_2^6}
-\frac{1345 C_A m_1 Y_s^4 \zeta_2}{2048 m_2^7}
+\frac{157 C_A m_1 Y_s^4 \zeta_2}{1024 \varepsilon m_2^7}
-\frac{15 C_A m_1 Y_s^4 \zeta_2}{256 \varepsilon ^2 m_2^7}+40 C_A C_F^2 \zeta_2
+\frac{5 C_A C_F^2 M_W^2 Y_s^2 \zeta_2}{8 m_- m_2^3}
+\frac{5 C_F M_W^2 Y_s^2 \zeta_2}{16 m_- m_2^3}
+\frac{5 C_A C_F^2 M_W^2 Y_s^2 \zeta_2}{4 m_2^4}
+\frac{5 C_F M_W^2 Y_s^2 \zeta_2}{8 m_2^4}
+\frac{5 C_A C_F^2 Y_s^2 \zeta_2}{16 m_2^2}
-\frac{65 C_A C_F Y_s^2 \zeta_2}{64 m_2^2}
+\frac{5 C_F Y_s^2 \zeta_2}{32 m_2^2}
-\frac{15 C_A C_F Y_s^2 \zeta_2}{4 \varepsilon m_2^2}
+\frac{5 C_A C_F^2 m_1 Y_s^2 \zeta_2}{16 m_2^3}
+\frac{255 C_A C_F m_1 Y_s^2 \zeta_2}{64 m_2^3}
+\frac{5 C_F m_1 Y_s^2 \zeta_2}{32 m_2^3}
+\frac{15 C_A C_F m_1 Y_s^2 \zeta_2}{8 \varepsilon m_2^3}
-\frac{2397 C_A C_F Y_s^2 \zeta_2}{64 m_2^4}
+\frac{79 C_A C_F Y_s^2 \zeta_2}{64 \varepsilon m_2^4}
-\frac{15 C_A C_F Y_s^2 \zeta_2}{16 \varepsilon ^2 m_2^4}
+\frac{65 C_F \zeta_2}{4}
-\frac{435 C_A C_F M_W Y_s \zeta_2}{32 m_2^2}
+\frac{15 C_A C_F M_W Y_s \zeta_2}{16 \varepsilon m_2^2}
-\frac{5 C_A C_F^2 \zeta_2}{\varepsilon }
-\frac{15 C_F \zeta_2}{2 \varepsilon }
+\frac{47 C_A Y_s^4 \mathcal{L}_{m_2}}{256 m_2^4}
-\frac{13 C_A Y_s^4 \mathcal{L}_{m_2}}{64 \varepsilon m_2^4}
-\frac{C_A Y_s^4 \mathcal{L}_{m_2}}{16 \varepsilon ^2 m_2^4}
-\frac{27 C_A m_1 Y_s^4 \mathcal{L}_{m_2}}{256 m_2^5}
+\frac{11 C_A m_1 Y_s^4 \mathcal{L}_{m_2}}{64 \varepsilon m_2^5}
+\frac{C_A m_1 Y_s^4 \mathcal{L}_{m_2}}{32 \varepsilon ^2 m_2^5}
+\frac{383 C_A Y_s^4 \mathcal{L}_{m_2}}{512 m_2^6}
-\frac{107 C_A Y_s^4 \mathcal{L}_{m_2}}{512 \varepsilon m_2^6}
+\frac{29 C_A Y_s^4 \mathcal{L}_{m_2}}{512 \varepsilon ^2 m_2^6}
-\frac{5 C_A Y_s^4 \mathcal{L}_{m_2}}{128 \varepsilon ^3 m_2^6}
-\frac{431 C_A m_1 Y_s^4 \mathcal{L}_{m_2}}{512 m_2^7}
+\frac{147 C_A m_1 Y_s^4 \mathcal{L}_{m_2}}{1024 \varepsilon m_2^7}
+\frac{11 C_A m_1 Y_s^4 \mathcal{L}_{m_2}}{512 \varepsilon ^2 m_2^7}
+\frac{3 C_A m_1 Y_s^4 \mathcal{L}_{m_2}}{128 \varepsilon ^3 m_2^7}
-\frac{377}{8} C_A C_F^2 \mathcal{L}_{m_2}
+\frac{5 C_A C_F^2 M_W^2 Y_s^2 \mathcal{L}_{m_2}}{8 m_- m_2^3}
+\frac{5 C_F M_W^2 Y_s^2 \mathcal{L}_{m_2}}{16 m_- m_2^3}
-\frac{C_A C_F^2 M_W^2 Y_s^2 \mathcal{L}_{m_2}}{4 \varepsilon m_- m_2^3}
-\frac{C_F M_W^2 Y_s^2 \mathcal{L}_{m_2}}{8 \varepsilon m_- m_2^3}
-\frac{C_A C_F^2 M_W^2 Y_s^2 \mathcal{L}_{m_2}}{2 m_2^4}
-\frac{C_F M_W^2 Y_s^2 \mathcal{L}_{m_2}}{4 m_2^4}
-\frac{C_A C_F^2 M_W^2 Y_s^2 \mathcal{L}_{m_2}}{2 \varepsilon m_2^4}
-\frac{C_F M_W^2 Y_s^2 \mathcal{L}_{m_2}}{4 \varepsilon m_2^4}
-\frac{19 C_A C_F^2 Y_s^2 \mathcal{L}_{m_2}}{8 m_2^2}
+\frac{49 C_A C_F Y_s^2 \mathcal{L}_{m_2}}{64 m_2^2}
-\frac{19 C_F Y_s^2 \mathcal{L}_{m_2}}{16 m_2^2}
-\frac{C_A C_F^2 Y_s^2 \mathcal{L}_{m_2}}{8 \varepsilon m_2^2}
+\frac{133 C_A C_F Y_s^2 \mathcal{L}_{m_2}}{32 \varepsilon m_2^2}
-\frac{C_F Y_s^2 \mathcal{L}_{m_2}}{16 \varepsilon m_2^2}
+\frac{3 C_A C_F Y_s^2 \mathcal{L}_{m_2}}{2 \varepsilon ^2 m_2^2}
+\frac{7 C_A C_F^2 m_1 Y_s^2 \mathcal{L}_{m_2}}{8 m_2^3}
-\frac{135 C_A C_F m_1 Y_s^2 \mathcal{L}_{m_2}}{64 m_2^3}
+\frac{7 C_F m_1 Y_s^2 \mathcal{L}_{m_2}}{16 m_2^3}
-\frac{C_A C_F^2 m_1 Y_s^2 \mathcal{L}_{m_2}}{8 \varepsilon m_2^3}
-\frac{111 C_A C_F m_1 Y_s^2 \mathcal{L}_{m_2}}{32 \varepsilon m_2^3}
-\frac{C_F m_1 Y_s^2 \mathcal{L}_{m_2}}{16 \varepsilon m_2^3}
-\frac{3 C_A C_F m_1 Y_s^2 \mathcal{L}_{m_2}}{4 \varepsilon ^2 m_2^3}
+\frac{347 C_A C_F Y_s^2 \mathcal{L}_{m_2}}{32 m_2^4}
+\frac{255 C_A C_F Y_s^2 \mathcal{L}_{m_2}}{16 \varepsilon m_2^4}
+\frac{C_A C_F Y_s^2 \mathcal{L}_{m_2}}{32 \varepsilon ^2 m_2^4}
+\frac{3 C_A C_F Y_s^2 \mathcal{L}_{m_2}}{8 \varepsilon ^3 m_2^4}
+\frac{15}{8} C_A C_F \mathcal{L}_{m_2}
+\frac{107}{16} C_F \mathcal{L}_{m_2}
-\frac{1}{2} C_A C_F n_f T_f \mathcal{L}_{m_2}
+\frac{361 C_A C_F M_W Y_s \mathcal{L}_{m_2}}{32 m_2^2}
+\frac{9 C_A C_F M_W Y_s \mathcal{L}_{m_2}}{2 \varepsilon m_2^2}
-\frac{3 C_A C_F M_W Y_s \mathcal{L}_{m_2}}{8 \varepsilon ^2 m_2^2}
-\frac{11 C_A C_F^2 \mathcal{L}_{m_2}}{\varepsilon }
+\frac{C_F \mathcal{L}_{m_2}}{\varepsilon }
+\frac{2 C_A C_F^2 \mathcal{L}_{m_2}}{\varepsilon ^2}
+\frac{3 C_F \mathcal{L}_{m_2}}{\varepsilon ^2}
+\frac{333 C_A C_F^2}{16 \varepsilon }
-\frac{15 C_A C_F}{16 \varepsilon }
-\frac{239 C_F}{32 \varepsilon }
+\frac{11 C_A C_F^2}{2 \varepsilon ^2}
-\frac{C_F}{2 \varepsilon ^2}
-\frac{C_A C_F^2}{\varepsilon ^3}
-\frac{3 C_F}{2 \varepsilon ^3}
\end{autobreak}
\end{align}
\end{tiny}

\begin{tiny}
\begin{align}
\begin{autobreak}
V^{(m_{1,2})}_6=
\frac{C_A C_F^2 Y_f Y_s \mathcal{L}_{m_2}^2 M_H^4}{128 m_- m_2^2 M_W^2}
+\frac{9 C_A Y_f Y_s \mathcal{L}_{m_2}^2 M_H^4}{128 m_- m_2^2 M_W^2}
+\frac{C_F Y_f Y_s \mathcal{L}_{m_2}^2 M_H^4}{256 m_- m_2^2 M_W^2}
-\frac{27 C_A C_F^2 Y_f Y_s M_H^4}{256 m_- m_2^2 M_W^2}
-\frac{243 C_A Y_f Y_s M_H^4}{256 m_- m_2^2 M_W^2}
-\frac{27 C_F Y_f Y_s M_H^4}{512 m_- m_2^2 M_W^2}
-\frac{C_A C_F^2 Y_f Y_s M_H^4}{128 \varepsilon m_- m_2^2 M_W^2}
-\frac{9 C_A Y_f Y_s M_H^4}{128 \varepsilon m_- m_2^2 M_W^2}
-\frac{C_F Y_f Y_s M_H^4}{256 \varepsilon m_- m_2^2 M_W^2}
+\frac{C_A C_F^2 Y_f Y_s M_H^4}{256 \varepsilon ^2 m_- m_2^2 M_W^2}
+\frac{9 C_A Y_f Y_s M_H^4}{256 \varepsilon ^2 m_- m_2^2 M_W^2}
+\frac{C_F Y_f Y_s M_H^4}{512 \varepsilon ^2 m_- m_2^2 M_W^2}
+\frac{5 C_A C_F^2 Y_f Y_s \zeta_2 M_H^4}{256 m_- m_2^2 M_W^2}
+\frac{45 C_A Y_f Y_s \zeta_2 M_H^4}{256 m_- m_2^2 M_W^2}
+\frac{5 C_F Y_f Y_s \zeta_2 M_H^4}{512 m_- m_2^2 M_W^2}
+\frac{C_A C_F^2 Y_f Y_s \mathcal{L}_{m_2} M_H^4}{64 m_- m_2^2 M_W^2}
+\frac{9 C_A Y_f Y_s \mathcal{L}_{m_2} M_H^4}{64 m_- m_2^2 M_W^2}
+\frac{C_F Y_f Y_s \mathcal{L}_{m_2} M_H^4}{128 m_- m_2^2 M_W^2}
-\frac{C_A C_F^2 Y_f Y_s \mathcal{L}_{m_2} M_H^4}{128 \varepsilon m_- m_2^2 M_W^2}
-\frac{9 C_A Y_f Y_s \mathcal{L}_{m_2} M_H^4}{128 \varepsilon m_- m_2^2 M_W^2}
-\frac{C_F Y_f Y_s \mathcal{L}_{m_2} M_H^4}{256 \varepsilon m_- m_2^2 M_W^2}
+\frac{357 C_A Y_f Y_s^2 M_H^2}{512 m_- m_2^2 M_W}
+\frac{3 C_A Y_f Y_s^2 M_H^2}{16 \varepsilon m_- m_2^2 M_W}
+\frac{3 C_A Y_f Y_s^2 M_H^2}{32 \varepsilon ^2 m_- m_2^2 M_W}
+\frac{3 C_A Y_f Y_s^2 M_H^2}{64 \varepsilon ^3 m_- m_2^2 M_W}
+\frac{17973 C_A Y_f Y_s^2 M_H^2}{8192 m_2^3 M_W}
+\frac{1509 C_A Y_f Y_s^2 M_H^2}{1024 \varepsilon m_2^3 M_W}
+\frac{513 C_A Y_f Y_s^2 M_H^2}{1024 \varepsilon ^2 m_2^3 M_W}
-\frac{309 C_A Y_f Y_s^2 M_H^2}{1024 \varepsilon ^3 m_2^3 M_W}
-\frac{16797 C_A m_1 Y_f Y_s^2 M_H^2}{8192 m_2^4 M_W}
-\frac{1185 C_A m_1 Y_f Y_s^2 M_H^2}{1024 \varepsilon m_2^4 M_W}
-\frac{405 C_A m_1 Y_f Y_s^2 M_H^2}{1024 \varepsilon ^2 m_2^4 M_W}
+\frac{189 C_A m_1 Y_f Y_s^2 M_H^2}{1024 \varepsilon ^3 m_2^4 M_W}
-\frac{3 C_A Y_f Y_s^2 \mathcal{L}_{m_2}^2 M_H^2}{64 m_- m_2^2 M_W}
+\frac{3 C_A Y_f Y_s^2 \mathcal{L}_{m_2}^2 M_H^2}{32 \varepsilon m_- m_2^2 M_W}
+\frac{2571 C_A Y_f Y_s^2 \mathcal{L}_{m_2}^2 M_H^2}{1024 m_2^3 M_W}
-\frac{309 C_A Y_f Y_s^2 \mathcal{L}_{m_2}^2 M_H^2}{512 \varepsilon m_2^3 M_W}
-\frac{1755 C_A m_1 Y_f Y_s^2 \mathcal{L}_{m_2}^2 M_H^2}{1024 m_2^4 M_W}
+\frac{189 C_A m_1 Y_f Y_s^2 \mathcal{L}_{m_2}^2 M_H^2}{512 \varepsilon m_2^4 M_W}
+\frac{3 C_A Y_f^2 Y_s \mathcal{L}_{m_2}^2 M_H^2}{32 m_- m_2 M_W}
-\frac{3 C_A Y_f^2 Y_s \mathcal{L}_{m_2}^2 M_H^2}{16 \varepsilon m_- m_2 M_W}
-\frac{75 C_A Y_f^2 Y_s \mathcal{L}_{m_2}^2 M_H^2}{256 m_2^2 M_W}
+\frac{21 C_A Y_f^2 Y_s \mathcal{L}_{m_2}^2 M_H^2}{128 \varepsilon m_2^2 M_W}
-\frac{357 C_A Y_f^2 Y_s M_H^2}{256 m_- m_2 M_W}
-\frac{3 C_A Y_f^2 Y_s M_H^2}{8 \varepsilon m_- m_2 M_W}
-\frac{3 C_A Y_f^2 Y_s M_H^2}{16 \varepsilon ^2 m_- m_2 M_W}
-\frac{3 C_A Y_f^2 Y_s M_H^2}{32 \varepsilon ^3 m_- m_2 M_W}
+\frac{1755 C_A Y_f^2 Y_s M_H^2}{2048 m_2^2 M_W}
+\frac{39 C_A Y_f^2 Y_s M_H^2}{256 \varepsilon m_2^2 M_W}
+\frac{15 C_A Y_f^2 Y_s M_H^2}{256 \varepsilon ^2 m_2^2 M_W}
+\frac{21 C_A Y_f^2 Y_s M_H^2}{256 \varepsilon ^3 m_2^2 M_W}
-\frac{15 C_A Y_f Y_s^2 \zeta_2 M_H^2}{128 m_- m_2^2 M_W}
+\frac{15 C_A Y_f Y_s^2 \zeta_2 M_H^2}{64 \varepsilon m_- m_2^2 M_W}
+\frac{12855 C_A Y_f Y_s^2 \zeta_2 M_H^2}{2048 m_2^3 M_W}
-\frac{1545 C_A Y_f Y_s^2 \zeta_2 M_H^2}{1024 \varepsilon m_2^3 M_W}
-\frac{8775 C_A m_1 Y_f Y_s^2 \zeta_2 M_H^2}{2048 m_2^4 M_W}
+\frac{945 C_A m_1 Y_f Y_s^2 \zeta_2 M_H^2}{1024 \varepsilon m_2^4 M_W}
+\frac{15 C_A Y_f^2 Y_s \zeta_2 M_H^2}{64 m_- m_2 M_W}
-\frac{15 C_A Y_f^2 Y_s \zeta_2 M_H^2}{32 \varepsilon m_- m_2 M_W}
-\frac{375 C_A Y_f^2 Y_s \zeta_2 M_H^2}{512 m_2^2 M_W}
+\frac{105 C_A Y_f^2 Y_s \zeta_2 M_H^2}{256 \varepsilon m_2^2 M_W}
-\frac{15 C_A Y_f Y_s^2 \mathcal{L}_{m_2} M_H^2}{128 m_- m_2^2 M_W}
-\frac{3 C_A Y_f Y_s^2 \mathcal{L}_{m_2} M_H^2}{16 \varepsilon m_- m_2^2 M_W}
-\frac{3 C_A Y_f Y_s^2 \mathcal{L}_{m_2} M_H^2}{32 \varepsilon ^2 m_- m_2^2 M_W}
-\frac{9435 C_A Y_f Y_s^2 \mathcal{L}_{m_2} M_H^2}{2048 m_2^3 M_W}
-\frac{513 C_A Y_f Y_s^2 \mathcal{L}_{m_2} M_H^2}{512 \varepsilon m_2^3 M_W}
+\frac{309 C_A Y_f Y_s^2 \mathcal{L}_{m_2} M_H^2}{512 \varepsilon ^2 m_2^3 M_W}
+\frac{6819 C_A m_1 Y_f Y_s^2 \mathcal{L}_{m_2} M_H^2}{2048 m_2^4 M_W}
+\frac{405 C_A m_1 Y_f Y_s^2 \mathcal{L}_{m_2} M_H^2}{512 \varepsilon m_2^4 M_W}
-\frac{189 C_A m_1 Y_f Y_s^2 \mathcal{L}_{m_2} M_H^2}{512 \varepsilon ^2 m_2^4 M_W}
+\frac{15 C_A Y_f^2 Y_s \mathcal{L}_{m_2} M_H^2}{64 m_- m_2 M_W}
+\frac{3 C_A Y_f^2 Y_s \mathcal{L}_{m_2} M_H^2}{8 \varepsilon m_- m_2 M_W}
+\frac{3 C_A Y_f^2 Y_s \mathcal{L}_{m_2} M_H^2}{16 \varepsilon ^2 m_- m_2 M_W}
+\frac{75 C_A Y_f^2 Y_s \mathcal{L}_{m_2} M_H^2}{512 m_2^2 M_W}
-\frac{15 C_A Y_f^2 Y_s \mathcal{L}_{m_2} M_H^2}{128 \varepsilon m_2^2 M_W}
-\frac{21 C_A Y_f^2 Y_s \mathcal{L}_{m_2} M_H^2}{128 \varepsilon ^2 m_2^2 M_W}
+\frac{3721 C_A Y_f Y_s^3}{2048 m_2^3}
+\frac{49 C_A Y_f Y_s^3}{64 \varepsilon m_2^3}
+\frac{3 C_A Y_f Y_s^3}{16 \varepsilon ^2 m_2^3}
-\frac{C_A Y_f Y_s^3}{256 \varepsilon ^3 m_2^3}
-\frac{3099 C_A m_1 Y_f Y_s^3}{2048 m_2^4}
-\frac{39 C_A m_1 Y_f Y_s^3}{64 \varepsilon m_2^4}
-\frac{3 C_A m_1 Y_f Y_s^3}{16 \varepsilon ^2 m_2^4}
+\frac{3 C_A m_1 Y_f Y_s^3}{256 \varepsilon ^3 m_2^4}
-\frac{27355 C_A Y_f Y_s^3}{4096 m_2^5}
-\frac{1555 C_A Y_f Y_s^3}{1024 \varepsilon m_2^5}
-\frac{1649 C_A Y_f Y_s^3}{4096 \varepsilon ^2 m_2^5}
+\frac{7 C_A Y_f Y_s^3}{4096 \varepsilon ^3 m_2^5}
+\frac{C_A Y_f Y_s^3}{1024 \varepsilon ^4 m_2^5}
+\frac{21391 C_A m_1 Y_f Y_s^3}{4096 m_2^6}
+\frac{1291 C_A m_1 Y_f Y_s^3}{1024 \varepsilon m_2^6}
+\frac{1491 C_A m_1 Y_f Y_s^3}{4096 \varepsilon ^2 m_2^6}
-\frac{175 C_A m_1 Y_f Y_s^3}{4096 \varepsilon ^3 m_2^6}
-\frac{C_A m_1 Y_f Y_s^3}{1024 \varepsilon ^4 m_2^6}
-\frac{3827}{64} C_A C_F^2
+\frac{245}{16} C_A C_F Y_f^2
+\frac{13 C_A C_F Y_f^2}{4 \varepsilon }
+\frac{7 C_A C_F Y_f^2}{4 \varepsilon ^2}
-\frac{C_A C_F Y_f^2}{2 \varepsilon ^3}
-\frac{1145 C_A Y_f^2 Y_s^2}{2048 m_2^2}
-\frac{145 C_A Y_f^2 Y_s^2}{256 \varepsilon m_2^2}
-\frac{15 C_A Y_f^2 Y_s^2}{256 \varepsilon ^2 m_2^2}
+\frac{9 C_A Y_f^2 Y_s^2}{256 \varepsilon ^3 m_2^2}
+\frac{1711 C_A m_1 Y_f^2 Y_s^2}{2048 m_2^3}
+\frac{65 C_A m_1 Y_f^2 Y_s^2}{128 \varepsilon m_2^3}
+\frac{33 C_A m_1 Y_f^2 Y_s^2}{256 \varepsilon ^2 m_2^3}
-\frac{7 C_A m_1 Y_f^2 Y_s^2}{256 \varepsilon ^3 m_2^3}
+\frac{1513 C_A Y_f^2 Y_s^2}{1024 m_2^4}
+\frac{209 C_A Y_f^2 Y_s^2}{512 \varepsilon m_2^4}
+\frac{27 C_A Y_f^2 Y_s^2}{256 \varepsilon ^2 m_2^4}
-\frac{7 C_A Y_f^2 Y_s^2}{1024 \varepsilon ^3 m_2^4}
+\frac{3 C_A Y_f^2 Y_s^2}{256 \varepsilon ^4 m_2^4}
-\frac{913 C_A C_F Y_s^2}{256 m_2^2}
-\frac{77 C_A C_F Y_s^2}{32 \varepsilon m_2^2}
-\frac{31 C_A C_F Y_s^2}{32 \varepsilon ^2 m_2^2}
+\frac{C_A C_F Y_s^2}{32 \varepsilon ^3 m_2^2}
+\frac{1059 C_A C_F m_1 Y_s^2}{256 m_2^3}
+\frac{69 C_A C_F m_1 Y_s^2}{32 \varepsilon m_2^3}
+\frac{39 C_A C_F m_1 Y_s^2}{32 \varepsilon ^2 m_2^3}
-\frac{3 C_A C_F m_1 Y_s^2}{32 \varepsilon ^3 m_2^3}
+\frac{539 C_A C_F Y_s^2}{128 m_2^4}
-\frac{65 C_A C_F Y_s^2}{64 \varepsilon m_2^4}
-\frac{643 C_A C_F Y_s^2}{256 \varepsilon ^2 m_2^4}
+\frac{15 C_A C_F Y_s^2}{64 \varepsilon ^3 m_2^4}
+\frac{101 C_A Y_f Y_s^3 \mathcal{L}_{m_2}^2}{256 m_2^3}
-\frac{C_A Y_f Y_s^3 \mathcal{L}_{m_2}^2}{128 \varepsilon m_2^3}
-\frac{111 C_A m_1 Y_f Y_s^3 \mathcal{L}_{m_2}^2}{256 m_2^4}
+\frac{3 C_A m_1 Y_f Y_s^3 \mathcal{L}_{m_2}^2}{128 \varepsilon m_2^4}
-\frac{63 C_A Y_f Y_s^3 \mathcal{L}_{m_2}^2}{64 m_2^5}
-\frac{C_A Y_f Y_s^3 \mathcal{L}_{m_2}^2}{2048 \varepsilon m_2^5}
+\frac{C_A Y_f Y_s^3 \mathcal{L}_{m_2}^2}{512 \varepsilon ^2 m_2^5}
+\frac{1065 C_A m_1 Y_f Y_s^3 \mathcal{L}_{m_2}^2}{1024 m_2^6}
-\frac{167 C_A m_1 Y_f Y_s^3 \mathcal{L}_{m_2}^2}{2048 \varepsilon m_2^6}
-\frac{C_A m_1 Y_f Y_s^3 \mathcal{L}_{m_2}^2}{512 \varepsilon ^2 m_2^6}
-\frac{23}{8} C_A C_F^2 \mathcal{L}_{m_2}^2+6 C_A C_F Y_f^2 \mathcal{L}_{m_2}^2
-\frac{C_A C_F Y_f^2 \mathcal{L}_{m_2}^2}{\varepsilon }
-\frac{75 C_A Y_f^2 Y_s^2 \mathcal{L}_{m_2}^2}{256 m_2^2}
+\frac{9 C_A Y_f^2 Y_s^2 \mathcal{L}_{m_2}^2}{128 \varepsilon m_2^2}
+\frac{101 C_A m_1 Y_f^2 Y_s^2 \mathcal{L}_{m_2}^2}{256 m_2^3}
-\frac{7 C_A m_1 Y_f^2 Y_s^2 \mathcal{L}_{m_2}^2}{128 \varepsilon m_2^3}
+\frac{199 C_A Y_f^2 Y_s^2 \mathcal{L}_{m_2}^2}{512 m_2^4}
-\frac{31 C_A Y_f^2 Y_s^2 \mathcal{L}_{m_2}^2}{512 \varepsilon m_2^4}
+\frac{3 C_A Y_f^2 Y_s^2 \mathcal{L}_{m_2}^2}{128 \varepsilon ^2 m_2^4}
-\frac{67 C_A C_F Y_s^2 \mathcal{L}_{m_2}^2}{32 m_2^2}
+\frac{C_A C_F Y_s^2 \mathcal{L}_{m_2}^2}{16 \varepsilon m_2^2}
+\frac{93 C_A C_F m_1 Y_s^2 \mathcal{L}_{m_2}^2}{32 m_2^3}
-\frac{3 C_A C_F m_1 Y_s^2 \mathcal{L}_{m_2}^2}{16 \varepsilon m_2^3}
-\frac{731 C_A C_F Y_s^2 \mathcal{L}_{m_2}^2}{128 m_2^4}
+\frac{15 C_A C_F Y_s^2 \mathcal{L}_{m_2}^2}{32 \varepsilon m_2^4}
+\frac{1}{4} C_A C_F \mathcal{L}_{m_2}^2
+\frac{117}{16} C_F \mathcal{L}_{m_2}^2
+\frac{1}{4} C_A C_F n_f T_f \mathcal{L}_{m_2}^2
-\frac{C_A C_F M_W Y_f \mathcal{L}_{m_2}^2}{2 m_-}
+\frac{C_A C_F M_W Y_f \mathcal{L}_{m_2}^2}{\varepsilon m_-}
-\frac{7 C_A Y_f^3 Y_s \mathcal{L}_{m_2}^2}{8 m_2}
+\frac{C_A Y_f^3 Y_s \mathcal{L}_{m_2}^2}{32 \varepsilon m_2}
+\frac{C_A m_1 Y_f^3 Y_s \mathcal{L}_{m_2}^2}{4 m_2^2}
+\frac{C_A m_1 Y_f^3 Y_s \mathcal{L}_{m_2}^2}{32 \varepsilon m_2^2}
+\frac{C_A C_F M_W Y_s \mathcal{L}_{m_2}^2}{4 m_- m_2}
-\frac{C_A C_F M_W Y_s \mathcal{L}_{m_2}^2}{2 \varepsilon m_- m_2}
-\frac{13 C_A C_F M_W Y_s \mathcal{L}_{m_2}^2}{2 m_2^2}
+\frac{3 C_A C_F M_W Y_s \mathcal{L}_{m_2}^2}{4 \varepsilon m_2^2}
+\frac{C_A C_F^2 M_W^2 Y_f Y_s \mathcal{L}_{m_2}^2}{2 m_- m_2^2}
+\frac{C_F M_W^2 Y_f Y_s \mathcal{L}_{m_2}^2}{4 m_- m_2^2}
+\frac{C_A C_F^2 Y_f Y_s \mathcal{L}_{m_2}^2}{2 m_2}
+\frac{55 C_A C_F Y_f Y_s \mathcal{L}_{m_2}^2}{32 m_2}
+\frac{C_F Y_f Y_s \mathcal{L}_{m_2}^2}{4 m_2}
-\frac{33 C_A C_F Y_f Y_s \mathcal{L}_{m_2}^2}{16 \varepsilon m_2}
-\frac{27 C_A C_F m_1 Y_f Y_s \mathcal{L}_{m_2}^2}{32 m_2^2}
+\frac{13 C_A C_F m_1 Y_f Y_s \mathcal{L}_{m_2}^2}{16 \varepsilon m_2^2}
-\frac{5 C_A C_F^2 \mathcal{L}_{m_2}^2}{4 \varepsilon }
-\frac{49 C_F \mathcal{L}_{m_2}^2}{8 \varepsilon }
+\frac{19 C_A C_F}{8}
-\frac{6167 C_F}{128}
+\frac{7}{4} C_A C_F n_f T_f
+\frac{C_A C_F n_f T_f}{2 \varepsilon }
+\frac{C_A C_F n_f T_f}{8 \varepsilon ^2}
+\frac{119 C_A C_F M_W Y_f}{16 m_-}
+\frac{2 C_A C_F M_W Y_f}{\varepsilon m_-}
+\frac{C_A C_F M_W Y_f}{\varepsilon ^2 m_-}
+\frac{C_A C_F M_W Y_f}{2 \varepsilon ^3 m_-}
-\frac{1245 C_A Y_f^3 Y_s}{512 m_2}
-\frac{15 C_A Y_f^3 Y_s}{32 \varepsilon m_2}
-\frac{51 C_A Y_f^3 Y_s}{128 \varepsilon ^2 m_2}
+\frac{C_A Y_f^3 Y_s}{64 \varepsilon ^3 m_2}
+\frac{403 C_A m_1 Y_f^3 Y_s}{512 m_2^2}
+\frac{5 C_A m_1 Y_f^3 Y_s}{32 \varepsilon m_2^2}
+\frac{21 C_A m_1 Y_f^3 Y_s}{128 \varepsilon ^2 m_2^2}
+\frac{C_A m_1 Y_f^3 Y_s}{64 \varepsilon ^3 m_2^2}
-\frac{119 C_A C_F M_W Y_s}{32 m_- m_2}
-\frac{C_A C_F M_W Y_s}{\varepsilon m_- m_2}
-\frac{C_A C_F M_W Y_s}{2 \varepsilon ^2 m_- m_2}
-\frac{C_A C_F M_W Y_s}{4 \varepsilon ^3 m_- m_2}
-\frac{791 C_A C_F M_W Y_s}{64 m_2^2}
-\frac{93 C_A C_F M_W Y_s}{16 \varepsilon m_2^2}
-\frac{37 C_A C_F M_W Y_s}{16 \varepsilon ^2 m_2^2}
+\frac{3 C_A C_F M_W Y_s}{8 \varepsilon ^3 m_2^2}
-\frac{13 C_A C_F^2 M_W^2 Y_f Y_s}{2 m_- m_2^2}
-\frac{13 C_F M_W^2 Y_f Y_s}{4 m_- m_2^2}
-\frac{5 C_A C_F^2 M_W^2 Y_f Y_s}{8 \varepsilon m_- m_2^2}
-\frac{5 C_F M_W^2 Y_f Y_s}{16 \varepsilon m_- m_2^2}
+\frac{C_A C_F^2 M_W^2 Y_f Y_s}{4 \varepsilon ^2 m_- m_2^2}
+\frac{C_F M_W^2 Y_f Y_s}{8 \varepsilon ^2 m_- m_2^2}
+\frac{81 C_A C_F^2 Y_f Y_s}{16 m_2}
-\frac{1347 C_A C_F Y_f Y_s}{256 m_2}
+\frac{81 C_F Y_f Y_s}{32 m_2}
+\frac{21 C_A C_F^2 Y_f Y_s}{16 \varepsilon m_2}
-\frac{295 C_A C_F Y_f Y_s}{64 \varepsilon m_2}
+\frac{21 C_F Y_f Y_s}{32 \varepsilon m_2}
+\frac{C_A C_F^2 Y_f Y_s}{4 \varepsilon ^2 m_2}
-\frac{55 C_A C_F Y_f Y_s}{32 \varepsilon ^2 m_2}
+\frac{C_F Y_f Y_s}{8 \varepsilon ^2 m_2}
-\frac{33 C_A C_F Y_f Y_s}{32 \varepsilon ^3 m_2}
-\frac{25 C_A C_F^2 m_1 Y_f Y_s}{16 m_2^2}
+\frac{135 C_A C_F m_1 Y_f Y_s}{256 m_2^2}
-\frac{25 C_F m_1 Y_f Y_s}{32 m_2^2}
-\frac{5 C_A C_F^2 m_1 Y_f Y_s}{16 \varepsilon m_2^2}
+\frac{187 C_A C_F m_1 Y_f Y_s}{64 \varepsilon m_2^2}
-\frac{5 C_F m_1 Y_f Y_s}{32 \varepsilon m_2^2}
+\frac{19 C_A C_F m_1 Y_f Y_s}{32 \varepsilon ^2 m_2^2}
+\frac{13 C_A C_F m_1 Y_f Y_s}{32 \varepsilon ^3 m_2^2}
+\frac{505 C_A Y_f Y_s^3 \zeta_2}{512 m_2^3}
-\frac{5 C_A Y_f Y_s^3 \zeta_2}{256 \varepsilon m_2^3}
-\frac{555 C_A m_1 Y_f Y_s^3 \zeta_2}{512 m_2^4}
+\frac{15 C_A m_1 Y_f Y_s^3 \zeta_2}{256 \varepsilon m_2^4}
+\frac{193 C_A Y_f Y_s^3 \zeta_2}{128 m_2^5}
-\frac{1465 C_A Y_f Y_s^3 \zeta_2}{4096 \varepsilon m_2^5}
+\frac{5 C_A Y_f Y_s^3 \zeta_2}{1024 \varepsilon ^2 m_2^5}
-\frac{495 C_A m_1 Y_f Y_s^3 \zeta_2}{2048 m_2^6}
+\frac{369 C_A m_1 Y_f Y_s^3 \zeta_2}{4096 \varepsilon m_2^6}
-\frac{5 C_A m_1 Y_f Y_s^3 \zeta_2}{1024 \varepsilon ^2 m_2^6}
-\frac{115}{16} C_A C_F^2 \zeta_2+15 C_A C_F Y_f^2 \zeta_2
-\frac{5 C_A C_F Y_f^2 \zeta_2}{2 \varepsilon }
-\frac{375 C_A Y_f^2 Y_s^2 \zeta_2}{512 m_2^2}
+\frac{45 C_A Y_f^2 Y_s^2 \zeta_2}{256 \varepsilon m_2^2}
+\frac{505 C_A m_1 Y_f^2 Y_s^2 \zeta_2}{512 m_2^3}
-\frac{35 C_A m_1 Y_f^2 Y_s^2 \zeta_2}{256 \varepsilon m_2^3}
-\frac{449 C_A Y_f^2 Y_s^2 \zeta_2}{1024 m_2^4}
+\frac{9 C_A Y_f^2 Y_s^2 \zeta_2}{1024 \varepsilon m_2^4}
+\frac{15 C_A Y_f^2 Y_s^2 \zeta_2}{256 \varepsilon ^2 m_2^4}
-\frac{335 C_A C_F Y_s^2 \zeta_2}{64 m_2^2}
+\frac{5 C_A C_F Y_s^2 \zeta_2}{32 \varepsilon m_2^2}
+\frac{465 C_A C_F m_1 Y_s^2 \zeta_2}{64 m_2^3}
-\frac{15 C_A C_F m_1 Y_s^2 \zeta_2}{32 \varepsilon m_2^3}
-\frac{4947 C_A C_F Y_s^2 \zeta_2}{256 m_2^4}
+\frac{107 C_A C_F Y_s^2 \zeta_2}{64 \varepsilon m_2^4}
+\frac{5}{8} C_A C_F \zeta_2
+\frac{585 C_F \zeta_2}{32}
+\frac{5}{8} C_A C_F n_f T_f \zeta_2
-\frac{5 C_A C_F M_W Y_f \zeta_2}{4 m_-}
+\frac{5 C_A C_F M_W Y_f \zeta_2}{2 \varepsilon m_-}
-\frac{35 C_A Y_f^3 Y_s \zeta_2}{16 m_2}
+\frac{5 C_A Y_f^3 Y_s \zeta_2}{64 \varepsilon m_2}
+\frac{5 C_A m_1 Y_f^3 Y_s \zeta_2}{8 m_2^2}
+\frac{5 C_A m_1 Y_f^3 Y_s \zeta_2}{64 \varepsilon m_2^2}
+\frac{5 C_A C_F M_W Y_s \zeta_2}{8 m_- m_2}
-\frac{5 C_A C_F M_W Y_s \zeta_2}{4 \varepsilon m_- m_2}
-\frac{65 C_A C_F M_W Y_s \zeta_2}{4 m_2^2}
+\frac{15 C_A C_F M_W Y_s \zeta_2}{8 \varepsilon m_2^2}
+\frac{5 C_A C_F^2 M_W^2 Y_f Y_s \zeta_2}{4 m_- m_2^2}
+\frac{5 C_F M_W^2 Y_f Y_s \zeta_2}{8 m_- m_2^2}
+\frac{5 C_A C_F^2 Y_f Y_s \zeta_2}{4 m_2}
+\frac{275 C_A C_F Y_f Y_s \zeta_2}{64 m_2}
+\frac{5 C_F Y_f Y_s \zeta_2}{8 m_2}
-\frac{165 C_A C_F Y_f Y_s \zeta_2}{32 \varepsilon m_2}
-\frac{135 C_A C_F m_1 Y_f Y_s \zeta_2}{64 m_2^2}
+\frac{65 C_A C_F m_1 Y_f Y_s \zeta_2}{32 \varepsilon m_2^2}
-\frac{25 C_A C_F^2 \zeta_2}{8 \varepsilon }
-\frac{245 C_F \zeta_2}{16 \varepsilon }
-\frac{795 C_A Y_f Y_s^3 \mathcal{L}_{m_2}}{512 m_2^3}
-\frac{3 C_A Y_f Y_s^3 \mathcal{L}_{m_2}}{8 \varepsilon m_2^3}
+\frac{C_A Y_f Y_s^3 \mathcal{L}_{m_2}}{128 \varepsilon ^2 m_2^3}
+\frac{657 C_A m_1 Y_f Y_s^3 \mathcal{L}_{m_2}}{512 m_2^4}
+\frac{3 C_A m_1 Y_f Y_s^3 \mathcal{L}_{m_2}}{8 \varepsilon m_2^4}
-\frac{3 C_A m_1 Y_f Y_s^3 \mathcal{L}_{m_2}}{128 \varepsilon ^2 m_2^4}
+\frac{8409 C_A Y_f Y_s^3 \mathcal{L}_{m_2}}{2048 m_2^5}
+\frac{1653 C_A Y_f Y_s^3 \mathcal{L}_{m_2}}{2048 \varepsilon m_2^5}
-\frac{7 C_A Y_f Y_s^3 \mathcal{L}_{m_2}}{2048 \varepsilon ^2 m_2^5}
-\frac{C_A Y_f Y_s^3 \mathcal{L}_{m_2}}{512 \varepsilon ^3 m_2^5}
-\frac{7137 C_A m_1 Y_f Y_s^3 \mathcal{L}_{m_2}}{2048 m_2^6}
-\frac{1495 C_A m_1 Y_f Y_s^3 \mathcal{L}_{m_2}}{2048 \varepsilon m_2^6}
+\frac{175 C_A m_1 Y_f Y_s^3 \mathcal{L}_{m_2}}{2048 \varepsilon ^2 m_2^6}
+\frac{C_A m_1 Y_f Y_s^3 \mathcal{L}_{m_2}}{512 \varepsilon ^3 m_2^6}
+\frac{377}{16} C_A C_F^2 \mathcal{L}_{m_2}
-\frac{37}{4} C_A C_F Y_f^2 \mathcal{L}_{m_2}
-\frac{7 C_A C_F Y_f^2 \mathcal{L}_{m_2}}{2 \varepsilon }
+\frac{C_A C_F Y_f^2 \mathcal{L}_{m_2}}{\varepsilon ^2}
+\frac{679 C_A Y_f^2 Y_s^2 \mathcal{L}_{m_2}}{512 m_2^2}
+\frac{15 C_A Y_f^2 Y_s^2 \mathcal{L}_{m_2}}{128 \varepsilon m_2^2}
-\frac{9 C_A Y_f^2 Y_s^2 \mathcal{L}_{m_2}}{128 \varepsilon ^2 m_2^2}
-\frac{597 C_A m_1 Y_f^2 Y_s^2 \mathcal{L}_{m_2}}{512 m_2^3}
-\frac{33 C_A m_1 Y_f^2 Y_s^2 \mathcal{L}_{m_2}}{128 \varepsilon m_2^3}
+\frac{7 C_A m_1 Y_f^2 Y_s^2 \mathcal{L}_{m_2}}{128 \varepsilon ^2 m_2^3}
-\frac{695 C_A Y_f^2 Y_s^2 \mathcal{L}_{m_2}}{512 m_2^4}
-\frac{3 C_A Y_f^2 Y_s^2 \mathcal{L}_{m_2}}{16 \varepsilon m_2^4}
+\frac{7 C_A Y_f^2 Y_s^2 \mathcal{L}_{m_2}}{512 \varepsilon ^2 m_2^4}
-\frac{3 C_A Y_f^2 Y_s^2 \mathcal{L}_{m_2}}{128 \varepsilon ^3 m_2^4}
+\frac{319 C_A C_F Y_s^2 \mathcal{L}_{m_2}}{64 m_2^2}
+\frac{31 C_A C_F Y_s^2 \mathcal{L}_{m_2}}{16 \varepsilon m_2^2}
-\frac{C_A C_F Y_s^2 \mathcal{L}_{m_2}}{16 \varepsilon ^2 m_2^2}
-\frac{309 C_A C_F m_1 Y_s^2 \mathcal{L}_{m_2}}{64 m_2^3}
-\frac{39 C_A C_F m_1 Y_s^2 \mathcal{L}_{m_2}}{16 \varepsilon m_2^3}
+\frac{3 C_A C_F m_1 Y_s^2 \mathcal{L}_{m_2}}{16 \varepsilon ^2 m_2^3}
+\frac{C_A C_F Y_s^2 \mathcal{L}_{m_2}}{m_2^4}
+\frac{643 C_A C_F Y_s^2 \mathcal{L}_{m_2}}{128 \varepsilon m_2^4}
-\frac{15 C_A C_F Y_s^2 \mathcal{L}_{m_2}}{32 \varepsilon ^2 m_2^4}
-\frac{5}{4} C_A C_F \mathcal{L}_{m_2}
+\frac{525}{32} C_F \mathcal{L}_{m_2}-C_A C_F n_f T_f \mathcal{L}_{m_2}
-\frac{C_A C_F n_f T_f \mathcal{L}_{m_2}}{4 \varepsilon }
-\frac{5 C_A C_F M_W Y_f \mathcal{L}_{m_2}}{4 m_-}
-\frac{2 C_A C_F M_W Y_f \mathcal{L}_{m_2}}{\varepsilon m_-}
-\frac{C_A C_F M_W Y_f \mathcal{L}_{m_2}}{\varepsilon ^2 m_-}
+\frac{131 C_A Y_f^3 Y_s \mathcal{L}_{m_2}}{128 m_2}
+\frac{51 C_A Y_f^3 Y_s \mathcal{L}_{m_2}}{64 \varepsilon m_2}
-\frac{C_A Y_f^3 Y_s \mathcal{L}_{m_2}}{32 \varepsilon ^2 m_2}
-\frac{29 C_A m_1 Y_f^3 Y_s \mathcal{L}_{m_2}}{128 m_2^2}
-\frac{21 C_A m_1 Y_f^3 Y_s \mathcal{L}_{m_2}}{64 \varepsilon m_2^2}
-\frac{C_A m_1 Y_f^3 Y_s \mathcal{L}_{m_2}}{32 \varepsilon ^2 m_2^2}
+\frac{5 C_A C_F M_W Y_s \mathcal{L}_{m_2}}{8 m_- m_2}
+\frac{C_A C_F M_W Y_s \mathcal{L}_{m_2}}{\varepsilon m_- m_2}
+\frac{C_A C_F M_W Y_s \mathcal{L}_{m_2}}{2 \varepsilon ^2 m_- m_2}
+\frac{219 C_A C_F M_W Y_s \mathcal{L}_{m_2}}{16 m_2^2}
+\frac{37 C_A C_F M_W Y_s \mathcal{L}_{m_2}}{8 \varepsilon m_2^2}
-\frac{3 C_A C_F M_W Y_s \mathcal{L}_{m_2}}{4 \varepsilon ^2 m_2^2}
+\frac{5 C_A C_F^2 M_W^2 Y_f Y_s \mathcal{L}_{m_2}}{4 m_- m_2^2}
+\frac{5 C_F M_W^2 Y_f Y_s \mathcal{L}_{m_2}}{8 m_- m_2^2}
-\frac{C_A C_F^2 M_W^2 Y_f Y_s \mathcal{L}_{m_2}}{2 \varepsilon m_- m_2^2}
-\frac{C_F M_W^2 Y_f Y_s \mathcal{L}_{m_2}}{4 \varepsilon m_- m_2^2}
-\frac{21 C_A C_F^2 Y_f Y_s \mathcal{L}_{m_2}}{8 m_2}
+\frac{227 C_A C_F Y_f Y_s \mathcal{L}_{m_2}}{64 m_2}
-\frac{21 C_F Y_f Y_s \mathcal{L}_{m_2}}{16 m_2}
-\frac{C_A C_F^2 Y_f Y_s \mathcal{L}_{m_2}}{2 \varepsilon m_2}
+\frac{55 C_A C_F Y_f Y_s \mathcal{L}_{m_2}}{16 \varepsilon m_2}
-\frac{C_F Y_f Y_s \mathcal{L}_{m_2}}{4 \varepsilon m_2}
+\frac{33 C_A C_F Y_f Y_s \mathcal{L}_{m_2}}{16 \varepsilon ^2 m_2}
+\frac{5 C_A C_F^2 m_1 Y_f Y_s \mathcal{L}_{m_2}}{8 m_2^2}
-\frac{231 C_A C_F m_1 Y_f Y_s \mathcal{L}_{m_2}}{64 m_2^2}
+\frac{5 C_F m_1 Y_f Y_s \mathcal{L}_{m_2}}{16 m_2^2}
-\frac{19 C_A C_F m_1 Y_f Y_s \mathcal{L}_{m_2}}{16 \varepsilon m_2^2}
-\frac{13 C_A C_F m_1 Y_f Y_s \mathcal{L}_{m_2}}{16 \varepsilon ^2 m_2^2}
+\frac{6 C_A C_F^2 \mathcal{L}_{m_2}}{\varepsilon }
-\frac{C_A C_F \mathcal{L}_{m_2}}{4 \varepsilon }
+\frac{8 C_F \mathcal{L}_{m_2}}{\varepsilon }
+\frac{5 C_A C_F^2 \mathcal{L}_{m_2}}{4 \varepsilon ^2}
+\frac{49 C_F \mathcal{L}_{m_2}}{8 \varepsilon ^2}
-\frac{27 C_A C_F^2}{2 \varepsilon }
+\frac{5 C_A C_F}{8 \varepsilon }
-\frac{133 C_F}{8 \varepsilon }
-\frac{3 C_A C_F^2}{\varepsilon ^2}
+\frac{C_A C_F}{8 \varepsilon ^2}
-\frac{4 C_F}{\varepsilon ^2}
-\frac{5 C_A C_F^2}{8 \varepsilon ^3}
-\frac{49 C_F}{16 \varepsilon ^3}
\end{autobreak}
\end{align}
\end{tiny}

\subsection{Matching at $\mu\sim m_2$: }
\begin{tiny}
\begin{align}
\begin{autobreak}
V^{(m_{2})}_1=
-\frac{2 n_f C_A C_F T_f \mathcal{L}_{m_2}}{\varepsilon }+2 n_f C_A C_F T_f \mathcal{L}_{m_2}^2-2 n_f C_A C_F T_f \mathcal{L}_{m_2}
+\frac{n_f C_A C_F T_f}{\varepsilon }
+\frac{n_f C_A C_F T_f}{\varepsilon ^2}+5 n_f \zeta_2 C_A C_F T_f+6 n_f C_A C_F T_f
+\frac{107 C_A C_F^2}{8 \varepsilon }
+\frac{9 C_A C_F}{16 \varepsilon }
-\frac{429 C_A C_F^2}{16 \varepsilon ^2}
+\frac{C_A C_F}{16 \varepsilon ^2}
+\frac{13 C_A C_F^2 \mathcal{L}_{m_2}^2}{2 \varepsilon }
+\frac{91 C_A C_F^2 \mathcal{L}_{m_2}}{8 \varepsilon }
-\frac{C_A C_F \mathcal{L}_{m_2}}{8 \varepsilon }
-\frac{91 Y_f^2 C_A C_F \mathcal{L}_{m_2}^2}{32 \varepsilon }
-\frac{Y_f^2 C_A C_F^2 \mathcal{L}_{m_2}}{2 \varepsilon }
+\frac{369 Y_f^2 C_A C_F \mathcal{L}_{m_2}}{64 \varepsilon }
+\frac{1}{2} Y_f^2 C_A C_F^2 \mathcal{L}_{m_2}^2
+\frac{561}{32} Y_f^2 C_A C_F \mathcal{L}_{m_2}^2
-\frac{7}{2} Y_f^2 C_A C_F^2 \mathcal{L}_{m_2}
-\frac{5157}{64} Y_f^2 C_A C_F \mathcal{L}_{m_2}
-\frac{723}{8} C_A C_F^2 \mathcal{L}_{m_2}^2
+\frac{3}{8} C_A C_F \mathcal{L}_{m_2}^2
+\frac{767}{2} C_A C_F^2 \mathcal{L}_{m_2}
-\frac{11}{4} C_A C_F \mathcal{L}_{m_2}
+\frac{135}{2} S_1 S_2 Y_f^2 C_A C_F
-\frac{15}{2} S_1 Y_f^2 C_A C_F
+\frac{7 Y_f^2 C_A C_F^2}{4 \varepsilon }
-\frac{919 Y_f^2 C_A C_F}{256 \varepsilon }
+\frac{Y_f^2 C_A C_F^2}{4 \varepsilon ^2}
+\frac{11 Y_f^2 C_A C_F}{4 \varepsilon ^2}
-\frac{251 Y_f^2 \zeta_2 C_A C_F}{64 \varepsilon }
+\frac{5}{4} Y_f^2 \zeta_2 C_A C_F^2
-\frac{23}{64} Y_f^2 \zeta_2 C_A C_F+5 i \pi Y_f^2 \zeta_2 C_A C_F-15 Y_f^2 \zeta_3 C_A C_F+8 Y_f^2 C_A C_F^2
+\frac{36443}{256} Y_f^2 C_A C_F
-\frac{3 \zeta_2 C_A C_F^2}{4 \varepsilon }
+\frac{849}{16} \zeta_2 C_A C_F^2
+\frac{15}{16} \zeta_2 C_A C_F
-\frac{23147}{32} C_A C_F^2
+\frac{177 C_A C_F}{32}
+\frac{7 Y_f^4 C_A \mathcal{L}_{m_2}^2}{32 \varepsilon }
-\frac{33 Y_f^4 C_A \mathcal{L}_{m_2}}{64 \varepsilon }
-\frac{133}{128} Y_f^4 C_A \mathcal{L}_{m_2}^2
+\frac{1515}{256} Y_f^4 C_A \mathcal{L}_{m_2}
+\frac{27}{8} S_1 S_2 Y_f^4 C_A
-\frac{9}{16} S_1 Y_f^4 C_A
-\frac{77 Y_f^4 C_A}{256 \varepsilon }
-\frac{Y_f^4 C_A}{8 \varepsilon ^2}
+\frac{35 Y_f^4 \zeta_2 C_A}{64 \varepsilon }
+\frac{339}{256} Y_f^4 \zeta_2 C_A
+\frac{1}{4} i \pi Y_f^4 \zeta_2 C_A
-\frac{3}{4} Y_f^4 \zeta_3 C_A
-\frac{11261 Y_f^4 C_A}{1024}
+\frac{3589 C_F}{32 \varepsilon }
-\frac{237 C_F}{32 \varepsilon ^2}
+\frac{11 C_F \mathcal{L}_{m_2}^2}{\varepsilon }
-\frac{823 C_F \mathcal{L}_{m_2}}{16 \varepsilon }
-\frac{Y_f^2 C_F \mathcal{L}_{m_2}}{4 \varepsilon }
+\frac{1}{4} Y_f^2 C_F \mathcal{L}_{m_2}^2
-\frac{7}{4} Y_f^2 C_F \mathcal{L}_{m_2}
-\frac{845}{16} C_F \mathcal{L}_{m_2}^2
+\frac{4295}{16} C_F \mathcal{L}_{m_2}-81 S_1 S_2 C_F+45 S_1 C_F
+\frac{7 Y_f^2 C_F}{8 \varepsilon }
+\frac{Y_f^2 C_F}{8 \varepsilon ^2}
+\frac{5}{8} Y_f^2 \zeta_2 C_F+4 Y_f^2 C_F
-\frac{13 \zeta_2 C_F}{2 \varepsilon }
-\frac{745 \zeta_2 C_F}{32}-6 i \pi \zeta_2 C_F+18 \zeta_3 C_F
-\frac{4085 C_F}{8}
\end{autobreak}
\end{align}
\end{tiny}

\begin{tiny}
\begin{align}
\begin{autobreak}
V^{(m_{2})}_2=
\frac{2 n_f C_A C_F T_f \mathcal{L}_{m_2}}{\varepsilon }-5 n_f C_A C_F T_f \mathcal{L}_{m_2}^2
+\frac{47}{2} n_f C_A C_F T_f \mathcal{L}_{m_2}
-\frac{2 n_f C_A C_F T_f}{\varepsilon }
-\frac{n_f C_A C_F T_f}{\varepsilon ^2}
-\frac{25}{2} n_f \zeta_2 C_A C_F T_f
-\frac{429}{8} n_f C_A C_F T_f
+\frac{53 C_A C_F^2}{2 \varepsilon }
-\frac{11 C_A C_F}{16 \varepsilon }
+\frac{115 C_A C_F^2}{16 \varepsilon ^2}
-\frac{3 C_A C_F}{16 \varepsilon ^2}
-\frac{C_A C_F^2 \mathcal{L}_{m_2}^2}{2 \varepsilon }
-\frac{89 C_A C_F^2 \mathcal{L}_{m_2}}{8 \varepsilon }
+\frac{3 C_A C_F \mathcal{L}_{m_2}}{8 \varepsilon }
-\frac{5 Y_f^2 C_A C_F \mathcal{L}_{m_2}^2}{8 \varepsilon }
-\frac{43 Y_f^2 C_A C_F \mathcal{L}_{m_2}}{16 \varepsilon }
+\frac{713}{96} Y_f^2 C_A C_F \mathcal{L}_{m_2}^2
-\frac{3}{2} Y_f^2 C_A C_F^2 \mathcal{L}_{m_2}
-\frac{5195}{192} Y_f^2 C_A C_F \mathcal{L}_{m_2}
+\frac{137}{8} C_A C_F^2 \mathcal{L}_{m_2}^2
-\frac{5}{8} C_A C_F \mathcal{L}_{m_2}^2
-\frac{305}{4} C_A C_F^2 \mathcal{L}_{m_2}+3 C_A C_F \mathcal{L}_{m_2}
-\frac{27}{4} S_1 S_2 Y_f^2 C_A C_F
+\frac{39}{8} S_1 Y_f^2 C_A C_F
+\frac{3 Y_f^2 C_A C_F^2}{4 \varepsilon }
+\frac{1029 Y_f^2 C_A C_F}{64 \varepsilon }
+\frac{3 Y_f^2 C_A C_F}{4 \varepsilon ^2}
-\frac{109 Y_f^2 \zeta_2 C_A C_F}{16 \varepsilon }
+\frac{313}{192} Y_f^2 \zeta_2 C_A C_F
-\frac{1}{2} i \pi Y_f^2 \zeta_2 C_A C_F
+\frac{3}{2} Y_f^2 \zeta_3 C_A C_F
+\frac{19}{4} Y_f^2 C_A C_F^2
+\frac{13805}{768} Y_f^2 C_A C_F
-\frac{17 \zeta_2 C_A C_F^2}{4 \varepsilon }
-\frac{131}{16} \zeta_2 C_A C_F^2
-\frac{25}{16} \zeta_2 C_A C_F
+\frac{5191}{32} C_A C_F^2
-\frac{289 C_A C_F}{32}
-\frac{23 Y_f^4 C_A \mathcal{L}_{m_2}}{64 \varepsilon }
+\frac{25}{384} Y_f^4 C_A \mathcal{L}_{m_2}^2
+\frac{1205}{768} Y_f^4 C_A \mathcal{L}_{m_2}
-\frac{27}{32} S_1 S_2 Y_f^4 C_A
-\frac{9}{64} S_1 Y_f^4 C_A
+\frac{235 Y_f^4 C_A}{256 \varepsilon }
+\frac{Y_f^4 C_A}{64 \varepsilon ^2}
-\frac{3 Y_f^4 \zeta_2 C_A}{8 \varepsilon }
+\frac{1373}{768} Y_f^4 \zeta_2 C_A
-\frac{1}{16} i \pi Y_f^4 \zeta_2 C_A
+\frac{3}{16} Y_f^4 \zeta_3 C_A
-\frac{17099 Y_f^4 C_A}{3072}
-\frac{651 C_F}{16 \varepsilon }
+\frac{131 C_F}{32 \varepsilon ^2}
-\frac{7 C_F \mathcal{L}_{m_2}^2}{4 \varepsilon }
+\frac{219 C_F \mathcal{L}_{m_2}}{16 \varepsilon }
-\frac{3}{4} Y_f^2 C_F \mathcal{L}_{m_2}
+\frac{399}{16} C_F \mathcal{L}_{m_2}^2
-\frac{2051}{16} C_F \mathcal{L}_{m_2}
-\frac{81}{2} S_1 S_2 C_F
+\frac{69 S_1 C_F}{2}
+\frac{3 Y_f^2 C_F}{8 \varepsilon }
+\frac{19 Y_f^2 C_F}{8}
+\frac{121 \zeta_2 C_F}{8 \varepsilon }
+\frac{451 \zeta_2 C_F}{32}-3 i \pi \zeta_2 C_F+9 \zeta_3 C_F
+\frac{8231 C_F}{32}
\end{autobreak}
\end{align}
\end{tiny}

\begin{tiny}
\begin{align}
\begin{autobreak}
V^{(m_{2})}_3=
-\frac{2 n_f C_A C_F T_f \mathcal{L}_{m_2}}{\varepsilon }+10 n_f C_A C_F T_f \mathcal{L}_{m_2}
+\frac{3 n_f C_A C_F T_f}{2 \varepsilon }
+\frac{n_f C_A C_F T_f}{\varepsilon ^2}
-\frac{89}{4} n_f C_A C_F T_f
-\frac{7853 C_A C_F^2}{32 \varepsilon }
+\frac{9 C_A C_F}{16 \varepsilon }
-\frac{45 C_A C_F^2}{16 \varepsilon ^2}
+\frac{C_A C_F}{16 \varepsilon ^2}
-\frac{65 C_A C_F^2 \mathcal{L}_{m_2}^2}{4 \varepsilon }
+\frac{445 C_A C_F^2 \mathcal{L}_{m_2}}{4 \varepsilon }
-\frac{C_A C_F \mathcal{L}_{m_2}}{8 \varepsilon }
-\frac{29 Y_f^2 C_A C_F \mathcal{L}_{m_2}^2}{16 \varepsilon }
-\frac{Y_f^2 C_A C_F^2 \mathcal{L}_{m_2}}{2 \varepsilon }
-\frac{229 Y_f^2 C_A C_F \mathcal{L}_{m_2}}{32 \varepsilon }
+\frac{1}{2} Y_f^2 C_A C_F^2 \mathcal{L}_{m_2}^2
+\frac{373}{32} Y_f^2 C_A C_F \mathcal{L}_{m_2}^2
-\frac{7}{2} Y_f^2 C_A C_F^2 \mathcal{L}_{m_2}
-\frac{2575}{64} Y_f^2 C_A C_F \mathcal{L}_{m_2}
+\frac{81}{8} C_A C_F^2 \mathcal{L}_{m_2}^2
+\frac{3}{8} C_A C_F \mathcal{L}_{m_2}^2
-\frac{377}{4} C_A C_F^2 \mathcal{L}_{m_2}
-\frac{11}{4} C_A C_F \mathcal{L}_{m_2}
+\frac{135}{4} S_1 S_2 Y_f^2 C_A C_F
-\frac{9}{4} S_1 Y_f^2 C_A C_F
+\frac{7 Y_f^2 C_A C_F^2}{4 \varepsilon }
+\frac{4611 Y_f^2 C_A C_F}{128 \varepsilon }
+\frac{Y_f^2 C_A C_F^2}{4 \varepsilon ^2}
+\frac{9 Y_f^2 C_A C_F}{4 \varepsilon ^2}
-\frac{541 Y_f^2 \zeta_2 C_A C_F}{32 \varepsilon }
+\frac{5}{4} Y_f^2 \zeta_2 C_A C_F^2
-\frac{127}{64} Y_f^2 \zeta_2 C_A C_F
+\frac{5}{2} i \pi Y_f^2 \zeta_2 C_A C_F
-\frac{15}{2} Y_f^2 \zeta_3 C_A C_F+8 Y_f^2 C_A C_F^2
+\frac{9617}{256} Y_f^2 C_A C_F
+\frac{223 \zeta_2 C_A C_F^2}{8 \varepsilon }
+\frac{53}{16} \zeta_2 C_A C_F^2
+\frac{15}{16} \zeta_2 C_A C_F
+\frac{6811}{32} C_A C_F^2
+\frac{177 C_A C_F}{32}
+\frac{33 Y_f^4 C_A \mathcal{L}_{m_2}^2}{128 \varepsilon }
-\frac{155 Y_f^4 C_A \mathcal{L}_{m_2}}{256 \varepsilon }
-\frac{69}{64} Y_f^4 C_A \mathcal{L}_{m_2}^2
+\frac{175}{32} Y_f^4 C_A \mathcal{L}_{m_2}
+\frac{27}{8} S_1 S_2 Y_f^4 C_A
-\frac{9}{16} S_1 Y_f^4 C_A
-\frac{339 Y_f^4 C_A}{1024 \varepsilon }
-\frac{Y_f^4 C_A}{8 \varepsilon ^2}
+\frac{177 Y_f^4 \zeta_2 C_A}{256 \varepsilon }
+\frac{117}{128} Y_f^4 \zeta_2 C_A
+\frac{1}{4} i \pi Y_f^4 \zeta_2 C_A
-\frac{3}{4} Y_f^4 \zeta_3 C_A
-\frac{1153 Y_f^4 C_A}{128}
+\frac{1951 C_F}{32 \varepsilon }
-\frac{109 C_F}{32 \varepsilon ^2}
+\frac{C_F \mathcal{L}_{m_2}^2}{4 \varepsilon }
-\frac{211 C_F \mathcal{L}_{m_2}}{16 \varepsilon }
-\frac{Y_f^2 C_F \mathcal{L}_{m_2}}{4 \varepsilon }
+\frac{1}{4} Y_f^2 C_F \mathcal{L}_{m_2}^2
-\frac{7}{4} Y_f^2 C_F \mathcal{L}_{m_2}
-\frac{185}{16} C_F \mathcal{L}_{m_2}^2
+\frac{819}{16} C_F \mathcal{L}_{m_2}-27 S_1 S_2 C_F
-\frac{75 S_1 C_F}{2}
+\frac{7 Y_f^2 C_F}{8 \varepsilon }
+\frac{Y_f^2 C_F}{8 \varepsilon ^2}
+\frac{5}{8} Y_f^2 \zeta_2 C_F+4 Y_f^2 C_F
-\frac{183 \zeta_2 C_F}{8 \varepsilon }
-\frac{333 \zeta_2 C_F}{32}-2 i \pi \zeta_2 C_F+6 \zeta_3 C_F
-\frac{875 C_F}{8}
\end{autobreak}
\end{align}
\end{tiny}

\begin{tiny}
\begin{align}
\begin{autobreak}
V^{(m_{2})}_4=
\frac{127 C_A C_F^2 Y_s^2 M_H^4}{256 m_2^4 M_W^2}
+\frac{1143 C_A Y_s^2 M_H^4}{256 m_2^4 M_W^2}
+\frac{127 C_F Y_s^2 M_H^4}{512 m_2^4 M_W^2}
-\frac{C_A C_F^2 Y_s^2 M_H^4}{32 \varepsilon m_2^4 M_W^2}
-\frac{9 C_A Y_s^2 M_H^4}{32 \varepsilon m_2^4 M_W^2}
-\frac{C_F Y_s^2 M_H^4}{64 \varepsilon m_2^4 M_W^2}
-\frac{C_A C_F^2 Y_s^2 M_H^4}{128 \varepsilon ^2 m_2^4 M_W^2}
-\frac{9 C_A Y_s^2 M_H^4}{128 \varepsilon ^2 m_2^4 M_W^2}
-\frac{C_F Y_s^2 M_H^4}{256 \varepsilon ^2 m_2^4 M_W^2}
+\frac{C_A C_F^2 Y_s^2 \mathcal{L}_{m_2}^2 M_H^4}{64 m_2^4 M_W^2}
+\frac{9 C_A Y_s^2 \mathcal{L}_{m_2}^2 M_H^4}{64 m_2^4 M_W^2}
+\frac{C_F Y_s^2 \mathcal{L}_{m_2}^2 M_H^4}{128 m_2^4 M_W^2}
+\frac{5 C_A C_F^2 Y_s^2 \zeta_2 M_H^4}{128 m_2^4 M_W^2}
+\frac{45 C_A Y_s^2 \zeta_2 M_H^4}{128 m_2^4 M_W^2}
+\frac{5 C_F Y_s^2 \zeta_2 M_H^4}{256 m_2^4 M_W^2}
-\frac{9 C_A C_F^2 Y_s^2 \mathcal{L}_{m_2} M_H^4}{64 m_2^4 M_W^2}
-\frac{81 C_A Y_s^2 \mathcal{L}_{m_2} M_H^4}{64 m_2^4 M_W^2}
-\frac{9 C_F Y_s^2 \mathcal{L}_{m_2} M_H^4}{128 m_2^4 M_W^2}
+\frac{C_A C_F^2 Y_s^2 \mathcal{L}_{m_2} M_H^4}{64 \varepsilon m_2^4 M_W^2}
+\frac{9 C_A Y_s^2 \mathcal{L}_{m_2} M_H^4}{64 \varepsilon m_2^4 M_W^2}
+\frac{C_F Y_s^2 \mathcal{L}_{m_2} M_H^4}{128 \varepsilon m_2^4 M_W^2}
-\frac{5373 C_A Y_s^3 M_H^2}{512 m_2^4 M_W}
+\frac{345 C_A Y_s^3 M_H^2}{64 \varepsilon m_2^4 M_W}
-\frac{45 C_A Y_s^3 \mathcal{L}_{m_2}^2 M_H^2}{64 m_2^4 M_W}
+\frac{3 C_A Y_s^3 \mathcal{L}_{m_2}^2 M_H^2}{8 \varepsilon m_2^4 M_W}
+\frac{63 C_A Y_s^3 \zeta_2 M_H^2}{128 m_2^4 M_W}
-\frac{3 C_A Y_s^3 \zeta_2 M_H^2}{8 \varepsilon m_2^4 M_W}
+\frac{603 C_A Y_s^3 \mathcal{L}_{m_2} M_H^2}{128 m_2^4 M_W}
-\frac{39 C_A Y_s^3 \mathcal{L}_{m_2} M_H^2}{16 \varepsilon m_2^4 M_W}
-\frac{5695 C_A Y_s^4}{2048 m_2^4}
+\frac{3457 C_A Y_s^4}{2048 \varepsilon m_2^4}
-\frac{C_A Y_s^4}{16 \varepsilon ^2 m_2^4}
+\frac{3289}{32} C_A C_F^2
+\frac{131 C_A C_F^2 M_W^2 Y_s^2}{4 m_2^4}
+\frac{131 C_F M_W^2 Y_s^2}{8 m_2^4}
-\frac{7 C_A C_F^2 M_W^2 Y_s^2}{4 \varepsilon m_2^4}
-\frac{7 C_F M_W^2 Y_s^2}{8 \varepsilon m_2^4}
-\frac{C_A C_F^2 M_W^2 Y_s^2}{2 \varepsilon ^2 m_2^4}
-\frac{C_F M_W^2 Y_s^2}{4 \varepsilon ^2 m_2^4}
+\frac{81 C_A C_F S_1 S_2 Y_s^2}{8 m_2^2}
+\frac{115 C_A C_F^2 Y_s^2}{16 m_2^2}
+\frac{11063 C_A C_F Y_s^2}{256 m_2^2}
+\frac{115 C_F Y_s^2}{32 m_2^2}
-\frac{3857 C_A C_F Y_s^2}{256 \varepsilon m_2^2}
+\frac{C_A C_F Y_s^2}{4 \varepsilon ^2 m_2^2}
-\frac{105 C_A Y_s^4 \mathcal{L}_{m_2}^2}{256 m_2^4}
+\frac{37 C_A Y_s^4 \mathcal{L}_{m_2}^2}{256 \varepsilon m_2^4}
+\frac{17}{4} C_A C_F^2 \mathcal{L}_{m_2}^2
+\frac{C_A C_F^2 M_W^2 Y_s^2 \mathcal{L}_{m_2}^2}{m_2^4}
+\frac{C_F M_W^2 Y_s^2 \mathcal{L}_{m_2}^2}{2 m_2^4}
+\frac{C_A C_F^2 Y_s^2 \mathcal{L}_{m_2}^2}{2 m_2^2}
+\frac{161 C_A C_F Y_s^2 \mathcal{L}_{m_2}^2}{32 m_2^2}
+\frac{C_F Y_s^2 \mathcal{L}_{m_2}^2}{4 m_2^2}
-\frac{61 C_A C_F Y_s^2 \mathcal{L}_{m_2}^2}{32 \varepsilon m_2^2}
+\frac{1}{4} C_A C_F \mathcal{L}_{m_2}^2
+\frac{57}{8} C_F \mathcal{L}_{m_2}^2+2 C_A C_F n_f T_f \mathcal{L}_{m_2}^2
-\frac{125 C_A C_F M_W Y_s \mathcal{L}_{m_2}^2}{16 m_2^2}
+\frac{19 C_A C_F M_W Y_s \mathcal{L}_{m_2}^2}{8 \varepsilon m_2^2}
-\frac{7 C_A C_F^2 \mathcal{L}_{m_2}^2}{\varepsilon }
-\frac{35 C_F \mathcal{L}_{m_2}^2}{4 \varepsilon }
+\frac{319 C_A C_F}{32}
+\frac{5913 C_F}{64}
-\frac{27 C_F S_1}{2}
+\frac{243}{4} C_F S_1 S_2+21 C_A C_F n_f T_f
+\frac{5 C_A C_F n_f T_f}{\varepsilon }
+\frac{C_A C_F n_f T_f}{\varepsilon ^2}
-\frac{11033 C_A C_F M_W Y_s}{128 m_2^2}
+\frac{2185 C_A C_F M_W Y_s}{64 \varepsilon m_2^2}
-\frac{57 C_A Y_s^4 \zeta_2}{512 m_2^4}
+\frac{5 C_A Y_s^4 \zeta_2}{512 \varepsilon m_2^4}
-\frac{35}{8} C_A C_F^2 \zeta_2
+\frac{5 C_A C_F^2 M_W^2 Y_s^2 \zeta_2}{2 m_2^4}
+\frac{5 C_F M_W^2 Y_s^2 \zeta_2}{4 m_2^4}
+\frac{5 C_A C_F^2 Y_s^2 \zeta_2}{4 m_2^2}
-\frac{15 C_A C_F Y_s^2 \zeta_2}{64 m_2^2}
+\frac{5 C_F Y_s^2 \zeta_2}{8 m_2^2}
-\frac{125 C_A C_F Y_s^2 \zeta_2}{64 \varepsilon m_2^2}
+\frac{3 i C_A C_F \pi Y_s^2 \zeta_2}{4 m_2^2}
+\frac{5}{8} C_A C_F \zeta_2
-\frac{59 C_F \zeta_2}{16}+5 C_A C_F n_f T_f \zeta_2
-\frac{25 C_A C_F M_W Y_s \zeta_2}{32 m_2^2}
+\frac{15 C_A C_F M_W Y_s \zeta_2}{16 \varepsilon m_2^2}
+\frac{29 C_A C_F^2 \zeta_2}{2 \varepsilon }
-\frac{35 C_F \zeta_2}{8 \varepsilon }
+\frac{9}{2} i C_F \pi \zeta_2
-\frac{9 C_A C_F Y_s^2 \zeta_3}{4 m_2^2}
-\frac{27 C_F \zeta_3}{2}
+\frac{833 C_A Y_s^4 \mathcal{L}_{m_2}}{512 m_2^4}
-\frac{375 C_A Y_s^4 \mathcal{L}_{m_2}}{512 \varepsilon m_2^4}
-\frac{315}{8} C_A C_F^2 \mathcal{L}_{m_2}
-\frac{19 C_A C_F^2 M_W^2 Y_s^2 \mathcal{L}_{m_2}}{2 m_2^4}
-\frac{19 C_F M_W^2 Y_s^2 \mathcal{L}_{m_2}}{4 m_2^4}
+\frac{C_A C_F^2 M_W^2 Y_s^2 \mathcal{L}_{m_2}}{\varepsilon m_2^4}
+\frac{C_F M_W^2 Y_s^2 \mathcal{L}_{m_2}}{2 \varepsilon m_2^4}
-\frac{13 C_A C_F^2 Y_s^2 \mathcal{L}_{m_2}}{4 m_2^2}
-\frac{1545 C_A C_F Y_s^2 \mathcal{L}_{m_2}}{64 m_2^2}
-\frac{13 C_F Y_s^2 \mathcal{L}_{m_2}}{8 m_2^2}
+\frac{551 C_A C_F Y_s^2 \mathcal{L}_{m_2}}{64 \varepsilon m_2^2}
+\frac{223}{8} C_A C_F \mathcal{L}_{m_2}
-\frac{707}{16} C_F \mathcal{L}_{m_2}-10 C_A C_F n_f T_f \mathcal{L}_{m_2}
-\frac{2 C_A C_F n_f T_f \mathcal{L}_{m_2}}{\varepsilon }
+\frac{1391 C_A C_F M_W Y_s \mathcal{L}_{m_2}}{32 m_2^2}
-\frac{247 C_A C_F M_W Y_s \mathcal{L}_{m_2}}{16 \varepsilon m_2^2}
+\frac{49 C_A C_F^2 \mathcal{L}_{m_2}}{\varepsilon }
+\frac{363 C_F \mathcal{L}_{m_2}}{8 \varepsilon }
-\frac{905 C_A C_F^2}{8 \varepsilon }
-\frac{59 C_A C_F}{4 \varepsilon }
-\frac{2655 C_F}{32 \varepsilon }
-\frac{7 C_A C_F^2}{4 \varepsilon ^2}
-\frac{17 C_F}{8 \varepsilon ^2}
\end{autobreak}
\end{align}
\end{tiny}

\begin{tiny}
\begin{align}
\begin{autobreak}
V^{(m_{2})}_5=
\frac{405 C_A C_F^2 Y_s^2 M_H^4}{512 m_2^4 M_W^2}
+\frac{3645 C_A Y_s^2 M_H^4}{512 m_2^4 M_W^2}
+\frac{405 C_F Y_s^2 M_H^4}{1024 m_2^4 M_W^2}
-\frac{C_A C_F^2 Y_s^2 M_H^4}{32 \varepsilon m_2^4 M_W^2}
-\frac{9 C_A Y_s^2 M_H^4}{32 \varepsilon m_2^4 M_W^2}
-\frac{C_F Y_s^2 M_H^4}{64 \varepsilon m_2^4 M_W^2}
+\frac{C_A C_F^2 Y_s^2 M_H^4}{256 \varepsilon ^2 m_2^4 M_W^2}
+\frac{9 C_A Y_s^2 M_H^4}{256 \varepsilon ^2 m_2^4 M_W^2}
+\frac{C_F Y_s^2 M_H^4}{512 \varepsilon ^2 m_2^4 M_W^2}
+\frac{7 C_A C_F^2 Y_s^2 \mathcal{L}_{m_2}^2 M_H^4}{128 m_2^4 M_W^2}
+\frac{63 C_A Y_s^2 \mathcal{L}_{m_2}^2 M_H^4}{128 m_2^4 M_W^2}
+\frac{7 C_F Y_s^2 \mathcal{L}_{m_2}^2 M_H^4}{256 m_2^4 M_W^2}
+\frac{35 C_A C_F^2 Y_s^2 \zeta_2 M_H^4}{256 m_2^4 M_W^2}
+\frac{315 C_A Y_s^2 \zeta_2 M_H^4}{256 m_2^4 M_W^2}
+\frac{35 C_F Y_s^2 \zeta_2 M_H^4}{512 m_2^4 M_W^2}
-\frac{31 C_A C_F^2 Y_s^2 \mathcal{L}_{m_2} M_H^4}{128 m_2^4 M_W^2}
-\frac{279 C_A Y_s^2 \mathcal{L}_{m_2} M_H^4}{128 m_2^4 M_W^2}
-\frac{31 C_F Y_s^2 \mathcal{L}_{m_2} M_H^4}{256 m_2^4 M_W^2}
-\frac{C_A C_F^2 Y_s^2 \mathcal{L}_{m_2} M_H^4}{128 \varepsilon m_2^4 M_W^2}
-\frac{9 C_A Y_s^2 \mathcal{L}_{m_2} M_H^4}{128 \varepsilon m_2^4 M_W^2}
-\frac{C_F Y_s^2 \mathcal{L}_{m_2} M_H^4}{256 \varepsilon m_2^4 M_W^2}
-\frac{3657 C_A Y_s^3 M_H^2}{1024 m_2^4 M_W}
+\frac{345 C_A Y_s^3 M_H^2}{128 \varepsilon m_2^4 M_W}
-\frac{9 C_A Y_s^3 \mathcal{L}_{m_2}^2 M_H^2}{128 m_2^4 M_W}
+\frac{3 C_A Y_s^3 \mathcal{L}_{m_2}^2 M_H^2}{16 \varepsilon m_2^4 M_W}
-\frac{189 C_A Y_s^3 \zeta_2 M_H^2}{256 m_2^4 M_W}
-\frac{9 C_A Y_s^3 \zeta_2 M_H^2}{16 \varepsilon m_2^4 M_W}
+\frac{303 C_A Y_s^3 \mathcal{L}_{m_2} M_H^2}{256 m_2^4 M_W}
-\frac{39 C_A Y_s^3 \mathcal{L}_{m_2} M_H^2}{32 \varepsilon m_2^4 M_W}
-\frac{3 C_A S_1 Y_s^4}{16 m_2^4}
-\frac{12611 C_A Y_s^4}{4096 m_2^4}
+\frac{3333 C_A Y_s^4}{4096 \varepsilon m_2^4}
-\frac{C_A Y_s^4}{64 \varepsilon ^2 m_2^4}
+\frac{78671}{192} C_A C_F^2
+\frac{413 C_A C_F^2 M_W^2 Y_s^2}{8 m_2^4}
+\frac{413 C_F M_W^2 Y_s^2}{16 m_2^4}
-\frac{17 C_A C_F^2 M_W^2 Y_s^2}{8 \varepsilon m_2^4}
-\frac{17 C_F M_W^2 Y_s^2}{16 \varepsilon m_2^4}
+\frac{C_A C_F^2 M_W^2 Y_s^2}{4 \varepsilon ^2 m_2^4}
+\frac{C_F M_W^2 Y_s^2}{8 \varepsilon ^2 m_2^4}
-\frac{69 C_A C_F S_1 Y_s^2}{16 m_2^2}
+\frac{135 C_A C_F S_1 S_2 Y_s^2}{4 m_2^2}
+\frac{219 C_A C_F^2 Y_s^2}{32 m_2^2}
+\frac{46149 C_A C_F Y_s^2}{1024 m_2^2}
+\frac{219 C_F Y_s^2}{64 m_2^2}
+\frac{C_A C_F^2 Y_s^2}{2 \varepsilon m_2^2}
-\frac{2523 C_A C_F Y_s^2}{1024 \varepsilon m_2^2}
+\frac{C_F Y_s^2}{4 \varepsilon m_2^2}
+\frac{C_A C_F Y_s^2}{4 \varepsilon ^2 m_2^2}
-\frac{197 C_A Y_s^4 \mathcal{L}_{m_2}^2}{512 m_2^4}
+\frac{57 C_A Y_s^4 \mathcal{L}_{m_2}^2}{512 \varepsilon m_2^4}
+\frac{731}{24} C_A C_F^2 \mathcal{L}_{m_2}^2
+\frac{7 C_A C_F^2 M_W^2 Y_s^2 \mathcal{L}_{m_2}^2}{2 m_2^4}
+\frac{7 C_F M_W^2 Y_s^2 \mathcal{L}_{m_2}^2}{4 m_2^4}
+\frac{C_A C_F^2 Y_s^2 \mathcal{L}_{m_2}^2}{4 m_2^2}
+\frac{691 C_A C_F Y_s^2 \mathcal{L}_{m_2}^2}{128 m_2^2}
+\frac{C_F Y_s^2 \mathcal{L}_{m_2}^2}{8 m_2^2}
-\frac{263 C_A C_F Y_s^2 \mathcal{L}_{m_2}^2}{128 \varepsilon m_2^2}
+\frac{1487}{24} C_A C_F \mathcal{L}_{m_2}^2
+\frac{503}{48} C_F \mathcal{L}_{m_2}^2
+\frac{1}{3} C_A C_F n_f T_f \mathcal{L}_{m_2}^2
-\frac{323 C_A C_F M_W Y_s \mathcal{L}_{m_2}^2}{32 m_2^2}
+\frac{45 C_A C_F M_W Y_s \mathcal{L}_{m_2}^2}{16 \varepsilon m_2^2}
-\frac{19 C_A C_F^2 \mathcal{L}_{m_2}^2}{2 \varepsilon }
-\frac{101 C_F \mathcal{L}_{m_2}^2}{8 \varepsilon }
+\frac{43073 C_A C_F}{192}
+\frac{68407 C_F}{384}
-\frac{108}{5} C_A C_F S_1
-\frac{117 C_F S_1}{2}
+\frac{405}{2} C_F S_1 S_2
+\frac{55}{6} C_A C_F n_f T_f
+\frac{23 C_A C_F n_f T_f}{2 \varepsilon }
+\frac{3 C_A C_F n_f T_f}{2 \varepsilon ^2}
-\frac{29295 C_A C_F M_W Y_s}{256 m_2^2}
+\frac{4791 C_A C_F M_W Y_s}{128 \varepsilon m_2^2}
-\frac{53 C_A Y_s^4 \zeta_2}{1024 m_2^4}
+\frac{153 C_A Y_s^4 \zeta_2}{1024 \varepsilon m_2^4}
+\frac{1975}{48} C_A C_F^2 \zeta_2
+\frac{35 C_A C_F^2 M_W^2 Y_s^2 \zeta_2}{4 m_2^4}
+\frac{35 C_F M_W^2 Y_s^2 \zeta_2}{8 m_2^4}
+\frac{5 C_A C_F^2 Y_s^2 \zeta_2}{8 m_2^2}
+\frac{99 C_A C_F Y_s^2 \zeta_2}{256 m_2^2}
+\frac{5 C_F Y_s^2 \zeta_2}{16 m_2^2}
-\frac{1447 C_A C_F Y_s^2 \zeta_2}{256 \varepsilon m_2^2}
+\frac{5 i C_A C_F \pi Y_s^2 \zeta_2}{2 m_2^2}
+\frac{1579}{48} C_A C_F \zeta_2
+\frac{3083 C_F \zeta_2}{96}
+\frac{5}{6} C_A C_F n_f T_f \zeta_2
+\frac{737 C_A C_F M_W Y_s \zeta_2}{64 m_2^2}
+\frac{81 C_A C_F M_W Y_s \zeta_2}{32 \varepsilon m_2^2}
+\frac{85 C_A C_F^2 \zeta_2}{4 \varepsilon }
-\frac{205 C_F \zeta_2}{16 \varepsilon }+15 i C_F \pi \zeta_2
-\frac{15 C_A C_F Y_s^2 \zeta_3}{2 m_2^2}-45 C_F \zeta_3
+\frac{1853 C_A Y_s^4 \mathcal{L}_{m_2}}{1024 m_2^4}
-\frac{499 C_A Y_s^4 \mathcal{L}_{m_2}}{1024 \varepsilon m_2^4}
-\frac{8789}{48} C_A C_F^2 \mathcal{L}_{m_2}
-\frac{61 C_A C_F^2 M_W^2 Y_s^2 \mathcal{L}_{m_2}}{4 m_2^4}
-\frac{61 C_F M_W^2 Y_s^2 \mathcal{L}_{m_2}}{8 m_2^4}
-\frac{C_A C_F^2 M_W^2 Y_s^2 \mathcal{L}_{m_2}}{2 \varepsilon m_2^4}
-\frac{C_F M_W^2 Y_s^2 \mathcal{L}_{m_2}}{4 \varepsilon m_2^4}
-\frac{21 C_A C_F^2 Y_s^2 \mathcal{L}_{m_2}}{8 m_2^2}
-\frac{6683 C_A C_F Y_s^2 \mathcal{L}_{m_2}}{256 m_2^2}
-\frac{21 C_F Y_s^2 \mathcal{L}_{m_2}}{16 m_2^2}
+\frac{1485 C_A C_F Y_s^2 \mathcal{L}_{m_2}}{256 \varepsilon m_2^2}
-\frac{4799}{48} C_A C_F \mathcal{L}_{m_2}
-\frac{7373}{96} C_F \mathcal{L}_{m_2}
-\frac{17}{3} C_A C_F n_f T_f \mathcal{L}_{m_2}
-\frac{3 C_A C_F n_f T_f \mathcal{L}_{m_2}}{\varepsilon }
+\frac{3737 C_A C_F M_W Y_s \mathcal{L}_{m_2}}{64 m_2^2}
-\frac{585 C_A C_F M_W Y_s \mathcal{L}_{m_2}}{32 \varepsilon m_2^2}
+\frac{103 C_A C_F^2 \mathcal{L}_{m_2}}{2 \varepsilon }
-\frac{107 C_A C_F \mathcal{L}_{m_2}}{2 \varepsilon }
+\frac{969 C_F \mathcal{L}_{m_2}}{16 \varepsilon }
-\frac{1971 C_A C_F^2}{16 \varepsilon }
+\frac{87 C_A C_F}{2 \varepsilon }
-\frac{6521 C_F}{64 \varepsilon }
+\frac{41 C_A C_F^2}{8 \varepsilon ^2}
+\frac{107 C_A C_F}{4 \varepsilon ^2}
-\frac{59 C_F}{16 \varepsilon ^2}
\end{autobreak}
\end{align}
\end{tiny}

\begin{tiny}
\begin{align}
\begin{autobreak}
V^{(m_{2})}_6=

-\frac{89 C_A Y_s \mathcal{L}_{m_2}^2 Y_f^3}{64 m_2}
+\frac{5 C_A Y_s \mathcal{L}_{m_2}^2 Y_f^3}{16 \varepsilon m_2}
+\frac{39 C_A S_1 Y_s Y_f^3}{32 m_2}
-\frac{81 C_A S_1 S_2 Y_s Y_f^3}{8 m_2}
-\frac{2413 C_A Y_s Y_f^3}{512 m_2}
-\frac{123 C_A Y_s Y_f^3}{128 \varepsilon m_2}
-\frac{3 C_A Y_s Y_f^3}{16 \varepsilon ^2 m_2}
-\frac{257 C_A Y_s \zeta_2 Y_f^3}{128 m_2}
+\frac{37 C_A Y_s \zeta_2 Y_f^3}{32 \varepsilon m_2}
-\frac{3 i C_A \pi Y_s \zeta_2 Y_f^3}{4 m_2}
+\frac{9 C_A Y_s \zeta_3 Y_f^3}{4 m_2}
+\frac{603 C_A Y_s \mathcal{L}_{m_2} Y_f^3}{128 m_2}
-\frac{11 C_A Y_s \mathcal{L}_{m_2} Y_f^3}{32 \varepsilon m_2}
-\frac{27 C_A S_1 Y_s^2 Y_f^2}{32 m_2^2}
+\frac{351 C_A S_1 S_2 Y_s^2 Y_f^2}{64 m_2^2}
+\frac{3277 C_A Y_s^2 Y_f^2}{512 m_2^2}
-\frac{345 C_A Y_s^2 Y_f^2}{128 \varepsilon m_2^2}
+\frac{19 C_A Y_s^2 \mathcal{L}_{m_2}^2 Y_f^2}{64 m_2^2}
-\frac{3 C_A Y_s^2 \mathcal{L}_{m_2}^2 Y_f^2}{16 \varepsilon m_2^2}
+\frac{89}{16} C_A C_F \mathcal{L}_{m_2}^2 Y_f^2
+\frac{69 C_A M_H^2 Y_s \mathcal{L}_{m_2}^2 Y_f^2}{64 m_2^2 M_W}
-\frac{33 C_A M_H^2 Y_s \mathcal{L}_{m_2}^2 Y_f^2}{64 \varepsilon m_2^2 M_W}
-\frac{5 C_A C_F \mathcal{L}_{m_2}^2 Y_f^2}{2 \varepsilon }
+\frac{4201}{128} C_A C_F Y_f^2
-\frac{309}{16} C_A C_F S_1 Y_f^2
+\frac{513}{8} C_A C_F S_1 S_2 Y_f^2
+\frac{6663 C_A M_H^2 Y_s Y_f^2}{512 m_2^2 M_W}
-\frac{3795 C_A M_H^2 Y_s Y_f^2}{512 \varepsilon m_2^2 M_W}
+\frac{129 C_A Y_s^2 \zeta_2 Y_f^2}{128 m_2^2}
+\frac{21 C_A Y_s^2 \zeta_2 Y_f^2}{32 \varepsilon m_2^2}
+\frac{13 i C_A \pi Y_s^2 \zeta_2 Y_f^2}{32 m_2^2}
+\frac{323}{32} C_A C_F \zeta_2 Y_f^2
-\frac{63 C_A M_H^2 Y_s \zeta_2 Y_f^2}{128 m_2^2 M_W}
+\frac{63 C_A M_H^2 Y_s \zeta_2 Y_f^2}{128 \varepsilon m_2^2 M_W}
-\frac{37 C_A C_F \zeta_2 Y_f^2}{4 \varepsilon }
+\frac{19}{4} i C_A C_F \pi \zeta_2 Y_f^2
-\frac{39 C_A Y_s^2 \zeta_3 Y_f^2}{32 m_2^2}
-\frac{57}{4} C_A C_F \zeta_3 Y_f^2
-\frac{319 C_A Y_s^2 \mathcal{L}_{m_2} Y_f^2}{128 m_2^2}
+\frac{39 C_A Y_s^2 \mathcal{L}_{m_2} Y_f^2}{32 \varepsilon m_2^2}
-\frac{763}{32} C_A C_F \mathcal{L}_{m_2} Y_f^2
-\frac{801 C_A M_H^2 Y_s \mathcal{L}_{m_2} Y_f^2}{128 m_2^2 M_W}
+\frac{429 C_A M_H^2 Y_s \mathcal{L}_{m_2} Y_f^2}{128 \varepsilon m_2^2 M_W}
+\frac{17 C_A C_F \mathcal{L}_{m_2} Y_f^2}{4 \varepsilon }
+\frac{87 C_A C_F Y_f^2}{16 \varepsilon }
+\frac{3 C_A C_F Y_f^2}{4 \varepsilon ^2}
-\frac{83 C_A Y_s^3 Y_f}{8 m_2^3}
+\frac{1273 C_A Y_s^3 Y_f}{256 \varepsilon m_2^3}
-\frac{C_A Y_s^3 Y_f}{32 \varepsilon ^2 m_2^3}
-\frac{3651 C_A M_H^2 Y_s^2 Y_f}{128 m_2^3 M_W}
+\frac{3795 C_A M_H^2 Y_s^2 Y_f}{256 \varepsilon m_2^3 M_W}
-\frac{3 C_A Y_s^3 \mathcal{L}_{m_2}^2 Y_f}{4 m_2^3}
+\frac{11 C_A Y_s^3 \mathcal{L}_{m_2}^2 Y_f}{32 \varepsilon m_2^3}
-\frac{15 C_A M_H^2 Y_s^2 \mathcal{L}_{m_2}^2 Y_f}{8 m_2^3 M_W}
+\frac{33 C_A M_H^2 Y_s^2 \mathcal{L}_{m_2}^2 Y_f}{32 \varepsilon m_2^3 M_W}
+\frac{C_A C_F^2 Y_s \mathcal{L}_{m_2}^2 Y_f}{m_2}
+\frac{21 C_A C_F Y_s \mathcal{L}_{m_2}^2 Y_f}{2 m_2}
+\frac{C_F Y_s \mathcal{L}_{m_2}^2 Y_f}{2 m_2}
-\frac{5 C_A C_F Y_s \mathcal{L}_{m_2}^2 Y_f}{2 \varepsilon m_2}
+\frac{99 C_A C_F S_1 Y_s Y_f}{8 m_2}
+\frac{189 C_A C_F S_1 S_2 Y_s Y_f}{8 m_2}
+\frac{141 C_A C_F^2 Y_s Y_f}{8 m_2}
+\frac{1595 C_A C_F Y_s Y_f}{32 m_2}
+\frac{141 C_F Y_s Y_f}{16 m_2}
+\frac{C_A C_F^2 Y_s Y_f}{2 \varepsilon m_2}
-\frac{193 C_A C_F Y_s Y_f}{32 \varepsilon m_2}
+\frac{C_F Y_s Y_f}{4 \varepsilon m_2}
-\frac{C_A Y_s^3 \zeta_2 Y_f}{16 m_2^3}
-\frac{29 C_A Y_s^3 \zeta_2 Y_f}{64 \varepsilon m_2^3}
+\frac{3 C_A M_H^2 Y_s^2 \zeta_2 Y_f}{8 m_2^3 M_W}
-\frac{87 C_A M_H^2 Y_s^2 \zeta_2 Y_f}{64 \varepsilon m_2^3 M_W}
+\frac{5 C_A C_F^2 Y_s \zeta_2 Y_f}{2 m_2}
-\frac{11 C_A C_F Y_s \zeta_2 Y_f}{4 m_2}
+\frac{5 C_F Y_s \zeta_2 Y_f}{4 m_2}
-\frac{31 C_A C_F Y_s \zeta_2 Y_f}{4 \varepsilon m_2}
+\frac{7 i C_A C_F \pi Y_s \zeta_2 Y_f}{4 m_2}
-\frac{21 C_A C_F Y_s \zeta_3 Y_f}{4 m_2}
+\frac{73 C_A Y_s^3 \mathcal{L}_{m_2} Y_f}{16 m_2^3}
-\frac{139 C_A Y_s^3 \mathcal{L}_{m_2} Y_f}{64 \varepsilon m_2^3}
+\frac{405 C_A M_H^2 Y_s^2 \mathcal{L}_{m_2} Y_f}{32 m_2^3 M_W}
-\frac{429 C_A M_H^2 Y_s^2 \mathcal{L}_{m_2} Y_f}{64 \varepsilon m_2^3 M_W}
-\frac{15 C_A C_F^2 Y_s \mathcal{L}_{m_2} Y_f}{2 m_2}
-\frac{329 C_A C_F Y_s \mathcal{L}_{m_2} Y_f}{8 m_2}
-\frac{15 C_F Y_s \mathcal{L}_{m_2} Y_f}{4 m_2}
+\frac{67 C_A C_F Y_s \mathcal{L}_{m_2} Y_f}{8 \varepsilon m_2}
+\frac{1413}{64} C_A C_F^2
+\frac{1403 C_A C_F Y_s^2}{32 m_2^2}
-\frac{1305 C_A C_F Y_s^2}{64 \varepsilon m_2^2}
-\frac{C_A C_F Y_s^2}{8 \varepsilon ^2 m_2^2}+C_A C_F^2 \mathcal{L}_{m_2}^2
+\frac{11 C_A C_F Y_s^2 \mathcal{L}_{m_2}^2}{4 m_2^2}
-\frac{11 C_A C_F Y_s^2 \mathcal{L}_{m_2}^2}{8 \varepsilon m_2^2}
+\frac{1}{4} C_A C_F \mathcal{L}_{m_2}^2
+\frac{159}{8} C_F \mathcal{L}_{m_2}^2+5 C_A C_F n_f T_f \mathcal{L}_{m_2}^2
-\frac{25 C_A C_F M_W Y_s \mathcal{L}_{m_2}^2}{2 m_2^2}
+\frac{5 C_A C_F M_W Y_s \mathcal{L}_{m_2}^2}{\varepsilon m_2^2}
-\frac{4 C_A C_F^2 \mathcal{L}_{m_2}^2}{\varepsilon }
-\frac{91 C_F \mathcal{L}_{m_2}^2}{8 \varepsilon }
+\frac{465 C_A C_F}{64}
+\frac{20327 C_F}{128}-36 C_F S_1+108 C_F S_1 S_2
-\frac{7}{4} C_A C_F n_f T_f
+\frac{6 C_A C_F n_f T_f}{\varepsilon }
+\frac{7 C_A C_F n_f T_f}{2 \varepsilon ^2}
-\frac{657 C_A C_F M_W Y_s}{4 m_2^2}
+\frac{575 C_A C_F M_W Y_s}{8 \varepsilon m_2^2}
-\frac{143}{4} C_A C_F^2 \zeta_2
-\frac{5 C_A C_F Y_s^2 \zeta_2}{8 m_2^2}
+\frac{29 C_A C_F Y_s^2 \zeta_2}{16 \varepsilon m_2^2}
+\frac{5}{8} C_A C_F \zeta_2
-\frac{107 C_F \zeta_2}{16}
+\frac{25}{2} C_A C_F n_f T_f \zeta_2
+\frac{33 C_A C_F M_W Y_s \zeta_2}{4 m_2^2}
-\frac{7 C_A C_F M_W Y_s \zeta_2}{2 \varepsilon m_2^2}
+\frac{6 C_A C_F^2 \zeta_2}{\varepsilon }
-\frac{315 C_F \zeta_2}{16 \varepsilon }+8 i C_F \pi \zeta_2-24 C_F \zeta_3
-\frac{65}{16} C_A C_F^2 \mathcal{L}_{m_2}
-\frac{151 C_A C_F Y_s^2 \mathcal{L}_{m_2}}{8 m_2^2}
+\frac{147 C_A C_F Y_s^2 \mathcal{L}_{m_2}}{16 \varepsilon m_2^2}
-\frac{49}{16} C_A C_F \mathcal{L}_{m_2}
-\frac{2991}{32} C_F \mathcal{L}_{m_2}+C_A C_F n_f T_f \mathcal{L}_{m_2}
-\frac{7 C_A C_F n_f T_f \mathcal{L}_{m_2}}{\varepsilon }
+\frac{77 C_A C_F M_W Y_s \mathcal{L}_{m_2}}{m_2^2}
-\frac{65 C_A C_F M_W Y_s \mathcal{L}_{m_2}}{2 \varepsilon m_2^2}
+\frac{263 C_A C_F^2 \mathcal{L}_{m_2}}{8 \varepsilon }
+\frac{C_A C_F \mathcal{L}_{m_2}}{8 \varepsilon }
+\frac{97 C_F \mathcal{L}_{m_2}}{2 \varepsilon }
-\frac{1309 C_A C_F^2}{16 \varepsilon }
+\frac{5 C_A C_F}{16 \varepsilon }
-\frac{4593 C_F}{64 \varepsilon }
-\frac{55 C_A C_F^2}{16 \varepsilon ^2}
-\frac{C_A C_F}{16 \varepsilon ^2}
-\frac{55 C_F}{32 \varepsilon ^2}
\end{autobreak}
\end{align}
\end{tiny}

\begin{tiny}
\begin{align}
\begin{autobreak}
V^{(m_{2})}_7=

-\frac{21 C_A Y_s \mathcal{L}_{m_2}^2 Y_f^3}{64 m_2}
+\frac{5 C_A Y_s \mathcal{L}_{m_2}^2 Y_f^3}{64 \varepsilon m_2}
-\frac{2741 C_A Y_s Y_f^3}{512 m_2}
+\frac{527 C_A Y_s Y_f^3}{512 \varepsilon m_2}
+\frac{79 C_A Y_s \zeta_2 Y_f^3}{128 m_2}
-\frac{11 C_A Y_s \zeta_2 Y_f^3}{128 \varepsilon m_2}
+\frac{299 C_A Y_s \mathcal{L}_{m_2} Y_f^3}{128 m_2}
-\frac{65 C_A Y_s \mathcal{L}_{m_2} Y_f^3}{128 \varepsilon m_2}
-\frac{3 C_A S_1 Y_s^2 Y_f^2}{32 m_2^2}
+\frac{27 C_A S_1 S_2 Y_s^2 Y_f^2}{64 m_2^2}
-\frac{119 C_A Y_s^2 Y_f^2}{512 m_2^2}
-\frac{C_A Y_s^2 \mathcal{L}_{m_2}^2 Y_f^2}{64 m_2^2}
+\frac{1}{2} C_A C_F \mathcal{L}_{m_2}^2 Y_f^2
+\frac{9 C_A M_H^2 Y_s \mathcal{L}_{m_2}^2 Y_f^2}{32 m_2^2 M_W}
-\frac{15 C_A M_H^2 Y_s \mathcal{L}_{m_2}^2 Y_f^2}{64 \varepsilon m_2^2 M_W}
-\frac{C_A C_F \mathcal{L}_{m_2}^2 Y_f^2}{4 \varepsilon }
+\frac{41}{4} C_A C_F Y_f^2
+\frac{267 C_A M_H^2 Y_s Y_f^2}{64 m_2^2 M_W}
-\frac{1725 C_A M_H^2 Y_s Y_f^2}{512 \varepsilon m_2^2 M_W}
+\frac{29 C_A Y_s^2 \zeta_2 Y_f^2}{128 m_2^2}
+\frac{i C_A \pi Y_s^2 \zeta_2 Y_f^2}{32 m_2^2}
-\frac{9}{16} C_A C_F \zeta_2 Y_f^2
-\frac{3 C_A M_H^2 Y_s \zeta_2 Y_f^2}{64 m_2^2 M_W}
+\frac{33 C_A M_H^2 Y_s \zeta_2 Y_f^2}{128 \varepsilon m_2^2 M_W}
+\frac{7 C_A C_F \zeta_2 Y_f^2}{8 \varepsilon }
-\frac{3 C_A Y_s^2 \zeta_3 Y_f^2}{32 m_2^2}
+\frac{13 C_A Y_s^2 \mathcal{L}_{m_2} Y_f^2}{128 m_2^2}
-\frac{33}{8} C_A C_F \mathcal{L}_{m_2} Y_f^2
-\frac{15 C_A M_H^2 Y_s \mathcal{L}_{m_2} Y_f^2}{8 m_2^2 M_W}
+\frac{195 C_A M_H^2 Y_s \mathcal{L}_{m_2} Y_f^2}{128 \varepsilon m_2^2 M_W}
+\frac{13 C_A C_F \mathcal{L}_{m_2} Y_f^2}{8 \varepsilon }
-\frac{113 C_A C_F Y_f^2}{32 \varepsilon }
+\frac{3 C_A S_1 Y_s^3 Y_f}{16 m_2^3}
-\frac{27 C_A S_1 S_2 Y_s^3 Y_f}{32 m_2^3}
-\frac{525 C_A Y_s^3 Y_f}{1024 m_2^3}
-\frac{95 C_A Y_s^3 Y_f}{1024 \varepsilon m_2^3}
+\frac{699 C_A M_H^2 Y_s^2 Y_f}{256 m_2^3 M_W}
-\frac{345 C_A M_H^2 Y_s^2 Y_f}{256 \varepsilon m_2^3 M_W}
-\frac{15 C_A Y_s^3 \mathcal{L}_{m_2}^2 Y_f}{128 m_2^3}
+\frac{5 C_A Y_s^3 \mathcal{L}_{m_2}^2 Y_f}{128 \varepsilon m_2^3}
+\frac{9 C_A M_H^2 Y_s^2 \mathcal{L}_{m_2}^2 Y_f}{32 m_2^3 M_W}
-\frac{3 C_A M_H^2 Y_s^2 \mathcal{L}_{m_2}^2 Y_f}{32 \varepsilon m_2^3 M_W}
+\frac{21 C_A C_F Y_s \mathcal{L}_{m_2}^2 Y_f}{8 m_2}
-\frac{5 C_A C_F Y_s \mathcal{L}_{m_2}^2 Y_f}{16 \varepsilon m_2}
-\frac{21 C_A C_F S_1 Y_s Y_f}{8 m_2}
+\frac{27 C_A C_F S_1 S_2 Y_s Y_f}{16 m_2}
+\frac{13 C_A C_F^2 Y_s Y_f}{4 m_2}
+\frac{1021 C_A C_F Y_s Y_f}{64 m_2}
+\frac{13 C_F Y_s Y_f}{8 m_2}
+\frac{C_A C_F^2 Y_s Y_f}{2 \varepsilon m_2}
+\frac{1435 C_A C_F Y_s Y_f}{128 \varepsilon m_2}
+\frac{C_F Y_s Y_f}{4 \varepsilon m_2}
-\frac{63 C_A Y_s^3 \zeta_2 Y_f}{256 m_2^3}
+\frac{37 C_A Y_s^3 \zeta_2 Y_f}{256 \varepsilon m_2^3}
-\frac{i C_A \pi Y_s^3 \zeta_2 Y_f}{16 m_2^3}
-\frac{27 C_A M_H^2 Y_s^2 \zeta_2 Y_f}{64 m_2^3 M_W}
-\frac{3 C_A M_H^2 Y_s^2 \zeta_2 Y_f}{64 \varepsilon m_2^3 M_W}
-\frac{29 C_A C_F Y_s \zeta_2 Y_f}{8 m_2}
-\frac{181 C_A C_F Y_s \zeta_2 Y_f}{32 \varepsilon m_2}
+\frac{i C_A C_F \pi Y_s \zeta_2 Y_f}{8 m_2}
+\frac{3 C_A Y_s^3 \zeta_3 Y_f}{16 m_2^3}
-\frac{3 C_A C_F Y_s \zeta_3 Y_f}{8 m_2}
+\frac{131 C_A Y_s^3 \mathcal{L}_{m_2} Y_f}{256 m_2^3}
-\frac{23 C_A Y_s^3 \mathcal{L}_{m_2} Y_f}{256 \varepsilon m_2^3}
-\frac{93 C_A M_H^2 Y_s^2 \mathcal{L}_{m_2} Y_f}{64 m_2^3 M_W}
+\frac{39 C_A M_H^2 Y_s^2 \mathcal{L}_{m_2} Y_f}{64 \varepsilon m_2^3 M_W}
-\frac{C_A C_F^2 Y_s \mathcal{L}_{m_2} Y_f}{m_2}
-\frac{189 C_A C_F Y_s \mathcal{L}_{m_2} Y_f}{16 m_2}
-\frac{C_F Y_s \mathcal{L}_{m_2} Y_f}{2 m_2}
-\frac{61 C_A C_F Y_s \mathcal{L}_{m_2} Y_f}{32 \varepsilon m_2}
-\frac{3177}{32} C_A C_F^2
-\frac{3 C_A C_F S_1 Y_s^2}{4 m_2^2}
+\frac{27 C_A C_F S_1 S_2 Y_s^2}{8 m_2^2}
+\frac{907 C_A C_F Y_s^2}{512 m_2^2}
+\frac{287 C_A C_F Y_s^2}{512 \varepsilon m_2^2}
+\frac{C_A C_F Y_s^2}{4 \varepsilon ^2 m_2^2}-12 C_A C_F^2 \mathcal{L}_{m_2}^2
+\frac{37 C_A C_F Y_s^2 \mathcal{L}_{m_2}^2}{64 m_2^2}
-\frac{5 C_A C_F Y_s^2 \mathcal{L}_{m_2}^2}{64 \varepsilon m_2^2}
+\frac{9}{8} C_A C_F \mathcal{L}_{m_2}^2
-\frac{157}{16} C_F \mathcal{L}_{m_2}^2
-\frac{1}{2} C_A C_F n_f T_f \mathcal{L}_{m_2}^2
+\frac{19 C_A C_F M_W Y_s \mathcal{L}_{m_2}^2}{8 m_2^2}
-\frac{C_A C_F M_W Y_s \mathcal{L}_{m_2}^2}{\varepsilon m_2^2}
+\frac{3 C_A C_F^2 \mathcal{L}_{m_2}^2}{4 \varepsilon }
+\frac{49 C_F \mathcal{L}_{m_2}^2}{16 \varepsilon }
+\frac{255 C_A C_F}{64}
-\frac{14029 C_F}{128}
-\frac{3 C_F S_1}{8}+27 C_F S_1 S_2
-\frac{309}{16} C_A C_F n_f T_f
+\frac{C_A C_F n_f T_f}{\varepsilon }
+\frac{C_A C_F n_f T_f}{2 \varepsilon ^2}
+\frac{2655 C_A C_F M_W Y_s}{64 m_2^2}
-\frac{115 C_A C_F M_W Y_s}{8 \varepsilon m_2^2}
+\frac{47}{4} C_A C_F^2 \zeta_2
+\frac{85 C_A C_F Y_s^2 \zeta_2}{128 m_2^2}
-\frac{37 C_A C_F Y_s^2 \zeta_2}{128 \varepsilon m_2^2}
+\frac{i C_A C_F \pi Y_s^2 \zeta_2}{4 m_2^2}
+\frac{13}{16} C_A C_F \zeta_2
+\frac{7 C_F \zeta_2}{32}
-\frac{5}{4} C_A C_F n_f T_f \zeta_2
-\frac{25 C_A C_F M_W Y_s \zeta_2}{16 m_2^2}
+\frac{2 C_A C_F M_W Y_s \zeta_2}{\varepsilon m_2^2}
-\frac{9 C_A C_F^2 \zeta_2}{8 \varepsilon }
-\frac{343 C_F \zeta_2}{32 \varepsilon }+2 i C_F \pi \zeta_2
-\frac{3 C_A C_F Y_s^2 \zeta_3}{4 m_2^2}-6 C_F \zeta_3
+\frac{447}{8} C_A C_F^2 \mathcal{L}_{m_2}
-\frac{181 C_A C_F Y_s^2 \mathcal{L}_{m_2}}{128 m_2^2}
-\frac{41 C_A C_F Y_s^2 \mathcal{L}_{m_2}}{128 \varepsilon m_2^2}
-\frac{1}{16} C_A C_F \mathcal{L}_{m_2}
+\frac{1747}{32} C_F \mathcal{L}_{m_2}
+\frac{31}{4} C_A C_F n_f T_f \mathcal{L}_{m_2}
-\frac{C_A C_F n_f T_f \mathcal{L}_{m_2}}{\varepsilon }
-\frac{281 C_A C_F M_W Y_s \mathcal{L}_{m_2}}{16 m_2^2}
+\frac{13 C_A C_F M_W Y_s \mathcal{L}_{m_2}}{2 \varepsilon m_2^2}
+\frac{33 C_A C_F^2 \mathcal{L}_{m_2}}{8 \varepsilon }
-\frac{C_A C_F \mathcal{L}_{m_2}}{\varepsilon }
-\frac{619 C_F \mathcal{L}_{m_2}}{32 \varepsilon }
-\frac{299 C_A C_F^2}{32 \varepsilon }
-\frac{3 C_A C_F}{8 \varepsilon }
+\frac{6197 C_F}{128 \varepsilon }
-\frac{9 C_A C_F^2}{2 \varepsilon ^2}
+\frac{C_A C_F}{2 \varepsilon ^2}
-\frac{9 C_F}{4 \varepsilon ^2}
\end{autobreak}
\end{align}
\end{tiny}

\subsection{Matching at $\mu\sim m_1$: }
The contributions from two-loop vertex corrections, $V^{(m_1)}_i$, with unevaluated MIs are too large to present here. We thus include the full expressions with description in an ancillary file.
\subsection{Matching at $\mu\sim M$: }
The contributions from two-loop vertex corrections, $V^{(M)}_i$, with unevaluated MIs are too large to present here. We thus include the full expressions with description in an ancillary file.
\section{Wave-function Corrections}
\label{sec:wfr2}
\subsection{Full theory field at $m=0$ and $M=0$:}

\begin{tiny}
\begin{align}
\begin{autobreak}
F^{(0,0)}_{\psi}=
\frac{2 \Delta_{M} Y_f^2 C_A C_F}{M_W \varepsilon _{\text{UV}}}
-\frac{\Delta_{M} Y_f^4 C_A}{4 M_W \varepsilon_{\text{UV}}}
+\frac{n_f C_A C_F T_f}{\varepsilon _{\text{IR}} \varepsilon _{\text{UV}}}
-\frac{n_f C_A C_F T_f}{2\varepsilon _{\text{UV}}}-\frac{n_f C_A C_F T_f}{2 \varepsilon _{\text{UV}}^2}
-\frac{8 C_A C_F^2}{\varepsilon _{\text{IR}}\varepsilon _{\text{UV}}}
+\frac{Y_f^2 C_A C_F}{2 \varepsilon _{\text{IR}} \varepsilon _{\text{UV}}}
+\frac{4 C_AC_F^2}{\varepsilon _{\text{UV}}^2}
-\frac{Y_f^2 C_A C_F}{4 \varepsilon _{\text{UV}}}-\frac{Y_f^2 C_A C_F}{4 \varepsilon_{\text{UV}}^2}
-\frac{Y_f^4 C_A}{8 \varepsilon _{\text{IR}} \varepsilon _{\text{UV}}}
+\frac{5 Y_f^4 C_A}{32 \varepsilon_{\text{UV}}}
+\frac{Y_f^4 C_A}{16 \varepsilon _{\text{UV}}^2}-\frac{4 C_F}{\varepsilon _{\text{IR}} \varepsilon_{\text{UV}}}+\frac{2 C_F}{\varepsilon _{\text{UV}}^2}
\end{autobreak}
\end{align}
\end{tiny}

\begin{tiny}
\begin{align}
\begin{autobreak}
F^{(0,0)}_{\chi}=
-\frac{10 \Delta_{M} Y_s^2 C_A C_F}{M_W^3 \varepsilon _{\text{IR}} \varepsilon _{\text{UV}}}-
\frac{14 \Delta_{M} Y_s C_A C_F}{M_W^2 \varepsilon _{\text{IR}} \varepsilon _{\text{UV}}}+
\frac{153 \Delta_{M} Y_s^2 C_A C_F^2}{16 M_W^3 \varepsilon _{\text{UV}}}+
\frac{37 \Delta_{M} Y_s^2 C_A C_F}{4 M_W^3 \varepsilon _{\text{UV}}}+
\frac{5 \Delta_{M} Y_s^2 C_A C_F}{M_W^3 \varepsilon _{\text{UV}}^2}+
\frac{11 \Delta_{M} Y_s C_A C_F}{M_W^2 \varepsilon _{\text{UV}}}+
\frac{7 \Delta_{M} Y_s C_A C_F}{M_W^2 \varepsilon _{\text{UV}}^2}+
\frac{\Delta_{M} C_A C_F}{M_W \varepsilon _{\text{UV}}}-
\frac{5 \Delta_{M} Y_s^4 C_A}{4 M_W^5 \varepsilon _{\text{IR}} \varepsilon _{\text{UV}}}+
\frac{7 \Delta_{M} Y_s^4 C_A}{4 M_W^5 \varepsilon _{\text{UV}}}+
\frac{5 \Delta_{M} Y_s^4 C_A}{8 M_W^5 \varepsilon _{\text{UV}}^2}-
\frac{3 \Delta_{M} Y_s^3 C_A}{4 M_W^4 \varepsilon _{\text{UV}}}+
\frac{9 \Delta_{M} Y_s^2 C_A}{16 M_W^3 \varepsilon _{\text{UV}}}+
\frac{2 n_f C_A C_F T_f}{\varepsilon _{\text{IR}} \varepsilon _{\text{UV}}}-
\frac{n_f C_A C_F T_f}{\varepsilon _{\text{UV}}}-
\frac{n_f C_A C_F T_f}{\varepsilon _{\text{UV}}^2}+
\frac{3 Y_s^2 C_A C_F}{M_W^2 \varepsilon _{\text{IR}} \varepsilon _{\text{UV}}}+
\frac{7 Y_s C_A C_F}{M_W \varepsilon _{\text{IR}} \varepsilon _{\text{UV}}}-
\frac{19 C_A C_F^2}{2 \varepsilon _{\text{IR}} \varepsilon _{\text{UV}}}-
\frac{7 C_A C_F}{2 \varepsilon _{\text{IR}} \varepsilon _{\text{UV}}}-
\frac{65 Y_s^2 C_A C_F^2}{32 M_W^2 \varepsilon _{\text{UV}}}-
\frac{2 Y_s^2 C_A C_F}{M_W^2 \varepsilon _{\text{UV}}}-
\frac{3 Y_s^2 C_A C_F}{2 M_W^2 \varepsilon _{\text{UV}}^2}-
\frac{2 Y_s C_A C_F}{M_W \varepsilon _{\text{UV}}}-
\frac{7 Y_s C_A C_F}{2 M_W \varepsilon _{\text{UV}}^2}+
\frac{49 C_A C_F^2}{2 \varepsilon _{\text{UV}}}-
\frac{5 C_A C_F}{2 \varepsilon _{\text{UV}}}+
\frac{19 C_A C_F^2}{4 \varepsilon _{\text{UV}}^2}+
\frac{7 C_A C_F}{4 \varepsilon _{\text{UV}}^2}+
\frac{5 Y_s^4 C_A}{16 M_W^4 \varepsilon _{\text{IR}} \varepsilon _{\text{UV}}}-
\frac{9 Y_s^4 C_A}{32 M_W^4 \varepsilon _{\text{UV}}}-
\frac{5 Y_s^4 C_A}{32 M_W^4 \varepsilon _{\text{UV}}^2}+
\frac{3 Y_s^3 C_A}{8 M_W^3 \varepsilon _{\text{UV}}}-
\frac{9 Y_s^2 C_A}{32 M_W^2 \varepsilon _{\text{UV}}}+
\frac{153 \Delta_{M} Y_s^2 C_F}{32 M_W^3 \varepsilon _{\text{UV}}}-
\frac{47 C_F}{4 \varepsilon _{\text{IR}} \varepsilon _{\text{UV}}}-
\frac{65 Y_s^2 C_F}{64 M_W^2 \varepsilon _{\text{UV}}}+
\frac{73 C_F}{4 \varepsilon _{\text{UV}}}+
\frac{47 C_F}{8 \varepsilon _{\text{UV}}^2}
\end{autobreak}
\end{align}
\end{tiny}

\subsection{Full theory field at $m=0$ and $M\neq0$ ($\Delta_{M}\equiv M_H-M_W$):}
\begin{tiny}
\begin{align}
\begin{autobreak}
F^{(0,M)}_{\psi}=
-\frac{4 \Delta_{M} Y_f^2 C_A C_F \mathcal{L}_{M_W}}{M_W}-
\frac{9 \Delta_{M} S_2 C_A C_F}{M_W}+
\frac{2 \Delta_{M} Y_f^2 C_A C_F}{M_W \varepsilon _{\text{UV}}}-
\frac{3 \Delta_{M} Y_f^2 \zeta_2 C_A C_F}{M_W}+
\frac{4 \Delta_{M} Y_f^2 C_A C_F}{M_W}+
\frac{2 \Delta_{M} C_A C_F}{M_W}+
\frac{\Delta_{M} Y_f^4 C_A \mathcal{L}_{M_W}}{2 M_W}-
\frac{\Delta_{M} Y_f^4 C_A}{4 M_W \varepsilon _{\text{UV}}}-
\frac{5 \Delta_{M} Y_f^4 C_A}{8 M_W}+
\frac{n_f C_A C_F T_f \mathcal{L}_{M_W}}{\varepsilon _{\text{UV}}}-n_f C_A C_F T_f \mathcal{L}_{M_W}^2+n_f C_A C_F T_f \mathcal{L}_{M_W}-
\frac{n_f C_A C_F T_f}{2 \varepsilon _{\text{UV}}}-
\frac{n_f C_A C_F T_f}{2 \varepsilon _{\text{UV}}^2}-
\frac{3}{2} n_f \zeta_2 C_A C_F T_f-
\frac{1}{2} n_f C_A C_F T_f-
\frac{8 C_A C_F^2 \mathcal{L}_{M_W}}{\varepsilon _{\text{UV}}}+
\frac{Y_f^2 C_A C_F \mathcal{L}_{M_W}}{2 \varepsilon _{\text{UV}}}-
\frac{1}{2} Y_f^2 C_A C_F \mathcal{L}_{M_W}^2+
\frac{1}{2} Y_f^2 C_A C_F \mathcal{L}_{M_W}+8 C_A C_F^2 \mathcal{L}_{M_W}^2+18 S_2 C_A C_F^2+
\frac{4 C_A C_F^2}{\varepsilon _{\text{UV}}^2}-
\frac{Y_f^2 C_A C_F}{4 \varepsilon _{\text{UV}}}-
\frac{Y_f^2 C_A C_F}{4 \varepsilon _{\text{UV}}^2}-
\frac{3}{4} Y_f^2 \zeta_2 C_A C_F-Y_f^2 C_A C_F+8 \zeta_2 C_A C_F^2-12 C_A C_F^2-
\frac{Y_f^4 C_A \mathcal{L}_{M_W}}{8 \varepsilon _{\text{UV}}}+
\frac{1}{8} Y_f^4 C_A \mathcal{L}_{M_W}^2-
\frac{5}{16} Y_f^4 C_A \mathcal{L}_{M_W}+
\frac{5 Y_f^4 C_A}{32 \varepsilon _{\text{UV}}}+
\frac{Y_f^4 C_A}{16 \varepsilon _{\text{UV}}^2}+
\frac{Y_f^4 C_A}{4}-
\frac{4 C_F \mathcal{L}_{M_W}}{\varepsilon _{\text{UV}}}+4 C_F \mathcal{L}_{M_W}^2+9 S_2 C_F+
\frac{2 C_F}{\varepsilon _{\text{UV}}^2}+4 \zeta_2 C_F-4 C_F
\end{autobreak}
\end{align}
\end{tiny}

\begin{tiny}
\begin{align}
\begin{autobreak}
F^{(0,M)}_{\chi}=
-\frac{5 C_A \mathcal{L}_{M_W}^2 Y_s^4}{16 M_W^4}+
\frac{5 C_A \Delta_{M} \mathcal{L}_{M_W}^2 Y_s^4}{4 M_W^5}-
\frac{3 C_A \zeta_2 Y_s^4}{32 M_W^4}+
\frac{3 C_A \Delta_{M} \zeta_2 Y_s^4}{8 M_W^5}+
\frac{9 C_A \mathcal{L}_{M_W} Y_s^4}{16 M_W^4}+
\frac{5 C_A \mathcal{L}_{M_W} Y_s^4}{16 \varepsilon _{\text{UV}} M_W^4}-
\frac{7 C_A \Delta_{M} \mathcal{L}_{M_W} Y_s^4}{2 M_W^5}-
\frac{5 C_A \Delta_{M} \mathcal{L}_{M_W} Y_s^4}{4 \varepsilon _{\text{UV}} M_W^5}-
\frac{17 C_A Y_s^4}{32 M_W^4}-
\frac{9 C_A Y_s^4}{32 \varepsilon _{\text{UV}} M_W^4}-
\frac{5 C_A Y_s^4}{32 \varepsilon _{\text{UV}}^2 M_W^4}+
\frac{13 C_A \Delta_{M} Y_s^4}{4 M_W^5}+
\frac{7 C_A \Delta_{M} Y_s^4}{4 \varepsilon _{\text{UV}} M_W^5}+
\frac{5 C_A \Delta_{M} Y_s^4}{8 \varepsilon _{\text{UV}}^2 M_W^5}+
\frac{81 C_A S_2 Y_s^3}{16 M_W^3}-
\frac{81 C_A \Delta_{M} S_2 Y_s^3}{8 M_W^4}-
\frac{3 C_A \mathcal{L}_{M_W} Y_s^3}{4 M_W^3}+
\frac{3 C_A \Delta_{M} \mathcal{L}_{M_W} Y_s^3}{2 M_W^4}-
\frac{3 C_A Y_s^3}{8 M_W^3}+
\frac{3 C_A Y_s^3}{8 \varepsilon _{\text{UV}} M_W^3}-
\frac{3 C_A \Delta_{M} Y_s^3}{4 M_W^4}-
\frac{3 C_A \Delta_{M} Y_s^3}{4 \varepsilon _{\text{UV}} M_W^4}-
\frac{3 C_A C_F \mathcal{L}_{M_W}^2 Y_s^2}{M_W^2}+
\frac{10 C_A C_F \Delta_{M} \mathcal{L}_{M_W}^2 Y_s^2}{M_W^3}+
\frac{1005 C_A C_F^2 S_2 Y_s^2}{64 M_W^2}+
\frac{189 C_A S_2 Y_s^2}{64 M_W^2}+
\frac{1005 C_F S_2 Y_s^2}{128 M_W^2}-
\frac{417 C_A C_F^2 \Delta_{M} S_2 Y_s^2}{4 M_W^3}-
\frac{27 C_A \Delta_{M} S_2 Y_s^2}{16 M_W^3}-
\frac{417 C_F \Delta_{M} S_2 Y_s^2}{8 M_W^3}-
\frac{5 C_A C_F \zeta_2 Y_s^2}{4 M_W^2}+
\frac{11 C_A C_F \Delta_{M} \zeta_2 Y_s^2}{2 M_W^3}+
\frac{65 C_A C_F^2 \mathcal{L}_{M_W} Y_s^2}{16 M_W^2}+
\frac{9 C_A \mathcal{L}_{M_W} Y_s^2}{16 M_W^2}+
\frac{4 C_A C_F \mathcal{L}_{M_W} Y_s^2}{M_W^2}+
\frac{65 C_F \mathcal{L}_{M_W} Y_s^2}{32 M_W^2}+
\frac{3 C_A C_F \mathcal{L}_{M_W} Y_s^2}{\varepsilon _{\text{UV}} M_W^2}-
\frac{153 C_A C_F^2 \Delta_{M} \mathcal{L}_{M_W} Y_s^2}{8 M_W^3}-
\frac{9 C_A \Delta_{M} \mathcal{L}_{M_W} Y_s^2}{8 M_W^3}-
\frac{37 C_A C_F \Delta_{M} \mathcal{L}_{M_W} Y_s^2}{2 M_W^3}-
\frac{153 C_F \Delta_{M} \mathcal{L}_{M_W} Y_s^2}{16 M_W^3}-
\frac{10 C_A C_F \Delta_{M} \mathcal{L}_{M_W} Y_s^2}{\varepsilon _{\text{UV}} M_W^3}-
\frac{213 C_A C_F^2 Y_s^2}{32 M_W^2}-
\frac{45 C_A Y_s^2}{32 M_W^2}-
\frac{3 C_A C_F Y_s^2}{M_W^2}-
\frac{213 C_F Y_s^2}{64 M_W^2}-
\frac{65 C_A C_F^2 Y_s^2}{32 \varepsilon _{\text{UV}} M_W^2}-
\frac{9 C_A Y_s^2}{32 \varepsilon _{\text{UV}} M_W^2}-
\frac{2 C_A C_F Y_s^2}{\varepsilon _{\text{UV}} M_W^2}-
\frac{65 C_F Y_s^2}{64 \varepsilon _{\text{UV}} M_W^2}-
\frac{3 C_A C_F Y_s^2}{2 \varepsilon _{\text{UV}}^2 M_W^2}+
\frac{645 C_A C_F^2 \Delta_{M} Y_s^2}{16 M_W^3}+
\frac{45 C_A \Delta_{M} Y_s^2}{16 M_W^3}+
\frac{45 C_A C_F \Delta_{M} Y_s^2}{4 M_W^3}+
\frac{645 C_F \Delta_{M} Y_s^2}{32 M_W^3}+
\frac{153 C_A C_F^2 \Delta_{M} Y_s^2}{16 \varepsilon _{\text{UV}} M_W^3}+
\frac{9 C_A \Delta_{M} Y_s^2}{16 \varepsilon _{\text{UV}} M_W^3}+
\frac{37 C_A C_F \Delta_{M} Y_s^2}{4 \varepsilon _{\text{UV}} M_W^3}+
\frac{153 C_F \Delta_{M} Y_s^2}{32 \varepsilon _{\text{UV}} M_W^3}+
\frac{5 C_A C_F \Delta_{M} Y_s^2}{\varepsilon _{\text{UV}}^2 M_W^3}-
\frac{7 C_A C_F \mathcal{L}_{M_W}^2 Y_s}{M_W}+
\frac{14 C_A C_F \Delta_{M} \mathcal{L}_{M_W}^2 Y_s}{M_W^2}+
\frac{261 C_A C_F S_2 Y_s}{4 M_W}-
\frac{192 C_A C_F \Delta_{M} S_2 Y_s}{M_W^2}-
\frac{4 C_A C_F \zeta_2 Y_s}{M_W}+
\frac{9 C_A C_F \Delta_{M} \zeta_2 Y_s}{M_W^2}+
\frac{4 C_A C_F \mathcal{L}_{M_W} Y_s}{M_W}+
\frac{7 C_A C_F \mathcal{L}_{M_W} Y_s}{\varepsilon _{\text{UV}} M_W}-
\frac{22 C_A C_F \Delta_{M} \mathcal{L}_{M_W} Y_s}{M_W^2}-
\frac{14 C_A C_F \Delta_{M} \mathcal{L}_{M_W} Y_s}{\varepsilon _{\text{UV}} M_W^2}-
\frac{35 C_A C_F Y_s}{2 M_W}-
\frac{2 C_A C_F Y_s}{\varepsilon _{\text{UV}} M_W}-
\frac{7 C_A C_F Y_s}{2 \varepsilon _{\text{UV}}^2 M_W}+
\frac{54 C_A C_F \Delta_{M} Y_s}{M_W^2}+
\frac{11 C_A C_F \Delta_{M} Y_s}{\varepsilon _{\text{UV}} M_W^2}+
\frac{7 C_A C_F \Delta_{M} Y_s}{\varepsilon _{\text{UV}}^2 M_W^2}+
\frac{81}{2} C_A C_F^2+
\frac{19}{2} C_A C_F^2 \mathcal{L}_{M_W}^2+
\frac{7}{2} C_A C_F \mathcal{L}_{M_W}^2+
\frac{47}{4} C_F \mathcal{L}_{M_W}^2-2 C_A C_F n_f T_f \mathcal{L}_{M_W}^2+
\frac{19 C_A C_F}{2}+
\frac{97 C_F}{4}-141 C_A C_F^2 S_2-6 C_A C_F S_2-
\frac{141 C_F S_2}{2}-
\frac{21 C_A C_F \Delta_{M} S_2}{M_W}-C_A C_F n_f T_f-
\frac{C_A C_F n_f T_f}{\varepsilon _{\text{UV}}}-
\frac{C_A C_F n_f T_f}{\varepsilon _{\text{UV}}^2}-
\frac{17}{4} C_A C_F^2 \zeta_2+
\frac{7}{4} C_A C_F \zeta_2+
\frac{39 C_F \zeta_2}{8}-3 C_A C_F n_f T_f \zeta_2-49 C_A C_F^2 \mathcal{L}_{M_W}+5 C_A C_F \mathcal{L}_{M_W}-
\frac{73}{2} C_F \mathcal{L}_{M_W}+2 C_A C_F n_f T_f \mathcal{L}_{M_W}+
\frac{2 C_A C_F n_f T_f \mathcal{L}_{M_W}}{\varepsilon _{\text{UV}}}-
\frac{19 C_A C_F^2 \mathcal{L}_{M_W}}{2 \varepsilon _{\text{UV}}}-
\frac{7 C_A C_F \mathcal{L}_{M_W}}{2 \varepsilon _{\text{UV}}}-
\frac{47 C_F \mathcal{L}_{M_W}}{4 \varepsilon _{\text{UV}}}-
\frac{2 C_A C_F \Delta_{M} \mathcal{L}_{M_W}}{M_W}+
\frac{49 C_A C_F^2}{2 \varepsilon _{\text{UV}}}-
\frac{5 C_A C_F}{2 \varepsilon _{\text{UV}}}+
\frac{73 C_F}{4 \varepsilon _{\text{UV}}}+
\frac{4 C_A C_F \Delta_{M}}{M_W}+
\frac{C_A C_F \Delta_{M}}{\varepsilon _{\text{UV}} M_W}+
\frac{19 C_A C_F^2}{4 \varepsilon _{\text{UV}}^2}+
\frac{7 C_A C_F}{4 \varepsilon _{\text{UV}}^2}+
\frac{47 C_F}{8 \varepsilon _{\text{UV}}^2}
\end{autobreak}
\end{align}
\end{tiny}

\subsection{Full theory field at $m\neq 0$ and $M=0$:}

\begin{tiny}
\begin{align}
\begin{autobreak}
F^{(m,0)}_{\psi}=
-\frac{47}{16} C_A \mathcal{L}_m^2 Y_f^4
+\frac{5 C_A \mathcal{L}_m^2 Y_f^4}{16 \varepsilon _{\text{IR}}}
-\frac{2889 C_A Y_f^4}{256}
-\frac{47}{16} C_A \zeta_2 Y_f^4
+\frac{37 C_A \zeta_2 Y_f^4}{32 \varepsilon _{\text{IR}}}
+\frac{543}{64} C_A \mathcal{L}_m Y_f^4
+\frac{77 C_A \mathcal{L}_m Y_f^4}{64 \varepsilon _{\text{IR}}}
-\frac{769 C_A Y_f^4}{256 \varepsilon _{\text{IR}}}
+\frac{5 C_A Y_f^4}{32 \varepsilon _{\text{UV}}}
-\frac{C_A Y_f^4}{8 \varepsilon _{\text{IR}} \varepsilon _{\text{UV}}}
-\frac{115 C_A Y_f^4}{128 \varepsilon _{\text{IR}}^2}
+\frac{C_A Y_f^4}{16 \varepsilon _{\text{UV}}^2}
-\frac{9131}{256} C_A^2 Y_f^2
-\frac{363}{32} C_A C_F^2 Y_f^2
-\frac{363}{32} C_A^2 \mathcal{L}_m^2 Y_f^2
-\frac{3}{4} C_A C_F^2 \mathcal{L}_m^2 Y_f^2
+\frac{3}{4} C_A^2 C_F \mathcal{L}_m^2 Y_f^2
+\frac{363}{8} C_A C_F \mathcal{L}_m^2 Y_f^2
-\frac{3}{8} C_F \mathcal{L}_m^2 Y_f^2
-\frac{5 C_A C_F \mathcal{L}_m^2 Y_f^2}{4 \varepsilon _{\text{IR}}}
+\frac{363}{32} \mathcal{L}_m^2 Y_f^2
+\frac{363}{32} C_A^2 C_F Y_f^2
+\frac{9131}{64} C_A C_F Y_f^2
-\frac{363 C_F Y_f^2}{64}
-\frac{699}{64} C_A^2 \zeta_2 Y_f^2
-\frac{15}{8} C_A C_F^2 \zeta_2 Y_f^2
+\frac{15}{8} C_A^2 C_F \zeta_2 Y_f^2
+\frac{699}{16} C_A C_F \zeta_2 Y_f^2
-\frac{15}{16} C_F \zeta_2 Y_f^2
-\frac{37 C_A C_F \zeta_2 Y_f^2}{8 \varepsilon _{\text{IR}}}
+\frac{699 \zeta_2 Y_f^2}{64}
+\frac{1501}{64} C_A^2 \mathcal{L}_m Y_f^2
+\frac{41}{8} C_A C_F^2 \mathcal{L}_m Y_f^2
-\frac{41}{8} C_A^2 C_F \mathcal{L}_m Y_f^2
-\frac{1501}{16} C_A C_F \mathcal{L}_m Y_f^2
+\frac{41}{16} C_F \mathcal{L}_m Y_f^2
-\frac{3 C_A C_F^2 \mathcal{L}_m Y_f^2}{4 \varepsilon _{\text{IR}}}
-\frac{53 C_A C_F \mathcal{L}_m Y_f^2}{4 \varepsilon _{\text{IR}}}
-\frac{3 C_F \mathcal{L}_m Y_f^2}{8 \varepsilon _{\text{IR}}}
-\frac{1501}{64} \mathcal{L}_m Y_f^2
+\frac{41 C_A C_F^2 Y_f^2}{16 \varepsilon _{\text{IR}}}
+\frac{61 C_A C_F Y_f^2}{4 \varepsilon _{\text{IR}}}
+\frac{41 C_F Y_f^2}{32 \varepsilon _{\text{IR}}}
-\frac{C_A C_F Y_f^2}{4 \varepsilon _{\text{UV}}}
+\frac{C_A C_F Y_f^2}{2 \varepsilon _{\text{IR}} \varepsilon _{\text{UV}}}
+\frac{3 C_A C_F^2 Y_f^2}{8 \varepsilon _{\text{IR}}^2}
+\frac{125 C_A C_F Y_f^2}{16 \varepsilon _{\text{IR}}^2}
+\frac{3 C_F Y_f^2}{16 \varepsilon _{\text{IR}}^2}
-\frac{C_A C_F Y_f^2}{4 \varepsilon _{\text{UV}}^2}
+\frac{9131 Y_f^2}{256}
-\frac{81 C_A^2}{64}
+\frac{8835}{32} C_A C_F^2
-\frac{1}{8} C_A^2 \mathcal{L}_m^2
+\frac{267}{4} C_A C_F^2 \mathcal{L}_m^2
-\frac{267}{4} C_A^2 C_F \mathcal{L}_m^2
+\frac{1}{2} C_A C_F \mathcal{L}_m^2
+\frac{375}{8} C_F \mathcal{L}_m^2
-\frac{1}{2} C_A^2 n_f T_f \mathcal{L}_m^2+2 C_A C_F n_f T_f \mathcal{L}_m^2
+\frac{1}{2} n_f T_f \mathcal{L}_m^2
+\frac{\mathcal{L}_m^2}{8}
-\frac{8835}{32} C_A^2 C_F
+\frac{81 C_A C_F}{16}
+\frac{10615 C_F}{64}
-\frac{59}{16} C_A^2 n_f T_f
+\frac{59}{4} C_A C_F n_f T_f
+\frac{11 C_A C_F n_f T_f}{4 \varepsilon _{\text{IR}}}
-\frac{C_A C_F n_f T_f}{2 \varepsilon _{\text{UV}}}
+\frac{C_A C_F n_f T_f}{\varepsilon _{\text{IR}} \varepsilon _{\text{UV}}}
-\frac{C_A C_F n_f T_f}{2 \varepsilon _{\text{UV}}^2}
+\frac{59 n_f T_f}{16}
-\frac{5 C_A^2 \zeta_2}{16}
+\frac{407}{8} C_A C_F^2 \zeta_2
-\frac{407}{8} C_A^2 C_F \zeta_2
+\frac{5}{4} C_A C_F \zeta_2
+\frac{659 C_F \zeta_2}{16}
-\frac{5}{4} C_A^2 n_f T_f \zeta_2+5 C_A C_F n_f T_f \zeta_2
+\frac{5}{4} n_f T_f \zeta_2
+\frac{5 \zeta_2}{16}
+\frac{11}{16} C_A^2 \mathcal{L}_m
-\frac{1169}{8} C_A C_F^2 \mathcal{L}_m
+\frac{1169}{8} C_A^2 C_F \mathcal{L}_m
-\frac{11}{4} C_A C_F \mathcal{L}_m
-\frac{1229}{16} C_F \mathcal{L}_m
+\frac{9}{4} C_A^2 n_f T_f \mathcal{L}_m-9 C_A C_F n_f T_f \mathcal{L}_m
-\frac{C_A C_F n_f T_f \mathcal{L}_m}{\varepsilon _{\text{IR}}}
-\frac{9}{4} n_f T_f \mathcal{L}_m
+\frac{267 C_A C_F^2 \mathcal{L}_m}{4 \varepsilon _{\text{IR}}}
-\frac{C_A C_F \mathcal{L}_m}{4 \varepsilon _{\text{IR}}}
+\frac{159 C_F \mathcal{L}_m}{8 \varepsilon _{\text{IR}}}
-\frac{11 \mathcal{L}_m}{16}
-\frac{1169 C_A C_F^2}{16 \varepsilon _{\text{IR}}}
+\frac{11 C_A C_F}{16 \varepsilon _{\text{IR}}}
-\frac{1109 C_F}{32 \varepsilon _{\text{IR}}}
-\frac{8 C_A C_F^2}{\varepsilon _{\text{IR}} \varepsilon _{\text{UV}}}
-\frac{4 C_F}{\varepsilon _{\text{IR}} \varepsilon _{\text{UV}}}
-\frac{235 C_A C_F^2}{8 \varepsilon _{\text{IR}}^2}
+\frac{C_A C_F}{8 \varepsilon _{\text{IR}}^2}
-\frac{127 C_F}{16 \varepsilon _{\text{IR}}^2}
+\frac{4 C_A C_F^2}{\varepsilon _{\text{UV}}^2}
+\frac{2 C_F}{\varepsilon _{\text{UV}}^2}
+\frac{81}{64}
\end{autobreak}
\end{align}
\end{tiny}

\begin{tiny}
\begin{align}
\begin{autobreak}
F^{(m,0)}_{\chi}=
\frac{15 Y_s^2 C_A^5}{8 m^2}+
\frac{Y_s^2 \mathcal{L}_m^2 C_A^5}{8 m^2}-
\frac{27}{8} \mathcal{L}_m^2 C_A^5-
\frac{27 \zeta_2 C_A^5}{16}-
\frac{13 Y_s^2 \mathcal{L}_m C_A^5}{16 m^2}+
\frac{191}{16} \mathcal{L}_m C_A^5-
\frac{883 C_A^5}{32}-
\frac{15 C_F Y_s^2 C_A^4}{4 m^2}+
\frac{8121 Y_s^2 C_A^4}{1024 m^2}-
\frac{C_F Y_s^2 \mathcal{L}_m^2 C_A^4}{4 m^2}+
\frac{239 Y_s^2 \mathcal{L}_m^2 C_A^4}{128 m^2}+
\frac{27}{4} C_F \mathcal{L}_m^2 C_A^4+6 \mathcal{L}_m^2 C_A^4+
\frac{883}{16} C_F C_A^4+
\frac{3}{2} n_f T_f C_A^4+
\frac{151 Y_s^2 \zeta_2 C_A^4}{256 m^2}+
\frac{27}{8} C_F \zeta_2 C_A^4+3 \zeta_2 C_A^4+
\frac{13 C_F Y_s^2 \mathcal{L}_m C_A^4}{8 m^2}-
\frac{1111 Y_s^2 \mathcal{L}_m C_A^4}{256 m^2}-
\frac{191}{8} C_F \mathcal{L}_m C_A^4-
\frac{1}{2} n_f T_f \mathcal{L}_m C_A^4-
\frac{1}{8} \mathcal{L}_m C_A^4+
\frac{175 C_A^4}{16}-
\frac{3341 Y_s^4 C_A^3}{8192 m^4}-
\frac{8121 C_F Y_s^2 C_A^3}{512 m^2}-
\frac{15 Y_s^2 C_A^3}{8 m^2}-
\frac{249 Y_s^4 \mathcal{L}_m^2 C_A^3}{1024 m^4}-
\frac{239 C_F Y_s^2 \mathcal{L}_m^2 C_A^3}{64 m^2}-
\frac{Y_s^2 \mathcal{L}_m^2 C_A^3}{8 m^2}-12 C_F \mathcal{L}_m^2 C_A^3+
\frac{21}{8} \mathcal{L}_m^2 C_A^3-
\frac{175}{8} C_F C_A^3-3 C_F n_f T_f C_A^3-
\frac{121 Y_s^4 \zeta_2 C_A^3}{2048 m^4}-
\frac{151 C_F Y_s^2 \zeta_2 C_A^3}{128 m^2}-6 C_F \zeta_2 C_A^3+
\frac{33 \zeta_2 C_A^3}{16}+
\frac{779 Y_s^4 \mathcal{L}_m C_A^3}{2048 m^4}+
\frac{1111 C_F Y_s^2 \mathcal{L}_m C_A^3}{128 m^2}+
\frac{13 Y_s^2 \mathcal{L}_m C_A^3}{16 m^2}+
\frac{1}{4} C_F \mathcal{L}_m C_A^3+C_F n_f T_f \mathcal{L}_m C_A^3-
\frac{47}{16} \mathcal{L}_m C_A^3+
\frac{463 C_A^3}{32}+
\frac{3341 C_F Y_s^4 C_A^2}{4096 m^4}+
\frac{15 C_F Y_s^2 C_A^2}{4 m^2}-
\frac{8121 Y_s^2 C_A^2}{1024 m^2}+
\frac{249 C_F Y_s^4 \mathcal{L}_m^2 C_A^2}{512 m^4}+
\frac{C_F Y_s^2 \mathcal{L}_m^2 C_A^2}{4 m^2}-
\frac{239 Y_s^2 \mathcal{L}_m^2 C_A^2}{128 m^2}-
\frac{21}{4} C_F \mathcal{L}_m^2 C_A^2-6 \mathcal{L}_m^2 C_A^2-
\frac{463}{16} C_F C_A^2-
\frac{3}{2} n_f T_f C_A^2+
\frac{121 C_F Y_s^4 \zeta_2 C_A^2}{1024 m^4}-
\frac{151 Y_s^2 \zeta_2 C_A^2}{256 m^2}-
\frac{33}{8} C_F \zeta_2 C_A^2-3 \zeta_2 C_A^2-
\frac{779 C_F Y_s^4 \mathcal{L}_m C_A^2}{1024 m^4}-
\frac{13 C_F Y_s^2 \mathcal{L}_m C_A^2}{8 m^2}+
\frac{1111 Y_s^2 \mathcal{L}_m C_A^2}{256 m^2}+
\frac{47}{8} C_F \mathcal{L}_m C_A^2+
\frac{1}{2} n_f T_f \mathcal{L}_m C_A^2+
\frac{1}{8} \mathcal{L}_m C_A^2-
\frac{175 C_A^2}{16}+
\frac{241 Y_s^4 C_A}{2048 \varepsilon _{\text{IR}} m^4}-
\frac{Y_s^4 C_A}{64 \varepsilon _{\text{IR}}^2 m^4}+
\frac{Y_s^4 C_A}{64 \varepsilon _{\text{IR}}^3 m^4}+
\frac{9 Y_s^4 C_A}{32 \varepsilon _{\text{IR}} M_W^4}+
\frac{5 Y_s^4 C_A}{16 \varepsilon _{\text{IR}} \varepsilon _{\text{UV}} M_W^4}-
\frac{9 Y_s^4 C_A}{32 \varepsilon _{\text{UV}} M_W^4}-
\frac{5 Y_s^4 C_A}{32 \varepsilon _{\text{IR}}^2 M_W^4}-
\frac{5 Y_s^4 C_A}{32 \varepsilon _{\text{UV}}^2 M_W^4}-
\frac{3 Y_s^3 C_A}{8 \varepsilon _{\text{IR}} M_W^3}+
\frac{3 Y_s^3 C_A}{8 \varepsilon _{\text{UV}} M_W^3}+
\frac{8121 C_F Y_s^2 C_A}{512 m^2}+
\frac{13 C_F^2 Y_s^2 C_A}{8 \varepsilon _{\text{IR}} m^2}+
\frac{733 C_F Y_s^2 C_A}{512 \varepsilon _{\text{IR}} m^2}+
\frac{C_F^2 Y_s^2 C_A}{4 \varepsilon _{\text{IR}}^2 m^2}+
\frac{15 C_F Y_s^2 C_A}{16 \varepsilon _{\text{IR}}^2 m^2}-
\frac{3 C_F Y_s^2 C_A}{16 \varepsilon _{\text{IR}}^3 m^2}+
\frac{65 C_F^2 Y_s^2 C_A}{32 \varepsilon _{\text{IR}} M_W^2}+
\frac{2 C_F Y_s^2 C_A}{\varepsilon _{\text{IR}} M_W^2}+
\frac{9 Y_s^2 C_A}{32 \varepsilon _{\text{IR}} M_W^2}-
\frac{65 C_F^2 Y_s^2 C_A}{32 \varepsilon _{\text{UV}} M_W^2}-
\frac{2 C_F Y_s^2 C_A}{\varepsilon _{\text{UV}} M_W^2}+
\frac{3 C_F Y_s^2 C_A}{\varepsilon _{\text{IR}} \varepsilon _{\text{UV}} M_W^2}-
\frac{9 Y_s^2 C_A}{32 \varepsilon _{\text{UV}} M_W^2}-
\frac{3 C_F Y_s^2 C_A}{2 \varepsilon _{\text{IR}}^2 M_W^2}-
\frac{3 C_F Y_s^2 C_A}{2 \varepsilon _{\text{UV}}^2 M_W^2}+
\frac{13 Y_s^4 \mathcal{L}_m^2 C_A}{256 \varepsilon _{\text{IR}} m^4}+
\frac{239 C_F Y_s^2 \mathcal{L}_m^2 C_A}{64 m^2}-
\frac{39 C_F Y_s^2 \mathcal{L}_m^2 C_A}{64 \varepsilon _{\text{IR}} m^2}+12 C_F \mathcal{L}_m^2 C_A+
\frac{3}{4} \mathcal{L}_m^2 C_A+
\frac{175 C_F C_A}{8}+3 C_F n_f T_f C_A+
\frac{3 C_F n_f T_f C_A}{2 \varepsilon _{\text{IR}}}-
\frac{C_F n_f T_f C_A}{\varepsilon _{\text{UV}}}+
\frac{2 C_F n_f T_f C_A}{\varepsilon _{\text{IR}} \varepsilon _{\text{UV}}}-
\frac{C_F n_f T_f C_A}{\varepsilon _{\text{IR}}^2}-
\frac{C_F n_f T_f C_A}{\varepsilon _{\text{UV}}^2}+
\frac{2 C_F Y_s C_A}{\varepsilon _{\text{IR}} M_W}-
\frac{2 C_F Y_s C_A}{\varepsilon _{\text{UV}} M_W}+
\frac{7 C_F Y_s C_A}{\varepsilon _{\text{IR}} \varepsilon _{\text{UV}} M_W}-
\frac{7 C_F Y_s C_A}{2 \varepsilon _{\text{IR}}^2 M_W}-
\frac{7 C_F Y_s C_A}{2 \varepsilon _{\text{UV}}^2 M_W}-
\frac{3 Y_s^4 \zeta_2 C_A}{512 \varepsilon _{\text{IR}} m^4}+
\frac{151 C_F Y_s^2 \zeta_2 C_A}{128 m^2}+
\frac{9 C_F Y_s^2 \zeta_2 C_A}{128 \varepsilon _{\text{IR}} m^2}+6 C_F \zeta_2 C_A-
\frac{3 \zeta_2 C_A}{8}-
\frac{7 Y_s^4 \mathcal{L}_m C_A}{512 \varepsilon _{\text{IR}} m^4}-
\frac{Y_s^4 \mathcal{L}_m C_A}{32 \varepsilon _{\text{IR}}^2 m^4}-
\frac{1111 C_F Y_s^2 \mathcal{L}_m C_A}{128 m^2}-
\frac{C_F^2 Y_s^2 \mathcal{L}_m C_A}{2 \varepsilon _{\text{IR}} m^2}-
\frac{171 C_F Y_s^2 \mathcal{L}_m C_A}{128 \varepsilon _{\text{IR}} m^2}+
\frac{3 C_F Y_s^2 \mathcal{L}_m C_A}{8 \varepsilon _{\text{IR}}^2 m^2}-
\frac{1}{4} C_F \mathcal{L}_m C_A-C_F n_f T_f \mathcal{L}_m C_A+
\frac{27 C_F^2 \mathcal{L}_m C_A}{2 \varepsilon _{\text{IR}}}-
\frac{12 C_F \mathcal{L}_m C_A}{\varepsilon _{\text{IR}}}-9 \mathcal{L}_m C_A-
\frac{387 C_F^2 C_A}{8 \varepsilon _{\text{IR}}}+
\frac{21 C_F C_A}{8 \varepsilon _{\text{IR}}}+
\frac{49 C_F^2 C_A}{2 \varepsilon _{\text{UV}}}-
\frac{5 C_F C_A}{2 \varepsilon _{\text{UV}}}-
\frac{19 C_F^2 C_A}{2 \varepsilon _{\text{IR}} \varepsilon _{\text{UV}}}-
\frac{7 C_F C_A}{2 \varepsilon _{\text{IR}} \varepsilon _{\text{UV}}}-
\frac{2 C_F^2 C_A}{\varepsilon _{\text{IR}}^2}+
\frac{31 C_F C_A}{4 \varepsilon _{\text{IR}}^2}+
\frac{19 C_F^2 C_A}{4 \varepsilon _{\text{UV}}^2}+
\frac{7 C_F C_A}{4 \varepsilon _{\text{UV}}^2}+
\frac{105 C_A}{8}+
\frac{13 C_F Y_s^2}{16 \varepsilon _{\text{IR}} m^2}+
\frac{C_F Y_s^2}{8 \varepsilon _{\text{IR}}^2 m^2}+
\frac{65 C_F Y_s^2}{64 \varepsilon _{\text{IR}} M_W^2}-
\frac{65 C_F Y_s^2}{64 \varepsilon _{\text{UV}} M_W^2}-
\frac{3}{2} C_F \mathcal{L}_m^2-
\frac{105 C_F}{4}+
\frac{3 C_F \zeta_2}{4}-
\frac{C_F Y_s^2 \mathcal{L}_m}{4 \varepsilon _{\text{IR}} m^2}+18 C_F \mathcal{L}_m+
\frac{33 C_F \mathcal{L}_m}{4 \varepsilon _{\text{IR}}}-
\frac{627 C_F}{16 \varepsilon _{\text{IR}}}+
\frac{73 C_F}{4 \varepsilon _{\text{UV}}}-
\frac{47 C_F}{4 \varepsilon _{\text{IR}} \varepsilon _{\text{UV}}}+
\frac{7 C_F}{4 \varepsilon _{\text{IR}}^2}+
\frac{47 C_F}{8 \varepsilon _{\text{UV}}^2}
\end{autobreak}
\end{align}
\end{tiny}

\subsection{Full theory field at $m\neq 0$ and $M\neq 0$ ($\Delta_{M}\equiv M_H-M_W$, $\Delta_{m,M}\equiv M_W-m$):}
\begin{tiny}
\begin{align}
\begin{autobreak}
F^{(m,M)}_{\psi}= 
\frac{5}{64} C_A \mathcal{L}_m^2 Y_f^4
-\frac{2229 C_A Y_f^4}{512}
+\frac{9}{4} C_A S_1 Y_f^4
-\frac{39}{64} C_A \log{(3)} S_1 Y_f^4
+\frac{39 C_A S_1 Y_f^4}{64 \varepsilon _{\text{UV}}}
+\frac{25 C_A \Delta_M S_1 Y_f^4}{12 m}
+\frac{25 C_A \Delta_{m,M} S_1 Y_f^4}{12 m}
-\frac{3 C_A \Delta_M \log{(3)} S_1 Y_f^4}{16 m}
-\frac{3 C_A \Delta_{m,M} \log{(3)} S_1 Y_f^4}{16 m}
+\frac{3 C_A \Delta_M S_1 Y_f^4}{16 \varepsilon _{\text{UV}} m}
+\frac{3 C_A \Delta_{m,M} S_1 Y_f^4}{16 \varepsilon _{\text{UV}} m}
+\frac{2295}{256} C_A S_2 Y_f^4
+\frac{81}{64} C_A S_1 S_2 Y_f^4
-\frac{117 C_A \Delta_M S_1 S_2 Y_f^4}{16 m}
-\frac{117 C_A \Delta_{m,M} S_1 S_2 Y_f^4}{16 m}
-\frac{351 C_A \Delta_M S_2 Y_f^4}{16 m}
-\frac{351 C_A \Delta_{m,M} S_2 Y_f^4}{16 m}
-\frac{25}{32} C_A \zeta_2 Y_f^4
+\frac{9 C_A \Delta_M \zeta_2 Y_f^4}{8 m}
+\frac{9 C_A \Delta_{m,M} \zeta_2 Y_f^4}{8 m}
-\frac{9}{32} C_A \zeta_3 Y_f^4
+\frac{13 C_A \Delta_M \zeta_3 Y_f^4}{8 m}
+\frac{13 C_A \Delta_{m,M} \zeta_3 Y_f^4}{8 m}
+\frac{271}{128} C_A \mathcal{L}_m Y_f^4
-\frac{39}{32} C_A S_1 \mathcal{L}_m Y_f^4
-\frac{3 C_A \Delta_M S_1 \mathcal{L}_m Y_f^4}{8 m}
-\frac{3 C_A \Delta_{m,M} S_1 \mathcal{L}_m Y_f^4}{8 m}
-\frac{5 C_A \mathcal{L}_m Y_f^4}{64 \varepsilon _{\text{UV}}}
+\frac{7 C_A \Delta_M \mathcal{L}_m Y_f^4}{8 m}
+\frac{7 C_A \Delta_{m,M} \mathcal{L}_m Y_f^4}{8 m}
-\frac{271 C_A Y_f^4}{256 \varepsilon _{\text{UV}}}
+\frac{37 C_A \Delta_M Y_f^4}{16 m}
+\frac{37 C_A \Delta_{m,M} Y_f^4}{16 m}
-\frac{7 C_A \Delta_M Y_f^4}{16 \varepsilon _{\text{UV}} m}
-\frac{7 C_A \Delta_{m,M} Y_f^4}{16 \varepsilon _{\text{UV}} m}
+\frac{5 C_A Y_f^4}{128 \varepsilon _{\text{UV}}^2}
-\frac{3 C_A Y_f^3}{8}
-\frac{3}{8} C_A S_1 Y_f^3
-\frac{13 C_A \Delta_M S_1 Y_f^3}{4 m}
-\frac{23 C_A \Delta_{m,M} S_1 Y_f^3}{8 m}
+\frac{27}{8} C_A S_1 S_2 Y_f^3
-\frac{27 C_A \Delta_{m,M} S_1 S_2 Y_f^3}{8 m}
+\frac{405 C_A \Delta_M S_2 Y_f^3}{32 m}
+\frac{405 C_A \Delta_{m,M} S_2 Y_f^3}{32 m}
+\frac{33 C_A \Delta_M \zeta_2 Y_f^3}{16 m}
+\frac{33 C_A \Delta_{m,M} \zeta_2 Y_f^3}{16 m}
-\frac{3}{4} C_A \zeta_3 Y_f^3
+\frac{3 C_A \Delta_{m,M} \zeta_3 Y_f^3}{4 m}
-\frac{3 C_A \Delta_M Y_f^3}{4 m}
-\frac{3 C_A \Delta_{m,M} Y_f^3}{8 m}
+\frac{861}{64} C_A C_F^2 Y_f^2
-\frac{1}{4} C_A C_F^2 \mathcal{L}_m^2 Y_f^2
-\frac{3}{8} C_A C_F \mathcal{L}_m^2 Y_f^2
-\frac{1}{8} C_F \mathcal{L}_m^2 Y_f^2
-\frac{45 C_A C_F \Delta_M \mathcal{L}_m^2 Y_f^2}{16 m}
-\frac{45 C_A C_F \Delta_{m,M} \mathcal{L}_m^2 Y_f^2}{8 m}
+\frac{135 C_A Y_f^2}{64}
+\frac{2327}{64} C_A C_F Y_f^2
+\frac{861 C_F Y_f^2}{128}
-\frac{521}{64} C_A C_F^2 S_1 Y_f^2
-\frac{105}{64} C_A S_1 Y_f^2
-\frac{19}{2} C_A C_F S_1 Y_f^2
-\frac{521}{128} C_F S_1 Y_f^2
+\frac{57}{32} C_A C_F^2 \log{(3)} S_1 Y_f^2
+\frac{9}{32} C_A \log{(3)} S_1 Y_f^2
+\frac{41}{8} C_A C_F \log{(3)} S_1 Y_f^2
+\frac{57}{64} C_F \log{(3)} S_1 Y_f^2
-\frac{57 C_A C_F^2 S_1 Y_f^2}{32 \varepsilon _{\text{UV}}}
-\frac{9 C_A S_1 Y_f^2}{32 \varepsilon _{\text{UV}}}
-\frac{41 C_A C_F S_1 Y_f^2}{8 \varepsilon _{\text{UV}}}
-\frac{57 C_F S_1 Y_f^2}{64 \varepsilon _{\text{UV}}}
+\frac{1891 C_A C_F^2 \Delta_M S_1 Y_f^2}{96 m}
-\frac{193 C_A \Delta_M S_1 Y_f^2}{32 m}
+\frac{5 C_A C_F \Delta_M S_1 Y_f^2}{6 m}
+\frac{1891 C_F \Delta_M S_1 Y_f^2}{192 m}
+\frac{C_A C_F^2 \Delta_{m,M} S_1 Y_f^2}{4 m}
-\frac{11 C_A \Delta_{m,M} S_1 Y_f^2}{4 m}
-\frac{245 C_A C_F \Delta_{m,M} S_1 Y_f^2}{36 m}
+\frac{C_F \Delta_{m,M} S_1 Y_f^2}{8 m}
+\frac{25 C_A C_F^2 \Delta_M \log{(3)} S_1 Y_f^2}{16 m}
+\frac{33 C_A \Delta_M \log{(3)} S_1 Y_f^2}{16 m}
+\frac{5 C_A C_F \Delta_M \log{(3)} S_1 Y_f^2}{2 m}
+\frac{25 C_F \Delta_M \log{(3)} S_1 Y_f^2}{32 m}
+\frac{9 C_A C_F^2 \Delta_{m,M} \log{(3)} S_1 Y_f^2}{2 m}
+\frac{3 C_A \Delta_{m,M} \log{(3)} S_1 Y_f^2}{2 m}
+\frac{15 C_A C_F \Delta_{m,M} \log{(3)} S_1 Y_f^2}{2 m}
+\frac{9 C_F \Delta_{m,M} \log{(3)} S_1 Y_f^2}{4 m}
-\frac{25 C_A C_F^2 \Delta_M S_1 Y_f^2}{16 \varepsilon _{\text{UV}} m}
-\frac{33 C_A \Delta_M S_1 Y_f^2}{16 \varepsilon _{\text{UV}} m}
-\frac{5 C_A C_F \Delta_M S_1 Y_f^2}{2 \varepsilon _{\text{UV}} m}
-\frac{25 C_F \Delta_M S_1 Y_f^2}{32 \varepsilon _{\text{UV}} m}
-\frac{9 C_A C_F^2 \Delta_{m,M} S_1 Y_f^2}{2 \varepsilon _{\text{UV}} m}
-\frac{3 C_A \Delta_{m,M} S_1 Y_f^2}{2 \varepsilon _{\text{UV}} m}
-\frac{15 C_A C_F \Delta_{m,M} S_1 Y_f^2}{2 \varepsilon _{\text{UV}} m}
-\frac{9 C_F \Delta_{m,M} S_1 Y_f^2}{4 \varepsilon _{\text{UV}} m}
+\frac{171}{32} C_A C_F^2 S_2 Y_f^2
+\frac{81}{32} C_A S_2 Y_f^2
-\frac{3501}{32} C_A C_F S_2 Y_f^2
+\frac{171}{64} C_F S_2 Y_f^2
-\frac{63}{4} C_A C_F S_1 S_2 Y_f^2
+\frac{81 C_A C_F \Delta_M S_1 S_2 Y_f^2}{4 m}
-\frac{9 C_A C_F \Delta_{m,M} S_1 S_2 Y_f^2}{4 m}
-\frac{10317 C_A C_F^2 \Delta_M S_2 Y_f^2}{64 m}
-\frac{513 C_A \Delta_M S_2 Y_f^2}{32 m}
-\frac{99 C_A C_F \Delta_M S_2 Y_f^2}{2 m}
-\frac{10317 C_F \Delta_M S_2 Y_f^2}{128 m}
-\frac{3051 C_A C_F^2 \Delta_{m,M} S_2 Y_f^2}{32 m}
-\frac{675 C_A \Delta_{m,M} S_2 Y_f^2}{32 m}
-\frac{639 C_A C_F \Delta_{m,M} S_2 Y_f^2}{4 m}
-\frac{3051 C_F \Delta_{m,M} S_2 Y_f^2}{64 m}
-\frac{57}{32} C_A C_F^2 \zeta_2 Y_f^2
-\frac{9}{32} C_A \zeta_2 Y_f^2
+\frac{5}{2} C_A C_F \zeta_2 Y_f^2
-\frac{57}{64} C_F \zeta_2 Y_f^2
+\frac{101 C_A C_F^2 \Delta_M \zeta_2 Y_f^2}{32 m}
-\frac{9 C_A \Delta_M \zeta_2 Y_f^2}{8 m}
-\frac{301 C_A C_F \Delta_M \zeta_2 Y_f^2}{32 m}
+\frac{101 C_F \Delta_M \zeta_2 Y_f^2}{64 m}
-\frac{33 C_A C_F^2 \Delta_{m,M} \zeta_2 Y_f^2}{16 m}
-\frac{9 C_A \Delta_{m,M} \zeta_2 Y_f^2}{16 m}
-\frac{439 C_A C_F \Delta_{m,M} \zeta_2 Y_f^2}{48 m}
-\frac{33 C_F \Delta_{m,M} \zeta_2 Y_f^2}{32 m}
+\frac{7}{2} C_A C_F \zeta_3 Y_f^2
-\frac{9 C_A C_F \Delta_M \zeta_3 Y_f^2}{2 m}
+\frac{C_A C_F \Delta_{m,M} \zeta_3 Y_f^2}{2 m}
-\frac{59}{16} C_A C_F^2 \mathcal{L}_m Y_f^2
-\frac{9}{16} C_A \mathcal{L}_m Y_f^2
-\frac{293}{16} C_A C_F \mathcal{L}_m Y_f^2
-\frac{59}{32} C_F \mathcal{L}_m Y_f^2
+\frac{57}{16} C_A C_F^2 S_1 \mathcal{L}_m Y_f^2
+\frac{9}{16} C_A S_1 \mathcal{L}_m Y_f^2
+\frac{41}{4} C_A C_F S_1 \mathcal{L}_m Y_f^2
+\frac{57}{32} C_F S_1 \mathcal{L}_m Y_f^2
+\frac{25 C_A C_F^2 \Delta_M S_1 \mathcal{L}_m Y_f^2}{8 m}
+\frac{33 C_A \Delta_M S_1 \mathcal{L}_m Y_f^2}{8 m}
+\frac{5 C_A C_F \Delta_M S_1 \mathcal{L}_m Y_f^2}{m}
+\frac{25 C_F \Delta_M S_1 \mathcal{L}_m Y_f^2}{16 m}
+\frac{9 C_A C_F^2 \Delta_{m,M} S_1 \mathcal{L}_m Y_f^2}{m}
+\frac{3 C_A \Delta_{m,M} S_1 \mathcal{L}_m Y_f^2}{m}
+\frac{15 C_A C_F \Delta_{m,M} S_1 \mathcal{L}_m Y_f^2}{m}
+\frac{9 C_F \Delta_{m,M} S_1 \mathcal{L}_m Y_f^2}{2 m}
+\frac{C_A C_F^2 \mathcal{L}_m Y_f^2}{4 \varepsilon _{\text{UV}}}
+\frac{3 C_A C_F \mathcal{L}_m Y_f^2}{8 \varepsilon _{\text{UV}}}
+\frac{C_F \mathcal{L}_m Y_f^2}{8 \varepsilon _{\text{UV}}}
-\frac{51 C_A C_F^2 \Delta_M \mathcal{L}_m Y_f^2}{4 m}
-\frac{27 C_A \Delta_M \mathcal{L}_m Y_f^2}{4 m}
-\frac{81 C_A C_F \Delta_M \mathcal{L}_m Y_f^2}{32 m}
-\frac{51 C_F \Delta_M \mathcal{L}_m Y_f^2}{8 m}
-\frac{149 C_A C_F^2 \Delta_{m,M} \mathcal{L}_m Y_f^2}{8 m}
-\frac{45 C_A \Delta_{m,M} \mathcal{L}_m Y_f^2}{8 m}
-\frac{225 C_A C_F \Delta_{m,M} \mathcal{L}_m Y_f^2}{16 m}
-\frac{149 C_F \Delta_{m,M} \mathcal{L}_m Y_f^2}{16 m}
+\frac{59 C_A C_F^2 Y_f^2}{32 \varepsilon _{\text{UV}}}
+\frac{9 C_A Y_f^2}{32 \varepsilon _{\text{UV}}}
+\frac{293 C_A C_F Y_f^2}{32 \varepsilon _{\text{UV}}}
+\frac{59 C_F Y_f^2}{64 \varepsilon _{\text{UV}}}
-\frac{29 C_A C_F^2 \Delta_M Y_f^2}{32 m}
+\frac{435 C_A \Delta_M Y_f^2}{32 m}
+\frac{23 C_A C_F \Delta_M Y_f^2}{128 m}
-\frac{29 C_F \Delta_M Y_f^2}{64 m}
+\frac{191 C_A C_F^2 \Delta_{m,M} Y_f^2}{8 m}
+\frac{75 C_A \Delta_{m,M} Y_f^2}{8 m}
+\frac{3989 C_A C_F \Delta_{m,M} Y_f^2}{192 m}
+\frac{191 C_F \Delta_{m,M} Y_f^2}{16 m}
+\frac{51 C_A C_F^2 \Delta_M Y_f^2}{8 \varepsilon _{\text{UV}} m}
+\frac{27 C_A \Delta_M Y_f^2}{8 \varepsilon _{\text{UV}} m}
+\frac{9 C_A C_F \Delta_M Y_f^2}{2 \varepsilon _{\text{UV}} m}
+\frac{51 C_F \Delta_M Y_f^2}{16 \varepsilon _{\text{UV}} m}
+\frac{149 C_A C_F^2 \Delta_{m,M} Y_f^2}{16 \varepsilon _{\text{UV}} m}
+\frac{45 C_A \Delta_{m,M} Y_f^2}{16 \varepsilon _{\text{UV}} m}
+\frac{27 C_A C_F \Delta_{m,M} Y_f^2}{2 \varepsilon _{\text{UV}} m}
+\frac{149 C_F \Delta_{m,M} Y_f^2}{32 \varepsilon _{\text{UV}} m}
-\frac{C_A C_F^2 Y_f^2}{8 \varepsilon _{\text{UV}}^2}
-\frac{3 C_A C_F Y_f^2}{16 \varepsilon _{\text{UV}}^2}
-\frac{C_F Y_f^2}{16 \varepsilon _{\text{UV}}^2}
+\frac{5 C_A C_F \Delta_M \mathcal{L}_m^2 Y_f}{2 m}
+\frac{5 C_A C_F \Delta_{m,M} \mathcal{L}_m^2 Y_f}{m}
-\frac{22 C_A C_F \Delta_M S_1 Y_f}{3 m}
+\frac{8 C_A C_F \Delta_{m,M} S_1 Y_f}{m}
-54 C_A C_F S_1 S_2 Y_f
+\frac{72 C_A C_F \Delta_M S_1 S_2 Y_f}{m}
-\frac{54 C_A C_F \Delta_{m,M} S_1 S_2 Y_f}{m}
+\frac{297 C_A C_F \Delta_M S_2 Y_f}{2 m}
+\frac{108 C_A C_F \Delta_{m,M} S_2 Y_f}{m}
+8 C_A C_F \zeta_2 Y_f
-\frac{197 C_A C_F \Delta_M \zeta_2 Y_f}{12 m}
-\frac{17 C_A C_F \Delta_{m,M} \zeta_2 Y_f}{6 m}
+12 C_A C_F \zeta_3 Y_f
-\frac{16 C_A C_F \Delta_M \zeta_3 Y_f}{m}
+\frac{12 C_A C_F \Delta_{m,M} \zeta_3 Y_f}{m}
-\frac{23 C_A C_F \Delta_M \mathcal{L}_m Y_f}{4 m}
-\frac{23 C_A C_F \Delta_{m,M} \mathcal{L}_m Y_f}{2 m}
-\frac{31 C_A C_F \Delta_M Y_f}{16 m}
-\frac{31 C_A C_F \Delta_{m,M} Y_f}{8 m}
-\frac{45485}{96} C_A C_F^2
-5 C_A C_F^2 \mathcal{L}_m^2
-3 C_F \mathcal{L}_m^2
-\frac{5 C_A C_F \Delta_M \mathcal{L}_m^2}{12 m}
-\frac{5 C_A C_F \Delta_{m,M} \mathcal{L}_m^2}{6 m}
-\frac{15 C_F \Delta_{m,M} \mathcal{L}_m^2}{2 m}
+\frac{1231 C_A C_F}{96}
-\frac{45233 C_F}{192}
+\frac{2945}{18} C_A C_F^2 S_1
-\frac{29}{18} C_A C_F S_1
+\frac{3461 C_F S_1}{36}
-\frac{1061}{12} C_A C_F^2 \log{(3)} S_1
+\frac{85}{36} C_A C_F \log{(3)} S_1
-\frac{1025}{24} C_F \log{(3)} S_1
+\frac{1061 C_A C_F^2 S_1}{12 \varepsilon _{\text{UV}}}
-\frac{85 C_A C_F S_1}{36 \varepsilon _{\text{UV}}}
+\frac{1025 C_F S_1}{24 \varepsilon _{\text{UV}}}
+\frac{79 C_A C_F \Delta_M S_1}{18 m}
+\frac{569 C_A C_F^2 \Delta_{m,M} S_1}{6 m}
+\frac{233 C_A C_F \Delta_{m,M} S_1}{54 m}
+\frac{2779 C_F \Delta_{m,M} S_1}{36 m}
-\frac{5 C_A C_F \Delta_M \log{(3)} S_1}{3 m}
-\frac{379 C_A C_F^2 \Delta_{m,M} \log{(3)} S_1}{3 m}
+\frac{19 C_A C_F \Delta_{m,M} \log{(3)} S_1}{9 m}
-\frac{93 C_F \Delta_{m,M} \log{(3)} S_1}{2 m}
+\frac{5 C_A C_F \Delta_M S_1}{3 \varepsilon _{\text{UV}} m}
+\frac{379 C_A C_F^2 \Delta_{m,M} S_1}{3 \varepsilon _{\text{UV}} m}
-\frac{19 C_A C_F \Delta_{m,M} S_1}{9 \varepsilon _{\text{UV}} m}
+\frac{93 C_F \Delta_{m,M} S_1}{2 \varepsilon _{\text{UV}} m}
+\frac{23163}{16} C_A C_F^2 S_2
-\frac{817}{16} C_A C_F S_2
+\frac{19167 C_F S_2}{32}
+144 C_A C_F^2 S_1 S_2
+72 C_F S_1 S_2
+\frac{450 C_A C_F^2 \Delta_{m,M} S_1 S_2}{m}
+\frac{225 C_F \Delta_{m,M} S_1 S_2}{m}
+\frac{27 C_A C_F \Delta_M S_2}{2 m}
+\frac{8769 C_A C_F^2 \Delta_{m,M} S_2}{4 m}
-\frac{427 C_A C_F \Delta_{m,M} S_2}{4 m}
+\frac{4113 C_F \Delta_{m,M} S_2}{8 m}
+\frac{377}{24} C_A C_F \text{nf} T_f
+\frac{55 C_A C_F \text{nf} T_f}{12 \varepsilon _{\text{UV}}}
+\frac{134 C_A C_F \Delta_{m,M} \text{nf} T_f}{9 m}
+\frac{6 C_A C_F \Delta_{m,M} \text{nf} T_f}{\varepsilon _{\text{UV}} m}
-5 C_A C_F \text{nf} S_1 T_f
+\frac{25}{9} C_A C_F \log{(3)} \text{nf} S_1 T_f
-\frac{25 C_A C_F \text{nf} S_1 T_f}{9 \varepsilon _{\text{UV}}}
+\frac{46 C_A C_F \Delta_{m,M} \text{nf} S_1 T_f}{27 m}
+\frac{28 C_A C_F \Delta_{m,M} \log{(3)} \text{nf} S_1 T_f}{9 m}
-\frac{28 C_A C_F \Delta_{m,M} \text{nf} S_1 T_f}{9 \varepsilon _{\text{UV}} m}
-\frac{175}{4} C_A C_F \text{nf} S_2 T_f
-\frac{49 C_A C_F \Delta_{m,M} \text{nf} S_2 T_f}{m}
-\frac{351}{8} C_A C_F^2 \zeta_2
-\frac{5}{24} C_A C_F \zeta_2
-\frac{287 C_F \zeta_2}{16}
-\frac{1}{2} C_A C_F \text{nf} T_f \zeta_2
-\frac{6 C_A C_F \Delta_{m,M} \text{nf} T_f \zeta_2}{m}
-\frac{55 C_A C_F \Delta_M \zeta_2}{72 m}
-\frac{559 C_A C_F^2 \Delta_{m,M} \zeta_2}{6 m}
-\frac{C_A C_F \Delta_{m,M} \zeta_2}{36 m}
-\frac{139 C_F \Delta_{m,M} \zeta_2}{3 m}
-32 C_A C_F^2 \zeta_3
-16 C_F \zeta_3
-\frac{100 C_A C_F^2 \Delta_{m,M} \zeta_3}{m}
-\frac{50 C_F \Delta_{m,M} \zeta_3}{m}
+\frac{2367}{8} C_A C_F^2 \mathcal{L}_m
-\frac{199}{24} C_A C_F \mathcal{L}_m
+\frac{2291}{16} C_F \mathcal{L}_m
-\frac{1061}{6} C_A C_F^2 S_1 \mathcal{L}_m
+\frac{85}{18} C_A C_F S_1 \mathcal{L}_m
-\frac{1025}{12} C_F S_1 \mathcal{L}_m
-\frac{10 C_A C_F \Delta_M S_1 \mathcal{L}_m}{3 m}
-\frac{758 C_A C_F^2 \Delta_{m,M} S_1 \mathcal{L}_m}{3 m}
+\frac{38 C_A C_F \Delta_{m,M} S_1 \mathcal{L}_m}{9 m}
-\frac{93 C_F \Delta_{m,M} S_1 \mathcal{L}_m}{m}
-\frac{55}{6} C_A C_F \text{nf} T_f \mathcal{L}_m
-\frac{12 C_A C_F \Delta_{m,M} \text{nf} T_f \mathcal{L}_m}{m}
+\frac{50}{9} C_A C_F \text{nf} S_1 T_f \mathcal{L}_m
+\frac{56 C_A C_F \Delta_{m,M} \text{nf} S_1 T_f \mathcal{L}_m}{9 m}
+\frac{5 C_A C_F^2 \mathcal{L}_m}{\varepsilon _{\text{UV}}}
+\frac{3 C_F \mathcal{L}_m}{\varepsilon _{\text{UV}}}
+\frac{167 C_A C_F \Delta_M \mathcal{L}_m}{24 m}
+\frac{449 C_A C_F^2 \Delta_{m,M} \mathcal{L}_m}{m}
-\frac{61 C_A C_F \Delta_{m,M} \mathcal{L}_m}{12 m}
+\frac{727 C_F \Delta_{m,M} \mathcal{L}_m}{4 m}
-\frac{2367 C_A C_F^2}{16 \varepsilon _{\text{UV}}}
+\frac{199 C_A C_F}{48 \varepsilon _{\text{UV}}}
-\frac{2291 C_F}{32 \varepsilon _{\text{UV}}}
-\frac{315 C_A C_F \Delta_M}{32 m}
-\frac{1328 C_A C_F^2 \Delta_{m,M}}{3 m}
+\frac{1757 C_A C_F \Delta_{m,M}}{144 m}
-\frac{9737 C_F \Delta_{m,M}}{48 m}
-\frac{3 C_A C_F \Delta_M}{\varepsilon _{\text{UV}} m}
-\frac{449 C_A C_F^2 \Delta_{m,M}}{2 \varepsilon _{\text{UV}} m}
+\frac{7 C_A C_F \Delta_{m,M}}{2 \varepsilon _{\text{UV}} m}
-\frac{329 C_F \Delta_{m,M}}{4 \varepsilon _{\text{UV}} m}
-\frac{5 C_A C_F^2}{2 \varepsilon _{\text{UV}}^2}
-\frac{3 C_F}{2 \varepsilon _{\text{UV}}^2}
\end{autobreak}
\end{align}
\end{tiny}

\begin{tiny}
\begin{align}
\begin{autobreak}
F^{(m,M)}_{\chi}=
\frac{5 C_A \Delta_{M} \mathcal{L}_m^2 Y_s^4}{64 m^5}
+\frac{5 C_A \Delta_{m,M} \mathcal{L}_m^2 Y_s^4}{64 m^5}
-\frac{5 C_A S_1 Y_s^4}{144 m^4}
-\frac{C_A \log{(3)} S_1 Y_s^4}{144 m^4}
+\frac{C_A S_1 Y_s^4}{144 \varepsilon _{\text{UV}} m^4}
+\frac{31 C_A \Delta_{M} S_1 Y_s^4}{216 m^5}
+\frac{31 C_A \Delta_{m,M} S_1 Y_s^4}{216 m^5}
-\frac{C_A \Delta_{M} \log{(3)} S_1 Y_s^4}{18 m^5}
-\frac{C_A \Delta_{m,M} \log{(3)} S_1 Y_s^4}{18 m^5}
+\frac{C_A \Delta_{M} S_1 Y_s^4}{18 \varepsilon _{\text{UV}} m^5}
+\frac{C_A \Delta_{m,M} S_1 Y_s^4}{18 \varepsilon _{\text{UV}} m^5}
+\frac{9 C_A S_1 S_2 Y_s^4}{32 m^4}
+\frac{31 C_A S_2 Y_s^4}{64 m^4}
+\frac{61 C_A \Delta_{M} S_2 Y_s^4}{32 m^5}
+\frac{61 C_A \Delta_{m,M} S_2 Y_s^4}{32 m^5}
-\frac{5 C_A \zeta_2 Y_s^4}{96 m^4}
+\frac{7 C_A \Delta_{M} \zeta_2 Y_s^4}{384 m^5}
+\frac{7 C_A \Delta_{m,M} \zeta_2 Y_s^4}{384 m^5}
-\frac{C_A \zeta_3 Y_s^4}{16 m^4}
-\frac{C_A S_1 \mathcal{L}_m Y_s^4}{72 m^4}
-\frac{C_A \Delta_{M} S_1 \mathcal{L}_m Y_s^4}{9 m^5}
-\frac{C_A \Delta_{m,M} S_1 \mathcal{L}_m Y_s^4}{9 m^5}
+\frac{C_A \mathcal{L}_m Y_s^4}{24 m^4}
-\frac{5 C_A \Delta_{M} \mathcal{L}_m Y_s^4}{384 m^5}
-\frac{5 C_A \Delta_{m,M} \mathcal{L}_m Y_s^4}{384 m^5}
-\frac{C_A Y_s^4}{24 m^4}
-\frac{C_A Y_s^4}{48 \varepsilon _{\text{UV}} m^4}
-\frac{919 C_A \Delta_{M} Y_s^4}{4608 m^5}
-\frac{919 C_A \Delta_{m,M} Y_s^4}{4608 m^5}
-\frac{C_A \Delta_{M} Y_s^4}{12 \varepsilon _{\text{UV}} m^5}
-\frac{C_A \Delta_{m,M} Y_s^4}{12 \varepsilon _{\text{UV}} m^5}
-\frac{5 C_A \Delta_{M} \mathcal{L}_m^2 Y_s^3}{64 m^4}
-\frac{5 C_A \Delta_{m,M} \mathcal{L}_m^2 Y_s^3}{64 m^4}
-\frac{C_A \Delta_{M} S_1 Y_s^3}{8 m^4}
-\frac{C_A \Delta_{m,M} S_1 Y_s^3}{8 m^4}
+\frac{27 C_A S_1 S_2 Y_s^3}{32 m^3}
+\frac{27 C_A \Delta_{M} S_1 S_2 Y_s^3}{16 m^4}
+\frac{27 C_A \Delta_{m,M} S_1 S_2 Y_s^3}{32 m^4}
-\frac{27 C_A \Delta_{M} S_2 Y_s^3}{16 m^4}
-\frac{27 C_A \Delta_{m,M} S_2 Y_s^3}{16 m^4}
-\frac{C_A \zeta_2 Y_s^3}{8 m^3}
-\frac{31 C_A \Delta_{M} \zeta_2 Y_s^3}{384 m^4}
+\frac{17 C_A \Delta_{m,M} \zeta_2 Y_s^3}{384 m^4}
-\frac{3 C_A \zeta_3 Y_s^3}{16 m^3}
-\frac{3 C_A \Delta_{M} \zeta_3 Y_s^3}{8 m^4}
-\frac{3 C_A \Delta_{m,M} \zeta_3 Y_s^3}{16 m^4}
+\frac{23 C_A \Delta_{M} \mathcal{L}_m Y_s^3}{128 m^4}
+\frac{23 C_A \Delta_{m,M} \mathcal{L}_m Y_s^3}{128 m^4}
+\frac{31 C_A \Delta_{M} Y_s^3}{512 m^4}
+\frac{31 C_A \Delta_{m,M} Y_s^3}{512 m^4}
+\frac{5 C_A \Delta_{M} \mathcal{L}_m^2 Y_s^2}{256 m^3}
-\frac{5 C_A C_F \Delta_{M} \mathcal{L}_m^2 Y_s^2}{64 m^3}
-\frac{35 C_A C_F^2 \Delta_{m,M} \mathcal{L}_m^2 Y_s^2}{2304 m^3}
+\frac{5 C_A \Delta_{m,M} \mathcal{L}_m^2 Y_s^2}{256 m^3}
-\frac{5 C_A C_F \Delta_{m,M} \mathcal{L}_m^2 Y_s^2}{32 m^3}
-\frac{35 C_F \Delta_{m,M} \mathcal{L}_m^2 Y_s^2}{4608 m^3}
-\frac{191 C_A C_F^2 S_1 Y_s^2}{96 m^2}
-\frac{5 C_A S_1 Y_s^2}{32 m^2}
-\frac{101 C_A C_F S_1 Y_s^2}{36 m^2}
-\frac{191 C_F S_1 Y_s^2}{192 m^2}
+\frac{7 C_A C_F^2 \log{(3)} S_1 Y_s^2}{32 m^2}
-\frac{C_A \log{(3)} S_1 Y_s^2}{32 m^2}
+\frac{3 C_A C_F \log{(3)} S_1 Y_s^2}{4 m^2}
+\frac{7 C_F \log{(3)} S_1 Y_s^2}{64 m^2}
-\frac{7 C_A C_F^2 S_1 Y_s^2}{32 \varepsilon _{\text{UV}} m^2}
+\frac{C_A S_1 Y_s^2}{32 \varepsilon _{\text{UV}} m^2}
-\frac{3 C_A C_F S_1 Y_s^2}{4 \varepsilon _{\text{UV}} m^2}
-\frac{7 C_F S_1 Y_s^2}{64 \varepsilon _{\text{UV}} m^2}
+\frac{59 C_A C_F^2 \Delta_{M} S_1 Y_s^2}{24 m^3}
-\frac{19 C_A \Delta_{M} S_1 Y_s^2}{24 m^3}
-\frac{35 C_A C_F \Delta_{M} S_1 Y_s^2}{18 m^3}
+\frac{59 C_F \Delta_{M} S_1 Y_s^2}{48 m^3}
-\frac{109 C_A C_F^2 \Delta_{m,M} S_1 Y_s^2}{144 m^3}
-\frac{23 C_A \Delta_{m,M} S_1 Y_s^2}{48 m^3}
-\frac{29 C_A C_F \Delta_{m,M} S_1 Y_s^2}{6 m^3}
-\frac{109 C_F \Delta_{m,M} S_1 Y_s^2}{288 m^3}
+\frac{9 C_A C_F^2 \Delta_{M} \log{(3)} S_1 Y_s^2}{8 m^3}
+\frac{C_A \Delta_{M} \log{(3)} S_1 Y_s^2}{8 m^3}
+\frac{C_A C_F \Delta_{M} \log{(3)} S_1 Y_s^2}{6 m^3}
+\frac{9 C_F \Delta_{M} \log{(3)} S_1 Y_s^2}{16 m^3}
+\frac{115 C_A C_F^2 \Delta_{m,M} \log{(3)} S_1 Y_s^2}{144 m^3}
+\frac{3 C_A \Delta_{m,M} \log{(3)} S_1 Y_s^2}{16 m^3}
+\frac{C_A C_F \Delta_{m,M} \log{(3)} S_1 Y_s^2}{2 m^3}
+\frac{115 C_F \Delta_{m,M} \log{(3)} S_1 Y_s^2}{288 m^3}
-\frac{9 C_A C_F^2 \Delta_{M} S_1 Y_s^2}{8 \varepsilon _{\text{UV}} m^3}
-\frac{C_A \Delta_{M} S_1 Y_s^2}{8 \varepsilon _{\text{UV}} m^3}
-\frac{C_A C_F \Delta_{M} S_1 Y_s^2}{6 \varepsilon _{\text{UV}} m^3}
-\frac{9 C_F \Delta_{M} S_1 Y_s^2}{16 \varepsilon _{\text{UV}} m^3}
-\frac{115 C_A C_F^2 \Delta_{m,M} S_1 Y_s^2}{144 \varepsilon _{\text{UV}} m^3}
-\frac{3 C_A \Delta_{m,M} S_1 Y_s^2}{16 \varepsilon _{\text{UV}} m^3}
-\frac{C_A C_F \Delta_{m,M} S_1 Y_s^2}{2 \varepsilon _{\text{UV}} m^3}
-\frac{115 C_F \Delta_{m,M} S_1 Y_s^2}{288 \varepsilon _{\text{UV}} m^3}
+\frac{9 C_A C_F \Delta_{M} S_1 S_2 Y_s^2}{2 m^3}
+\frac{27 C_A C_F \Delta_{m,M} S_1 S_2 Y_s^2}{4 m^3}
+\frac{855 C_A C_F^2 S_2 Y_s^2}{128 m^2}
+\frac{279 C_A S_2 Y_s^2}{128 m^2}
-\frac{63 C_A C_F S_2 Y_s^2}{4 m^2}
+\frac{855 C_F S_2 Y_s^2}{256 m^2}
-\frac{1467 C_A C_F^2 \Delta_{M} S_2 Y_s^2}{32 m^3}
+\frac{63 C_A \Delta_{M} S_2 Y_s^2}{16 m^3}
+\frac{27 C_A C_F \Delta_{M} S_2 Y_s^2}{8 m^3}
-\frac{1467 C_F \Delta_{M} S_2 Y_s^2}{64 m^3}
-\frac{523 C_A C_F^2 \Delta_{m,M} S_2 Y_s^2}{64 m^3}
-\frac{27 C_A \Delta_{m,M} S_2 Y_s^2}{64 m^3}
+\frac{147 C_A C_F \Delta_{m,M} S_2 Y_s^2}{8 m^3}
-\frac{523 C_F \Delta_{m,M} S_2 Y_s^2}{128 m^3}
-\frac{73 C_A C_F^2 \zeta_2 Y_s^2}{192 m^2}
-\frac{3 C_A \zeta_2 Y_s^2}{64 m^2}
+\frac{3 C_A C_F \zeta_2 Y_s^2}{4 m^2}
-\frac{73 C_F \zeta_2 Y_s^2}{384 m^2}
+\frac{5 C_A C_F^2 \Delta_{M} \zeta_2 Y_s^2}{16 m^3}
-\frac{305 C_A \Delta_{M} \zeta_2 Y_s^2}{1536 m^3}
-\frac{431 C_A C_F \Delta_{M} \zeta_2 Y_s^2}{384 m^3}
+\frac{5 C_F \Delta_{M} \zeta_2 Y_s^2}{32 m^3}
-\frac{8089 C_A C_F^2 \Delta_{m,M} \zeta_2 Y_s^2}{13824 m^3}
-\frac{161 C_A \Delta_{m,M} \zeta_2 Y_s^2}{1536 m^3}
-\frac{415 C_A C_F \Delta_{m,M} \zeta_2 Y_s^2}{192 m^3}
-\frac{8089 C_F \Delta_{m,M} \zeta_2 Y_s^2}{27648 m^3}
-\frac{C_A C_F \Delta_{M} \zeta_3 Y_s^2}{m^3}
-\frac{3 C_A C_F \Delta_{m,M} \zeta_3 Y_s^2}{2 m^3}
+\frac{7 C_A C_F^2 S_1 \mathcal{L}_m Y_s^2}{16 m^2}
-\frac{C_A S_1 \mathcal{L}_m Y_s^2}{16 m^2}
+\frac{3 C_A C_F S_1 \mathcal{L}_m Y_s^2}{2 m^2}
+\frac{7 C_F S_1 \mathcal{L}_m Y_s^2}{32 m^2}
+\frac{9 C_A C_F^2 \Delta_{M} S_1 \mathcal{L}_m Y_s^2}{4 m^3}
+\frac{C_A \Delta_{M} S_1 \mathcal{L}_m Y_s^2}{4 m^3}
+\frac{C_A C_F \Delta_{M} S_1 \mathcal{L}_m Y_s^2}{3 m^3}
+\frac{9 C_F \Delta_{M} S_1 \mathcal{L}_m Y_s^2}{8 m^3}
+\frac{115 C_A C_F^2 \Delta_{m,M} S_1 \mathcal{L}_m Y_s^2}{72 m^3}
+\frac{3 C_A \Delta_{m,M} S_1 \mathcal{L}_m Y_s^2}{8 m^3}
+\frac{C_A C_F \Delta_{m,M} S_1 \mathcal{L}_m Y_s^2}{m^3}
+\frac{115 C_F \Delta_{m,M} S_1 \mathcal{L}_m Y_s^2}{144 m^3}
-\frac{5 C_A C_F^2 \mathcal{L}_m Y_s^2}{16 m^2}
+\frac{3 C_A \mathcal{L}_m Y_s^2}{16 m^2}
-\frac{3 C_A C_F \mathcal{L}_m Y_s^2}{m^2}
-\frac{5 C_F \mathcal{L}_m Y_s^2}{32 m^2}
-\frac{43 C_A C_F^2 \Delta_{M} \mathcal{L}_m Y_s^2}{8 m^3}
-\frac{215 C_A \Delta_{M} \mathcal{L}_m Y_s^2}{512 m^3}
-\frac{105 C_A C_F \Delta_{M} \mathcal{L}_m Y_s^2}{128 m^3}
-\frac{43 C_F \Delta_{M} \mathcal{L}_m Y_s^2}{16 m^3}
-\frac{15583 C_A C_F^2 \Delta_{m,M} \mathcal{L}_m Y_s^2}{4608 m^3}
-\frac{407 C_A \Delta_{m,M} \mathcal{L}_m Y_s^2}{512 m^3}
-\frac{41 C_A C_F \Delta_{m,M} \mathcal{L}_m Y_s^2}{64 m^3}
-\frac{15583 C_F \Delta_{m,M} \mathcal{L}_m Y_s^2}{9216 m^3}
+\frac{33 C_A C_F^2 Y_s^2}{16 m^2}
-\frac{3 C_A Y_s^2}{16 m^2}
+\frac{20 C_A C_F Y_s^2}{3 m^2}
+\frac{33 C_F Y_s^2}{32 m^2}
+\frac{5 C_A C_F^2 Y_s^2}{32 \varepsilon _{\text{UV}} m^2}
-\frac{3 C_A Y_s^2}{32 \varepsilon _{\text{UV}} m^2}
+\frac{3 C_A C_F Y_s^2}{2 \varepsilon _{\text{UV}} m^2}
+\frac{5 C_F Y_s^2}{64 \varepsilon _{\text{UV}} m^2}
+\frac{83 C_A C_F^2 \Delta_{M} Y_s^2}{16 m^3}
+\frac{1377 C_A \Delta_{M} Y_s^2}{2048 m^3}
+\frac{4189 C_A C_F \Delta_{M} Y_s^2}{1536 m^3}
+\frac{83 C_F \Delta_{M} Y_s^2}{32 m^3}
+\frac{65881 C_A C_F^2 \Delta_{m,M} Y_s^2}{18432 m^3}
+\frac{2145 C_A \Delta_{m,M} Y_s^2}{2048 m^3}
+\frac{2653 C_A C_F \Delta_{m,M} Y_s^2}{768 m^3}
+\frac{65881 C_F \Delta_{m,M} Y_s^2}{36864 m^3}
+\frac{43 C_A C_F^2 \Delta_{M} Y_s^2}{16 \varepsilon _{\text{UV}} m^3}
+\frac{3 C_A \Delta_{M} Y_s^2}{16 \varepsilon _{\text{UV}} m^3}
+\frac{C_A C_F \Delta_{M} Y_s^2}{2 \varepsilon _{\text{UV}} m^3}
+\frac{43 C_F \Delta_{M} Y_s^2}{32 \varepsilon _{\text{UV}} m^3}
+\frac{41 C_A C_F^2 \Delta_{m,M} Y_s^2}{24 \varepsilon _{\text{UV}} m^3}
+\frac{3 C_A \Delta_{m,M} Y_s^2}{8 \varepsilon _{\text{UV}} m^3}
+\frac{C_A C_F \Delta_{m,M} Y_s^2}{2 \varepsilon _{\text{UV}} m^3}
+\frac{41 C_F \Delta_{m,M} Y_s^2}{48 \varepsilon _{\text{UV}} m^3}
+\frac{5 C_A C_F \Delta_{m,M} \mathcal{L}_m^2 Y_s}{8 m^2}
+\frac{43 C_A C_F S_1 Y_s}{6 m}
-\frac{2 C_A C_F \log{(3)} S_1 Y_s}{m}
+\frac{2 C_A C_F S_1 Y_s}{\varepsilon _{\text{UV}} m}
-\frac{17 C_A C_F \Delta_{M} S_1 Y_s}{18 m^2}
+\frac{107 C_A C_F \Delta_{m,M} S_1 Y_s}{9 m^2}
-\frac{2 C_A C_F \Delta_{M} \log{(3)} S_1 Y_s}{3 m^2}
-\frac{8 C_A C_F \Delta_{m,M} \log{(3)} S_1 Y_s}{3 m^2}
+\frac{2 C_A C_F \Delta_{M} S_1 Y_s}{3 \varepsilon _{\text{UV}} m^2}
+\frac{8 C_A C_F \Delta_{m,M} S_1 Y_s}{3 \varepsilon _{\text{UV}} m^2}
-\frac{63 C_A C_F S_1 S_2 Y_s}{8 m}
+\frac{27 C_A C_F \Delta_{M} S_1 S_2 Y_s}{4 m^2}
+\frac{99 C_A C_F \Delta_{m,M} S_1 S_2 Y_s}{8 m^2}
+\frac{18 C_A C_F S_2 Y_s}{m}
+\frac{24 C_A C_F \Delta_{M} S_2 Y_s}{m^2}
+\frac{39 C_A C_F \Delta_{m,M} S_2 Y_s}{2 m^2}
+\frac{C_A C_F \zeta_2 Y_s}{m}
-\frac{7 C_A C_F \Delta_{M} \zeta_2 Y_s}{3 m^2}
-\frac{289 C_A C_F \Delta_{m,M} \zeta_2 Y_s}{48 m^2}
+\frac{7 C_A C_F \zeta_3 Y_s}{4 m}
-\frac{3 C_A C_F \Delta_{M} \zeta_3 Y_s}{2 m^2}
-\frac{11 C_A C_F \Delta_{m,M} \zeta_3 Y_s}{4 m^2}
-\frac{4 C_A C_F S_1 \mathcal{L}_m Y_s}{m}
-\frac{4 C_A C_F \Delta_{M} S_1 \mathcal{L}_m Y_s}{3 m^2}
-\frac{16 C_A C_F \Delta_{m,M} S_1 \mathcal{L}_m Y_s}{3 m^2}
+\frac{6 C_A C_F \mathcal{L}_m Y_s}{m}
+\frac{4 C_A C_F \Delta_{M} \mathcal{L}_m Y_s}{m^2}
+\frac{137 C_A C_F \Delta_{m,M} \mathcal{L}_m Y_s}{16 m^2}
-\frac{23 C_A C_F Y_s}{2 m}
-\frac{3 C_A C_F Y_s}{\varepsilon _{\text{UV}} m}
-\frac{8 C_A C_F \Delta_{M} Y_s}{3 m^2}
-\frac{2813 C_A C_F \Delta_{m,M} Y_s}{192 m^2}
-\frac{2 C_A C_F \Delta_{M} Y_s}{\varepsilon _{\text{UV}} m^2}
-\frac{5 C_A C_F \Delta_{m,M} Y_s}{\varepsilon _{\text{UV}} m^2}
+\frac{1521}{32} C_A C_F^2
+\frac{73}{4} C_A C_F^2 \mathcal{L}_m^2
-\frac{1}{4} C_A C_F \mathcal{L}_m^2
+\frac{105}{8} C_F \mathcal{L}_m^2-C_A C_F n_f T_f \mathcal{L}_m^2
-\frac{5 C_A C_F^2 \Delta_{m,M} \mathcal{L}_m^2}{8 m}
+\frac{5 C_A C_F \Delta_{m,M} \mathcal{L}_m^2}{2 m}
+\frac{5 C_F \Delta_{m,M} \mathcal{L}_m^2}{8 m}
+\frac{659 C_A C_F}{96}
+\frac{2045 C_F}{64}
-\frac{493}{12} C_A C_F^2 S_1
-\frac{89}{12} C_A C_F S_1
-\frac{493 C_F S_1}{24}
+\frac{53}{4} C_A C_F^2 \log{(3)} S_1
+\frac{41}{12} C_A C_F \log{(3)} S_1
+\frac{53}{8} C_F \log{(3)} S_1
-\frac{53 C_A C_F^2 S_1}{4 \varepsilon _{\text{UV}}}
-\frac{41 C_A C_F S_1}{12 \varepsilon _{\text{UV}}}
-\frac{53 C_F S_1}{8 \varepsilon _{\text{UV}}}
-\frac{5 C_A C_F \Delta_{M} S_1}{3 m}
-\frac{841 C_A C_F^2 \Delta_{m,M} S_1}{9 m}
+\frac{16 C_A C_F \Delta_{m,M} S_1}{3 m}
-\frac{1243 C_F \Delta_{m,M} S_1}{18 m}
+\frac{C_A C_F \Delta_{M} \log{(3)} S_1}{2 m}
+\frac{179 C_A C_F^2 \Delta_{m,M} \log{(3)} S_1}{3 m}
+\frac{29 C_A C_F \Delta_{m,M} \log{(3)} S_1}{9 m}
+\frac{185 C_F \Delta_{m,M} \log{(3)} S_1}{6 m}
-\frac{C_A C_F \Delta_{M} S_1}{2 \varepsilon _{\text{UV}} m}
-\frac{179 C_A C_F^2 \Delta_{m,M} S_1}{3 \varepsilon _{\text{UV}} m}
-\frac{29 C_A C_F \Delta_{m,M} S_1}{9 \varepsilon _{\text{UV}} m}
-\frac{185 C_F \Delta_{m,M} S_1}{6 \varepsilon _{\text{UV}} m}
-\frac{441}{16} C_A C_F^2 S_2
-\frac{609}{16} C_A C_F S_2
-\frac{1089 C_F S_2}{32}
+\frac{63}{4} C_A C_F^2 S_1 S_2
+\frac{63}{8} C_F S_1 S_2
-\frac{9 C_A C_F^2 \Delta_{m,M} S_1 S_2}{m}
-\frac{9 C_F \Delta_{m,M} S_1 S_2}{2 m}
+\frac{3 C_A C_F \Delta_{M} S_2}{8 m}
-\frac{2985 C_A C_F^2 \Delta_{m,M} S_2}{4 m}
-\frac{53 C_A C_F \Delta_{m,M} S_2}{4 m}
-\frac{2787 C_F \Delta_{m,M} S_2}{8 m}
-\frac{55}{24} C_A C_F n_f T_f
-\frac{3 C_A C_F n_f T_f}{4 \varepsilon _{\text{UV}}}
-\frac{8 C_A C_F \Delta_{m,M} n_f T_f}{m}
-\frac{4 C_A C_F \Delta_{m,M} n_f T_f}{3 \varepsilon _{\text{UV}} m}
-\frac{C_A C_F n_f T_f}{2 \varepsilon _{\text{UV}}^2}
+\frac{4}{3} C_A C_F n_f S_1 T_f
-\frac{1}{3} C_A C_F \log{(3)} n_f S_1 T_f
+\frac{C_A C_F n_f S_1 T_f}{3 \varepsilon _{\text{UV}}}
+\frac{8 C_A C_F \Delta_{m,M} n_f S_1 T_f}{3 m}
-\frac{16 C_A C_F \Delta_{m,M} \log{(3)} n_f S_1 T_f}{9 m}
+\frac{16 C_A C_F \Delta_{m,M} n_f S_1 T_f}{9 \varepsilon _{\text{UV}} m}
+\frac{21}{4} C_A C_F n_f S_2 T_f
+\frac{28 C_A C_F \Delta_{m,M} n_f S_2 T_f}{m}-C_A C_F^2 \zeta_2
+\frac{1}{4} C_A C_F \zeta_2-2 C_A C_F n_f T_f \zeta_2
+\frac{C_A C_F \Delta_{M} \zeta_2}{12 m}
+\frac{115 C_A C_F^2 \Delta_{m,M} \zeta_2}{16 m}
+\frac{25 C_A C_F \Delta_{m,M} \zeta_2}{12 m}
+\frac{153 C_F \Delta_{m,M} \zeta_2}{16 m}
-\frac{7}{2} C_A C_F^2 \zeta_3
-\frac{7 C_F \zeta_3}{4}
+\frac{2 C_A C_F^2 \Delta_{m,M} \zeta_3}{m}
+\frac{C_F \Delta_{m,M} \zeta_3}{m}
-\frac{307}{8} C_A C_F^2 \mathcal{L}_m
+\frac{21}{8} C_A C_F \mathcal{L}_m
-\frac{343}{16} C_F \mathcal{L}_m
+\frac{53}{2} C_A C_F^2 S_1 \mathcal{L}_m
+\frac{41}{6} C_A C_F S_1 \mathcal{L}_m
+\frac{53}{4} C_F S_1 \mathcal{L}_m
+\frac{C_A C_F \Delta_{M} S_1 \mathcal{L}_m}{m}
+\frac{358 C_A C_F^2 \Delta_{m,M} S_1 \mathcal{L}_m}{3 m}
+\frac{58 C_A C_F \Delta_{m,M} S_1 \mathcal{L}_m}{9 m}
+\frac{185 C_F \Delta_{m,M} S_1 \mathcal{L}_m}{3 m}
+\frac{3}{2} C_A C_F n_f T_f \mathcal{L}_m
+\frac{C_A C_F n_f T_f \mathcal{L}_m}{\varepsilon _{\text{UV}}}
+\frac{8 C_A C_F \Delta_{m,M} n_f T_f \mathcal{L}_m}{3 m}
-\frac{2}{3} C_A C_F n_f S_1 T_f \mathcal{L}_m
-\frac{32 C_A C_F \Delta_{m,M} n_f S_1 T_f \mathcal{L}_m}{9 m}
-\frac{73 C_A C_F^2 \mathcal{L}_m}{4 \varepsilon _{\text{UV}}}
+\frac{C_A C_F \mathcal{L}_m}{4 \varepsilon _{\text{UV}}}
-\frac{105 C_F \mathcal{L}_m}{8 \varepsilon _{\text{UV}}}
-\frac{2601 C_A C_F^2 \Delta_{m,M} \mathcal{L}_m}{16 m}
-\frac{565 C_A C_F \Delta_{m,M} \mathcal{L}_m}{12 m}
-\frac{1047 C_F \Delta_{m,M} \mathcal{L}_m}{16 m}
+\frac{307 C_A C_F^2}{16 \varepsilon _{\text{UV}}}
-\frac{21 C_A C_F}{16 \varepsilon _{\text{UV}}}
+\frac{343 C_F}{32 \varepsilon _{\text{UV}}}
+\frac{C_A C_F \Delta_{M}}{2 m}
+\frac{48925 C_A C_F^2 \Delta_{m,M}}{192 m}
+\frac{497 C_A C_F \Delta_{m,M}}{16 m}
+\frac{29123 C_F \Delta_{m,M}}{192 m}
+\frac{82 C_A C_F^2 \Delta_{m,M}}{\varepsilon _{\text{UV}} m}
+\frac{62 C_A C_F \Delta_{m,M}}{3 \varepsilon _{\text{UV}} m}
+\frac{32 C_F \Delta_{m,M}}{\varepsilon _{\text{UV}} m}
+\frac{73 C_A C_F^2}{8 \varepsilon _{\text{UV}}^2}
-\frac{C_A C_F}{8 \varepsilon _{\text{UV}}^2}
+\frac{105 C_F}{16 \varepsilon _{\text{UV}}^2}
\end{autobreak}
\end{align}
\end{tiny}
\subsection{HPET field at $M\neq 0$:}
The contributions from two-loop wavefunction corrections, $F_h^{(M)}$, with unevaluated MIs are too large to present here. We thus include the full expressions with description in an ancillary file.
\subsection{Parametric Integrals}
\label{sec:PIF}
\begin{align}
    P(z)&=-\int_0^1 dx\left\lbrace 2(1-x)\log{\left(\frac{1-x+z^2x^2}{1-x}\right)}+\frac{4z^2x(1-x^2)}{1-x+z^2x^2} \right\rbrace \nonumber\\&
    =\frac{3}{z^2}+\left\lbrace\frac{3}{2z^4}-3\right\rbrace\log{z^2}+\frac{(3-6z^2-12z^4)}{z^4\sqrt{1-4z^2}}\tanh^{-1}{(\sqrt{1-4z^2})}
\end{align}
\begin{align}
    P^{'}(z)&=-\int_0^1 dx\left\lbrace (1-x)\log{\left(\frac{1-x+z^2x^2}{1-x}\right)}-\frac{2z^2x(1-x)(2-x)}{1-x+z^2x^2} \right\rbrace \nonumber\\&=\frac{3}{2z^2}-\left\lbrace\frac{3}{z^2}-\frac{3}{4z^4}-\frac{3}{2}\right\rbrace\log{z^2}+\frac{(3-6z^2)\sqrt{1-4z^2}}{2z^4}\tanh^{-1}{(\sqrt{1-4z^2})}
\end{align}

\begin{align}
    S(z)&=\int_0^1 dx\left\lbrace (3x^2-6x+4)\log{\left(\frac{1-x+z^2x^2}{1-x}\right)}-\frac{z^2x(1-x^2)}{(1-x)(2-x)^2} \right\rbrace \nonumber\\&=-\frac{1}{z^2}+\left\lbrace\frac{3}{2z^2}-\frac{1}{2z^4}\right\rbrace\log{z^2}+\frac{(z^2-1)\sqrt{1-4z^2}}{z^4}\tanh^{-1}{(\sqrt{1-4z^2})}
\end{align}
\begin{align}
    S^{'}(z)&=-\int_0^1 dx\left\lbrace\frac{z^2x^3}{1-x+z^2x^2} \right\rbrace \nonumber\\&=-\frac{1}{z^2}+\left\lbrace\frac{1}{2z^2}-\frac{1}{2z^4}\right\rbrace\log{z^2}+\frac{3z^2-1}{z^4\sqrt{1-4z^2}}\tanh^{-1}{(\sqrt{1-4z^2})}
\end{align}
The integrals ma be analytically continued in the regime, $4z^2\geq 1$, using $\sqrt{1-4z^2}\mapsto i\sqrt{4z^2-1}$ and therefore $\tanh^{-1}(\sqrt{1-4z^2})\mapsto i\tanh^{-1}(\sqrt{4z^2-1})$, and plugging this back into the integrals one can verify that the integral remains real.
\acknowledgments
We are indebted to O. Veretin for the many discussions we had on the subject and technical details of the calculation. 
\printbibliography

@article{40,
  title={The pole mass in perturbative QCD},
  author={Tarrach, Rolf},
  journal={Nuclear Physics B},
  volume={183},
  number={3},
  pages={384--396},
  year={1981},
  publisher={Elsevier}
}

@article{HQET,
  title={Heavy quark effective theory beyond perturbation theory: Renormalons, the pole mass and the residual mass term},
  author={Beneke, Martin and Braun, Vladimir M},
  journal={Nuclear Physics B},
  volume={426},
  number={2},
  pages={301--343},
  year={1994},
  publisher={Elsevier}
}

@article{higgs,
  title={Combined search for the Standard Model Higgs boson using up to 4.9 fb- 1 of pp collision data at s= 7 TeV with the ATLAS detector at the LHC},
  author={Aad, Georges and Abbott, B and Abdallah, J and Khalek, S Abdel and Abdelalim, AA and Abdesselam, A and Abdinov, O and Abi, B and Abolins, M and AbouZeid, OS and others},
  journal={Physics Letters B},
  volume={710},
  number={1},
  pages={49--66},
  year={2012},
  publisher={Elsevier}
}

@article{vermaseren2000new,
  title={New features of FORM},
  author={Vermaseren, Jos AM},
  journal={arXiv preprint math-ph/0010025},
  year={2000}
}

@article{nogueira1993automatic,
  title={Automatic Feynman graph generation},
  author={Nogueira, P},
  journal={Journal of Computational Physics},
  volume={105},
  number={2},
  pages={279--289},
  year={1993},
  publisher={Elsevier}
}

@article{bigi1994pole,
  title={Pole mass of the heavy quark: perturbation theory and beyond},
  author={Bigi, Ikaros I and Shifman, Mikhail A and Uraltsev, NG and Vainshtein, AI},
  journal={Physical Review D},
  volume={50},
  number={3},
  pages={2234},
  year={1994},
  publisher={APS}
}

@article{manohar1997heavy,
  title={Heavy quark effective theory and nonrelativistic QCD Lagrangian to order $\alpha$ s/m 3},
  author={Manohar, Aneesh V},
  journal={Physical Review D},
  volume={56},
  number={1},
  pages={230},
  year={1997},
  publisher={APS}
}

@article{denner1993techniques,
  title={Techniques for the Calculation of Electroweak Radiative Corrections at the One-Loop Level and Results for W-physics at LEP 200},
  author={Denner, Ansgar},
  journal={Fortschritte der Physik/Progress of Physics},
  volume={41},
  number={4},
  pages={307--420},
  year={1993},
  publisher={Wiley Online Library}
}

@article{martinez2003multi,
  title={Multi-parameter fits to threshold observables at a future ee linear collider},
  author={Martinez, Manel and Miquel, Ramon},
  journal={The European Physical Journal C-Particles and Fields},
  volume={27},
  number={1},
  pages={49--55},
  year={2003},
  publisher={Springer}
}

@article{simon2016impact,
  title={Impact of Theory Uncertainties on the Precision of the Top Quark Mass in a Threshold Scan at future $e^+ e^-$-Colliders},
  author={Simon, Frank},
  journal={arXiv preprint arXiv:1611.03399},
  year={2016}
}

@article{chiu2009factorization,
  title={Factorization structure of gauge theory amplitudes and application to hard scattering processes at the LHC},
  author={Chiu, Jui-yu and Fuhrer, Andreas and Kelley, Randall and Manohar, Aneesh V},
  journal={Physical Review D},
  volume={80},
  number={9},
  pages={094013},
  year={2009},
  publisher={APS}
}

@article{chiesa2013electroweak,
  title={Electroweak Sudakov corrections to new physics searches at the LHC},
  author={Chiesa, Mauro and Montagna, Guido and Barze, Luca and Moretti, Mauro and Nicrosini, Oreste and Piccinini, Fulvio and Tramontano, Francesco},
  journal={Physical review letters},
  volume={111},
  number={12},
  pages={121801},
  year={2013},
  publisher={APS}
}

@article{dittmaier1996integrating,
  title={Integrating out the standard Higgs field in the path integral},
  author={Dittmaier, Stegan and Grosse-Knetter, Carsten},
  journal={Nuclear Physics B},
  volume={459},
  number={3},
  pages={497--536},
  year={1996},
  publisher={Elsevier}
}

@article{denner2020electroweak,
  title={Electroweak radiative corrections for collider physics},
  author={Denner, Ansgar and Dittmaier, Stefan},
  journal={Physics Reports},
  year={2020},
  publisher={Elsevier}
}

@article{ovanesyan2015heavy,
  title={Heavy dark matter annihilation from effective field theory},
  author={Ovanesyan, Grigory and Slatyer, Tracy R and Stewart, Iain W},
  journal={Physical Review Letters},
  volume={114},
  number={21},
  pages={211302},
  year={2015},
  publisher={APS}
}

@article{ciafaloni2011weak,
  title={Weak corrections are relevant for dark matter indirect detection},
  author={Ciafaloni, Paolo and Comelli, Denis and Riotto, Antonio and Sala, Filippo and Strumia, Alessandro and Urbano, Alfredo},
  journal={Journal of Cosmology and Astroparticle Physics},
  volume={2011},
  number={03},
  pages={019},
  year={2011},
  publisher={IOP Publishing}
}

@article{ablinger2018heavy,
  title={Heavy quark form factors at three loops in the planar limit},
  author={Ablinger, J and Bl{\"u}mlein, J and Marquard, P and Rana, N and Schneider, C},
  journal={Physics Letters B},
  volume={782},
  pages={528--532},
  year={2018},
  publisher={Elsevier}
}

@article{brambilla2005effective,
  title={Effective-field theories for heavy quarkonium},
  author={Brambilla, Nora and Pineda, Antonio and Soto, Joan and Vairo, Antonio},
  journal={Reviews of Modern Physics},
  volume={77},
  number={4},
  pages={1423},
  year={2005},
  publisher={APS}
}

@article{ciafaloni2000bloch,
  title={Bloch-Nordsieck violating electroweak corrections to inclusive TeV scale hard processes},
  author={Ciafaloni, M and Ciafaloni, P and Comelli, D},
  journal={Physical review letters},
  volume={84},
  number={21},
  pages={4810},
  year={2000},
  publisher={APS}
}

@article{ciafaloni2000electroweak,
  title={Electroweak Sudakov form factors and nonfactorizable soft QED effects at NLC energies},
  author={Ciafaloni, P and Comelli, D},
  journal={Physics Letters B},
  volume={476},
  number={1-2},
  pages={49--57},
  year={2000},
  publisher={Elsevier}
}

@article{fadin2000resummation,
  title={Resummation of double logarithms in electroweak high energy processes},
  author={Fadin, Victor S and Lipatov, LN and Martin, Alan D and Melles, M},
  journal={Physical Review D},
  volume={61},
  number={9},
  pages={094002},
  year={2000},
  publisher={APS}
}

@article{kuhn2000summing,
  title={Summing up subleading Sudakov logarithms},
  author={K{\"u}hn, Johann H and Penin, Aleksandr A and Smirnov, Vladimir A},
  journal={The European Physical Journal C-Particles and Fields},
  volume={17},
  number={1},
  pages={97--105},
  year={2000},
  publisher={Springer}
}

@article{feucht2004two,
  title={Two-loop Sudakov form factor in a theory with a mass gap},
  author={Feucht, Bernd and K{\"u}hn, Johann H and Penin, Alexander A and Smirnov, Vladimir A},
  journal={Physical review letters},
  volume={93},
  number={10},
  pages={101802},
  year={2004},
  publisher={APS}
}

@article{jantzen2005two,
  title={Two-loop high-energy electroweak logarithmic corrections in a spontaneously broken SU(2) gauge model},
  author={Jantzen, Bernd and K{\"u}hn, Johann H and Penin, Alexander A and Smirnov, Vladimir A},
  journal={Physical Review D},
  volume={72},
  number={5},
  pages={051301},
  year={2005},
  publisher={APS}
}

@article{denner2001one,
  title={One-loop leading logarithms in electroweak radiative corrections},
  author={Denner, Ansgar and Pozzorini, Stefano},
  journal={The European Physical Journal C-Particles and Fields},
  volume={18},
  number={3},
  pages={461--480},
  year={2001},
  publisher={Springer}
}

@article{hori2000electroweak,
  title={Electroweak Sudakov at two loop level},
  author={Hori, Masaki and Kawamura, H and Kodaira, J},
  journal={Physics Letters B},
  volume={491},
  number={3-4},
  pages={275--279},
  year={2000},
  publisher={Elsevier}
}

@article{jantzen2006two,
  title={The two-loop vector form factor in the Sudakov limit},
  author={Jantzen, Bernd and Smirnov, Vladimir A},
  journal={The European Physical Journal C-Particles and Fields},
  volume={47},
  number={3},
  pages={671--695},
  year={2006},
  publisher={Springer}
}

@article{von2017quark,
  title={Quark and gluon form factors to four-loop order in QCD: the $N_f^3$ contributions},
  author={von Manteuffel, Andreas and Schabinger, Robert M},
  journal={Physical Review D},
  volume={95},
  number={3},
  pages={034030},
  year={2017},
  publisher={APS}
}

@article{gehrmann2010calculation,
  title={Calculation of the quark and gluon form factors to three loops in QCD},
  author={Gehrmann, T and Glover, EWN and Huber, T and Ikizlerli, N and Studerus, Cedric},
  journal={Journal of High Energy Physics},
  volume={2010},
  number={6},
  pages={94},
  year={2010},
  publisher={Springer}
}

@article{blumlein2019heavy,
  title={Heavy quark form factors at two loops},
  author={Ablinger, J and Behring, A and Bl{\"u}mlein, J and Falcioni, G and De Freitas, A and Marquard, P and Rana, N and Schneider, C},
  journal={Physical Review D},
  volume={97},
  number={9},
  pages={094022},
  year={2018},
  publisher={APS}
}

@article{bernreuther2005two,
  title={Two-loop QCD corrections to the heavy quark form factors: Anomaly contributions},
  author={Bernreuther, Werner and Bonciani, R and Gehrmann, T and Heinesch, R and Leineweber, T and Remiddi, E},
  journal={Nuclear Physics B},
  volume={723},
  number={1-2},
  pages={91--116},
  year={2005},
  publisher={Elsevier}
}

@article{chiu2008electroweak,
  title={Electroweak corrections to high energy processes using effective field theory},
  author={Chiu, Jui-yu and Golf, Frank and Kelley, Randall and Manohar, Aneesh V},
  journal={Physical Review D},
  volume={77},
  number={5},
  pages={053004},
  year={2008},
  publisher={APS}
}

@article{bauer2001effective,
  title={An Effective field theory for collinear and soft gluons: Heavy to light decays},
  author={Bauer, Christian W and Fleming, Sean and Pirjol, Dan and Stewart, Iain W},
  journal={Physical Review D},
  volume={63},
  number={11},
  pages={114020},
  year={2001},
  publisher={APS}
}

@article{bauer2001invariant,
  title={Invariant operators in collinear effective theory},
  author={Bauer, Christian W and Stewart, Iain W},
  journal={Physics Letters B},
  volume={516},
  number={1-2},
  pages={134--142},
  year={2001},
  publisher={Elsevier}
}

@article{chiu2008electroweak0,
  title={Electroweak Sudakov corrections using effective field theory},
  author={Chiu, Jui-yu and Golf, Frank and Kelley, Randall and Manohar, Aneesh V},
  journal={Physical review letters},
  volume={100},
  number={2},
  pages={021802},
  year={2008},
  publisher={APS}
}

@article{chiu2008electroweak2,
  title={Electroweak corrections using effective field theory: Applications to the CERN LHC},
  author={Chiu, Jui-yu and Kelley, Randall and Manohar, Aneesh V},
  journal={Physical Review D},
  volume={78},
  number={7},
  pages={073006},
  year={2008},
  publisher={APS}
}

@article{czakon2013total,
  title={Total top-quark pair-production cross section at hadron colliders through  $\mathcal{O}(\alpha_s^4)$},
  author={M Czakon, P Fiedler, A Mitov},
  journal={Physical Review Letters},
  volume={110},
  number={25},
  pages={252004},
  year={2013},
  publisher={APS}
}

@article{beenakker2010supersymmetric,
  title={Supersymmetric top and bottom squark production at hadron colliders},
  author={Beenakker, Wim and Brensing, Silja and Kr{\"a}mer, Michael and Kulesza, Anna and Laenen, Eric and Niessen, Irene},
  journal={Journal of High Energy Physics},
  volume={2010},
  number={8},
  pages={98},
  year={2010},
  publisher={Springer}
}

@article{georgi1990effective,
  title={An effective field theory for heavy quarks at low energies},
  author={Georgi, Howard},
  journal={Physics Letters B},
  volume={240},
  number={3-4},
  pages={447--450},
  year={1990},
  publisher={Elsevier}
}

@article{georgi1991heavy,
  title={Heavy quark effective field theory},
  author={Georgi, Howard},
  journal={Proc. of the Theoretical Advanced Study Institute},
  pages={589},
  year={1991}
}

@book{manohar2007heavy,
  title={Heavy quark physics},
  author={Manohar, Aneesh V and Wise, Mark B},
  volume={10},
  year={2007},
  publisher={Cambridge university press}
}

@book{schwartz2014quantum,
  title={Quantum field theory and the standard model},
  author={Schwartz, Matthew D},
  year={2014},
  publisher={Cambridge University Press}
}

@article{guth2015dark,
  title={Do dark matter axions form a condensate with long-range correlation?},
  author={Guth, Alan H and Hertzberg, Mark P and Prescod-Weinstein, Chanda},
  journal={Physical Review D},
  volume={92},
  number={10},
  pages={103513},
  year={2015},
  publisher={APS}
}

@article{braaten2018classical,
  title={Classical nonrelativistic effective field theories for a real scalar field},
  author={Braaten, Eric and Mohapatra, Abhishek and Zhang, Hong},
  journal={Physical Review D},
  volume={98},
  number={9},
  pages={096012},
  year={2018},
  publisher={APS}
}

@article{namjoo2018relativistic,
  title={Relativistic corrections to nonrelativistic effective field theories},
  author={Namjoo, Mohammad Hossein and Guth, Alan H and Kaiser, David I},
  journal={Physical Review D},
  volume={98},
  number={1},
  pages={016011},
  year={2018},
  publisher={APS}
}

@article{pineda1998effective,
  title={Effective field theory for ultrasoft momenta in NRQCD and NRQED},
  author={Pineda, Antonio and Soto, Joan},
  journal={Nuclear Physics B-Proceedings Supplements},
  volume={64},
  number={1-3},
  pages={428--432},
  year={1998},
  publisher={Elsevier}
}

@article{luke2000renormalization,
  title={Renormalization group scaling in nonrelativistic QCD},
  author={Luke, Michael E and Manohar, Aneesh V and Rothstein, Ira Z},
  journal={Physical Review D},
  volume={61},
  number={7},
  pages={074025},
  year={2000},
  publisher={APS}
}

@article{bauer2003enhanced,
  title={Enhanced nonperturbative effects in jet distributions},
  author={Bauer, Christian W and Manohar, Aneesh V and Wise, Mark B},
  journal={Physical review letters},
  volume={91},
  number={12},
  pages={122001},
  year={2003},
  publisher={APS}
}

@article{hiller2004more,
  title={More model-independent analysis of $b\rightarrow s$ processes},
  author={Hiller, Gudrun and Kr{\"u}ger, Frank},
  journal={Physical Review D},
  volume={69},
  number={7},
  pages={074020},
  year={2004},
  publisher={APS}
}

@article{georgi1992d,
  title={$D$-$\backslash\bar{D} $ Mixing in Heavy Quark Effective Field Theory},
  author={Georgi, Howard},
  journal={arXiv preprint hep-ph/9209291},
  year={1992}
}

@article{jenkins1991baryon,
  title={Baryon chiral perturbation theory using a heavy fermion Lagrangian},
  author={Jenkins, Elizabeth and Manohar, Aneesh V},
  journal={Physics Letters B},
  volume={255},
  number={4},
  pages={558--562},
  year={1991},
  publisher={Elsevier}
}

@article{hoang2000charm,
  title={Charm effects in the MS bottom quark mass from $\Upsilon$ mesons},
  author={Hoang, AH and Manohar, AV},
  journal={Physics Letters B},
  volume={483},
  number={1-3},
  pages={94--98},
  year={2000},
  publisher={Elsevier}
}

@incollection{manohar1977effective,
  title={Effective field theories},
  author={Manohar, Aneesh V},
  booktitle={Perturbative and nonperturbative aspects of quantum field theory},
  pages={311--362},
  year={1977},
  publisher={Springer}
}

@article{manohar2003deep,
  title={Deep inelastic scattering as $x \rightarrow 1$ using soft-collinear effective theory},
  author={Manohar, Aneesh V},
  journal={Physical Review D},
  volume={68},
  number={11},
  pages={114019},
  year={2003},
  publisher={APS}
}

@article{machacek1985two,
  title={Two-loop renormalization group equations in a general quantum field theory:(III). Scalar quartic couplings},
  author={Machacek, Marie E and Vaughn, Michael T},
  journal={Nuclear Physics B},
  volume={249},
  number={1},
  pages={70--92},
  year={1985},
  publisher={Elsevier}
}

@article{bauer2004shape,
  title={Shape function effects in $B\rightarrow X s \gamma$ and $B\rightarrow X u l \nu$ decays},
  author={Bauer, Christian W and Manohar, Aneesh V},
  journal={Physical Review D},
  volume={70},
  number={3},
  pages={034024},
  year={2004},
  publisher={APS}
}

@book{becher2015introduction,
  title={Introduction to soft-collinear effective theory},
  author={Becher, Thomas and Broggio, Alessandro and Ferroglia, Andrea},
  volume={896},
  year={2015},
  publisher={Springer}
}

@article{leibovich2003comment,
  title={Comment on quark masses in SCET},
  author={Leibovich, Adam K and Ligeti, Zoltan and Wise, Mark B},
  journal={Physics Letters B},
  volume={564},
  number={3-4},
  pages={231--234},
  year={2003},
  publisher={Elsevier}
}

@article{fleming2008jets,
  title={Jets from massive unstable particles: Top-mass determination},
  author={Fleming, Sean and Hoang, Andre H and Mantry, Sonny and Stewart, Iain W},
  journal={Physical Review D},
  volume={77},
  number={7},
  pages={074010},
  year={2008},
  publisher={APS}
}

@article{chetyrkin1980new,
  title={New approach to evaluation of multiloop Feynman integrals: The Gegenbauer polynomial x-space technique},
  author={Chetyrkin, KG and Kataev, AL and Tkachov, FV},
  journal={Nuclear Physics B},
  volume={174},
  number={2-3},
  pages={345--377},
  year={1980},
  publisher={Elsevier}
}

@article{kotikov1991differential,
  title={Differential equations method. New technique for massive Feynman diagram calculation},
  author={Kotikov, Anatoli V},
  journal={Physics Letters B},
  volume={254},
  number={1-2},
  pages={158--164},
  year={1991},
  publisher={Elsevier}
}

@article{remiddi1997differential,
  title={Differential equations for Feynman graph amplitudes},
  author={Remiddi, Ettore},
  journal={Il Nuovo Cimento A (1971-1996)},
  volume={110},
  number={12},
  pages={1435--1452},
  year={1997},
  publisher={Springer}
}

@article{lee2012presenting,
  title={Presenting LiteRed: a tool for the loop integrals reduction},
  author={Lee, RN},
  journal={arXiv preprint arXiv:1212.2685},
  year={2012}
}

@article{van1986dimensional,
  title={Dimensional regularization of mass and infrared singularities in two-loop on-shell vertex functions},
  author={Van Neerven, WL},
  journal={Nuclear Physics B},
  volume={268},
  number={2},
  pages={453--488},
  year={1986},
  publisher={Elsevier}
}

@article{fleischer2000shell2,
  title={ON-SHELL2: FORM based package for the calculation of two-loop self-energy single scale Feynman diagrams occurring in the Standard Model},
  author={Fleischer, Jochem and Kalmykov, M Yu},
  journal={Computer physics communications},
  volume={128},
  number={3},
  pages={531--549},
  year={2000},
  publisher={Elsevier}
}

@misc{assi,
  title={Evaluating Heavy Particle Effective Theory Master Integrals by Differential Equations},
  author={B. Assi and B. A. Kniehl},
  year={In Preparation}
}
\end{document}